\documentclass[twoside,a4paper,10pt]{report}
\pdfoutput=1
\usepackage[frenchb,american]{babel}
\usepackage[utf8]{inputenc}
       \usepackage{amsmath}
       \usepackage{amsfonts}
       \usepackage{amssymb}
       \usepackage{graphicx}
       \usepackage{array}
       \usepackage{authblk}
       \usepackage{appendix}
       \usepackage{subfigure}
       \usepackage{multirow}
       \usepackage{tabulary}
       \usepackage{booktabs}
       \usepackage{color}
       \usepackage{paralist}
       \usepackage{url}
       \usepackage{float}

       \usepackage[font=small]{caption}

\usepackage[toc]{glossaries}

       \usepackage[citestyle=numeric]{biblatex}
 \numberwithin{equation}{chapter}
\numberwithin{figure}{chapter}
\usepackage{xcolor}
\usepackage{wallpaper}

\newcommand{\titlecommand}{
\ThisULCornerWallPaper{0.2}{ParisDiderotLogo}
\flushleft{\large
\textcolor{red}{UNIVERSITE PARIS.
DIDEROT (Paris 7)\\
SORBONNE PARIS CITE}}

\vspace{0.04\textheight}
\centering
{\large Ecole Doctoral 517 Particules, Noyaux et Cosmos}\par
\vspace{0.4\baselineskip}
{\large UMR7164  Astroparticule et Cosmologie (APC)}\par
\vspace{1\baselineskip}

{\large DOCTORAT}\par
{\large Physique}\par
\vspace{0.06\textheight}

{\Large \bfseries Carl BLAKSLEY \\ \vspace{0.04\textheight}}\par

{\Large\bfseries Photodetection Aspects of JEM-EUSO and Studies of the Ultra-High Energy Cosmic Ray Sky}\par
 \vspace{2.0 \baselineskip}
 
{\large\bfseries \'{E}tudes de Photod\'{e}tection pour l'Observatoire JEM-EUSO et Ph\'{e}nom\'{e}nologie des Rayons Cosmique d'Ultra-Haute \'{E}nergie}\par
 \vspace{1.6 \baselineskip}

 \rule{22em}{0.2pt}\\

{Th\`{e}se dirig\'{e}e par Professeur Etienne PARIZOT
\\[0.5\baselineskip]
Soutenue le 19 Novembre 2013 }\par

\flushleft
{devant le jury compos\'{e} de:}\par
\vspace{1.0\baselineskip}
 \textit{Pr\'{e}sident} \\
Mme. Tiina Suomij\"{a}rvi, Professeur Universit\'{e} Paris-Sud (Paris XI) \\[1\baselineskip]
\vspace{0.18\baselineskip}
\textit{Rapporteurs}\\
M. Alan Watson, Professeur University of Leeds \\ [1\baselineskip] 
M. Marco Casolino,  Professeur RIKEN \\[1\baselineskip] 
\vspace{0.18\baselineskip}
\textit{Examinateurs}\\
M. Angela Olinto,  Professeur University of Chicago  \\[1\baselineskip]
M. Andreas Haungs, Professeur Karlsruhe Institute of Technology \\[1\baselineskip]
M. Beno\^{i}t Revenu, Charge de Recherche SUBATECH \\[1\baselineskip]
\vspace{0.15\baselineskip}
\textit{Examinateur \& Co-directeur du th\`{e}se}\\
M. Philippe Gorodetzky, DREM Astroparticule et Cosmologie \par
\centering
\vfill
}

\def\fig{Fig.~}
\def\eq{Eq.~(}

\def\M64{M64}

\title{Photodetection Aspects of JEM-EUSO and Studies of the Ultra-High Energy Cosmic Ray Sky}
\author{Carl J. Blaksley}

\begin{document}

 \begin{titlepage}
\titlecommand
 \end{titlepage}

\makeglossaries

\newacronym{UVB}{UVB}{Ultraviolet B}

\newacronym{CLOUDS}{CLOUDS}{Cosmics Leaving Outdoor Droplets}

\newacronym{eV}{eV}{electron Volt}

\newacronym{CERN}{CERN}{Organisation Europ\'{e}enne pour la Recherche Nucl\'{e}aire (European Organization for Nuclear Research)}

\newacronym{DNA}{DNA}{Deoxyribonucleic acid}

\newacronym{EF}{EF}{Exposure Facility}

\newacronym{CNRS}{CNRS}{Centre national de la recherche scientifique (French National Center for Scientific Research)}

 \newacronym{MIT}{MIT}{Massachusetts Institute of Technology}

\newacronym{JEM-EUSO}{JEM-EUSO}{Extreme Universe Space Observatory on-board the Japanese Experiment Module}

\newacronym{ASIC}{ASIC}{Application Specific Integrated Circuit}

\newacronym{SPACIROC}{SPACIROC}{Spatial Photomultiplier Array Counting and
Integrating Readout Chip}

\newacronym{TA}{TA}{Telescope Array}

\newacronym{Auger}{Auger}{Pierre Auger Observatory}

\newacronym{PMT}{PMT}{Photomultiplier Tube}

\newacronym{MAPMT}{MAPMT}{Multi-anode Photomultiplier Tube}

\newacronym{DAQ}{DAQ}{Data Acquisition System}

\newacronym{RMS}{RMS}{Root Mean Square}

\newacronym{DAC}{DAC}{Digital-to-Analog Convertor}

\newacronym{SFR}{SFR}{Stellar Formation Rate}

\newacronym{FR-II}{FR-II}{Fanaroff-Riley Class II}

\newacronym{Mpc}{Mpc}{MegaParsec}

\newacronym{ODB}{ODB}{Online DataBase}

\newacronym{FSR}{FSR}{Full Scale Range}

\newacronym{LSB}{LSB}{Least Significant Bit}

\newacronym{DNL}{DNL}{Differential Non-Linearity}

\newacronym{INL}{INL}{Integral Non-Linearity}

\newacronym{NIM}{NIM}{Nuclear Instrumentation Module}

\newacronym{QDC}{QDC}{Charge-to-Digital Convertor}

\newacronym{ADC}{ADC}{Analog-to-Digital Convertor}

\newacronym{TDC}{TDC}{Time-to-Digital Convertor}

\newacronym{CAMAC}{CAMAC}{Computer Automated Measurement And Control}

\newacronym{PDM}{PDM}{Photodetection Module}

\newacronym{EC}{EC}{Elementary Cell}

\newacronym{LIDAR}{LIDAR}{Light Detection And Ranging}

\newacronym{CLF}{CLF}{Central Laser Facility}

\newacronym{ELS}{ELS}{Electron Light Source}

\newacronym{TALE}{TALE}{Telescope Array Low Energy Extension}

\newacronym{XLF}{XLF}{eXtreme Laser Facility}

\newacronym{AMIGA}{AMIGA}{Auger Muons and Infill for the Ground Array}

\newacronym{HEAT}{HEAT}{High Elevation Auger Telescopes}

\newacronym{AERA}{AERA}{Auger Engineering Radio Array}

\newacronym{IR}{IR}{Infrared}

\newacronym{PCB}{PCB}{Printed Circuit Board}

\newacronym{UV}{UV}{Ultraviolet}

\newacronym{CW-HVPS}{CW-HVPS}{Cockcroft-Walton High Voltage Power Supply}

\newacronym{HVPS}{HVPS}{High Voltage Power Supply}

\newacronym{PS}{PS}{Power Supply}

\newacronym{pe}{pe}{photoelectron}

\newacronym{SNO}{SNO}{Sudbury Neutrino Observatory}

\newacronym{spe}{spe}{single photoelectron}

\newacronym{FD}{FD}{Fluorescence Detector}

\newacronym{FoV}{FoV}{Field of View}

\newacronym{SD}{SD}{Surface Detector}

\newacronym{UHECR}{UHECR}{Ultra-High Energy Cosmic Ray}

\newacronym{EAS}{EAS}{Extensive Air Shower}

\newacronym{PMMA}{PMMA}{Poly(methyl methacrylate)}

\newacronym{BBM}{BBM}{Bread Board Model}

\newacronym{TLE}{TLE}{Transient Luminous Events}

\newacronym{HTV}{HTV}{H-IIB Transfer Vehicle}

\newacronym{TDRS}{TDRS}{Tracking and Data Relay Satellite}

\newacronym{JAXA}{JAXA}{Japan Aerospace Exploration Agency}

\newacronym{GLS}{GLS}{Global Light System}

\newacronym{PTT}{PTT}{Persistent Track Trigger}

\newacronym{LTT}{LTT}{Linear Track Trigger}

\newacronym{TCU}{TCU}{Telemetry Command Unit} 

\newacronym{IDAQ}{IDAQ}{Data Acquisition Interface}

\newacronym{MEM}{MEM}{MicroElectroMechanical}

\newacronym{DMSPac}{DMSP}{Defense Meteorological Satellite Program}

\newacronym{MDP}{MDP}{Mission Data Processor}

\newacronym{CCB}{CCB}{Cluster Control Board}

\newacronym{GLS-X}{GLS-X}{Global Light System - Xenon Flashers}

\newacronym{GLS-XL}{GLS-XL}{Global Light System - Xenon Flashers and Lasers}

\newacronym{FSU}{FSU}{Fast Shaper Unit}

\newacronym{NIST}{NIST}{National Institute of Standards and Technology}

\newacronym{LED}{LED}{Light-Emitting Diodes}

\newacronym{VME}{VME}{Versa Module European bus}

\newacronym{ISS}{ISS}{International Space Station}

\newacronym{AMS}{AMS}{Atmospheric Monitoring System}

\newacronym{GTU}{GTU}{Gate Time Unit}

\newacronym{MIDAS-UK}{MIDAS-UK}{Multi Instance Data Acquisition System}

\newacronym{MIDAS}{MIDAS}{Maximum Integrated Data Acquisition System}

\newacronym{LHC}{LHC}{Large Hadron Collider}

\newacronym{HiRes}{HiRes}{High Resolution Fly's Eye}

\newacronym{CMB}{CMB}{Cosmic Microwave Background}

\newacronym{GRB}{GRB}{Gamma Ray Burst}

\newacronym{GPIB}{GPIB}{General Purpose Interface Bus}

\newacronym{USB}{USB}{Universal Serial Bus}

\newacronym{CCD}{CCD}{Charge-Coupled Device}

\newacronym{GZK}{GZK}{Greisen, Zatsepin, and Kuzmin}

\newacronym{EM}{EM}{Electromagnetic}

\newacronym{LPM}{LPM}{Landau-Pomeranchuk-Migdal}

\newacronym{DICE}{DICE}{Dual Imaging Cherenkov Experiment}

\newacronym{KASCADE}{KASCADE}{Karlsruhe Shower Core and Array Detector}

\newacronym{AGASA}{AGASA}{Akeno Giant Air Shower Array} 

\newacronym{UHE}{UHE}{Ultra High Energy}

\newacronym{CNES}{CNES}{Centre national d'\'{e}tudes spatiales} 
 
\newacronym{FWHM}{FWHM}{Full Width at Half Maximum} 

\newacronym{NKG}{NKG}{Nishimura-Kamata-Greisen} 

%
\pagenumbering{roman}

\begin{abstract}
The Earth is constantly bathed in a sea of particles from space. These particles,
known as cosmic rays, were fundamental in early particle physics, and continue to be a source of unanswered questions. 
Cosmic rays with energies of more than $10^{20}~$eV have been observed, making these the most energetic particles known in the universe.
These so-called Ultra-High Energy Cosmic Rays (UHECRs) are incredibly rare, hitting the earth at a rate of less than 
1 per km$^{2}$ per century. It is still unknown where or how these particles are accelerated up to such energies.

Because of this extremely low rate, UHECRs are observed indirectly through the Extensive Air Shower (EAS) which they create in the atmosphere using huge ground arrays. 
These ground arrays sample the shower at the ground using an array of surface detectors, and also observe the light track created by the EAS. 
This second technique is known as the air fluorescence method, as the emitted light originates from the fluorescence of nitrogen which is excited by the charged particles in the EAS.
Because the amount of fluorescence light emitted is proportional to the energy deposited in the air, the air fluorescence technique allows a calormetric measurement of the energy of the UHECR which created the EAS.
A major obstacle in the study of UHECRs is the low number of UHECR events observed, despite the fact that the current generation of ground arrays cover surface areas on the order of a thousand 
km$^{2}$.

JEM-EUSO is a proposed next generation UHECR observatory which would give an order of magnitude increase in the total exposure. This large increase in exposure would be achieved by observing EAS from space using the air fluorescence technique.
The JEM-EUSO instrument is a fluorescence telescope, sensitive in the near UV, with an aperture of several m$^{2}$ and a full field of view of $60^{\circ}$. This telescope would be attached to the 
JEM module of the International Space Station (ISS), orbiting the Earth at an altitude of $\sim 400~$km. From this height, JEM-EUSO would observe a ground-area of $\approx 10^{5}~$km$^{2}$. 
The JEM-EUSO focal surface is made of 137 Photodetection Modules (PDM), each composed of 9
Elementary Cells (EC) with each EC built from 4 Hamamatsu R11265-M64 Multi-Anode Photomultiplier Tubes (PMTs). Each M64 PMT contains 64 pixels.
JEM-EUSO’s daughter experiment, EUSO-Balloon, is a path-finder mission composed of a single JEM-EUSO
PDM with optics in a balloon-borne gondola. 

Vacuum photomultiplier tubes, such as the Hamamatsu R11265-M64 used in JEM-EUSO, convert incident photons into a small shower of electrons, giving a signal pulse which can be detected by read-out electronics. 
In JEM-EUSO the flux of photons arriving from an EAS is low enough that individual electron showers in the PMT can be separated,  allowing single photons to be counted. This is known as single photoelectron counting.
The efficiency of a PMT, that is the number of single photoelectron counts at the PMT output compared to the number of photons incident on the PMT input window, is a critical parameter for
determining the energy of an observed UHECR from the number of single photoelectron counts on the focal surface. Measuring the absolute efficiency of PMTs is an experimentally challenging task, however, and
most methods give an uncertainty on the order of 10\%. By comparing the PMT to an absolutely calibrated NIST photodiode and using an integrating sphere as a stable splitter, we measure the absolute efficiency 
with an uncertainty on the order of a few percent.

In this thesis, a general introduction to the UHECR field is given, including both EAS physics, UHECR astrophysics, and experimental techniques.
The current theoretical questions in UHECR physics are introduced, and the experimental challenges encountered in the field, mostly related to understanding the results of the main experiments, are also discussed.
The physics of air fluorescence is also presented, as it is an important element in working with the air fluorescence technique and motives part of the experimental work presented. 
The JEM-EUSO experiment is introduced in detail to set the backdrop for the instrumental work presented later.   

The original contributions in this thesis are divided into experimental work on photodetection aspects of JEM-EUSO and phenomenological studies of UHECRs composition and source statistics.
The experimental part starts with a comprehensive introduction to photomultiplier tubes and their calibration. Single photoelectron detection is explained, and the calibration technique is 
discussed in detail. An experimental setup for measuring the air fluorescence yield is also presented based on the absolute calibration of PMTs using our method. The preliminary
calibration of two PMTs for this setup is shown to illustrate the application of the PMT calibration technique.

The instrumental work directly connected to JEM-EUSO begins with the testing of the JEM-EUSO high voltage power supply and switch system. 
As PMTs are electrostatic devices, the properties of the high voltage power supply directly affect both their gain and efficiency. The limited power budget of JEM-EUSO
requires a high power supply with a low power consumption, here achieved using a design based on a Cockcroft-Walton circuit. 
At the same time, JEM-EUSO will observe atmospheric phenomena which cover a dynamic range in light of $10^{6}$. 
A system of fast switches is needed in order to match the dynamic range of the PMTs to this range of light.
These switches reduce the number of electrons reaching the output of the PMT by 
changing the voltage of the PMT photocathode in a few microseconds. 
Both preliminary design tests and a test of the final EUSO-Balloon high voltage power supply prototype were performed, and this work was included in the successful CNES phase B review of EUSO-Balloon.

One Cockcroft-Walton high voltage power supply is used per EC, and so the gain of the EC can be adjusted as a unit by changing the power supply
output. The JEM-EUSO read-out Application Specific Integrated Circuit (ASIC) includes a preamplifier which allows the gain of each pixel within the PMT to
be equalized. There is up to a factor of 4 variation in gain between PMTs, however, and around a 20\% variation in gain from
pixel to pixel within a PMT. The gain and efficiency of each PMT is measured in single photon electron mode, and
they are sorted so that each EC can be build from PMTs with a similar enough gain that all 256 pixels can be
equalized using the dynamic range of the ASIC preamplifier. Sorting the PMTs in this way also allows a rejection
of defective PMTs. For JEM-EUSO the PMT sorting requires measuring the gain and efficiency of 64 pixels for
over 5,000 photomultiplier tubes. The development of a PMT sorting setup included the building and calibration of a data acquisition system using CAMAC Charge-to-Digital Conversion (QDC) electronics,
the development of data acquisition software, and the creation of routines to perform the analysis of 64 spectra for each PMT.


This system was then used to perform a first absolute calibration of the entire focal surface of EUSO-balloon. These calibration measurements were performed on the assembled and potted EC units. The goal
was to check that each EC function correctly, and, at the same time, measure the absolute efficiency.
Due to the difference in sensitivity between the QDC system and the ASIC read-out electronics of EUSO-Balloon, these preliminary results can only serve as a cross check.  
Measurements of the pixel width and the dead-space between photomultiplier tubes within an EC, again using the capabilities of the developed test bench, are also presented.
An extension of these measurements to a final absolute calibration of the EUSO-Balloon PDM is also discussed.

In the phenomenological part of this thesis two different bodies of work are presented. In the first,
a generic class of models for ultra-high energy cosmic ray (UHECR) phenomenology are studied.
In these models the sources accelerate protons and nuclei with a power-law spectrum having the same index, but with
different values for the maximum proton energies, distributed according to a power-law. 
It is shown that, for energies sufficiently lower than the maximum proton energy, such models are equivalent to single-type source models, but with a larger effective power law index and a heavier composition at the source.
The resulting enhancement of the abundance of nuclei is calculated, and typical values of a factor
2–10 are found for Fe nuclei. At the highest energies, the heavy nuclei enhancement ratios become larger, and
the granularity of the sources must also be taken into account. 
This shows that the effect of a distribution of maximum energies among sources must be considered in order to understand both the energy
spectrum and the composition of UHECRs as measured on Earth.

The second phenomenological study focuses on the number of sources which can be expected to contribute to the UHECR sky.
The GZK effect, the interaction of UHECR protons and nuclei with the intergalactic photon background, results
in a drastic reduction of the number of sources contributing to the observed flux above $\sim 6~10^{19}~$eV.  
The source statistics are studied quantitatively as a function of energy for a range of models compatible with the current data.
The source composition and injection spectrum, as well as the source density and luminosity distribution are varied, and various realizations of the
source distribution are explored. It is found that, in typical cases, the brightest source in the sky contributes more than 20\% of the total flux
above $8~10^{19}~$eV, and about 1/3 of the total flux at $10^{20}~$eV. 

It is also shown that typically between 2 and 5 sources contribute
more than half of the UHECR flux at $10^{20}~$eV. With such low source numbers, the isolation of the few brightest sources in the
sky may be possible for experiments collecting sufficient statistics at the highest energies, even in the event of relatively large
particle deflections. This last point loops the work presented in this thesis back to JEM-EUSO, showing that it is natural to expect that next-generation experiments with large exposure
will begin to answer the most fundamental questions about UHECRs.

\end{abstract}

\renewcommand{\abstractname}{r\'{e}sum\'{e}}
\begin{otherlanguage}{frenchb}
  \begin{abstract}
   La Terre est constamment soumise \`{a} un flux de particules provenant de l'espace. 
Ces particules, connues sous le nom de rayons cosmiques, ont jou\'{e} un r\^{o}le fondamental dans les premiers d\'{e}veloppements de la physique des particules, et de nombreuses questions les concernants demeurent sans r\'{e}ponse.
Des rayons cosmiques ont \'{e}t\'{e} observ\'{e}s avec des \'{e}nergies sup\'{e}rieures \`{a} $10^{20}~$eV, ce qui fait d'eux les particules les plus \'{e}nerg\'{e}tiques connues dans l'univers.
Ces Rayons Cosmiques d'Ultra-Haute \'{E}nergie (RCUHE) sont extr\^{e}mement rares et frappent la terre \`{a} raison de moins d'une particule par km$^{2}$ par si\`{e}cle. 
O\`{u} et comment ces particules ont-elles \'{e}l\'{e} acc\'{e}l\'{e}r\'{e}es jusqu'\`{a} de telles \'{e}nergies? Cela demeure \`{a} ce jour un myst\`{e}re.

En raison de leur flux extr\^{e}mement faible, les RCUHE ne peuvent \^{e}tre observ\'{e}s que de mani\`{e} indirecte, via les Gerbes Atmosph\'{e}riques qu'ils cr\'{e}ent dans l'atmosph\`{e}re, au moyen de d\'{e}tecteurs d\'{e}ploy\'{e}s sur d'immenses surfaces au sol.
Ces observatoires \'{e}chantillonnent les particules de la gerbe \`{a} l'aide d'une matrice de d\'{e}tecteurs de surface, mais peuvent \'{e}galamment d\'{e}tecter la gerbe par voie lumineuse.
Cette deuxi\`{e}me technique est connue sous le nom de m\'{e}thode de fluorescence de l'air, car la lumi\`{e}re \'{e}mise provient de la fluorescence des mol\'{e}cules d'azote excit\'{e}es par les particules charg\'{e}es de la gerbe.
La quantit\'{e} de lumi\`{e}re de fluorescence \'{e}mise \'{e}tant proportionnelle \`{a} l'\'{e}nergie d\'{e}pos\'{e}e dans l'air, la technique de fluorescence de l'air permet une mesure calorim\'{e}trique de l'\'{e}nergie des RCUHE incidents.
Un obstacle majeur \`{a} l'\'{e}tude des RCUHE est le faible nombre d'\'{e}v\'{e}nements ultra-\'{e}nerg\'{e}tiques observ\'{e}s, malgr\'{e} le fait que la g\'{e}n\'{e}ration actuelle de r\'{e}seaux de d\'{e}tecteurs couvrent des surfaces du sol de l'ordre du millier de km$^{2}$.

JEM-EUSO est un projet d'observatoire de RCUHE de prochaine g\'{e}n\'{e}ration, qui conduirait \`{a} une augmentation de l'exposition totale du ciel d'un ordre de grandeur. 
Ce gain important sera rendu possible par l'utilisation de la technique de fluorescence de l'azote depuis l'espace.
L'instrument JEM-EUSO est un t\'{e}lescope fluorescence, sensible dans le proche UV, avec une ouverture de plusieurs m$^{2}$ et un champ de vue de $60^{\circ}$. 
Ce t\'{e}l\'{e}scope serait fix\'{e} au module JEM de la Station Spatiale Internationale (ISS), en orbite autour de la Terre \`{a} une altitude d'environ 400~km. 
Depuis cette altitude, JEM-EUSO observe une surface au sol d'environ $2\,10^{5}~$km$^{2}$, avec un cycle utile de l'ordre de 14\%.
La surface focale de JEM-EUSO est compos\'{e}e de 137 Modules de PhotoD\'{e}tection (PDM), chacun compos\'{e} de 9
Cellules \'{E}l\'{e}mentaires (EC), regroupant chacune 4 Tubes PhotoMultiplicateurs Multi-Anode (MA-PMT) Hamamatsu R11265-M64, de 64 pixels chacun.
L'exp\'{e}rience EUSO-Ballon, \'{e}galamment port\'{e}e par la collaboration JEM-EUSO, est une mission Pathfinder compos\'{e}e d'un seul PDM, identique \`{a} ceux de JEM-EUSO, et d'une optique de Fresnel de m\^{e}me type, destin\'{e}e \`{a} un vol en ballon statosph\'{e}rique devant avoir lieu en 2014.

Les tubes photomultiplicateurs \`{a} vide, comme le Hamamatsu R11265-M64 utilis\'{e} pour JEM-EUSO, convertissent les photons incidents en une petite gerbes d'\'{e}lectrons,
ce qui donne une impulsion de signal qui peut \^{e}tre d\'{e}tect\'{e}e par l'\'{e}lectronique de lecture.
Dans JEM-EUSO, le flux de photons en provenance d'une gerbe atmosph\'{e}rique est suffisamment faible pour que les gerbes d'\'{e}lectrons individuels se d\'{e}veloppant dans le PMT puissent \^{e}tre s\'{e}par\'{e}es. Il est ainsi possible de compter les photons un par un. 
C'est ce qu'on appelle le comptage de photo\'{e}lectron unique.
L'efficacit\'{e} d'un PMT, d\'{e}finie comme le rapport entre nombre de photo\'{e}lectrons uniques \`{a} la sortie du PMT et le nombre de photons incidents sur la fen\^{e}tre d'entr\'{e}e du PMT, est un param\`{e}tre critique pour
la d\'{e}termination de l'\'{e}nergie d'un RCUHE, qui se d\'{e}duit du nombre de coups de photo\'{e}lectrons enregistr\'{e}s sur la surface focale. 
Cependant, mesurer l'efficacit\'{e} absolue du PMT est une t\^{a}che difficile exp\'{e}rimentalement, et la plupart des m\'{e}thodes donnent une incertitude de l'ordre de 10\%. 
En comparant le PMT \`{a} une photodiode NIST calibr\'{e}e de mani\`{e}re absolue, et en utilisant une sph\`{e}re d'int\'{e}gration comme diviseur stable de lumi\`{e}re, nous sommes capables de mesurer l'efficacit\'{e} absolue
avec une incertitude de l'ordre de quelques pour cent.

Dans cette th\'{e}se, nous proposons d'abord une introduction g\'{e}n\'{e}rale au domaine des RCUHE, ainsi qu'\`{a} la physique des gerbes, \'{a} l'astrophysique des RCUHE et aux techniques exp\'{e}rimentales associ\'{e}es.
Les questions th\'{e}oriques actuelles de la physique des RCUHE sont introduites, et les difficult\'{e}s exp\'{e}rimentales rencontr\'{e}es dans le domaine, principalement 
li\'{e}es \`{a} la compr\'{e}hension des r\'{e}sultats des principales exp\'{e}riences, sont \'{e}galement discut\'{e}es.
La physique de la fluorescence de l'air est \'{e}galement pr\'{e}sent\'{e}e, car c'est un \'{e}l\'{e}ment important de la d\'{e}tection des gerbes par la technique de fluorescence, et 
c'est ce qui motive une partie du travail exp\'{e}rimental pr\'{e}sent\'{e}.
L'exp\'{e}rience JEM-EUSO est pr\'{e}sent\'{e}e en d\'{e}tail, comme toile de fond pour le travail instrumental pr\'{e}sent\'{e} ensuite.

Les contributions originales de cette th\`{e}se sont divis\'{e}es en des travaux exp\'{e}ri-mentaux sur les aspects de photod\'{e}tection de JEM-EUSO et des \'{e}tudes ph\'{e}nom\'{e}-nologiques portant sur la composition des RCUHEs et la statistique de leurs sources en fonction de l'\'{e}nergie.
La partie exp\'{e}rimentale commence par une introduction compl\`{e}te aux tubes photomultiplicateurs et \`{a} leur \'{e}talonnage. 
Apr\`{e}s cela, la d\'{e}tection de photo\'{e}lectrons uniques est expliqu\'{e}e, et la technique de calibration est
discut\'{e}e en d\'{e}tail. Un dispositif exp\'{e}rimental devant permettre la mesure du rendement de fluorescence de l'air est \'{e}galement pr\'{e}sent\'{e}, sur la base de l'\'{e}talonnage absolu des PMTs, en utilisant la m\'{e}thode que nous avons d\'{e}velopp\'{e}e. 
Le pr\'{e}-\'{e}talonnage de deux PMTs selon cette configuration est indiqu\'{e} pour illustrer l'application de la technique de calibration.

Le travail instrumental directement connect\'{e} \`{a} JEM-EUSO commence avec le test de l'alimentation haute tension de l'instrument, ainsi que de son syst\`{e}me de commutation.
Comme les PMTs sont des dispositifs \'{e}lectrostatiques, les propri\'{e}t\'{e}s de l'alimentation haute tension affectent directement \'{a} la fois leur gain et leur efficacit\'{e}. 
Le budget de puissance limit\'{e} de JEM-EUSO n\'{e}cessite une alimentation de haute tension avec une faible consommation d'\'{e}nergie. 
Ceci est obtenu gr\^{a} \`{a} la mise en oeuvre d'un concept original, bas\'{e} sur un circuit Cockcroft-Walton.
Dans le m\^{e}me temps, JEM-EUSO doit pouvoir observer des ph\'{e}nom\`{e}nes atmosph\'{e}riques couvrant une vaste gamme dynamique, de l'ordre de 10$^{6}$.
Un syst\`{e}me de commutateurs rapides est donc n\'{e}cessaire afin de faire correspondre la gamme dynamique des PMT \`{a} de tels \'{e}carts d'intensit\'{e} lumineuse.
Ces commutateurs permettent de r\'{e}duire le nombre d'\'{e}lectrons atteignant la sortie du PMT en modifiant la tension de la photocathode du PMT en quelques micro-secondes.
Un test de conception pr\'{e}liminaire et un test du prototype du circuit haute tension final de l'instrument EUSO-Ballon ont \'{e}t\'{e} men\'{e}s avec succ\`{e}s.

Une alimentation \`{a} haute tension Cockcroft-Walton est utilis\'{e}e par les ECs, de sorte que le gain des ECs puisse \^{e}tre r\'{e}gl\'{e} comme une unit\'{e} en changeant la tension d'alimentation. 
Le syst\`{e}me de lecture de JEM-EUSO est un ASIC (Application Specific Integrated Circuit) qui comprend un pr\'{e}amplificateur permettant au gain de chaque pixel dans le PMT d'\^{e}tre \'{e}galis\'{e}.
Cependant, il y a une variation de gain pouvant aller jusqu'\`{a} un facteur 4 entre les PMT et une variation d'environ 20\% du gain de pixel \`{a} pixel au sein d'un PMT. 
Le gain et l'efficacit\'{e} de chaque PMT est mesur\'{e} en mode photo\'{e}lectron unique, et les PMT sont tri\'{e}s afin que chaque EC puisse \^{e}tre construite \`{a} partir de PMT ayant un gain suffisamment similaire pour que l'ensemble des 256 pixels puissent \^{e}tre
\'{e}galis\'{e}s \`{a} l'aide de la gamme dynamique du pr\'{e}amplificateur de l'ASIC. 
Le tri des PMT de cette fa\c{c}on permet \'{e}galement un rejet des PMTs d\'{e}fectueux. 
Pour JEM-EUSO, le tri des PMT n\'{e}cessite de mesurer le gain et l'efficacit\'{e} de 64 pixels pour plus de 5000 tubes photomultiplicateurs. 
Le d\'{e}veloppement d'une configuration de tri des PMT inclus la construction et l'\'{e}talonnage d'un syst\`{e}me d'acquisition de donn\'{e}es en utilisant un QDC (Charge-to-Digital Convertor) \'{e}lectronique CAMAC,
le d\'{e}veloppement de logiciels d'acquisition de donn\'{e}es et la cr\'{e}ation de routines pour effectuer l'analyse des 64 spectres pour chaque PMT.

Ce syst\`{e}me a ensuite \'{e}t\'{e} utilis\'{e} pour effectuer un premier \'{e}talonnage absolu de toute la surface focale d'EUSO-ballon. 
Ces mesures d'\'{e}talonnage ont \'{e}t\'{e} effectu\'{e}es sur les ECs assembl\'{e}s et encapsul\'{e}s. 
L'objectif \'{e}tait de v\'{e}rifier que chaque EC fonctionnait correctement et, en m\^{e}me temps, de mesurer son efficacit\'{e} absolue.
En raison de la diff\'{e}rence de sensibilit\'{e} entre le syst\`{e}me QDC et le syst\`{e}me de lecture \'{e}lectronique ASIC d'EUSO-Ballon, ces r\'{e}sultats pr\'{e}liminaires ne peuvent servir que de recoupement.
Les mesures de la largeur en pixels et de l'espace mort entre les tubes photomultiplicateurs au sein d'une EC, en utilisant une fois encore les capacit\'{e}s du banc d'essai d\'{e}velopp\'{e}, sont \'{e}galement pr\'{e}sent\'{e}es.
Une extension de ces mesures \`{a} un \'{e}talonnage absolu final du PDM d'EUSO-Ballon est \'{e}galement discut\'{e}e.

Dans la partie ph\'{e}nom\'{e}nologique de cette th\`{e}se, deux travaux diff\'{e}rents sont pr\'{e}sent\'{e}s. 
Dans la premi\`{e}re partie, une classe g\'{e}n\'{e}rique du mod\`{e}les ph\'{e}nom\'{e}-nologiques de RCUHE est \'{e}tudi\'{e}.
Dans ces mod\`{e}les, les sources acc\'{e}l\`{e}rent des protons et des noyaux avec un spectre en loi de puissance ayant le m\^{e}me indice spectral, mais des valeurs diff\'{e}rentes de l'\'{e}nergies maximale atteinte par les protons, qui se distribuent selon une loi de puissance.
Il est d\'{e}montr\'{e} que, pour des \'{e}nergies suffisamment inf\'{e}rieures \`{a} l'\'{e}nergie maximale des protons, ces mod\`{e}les sont \'{e}quivalents aux mod\`{e}les supposant que toutes les sources sont identiques, 
mais l'indice de la loi de puissance du spectre source effectif r\'{e}sultant est alors plus \'{e}lev\'{e}, et la composition source effective \'{e}quivalente est quant \`{a} elle plus lourde.
L'augmentation effective de l'abondance des noyaux qui en r\'{e}sulte est calcul\'{e}e, et des valeurs typiques d'un facteur
2-10 sont trouv\'{e}es pour les noyaux de fer. Aux plus hautes \'{e}nergies, les facteurs d'augmentation de l'abondance des noyaux lourds deviennent plus grands, et la granularit\'{e} des sources doit alors \^{e}tre prise en compte.
Cela montre que l'effet d'une distribution d'\'{e}nergies maximales entre les diff\'{e}rentes sources doit \^{e}tre pris en compte pour comprendre \`{a} la fois le spectre d'\'{e}nergie
et la composition de RCUHEs, tels que mesur\'{e}s sur Terre.

La deuxi\`{e}me \'{e}tude ph\'{e}nom\'{e}nologique met l'accent sur le nombre de sources que l'on peut voir contribuer au ciel RCUHE.
L'effet GZK, l'interaction des protons et des noyaux RCUHE avec le fond de photons intergalactiques, entra\^{i}ne une r\'{e}duction drastique du nombre de sources qui contribuent au flux observ\'{e} au-dessus de $6\,10^{19}~$eV environ.
La statistique des sources est \'{e}tudi\'{e}e quantitativement en fonction de l'\'{e}nergie pour toute une gamme de mod\`{e}les compatibles avec les donn\'{e}es actuelles.
Diverses hypoth\`{e}ses sont explor\'{e}es concernant la composition et le spectre d'injection des sources, ainsi que la densit\'{e} des sources et leur distribution en luminosit\'{e}. Enfin, diverses r\'{e}alisations de la
distribution des sources sont \'{e}tudi\'{e}es. On constate que, dans les cas typiques, la source la plus lumineuse du ciel contribue \`{a} plus de 20\% du flux total
au-dessus $8~10^{19}~$eV, et \`{a} environ 1/3 du flux total \`{a} 10$^{20}~$eV.
Il est \'{e}galement montr\'{e} que, g\'{e}n\'{e}ralement, entre 2 et 5 sources contribuent
\`{a} plus de la moiti\'{e} du flux RCUHE \`{a} $10^{20}~$eV. Compte tenu de ces nombres tr\`{e}s faibles, des exp\'{e}riences capables de collecter une statistique appr\'{e}ciable aux plus hautes \'{e}nergies devraient \^{e}tre en mesure d'isoler les quelques rares sources lumineuses visibles dans le ciel, m\^{e}me en cas de relativement grandes
d\'{e}flexions angulaires des particules. Ce dernier point boucle le travail pr\'{e}sent\'{e} dans cette th\`{e}se relativement \`{a} JEM-EUSO, en montrant que l'on peut s'attendre de mani\`{e}re naturelle \`{a} ce que les exp\'{e}riences de la prochaine g\'{e}n\'{e}ration, telles que JEM-EUSO, commencent \`{a} r\'{e}pondre aux questions les plus fondamentales sur le RCUHEs.
\\
\\
\textbf{mots-cl\'{e}s}: RCUHE, JEM-EUSO,  EUSO-Ballon, instrumentation, photod\'{e}tection, photomultiplicateur, \'{e}talonnage, efficacit\'{e} 

  \end{abstract}
\end{otherlanguage}

\chapter*{Acknowledgements and Foreword}

\begin{quote}
 HORATIO. \begin{quote}O day and night, but this is wondrous strange!\end{quote}
 HAMLET. \begin{quote} And therefore as a stranger give it welcome.
 There are more things in heaven and earth, Horatio,
 Than are dreamt of in your philosophy. \end{quote}

\hspace*\fill{--- William Shakespeare, Hamlet, I.v}
\end{quote}
The history of science is one of welcoming the stranger, albeit often only because he already had his foot in the door.
Despite what Eugene Wigner called the ``unreasonable effectiveness of mathematics in the natural sciences'' we must never forget that the sole arbiter of truth in science
is to ask nature a question through a well designed and properly conducted experiment. Understanding the answer, that is another question.


In our time it is often said by those outside the physical sciences that removing the mystery of the world somehow reduces the beauty, as though ignorance were a perquisite for wonder or appreciation. 
I would prefer to agree with those who are of a mind that the truth of the universe is more amazing and beautiful than the human scale of mythology, as  
\begin{quote}
 [n]othing is ``mere''. I too can see the stars on a desert night, and feel them. But do I see less or more? 
The vastness of the heavens stretches my imagination -- stuck on this carousel my little eye can catch one-million-year-old light. 
A vast pattern -- of which I am a part... What is the pattern, or the meaning, or the why? It does not do harm to the mystery to know a little about it. 
For far more marvelous is the truth than any artists of the past imagined it.

 \hspace*\fill{---Richard Feynman}
\end{quote}

In this quest to understand the workings of the world,
I have been privileged to work for the last three years in a city which is full of the humane disciplines of art and culture, on a scientific project which is interesting, and with people who are wonderful. 
While I have certainly learned a small part of the vast field of physics, I have also had the opportunity to see much of life.

I would first like to thank Etienne Parizot and Philippe Gorodetzky, my thesis supervisors. I have learned many things 
from them about physics, about being a scientist, and about life. It has been my pleasure to discuss many topics with them. I consider the two of them to be both role-models and friends.
I cannot help but also mention the many colleagues who I have had the opportunity to work with over these years. These are people with whom I have spent hours in the lab, and shared drinks with on the far corners of the Earth. 
There are, of course, also numerous other people who deserve my appreciation for making my world what it is; I feel that they know who they are. I am especially grateful to those who have 
bore with me (and in some cases bared with me) while I have been writing this thesis. 
With that out of the way, I hope that you, the reader, will find this work interesting and worthwhile.

  \tableofcontents
     \listoffigures
     \listoftables

\part{Ultra-High Energy Cosmic Rays: Experiment and Theory}
\pagenumbering{arabic}

  \label{PART1}
    \begin{refsection}
  \chapter{Introduction to Cosmic Rays}
    
The Earth is constantly bathed in a sea of particles from space. 
These particles, known as cosmic rays, were fundamental in early particle physics, and continue to be a source of unanswered questions. 
The first evidence of cosmic rays came from measurements of atmospheric ionization at high altitude by Victor Hess~\cite{hess}.
In the course of multiple balloon flights between 1911 and 1913, Hess measured the ionization of the atmosphere using an electrometer. 
He found that the ionization decreased with altitude up to 1 km, and then increased at higher altitude.
This was particularity surprising, as the motivation of Hess's flights was to study the ionization which was presumed to be caused by the
radioactivity of the Earth. The belief at the time was that the atmosphere ionization should decrease with altitude.
Having observed the opposite, Hess's conclusion was that the particles responsible for the ionization at high altitudes had their origin in outer space.

The name ``cosmic ray'' was given to these particles due to their presumed origin in the cosmos and early theories proposed by Robert Millikan that they where composed of electromagnetic radiation.
Further studies in the late 1920s by W.\ Bothe and W.\ Kolh\"{o}rster using two Geiger-M\"{u}ller counters with an absorber in between showed that cosmic rays are composed of discrete particles~\cite{1929ZPhy...56..751B}, and further results 
by J.\ Clay showing that the flux of cosmic rays depends on magnetic latitude led to the conclusion that a significant fraction of cosmic rays must be charged particles~\cite{JClay}.
 
Cosmic rays provided a source of many early discoveries in particle physics due to their constant availability and relatively large energies. Most notable of 
these early discoveries was that of the positron by Carl Anderson in the early 1930s. 
Using a vertically oriented Wilson chamber with an applied magnetic field, he observed positively charged particles with the same radius of curvature as electrons in 15 out of 1300 photographed events~\cite{Anderson1933}.
Based on the track length, energy loss, and resulting charge-to-mass ratio  Anderson concluded that this new particle was a positive electron, or positron.  
In similar cloud chamber experiments, the mu-meson, so called because it was at the time thought to be the mediator of the strong force proposed by Yukawa~\cite{Yukawa:1935xg}, was discovered by
Neddermeyer and Anderson~\cite{PhysRev.51.884}, and later confirmed by Street and Stevenson~\cite{PhysRev.52.1003}.

The mediating particle theorized by Yukawa, the pion, was itself discovered in 1947 in nuclear emulsions exposed to cosmic rays on a mountain top~\cite{Lattes:1947mx}.
The mu-meson, on the other hand, is now known as the muon. The majority of muons which reach the ground are secondaries produced in the decay of pions. 
The muon flux at sea level is roughly $1/\text{cm}^{2}/\text{min}$, and knowledge of this flux is extremely important.
This is true both as a background in precision high-energy particle physics experiments, and as a tool for muon tomography in the fields of archeology, geoscience, and volcanology~\cite{Alvarez1970832, Marteau:2012zv}.
The muon flux can also be put to use for detector calibration and performance tests, such as those done by the Atlas collaboration at the \gls{LHC}~\cite{2011EPJC...71.1593A}, 
or the ground array of the Pierre Auger Observatory \cite{Abraham200450}.
Background events caused by cosmic-ray muons can even be seen in the output signal of the photodetectors which will be
studied in this thesis~\cite{Tubs100}.

Each muon which is created by pion decay is accompanied by a muon neutrino. This flux of ``atmospheric'' neutrinos is calculable (cf.~\cite{PhysRevD.39.3532})  
and was studied, for example, by the Kamiokande experiment~\cite{Hirata:1988uy}, which found a deficit of muon neutrinos. This led to further activity in the field,
 and the eventual discovery of atmospheric neutrino oscillation by Super Kamiokande~\cite{PhysRevLett.81.1562}. In this sense atmospheric neutrinos, originating from cosmic rays, played an important 
role in unraveling the phenomena of neutrino oscillation, which is one of the first phenomena discovered ``outside'' the standard model of particle physics.
 
In addition to these key discoveries in particle physics, there are also hints that cosmic rays play an important role in the overall Earth system. 
The ionization of the atmosphere is thought to influence the creation of lightning by forming an electron avalanche which
leads to relativistic runaways, resulting in an abrupt discharge~\cite{Gurevich2001180}. This theory has been studied in both a laboratory setting~\cite{Gurevich20112845}, and in collaboration with 
cosmic ray observatories~\cite{Gurevich2004348}.
  
Cosmic rays may also impact the formation of aerosols in the atmosphere, which are a prerequisite to cloud formation. This idea was first proposed by E.P. Ney, who
argued for the existence of large tropospheric and stratospheric effects produced by the solar-cycle modulation of cosmic rays~\cite{Ney1959451}.
Newer results seemed to show both a correlation between the global cloud cover and cosmic ray flux and a dependence on ionization in aerosol formation~\cite{Svensmark19971225, Svensmark20132343}.
These controversial results have led to a controlled study of aerosol formation using an accelerator beam in the \acrshort{CLOUDS} experiment at \acrshort{CERN}~\cite{Fastrup:2000tm}.

The cosmic ray flux may also impact the biosphere. One example from biology is the role of lightning in the formation of organic compounds.
This was shown by Miller-Urey experiment, in which amino acids are produced in an ``primitive earth-like'' atmosphere subjected to electrical discharges~\cite{Miller15051953}.
As lightning is possibly influenced by cosmic rays, this could connect cosmic ray phenomenology with the formation of life.

In addition, atmospheric ionization induced by cosmic rays can disintegrate N$_{2}$ and O$_{2}$ molecules, changing the chemistry of the atmosphere. Free oxygen and nitrogen
atoms bond with the ozone in the upper atmosphere, depleting the ozone layer and forming nitrates~\cite{GRL:GRL50222}. 
The nitrates formed can find their way to the Earth's surface through rain and act
as fertilizers for plant life~\cite{2008arXiv0804.3604T, GRL:GRL19929}. 
In the same way, cosmic rays continuously produce various unstable isotopes in the Earth's atmosphere, such as carbon-14. The cosmic ray flux has kept the level of carbon-14 in
the atmosphere roughly constant for the last 100,000 years, which makes possible the use of radiocarbon dating~\cite{ANDERSON30051947, Libby1949}. 

The decrease in ozone, on the other hand, leads to an increase of \acrshort{UVB} radiation, which is damaging to \acrshort{DNA}. This increase, combined with the direct flux of secondary muons and neutrons from cosmic rays, may lead to genetic mutations, acting as a catalyst 
for evolution. Both the increase of UVB radiation and the formation of nitrates may also have effects on biodiversity~\cite{JGRE:JGRE2567}, and a connection between cosmic rays and evolutionary process was first 
proposed in the late 1920s by J.\ Joly~\cite{JolyJ}. 
As a more recent corollary to this, interesting evidence for a 60-Myr and 140-Myr cycle in fossil diversity the during the course of the Phanerozoic era was found by Rohde and Muller~\cite{Rohde2005}.
Although the 140-Myr cycle which they found was less significant than the 60-Myr cycle, the authors suggest that the fact that it is consistent with the periods of other cycles reported in the climate and cosmic ray flux potentially warrants further investigation.

All of these topics are, of course, parallel to the physical interest of cosmic ray phenomena themselves. 
In the late 1930s, W.\ Kolh\"{o}ster and Pierre Auger separately began to experiment with correlated detectors. 
This work built on the previous refinement of the coincidence technique by B. Rossi \cite{Rossi1930},
whose circuit gave an improved time resolution compared to the first coincidence circuits developed by W. Bothe \cite{Bothe1929}.
A further key aspect of Rossi's coincidence circuit was the possibly of multifold coincidences, which greatly reduced the rate of random triggers and allowed the study of rare cosmic-ray events \cite{Rossi1933}. 
Putting these techniques to use with Geiger-M\"{u}ller counters, Kolh\"{o}ster found coincidence signals in detectors up to 75 m apart \cite{Kolhorster1938}, while
Auger performed similar experiments in the Swiss Alps using Wilson chambers and Geiger-M\"{u}ller counters separated by large distances \cite{Auger1938}. 
From the experiments of these pioneers, it was concluded that the detected signals were secondary particles created in an extensive air shower initiated by a single primary cosmic ray. 

The discovery of extensive air showers lead to the indirect detection of cosmic rays using ground arrays consisting of spaced detector elements which sample the shower.
In the 1960s a detector composed of scintillation counters split into 20 stations was built at Volcano Ranch, New Mexico by a group from MIT led by John Linsley.
This detector recorded the first extensive air shower coming from a primary particle with an energy of $10^{20}~$eV~\cite{PhysRevLett.10.146}. 

The observation of a single particle with such a huge energy (a $10^{20}~$\acrshort{eV} proton has the same energy as a tennis ball thrown at 86 km/h, confined in a subatomic particle with a radius of less than $9~10^{-16}~$m) 
raises an intriguing question, and perhaps the one which most captures the imagination in all of cosmic ray physics: What in the universe as is understood by modern astrophysics could accelerate
particles up to such energies?

\section{The Physics of Cosmic Rays}
The next few sections will give a very brief summary of the field of cosmic ray physics, focusing on so-called ultra-high-energy cosmic rays, or UHECRs.
The term UHECR is generally used for those cosmic rays with energies above $10^{18}~$eV, which represents the limit of the cosmic ray spectrum, 
and, as is the case with the most interesting science, this limit pushes against the boundaries of what is well-understood. 

Naturally, the overview presented will be brief, as several theses could be written (as many have) on the knowledge gathered in the 100 years since cosmic rays were first discovered.
The basic stage will be set by introducing the cosmic ray spectrum and what is known about the composition and sources of cosmic rays at the highest energies. After that, 
a brief discussion of the astrophysics and particle physics involved in cosmic ray phenomenology will be presented. A discussion of extensive air showers and 
the experiments which observe the highest energy particles known in the universe will be presented in the next chapter.

Curious readers who wish for a more in-depth introduction are recommended towards the review of M.\ Walter and A.\ Wolfendale on the early history of cosmic ray physics~\cite{Walter:2012zz}, the further historical review of the field by
Kampert et al.\ \cite{Kampert:2012vi},
the review of  Bl\"{u}mer et al.\ for a general overview of cosmic ray physics above $10^{15}~$eV \cite{Bluemer:2009zf}, 
the review of D.\ Allard on the extragalactic propagation of UHECR in the universe~\cite{Allard:2011aa}, the review of Olinto and Kotera~\cite{2011ARA&A..49..119K} 
on the subject of cosmic ray astrophysics, and the experimental and theoretical summaries of the UHECR 2012 symposium \cite{2013EPJWC..5302001O,2013EPJWC..5302002F}. 
These texts, and the many other articles which they reference were heavily and shamelessly referenced during the writing of this introduction.

\section{The Cosmic Ray Spectrum}

\begin{figure}[h]
  \centering
  \includegraphics[angle=270, width=1.0\textwidth]{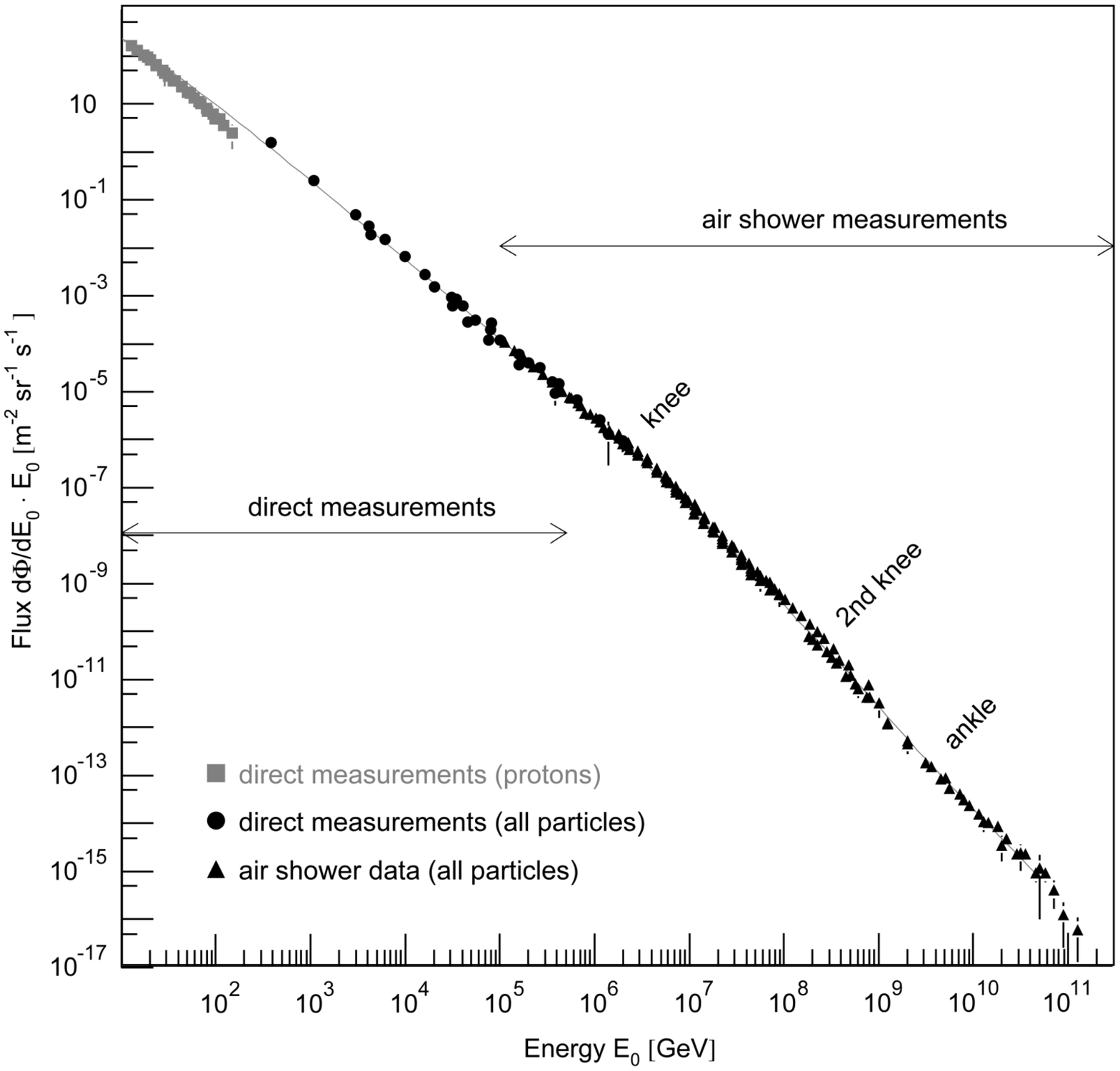}
   \caption[The Cosmic Ray Spectrum]{\label{fig:CosmicRaySpectrum:Overall} A plot of the total cosmic ray spectrum from the lowest energies up to the highest energy particles observed.
The overall spectrum extends more than 12 decades in energy and 18 decades in flux.  
Above $\sim 100~$GeV the total cosmic ray flux is shown, below this energy only the proton flux is plotted.
Below $10^{5}~$GeV the flux is high enough that cosmic rays can be studied by direct observation. Above this energy, cosmic rays are observed primarily by indirect observation of extensive air showers.
The figure itself is taken from ref.~\cite{Bluemer:2009zf}.}
 \end{figure}

The cosmic ray spectrum, that is the number of particles which reach earth per unit of energy per square meter per steradian per second, is shown in \fig\ref{fig:CosmicRaySpectrum:Overall}. 
A salient feature of the cosmic ray spectrum is that it extends from several MeV up to at least $10^{20}~$eV.
It should be remembered that the single plot shown in \fig\ref{fig:CosmicRaySpectrum:Overall} represents an enormous experimental effort which cannot be truly appreciated in these few short pages. 

Some fraction of cosmic rays with energies of up to several GeV originate from the sun, accelerated by solar flares and coronal mass ejections. 
The flux of lower energy galactic cosmic rays is influenced by solar winds~\cite{1968ApJ...154.1011G},
in addition to this, the Earth's magnetic field deflects cosmic rays away from the surface.  This causes the flux measured at earth at low energy to be dependent on latitude, longitude, and azimuth angle. 
The magnetic field lines of the Earth sweep low energy cosmic rays towards the poles, giving rise to aurorae. 

As can be seen in \fig\ref{fig:CosmicRaySpectrum:Overall}, the flux decreases with increasing energy, by about a factor of 500 per decade in energy. 
This results in the flux going from more than 1000 particles per second and m$^{2}$ at GeV energies to about one particle per m$^{2}$
per year at a $10^{15}~$eV, and further, to less than one particle per km$^{2}$ per century at $10^{20}~$eV.

Because of this rapid decrease in flux, our ability to detect cosmic rays also decreases with energy. At energies in the range of a GeV to TeV cosmic rays can be detected directly
using balloon borne detectors, or detectors in space (i.e.\ detectors with an area on the order of several m$^{2}$). 
Above 100 TeV, larger and larger collection areas are needed, generally in the form of ground arrays, the largest of which span an effective area of several thousand km$^{2}$.  
These ground observatories detect the cosmic rays indirectly, by sampling the secondary particles in the \emph{Extensive Air Shower} (\acrshort{EAS}) created by the interaction of the cosmic ray with the 
atmosphere.

\begin{figure}[h]
  \centering
  \includegraphics[angle=270, width=1.0\textwidth]{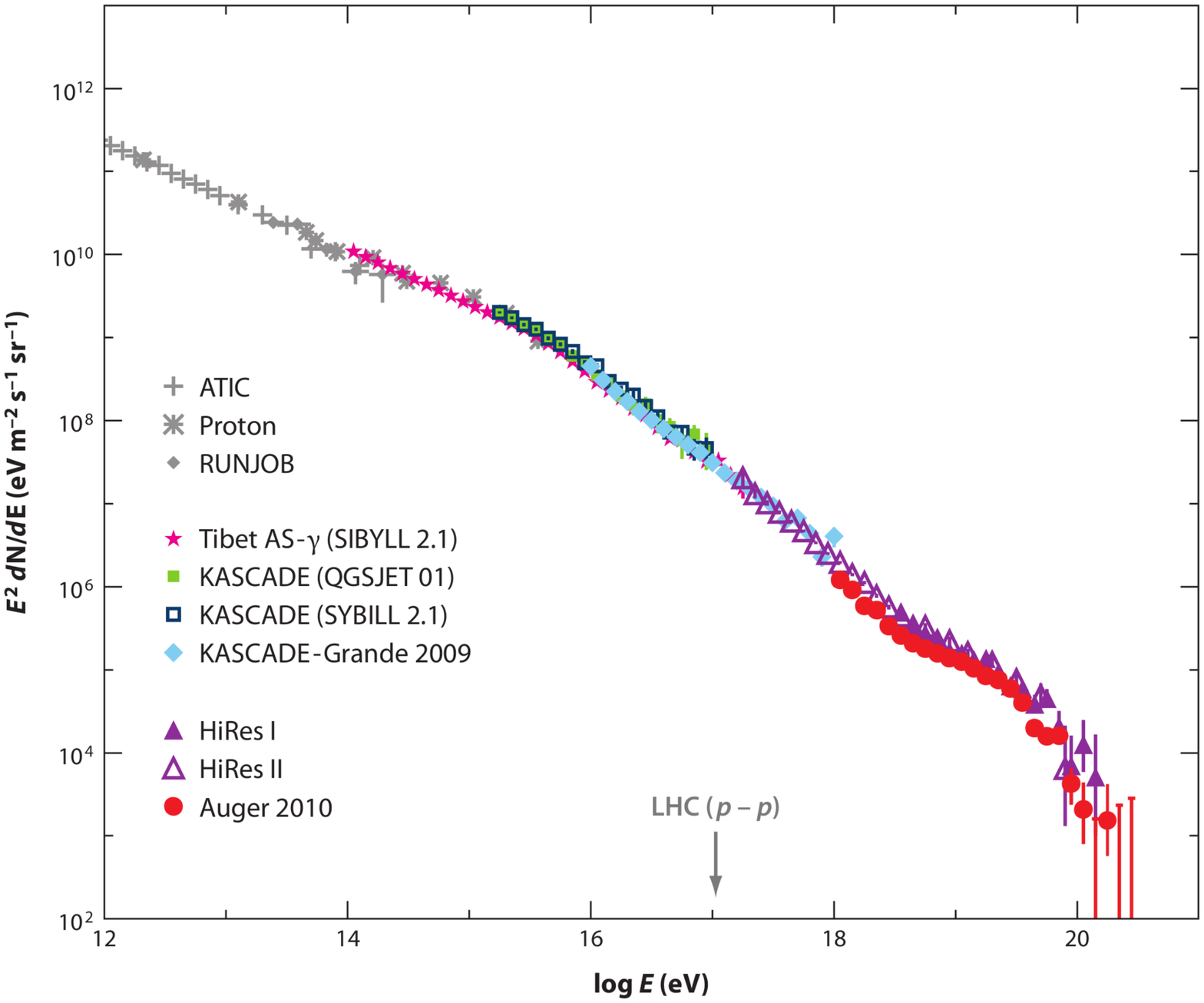}
   \caption[The Spectrum of Cosmic Rays Above $10^{12}~$eV]{\label{fig:UHECRspectrum} A plot of the cosmic ray spectrum above $10^{12}~$eV taken from ref.~\cite{2011ARA&A..49..119K}.
Here the flux is multiplied by $E^{2}$ in order to make structures in the spectrum more visible. Several breaks in the spectrum can be seen, such as the knee, and the ankle (see text).
The measured data are from ATIC \cite{Ahn20061950}, Proton \cite{Akimov:1971gq}, RUNJOB \cite{Apanasenko:2000xj}, Tibet-AS-$\gamma$ \cite{Amenomori:2008aa}, KASCADE \cite{Kampert:2004rz} and KASCADE-Grande \cite{Apel:2009uq}, HiRes I \cite{Abbasi:2009ix} and HiRes II \cite{Abbasi:2007sv}, and the Pierre Auger Observatory
\cite{Abraham:2010mj}.
The equivalent LHC ($pp$ fixed target) energy is shown for comparison.
}
 \end{figure}

At high energy the spectrum shows several features which can give us information about the underlying physics of cosmic rays.
The \gls{UHECR} spectrum is shown, multiplied by a factor of $E^{2}$, in \fig\ref{fig:UHECRspectrum}. 
Within the energy range shown in \fig\ref{fig:UHECRspectrum}, the flux decreases by 24 orders of magnitude over the course of 8 decades of energy. 
This plot is a compilation of 
published results from several past and ongoing experiments, and it should be noted that multiplying the 
flux by a factor of $E$, while useful visually, is dangerous as it mixes the systematic and statistical uncertainties of the energy scales of the different experiments~\cite{Bluemer:2009zf}.

The first feature which can be seen in \fig\ref{fig:UHECRspectrum} is the steepening of the UHECR spectrum between $10^{15}~$eV and $10^{16}~$eV, known as the ``knee.'' 
After the knee is the still-debated so-called ``second-knee,'' a further steepening around $3~10^{17}$ eV, followed by a recovery in the slope of the spectrum known as the ``ankle,'' which appears
between $10^{18}$ and $10^{19}~$eV. Also seen in \fig\ref{fig:UHECRspectrum} is the equivalent LHC energy (proton-proton fixed target), which is $\sim 10^{17}~$eV. This shows clearly the fact that particle physics above the knee
is no longer directly constrained by data from accelerator experiments.

Above the ankle is the cutoff. The overall low flux above $10^{18}~$eV makes the observation of a cutoff non-trivial, but the jump in statistics given by 
\gls{HiRes}, \gls{Auger}, and the \gls{TA} have made the observation of the cutoff statistically significant \cite{Abbasi:2007sv, Abbasi:2009ix, Abbasi+04, Abraham:2008ru, Abraham:2010mj, AbuZayyad:2012ru}.
Such a cut-off in the UHECR spectrum was predicted by Greissen, Zatsepin, and Kuzmin (\acrshort{GZK})~\cite{Greisen1966,Zatsepin:1966jv} just after the discovery of the \gls{CMB} radiation.
They predicted that the interaction of extremely high energy cosmic rays with the photons of the CMB would lead to a decrease in the average propagation length of UHECR, known as the GZK-effect.

Cosmic ray protons are affected mainly by the pair production mechanism, which has an energy threshold with CMB photons of around $10^{18}~$eV, and pion production,
which dominates above $\simeq 7~10^{19}~$eV. The same interactions can occur with infrared, optical, and ultraviolet backgrounds in intergalactic space, but this contribution is 
almost irrelevant over the entire energy range. Cosmic ray nuclei, such as iron, on the other hand, interact with the CMB through the giant dipole resonance. 
This type of interaction leads to the photo-disintegration of the nucleus. 
The photo-disintegration threshold energy is proportional to the atomic number, in the laboratory frame of the cosmic ray.

While the observed cutoff could be due to the GZK-effect, the cutoff could also be the result of the maximum energy of the sources of UHECR, whatever they might be. 
This makes the question of the absolute energy scale in the measured spectrum an important problem, as if the absolute energy scale of spectral features are known this information can help constrain different UHECR models.
Unfortunately, the two largest experiments (Auger and TA/HiRes) have an energy scale which differs by approximately 20\%, and 
this difference in energy scale is the object of continuing study~\cite{2013EPJWC..5301005D}.

\begin{figure}[h]
  \centering
  \includegraphics[angle=270, width=1.0\textwidth]{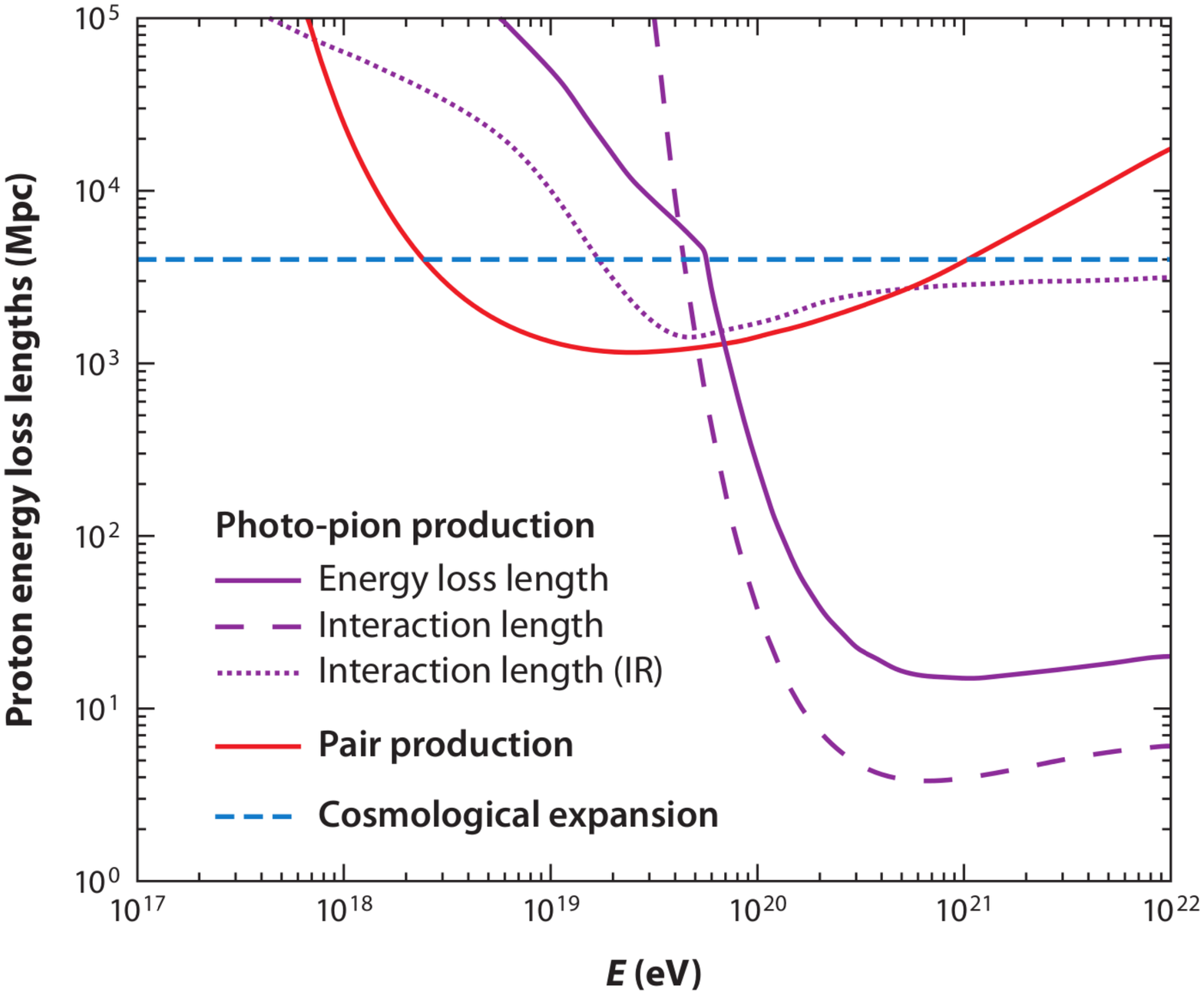}
   \caption[Energy Loss Length of Cosmic Ray Protons]{\label{fig:energyLosses}  A plot of the energy loss length of protons taken from ref.~\cite{2011ARA&A..49..119K}.
The purple lines show the energy loss (solid line) and interaction lengths for photo-pion production on the CMB (dashed line), and IR-UV photons (dotted line).
The solid red line shows loss length for pair production on CMB photons, using the background of ref.~\cite{2006ApJ...648..774S}. The dashed blue line shows energy loss due to cosmological expansion.
As can be seen, the interaction length for photo-pion production on the CMB becomes $\lesssim 10~$Mpc for protons with energies above $10^{20}~$eV.
}
 \end{figure}

Energy-loss mechanisms such as photo-disintegration and pion production are a natural extension of known particle physics. 
The propagation effects which UHECRs experience as they propagate away from their respective sources can be fit into two general categories: 
\begin{inparaenum}[i\upshape)]
 \item effects which modify the UHECR direction, but not their energy or composition, such as deflection in magnetic fields, and
 \item effects which modify the UHECR composition and or energy, but not their direction.
\end{inparaenum}

The second category is well-represented by the GZK effect and adiabatic losses due to the expansion of the universe. Magnetic deflections, on the other hand, fall into the first category, and change the trajectory of the UHECR, but not 
their energy. 
The actual deflection of a given cosmic ray depends on its charge $Z$ and on the magnetic fields through which it propagates.
It is known\footnote{This comes from the overabundance of certain elements in the composition of galactic cosmic rays, which will be discussed in the next section} that lower energy cosmic rays, those which are thought to originate from 
within the galaxy,  must propagate an average distance of $\sim 1~$Mpc.
This implies that galactic cosmic rays diffuse through the galaxy and so arrive isotropically at the Earth.

At ultra-high energies, on the other hand, cosmic rays are most likely extragalactic in origin, as their Larmor radius exceeds the size of the galactic disk. 
The Larmor radius for a particle of charge $Z$ and with energy $E$ is 
\begin{equation}
\label{eq:LarmorRadius}
 r_{\text{Larmor}} = \frac{E}{ZeB} \approx \frac{110~\text{kpc}}{Z}\left(\frac{\mu\text{G}}{B}\right)\left(\frac{E}{10^{20}~\text{eV}}\right)
\end{equation}
From this equation it can be seen that the deflection of a cosmic ray in a given magnetic field is proportional to the energy of the cosmic ray. This implies that 
the cosmic ray sky should become more anisotropic with increasing energy if UHECRs come from discrete sources. 

The expected deflection in the Galactic magnetic field for UHECR protons with energies greater than $\simeq 10^{20}~$eV is a few degrees. The correlation between the arrival directions of UHECR and some manner 
of astrophysical object is then a question of statistics, which is highly limited at these energies by the low flux.
At the same time, the typical deflection of a UHECR will increase with increasing charge, washing out the anisotropy. 
These two facts make the isolation of cosmic ray sources dependent on the composition, the source density, and the number of observed UHECR events.
The second point, the number of UHECR sources in the sky, will be studied in chapter~\ref{CHAPTER:UHECR Source Statistics}, while the number of observed UHECR events is a strong motivation for experimental advancement in the field, which will be 
discussed in chapters~\ref{CHAPTER:UHECRobservation} and~\ref{CHAPTER:JEMEUSO}.

\section{Cosmic Ray Composition}
\begin{figure}[h]
  \centering
  \includegraphics[angle=270, width=1.0\textwidth]{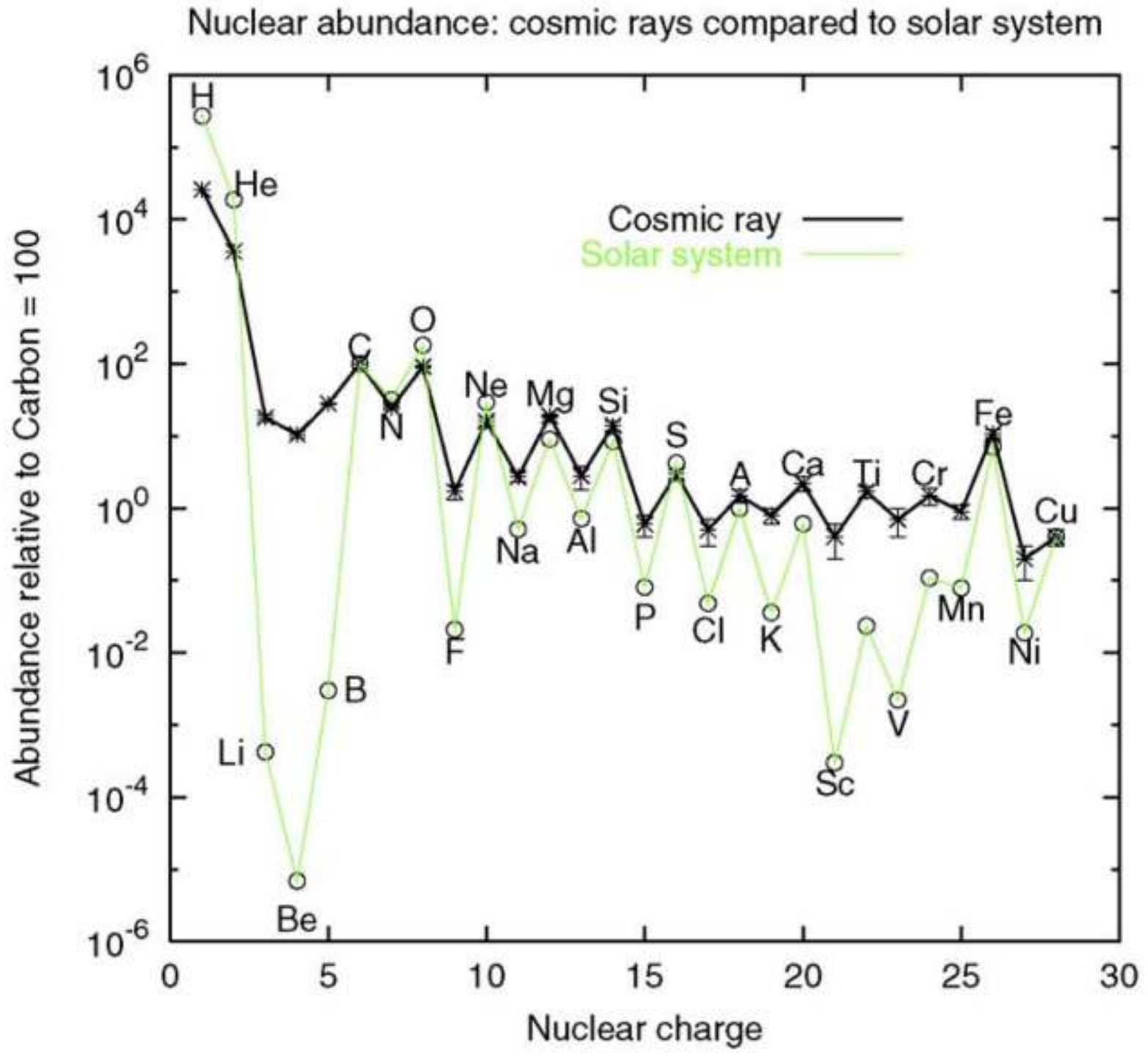}
   \caption[Cosmic Ray Composition]{\label{fig:CRcomposition} A plot of the cosmic ray composition by nuclear species, outside the heliosphere, for $E < 10^{14}~$eV cosmic rays, taken from ref.~\cite{Gaisser200698}.
The shown nuclear abundances are from \cite{1990A&A...233...96E}, and the proton and helium abundances are from 
\cite{Alcaraz200027,0004-637X-545-2-1135} and \cite{Alcaraz2000193}.
}
 \end{figure}

After the cosmic ray spectrum, the next property which gives us information on cosmic ray phenomena is their composition.
The composition of cosmic rays can be directly measured up to energies of $\simeq 100~$TeV, and this is shown in \fig\ref{fig:CRcomposition}, compared to the abundance of nuclei in the 
solar system. As can be seen, the abundance of many elements in the measured cosmic ray flux matches well with their abundance in the solar system.

For some elements, however, such as lithium, beryllium, and boron, the abundance in cosmic rays is several orders of magnitude higher than in the solar system.
This can be explained by the phenomena of ``primary'' versus ``secondary'' cosmic rays.
Primary cosmic rays are those particles which are accelerated by some astrophysical source, whereas secondary cosmic rays are created by the spallation of
primary cosmic rays. Spallation is the emission of a small number of nucleons as the result of a heavier nucleus being hit by a high-energy particle. This process is a natural result of 
both low energy interactions with the Galactic medium, and GZK-type energy loss mechanisms like photo-disintegration of nuclei.
The lithium, beryllium, and boron overabundance can be easily explained by the spallation of carbon and oxygen if cosmic rays transverse at least $\simeq 5~$g/cm$^{2}$ of matter. 
The same mechanism can also account for the overabundance of elements below iron in \fig\ref{fig:CRcomposition}. 

Further information about cosmic rays can be gleamed from the composition by looking at the ratio of unstable to stable isotopes in the cosmic ray flux.
One example is the ratio $R_{10}$ of unstable $^{10}$Be to $^{9}$Be. The two isotopes of beryllium are known to be produced in roughly equal amounts by spallation, and 
the half-life of $^{10}$Be is known to be $1.5~10^{6}~$years. The measurement of  $R_{10}$ can therefore constrain the escape time of cosmic rays in the Galaxy, giving a
value of $\tau_{\text{escape}} \approx 2~10^{7}~$years. This result can be used to estimate the density of matter through which the cosmic rays propagate, and a comparison to the matter density
in the Galactic disk and halo shows that Galactic cosmic rays must spend a significant fraction of this time in the halo. 

These estimated values of $\tau_{\text{escape}}$ imply that cosmic ray nuclei must spend a significant length of time diffusing in low-density regions of the galaxy. 
The ratio of primary to secondary cosmic rays is also known to be energy dependent, which in turn implies that $\tau_{\text{escape}}$ decreases with increasing energy, implying energy dependent diffusion of cosmic rays
in the galaxy. This is expected theoretically, as the Larmor radius, as well as the diffusion coefficient, of the cosmic rays will increase with energy.
Above 100 TeV a direct measurement of the cosmic ray composition is more difficult, and in this energy range the composition is derived from the 
observation of the depth of shower maximum of the extensive air shower created by the primary cosmic ray, which will be discussed in chapter~\ref{CHAPTER:UHECRobservation}. 

\begin{figure}[h]
  \centering
  \includegraphics[angle=270, width=1.0\textwidth]{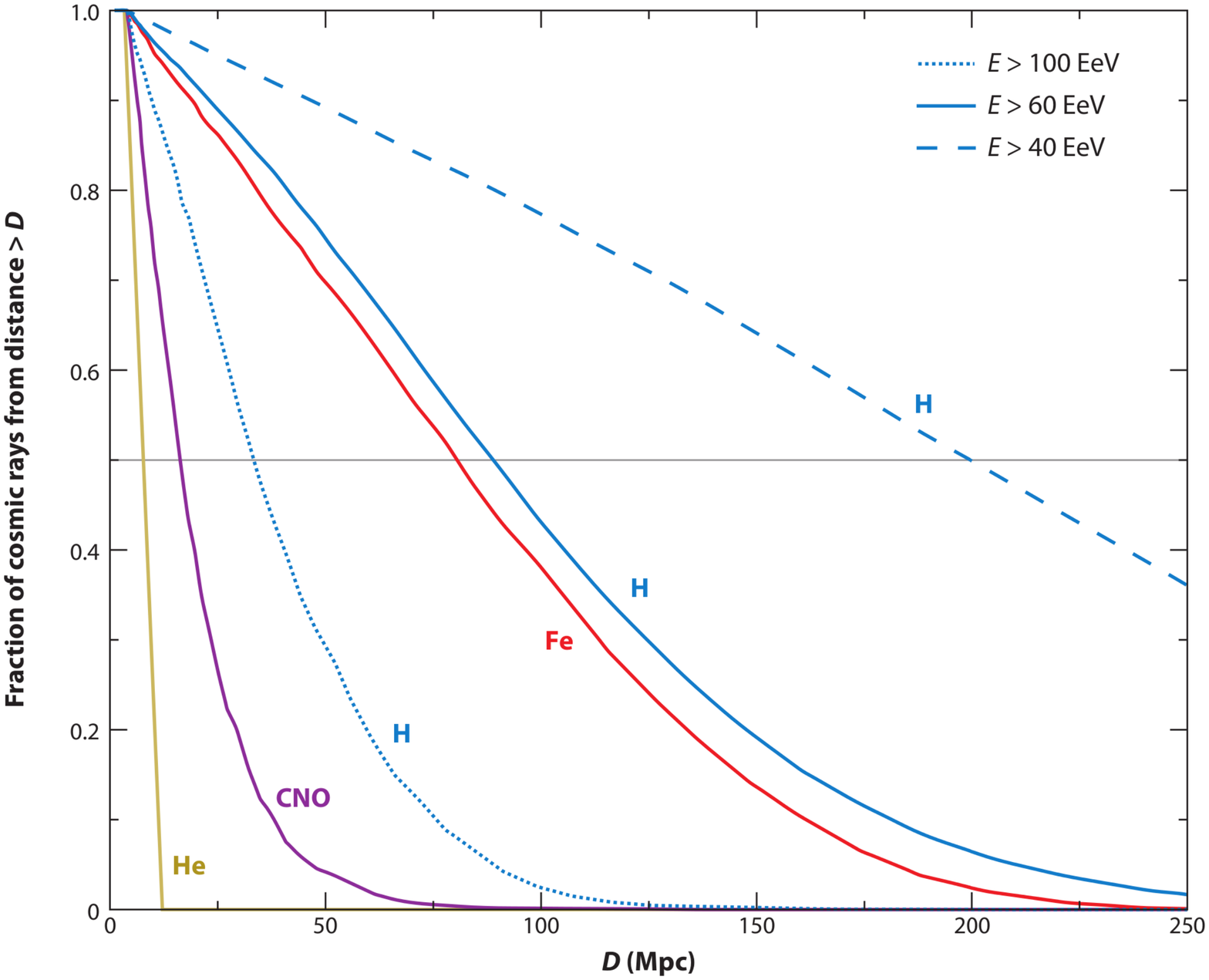}
   \caption[Cosmic Ray Extragalactic Propagation]{\label{fig:UHECRlosslength} A plot of the fraction of cosmic rays which survive propagation over a distance greater than $D$, taken from ref.~\cite{2011ARA&A..49..119K}. The fraction is
shown for protons above 40, 60, and $100~10^{18}~$eV, and for He, CNO, and Fe above $60~10^{18}~$eV. The gray line shows the distance from which 50\% of a given species can originate for a given atomic mass and energy.
Above $\simeq 60~10^{18}~$eV only protons and iron survive propagation over distances greater than $\simeq 50~$Mpc.}
 \end{figure}

One interesting point regarding the composition at the highest energies can be understood by considering the previously mentioned energy loss mechanisms. 
The horizon structure at UHECR energies is shown in \fig\ref{fig:UHECRlosslength} as the percentage of 
cosmic rays of a given nuclear species which survive propagation over a distance greater than $D$. 
This horizon is due to the energy and atomic number dependence of the interaction cross sections for processes such as the giant-dipole resonance, photo-pion production on CMB photons, and interactions with the \gls{IR}/\gls{UV}/optical photon background, as shown for protons
in \fig\ref{fig:energyLosses}.
Due to these energy losses, at energies above $60~10^{18}~$eV only protons and nuclei with an atomic number near iron survive a propagation distance
of greater than $50~$Mpc. This implies that the entirety of the UHECR flux at the highest energies is dominated by some combination of protons or nuclei near iron.

Observations of cosmic rays with energies from just above the knee up to the ankle show a trend from a light composition (protons) at the knee to a heavier composition up to $\sim 10^{17}~$eV \cite{Bluemer:2009zf}. This follows the 
general expectation that the knee is created by the end of the major Galactic cosmic ray sources and that the maximum acceleration energy is proportional to the cosmic ray charge.
Just after the ankle, the composition, as observed by both Auger \cite{Abraham:2010mj} and HiRes \cite{HiResComp2010}, 
appears to reverse back towards a lighter composition.

Above $10^{19}~$eV, however, it appears that the UHECR composition again changes back towards a heavy composition, as measured by both the Auger average
depth of shower maximum $\langle X_{\text{max}}\rangle$ and the root-mean-square of $X_{\text{max}}$ \cite{1742-6596-375-5-052003}.  
This shift in the depth of shower maximum could, however, also be due to a change in particle interactions at center-of-mass energies above 100 TeV.
At the same time, the measurement of $\langle X_{\text{max}}\rangle$ and $X_{\text{max}}$ root-mean-square by HiRes and Telescope array are consistent with a proton composition at the highest energies \cite{2013EPJWC..5304005T}. 
This potential inconsistency is unclear, as the HiRes and TA results are compatible with both heavy nuclei and protons,
difficult to resolve, due the use of different analysis techniques and experimental methods by the two collaborations, and a point of ongoing investigation \cite{Barcikowski:2013nfa}.

\section{Cosmic Ray Sources and Acceleration Mechanisms}

\begin{figure}[h]
  \centering
  \includegraphics[angle=270, width=1.0\textwidth]{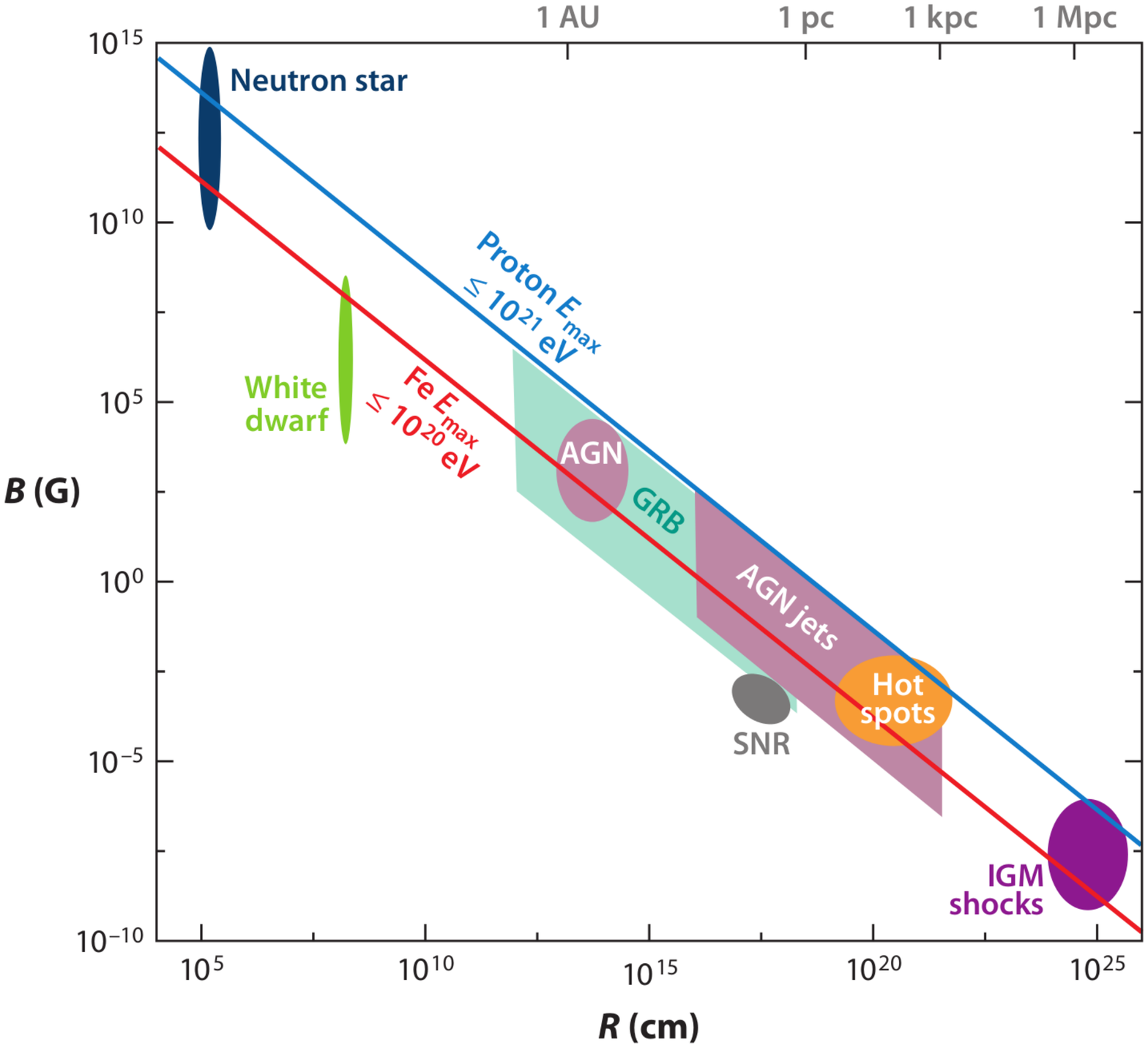}
   \caption[Hillas Diagram]{\label{fig:HillasDiagram} A Hillas diagram showing the possible classes of astrophysical objects versus their size and magnetic field strength taken from ref.~\cite{2011ARA&A..49..119K}
Above the dark blue line is the region of parameters which could confine protons above $E_{\text{max}} = 10^{21}~$eV, 
while above the red line are those combinations of parameters which would allow acceleration of iron up to  $E_{\text{max}} = 10^{20}~$eV.
The region occupied by each source type indicates the uncertainties in their parameters. 
The abbreviations in the diagram are
\begin{inparaenum}[i\upshape)]
 \item AGN: Active Galactic Nuclei,
 \item GRB: Gamma-Ray Burst,
 \item IGM: InterGalactic Medium, and
 \item SNR: Supernova Remnant.
\end{inparaenum}
As can quickly be seen in this plot, the types of objects which could accelerate cosmic rays up to ultra-high energies is limited.
}
\end{figure}

The simplest and yet largest question in cosmic ray physics is ``Where do they come from?,'' and this question has yet to be answered.
Any electric field can easily accelerate charged particles, but large-scale electric fields are limited in the universe due to the presence of highly conductive astrophysical plasmas. 
Magnetic fields, on the other hand are ever present in the universe. 
At the low-energy side of the cosmic ray spectrum, cosmic rays can originate from any of a large number of bodies which possess a spatially or temporally varying magnetic field, the sun being one such example, 
and the main question in this energy region is that of total power cosmic ray power.    
In the UHECR energy region, however, the possible acceleration mechanisms are more constrained, due to the energy scale involved. The two best understood mechanisms which have been proposed 
are shock acceleration and unipolar induction. There are additional models which have been discussed in the literature (see section 5.3 of ref.~\cite{2011ARA&A..49..119K} for examples), but which will not be discussed here. 

The basic principle behind shock acceleration (also known as first-order Fermi acceleration, as the second-order version is originally due to Enrico Fermi)
is the transfer of energy from macroscopic motion to microscopic particles through their interaction with magnetic inhomogeneities.
In second-order Fermi acceleration the acceleration is due to the random velocities of magnetic scattering centers and leads to an energy gain of $\Delta E/E \propto \beta^{2}$, where $\beta$ is the average velocity of the scattering centers \cite{Fermi:1949ee}.
This is in contrast to first-order Fermi acceleration, in which the acceleration is due to a coherent shock wave such that the accelerated particles gain energy as they bounce back and forth. This results is an energy gain of $\Delta E/E \propto \beta$
\cite{1978ApJ...221L..29B,Bell:1978zc}.
Such shock waves are frequent in the universe, arising wherever supersonic ejecta interact with the interstellar medium. This includes supernova remnants, gamma ray burst shocks, active galactic nuclei jets, and gravitational accretion shocks.

Supernova remnants bear special mention in the area of galactic cosmic rays, and were first discussed as extragalactic sources of cosmic rays by Baade and Zwicky \cite{Baade01051934,1934PhRv...46...76B,}. The energy density of cosmic rays in the galaxy, $\simeq1~$eV /cm$^{3}$, is the same order of magnitude as the magnetic field energy density and thermal gas energy density.
Given the typical cosmic ray residence time in the Galaxy, this gives a cosmic ray power in the Galaxy of approximately $3~10^{40}~$erg/s, which can be compared to the power emitted by supernovae in the galaxy of $3~10^{41}~$erg/s, given the expected supernova rate.
This implies that supernovae alone could maintain the cosmic ray population provided that about 10\% of their kinetic energy is converted into cosmic rays, and
supernova shock acceleration has been shown to fit the spectrum up to $10^{15}~$eV, that is up to the knee in the cosmic ray spectrum.

Unipolar induction, on the other hand, is due to bodies such as neutron stars or other relativistic magnetic rotators, such as magnetized black holes, which lose rotational energy in jets (see for example, refs.~\cite{Shapiro:1983du} and \cite{Ginzburg:1990sk}). 
These rapidly rotating magnetized bodies create relativistic winds which, combined with the magnetic field, produce an electric field $E = - v \times B/c$, where $v$ and $B$ are the velocity and magnetic field of the out-flowing plasma. 
This creates a large voltage drop, which can accelerate particles to high energy.

The basic ability of an accelerator to accelerate particles to a given energy is limited by the ability of the accelerating object to contain the particles inside the acceleration region. 
This can be parametrized by the Larmor radius, given in \eq\ref{eq:LarmorRadius}), which sets the scale for 
\begin{equation}
\label{eq:CRsourceEmax}
E_{\max} \simeq 10^{18}~\text{eV} \left(\frac{B}{1~\mu\text{G}}\right)\left(\frac{R}{1~\text{kpc}}\right)
\end{equation}
Based on \eq\ref{eq:CRsourceEmax}), the astrophysical objects which could possibly accelerate charged particles up to UHECR energies are shown in \fig\ref{fig:HillasDiagram}, the so-called ``Hillas diagram''.
In the Hillas diagram, the known classes of astrophysical objects are plotted versus their radius and magnetic field \cite{1984ARA&A..22..425H}. The size of the region for each object in \fig\ref{fig:HillasDiagram} accounts for the 
uncertainty on $B$ and $R$ for that class of object. 

Because the Larmor radius is proportional to the charge of the accelerated particle, the ability of a given accelerator to reach a certain energy depends on the nuclei being accelerated.
The (lower) red line in \fig\ref{fig:HillasDiagram} indicates the combinations of magnetic field and size, according to \eq\ref{eq:CRsourceEmax}), which are capable of accelerating
iron nuclei up to a maximum energy of $10^{20}~$eV. The blue (upper) line shows the same for protons at $10^{21}~$eV.  
As can be seen, there are only a limited number of astrophysical objects which could potentially accelerate protons up to the highest energies. 

This diagram also does not take into account the acceleration efficiency of the source(s) or corrections due to relativistic effects. 
Accounting for acceleration efficiency will decrease the actual reach of an accelerator, bringing down $E_{\text{max}}$ further, while relativistic effects could increase an accelerator's reach. 
Other details of the acceleration also come into play, such as the required acceleration time compared to the average age of a given class of objects and the energy loss time. 

In addition to the acceleration mechanisms discussed above, alternative non-acceleration scenarios have been proposed. In these ``top-down'' models the highest energy cosmic rays are theorized to be the decay products of some super-heavy ``particle''.
The super-heavy candidate ranges from dark matter~\cite{Berezinsky:1997hy}, cryptons~\cite{Ellis:2004cj, Ellis:2005jc}, or topological defects~\cite{Hill:1982iq,Berezinsky:1998ft}. 
Top-down models as a class include a postulation of new particle physics and generally predict a high flux of gamma rays at UHECR energies.
Due to these predictions, the non-observation of UHECR gamma-rays by Auger and TA has put strong constrains on this type of UHECR source.
Results from Auger has placed an upper limit on the photon fraction in the UHECR flux, above $10^{19}~$eV, of less than 11.7\% using hybrid events \cite{Abraham:2009qb} and less than 2.0\% using surface detector events \cite{Aglietta:2007yx} (both at 95\% c.l.).  
The corresponding upper limit from Telescope Array is 6.2\% photons above $10^{19}~$eV \cite{Abu-Zayyad:2013dii}.

\section{Phenomenology of UHE Cosmic Rays}

We now turn to a brief overview of the relationship between the composition, sources, and spectra of cosmic rays, and the open questions in the field.
In the UHECR energy range several different astrophysical models have been proposed which account for the different features of the spectrum and composition at the highest energies.
In mixed-composition models and models dominated by iron, the ankle is the signature of the transition from Galactic to extragalactic cosmic rays. In these types of models, the composition would be expected to 
be heavy up to the ankle, where it would transition to a light composition, followed by a re-transition to a heavy composition due to the charge dependence of the 
maximum energy of the sources.

In the so-called ``dip-model'', on the other hand, the energy break points and shape of the UHECR spectrum is explained by GZK processes.
The ankle structure is due to $e^{+}e^{-}$ pair production interaction of UHECR protons on the CMB at around $5~10^{18}~$eV. The cut-off is then taken to be the result of photo-pion production at around $4~10^{19}~$eV.
If compared to the HiRes spectrum measurement, dip models are consistent,  within the energy scale uncertainty of HiRes, with theoretical calculations of the GZK cutoff energy \cite{2013EPJWC..5301003B}. The dip model is also consistent
with the Telescope Array and HiRes observation of a light composition above $10^{18}~$eV.  
At this moment, however, the observed GZK-like feature in the UHECR spectrum does not distinguish between propagation energy loss (i.e.\ the true GZK effect) and source maximum energy.

At the same time, the continued isotropy of the UHECR sky is a ongoing puzzle. 
Both Auger \cite{Kampert:2012vh} and Telescope Array \cite{AbuZayyad:2012hv} have reported a small, seemingly growing, number of correlated UHECR events, but no clear signal of anisotropy or correlation has yet been established with certainty. 
The expected correlation at a given energy depends on several factors, the composition being one example.
If UHECR are primarily protons, then the UHECR should display some anisotropy above $6~10^{19}~$eV as the \gls{UHE} protons would not be deflected as much as heavy nuclei by magnetic fields.
On the other hand, if the composition is indeed dominated by iron at the highest energies, as reported by Auger, then any anisotropy could be washed out by galactic magnetic fields. 
This could also be true if the intergalactic magnetic field is stronger than expected.
These questions will be discussed further in chapters~\ref{CHAPTER:UHECR Source Maximum Energy and the UHECR Spectrum} 
and~\ref{CHAPTER:UHECR Source Statistics}, which present some original contributions to studies of UHECR phenomenology. 

These questions are influenced by ongoing experimental and/or instrumental problems, and this leads to several key areas for future work. One such
area is the energy scale uncertainty of both Auger and Telescope Array. This uncertainty complicates energy cuts for comparative anisotropy analysis and for
large-scale anisotropy studies, and also leads to ambiguity in associating spectral features to physical phenomena.
This transition is an important element in models of the UHECR spectrum, and the transition should be signaled by a composition change from heavy to light, some manner of deformation in the spectrum, and an energy dependent anisotropy \cite{2013EPJWC..5301003B}.

The study of the anisotropy, and the associated search for the sources of UHECRs, would also be aided by having a single experiment with full-sky coverage. This would introduce the minimal exposure distortion in the correlation analysis, and remove the 
need for additional assumptions about the unobserved portion of the sky when performing spherical harmonic or multipole analysis of the anisotropy. 
Above all, the hints of anisotropy observed by Auger and Telescope array will be clarified with the observation of more UHECR events, and it is generally held that 
an order of magnitude increase in statistics is needed to find anisotropy and characterize its cause. These points are the primary motivations for 
the next generation UHECR experiment JEM-EUSO, which will be introduced in chapter~\ref{CHAPTER:JEMEUSO}.
 
In addition to the observation of UHECRs themselves, multimessenger information can also be used to constrain UHECR phenomenology. 
One previously mentioned example is the limits placed on top-down models by the non-observation of ultra-high energy gamma rays and neutrinos. 
Some number of ultra-high energy gamma rays and neutrinos are also expected as the result of pion decay from GZK interactions. 

Transient Large Luminosity events (\glspl{GRB}, etc.) may account for anisotropy for larger source densities. For these, source densities and transient time profiles can be used to constrain source parameters \cite{2013EPJWC..5306008T}.
The distribution of parameters of UHECR source candidates (AGN black hole masses, for example) affects predictions of UHECR observables \cite{2013EPJWC..5306003K}, and the observation of gamma rays from Blazars may 
\emph{require} UHECR acceleration in AGN \cite{2013EPJWC..5306006K}. The observation of neutrinos with energies in the range of $10^{15}$ to $10^{21}~$eV can also strongly constrain models for the origin of UHECRs \cite{2013EPJWC..5301013S}.

The current state of UHECR physics can be summarized by a series of open questions. First, is the observed cut-off in the UHECR spectrum around $10^{19}~$eV due to the particle energy loss, i.e.\ the GZK effect, or to
the maximum energy of the UHECR accelerators?
The next question is whether the composition does in fact change above $10^{19}~$eV, as reported by Auger, or if the observed depth of shower maximum distributions argue for a change in particle interaction at these high energies.
A continued question from an astrophysical perspective is at what energy does the cosmic ray flux transition from being of galactic to extra-galactic origin, and finally, the greatest question of the lot still remains: ``What are the sources of 
UHECR cosmic rays?''

%
%
%
    \printbibliography[heading=subbibliography]
    \end{refsection}
    \begin{refsection}
  \chapter{Observation of Ultra-High Energy Cosmic Rays}
    \label{CHAPTER:UHECRobservation}
    \label{chapt:ObservationOfUHECR}
This chapter will discuss the actual observation of UHECRs. 
The first sections present the physics of extensive air showers and a very basic analytic derivation of some of their properties.
This is followed by a discussion of air shower simulations and 
the interplay between extensive air shower physics and data from accelerator experiments.

After that, a very brief overview of detection techniques and the two main operating UHECR observatories, the Pierre Auger Observatory and the Telescope Array, will be presented.
This is followed by a discussion of some of the experimental challenges encountered in the field,  mostly related to understanding the results of the main experiments.

\section{The Physics of Extensive Air Showers}

\begin{figure}[ht!]
  \centering
 \subfigure[$\gamma$-ray]{ \label{fig:EASpictures:photon} \includegraphics[angle=0, width=0.48\textwidth]{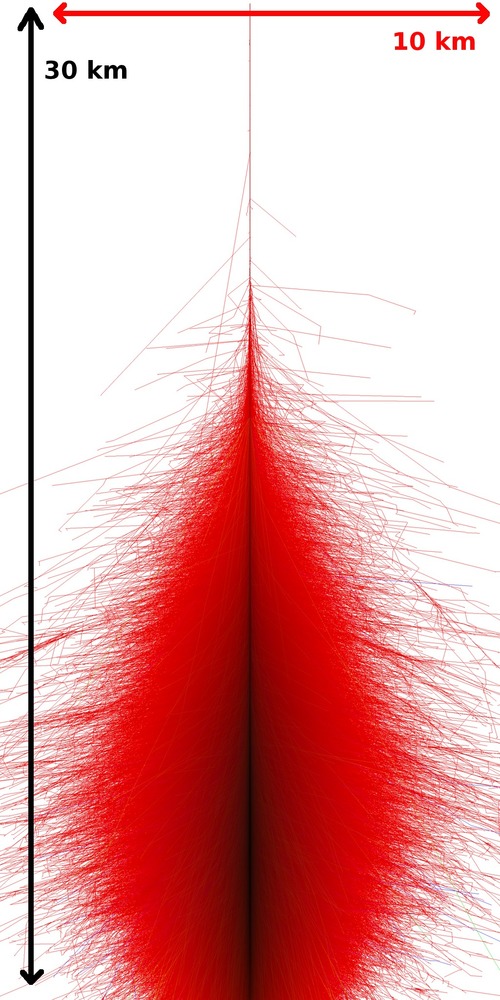}}
 \subfigure[proton]{\label{fig:EASpictures:proton} \includegraphics[angle=0, width=0.48\textwidth]{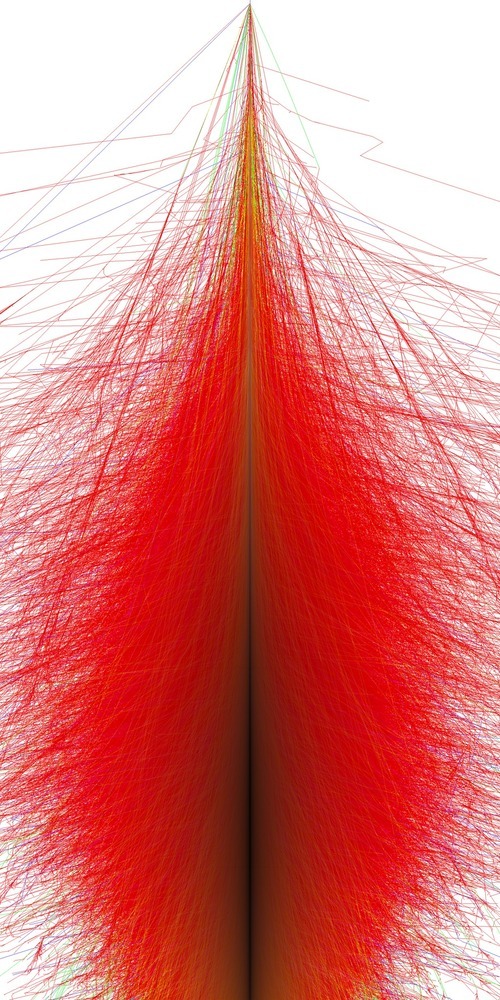}}
 \caption[Extensive Air Showers Simulated by CORSIKA]{\label{fig:EASpictures} 
Two pictures of EAS simulated by CORSIKA \cite{CorsikaShowerImages, CORSIKA}. \fig\ref{fig:EASpictures:photon} shows an EAS generated by a $10^{15}~$eV photon, while \fig\ref{fig:EASpictures:proton} shows a shower created by a 
proton of the same energy. In both simulations, the primary cosmic ray arrives at a zenith angle of $0^{\circ}$, and the first interaction is fixed at a height of 30 km. 
The range of the vertical axis is from 0 to 30.1 km, and the X/Y axis range is $\pm 5~$km around the shower core.
The color of the particle tracks indicates type:  red = electrons, positrons, gammas; green = muons; and blue = hadrons.
The color scale is logarithmic, with dark color corresponding to high track density.
Only particles with kinetic energies above 0.1 MeV (for $e^{\pm}$ and $\gamma$) and 0.1 GeV ($\mu$ and hadrons) are shown, and neutrinos are not plotted.
As can be seen, the maximum of the photon shower is deeper in the atmosphere than the proton shower, and its core is smaller in radius. }
 \end{figure}

Above approximately $10^{15}$ eV, the flux of cosmic rays is so low, on the order of one particle/m$^{2}$/year, that direct detection is no longer feasible, as the probability of having an event in a typical detector is too low.
At such energies, the primary cosmic ray can be detected through its interaction with the Earth's atmosphere. The huge energy of UHECR cosmic rays, released on their impact with a nucleus in the air, generates 
a cascade of secondary particles, known as an Extensive Air Shower (EAS).
The properties of the primary cosmic ray, namely its type, mass, and energy can be inferred from the properties of the generated air shower.
Extensive air showers can be characterized by several parameters:
\begin{itemize}
 \item  $N_{\text{max}}$, the maximum size, in number of particles, of the shower. The size can be divided into components, such as the 
        number of muons $N_{\mu}$ or electrons $N_{e}$ at the shower maximum.
 \item $X_{\text{max}}$, the depth in the atmosphere of the shower maximum. 
 \item  $\Lambda$, the elongation rate, that is the rate of increase of $X_{\text{max}}$ with the energy of the primary cosmic ray.
 \item $N_{e}(X)$, the mean longitudinal shower size profile, in other words the number of charged particles as a function of the shower depth.
 \item The lateral particle distribution  $dN_{e}/rdrd\varphi$, the distribution of particles in the shower as a function of angle and radial distance to the shower core.
\end{itemize}
A detailed understanding and modeling of the development of EAS are complicated by the large number of particles involved, and, in the case of a hadronic primary particle, the lack of an analytical description of QCD.
These two factors can be treated using detailed numerical simulations, but such simulations involve a large extrapolation of interaction cross sections and particle production mechanisms to extremely high energies 
where little or no data is available.
The basic properties of EAS can be understood, however, by using simple arguments, starting with the properties of purely electromagnetic showers. The EAS models derived below are due to Heitler \cite{heitlerModel},
Matthews \cite{Matthews:2005sd}, and the discussion on EAS properties in \cite{Bluemer:2009zf}.

\subsection{Electromagnetic Showers}

A simple model for electromagnetic cascades which reproduces well the basic characteristics of EAS was developed by W. Heitler in the 1950s \cite{heitlerModel}.
The Heitler model describes EAS which are created by an electromagnetic primary ($\gamma$-rays), and the evolution of an \gls{EM} EAS is controlled by
the processes which produce additional particles: bremsstrahlung and pair production.

Pair production is the creation of an electron-positron pair from an incident photon in the coulomb field of a nucleus ($\gamma \gamma \rightarrow e^{+}e^{-}$).
This interaction is threshold dependent and requires an energy of greater than $2m_{e}c^{2}$.
Bremsstrahlung, on the other hand, occurs when a incident charged particle is deflected in the coulomb field of a nucleus in the material through which it is passing ($\gamma e^{\pm}  \rightarrow \gamma e^{\pm}$). Acceleration of a charge
produces radiation, and the charged particle will lose an amount of energy proportional to $(E/mc^{2})^{4}$ in the creation of photons. 
The $1/m^{4}$ dependence of bremsstrahlung makes it an extremely important energy loss mechanism for electrons and positrons, but less so for heavier charged particles such as muons, pions, and protons.

The electrons and positrons in the EAS lose energy through bremsstrahlung and ionization, and the total energy loss can be written as
\begin{equation}
\label{eq:TotalEMeloss}
 \frac{dE}{dX} = -\frac{E}{\lambda_{\text{r}}} - \left.\frac{dE}{dX}\right|_{\text{ion}}
\end{equation}
The first term is the radiative energy loss due to bremsstrahlung, which feeds the shower by leading to the creation of new photons.
The second term in \eq\ref{eq:TotalEMeloss}) is the ionization energy loss, which is given by the Bethe-Bloch formula\footnote{This is an approximate non-relativistic form for the case of a heavy 
charged particle, and it does not include corrections for electron indistinguishably, the shell correction for the motion of atomic electrons, and higher order terms in the perturbative expansion.}:
\begin{equation}
\label{eq:Bethe-Bloch}
  \left.\frac{dE}{dX}\right|_{\text{ion}} = \frac{4\pi n z^{2} e^{4}}{m_{e}v^{2}}\left[\ln\frac{2m_{e}v^{2}}{I[1-\left(v/c\right)^{2}]}  -\left(\frac{v}{c}\right)^{2} \right]
\end{equation}
where $n$ is the number of electrons per cm$^{3}$ in the absorbing material, $m_{e}$ is the electron mass, $ze$ is the charge of the particle ($z = 1$ in this case), $v$ is the velocity of the particle,
and $I$ is the mean excitation potential of the atoms of the absorber ($I \approx Z \times 10~\text{eV}$). 
The ionization energy loss transfers energy from the shower electrons and positrons to the atmosphere.

As the average energy of the charged particles in the EAS decreases, the relative importance of ionization energy loss (which transfers the shower energy to the atmosphere) and bremsstrahlung (which adds particles to the shower)
changes. The energy at which the energy loss due to ionization and bremsstrahlung are equal is known as the critical energy $E_{\text{c}}$.
The critical energy depends on the properties of the absorbing material, and is given approximately by the relation:
\begin{equation}
 E_{\text{c}} \approx \frac{600 \text{MeV}}{Z}
\end{equation}
where $Z$ is the charge number of the atoms in the absorbing material. In air the critical energy is $E_{\text{c}} = 85~$MeV.

\begin{figure}[h]
  \centering
  \includegraphics[angle=0, width=1.0\textwidth]{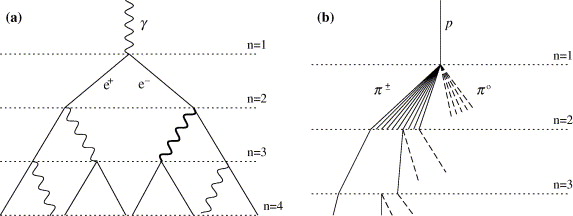}
   \caption[Diagram of the Heitler Model]{\label{fig:HeitlerModeldiagram} A diagram of the Heitler model (a) and a hadronic extension (b), taken from ref.~\cite{Matthews:2005sd}.
In Fig.$~$(a), an EM shower is shown. At each interaction length each particle in the shower is assumed to split into 2 new particles, with each electron emitting a photon through Bremsstrahlung, and each photon producing a $e^{+}e^{-}$ pair.
In Fig.$~$(b), a similar model is shown for a hadronic shower. At each interaction length a number $N_{\text{ch}}$ of charged pions and a number $\frac{1}{2}N_{\text{ch}}$ neutral pions are created. 
$\pi^{0}$ are assumed to decay to $\gamma\gamma$ pairs, creating EM sub-showers. $\pi^{\pm}$ are assumed to continue to split until the average pion energy reaches $E_{c}^{\pi}$, at which point all $\pi^{\pm}$ decay to muons and neutrinos. 
}
 \end{figure}

In the Heitler model, a simple picture of an EAS is created by assuming that the electrons and positrons created in the initial interaction undergo repeated 2-body splitting, in either single photon bremsstrahlung or $e^{+}e^{-}$ pair production interactions.
This is shown in part (a) of \fig\ref{fig:HeitlerModeldiagram}.  
On average, each particle in the shower is assumed to undergo an interaction after traveling a fixed distance $d$ related to the radiation length $\lambda_{\text{r}}$ 
as $d=\lambda_{\text{r}}\ln 2$ (the radiation length in air is $\lambda^{\text{air}}_{\text{r}} \approx 37~\text{g/cm}^{2}$).
In this definition $d$ is the average distance over which an $e^{\pm}$ loses one half of its energy by radiation.

After $n$ interactions there are $2^{n}$ total particles in the shower. The distance traveled by the shower is then 
$x = n\lambda_{\text{r}}\ln 2$, so that the total number of particles is $N = e^{x/\lambda_{\text{r}}}$. The multiplication of particles is assumed to stop when the average 
particle energy, given by $E_{o}/N$, is too low for continued pair production or bremsstrahlung. 
This is assumed to be equal to the critical energy $E_{\text{c}}$.

Using these assumptions, the maximum size of the shower is simply given by the relation:
\begin{equation}
\label{eq:EAS:EM:MaxSize}
 N_{\text{max}} = E_{o}/E_{\text{c}}
\end{equation}
The penetration depth at which the shower reaches maximum size is given by the number of interaction lengths needed for the average energy per particle to reach $E_{\text{c}}$, beyond which point no further particles are produced (by assumption).
Using \eq\ref{eq:EAS:EM:MaxSize}), the depth of shower maximum is
\begin{equation}
\label{eq:EAS:EM:depthMax}
X_{\text{max}}^{\text{em}} = \lambda_{\text{r}}\ln\left(E_{o}/E_{\text{c}}\right)
\end{equation}
The radiation length, and thus the depth of shower maximum, are most conveniently measured as an absorber thickness in g/cm$^{2}$, which accounts for the density profile of the atmosphere.
The shower depth X is then related to the distance of the shower inside the atmosphere $x$ and the density of the atmosphere $\rho$ as $X = \int\rho(x) dx$.

The elongation rate is defined as the rate of increase of $X_{\text{max}}$ per decade of primary particle energy:
\begin{equation}
 \Lambda = \frac{dX_{\text{max}}}{d\log_{10}E_{o}}
\end{equation}
and using \eq\ref{eq:EAS:EM:depthMax}) the elongation rate for electromagnetic EAS in the Heitler model is $\Lambda^{\text{em}} = 2.3 \lambda_{\text{r}} = 85 \text{g}/\text{cm}^{2}/$decade.

This simple model accounts well for two basic features of electromagnetic showers:
\begin{inparaenum}[i\upshape)]
 \item The maximum number of electrons, positrons, and photons in the shower is proportional to $E_{o}$, and
 \item the depth of shower maximum is proportional to the logarithm of $E_{o}$ and scales at a rate of 85 g/cm$^{2}$ per decade of $E_{o}$.
\end{inparaenum}
The longitudinal shower profile of an electromagnetic EAS can be calculated from cascade theory, and a related parametrization due to Gaisser and Hillas \cite{GaisserHillasFunc} is often used to fit measured shower profiles 
\begin{equation}
\label{eq:GaisserHillasEq}
 N(x) = N_{\text{max}}\left( \frac{X -X_{0}}{X_{\text{max}} - X_{0}}\right)^{\left(X_{\text{max}} - X\right)/\upsilon} \text{exp}\left[ \frac{X_{\text{max}} -X}{\upsilon}\right]
\end{equation}
where $X_{0}$ and $\upsilon$ are shower shape parameters.

The dependence of the particle density on the distance to the shower core, i.e.\ the lateral distribution, is determined mainly by the multiple Coulomb scattering of electrons.  
Detailed calculations of the lateral shower profile by Nishimura and Kamata were parametrized by Greisen in the so-called \acrshort{NKG} function \cite{Kamata01021958,NKGfunc}:
\begin{equation}
\label{eq:NKGfunct}
 \frac{dN_{e}}{rdrd\varphi} = C(s)N_{e}(X)\left(\frac{r}{r_{1}}\right)^{s-2} \left(1 + \frac{r}{r_{1}}\right)^{s-4.5}
\end{equation}
where $s$ is the shower age parameter (often defined as $s \approx 3X/[X+2X_{\text{max}}]$), $C(s) = \Gamma(4.5 - s)/[2\pi r_{1}^{2}\Gamma(s)\Gamma(4.5-2s)]$ is a normalization constant.
The quantity $r_{1}$ is the Moli\`{e}re radius 
$r_{1} \propto \lambda_{\text{r}}/E_{\text{c}} \approx 9.3~$g/cm$^{2}$.  The lateral density of the shower
depends on the air density, due to the dependence of the lateral shower profile on the Moli\`{e}re radius.

The Heitler model predicts that the number of electrons approaches $N_{e} \approx \frac{2}{3}N_{\text{max}}$, which overestimates the true ratio of electrons and positrons to photons. 
This is primarily because the model does not account for multiple photons being radiated through bremsstrahlung or for the range out of electrons and positrons.
These details of shower development past its maximum require a more careful treatment of particle production and energy loss than is provided by the Heitler model.
To account for these shortcomings, the number of electrons and positrons can be corrected by some factor $g$, so that the 
electron size is related to the overall shower size as $N_{e} = N_{\text{max}}/g$.
A comparison to simulations shows that the actual correction factor is $g=10$, so that the number of electrons (as an order of magnitude estimate) is given by $N_{e} = N_{\text{max}}/10$ \cite{Matthews:2005sd}.

Two further effects in UHE electromagnetic showers should be mentioned before moving to hadronic EAS. The first is the so-called \gls{LPM} effect. The LPM effect suppresses particle production in certain 
kinetic regions due to the coherent addition of the interactions of photons and electrons when the interaction length is comparable to the separation between subsequent interactions. This effect becomes important above $10^{18}~$eV
and increases shower-to-shower fluctuations while pushing $X_{\text{max}}$ deeper into the atmosphere.

The second effect is that of geomagnetic pair production and bremsstrahlung, which is due to photons with energies above $3~10^{19}~$eV interacting with the magnetic field of the Earth. 
This causes a pre-shower in which the primary photon interacts high above the atmosphere, creating hundreds of simultaneous sub-showers. Due to the division of the primary particle energy among numerous sub-showers the LPM effect is not important and the 
superposition of the many lower-energy showers reduces the overall shower-to-shower fluctuations. The dependence of this geomagnetic pre-shower effect on the arrival direction allows a model-independent search of UHE photons (e.g.~\cite{Bertou2000121}).

\subsection{Hadronic Showers}
A model for hadronic showers, i.e.\ those EAS initiated by a hadronic primary cosmic ray, can be built with an approach similar to the Heitler model. 
The model present here is from J. Matthews~\cite{Matthews:2005sd}.
In this model, the atmosphere is considered in layers of fixed thickness $\lambda_{\text{I}}\ln2$, where $\lambda_{\text{I}}$ is the interaction length of strongly interacting particles. 
Here $\lambda_{\text{I}}$ will be assumed to be constant. This is a good approximation in the energy range of 10 to 1000 GeV, where for pions in air $\lambda_{\text{I}} \approx 120~$g/cm$^{2}$.

As they transverse each layer, the hadrons are assumed to interact, producing $N_{\text{ch}}$ charged pions and $\frac{1}{2}N_{\text{ch}}$ neutral pions. 
Each $\pi^{0}$ is considered to immediately decay, yielding two photons, which create electromagnetic showers. The $\pi^{\pm}$, on the other hand, are assumed to continue to the next interaction layer, where they interact. 
This processes continues until the average energy per pion decreases below some critical energy $E^{\pi}_{\text{c}}$, at which point all the charged pions are assumed to decay to muons and muon neutrinos.
The critical pion energy $E_{\text{c}}^{\pi}$ is a slowly decreasing function of the primary particle energy, passing from 30 GeV for a primary proton of $E_{o}=10^{14}~$eV to 10~GeV for $E_{o}=10^{17}~$eV.
That being said, a constant value of $E_{\text{c}}^{\pi} = 20~$GeV will be used during the rest of this discussion.

The multiplicity $N_{\text{ch}}$ of charged particles produced in hadron interactions is a slowly increasing function of the interaction energy in the laboratory frame, and grows as $E^{1/5}$ in $pp$ and $p\bar{p}$ interactions.
A useful working value of $N_{\text{ch}}$ is $N_{\text{ch}}=10$, which is the correct order of magnitude for pions with kinetic energies from $\sim 1~$GeV through $10~$TeV. This approximation is reasonable because the majority of pion
interactions within an EAS occur at energies of around 100~GeV, as opposed to higher energies\footnote{This is not true, however, for studies of some quantities, $X_{\text{max}}$ being an example, where the first interaction (at high energy) is more important}. 
For example, the center of mass energy of 250 GeV pions colliding with stationary air nuclei is 22 GeV. The mean $pp$ charged multiplicity at this energy is 
approximately 8. This implies that a value of $N_{\text{ch}}=10$ is reasonable, considering that it allows for multiple interactions of pions with air nuclei.


Using the Matthews model, we can then consider a primary cosmic ray proton entering the atmosphere with an energy $E_{o}$. Analogous to the electromagnetic Heitler model, the number of charged particles after $n$ interactions is 
$N_{\pi}=(N_{\text{ch}})^{n}$. The total energy of the charged pions is assumed to be $(2/3)^{n}E_{o}$, with the remaining one third of the energy going into electromagnetic showers through the decay of neutral pions. 
The average energy per charged pion is
\begin{equation}
\label{eq:EAS:HAD:pionE}
 E_{\pi} = \frac{E_{o}}{(\frac{3}{2} N_{\text{ch}})^{n}}
\end{equation}
and the number 
of interactions needed to reach $E_{\pi}=E_{\text{c}}^{\pi}$ is then
\begin{equation}
\label{eq:EAS:HAD:Interactions}
 n_{\text{c}} = \frac{\ln[E_{o}/E_{\text{c}}^{\pi}]}{\ln[\frac{3}{2}N_{\text{ch}}]} = 0.85 \log_{10}\left[E_{o}/E_{\text{c}}^{\pi}\right]
\end{equation}
which gives $n_{\text{c}}=6,7,8,9\ldots$ for $E_{o}=10^{17}, 10^{18},10^{19},~\text{and}~10^{20}~$eV.
The number of interaction lengths to reach $E_{\text{c}}^{\pi}$, \eq\ref{eq:EAS:HAD:Interactions}), is not highly sensitive to the above assumption for the value of $N_{\text{ch}}$, reducing by 1 if $N_{\text{ch}}=20$ and $E_{o}<10^{16}~$eV, for example. 

The energy of the hadronic primary is divided between $N_{\pi}$ pions and $N_{\text{max}}$ electromagnetic particles in sub-showers, and the number of muons at the end of the shower is equal to the total number of charged pions. 
From the same logic as \eq\ref{eq:EAS:EM:MaxSize}), the total energy in the shower is 
\begin{equation}
\label{eq:EAS:HAD:PrimaryE}
 E_{o} = E_{\text{c}}N_{\text{max}} + E_{\text{c}}^{\pi}N_{\mu}
\end{equation}
which can be scaled to the ``electron size'', to account for the overestimation of the $e^{\pm}$ number (as in the case of an EM shower):
\begin{equation}
\label{eq:PrimaryEfromNumber}
 E_{o} =  gE_{\text{c}}N_{\text{max}} + \frac{E_{\text{c}}^{\pi}}{gE_{\text{c}}}N_{\mu} \approx 0.85~\text{GeV} \left(N_{e} + 24 N_{\mu}\right)
\end{equation}
The relative magnitude of the contribution from $N_{\mu}$ and $N_{e}$ are determined by the critical energies. The basic feature of \eq\ref{eq:PrimaryEfromNumber}) is that the primary cosmic ray energy is 
calculable in a simple way if the number of muons and electrons is known at the shower maximum. The actual application of this relationship, however, requires corrections for experimental details such as 
the relative sensitivity of the detectors to muons and electrons and the fact that the shower is not viewed at the shower maximum.

Using \eq\ref{eq:EAS:HAD:Interactions}), the energy dependence of the muon number is given by 
\begin{equation}
 \ln N_{\mu} = \ln N_{\pi} = n_{\text{c}}\ln N_{\text{ch}} = \beta \ln[E_{o}/E_{\text{c}}^{\pi}]
\end{equation}
The quantity $\beta$ is itself given by 
\begin{equation}
 \beta \approx 1 - \frac{\kappa}{3\ln[N_{\text{ch}}]} = 1 - 0.14 \kappa
\end{equation}
where $\kappa$ is the inelasticity parameter, which accounts for the energy which is not available for particle production because it is carried away by a single leading particle.
Values of $\beta$ from Monte Carlo studies range from $\beta=0.85~$to$~0.92$, depending on the muon energy threshold and the hadronic interaction model \cite{AlvarezMuniz:2002xs}.  
The muon number at the shower maximum is then 
\begin{equation}
 N_{\mu} = \left(\frac{E_{o}}{E_{\text{c}}^{\pi}}\right)^{\beta} \approx 10^{4}\left(\frac{E_{o}}{1~\text{PeV}}\right)^{\beta}
\end{equation}

The energy of the shower is split between the hadronic and electromagnetic part of the total shower and energy conservation implies that the fraction of the total shower energy which is in the electromagnetic
component is
\begin{equation}
 \frac{E_{\text{em}}}{E_{o}} = \frac{E_{o} - N_{\mu}E_{\text{c}}^{\pi}}{E_{o}} = 1 - \left(\frac{E_{o}}{E_{\text{c}}}^{\pi}\right)^{\beta-1}
\end{equation}
From this relation, it can be determined that the electromagnetic component of the EAS is about 70-80\% of the total at $E_{o}= 10^{15}~$eV  up to  90-95\% at $10^{20}$~eV.
In a more detailed treatment, this percentage is known, but weakly dependent on the parameters of hadronic interaction models.

Unlike the electromagnetic case, the depth of shower maximum for a hadronic shower must account for each sub-shower from its respective point of origin and account for its possible attenuation 
at or after its maximum. A simple estimate can be made, however, by using only the first interaction. This method will tend to underestimate $X_{\text{max}}$, but will account well for the elongation rate.
The interaction cross section of the primary particle rises with energy, as does the multiplicity $N_{\text{ch}}$. This will tend to raise the altitude of $X_{\text{max}}$ with increasing $E_{o}$, and causes the 
electromagnetic sub-showers to have shorter development depths.

To estimate $X_{\text{max}}$, we assume that the first interaction occurs at an atmospheric depth of $\lambda_{\text{I}}\ln2$ (i.e.\ in the first interaction layer). 
The first interaction yields, on average, $\frac{1}{2}N_{\text{ch}}$ neutral pions which yield $N_{\text{ch}}$ photons.
The same logic as the derivation of \eq\ref{eq:EAS:EM:depthMax}) in the electromagnetic case can then be applied to find
\begin{equation}
\label{eq:EAS:HAD:depthMax}
 X_{\text{max}}^{p} \approx \lambda_{\text{I}}\ln2 + \lambda_{\text{r}}\ln\left(\frac{E_{o}}{3N_{\text{ch}}E_{\text{c}}}\right) \approx \lambda_{\text{I}}\ln2 + X_{\text{max}}^{e}\left(E_{o}/3N_{\text{ch}}\right)
\end{equation}
which gives an elongation rate of
\begin{equation}
 \Lambda = \Lambda^{\gamma} + \frac{d}{d\log_{10} E_{o}}\left[\lambda_{\text{I}}\ln2 - \lambda_{\text{r}}\ln(3N_{\text{ch}})\right] \approx 58~\text{g}/\text{cm}^{2}
\end{equation}
This leads to the \emph{elongation rate theorem}, which says that the elongation rate of hadronic showers is always less than or equal to that of electromagnetic showers.
This is due to the increasing multiplicity $N_{\text{ch}}$ with primary particle energy, and the decreasing depth of the first interaction due to the increase in cross section with energy.
As pointed out by J. Linsley \cite{1977ICRC...12...89L}, the elongation rate theorem can be used to estimate the properties of rare high-energy showers thanks to the relation between depth dependence and energy dependence.
For any shower parameter which is a linear function of the shower depth, such as the shower age, the change in the parameter per decade of $E_{o}$, evaluated at a given depth, is proportional to the elongation rate.

\subsection{The Superposition Model}
The results derived so far assume that the shower is initiated by a proton. The interaction of a nucleus with the atmosphere can be treated in a simplified way using the \emph{superposition model}.
In the superposition model, the shower generated by a nucleus with atomic number $A$ and energy $E_{o}$ is modeled as the sum of $A$ independent sub-showers, each with energy $E_{o}/A$. Simple substitution into
previous results gives expressions for the number of charged particles at shower maximum:
\begin{equation}
 N_{\text{max}}^{(A)} = A \frac{E_{o}}{AE_{\text{c}}} = N_{\text{max}}^{(\text{proton})}
\end{equation}
the number of muons at shower maximum
\begin{equation}
 N_{\mu}^{A} = A\left(\frac{E_{o}/A}{E_{\text{c}}^{\pi}}\right)^{\beta} = A^{1-\beta}N_{\mu}
\end{equation}
and the depth of shower maximum
\begin{equation}
  X_{\text{max}}^{A} = X^{p}_{\text{max}}\left(E_{o}/A\right)
\end{equation}
As can be seen, the number of charged particles is independent of the primary hadron, but $X_{\text{max}}$ and $N_{\mu}$ are sensitive to the composition.
Air showers initiated by nuclei produce a larger number of muons than proton EAS, and an iron EAS will give
$(56)^{0.1} =  1.5$ times the number of muons than a proton EAS of the same energy (using a value of $\beta = 0.9$).
At the same time, the lower energy of each sub-shower means that the overall shower will not penetrate as deeply into the atmosphere.

\subsection{Simulations and Hadronic Interaction Modeling}
The basic properties of EAS described above are confirmed by detailed simulation studies. The exact results from simulations are, however, dependent on the modeling of hadronic interactions.
Modeling hadronic multiparticle production is difficult because, unlike QED, multiparticle hadronic interactions can not be calculated analytically. Typically, the final
states of hadronic interactions are modeled phenomenologically, with the model parameters
being tuned to accelerator measurements. 

At high interaction energies, greater than $\sim100~$GeV in the laboratory frame, hadronic interaction simulation programs such as DPMJET-III~\cite{2008PhRvC..77a4904B, Roesler:2001mn}, 
EPOS \cite{2006PhRvC..74d4902W, 2009NuPhS.196..102P, 2011PhRvL.106l2004W}, QGSJET II \cite{2006PhRvD..74a4026O, 2006NuPhS.151..143O, 2007AIPC..928..118O}, and SIBYLL 2.1 \cite{2009PhRvD..80i4003A, Engel:1992vf, Fletcher:1994bd} are often used. 
These interaction models
reproduce accelerator data well, but use different extrapolations above a center-of-mass energy of $\sim~$1.8 TeV, which leads to very different predictions on shower properties at high energy.
At lower energies, the main difficulty is the extrapolation of accelerator data to the very forward region near the beam axis, where most accelerator detectors have limited acceptance.


The Auger, Telescope Array, and Yakutsk air shower arrays each respectively use  AIRES~\cite{1999astro.ph.11331S, 2001ICRC....1..237S}, CORSIKA~\cite{CORSIKA}, and COSMOS~\cite{COSMOS} for EAS simulations.
There are a number of comparisons between the predictions of shower simulation packages for the same interaction models available in the literature (e.g.~\cite{2003APh....19...77K}), and
in general the differences between the predictions
for muon multiplicities and lateral distributions of different simulation packages are on the order of 5\%, increasing in some regions of the phase space up to 10\%. 
For example, the muon number measured at ground is typically a factor of $\sim2$ larger than predictions from air shower simulations, depending on the interaction model, 
and data from inclined showers implies that the discrepancy is closely related to a deficiency in simulating the muon component of air showers \cite{2013EPJWC..5301007A}.

An interesting point is that the muon deficit is lower in EPOS 1.99 compared to QGSJET II, as EPOS accounts for baryon-antibaryon pairs, which lead to higher muon multiplicities. It also 
accounts for the possibility of leading $\rho^{0}$ replacing neutral pions at large values of Feynman-$x$. This increases the muon number, as rho-mesons decay into charged pions (giving muons) rather than photons as is the case for 
neutral pions.
Another effect, which is not included in EPOS, is a possible increase in kaon production, which would lead to a higher muon multiplicity and a harder muon spectrum \cite{AlvarezMuniz:2012dd,2008PhRvD..77e6003D}. 
Other possibilities include a drastic change in interaction properties due to chiral symmetry
restoration \cite{2013EPJWC..5307007F}, as an example.

In addition to the muon deficiency, each interaction model gives different results for longitudinal shower development and differing distributions of $X_{\text{max}}$. 
This results in different predictions of $\langle X_{\text{max}}\rangle$, RMS($ X_{\text{max}}$), and the elongation rate, and thus these quantities are not well produced by simulations.
The discrepancies are larger for protons, with the same predictions for iron showing little difference between models.
The lateral shower development also seems to be poorly predicted by simulations. For example, the energy of UHECRs as measured by the Telescope Array surface detector is 27\% higher than the energy measured by the Telescope Array fluorescence detector.
As the systematic uncertainty of these two energy methods is 21\% and 20\%, respectively (in TA), the difference in energy is at the limit of the systematics. 
This energy difference translates into a situation in which the electromagnetic shower energy measured
$\sim 700~$m from the shower core by the surface detector is larger than the electromagnetic shower energy measured near the shower axis by the fluorescence detector. 
This would suggest that the physical EAS have a larger lateral development than predicted by shower simulations \cite{2013EPJWC..5302002F}.

The agreement between models for the electromagnetic component of EAS, on the other hand, is very good. 
The dominating source of systematic uncertainties in air shower prediction are a limited theoretical understanding of hadronic multiparticle production and the
limitations of accelerator measurements in energy, phase space coverage, and projectile-target combinations \cite{2013EPJWC..5301007A}.

\subsubsection{Impact of LHC data and \textit{Vice Versa}}
The recent availability of data from the Large Hadron Collider has had an interesting impact on the area of EAS simulations. 
This is especially so because the LHC is the first accelerator which has provided data at interaction energies above the cosmic ray knee ($\sim3$-$5~10^{15}~$eV).

Overall, the predictions of cosmic ray high energy interaction models (such as QGSJET, EPOS, etc.) bracket the LHC data, but no single model
reproduced the center of mass energy ($\sqrt{s}$) evolution of all observables consistently \cite{2011APh....35...98D}. In fact, the bracketing of the LHC data 
by the cosmic ray interaction models was found to be more natural than that of PYTHIA. The result of this is that alternative interpretations of the 
knee in the cosmic ray spectrum as being a consequence of rapidly changing hadronic interactions above $\sqrt{s}\simeq 2~$TeV are disfavored,
and the LHC data at 7 TeV has reduced the uncertainty in extrapolating to the highest energies by a factor $\sim2$ in average $X_{\text{max}}$ when comparing QGSJET II and EPOS \cite{2013EPJWC..5302001O}.

At the same time, air shower measurements are being used to study particle interactions at energies outside the reach of the LHC. 
One example is the measurement of the proton-air cross section at a center of mass energy per nucleon of 57 TeV \cite{2013EPJWC..5307005U}.
This measurement can be converted to a proton-proton cross section, which can then be compared to accelerator data and model predictions.
Despite the fact that the systematic uncertainties of this type of air shower measurement are higher than studies at colliders, this is the only way to study particle 
production at interaction energies beyond the reach of current accelerators.

\section{Detection Techniques}

As cosmic rays are fundamentally nothing more that particles from space, they can be studied using any typical particle detection technique. 
As the energy of the primary cosmic ray increases, however, several considerations come into play which allow for, or require, novel detection techniques.
The next few sections will introduce in a very general way the detection techniques currently used to register extensive air showers. Detection techniques which 
are still in the development stages, such as radio detection, will not be discussed in 
detail\footnote{Some current experiments working with radio detection include CODALEMA~\cite{Ardouin:2006gj}, and on-going radio detection tests at the Pierre Auger Observatory \cite{2008ICRC....5..885V}. See \cite{2008arXiv0804.0548F} for a review of this method.}.

\subsection{Surface Arrays}
An array of scintillation detectors on the ground is the classic EAS detection method. The surface coverage needed for ground arrays is low, due to the very large number of particles generated by the shower. 
This can range from ~1\% of the total array area in KASCADE, down to $5~10^{-6}$ in the Pierre Auger Observatory surface array. The duty cycle of a ground array is typically close to 100\%, as it is not affected by atmospheric conditions
or light levels.
 
A scintillation detector is made up of some material which scintillates, that is a material which emits photons when energy is deposited in it by the passage of a particle, paired with some readout
for this light, typically a photomultiplier tube coupled to the scintillating material by a light guide.
The number of photons created in the scintillator is given by the scintillation yield and the energy deposited\footnote{In plastic scintillators the number of scintillation photons
is linear with the energy deposit only at low energies, and saturates at high energies according to Birks' formula \cite{PhysRevD.86.010001}}.
The pulse height on the anode of the photomultiplier is proportional to the number of scintillation photons, with the constant of proportionality given by the light collection
efficiency, and the photomultiplier gain.

A method similar to scintillation detectors are water Cherenkov detectors. These are made up of a volume of water which is viewed by one or more photomultiplier tubes. 
Particles with high enough energies emit Cherenkov photons as they pass through the water. This light is reflected off the walls of the water tank and is detected by one or more photomultiplier tubes.
Water Cherenkov detectors have several benefits over scintillator counters,
and were first used for EAS detection in the Haverah Park observatory \cite{0370-1328-92-3-315}, which was the precursor of the Pierre Auger Observatory surface detector array.
These benefits will be mentioned later in a comparison between current observatories.

The muon component of a shower can be isolated by placing absorbers with thickness of several tens of radiation lengths above scintillation detectors. 
Muons can also be separated in tracking detectors, such as the HEGRA CRT detector \cite{Bernlohr:1997qj} or the muon tracking detector of KASCADE \cite{Doll:2002js}.
Hadron calorimeters, such as those in KASCADE \cite{Engler:1999pb} or EAS-TOP \cite{AdinolfiFalcone:1998tf} can also be employed in a surface array to measure the energy of the hadrons in the shower.

In scintillator and water Cherenkov ground arrays each detector samples the density of charged particles in the shower and 
allows a mapping of the lateral distribution of the electromagnetic component of the shower. 
If the arrival times of particles at each sub-detector are known with a resolution on the order of a few nanoseconds, 
then the orientation of the shower plane can be found with an accuracy on the order of a degree, and 
the location of the shower core can then be determined from this information.

The position of the shower plane is determined by fitting 
a lateral distribution function, such as the NKG function, \eq\ref{eq:NKGfunct}), to the measured particle densities. The shower size, in number of particles, is found by integration of the 
measured lateral distribution. The position of the shower core is generally determined with an uncertainty on the order of several meters, depending on the number of electrons in the shower \cite{Bluemer:2009zf}.
The energy of the shower can be estimated in a similar manner by using the lateral distribution of muons or the correlation between the muon and electron number (using \eq\ref{eq:PrimaryEfromNumber}), for example).

\subsection{Cherenkov Detection}

Another possible EAS detection method is the (direct) use of Cherenkov radiation. Many particles in the shower ``disc'' travel with relativistic energies  and approximately 
one third of charged particles in the shower emit Cherenkov radiation in the forward direction \cite{Giller:2004cf}.
The threshold energy for Cherenkov radiation by electrons (at sea level) is 21 MeV, and the Cherenkov angle in the air (also at sea level) is $1.3^{\circ}$. 
Because of this low energy threshold and the large number of electrons, positrons, and photons in the shower, the majority of Cherenkov light in an EAS is due to the EM component of the shower. 

This Cherenkov light is detected in one of two methods:
\begin{inparaenum}[i\upshape)]
 \item integrating detectors, and
 \item imaging detectors or telescopes.
\end{inparaenum}
Integrating Cherenkov detectors consists of an array of photomultipliers with some light collection optics (Winston Cones) looking up towards the oncoming showers. 
Examples of experiments using this technique are the Yatkusk Air Shower Array \cite{Anatoly:2013lya}.
These detectors measure the lateral density of Cherenkov photons, which allows a determination of both 
the energy and mass of the primary particle. 

Imaging Cherenkov detectors, on the other hand, produce a focal plane image which gives the direction and intensity of the incoming Cherenkov light. 
If the direction of the air shower core and the distance of the shower axis to the telescope are known, then the light received from the shower as a function of altitude can be 
reconstructed geometrically. This can be used to estimate the number of electrons in the shower, and thus the electromagnetic shower size as a function of atmospheric depth. 
Imaging Cherenkov detectors are used for the observation of TeV energy cosmic rays in experiments such as 
HESS~\cite{Hinton:2004eu}, MAGIC~\cite{Ferenc:2005fd}, and VERITAS~\cite{Weekes:2001pd}, and for the reconstruction of hadronic EAS in experiments such as
the \gls{DICE}~\cite{Swordy:1999um} at the CASA-MIA array.
 
\subsection{Air Fluorescence}
\label{sec:AirFluorEAS}
The charged particles in an EAS also excite nitrogen in the atmosphere, which then emits ultraviolet fluorescence light isotropically (in the 290 to 430 nm wavelength range). 
This phenomenon allows the detection of EAS with energies higher than $\simeq 10^{17}~$eV by
the so-called \emph{air fluorescence method}. The air fluorescence technique was first proposed by K.\ Suga et al.~\cite{AFt1, AFt2}, and first
attempted by K.\ Greisen at Cornell University in the mid-1960s~\cite{Greisen:1972zz}. The first observation of a cosmic ray by air fluorescence was
achieved by G.\ Tanahashi in 1968~\cite{AFt3}, and
the first completely successful fluorescence detector was the Fly's Eye detector in Utah, which started taking data in 1982 \cite{Baltrusaitis:1985mx}.
The discussion of air fluorescence will be slightly more detailed than that for Cherenkov detection or surface arrays, as this method is used by the JEM-EUSO experiment, which will be presented in chapter~\ref{CHAPTER:JEMEUSO}.

\begin{figure}[h]
  \centering
  \includegraphics[angle=0, width=1.0\textwidth]{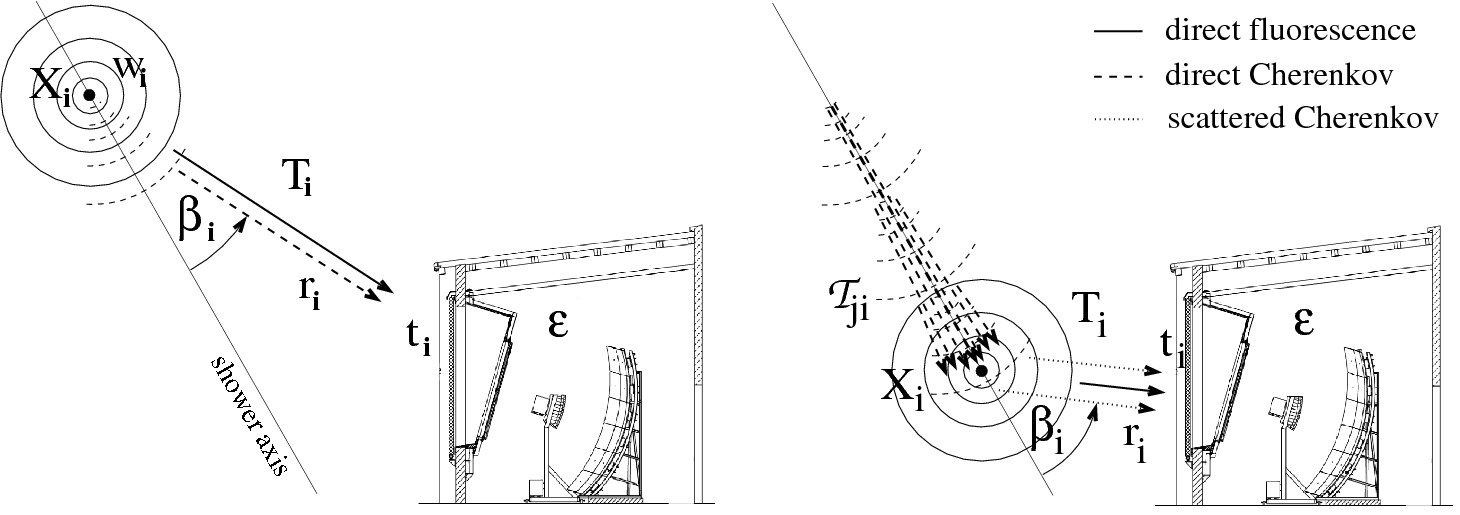}
   \caption[Air Fluorescence Detection]{\label{fig:sketch_of_fluorescencedetection} A sketch of air fluorescence detection, taken from \cite{Unger:2008uq}. The first figure shows the contribution of direct Cherenkov light and fluorescence, while 
the second figure shows the contribution of the fluorescence light and the scattered Cherenkov photons. The parameters shown in the two figures are described in the text.}
 \end{figure}

The number of emitted fluorescence photons is proportional to the energy deposited in the atmosphere, and the constant of proportionality is known as the fluorescence yield. This yield depends on the 
pressure, temperature, and composition of the atmosphere.
The physics of air fluorescence itself, and experimental determinations of the yield will be discussed in chapter~\ref{sec:INTROtoAFY}. 
The impact of uncertainties in the fluorescence yield on the reconstruction of EAS is rather large, with the uncertainty on the absolute fluorescence yield being in the range of 10-15\% (before AirFly, see chapter \ref{sec:INTROtoAFY}).
This impact is well-known and is discussed in ref.~\cite{2013EPJWC..5310002V}, for example.

Because the number of fluorescence photons is proportional to the energy deposit, the air fluorescence technique provides an almost calorimetric measurement of the energy of the primary UHECR.
This measurement is almost independent of hadronic interaction models, as the fraction of the UHECR energy which is deposited in the atmosphere is well-known.

A diagram of a typical ground \gls{FD} is shown in \fig\ref{fig:sketch_of_fluorescencedetection} (The diagram here is of an Auger FD).
A fluorescence detector consists of some light collection optics, in this case a parabolic mirror, which focuses the incoming light on a segmented ``camera'' made of photomultiplier tubes.
The entrance window of the fluorescence detector generally contains a filter in order to restrict the bandwidth of the telescope to the range of fluorescence emission.
Both air fluorescence and Cherenkov detection have a duty cycle of $\sim10\%$, as optical detectors can only operate on clear, moon-less nights. 

The reconstruction of the shower properties using a fluorescence telescope requires the geometrical determination of the shower axis, an estimation of the amount of Cherenkov light, and a correction
for the transmission properties of the atmosphere (cf.\ \cite{Tonachini:2011zza} for a discussion on the impact of atmospheric monitoring on EAS observables). Using a single fluorescence telescope, the shower arrival direction can be determined from the image of the light track and the timing information of the light signal.
This can in principle give an angular resolution on the order of $1^{\circ}$.
With multiple fluorescence telescopes, on the other hand, the shower arrival direction can be determined through stereo reconstruction, which can give reconstruction resolutions on the order of $0.6^{\circ}$ \cite{Bluemer:2009zf}.

If the fluorescence yield is parametrized as the number of photons emitted per unit track length per charged particle, then only a weak pressure and temperature dependence must be taken into account,
 and the longitudinal shower development is given by the number of charged particles as a function of atmospheric depth. 
This description is characterized by several shortcomings, however, such as the fact that the tracks of secondary particles are not parallel to the shower axis, and the dependence of the ionization energy deposit on particle energy (which changes 
with shower development). These issues can be avoided by using the energy deposit as the primary quantity for the shower profile reconstruction \cite{Unger:2008uq}.

As discussed in the last section, a significant amount of Cherenkov emission will also be present. For fluorescence detectors, the Cherenkov light can either be considered as a background contribution to be removed (generally in an iterative subtraction procedure)
or as a signal to be exploited in the analysis.
Consider a fluorescence detector which is observing an incoming shower as in \fig\ref{fig:sketch_of_fluorescencedetection}.
The directly observed fluorescence signal at a slant depth $X_{i}$ is measured at the fluorescence detector at a time $t_{i}$. For some value of the fluorescence yield (at this point in the atmosphere) $Y^{\text{f}}_{i}$, 
which is a non-trivial input from dedicated measurements, the number of 
photons produced in the shower at a slant depth interval $\Delta X_{i}$ is 
\begin{equation}
\label{eq:NfluorPhotons}
 N_{\gamma}^{\text{f}}(X_{i}) = Y^{\text{f}}_{i} w_{i} \Delta X_{i}
\end{equation}
where $w_{i}$ is the energy deposited per unit depth at a slant depth $X_{i}$, which is defined as
\begin{equation}
 w_{i} = \frac{1}{\Delta X_{i}} \int_{0}^{2\pi}d\phi \int_{0}^{\infty}rdr\int_{\Delta z_{i}}dz \frac{dE_{\text{dep}}}{dV}
\end{equation}
$dE_{\text{dep}}/dV$ is the energy deposited per unit volume, and $(\phi,R,z)$ are cylindrical coordinates with the shower axis at $r=0$.
The distance interval $\Delta z_{i}$ is given by the slant depth interval $\Delta X_{i}$. The fluorescence photons \eq\ref{eq:NfluorPhotons}) are distributed over a sphere of area
$4\pi r_{i}^{2}$, where $r_{i}$ is the distance to the detector. Only some fraction $T_{i}$ of these photons reach the fluorescence detector due to 
atmospheric attenuation.  If the fluorescence detector has some aperture area $A$, and efficiency $\epsilon$, then the measured photon flux at the detector (from fluorescence) will be
\begin{equation}
\label{eq:AF:AF}
 y^{\text{f}}_{i} = Y^{\text{f}}_{i}w_{i}\Delta X_{i} \frac{\epsilon T_{i}A}{4\pi r^{2}_{i}}
\end{equation}
The detector and atmospheric parameters can be grouped together in the quantity $d_{i} = \epsilon T_{i}A/4\pi r^{2}_{i}$.
The direct and scattered Cherenkov contribution to the photon flux can be determined in a similar way, and the calorimetric energy deposit $E_{\text{cal}}$ can be found using 
an analytic least-squares solution for the estimation of the longitudinal shower profile, as is done by Unger et al.~\cite{Unger:2008uq}.

The spectra of electrons in the shower, the Cherenkov yield (which replaces the fluorescence yield in the direct and scattered Cherenkov terms for the photon flux at the detector), and the average electron energy deposit $\alpha_{i}$ depend on the depth of shower maximum, which is not known before the analysis.
As the dependence of these quantities on the shower depth is small, however, a good first estimate of each can be obtained by taking the shower maximum at the position of the maximum light signal. After the shower profile is
calculated with these estimates, the shower maximum can be determined from the energy deposit profile, and the profile can be recalculated iteratively. 
If only part of the shower profile is observed due to a limited field of view, then a extrapolation to unobserved depths can be performed using, for example, the Gaisser Hillas function, \eq\ref{eq:GaisserHillasEq}).

Several sources of uncertainty exist for the determination of the energy deposit with this method. These include the flux uncertainty on the calorimetric energy, the 
geometrical uncertainties on the shower geometry, atmospheric uncertainties, and the loss of invisible energy. 
The geometrical uncertainties are due to the limited precision with which the distance to each shower point is known, 
and the propagation of this uncertainty through the transmission factors $T_{i}(r_{i})$ and geometry factors $1/(4\pi r^{2}_{i})$ of each point. 

The atmospheric uncertainties come from the uncertainty on the molecular density profiles and aerosol content of the atmosphere. These uncertainties require a constant monitoring of atmospheric
conditions around the fluorescence detector, in particular the measurement of Mie scattering and detection of clouds \cite{Abraham:2010pf}. For example, the variation of the atmospheric density profile over time can lead to systematic uncertainties
in the measurement of the depth of shower maximum.

The invisible energy is due to the fact that less than 100\% of the shower energy ends up in the electromagnetic component of the shower. 
The muon component of the shower requires a long path length to fully deposit its energy, and a number of muons carry energy past ground level. 
The neutrino flux created by pion decay also escapes. Both of these components can be accounted for by introducing a correction factor $f_{\text{inv}}$. The total energy of the shower is
then given by
\begin{equation}
 E_{\text{tot}} = f_{\text{inv}}E_{\text{cal}}
\end{equation}
The actual correction factor $f_{\text{inv}}$ is determined from shower simulations, and depends on the energy and mass of the primary UHECR and on the angle of EAS in the atmosphere. 
A mean parametrization \cite{Barbosa:2003dc} for showers at zenith angles of $45^{\circ}$ is given by:
\begin{equation}
 1/f_{\text{inv}} = 0.967 - 0.078\left(\frac{E_{\text{cal}}}{10^{18}~\text{eV}}\right)^{-0.140}
\end{equation}
resulting in a value of $f_{\text{inv}} = 1.08$ at $E_{\text{cal}}= 10^{20}~$eV.
The systematic error introduced by using this mean parametrization is on the order of $\pm3\%$ at $10^{18}~$eV 
and reduces with shower energy. Similarly,
shower-to-shower fluctuations introduce a systematic error of $\sim1\%$, and there is also an uncertainty between hadronic interaction models on the order of a few percent.

\subsection{Hybrid Detection}
It is worth  separately mentioning the so-called ``Hybrid'' detection of EAS. Hybrid detection is simply the observation of 
the same EAS by several different detector types, for example both air fluorescence telescopes and a surface array. 
This technique was pioneered on large scales by the Pierre Auger Observatory and is also used by Telescope Array, both of which will be briefly presented in the next section.

\begin{figure}
  \centering
  \includegraphics[angle=0, width=0.8\textwidth]{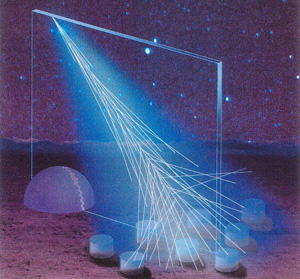}
   \caption[Hybrid Detection]{\label{fig:hybriddetectiondetection} A sketch of hybrid detection taken from \cite{Augerwebsite}. The same extensive air shower is viewed
by both a fluorescence detector (the dome) and a surface array. The fluorescence observation gives the shower track, while the surface detection gives the footprint of the shower on the ground.
The shower location is known to be somewhere inside the plane based on observation by a single fluorescence telescope, but the location of the shower within the plane is more uncertain. The signature 
of the shower on the ground can be used to further constrain the actual shower path. 
}
 \end{figure}

As the \gls{SD} array and fluorescence detector (FD) measure the EAS in a different manner, they provide different information on the shower. The SD provides the lateral development and the location of the shower core
from the EAS footprint on the ground, while the fluorescence telescope provides an accurate determination of the longitudinal shower development and the arrival direction of the shower within a plane from the fluorescence track.
This is shown as a diagram in \fig\ref{fig:hybriddetectiondetection}. 
For an EAS viewed with only a single fluorescence telescope, the angular resolution is often elliptical, that is worse within the shower plane.
One way to reduce this uncertainty is to view the same EAS with two FD placed some distance apart, known as stereo observation, which was previously mentioned.
The angular resolution can also be improved using a hybrid approach, as the footprint of the EAS measured by the surface detector constrains the shower axis.

Similarly, the energy determination from the ground array has a relativity high uncertainty due to the facts that only a fraction of the shower is sampled and that the energy reconstruction is dependent on EAS models.  
The fluorescence detector, on the other hand, gives a (nearly) calorimetric determination of the shower energy with a lower overall uncertainty than the surface array. 
At the same time, however, the number of true hybrid events is limited by the $\sim 10\%$ duty cycle of the FD.
A key idea in hybrid detection is thus to cross calibrate the energy reconstruction of surface array with the FD energy measurement using the subset of hybrid events. 
This calibration can then be applied to the events which are viewed only by the SD, which has a $\sim 100\%$ duty cycle, in principle 
greatly increasing the number of high quality events compared to a FD or SD only approach.

\section{Current UHECR Observatories}

There have been numerous  experiments which use these techniques to study UHECRs, since the pioneering measurements of EAS. 
Surface detector experiments in the recent past include the \gls{KASCADE} \cite{Antoni:2005wq}, its extension to higher energies
KASCADE-Grande \cite{Apel:2011mi}, and the \gls{AGASA} \cite{Shinozaki:2006kk}. 
The High-Resolution Fly's Eye (HIRES) \cite{Sokolsky:2011zz} experiment was the first UHECR observatory to successfully use the air fluorescence technique.

The main UHECR observatories which are still taking data are the Pierre Auger Observatory (Auger) \cite{forthePierreAuger:2013bha}, the Telescope Array (TA) \cite{Ogio:2013pka}, and the Yatkusk Air Shower Array \cite{Anatoly:2013lya}.
Each of these experiments are characterized by the use of both a ground array and some form of optical detection.
The Pierre Auger Observatory and the Telescope Array are particularly noteworthy, as they increased the number of observed UHECR by more than an order of magnitude compared to past experiments, and these two observatories will be described in the next sections.

\subsection{Telescope Array}

\begin{figure}[]
  \centering
\subfigure[ELS and Black Rock Mesa]{\label{fig:TAphotos:ELS}  \includegraphics[angle=270, width=1.0\textwidth]{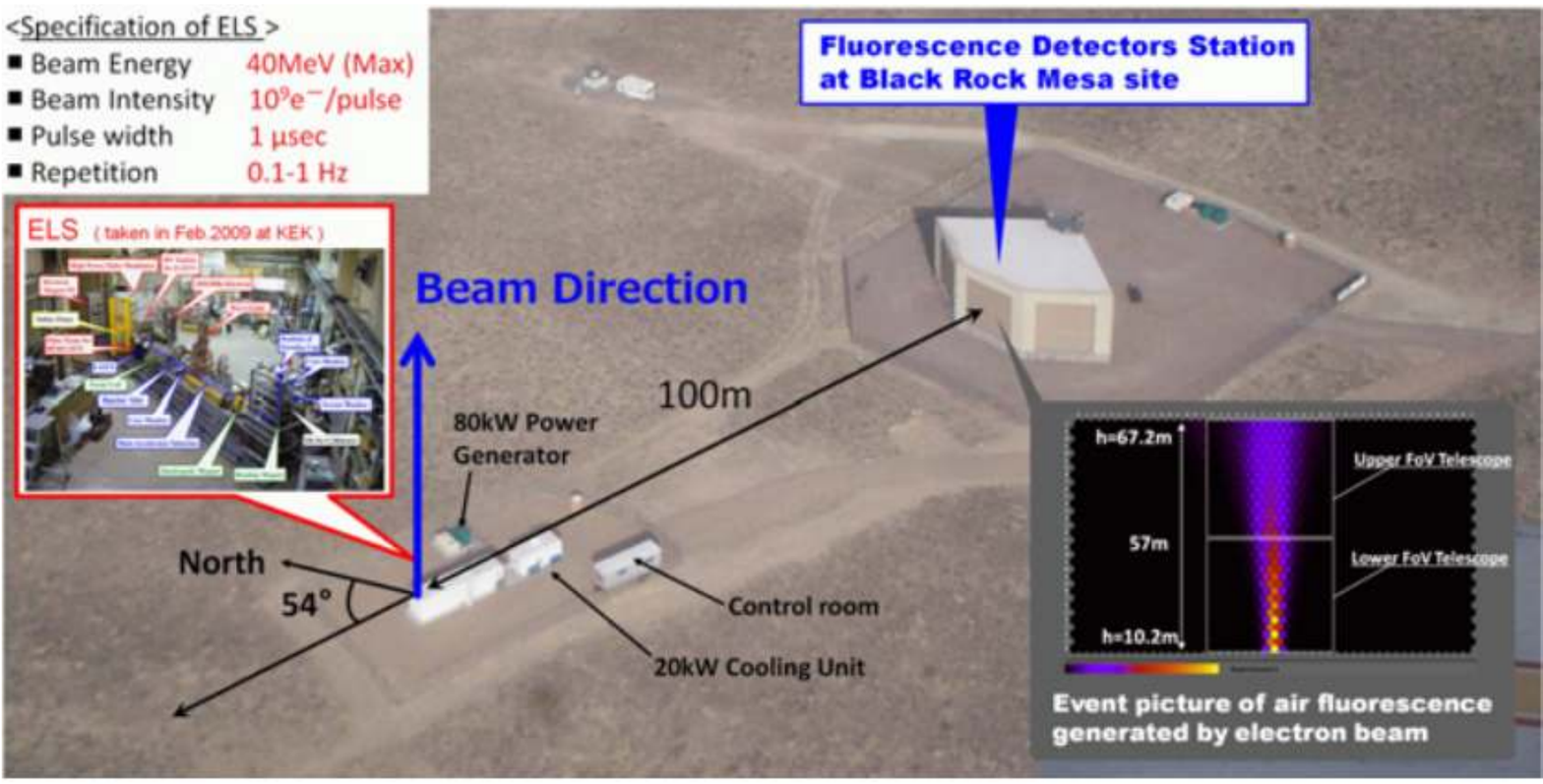}}\\
\subfigure[Fluorescence Telescope]{\label{fig:TAphotos:BRM}    \includegraphics[angle=0, width=0.495\textwidth]{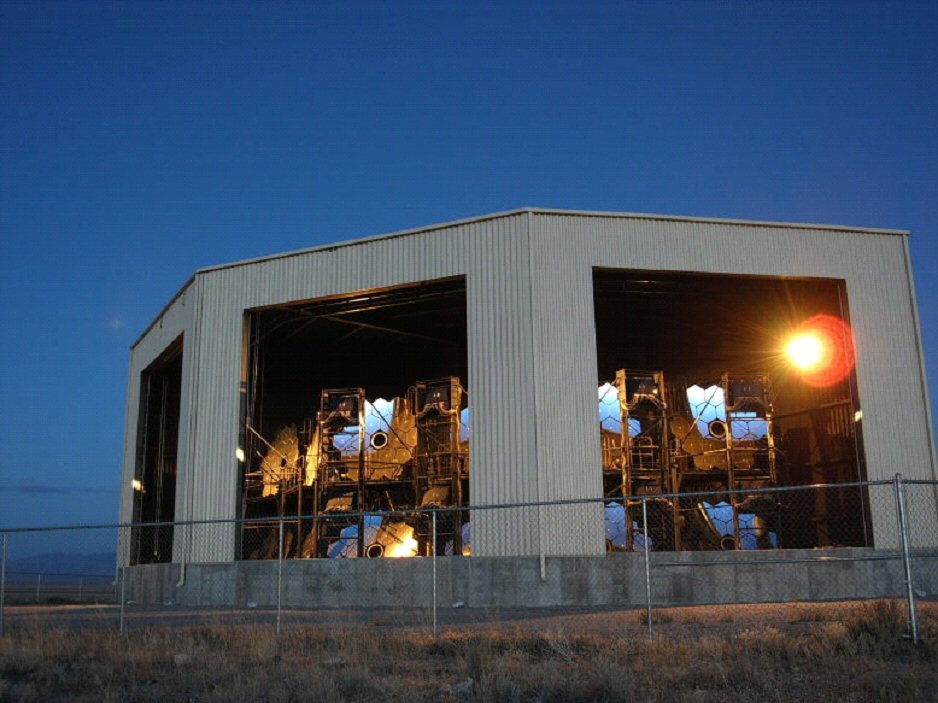}}
\subfigure[Scintillator Counter]{\label{fig:TAphotos:SD}    \includegraphics[angle=0, width=0.43\textwidth]{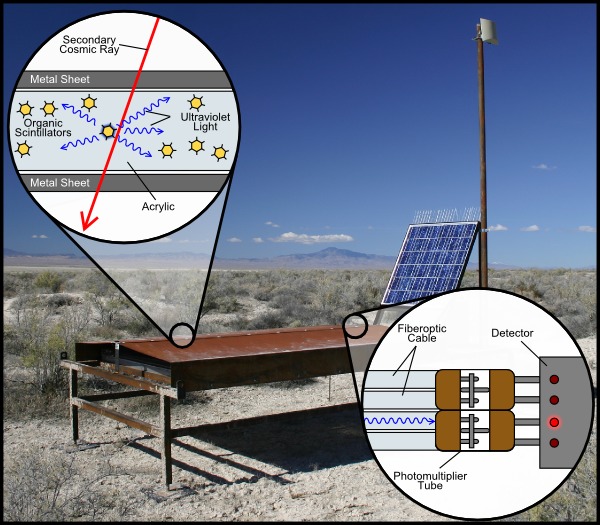}}
\caption[Photographs of the Telescope Array SD and FD]{\label{fig:TAphotos}
Several photographs of the Telescope Array (TA), taken from  \cite{Shibata:2013xka, TAwebsite}. \fig\ref{fig:TAphotos:ELS} shows an overview of the ELS calibration system, which uses an electron beam to generate test air showers. The ELS itself is located in front of the Black Rock Mesa
FD detector.  \fig\ref{fig:TAphotos:BRM} shows the fluorescence telescopes at Black Rock Mesa, and \fig\ref{fig:TAphotos:SD} shows a photograph of a TA scintillator counter, $\sim500$ of which make up the TA surface array. 
}
 \end{figure}

The Telescope Array (TA) is situated in the dessert of Millard County, Utah, USA at a latitude of $39.4^{\circ}$ North.
TA is made up of a surface array of scintillator detectors and several batteries of fluorescence telescopes, together covering a 
total area of 700 km$^{2}$.

The surface detector array (SD) is made of 507 scintillation counters deployed in a grid with a spacing of 1.2 km. A photograph of a SD detector unit is shown in \fig\ref{fig:TAphotos:SD}. 
Each individual scintillating counter has two stacked layers of plastic scintillator 1.2 cm thick and 3 m$^{2}$ in area. 
The scintillator plates are connected to photomultiplier tubes by 96 wavelength-shifting fibers.

The TA has three fluorescence detectors, Black Rock Mesa, Long Ridge, and Middle Drum, located on a triangle approximately 35 km apart. 
There are between 12 and 14 telescopes in each station, with a field of view from $3^{\circ}$ to $33^{\circ}$ in elevation. One TA FD station is shown 
in \fig\ref{fig:TAphotos:BRM}.
Each telescope of the FD array is made of a primary mirror, consisting of 18 hexagonal mirror segments, and a read-out camera. The cameras are made of 256 Hamamatsu R9508 photomultiplier tubes with attached BG3 filters.
This model of PMT is sensitive in the UV range and has a hexagonal shape with a length between sides of 60 mm. A subset of these photomultiplier tubes were calibrated with a systematic uncertainty on the absolute efficiency of 10\% using a 
comparison to Rayleigh scattered photons from a pulsed nitrogen laser beam \cite{Kawana201268}.
In connection to the FD array TA operates various atmosphere monitoring equipment, which includes infrared cameras to monitor clouds \cite{Shibata:2011zza}, and a \gls{LIDAR} system and a \gls{CLF} to measure the aerosol content and molecular profiles of the atmosphere \cite{Kobayashi:2011zz}.

A notable feature of TA is the calibration of the FD using artificial air showers generated by an on-site electron accelerator \cite{Shibata:2008zzg,Shibata:2013xka}.
This \gls{ELS}, shown in \fig\ref{fig:TAphotos:ELS}, is placed 100 meters from the Black Rock Mesa site. 
As the ELS is a 40 MeV linear electron accelerator with a bunch size of $\simeq 10^{9}~e^{-}$, it gives an energy deposit in the atmosphere equivalent to a $10^{16}~$eV UHECR.
At 100 m from the FD, this gives the same detector signal as a $\simeq 10^{20}~$eV shower at 10 km.

A comparison between the ELS air showers and true UHECR air showers allows a simultaneous calibration of all detector parameters such as fluorescence yield, mirror reflectivity, the transparencies of filters
and windows, photomultiplier tube quantum efficiency, and  photomultiplier tube gain. The atmospheric transparency and wavelength dependence of the
detector response can not be calibrated by this method, however, and
the TA collaboration estimates that the systematic error on the energy measurement is reduced from $\simeq23\%$ to $\simeq 17\%$ by the use of this end-to-end energy calibration.

TA is currently working on several extensions of their detector, known as the \gls{TALE}, designed to 
observe cosmic rays in the energy range between $3~10^{16}~$eV and $10^{19}~$eV \cite{Martens:2007ef}. This is accomplished by adding 10 new telescopes to the Middle Drum site, in order to extend 
the vertical field of view of the FD to the range of $3^{\circ}$ to $59^{\circ}$ in elevation. This field of view extension is intended to allow the observation of 
the shower development up to the shower maximum for lower energy EAS. In addition to FD upgrades, the TA SD will be extended by way of a graded infill of the ground array with a spacing of 400 m and 600 m.

\subsection{The Pierre Auger Observatory}

\begin{figure}[]
  \centering
  \includegraphics[angle=0, width=0.6\textwidth]{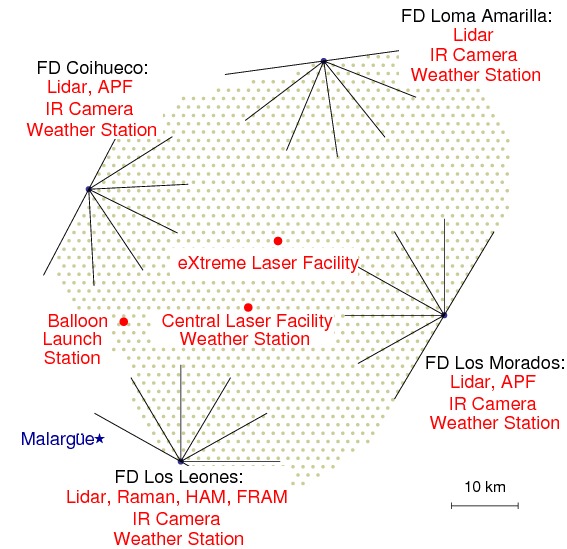}
   \caption[Map of the Pierre Auger Observatory]{\label{fig:PAOmap}  A map of the Pierre Auger Observatory, taken from \cite{Abraham:2010pf}, showing the 
layout of the four FD stations. The dots represent the individual water Cherenkov tanks of the SD detector. Each element is described further in the text.}
 \end{figure}

The Pierre Auger Observatory (Auger) is located in western Mendoza Province, Argentina, near the Andes mountains.
This places Auger in the southern hemisphere at a latitude of $35.3^{\circ}$, in contrast to TA.
Like TA, Auger is a hybrid observatory consisting of both a ground array and several batteries of fluorescence detectors.
A schematic of Auger is shown in \fig\ref{fig:PAOmap}.

\begin{figure}[]
  \centering
\subfigure[Water Cherenkov Tank]{\label{fig:PAOphotos:SD}    \includegraphics[angle=270, width=0.51\textwidth]{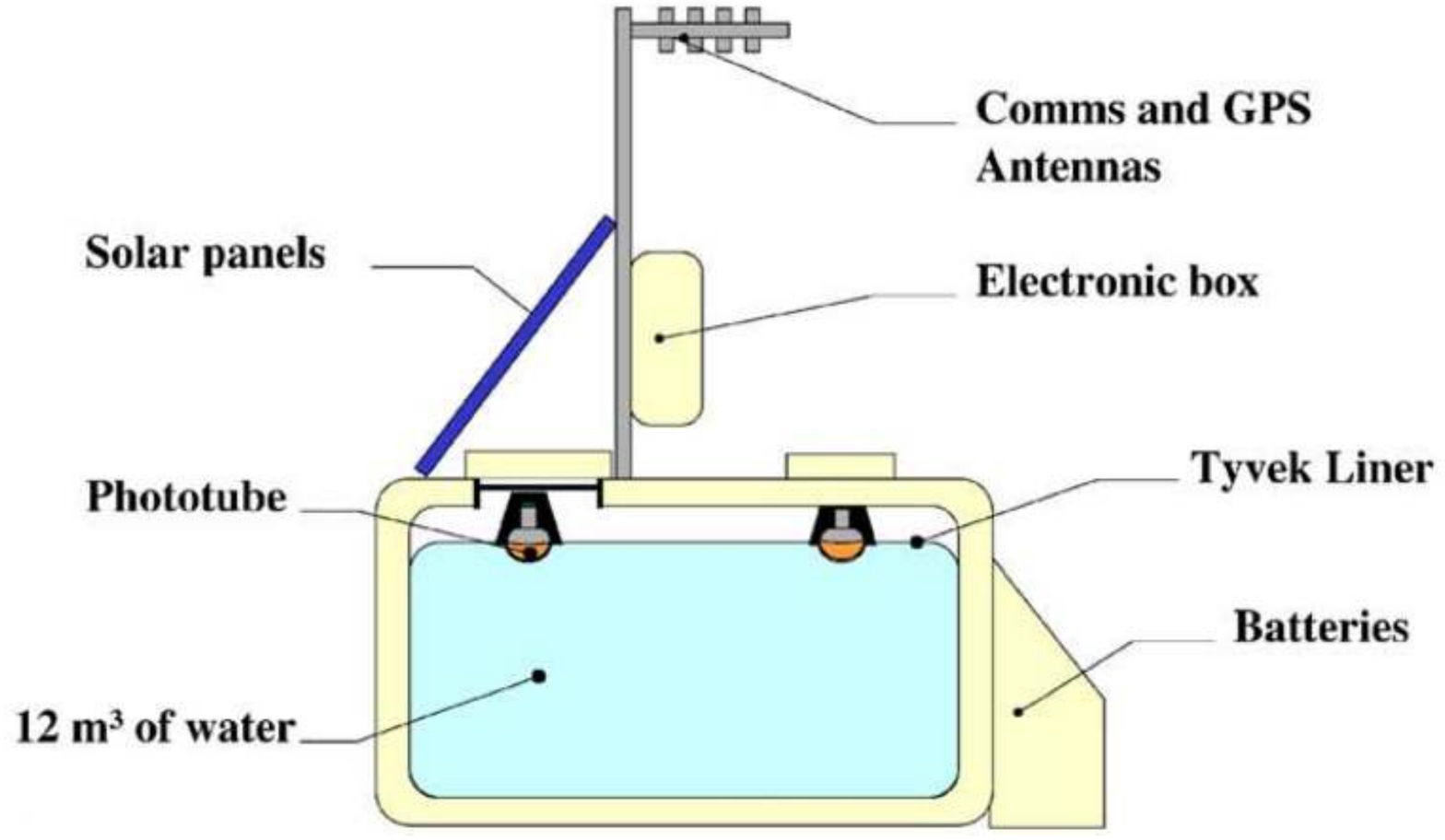}}
\subfigure[Fluorescence Telescope]{\label{fig:PAOphotos:FD}  \includegraphics[angle=270, width=0.46\textwidth]{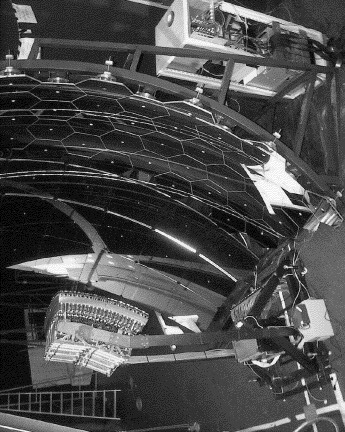}}
   \caption[Photographs of the Auger SD and FD]{\label{fig:PAOphotos} The Pierre Auger Observatory SD and FD arrays (taken from \cite{Abraham200450}).  
\fig\ref{fig:PAOphotos:SD} shows a diagram of the Auger water Cherenkov detector which makes up the SD. The main elements labeled in the diagram are described in the text. \fig\ref{fig:PAOphotos:FD} shows
a photograph of the Auger prototype FD, showing the spherical collection mirror and the camera consisting of 440 photomultiplier tubes.}
 \end{figure}

The Auger SD is composed of $\sim 1600$ water Cherenkov tanks deployed in a triangular grid with a spacing of 1.5 km. 
The total ground array covers an area of 3000 km$^{2}$, which is the same order of size as the US state of Rhode Island or the country of Luxembourg. 
A diagram of the  Auger water Cherenkov detector is shown in \fig\ref{fig:PAOphotos:SD}. Each cylindrical water Cherenkov detector has a footprint of 10 m$^{2}$, stands 1.2 m high, and is made of polyethylene resin\cite{Abraham200450}.
The water is held inside a cylindrical polyolefin bag inside the tank. This water bag has a $140~\mu$m thick lining of DuPont Tyvek$^{\text{\textregistered}}$ to diffusively reflect
the Cherenkov light to the three photomultiplier tubes installed in the top of the tank. The three photomultiplier tubes are installed symmetrically at a distance of 1.2 m from the center of the tank lid.

The Auger FD has four sites each with six fluorescence telescopes. A photograph of the first Auger fluorescence telescope can see in \fig\ref{fig:PAOphotos:FD}.
In the Auger FD telescopes, the light is focused by a 3.5 m by 3.5 m spherical mirror into a camera made of 440 Photonis XP3062 photomultiplier tubes. The photomultiplier array is made of 22
rows and 20 columns, and the Photonis XP3062 photomultiplier tube has a hexagonal photocathode with a 40 mm side-to-side length.

A circular diaphragm  at the center of curvature of the spherical mirror defines the aperture of the Schmidt optical system.
UV transmitting filters are installed at the entrance of the aperture, and just inside the filter is a ring of Schmidt corrector elements.
Each telescope has a field of view of $30^{\circ}$ in azimuthal angle and from $0^{\circ}$ to $28.6^{\circ}$ in elevation, and each pixel has a viewing angle of $\approx1.5^{\circ}$. 
The FD are calibrated using a ``drum calibration'' method with an uncertainty on the absolute efficiency on the order of 10\% \cite{Abraham200450}.
The systematic uncertainty on EAS energy measurements is $\approx 22\%$ for the Auger FD. 

As in the case of TA, the Pierre Auger Observatory runs an entire battery of atmospheric monitoring activities \cite{Abreu:2013qtw, Rizi:2012dn, Keilhauer:2012yp, Abreu:2012oza}. 
These include a Central Laser Facility (CLF) station, and an \gls{XLF}. In addition, 
each FD station operates a LIDAR, an IR camera for cloud detection, and a weather station. Two out of the four FD stations also operate aerosol phase function monitors.
During the early years of Auger operation, a balloon flight was also conducted after notable EAS events to record atmospheric data up to 23 km in altitude. 

Auger has also implemented several new projects beyond its original plans. These include the \gls{AMIGA} \cite{2011arXiv1107.4809T}, and the \gls{HEAT} \cite{Meurer:2011ms} extensions.
Both AMIGA and HEAT are similar to the TALE extensions of Telescope Array.
AMIGA is a infill of the SD array with more water Cherenkov detectors at a spacing of 750 m, which extends the energy range of the SD array down to $3~10^{17}~$eV.
HEAT, on the other hand, is a new array of tilt-able fluorescence detectors which extend the elevation range of the FD up to $60^{\circ}$. The idea of HEAT is to allow 
the observation of lower energy showers, by enabling the FD array to view the shower maximum for showers which are close by.
In addition to AMIGA and HEAT, Auger is also home to a prototype radio telescope array, the \gls{AERA} for detecting radio emission from the shower cascades in the frequency range 30-80 MHz \cite{Kleifges:2013fwa}.

\subsection{A Comparison between TA and Auger}
A short experimental comparison between Pierre Auger Observatory and the Telescope Array is instructive in understanding the difficulty of directly comparing 
results from the two experiments.
Firstly, TA and Auger use two different detection methods in their surface array.
The depth of the water Cherenkov tank, 1.2 m, compared to the thickness of the TA scintillator counters, several centimeters, extends the range of detectable zenith angles closer to the horizon.
This gives Auger a  factor $\approx2$ greater sky coverage compared to TA~\cite{Abraham200450}.
The water Cherenkov and scintillator counters are sensitive to both the electromagnetic and hadronic component of an EAS, but the signal in a Cherenkov tank is higher for 
hard muons than for less penetrating electrons and photons. The plastic scintillator, on the other hand, samples both equally. The result of this is that the Auger SD is more sensitive to the hadronic component of a EAS, 
whereas in TA this signal is dominated by the more numerous electrons and photons in the EAS.

Both TA and Auger use a sub-set of the events which they collect in their analysis.
Auger uses hybrid events, viewed by both FD and SD in coincidence, whereas TA uses stereo FD events in coincidence. 
Both experiments make use of the so-called ``bracket'' cut, which requires that the $X_{\text{max}}$ of individual events be inside the field of view of the FD telescope. 
A tight bracket cut will introduce some level of bias into $\langle X_{\text{max}}\rangle$ and RMS($ X_{\text{max}}$), as for a given event geometry it is difficult to know with certainty how many unbracketed
events lie outside the field of view \cite{2013EPJWC..5302002F}.

The two experiments follow a different philosophy in terms of applying acceptance cuts.
TA uses loser acceptance cuts and performs detailed simulations to estimate the acceptance bias of their detectors. 
This is done primarily to keep a larger percentage of their observed events.
Auger, on the other hand, uses tight acceptance cuts, requiring that both the $X_{\text{max}}$ of the observed shower 
and the entire range of probable $X_{\text{max}}$ be in the field of view. This choice is intended to minimize the field of view acceptance bias, and results in about half of their observed events being rejected above $1.6~10^{18}~$eV.
Auger can afford to do this, as their detection area is much larger than TA's (3000 km$^{2}$ vs. 700 km$^{2}$) and Auger has more events above $10^{19}~$eV than all other UHECR experiments combined, even after their strict
acceptance cuts. 

The two collaborations are currently working to understand the effect of these differences on their respective results.
This effort has resulted in several inter-collaboration working groups \cite{2013EPJWC..5302001O, 2013EPJWC..5302002F} and 
a program of cross calibration \cite{2013arXiv1310.0647T}. The hope is that these activities will help in making
sense of the conflicting results on the composition at the highest energies, and the overall disagreement in energy scale between the two experiments.
     \printbibliography[heading=subbibliography]
    \end{refsection}
    \begin{refsection}
  \chapter{Introduction to the Physics of Air Fluorescence}
    \label{sec:INTROtoAFY}
    As shown in section \ref{sec:AirFluorEAS}, fluorescence telescopes provide a calorimetric measurement of the electromagnetic part of an extensive air shower. This gives an almost model-independent measurement of the 
energy of the primary UHECR particle. The fluorescence track also records the longitudinal development of the air shower and allows an accurate geometrical reconstruction of the 
arrival direction of the incoming (primary) particle. 

The fluorescence yield is the proportionality between the energy deposited in the atmosphere by the charged particles in the EAS and the number of fluorescence photons emitted. 
This parameter must be known in order to reconstruct the primary energy, and the uncertainty
on the fluorescence yield is a major source of systematic error in the overall energy calibration. 
A complete understanding of the fluorescence yield can be considered as an extension of the calibration
for instruments based on the air fluorescence technique, such as JEM-EUSO, and the Telescope Array and Pierre Auger Observatory fluorescence detectors.

Section \ref{sec:Physics of Air Fluorescence} gives an overview of the physics of air fluorescence based on a large body of theoretical and experimental work by Arqueros et al. 
The main focus is on the characteristics of the fluorescence emission that are of particular importance in an accurate
measurement, and section \ref{sec:ExperimentalMeasurementsofAirFluorescence} discusses currently available experimental results and their interpretation.
This introduction will be used to motivate a new experimental setup for the absolute measurement of the air fluorescence yield and spectrum in all conditions relevant to air shower physics, which will
be presented later in chapter~\ref{CHAPTER:AIRFLUOR}.

\section{Physics of Air Fluorescence}
\label{sec:Physics of Air Fluorescence}
In the case of extensive air showers, atmospheric air fluorescence is due to the de-excitation of molecular nitrogen which has been excited by the charged particles in the air shower.
Electrons and other charged particles passing through the air lose energy through inelastic scattering. 
These collisions excite and/or ionize molecular nitrogen (N$_{2}$) in the ground state to the upper electron states of $N_{2}$ and $N^{+}_{2}$. 
Each electron state is split into vibrational energy levels $\nu$. The set of all transitions $\nu$-$\nu^{\prime}$ is called a band system.

\begin{figure}[h]
\centering
\includegraphics[angle=270,width=0.95\textwidth]{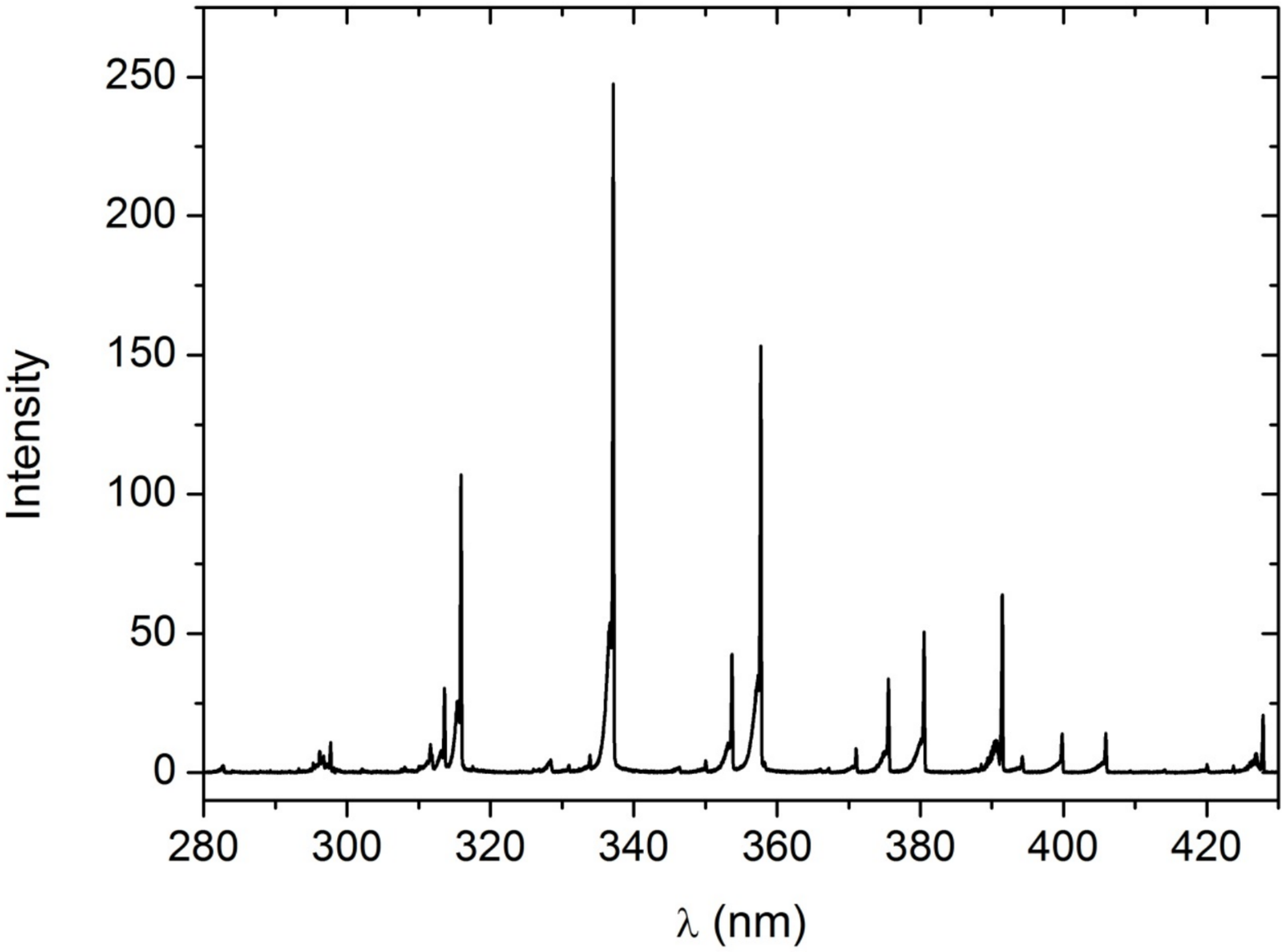}
\caption[Air Fluorescence Spectrum]{ \label{fig:FYSpectrum} 
The spectrum of artificial air (80\% $N_{2}$, 20\% $O_{2}$) measured by Dandl et al.~\cite{doi:1606299} at 800 hPa. The air mixture was excited using a 12 keV electron beam with a current of 1 to 5 $\mu$A. 
The vertical scale is not absolute, but is proportional to the photon flux. The dominating 337 line of the 2P can be clearly seen.  
}
\end{figure}

UV fluorescence in the spectral range 290 to 430 nm is dominated by the 
first negative (1N) and the second positive (2P) band systems \cite{Arqueros:2009zz}.
Once excited, the nitrogen de-excites through the emission of a photon with one of the allowed wavelengths of the band system. 
A measurement of the fluorescence spectrum for pure air in this wavelength region is 
shown in \fig\ref{fig:FYSpectrum}.  The dominant line is clearly the 337 nm line for the (0,0) band of the 2P system.

The yield $\varepsilon_{\nu \nu}$ is defined as the number of photons emitted from the $\nu$-$\nu^{\prime}$ band per unit length and per incident electron, and is given by
\begin{equation}
\label{eq:OpticalXsection}
 \varepsilon_{\nu\nu} = N \sigma_{\nu} \frac{A_{\nu\nu^{\prime}}}{A_{\nu}} = N \sigma_{\nu\nu^{\prime}}
\end{equation}
where $N$ is the number of nitrogen molecules per unit volume, $A_{\nu\nu^{\prime}}$ is the partial and $A_{\nu}$ the total radiative transition probability, and $\sigma_{\nu}$ is the cross section for excitation to the $\nu$ excited state. 

In addition to de-excitation by photon emission, the nitrogen can also be de-excited by collision with other molecules in the medium. This is known as collisional quenching.
 Including quenching, the total de-excitation probability is given by $A_{\nu}+A^{\text{c}}_{\nu}$. For a given temperature, $A_{\nu}^{\text{c}}(P)$ is proportional to the collision frequency and, therefore, to the pressure.
$A^{\text{c}}_{\nu}$ can be parametrized as $A^{\text{c}}_{\nu} = A_{\nu} P / P^{'}_{\nu}$, where $P^{'}_{\nu}$ is the characteristic pressure at which $A^{\text{c}}_{\nu}(P^{'}_{\nu}) = A_{\nu}$. 
Collisional quenching becomes important at moderate pressures and adds an additional pressure dependence to the fluorescence emission, which can be expressed by the Stern-Volmer correction factor:  
\begin{equation}
\label{eq:PrimaryFluor}
  \varepsilon_{\nu\nu^{\prime}}(P) = N \sigma_{\nu\nu^{\prime}} \frac{1}{1 + P/P^{'}_{\nu}}
\end{equation}
This yield accounts only for photons from primary interactions.
The characteristic pressure can be determined from the slope of the Stern-Volmer plot (due to shortening of the effective lifetime of the excited state).

Every component of the atmosphere, including oxygen, water vapor, aerosols, and other nitrogen molecules can quench excited N$_{2}$. The characteristic pressure for any mixture of gases can be determined as the sum of the ratio of
the fraction of each component over the partial characteristic pressure of that component. The partial characteristic pressure of any molecule can be expressed in terms of elementary molecular parameters and is proportional to
the square root of the temperature. There is also an additional temperature effect on the characteristic pressure through the collisional cross section $\sigma_{Ni}$. This dependence can be model as $\sigma_{Ni}\propto T^{\alpha}$ where $\alpha$ depends on
the dominating interaction for each quencher. This simple treatment of the quenching has been shown to break down when the concentration of electronegative O$_{2}$ molecules is low \cite{doi:1606299}.

\begin{figure}[h]
 \includegraphics[angle=270,width=1.0\textwidth]{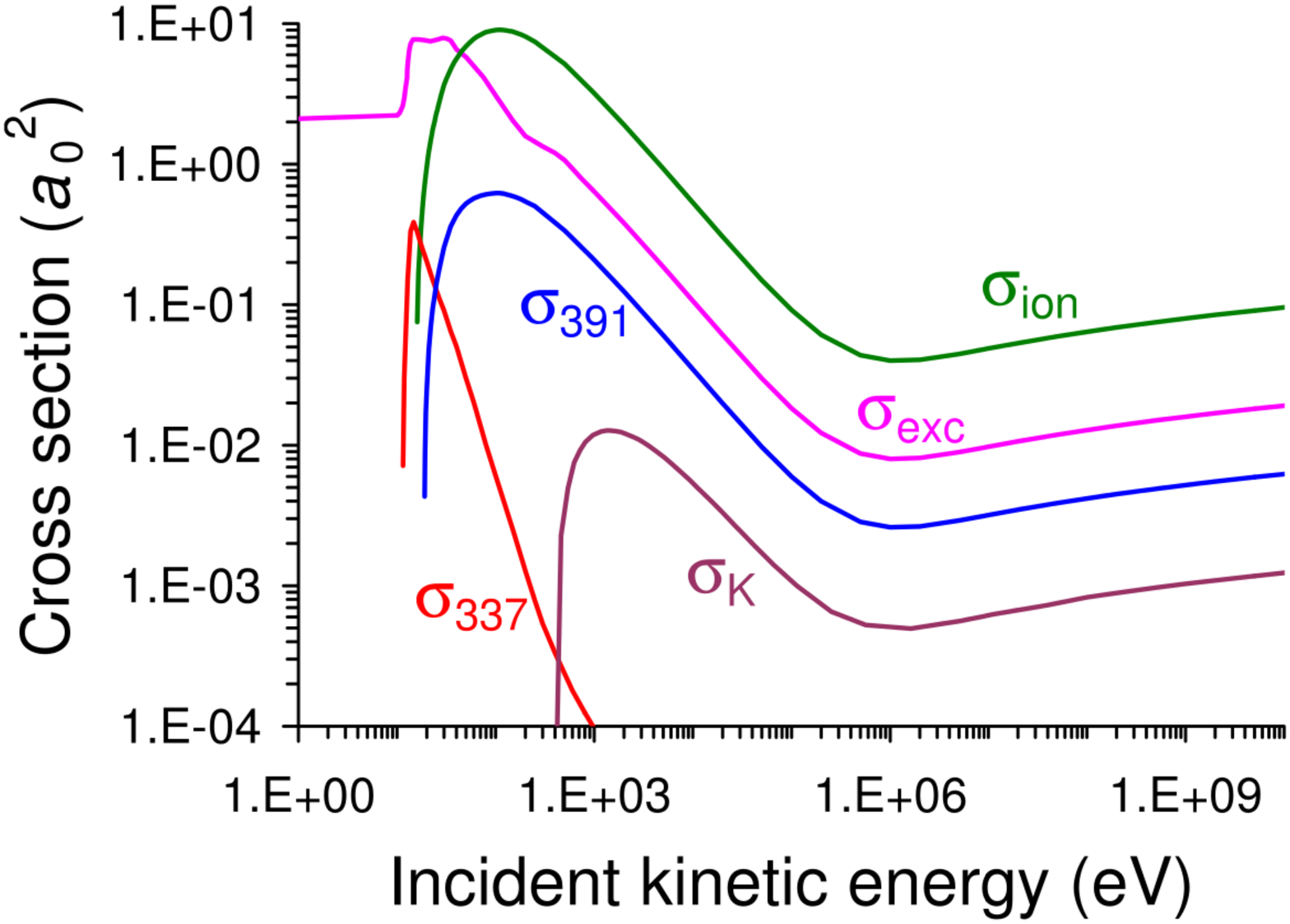}
\caption[Cross Sections for Air Fluorescence Processes]{\label{fig:AirFluorCrossSections} 
A plot of the cross section for several processes important in air fluorescence, taken from ref.~\cite{Arqueros:2009zz}.  
Each cross section is given in units of the Bohr radius squared $a_{o}^{2}$. $\sigma_{\text{ion}}$ is the cross section for all processes leading to the emission of a secondary electron,
and $\sigma_{\text{exc}}$ is the total cross section for all excitation processes which do not lead to the emission of a secondary electron.
$\sigma_{\text{K}}$  is the cross section for the ejection of a K shell electron, shown for comparison. $\sigma_{337}$ and $\sigma_{391}$ are the optical cross sections for excitation and emission of the (0,0) band of the 1P and 2N system, respectively.
The steep decrease of $\sigma_{337}$ with energy would seem to imply that its contribution to the total fluorescence yield is negligible, but instead it dominates due to the large number of low-energy secondary electrons. 
}
\end{figure}

The energy dependence of inelastic cross sections and the dominate 2P and 1N bands of air fluorescence excitation cross sections are shown in \fig\ref{fig:AirFluorCrossSections}. The cross sections for the
(0,0) band of the 2P system at 337 nm and the 1N system at 391 nm are shown separately, along with the total excitation and ionization cross sections.
The difference in energy dependance in 2P $\left(E^{-2}\right)$ and 1N $\left(\log E/E\right)$ systems would seem to imply that the 2P band system is negligible above $\approx10~$keV, 
but fluorescence in both air and pure N$_{2}$ at pressures over a few hPa is in fact dominated by the 2P system \cite{Arqueros:2009zz}.
Due to their different behavior with pressure, the 1N system is more important at low pressure, and the total fluorescence yield is approximately uniform at all pressures and so at all altitudes.  

It was been shown quantitatively that the large number of 2P fluorescence photons seen at high pressure come from the excitation by low-energy secondary electrons \cite{Blanco2005355}.
Because of this, the fluorescence yield is approximately constant at energies larger than 100 MeV and shows an increase of 2.5\% at 1 MeV and 10\% at 1 keV \cite{RosadoICRC0377}.

 The contribution of secondary electrons to the fluorescence emission
depends on their path length, and so on the pressure of the air and the geometry of the observation region. 
In addition, the average energy of the secondary electrons and the shape of their distribution is pressure dependent due to the dependence of the total cross section on pressure (in the GeV region).
Because of electron indistinguishably, the secondary electron spectrum extends up to a maximum energy of $E_{\text{secondary}}= \left(E_{\text{primary}} - I\right)/2$, where $I$ is the ionization potential of the target molecule.

If the pressure dependence of the ionization cross section is neglected,
the average number of $\nu$-$\nu^{\prime}$ photons produced in pure nitrogen inside a given region per secondary electron coming from ionization interactions of the primary electron is a function of the thickness of the interaction region.
This number of photons is denoted by $\alpha_{\nu\nu^{\prime}}$ and the ``thickness'' of the interaction region can be parametrized by the product $P\cdot R$ of the radius of the interaction volume and the pressure \cite{Blanco2005355}. 

The effect of the ionization cross section pressure dependence is to keep $\alpha_{\nu\nu^{\prime}}$ nearly constant at energies above a minimum pressure dependent value given by $E_{\min}(\text{GeV})=3.2/\sqrt{P(\text{hPa})}$ \cite{Arqueros:2009zz}. 
The relative intensities of bands within a system are also pressure dependent \cite{Blanco2005355, Arqueros:2009zz}.

In practice, the fluorescence yield is often defined as the number of $\nu$-$\nu^{\prime}$ photons per unit energy deposited in the medium:
\begin{equation}
\label{eq:AirFlYperEdep}
 Y_{\nu\nu^{\prime}} = \frac{\varepsilon_{\nu\nu^{\prime}}}{\left(dE / dX\right)_{\text{dep}}}
\end{equation}
where $\left(dE / dX\right)_{\text{dep}}$ is the energy deposited per primary electron and unit path length. This relationship requires that the emitted photons and the deposited energy correspond to the same air volume. 
Above 1 keV, the energy loss per unit path length of a primary electron is given by the Bethe-Bloch formula. A significant fraction of this energy can 
be carried out of the interaction region by high energy secondary electrons or radiation. It is the locally deposited energy, which depends on the interaction region geometry, size, and air pressure, which must be used in the calculation of the yield.

\begin{figure}[h]
 \includegraphics[angle=270,width=1.0\textwidth]{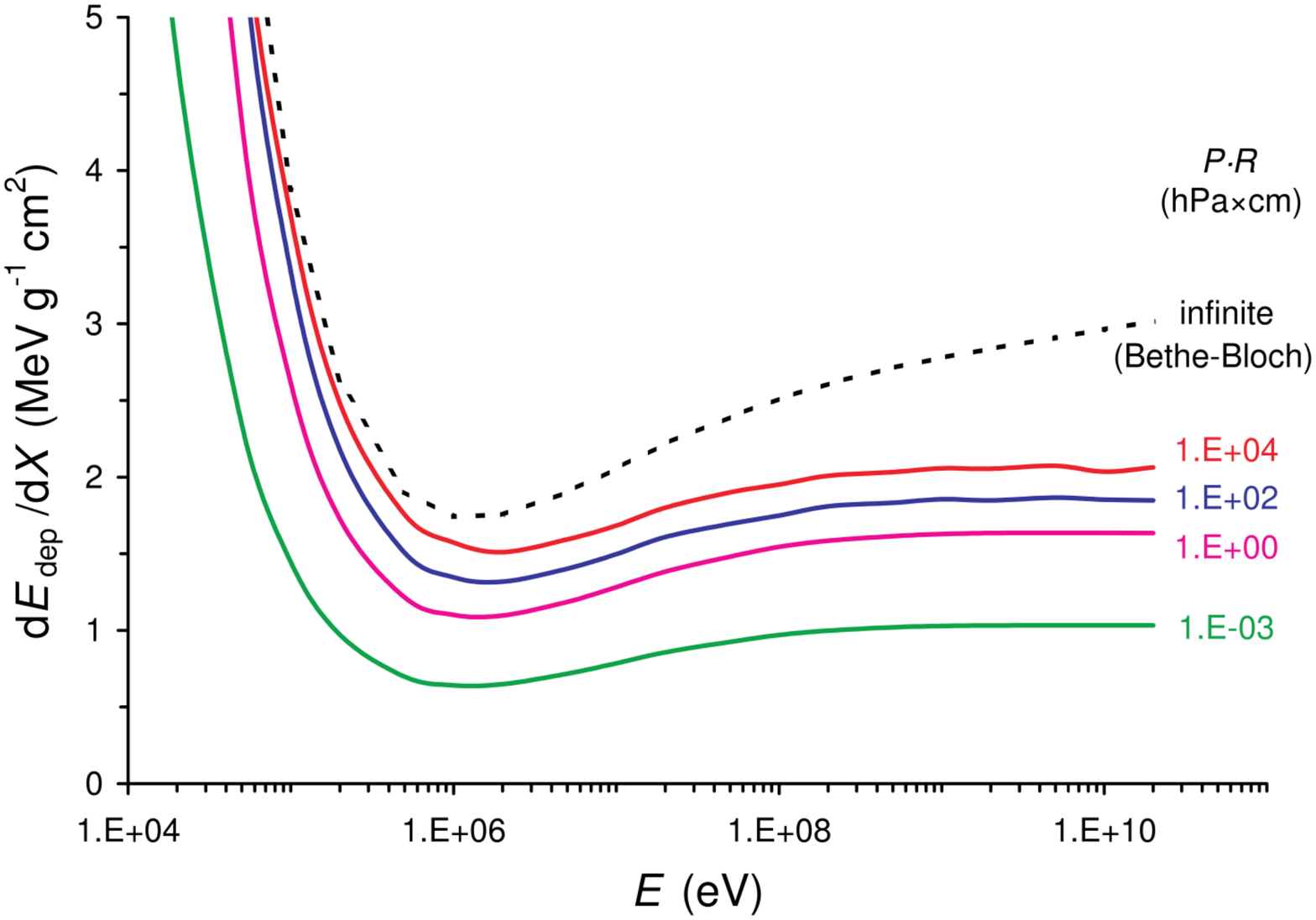}
\caption[Plot of the Local Energy Deposit]{\label{fig:AirFluorEnergyDepositPR}  A plot from ref.~\cite{Arqueros:2009zz} showing the local energy deposit $dE_{\text{dep}}/dX$ as a function of primary particle energy in eV and the thickness
parameter $P\cdot R$ in units of hPa$\times$cm. The dashed line shows the energy deposit calculated using the Bethe-Bloch formula, which does not take into account the escape of secondary particles from a ``thin'' interaction region.
As an example, there is a factor of $\approx2.5$ difference between the Bethe-Bloch formula and the actual energy deposited in the interaction region for 100 MeV primaries at $P\cdot R = 1~10^{-3}$ hPa$\times$cm. 
}
\end{figure}

Figure \ref{fig:AirFluorEnergyDepositPR} shows the dependence of the energy deposited locally per unit path length on the primary particle energy and on the product $P\cdot R$ for a spherical interaction volume of radius R. 
For $P\cdot R \rightarrow \infty$ the locally deposited energy is given by the Bethe-Bloch formula. The $P\cdot R$ dependence of the deposited energy shows a smooth logarithmic growth with energy due to very energetic, but rare, secondary electrons with a 
large range and a high probability of escaping the interaction region. 

\begin{figure}[h]
\centering
 \subfigure[2P emission]{\label{fig:AirFluorSpatial2P} \includegraphics[angle=270,width=0.47\textwidth]{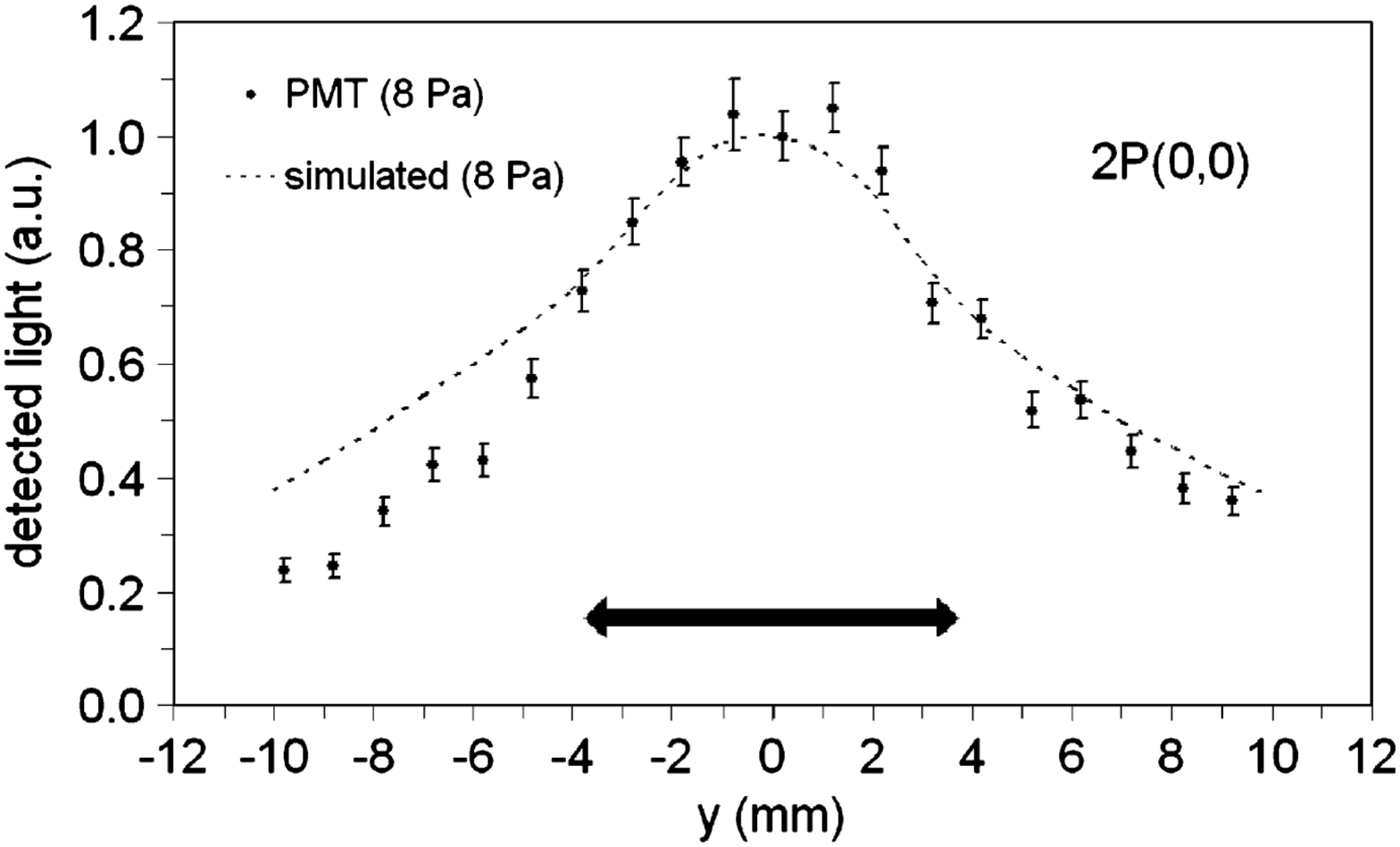}}
  \subfigure[1N emission]{\label{fig:AirFluorSpatial1N} \includegraphics[angle=270,width=0.48\textwidth]{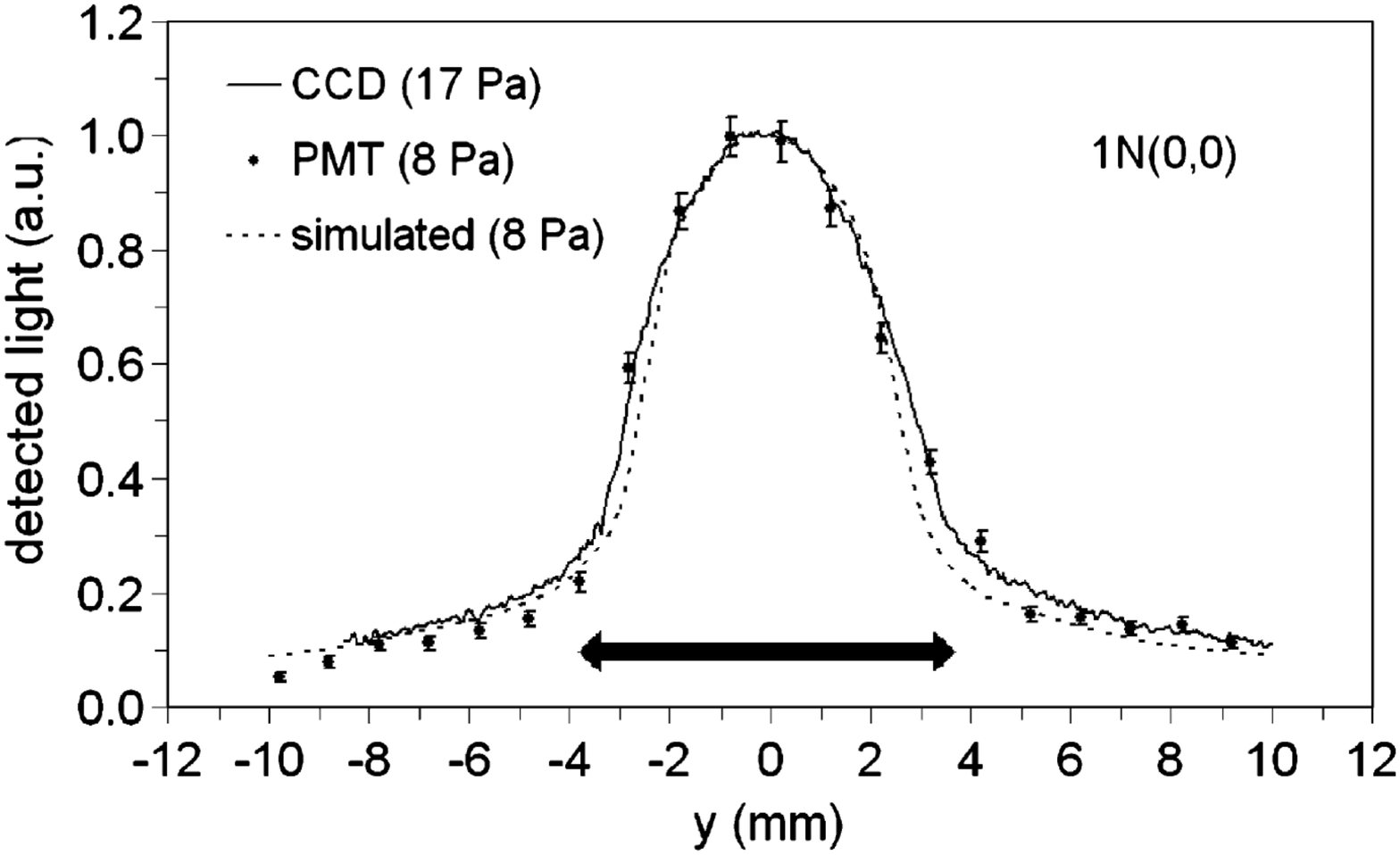}}
\caption[Air Fluorescence Emission Spatial Distribution]{ \label{fig:AirFluorSpatial} Two plots from ref.~\cite{Rosado:2008zz} showing their measurement of the spatial distribution of the fluorescence emission. 
\fig\ref{fig:AirFluorSpatial2P} shows the distribution for the 2P(0,0) line at 337 nm, the cross section for which is less than $\sim1~10^{-4}~a_{o}^{2}$ above 1 keV (see \fig\ref{fig:AirFluorCrossSections}). 
This was measured using a monochromator and photomultiplier tube (PMT), moving the focusing lens of the monochromator across the $y$ direction.
\fig\ref{fig:AirFluorSpatial1N} shows the distribution for the 1N(0,0) line at 391 nm measured using a PMT and monochromator as before (circles) and also a CCD image (solid line).
The black arrow in each plot shows the actual size of the primary electron beam. As can be seen by comparing the two plots, the 
1N emission is confined mostly to the beam, while 2P emission has a wider distribution. This corresponds to the fact that the 2P emission is dominated by the interactions of lower energy secondary electrons,
while the 1N  emission comes from interactions of the primary electrons. The dashed lines show predictions from simulation. 
}
 \end{figure}

The spatial distribution of the fluorescence emission is different in the case of the 2P and 1N system, due to their different energy dependencies. A measurement of the distribution by Rosado et al.\ \cite{Rosado:2008zz} is shown in \fig\ref{fig:AirFluorSpatial} for both the 2P (0,0) and 1N (0,0) bands.
The large black arrows in both plots shows the actual size of the primary electron beam. As can clearly be seen, the emission of the 2P band is significantly larger than the beam, while the 1N emission drops sharply outside the beam region.
This is easily understood as the 2P contribution comes mainly from low-energy secondary electrons, while the 1N contribution is dominated by the interactions of the primary electrons.

The physics of air fluorescence yield is, in sum, not trivial. 
A brief summary of important aspects includes:
\begin{enumerate}[i\upshape)]
\item  The observed fluorescence yield is dependent on the pressure and size of the observation volume due to both the average number of secondary electrons per primary electron in a given region 
and the deviation of the local energy deposit from the Bethe-Bloch formula. 
\item  The relative intensities of the fluorescence spectrum are dependent on the pressure, temperature, and composition of the air, and the total fluorescence yield is temperature dependent due to the dependence on temperature of collisional quenching.
\item  Because of quenching, the yield is also dependent on the composition of the air regarding water vapor, aerosols, and other pollutants.
\end{enumerate}
All of these aspects must be accounted for in a measurement of the air fluorescence yield.

\section{Experimental Results on the\\ Air Fluorescence Yield}
\label{sec:ExperimentalMeasurementsofAirFluorescence}

The laboratory measurement of the properties of air fluorescence has a long history. Most of these early measurements did not, however, reach the precision needed for the fluorescence observation of extensive air showers. 
See the review of Arqueos et al.\ \cite{Arqueros:2008en} for an overview of early air fluorescence experimental results.
Several more recent measurements have been performed since the advent of observing extensive air showers through fluorescence.
All of these experiments essentially consist of charged particles from some source crossing an air-filled chamber with an optical and electronic system for detecting the generated fluorescence light. 

The yield results of Kakimoto et al.\ \cite{Kakimoto1996527} are of particular note as these are used in the shower reconstruction of the HiRes and Telescope Array experiments.
The Pierre Auger Observatory has been using the results of Nagano et al.\ \cite{Nagano:2004am}, but has recently begun a transition to the use of the newer AirFly \cite{2008NIMPA.597...55A} results in their analysis.

Many of these measurements \cite{Kakimoto1996527,  Nagano:2004am, Lefeuvre:2007jq, Colin:2006nj, Waldenmaier:2007am} have used a $\beta$ emitting radioactive source, such as $^{90}$Sr, which gives electrons in a $\beta$-spectrum centered at 
1 MeV with an endpoint at 2.2 MeV, as a source of charged particles.
In this technique the detection of fluorescence photons is made using electron-photon coincidences. While radioactive sources have the advantage of simplicity, they have a large disadvantage due to the low rate of electron-photon coincidences and the 
fact that the $\beta$-emission is in $4\pi$. The low counting
rate results in a large uncertainty on spectral measurements, and the $4\pi$ emission requires one to setup an experimentally well-defined solid angle.
Several other measurements have used keV electron guns \cite{doi:1606299}, higher energy electrons from accelerators \cite{Colin:2006nj,Abbasi:2007am}, or even a 120 GeV proton beam \cite{Ave:2012ifa}
(using the fact that the air fluorescence is nearly independent of the type of primary charged particle) as the primary charged particles.
These approaches naturally require much greater infrastructure, but have the advantage of a large increase in fluorescence signal and range of possible primary particle energies.

Table~\ref{tab:SummaryOfAirFlResults} shows a summary of the parameters and results of these recent measurements.
A direct comparison between the results is difficult as the air pressure, air temperature, air mixture, and observation bandwidth in each must be taken into account. 

%
%
%

\begin{table}
\begin{tabulary}{1.125\textwidth}{LCCCCCC}

\toprule
\small Experiment & \small $\Delta\lambda$ & \small P & \small T & \small E & \small Experimental & \small Error\\
 &\small (nm)  &\small (hPa) &\small (K)  &\small (MeV) &\small Result&  (\%)\\
\cmidrule(l){1-7}

\multirow{5}{*}{\small Kakimoto et al.\ \cite{Kakimoto1996527} }& 337 & 800 & 288 & 1.4 & 5.7 \footnotesize ph/MeV & 10 \\
\cmidrule(r){2-7}
& \multirow{4}{*}{300 - 400 }& \multirow{4}{*}{1013} & \multirow{4}{*}{288} & 1.4 & 3.3 \footnotesize ph/m & \multirow{4}{*}{10} \\
&  &  &  & 300 & 4.9 \footnotesize ph/m &  \\
&  &  &  & 650 & 4.4 \footnotesize ph/m &  \\
&  &  &  & 1000 & 5.0 \footnotesize ph/m &  \\
\cmidrule(r){1-7}

\small Nagano et al.\ \cite{Nagano:2004am} & 337 & 1013 & 293 & 0.85 & 1.021 \footnotesize ph/m & 13 \\
\cmidrule(r){1-7}

\multirow{2}{*}{\small Lefeuvre et al.\ \cite{Lefeuvre:2007jq} }& \multirow{2}{*}{300 - 430 }& \multirow{2}{*}{1005} & \multirow{2}{*}{296} & 1.1  & 3.95 \footnotesize ph/m  & \multirow{2}{*}{5} \\
 & & & & 1.5 & 4.34 \footnotesize ph/m & \\
\cmidrule(r){1-7}

\multirow{3}{*}{\small  MACFLY  \cite{Colin:2006nj} }&\multirow{3}{*}{ 290 - 440} & \multirow{3}{*}{1013} & \multirow{3}{*}{296} & 1.5 & 17.0 \footnotesize ph/MeV & \multirow{3}{*}{13}\\
&&&&$20 \cdot 10^{3}$&17.4 \footnotesize ph/MeV& \\
&&&&$50 \cdot 10^{3}$&18.2 \footnotesize ph/MeV& \\
\cmidrule(r){1-7}

\small Dandl \cite{doi:1606299} & 337 & 800 & 293 & $\sim 0.01$ & 8.1 \footnotesize ph/MeV & 10\\
\cmidrule(r){1-7}
\small FLASH \cite{Abbasi:2007am} & 300 - 420 & 1013 & 304 & $28.5 \cdot 10^{3}$ & 20.8 \footnotesize ph/MeV & 7.5\\
\cmidrule(r){1-7}
\small Airlight \cite{Waldenmaier:2007am} & 337 & - & - & 0.2 - 2 & \footnotesize$Y^{0} = 384$ \footnotesize ph/MeV & 16 \\
\cmidrule(r){1-7}
\small AirFly \cite{Ave:2011ub} & 337 & 1013 & 293 & $120 \cdot 10^{3}$ & 5.60 \footnotesize ph/MeV & 4\\
\bottomrule
\end{tabulary}
\caption[Summary of Air Fluorescence Yield Experimental Results]{
\label{tab:SummaryOfAirFlResults} A summary and comparison of results on the absolute value of the fluorescence yield from the experiments discussed in the text.
The experimental results given by the authors are shown in column 6 and their quoted uncertainties in the 7th column.
The spectral range $\Delta\lambda$, the pressure P, the temperature T, and the energy of the primary particles E are given for each measurement. 
The primary particles in each case are electrons, with the exception of the AirFly group: who used 120 GeV protons.
Airlight reports their yield extrapolated to null pressure. The results from each group are listed in either photons per MeV of deposited energy or
photons per meter per electron (depending on how it was originally reported).
As can be seen the experimental conditions of the measurement vary from group to group, making a direct comparison not trivial.
}
\end{table}

% ...
%

Aside from differences in air conditions and observation bandwidth, there can be a large systematic error from the estimation of the deposited energy due to geometry \cite{Rosado:2012cy}.
As was discussed in the previous section, the true deposited energy is geometry dependent, and in any case less than the total energy loss given by the Bethe-Bloch formula due to the escape of high energy secondary electrons.
Nagano et al.\ \cite{Nagano:2004am}, Kakimoto et al.\ \cite{Kakimoto1996527}, and Lefeuvre et al.\ \cite{Lefeuvre:2007jq} each assumed that the energy deposited in the observation region was equal to the total energy loss. 
Rosado et al.\ calculated a correction to their reported values of the fluorescence yield using a custom simulation and found that a correction as high as $ 29\%$ (for 1000 MeV electrons in the geometry of Kakimoto et al.)
is needed to account for the true local energy deposit.

MACFLY \cite{Colin:2006nj} and Airlight \cite{Waldenmaier:2007am} calculated their energy deposit in the observation region using GEANT4, and 
FLASH \cite{Abbasi:2007am} used EGS4 to perform their own estimate of the local energy deposit. 
In each of these cases, Rosado et al.\ nonetheless found corrections on the order of several percent due to the more detailed treatment in their simulation of primary and 
secondary electron interactions with the molecules of the medium.

The absolute measurement of the yield requires an absolutely calibrated photodetector. 
The first experiments used the absolute calibration given by the manufacturers, and their experimental uncertainties reflect this.
The FLASH group began to address this issue, and performed an absolute calibration of their fluorescence bench using a comparison with the measurement of Rayleigh-scattered light from a nitrogen laser.
The decrease from an uncertainty of $\geq 10\%$ in previous measurements to $7.5\%$  in their result reflects this. 
A even greater attention to absolute calibration can be seen in the reported uncertainties of Lefeuvre et al.\ and AirFly, who focused a large effort on reducing this source of systematic error.
In the case of AirFly, the absolute measurement was done by measuring the fluorescence yield relative to the Cherenkov emission of the 120 GeV proton beam. 
Lefeuvre et al.\ used a different philosophy, preferring to do an absolute calibration of their photomultiplier tubes with high precision. 
To this end, they developed the calibration technique which will be presented in section~\ref{sec:Measuring Absolute Detection Efficiency}. 

Even with a low systematic uncertainty on the absolute calibration of the photodetector(s), however, it is not enough to measure the fluorescence yield at only one combination of pressure, temperature, and
gas mixture. As an example, neglecting the temperature dependence appears to result in an overestimation of the fluorescence yield by an amount of up to 20\% for the 391.4 nm band \cite{Ave:2007em}.
The most recent attempt to address this point has been the measurement campaign of the AirFly group.
Using various experimental setups, they have separately measured: 
\begin{itemize}
\item The fluorescence spectrum between 84–369 nm and 344–429 nm in dry air at 800 hPa and 293 K \cite{Ave:2008zza}.
\item The temperature and humidity dependence between 240 and 310 K in dry air  for the 313.6, 337.1, 353.7, and 391.4 nm bands \cite{Ave:2007em}.
\item The pressure dependence of the 337 nm band \cite{Ave:2008zza}.
\item Relative intensities of 34 fluorescence bands in the wavelength range from 284 to 429 nm at pressures from a few hPa to atmospheric pressure \cite{Ave:2007xh}.
\item The energy dependence \cite{Ave:2008zz}. 
\item The absolute yield of the 337 nm line \cite{Ave:2011ub, Ave:2012ifa} and \cite{Ave:2012ifa}.
\end{itemize}

\begin{figure}[h]
\centering
\includegraphics[angle=270,width=0.90\textwidth]{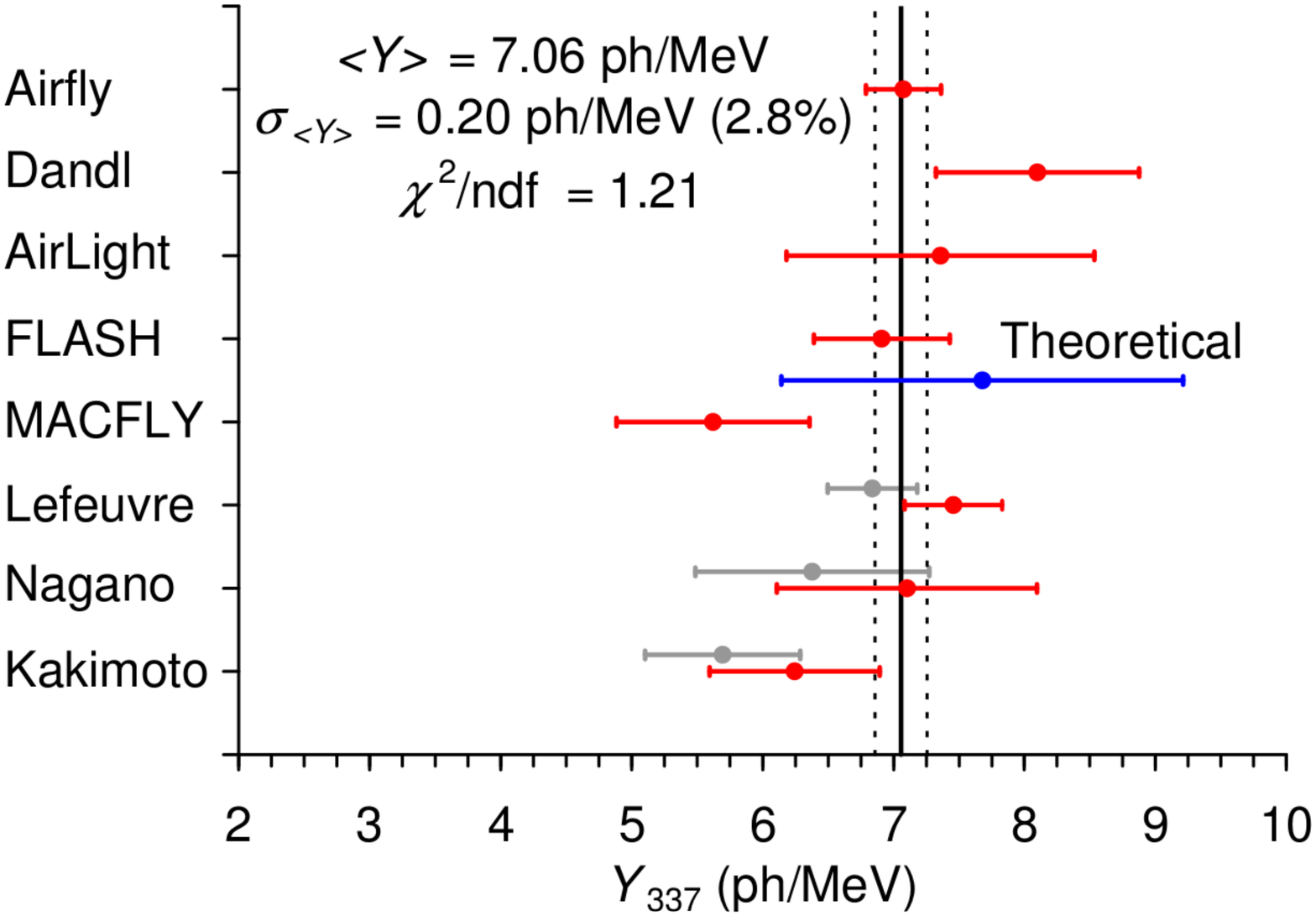}
\caption[Global Average of the Fluorescence Yield]{ \label{fig:FYaverage} 
This plot by Rosado et al.\ \cite{RosadoICRC0377} shows a compilation of results on the absolute fluorescence yield 
normalized to 800 hPa and 293 K in dry air (red markers). Also shown is their calculation of a global average value from this data set. 
The gray bars show the impact of their corrections on the original results. The weighted average and associated uncertainty
are represented by solid and dashed vertical lines. Rosado et al.\ suggest in their text a more conservative uncertainty on the average value of 3.5\%. 
The blue bar shows the theoretical absolute yield predicted by their MC algorithm 
}
\end{figure}

In addition to this large experimental effort, Rosado et al.\ \cite{RosadoICRC0377, Rosado:2011qi, Rosado2010164} have performed a systematic analysis of all the recent fluorescence results in order to determine a global average value.
This includes corrections for the true energy deposit and a normalization of each result to the same temperature pressure and bandwidth conditions. 
Their results are shown in \fig\ref{fig:FYaverage} and give a global average of $Y_{337} = 7.06 \pm 0.25~$ph/MeV.
Due to the small uncertainty on the AirFly absolute yield, it and the global average can be considered complementary measurements.  
 
Even considering the numerous results from the AirFly group, however, there are still gaps as far as a systematic exploration of pressure, temperature, and air mixture are concerned. 
In particular, the AirFly measurements were done with several different setups, and their absolute yield measurement 
was performed only for the 337 nm line using a narrow-band filter. 

There also appear to be strong effects due to residual oxygen and other impurities in measurements using nitrogen or artificial air mixtures. These effects are extremely important when a narrow band measurement of the 337 line in nitrogen
is used to derive the yield in dry air, as is done in many absolute yield measurements.  
Dandl et al.\ \cite{doi:1606299} found evidence for an increase in the intensity ratio between the 337 line in a nitrogen oxygen mixture and in air at very low oxygen content. 
This effect cannot be completely accounted for by collisional quenching, and so they investigated further by studying the 
nitrogen afterglow using a pulsed electron beam with a pulse length of 10$~\mu$s. They found a long afterglow phase in pure nitrogen which can be attributed to recombination processes. It is particularly important to note that Dandl et al.'s
results on the intensity ratio versus pressure are not consistent with the results of AirFly \cite{Ave:2007xh} or Nagano et al.\ \cite{Nagano:2004am} who did not pay the same level of attention to nitrogen purity.

In order to reach the best precision for the application of
air fluorescence to the reconstruction of UHECR air showers, it is best to go around these problems by measuring the absolute yield directly in air for all combinations of temperature, pressure, and humidity (and also 
possibly in the presence of aerosols and other impurities).
For these reasons, we propose a measurement of the absolute spectrum and integrated yield in all conditions using the same setup with 5\% uncertainty. This measurement would be strongly complementary to both the 
average value of Rosado et al.\ and the absolute yield measurement of AirFly. The latter point is particularly true as our absolute 
calibration will use an entirely different technique from AirFly, which would result in limited correlated systematic uncertainties between the two measurements.
Our proposed setup and some preliminary laboratory work will be presented in chapter~\ref{CHAPTER:AIRFLUOR}.
     \printbibliography[heading=subbibliography]
    \end{refsection}

    \begin{refsection}
  \chapter{JEM-EUSO}
     \label{CHAPTER:JEMEUSO}
     
\label{sec:INTROtoJEMEUSO}

The \acrshort{JEM-EUSO} (Extreme Universe Space Observatory on-board the Japanese Experiment Module) mission is a proposed Ultra-High Energy Cosmic Ray (UHECR) observatory which would be placed on the International Space Station.
JEM-EUSO would use a large volume of atmosphere to observe cosmic ray Extensive Air Showers (EAS) by looking down at the Earth,  a concept first proposed
by Benson and Linsley \cite{1981ICRC....8..145B} in the early 1980s, with the goal of increasing the 
detector exposure by almost an order of magnitude compared to previous UHECR observatories.

As JEM-EUSO would be attached to the ISS, it would be at an altitude of approximately 400 km, orbiting the Earth at a speed of $\sim7~$km/s with an orbital inclination of $51.6^{\circ}$. 
Extensive air showers in the telescope's field of view are observed through air fluorescence emission and diffused Cherenkov light. 
The primary scientific objectives of JEM-EUSO are charged-particle astronomy and astrophysics, as well as
a number of exploratory objectives in physics and atmospheric science. 

Research and development work for JEM-EUSO is ongoing.
At the same time, the JEM-EUSO collaboration is completing several pathfinder experiments, including EUSO-TA, EUSO-Balloon, and Mini-EUSO.
The EUSO-TA project is a joint effort of the JEM-EUSO and Telescope Array collaborations with the aim of placing a small scale JEM-EUSO prototype at the Black Rock Mesa site of Telescope Array.
EUSO-TA's field of view will overlap with the TA fluorescence detector, allowing the simultaneous observation of extensive air showers by both experiments. EUSO-TA will have a finer granularity than the Telescope Array 
Fluorescence detector, allowing it ``see inside'' the observed shower.

The EUSO-Balloon instrument is a similar reduced-scale prototype of JEM-EUSO, which will be flown in a stratospheric balloon. EUSO-Balloon is a project of the French space agency, \gls{CNES}, and
is being built as a technology demonstrator for the hardware and methods to be used in the full space mission. EUSO-Balloon will also measure the UV background looking down from an altitude of 40 km.
Finally, Mini-EUSO is a special project to place a very small version of the JEM-EUSO detector
inside the ISS, observing the atmosphere through a UV transparent window. Mini-EUSO will later be placed outside the ISS to test its operation in vacuum.

This chapter will introduce the JEM-EUSO mission, instrument, and pathfinders. The first sections will present the JEM-EUSO mission
and its science goals, and explain the overall plan of the mission. After that, a more technical introduction to the instrument will be presented.
The last sections will discuss the design and progress of the JEM-EUSO pathfinders. 
The information presented here on JEM-EUSO is taken from work done by the entire collaboration, which can be found in \cite{Takahashi:2009zzc,PurpleBook,Adams:2012tt,TheJEM-EUSO:2013vea,AdamsJr201376} and the references therein.
Specific publications are referenced only in a few instances or in those cases were work outside of JEM-EUSO is cited.

\section{The JEM-EUSO Mission}
The JEM-EUSO instrument is planned to be hosted on-board the \gls{EF} of the Japanese Experiment Module \textit{Kibo} of the \gls{ISS}. 
JEM-EUSO consists of
\begin{inparaenum}[i\upshape)]
\item a main fluorescence telescope sensitive in the near-UV band, and
\item an \gls{AMS}.
\end{inparaenum}
The main telescope will have three Fresnel lenses and a focal surface which is a high-speed digital camera with a 
high pixel density and a large-aperture wide-\gls{FoV}. 
The primary technical achievement of JEM-EUSO would be an order of magnitude increase in total exposure compared to the current generation of UHECR observatories, and the first observation of extensive air showers from space using the 
fluorescence technique.

\begin{figure}
\centering
 \subfigure[UHECR Exposure]{\label{fig:JEMEUSOExposureComparison} \includegraphics[angle=270,width=0.8\textwidth]{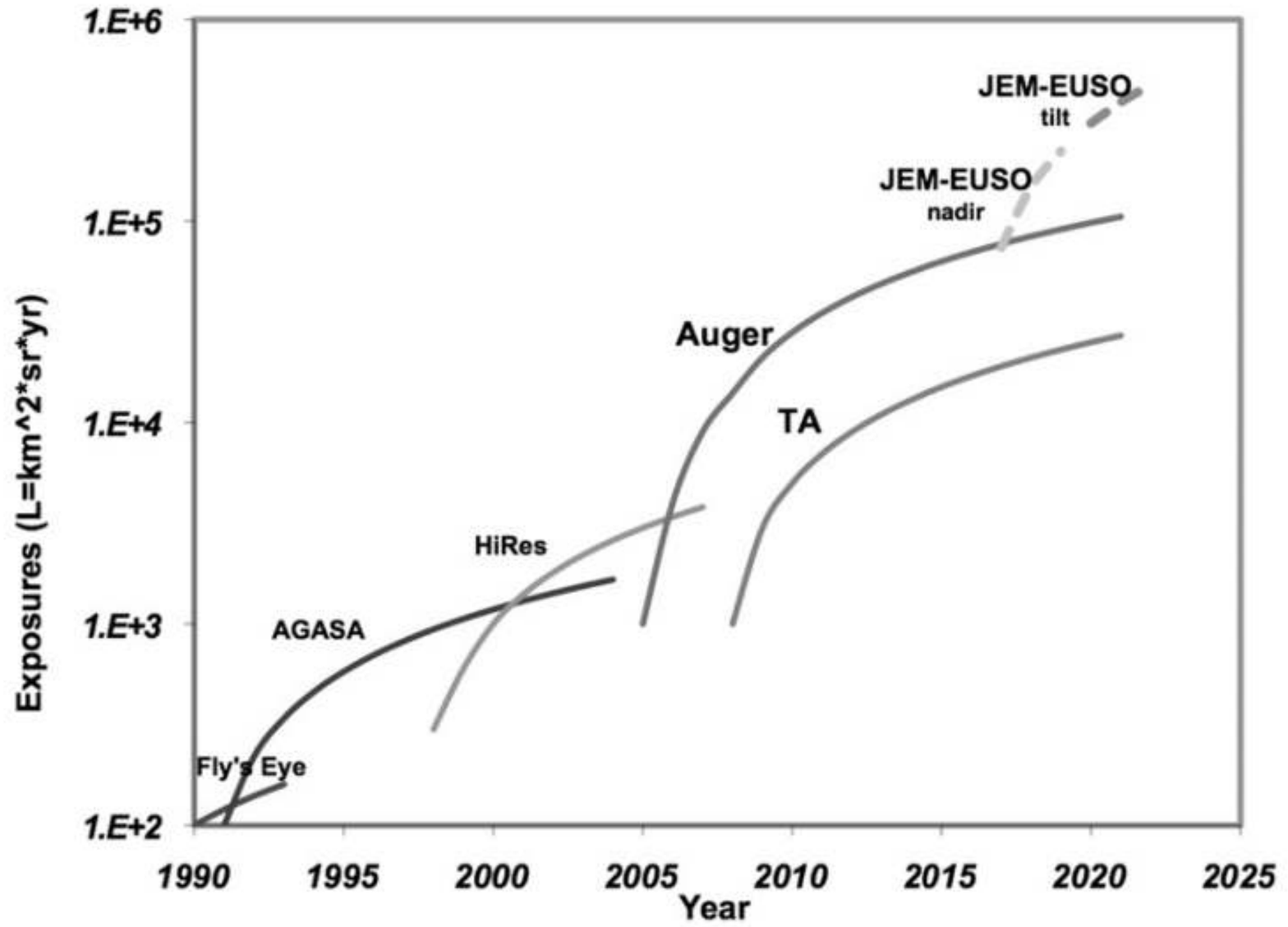}}
 \subfigure[Accelerator Energy]{\label{fig:PanofskyAcceleratorPlot}\includegraphics[width=0.6\textwidth]{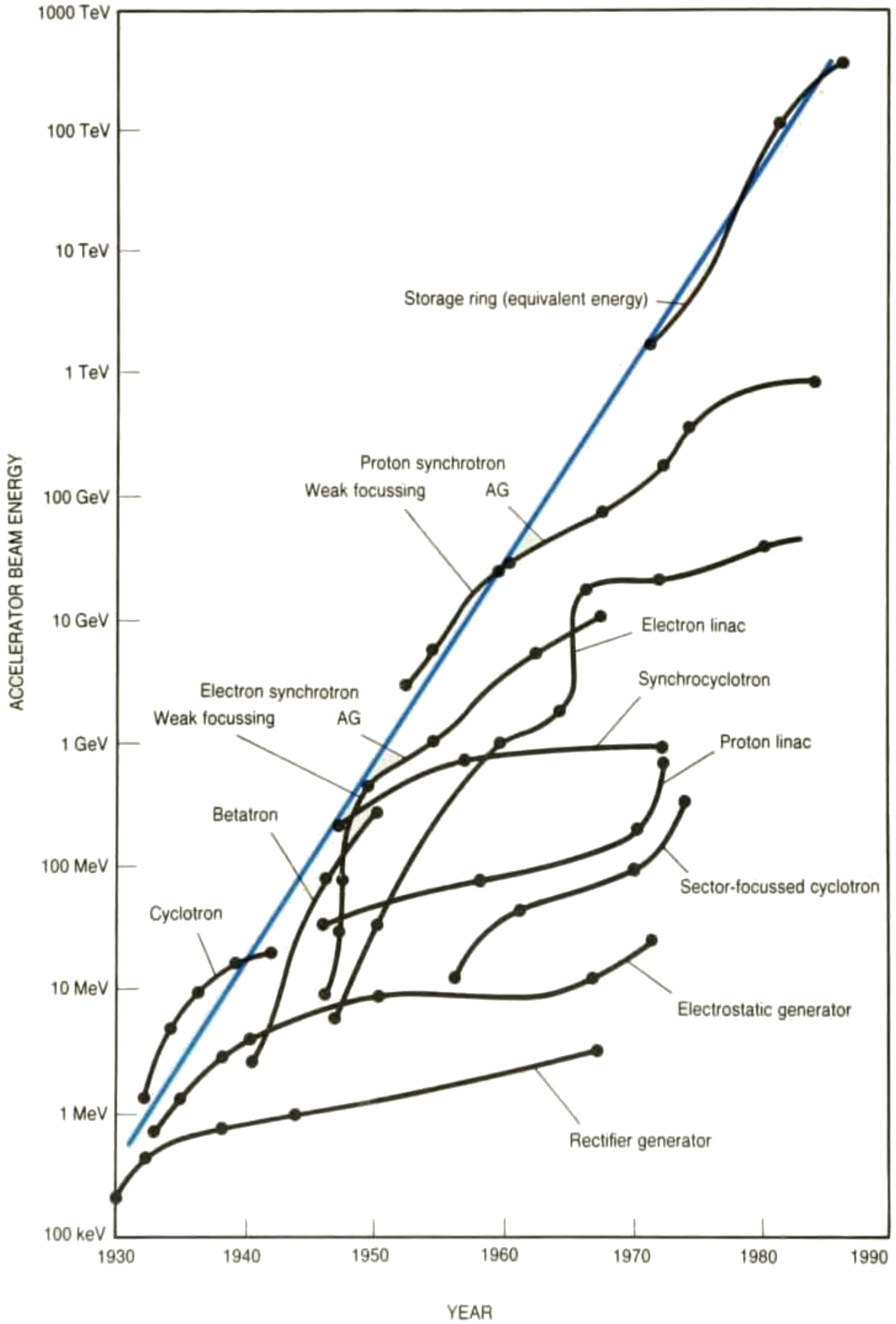}}
\caption[The Exposure of JEM-EUSO Compared to Current UHCER Observatories]{Two figures with no actual relation. 
\fig\ref{fig:JEMEUSOExposureComparison} (from \cite{2013EPJWC..5302001O}) shows a comparison of the estimated JEM-EUSO exposure to that of current and past UHECR observatories.
The expectation is that JEM-EUSO will deliver an order of magnitude increase in total exposure compared to both Auger and TA. 
The second figure, \fig\ref{fig:PanofskyAcceleratorPlot} (from \cite{panofsky:24}) is merely to show an interesting comparison between
the increase in UHECR exposure and the increased of particle accelerator energy. Both quantities increase with successive generations, with new techniques (or observatories in the case of UHECR physics) 
bringing large, often order of magnitude, increases. 
}
\end{figure}

A comparison between the exposure of JEM-EUSO and current or past cosmic ray observatories is shown in \fig\ref{fig:JEMEUSOExposureComparison}. 
As discussed in the last chapters, the Pierre Auger Observatory and Telescope Array have already increased the total exposure by an order of magnitude compared to past ground arrays.
An interesting comparison can be made between \fig\ref{fig:JEMEUSOExposureComparison}, showing the exposure of UHECR observatories, and \fig\ref{fig:PanofskyAcceleratorPlot}, which shows the increase of accelerator energy by year and technology. 
While the energy available using a given accelerator method increases slowly with time,
large leaps are made by the introduction of a new method. The argument can be made that, while ground-based arrays can continue to increase in overall exposure using ever larger surface areas, 
as the Pierre Auger Observatory already extends over an area the size of Luxembourg it is perhaps time to also consider new strategies.

\subsection{Science Goals}
The main scientific goals of JEM-EUSO are related to UHECR astronomy and astrophysics. 
The increase in exposure offered by JEM-EUSO, would result in high statistics above $10^{20}~$eV, the region of the cosmic ray spectrum where only a few sources are expected to dominate due to the GZK-effect.
This would allow:
\begin{enumerate}[1\upshape)]
 \item The identification of sources by high-statistics arrival direction analysis and possibly the measurement of the individual energy spectra in a few sources with high event multiplicity,
 \item The study of the anisotropies of the UHECR sky (at all angular scales), and
 \item A high statistics measurement of the trans-GZK spectrum.
\end{enumerate}
Aside from these primary scientific objectives, several exploratory objectives in astrophysics have been set for the JEM-EUSO mission.
These are areas to which the experiment may contribute, depending on the actual nature of the UHECR flux and include:
\begin{enumerate}[i\upshape)]
 \item The study of the UHE neutrino component which can be achieved by discriminating weakly interacting events through the position of the first visible point and the shower maximum,
 \item The discovery of UHE Gamma-rays, whose shower maximum is strongly affected by the geomagnetic and LPM effects, and
 \item The study of the Galactic and local extragalactic magnetic fields, through the analysis of the ``magnetic point spread function'' of one or more identified sources.
\end{enumerate}

In addition to the astrophysics objectives of the mission, JEM-EUSO would also be 
able to contribute in several areas of atmospheric science due to its continuous monitoring of 
the Earth's atmosphere in the UV band (290-430 nm). This fact would allow a  characterization of atmospheric night-glow and of \gls{TLE}. 
JEM-EUSO would also be able to detect the slow (compared to EAS) UV tracks created by meteors and meteoroids, with important scientific outcomes.


In order to meet these scientific goals, a list of instrumental requirements have been determined.
The statistics collected by the experiment should be on the order of several hundred events above $7~10^{19}~$eV, which implies a total exposure over three years of more than $10^{5}~$km$^{2}$ sr yr.
The angular resolution of the telescope should be better than $3^{\circ}$ for EAS of energies greater than $8~10^{19}~$eV.
Similarly, the energy resolution, as a 68\% interval, should be better than 30\% for EAS of $E_{o} = 8~10^{19}~$eV. 
It would also be preferable if the experiment has some ability to distinguish between nuclei, gamma rays, and neutrinos, which would require a determination of $X_{\text{max}}$ with an uncertainty of less than $120~$g/cm$^{2}$
at $E_{o} = 10^{20}~$eV for showers at zenith angles of 60 degrees. 
As JEM-EUSO will be placed on the ISS, it will observe the full sky with much less than 30\% non-uniformity between hemispheres, which is ideal for anisotropies studies.
The current design of the JEM-EUSO instrument is foreseen to meet or exceed all these requirements. 

\subsection{The Mission}
\begin{figure}[h]
\centering
\includegraphics[angle=270,width=0.9\textwidth]{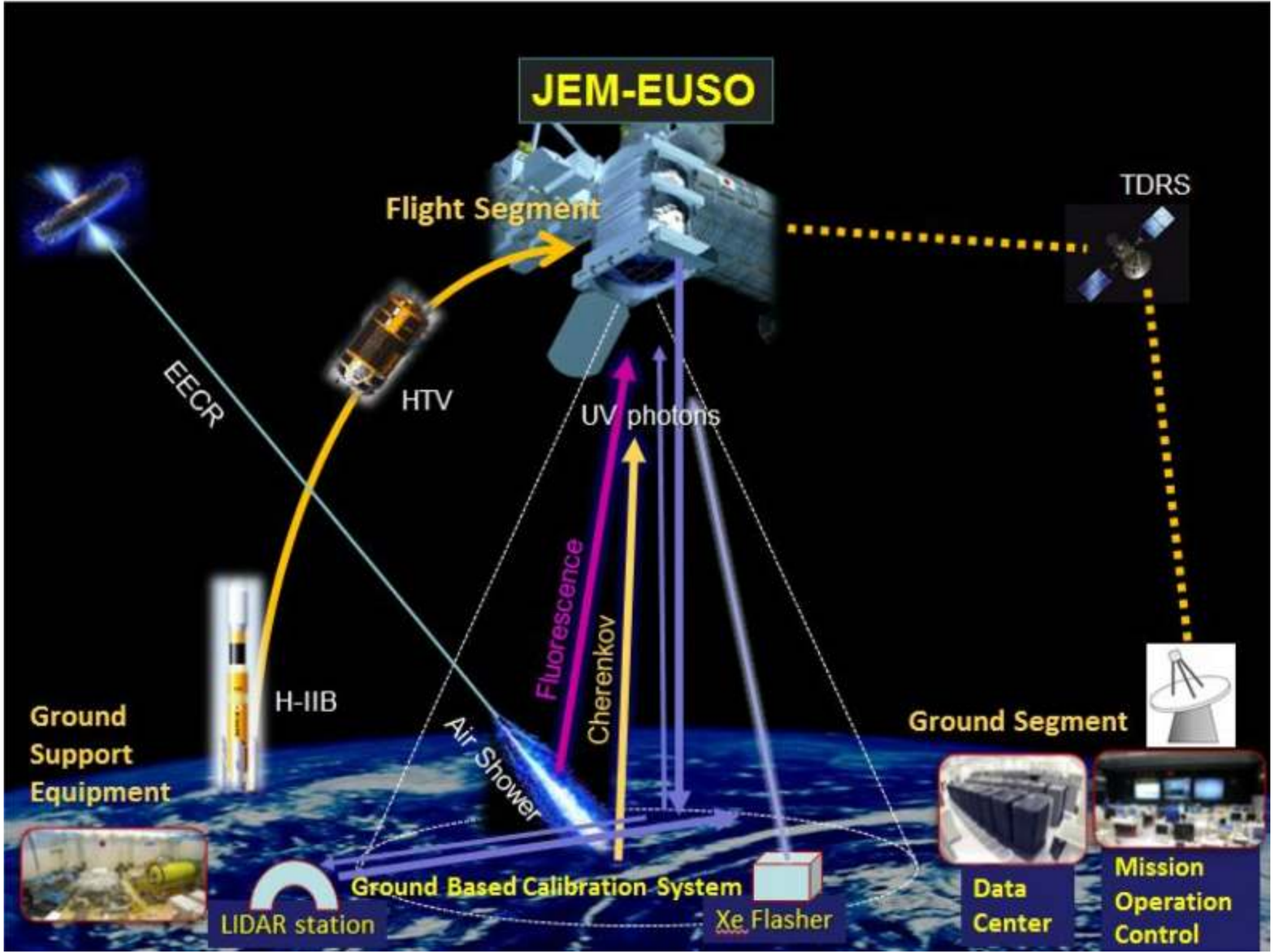}
\caption[Overview of the JEM-EUSO Mission]{\label{fig:JEMEUSOmissionOverview} An overview of the JEM-EUSO mission from launch by H-IIB (or SpaceX) to attachment on the 
International Space Station. See the text for a description of each element.}
\end{figure}

An outline of the JEM-EUSO mission is shown in \fig\ref{fig:JEMEUSOmissionOverview}.
JEM-EUSO directly inherits from the ESA-EUSO proposal, which defined the EUSO concept.
At this time, JEM-EUSO is designed to meet a launch date in 2017.
The current baseline mission plan is that JEM-EUSO will be launched by a Japanese H-IIB rocket and transported to the ISS by an unmanned \gls{HTV}.
The accommodation of JEM-EUSO in the HTV requires that the telescope be cut on two sides so that the maximum diameter of the lenses is 2.65 m and the minimum diameter is 1.9 m.
Another recent launch possibility, however, is the use of the SpaceX Dragon as the transfer vehicle. 
The accommodation of JEM-EUSO in the trunk section of the SpaceX Dragon Spacecraft will require slight modifications to the instrument and an optimization of the instrument shape and size.
This would include a return to a circular shape.
 
In either case, the instrument will then be attached, using the Canadian and Japanese robotic arms, to one of the ports for non-standard payloads of the Exposure Facility (EF) of the JEM.
In order to be accommodated in either the HTV or the Dragon, the JEM-EUSO instrument will be stored in a folded configuration during launch and transport, and the telescope
will be deployed after the attachment procedure is completed. 
Data from the JEM-EUSO mission will be transmitted via \gls{TDRS} to a Mission Operation Center hosted by \acrshort{JAXA} in the Tsukuba Space Center and managed by RIKEN with the support of the collaboration. 
It is also planned to establish several data centers in the major participating countries.

According to the currently planned mission profile, JEM-EUSO should be operated for one to three years in a Nadir configuration -- i.e.\ looking directly towards the ground.
Because the acceptance for low energy events is better in Nadir configuration, this would be done to maximize the number of events at the lowest energies
so that the cosmic ray spectrum measured by JEM-EUSO overlaps well with that measured by the current generation of ground-based detectors (when they agree). 
The instrument will then be tilted (by up to 35 degrees) with respect to Nadir, to increase the viewed volume of atmosphere and maximize the event statistics at the highest energies.

During flight, JEM-EUSO will be calibrated by an on-board calibration system, which will check for changes in photodetection efficiency and lens through-put relative to the 
precise on-ground calibration. A ground-based \gls{GLS} is planned in addition.
The GLS would be a world-wide network combining ground-based Xenon flash lamps and steered UV lasers.
These will be used to generate known optical signatures in the atmosphere with similar optical characteristics to EAS and with a known  event energy, time, and direction. 
There are planned to be 12 ground based units placed at sites around the world, with six locations equipped with Xenon flashers only (\acrshort{GLS-X}) and six equipped with both
flashers and remotely operated steerable lasers (\acrshort{GLS-XL}). The future GLS sites will be chosen for low background light and an altitude above the planetary boundary layer.

\section{The Observation Principle}
  \begin{figure}
\centering
\includegraphics[angle=270,width=0.8\textwidth]{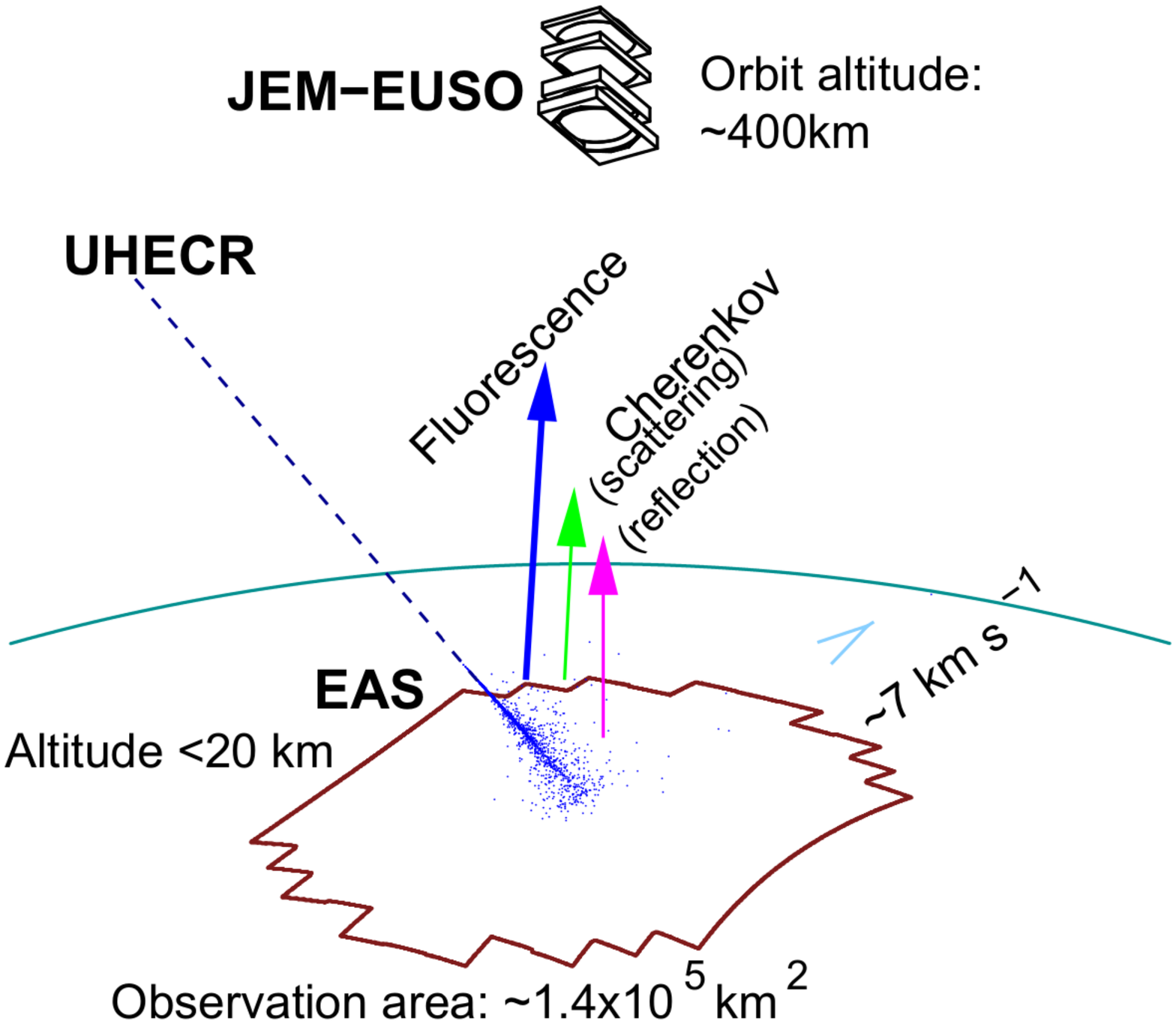}
\caption[The JEM-EUSO Observational Principle]{\label{fig:JEMEUSOprinciple} A diagram of the observational principle behind JEM-EUSO. Shown is an EAS in the field of view of JEM-EUSO, which is orbiting at a height of 400 km and at a speed of 7 km/s.
The field of view of the optics translates into an observed area on ground of $1.4~10^{5}~$km$^{2}$. The EAS is viewed directly, through fluorescence emission, and indirectly through scattered or reflected Cherenkov light.
See the text for a more detailed explanation.  }
\end{figure}

The basic observation principle of the telescope is shown in \fig\ref{fig:JEMEUSOprinciple}.
From its orbit on the ISS, at an altitude of $\sim400~$km, JEM-EUSO
would detect the fluorescence light and Cherenkov emission from EAS occurring in its FoV. 
The fluorescence light from the EAS is emitted isotropically along the shower track, tracing the longitudinal development of the EAS. Because it is emitted isotropically,
this light is observed directly. The Cherenkov radiation, on the other hand, is emitted in a cone around the axis of the EAS, 
and the Cherenkov light which would be seen by JEM-EUSO would be that which was either scattered by the atmosphere or reflected off the ground or a cloud top.
This reflected signal is known as the ``Cherenkov mark'', and can be used as a tool to gain additional information about the geometry of the shower. 

A UHECR produces an EAS with approximately $10^{11}$ particles in the region of the shower maximum. These particles cause the emission of fluorescence 
photons as they deposit their energy in the atmosphere, allowing the detection of EAS by the air fluorescence technique, which was discussed in some detail in chapter~\ref{chapt:ObservationOfUHECR}. 
The experimental determination of the air fluorescence yield was discussed in chapter~\ref{sec:INTROtoAFY}, and the yield can be taken as being approximately 4 photons per meter per charged particle. 
This results in the emission of $\sim 10^{15}$ photons during the development of an UHECR EAS. 

As JEM-EUSO views the shower development from $\sim400~$km away, the number of photons which actually reach the telescope will be strongly attenuated.
The solid angle subtended by telescope with a collection area of about 5 m$^{2}$ at a distance of 400 km is on the order of $10^{-11}~$sr.
Thus several thousand photons will reach the aperture of JEM-EUSO from each EAS in the field of view in a time window of about one hundred $\mu$s, depending on the atmospheric conditions.
Given the throughput of the JEM-EUSO optics and the quantum efficiency of the photodetector (c.f.\ table~\ref{tab:JEMEUSOCharacteristics}), these several thousand photons will yield several hundred photoelectrons. 
As each photoelectron registered occupies $\sim 8~$ns, the focal surface of JEM-EUSO will operate in \emph{single photoelectron counting} mode.

\begin{figure}
\centering
\includegraphics[angle=270,width=0.8\textwidth]{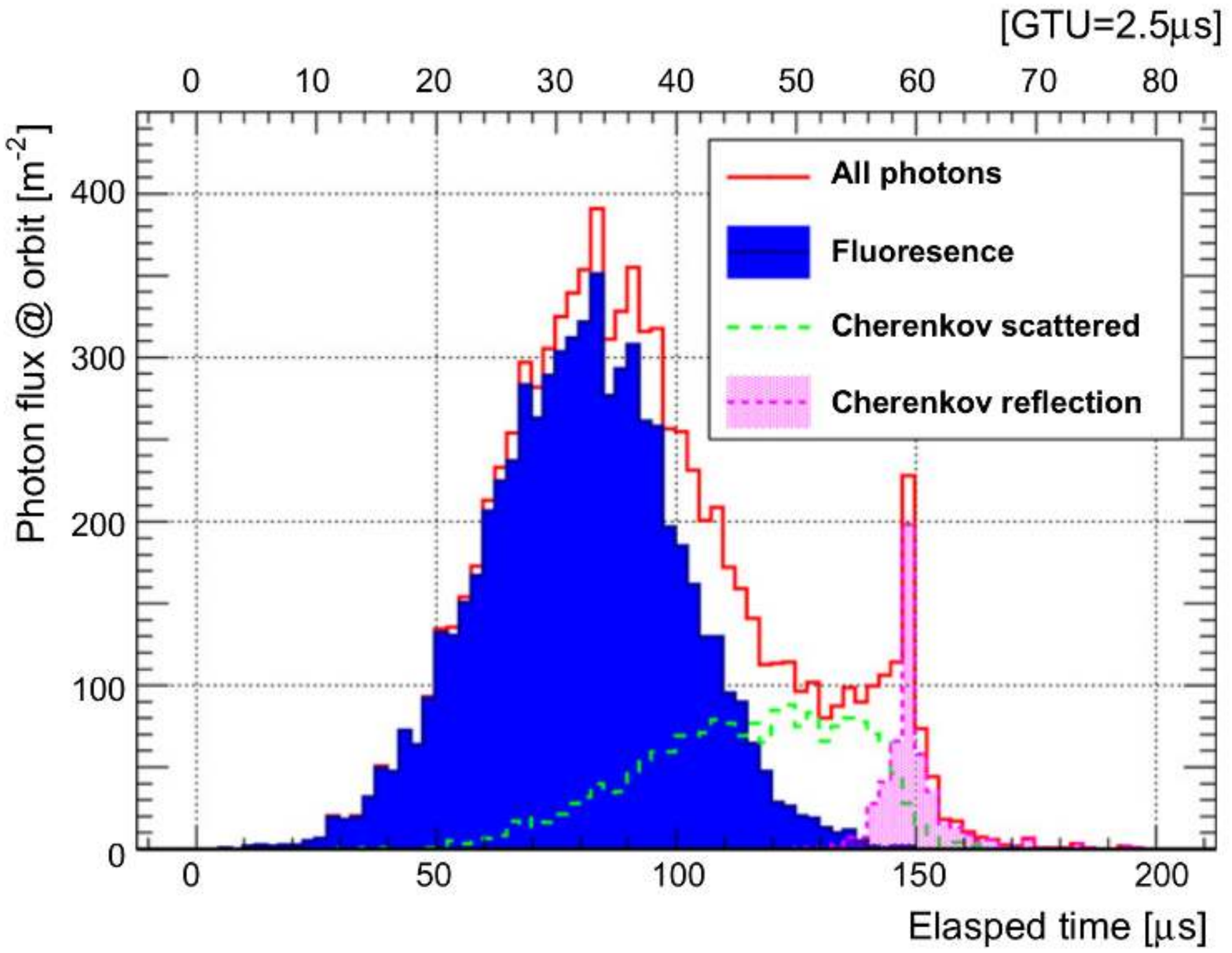}
\caption[Photon Flux at the JEM-EUSO Aperture]{\label{fig:PhotonArrival}  A plot of the number of photons arriving at the entrance aperture of JEM-EUSO versus time for an EAS with $E_{o}=10^{20}~$eV with a zenith
angle of $\theta=60^{\circ}$ (from simulations). The photon flux is broken down by origin. The fluorescence component dominates the overall signal. The end of the shower (on the ground or a cloud top) 
is marked by reflected Cherenkov light. As can be seen, the development time of the EAS is on the order $200~\mu$s.}
\end{figure}

The distribution of photons from an EAS according to arrival time is shown in \fig\ref{fig:PhotonArrival}. 
As can be seen, the shower develops over a time span of $\sim 200~\mu$s.
 The fluorescence light, shown by the solid blue, is the dominant contribution to the signal. 
It is also the most prompt part of the signal to arrive at the telescope. The Cherenkov contribution, on the other hand, is lower and arrives later in time. 
The Cherenkov reflection, shown in \fig\ref{fig:PhotonArrival} by the red area provides a time mark for the end of the shower and thus gives information on the altitude of the reflection. 
The domination of the fluorescence signal (for low albedo) allows the energy of the UHECR to be determined with only small corrections for the Cherenkov component.
At the same time, the Cherenkov light can be used as an additional signal using the technique of \cite{Unger:2008uq}, for example.

As JEM-EUSO is at an altitude of around 400 km, the shower maximum, which generally occurs below 20 km altitude ($\sim 10~$km in \fig\ref{fig:PhotonArrival}), will be at an almost constant distance for a given location in the field of view.
This greatly reduces the importance of proximity effects in the shower reconstruction. 
This, and the domination of the fluorescence component, are beneficial characteristics of observation from space, and
because of them JEM-EUSO can be understood as essentially turning a volume of atmosphere into a Time Projection Chamber in which the drift velocity is that of light. 

\section{The JEM-EUSO Instrument}
\label{JEMEUSOInstrument}
\begin{figure}
\centering
\includegraphics[angle=270,width=0.8\textwidth]{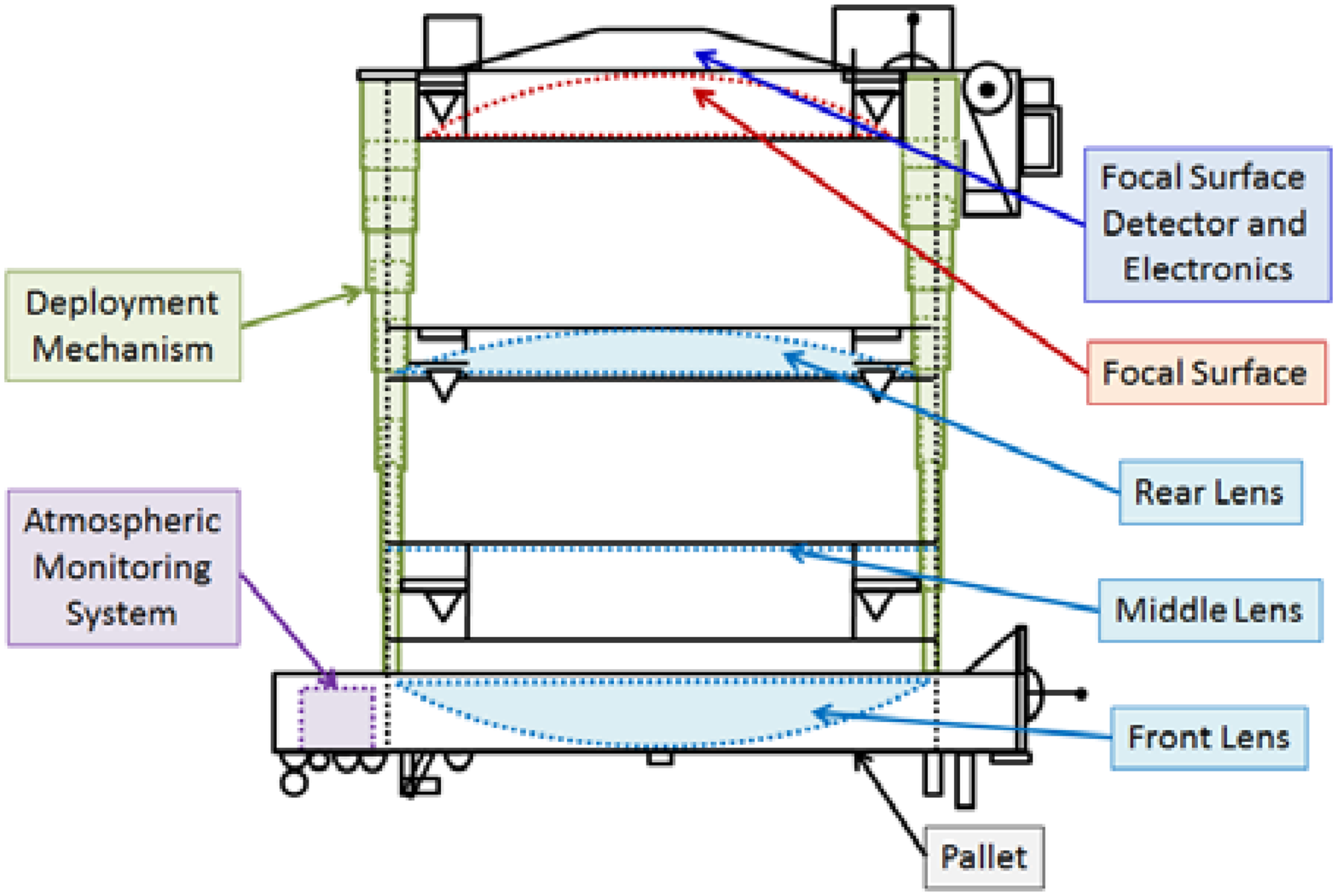}
\caption[Overview of the JEM-EUSO Instrument]{\label{fig:JEM-EUSOConceptOverView}  A conceptual overview of the JEM-EUSO instrument after deployment. The telescope is composed of three lenses, a focal surface with electronics, and an atmospheric monitoring system. 
All of these are mounted on a transport palette. The instrument is transported in a stowed (collapsed vertically) configuration, and deployed (expanded) after attachment.}
\end{figure}

The actual JEM-EUSO instrument consists of three main parts:
\begin{inparaenum}[i\upshape)]
 \item the main telescope,
 \item a calibration system, and
 \item an atmospheric monitoring system.
\end{inparaenum}
A conceptual overview of the JEM-EUSO telescope is shown in \fig\ref{fig:JEM-EUSOConceptOverView}. 
The telescope itself is made up of four main components,
\begin{inparaenum}[a\upshape)]
 \item light collecting optics consisting of three lenses,
 \item the focal surface,
 \item focal surface electronics, and 
 \item the support structure.
\end{inparaenum}
Each of these components will be presented in turn.
The mission parameters give strict limits on the weight and power budget of the instrument, which heavily influence the chosen designs.

\subsection{Optics}
The main telescope has a total of three lenses. The first and third lens are each curved double-sided Fresnel lenses with an external diameter of 2.65 m.
The middle lens has a curved precision Fresnel lens on one side with a diffractive optical element on the other.
The combination of these three lenses gives a full angle FoV of $60^{\circ}$ and an angular resolution of $0.07^{\circ}$.
This resolution corresponds to 550 m on the ground at an altitude of 400 km and facing Nadir.

A system of plastic Fresnel lenses was chosen because a light-weight design is a necessity for a space mission.
In addition, the Fresnel-type lens have the advantage of a higher transmission factor, which is 
critical when observing only several thousands of photons from a given EAS. The central diffractive lens was added to the design in order 
to correct the chromatic aberration, and also acts as a field lens due to its location near the Aperture Stop.

\begin{figure}
\centering
\subfigure[Baseline]{\label{fig:JEM-EUSOlensdesigns:Baseline} \includegraphics[angle=270,width=0.47\textwidth]{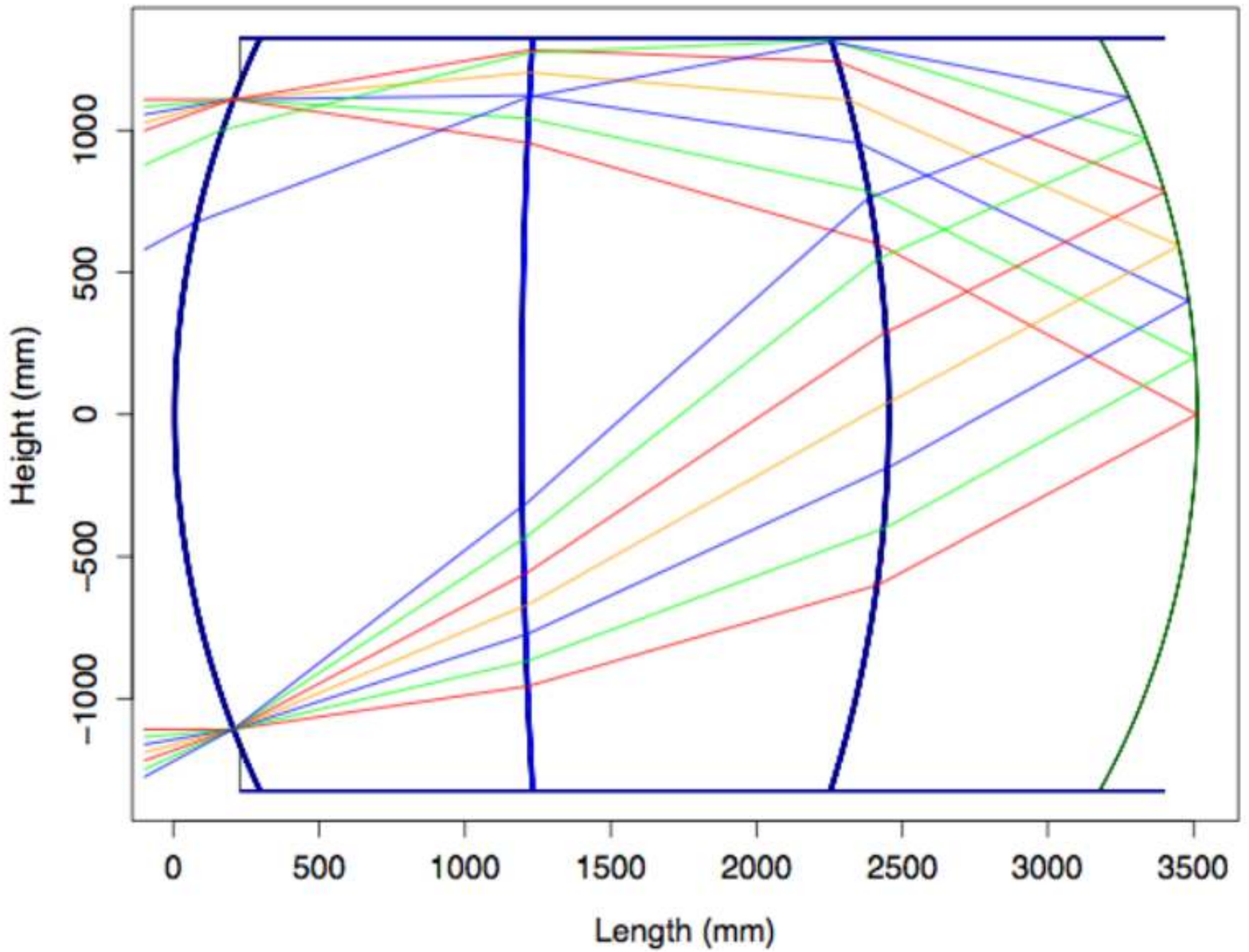}}
\subfigure[Advanced]{\label{fig:JEM-EUSOlensdesigns:Advanced} \includegraphics[angle=270,width=0.48\textwidth]{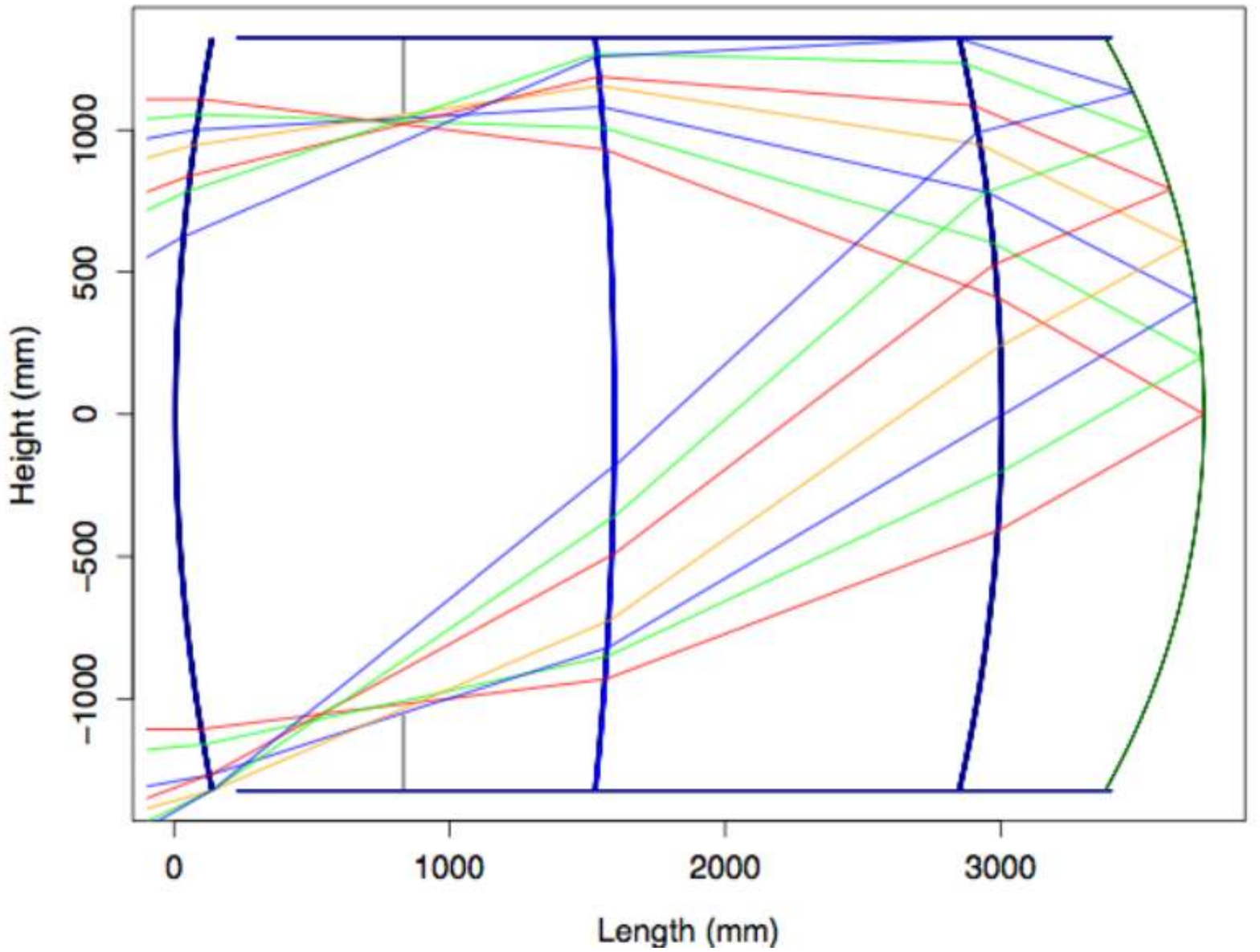}}
\caption[JEM-EUSO Optics Designs]{\label{fig:JEM-EUSOlensdesigns}  Ray tracing simulations of the Baseline, \fig\ref{fig:JEM-EUSOlensdesigns:Baseline}, and Advanced, \fig\ref{fig:JEM-EUSOlensdesigns:Baseline}, optics designs.
The baseline design is a three lens system with each lens made of PMMA. The advanced design uses the same number of lenses, but with the first lens being made of CYTOP.
See the text for more details.}
\end{figure}

\begin{figure}
\centering
\subfigure[Photograph of the Bread Board Model]{\includegraphics[angle=270,width=0.8\textwidth]{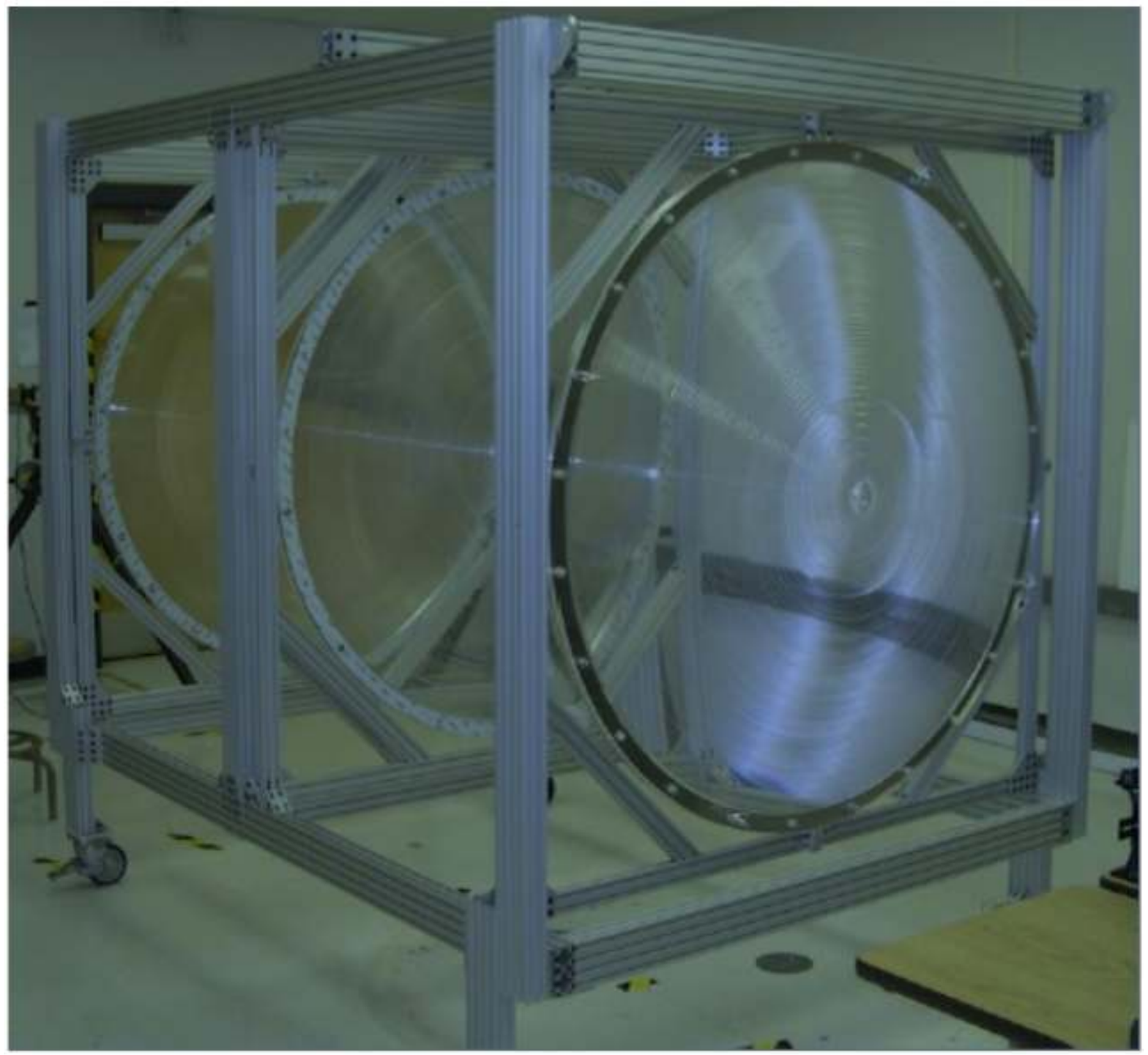}}
\subfigure[Optics Requirements]{\label{fig:JEM-OpticsRequirements}\includegraphics[angle=270,width=0.8\textwidth]{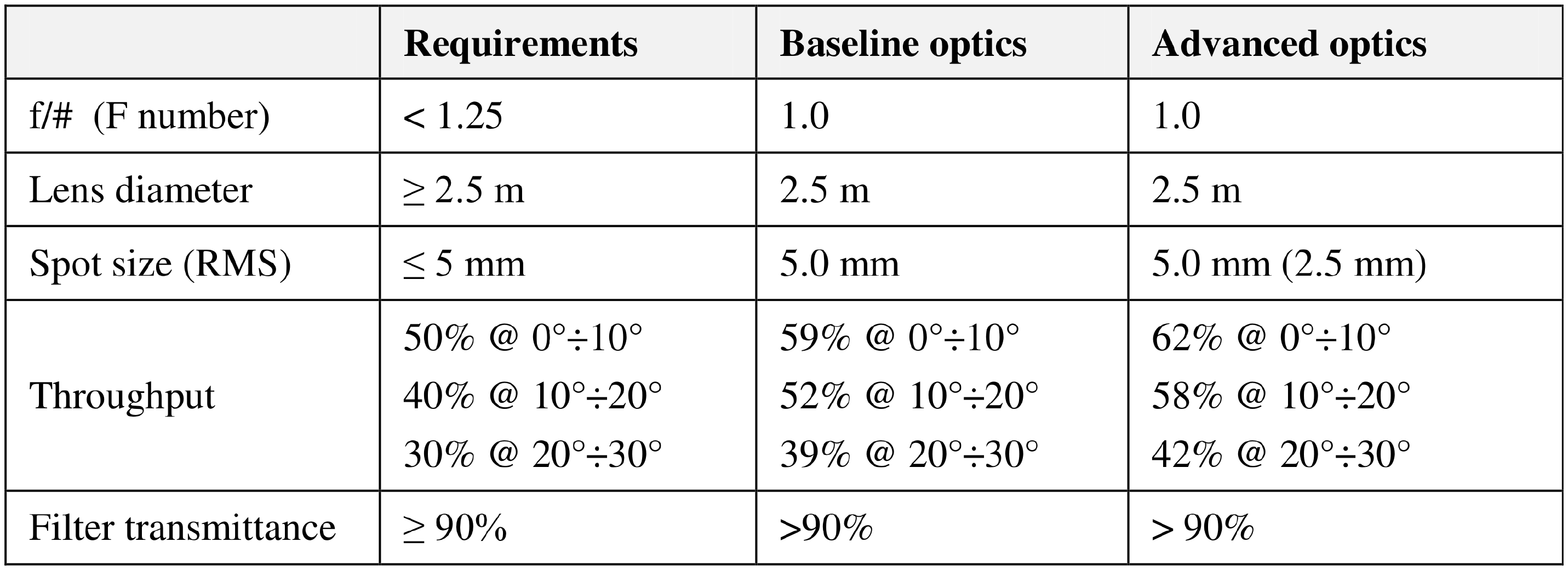}}
\caption[Optics Requirements]{\label{fig:JEM-EUSOlensBBM} A photograph of the Bread Board Module (BBM) of the JEM-EUSO optics, and a table of the optical requirements. The table shows the requirements 
and the specifications of both the baseline and advanced optics designs. This photograph of the BBM was taken 
at UAH, Huntsville, where the lenses underwent testing. }
\end{figure}

Two possible lens materials have been studied, CYTOP and \acrshort{PMMA}-000. CYTOP (basically transparent Teflon) provides less dispersion, higher transmittance, and a generally lower refractive index than PMMA-000. 
CYTOP also has a greater resistance to damage caused by exposure to atomic oxygen in orbit.
At the same time, CYTOP is more expensive and heavier than PMMA. The baseline design is for all three lenses to be made from PMMA, while the ``advanced design'' is
for the front lens to made of CYTOP with the inner two lenses made of PMMA. Each of these designs is shown in \fig\ref{fig:JEM-EUSOlensdesigns}.

A full-scale set of JEM-EUSO lenses, i.e.\ a so-called \gls{BBM}, has been manufactured at RIKEN using a large ultra-precision ($\sim 4~$nm) diamond turning
machine.
A photograph of the BBM lens is shown in \fig\ref{fig:JEM-EUSOlensBBM}. 
It is worth mentioning that the expertise in lens manufacturing at RIKEN is an invaluable asset for the JEM-EUSO collaboration.
The optics requirements are shown in \fig\ref{fig:JEM-OpticsRequirements}, and include a point-spread function of less than 5 mm, to be compared to a pixel size of 2.88 mm, and a throughput of 50\% at normal incidence.
The BBM shown in \fig\ref{fig:JEM-EUSOlensBBM} underwent testing at Marshall Space Flight Center in Huntsville, Alabama, and was found to meet or exceed all the optics system requirements. 

\subsection{Focal Surface}

\begin{figure}
\centering
\includegraphics[angle=270,width=0.8\textwidth]{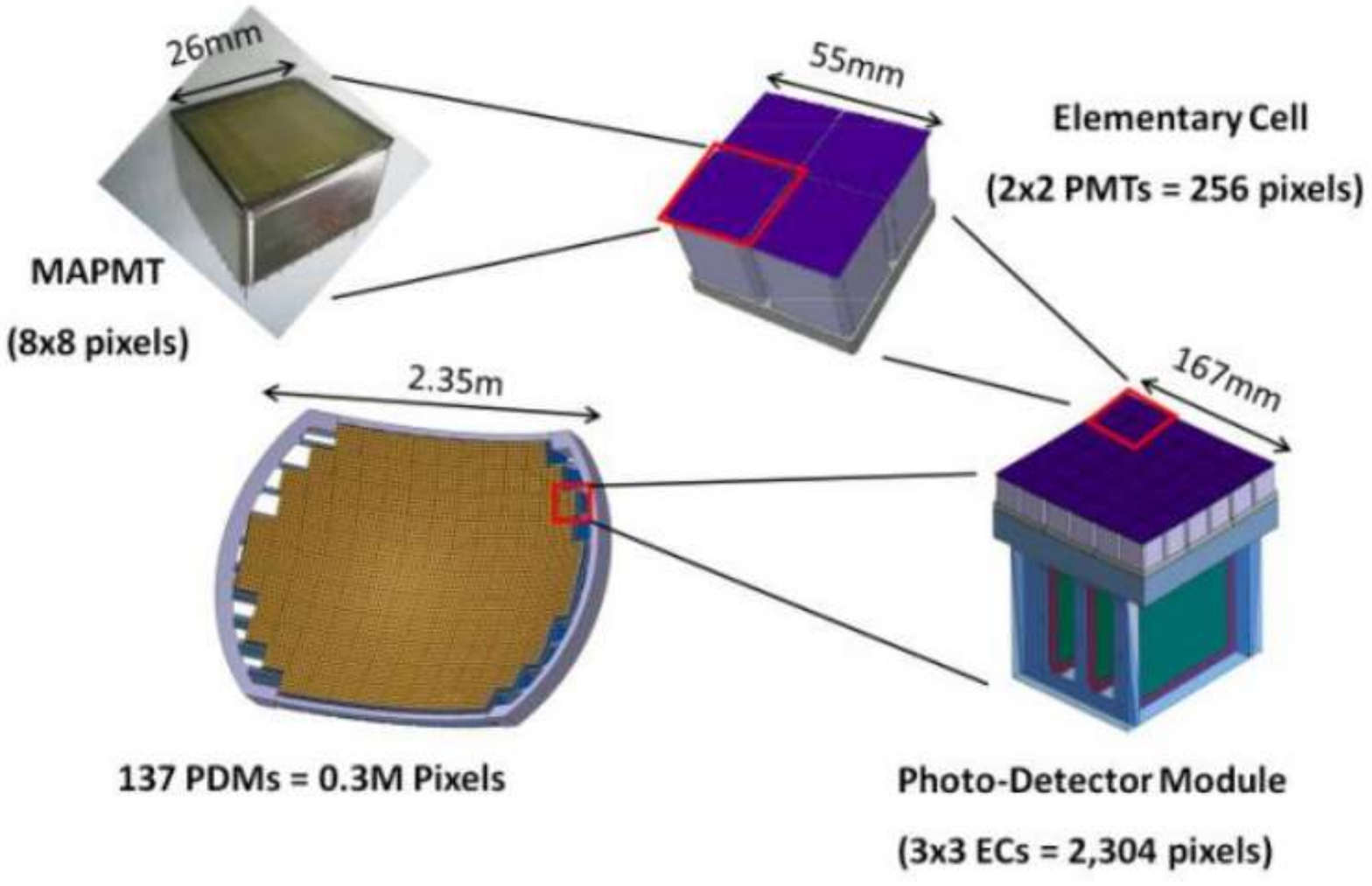}
\caption[Breakdown of the Focal Surface]{\label{fig:JEM-EUSOfocalsurfacebreakdown}  A conceptual breakdown of the JEM-EUSO focal surface. The total focal surface is made of 137 Photodetection Modules (PDM).
Each PDM is made of 9 Elementary Cells (EC) and their read-out electronics. Every EC is composed of 4 multi-anode photomultiplier tubes, each having 64 pixels.}
\end{figure}

The heart of the JEM-EUSO instrument is the focal surface, shown in \fig\ref{fig:JEM-EUSOfocalsurfacebreakdown}.
The focal surface is composed of 137 \gls{PDM}, the largest self-triggering element of JEM-EUSO. 
Each PDM is built from 9 \gls{EC}, the smallest flat surface, with each EC composed of 4 Hamamatsu M64 \glspl{MAPMT}.
This gives a total of $\sim 5000$  64-pixel PMTs, making the JEM-EUSO focal surface a $0.3~$Mpixel camera. The focal surface is approximately 2.3 m in diameter, with a 2.5 m radius of curvature.

Each MAPMT is read-out through a dedicated \gls{ASIC}.
The ASIC used in JEM-EUSO is the \textit{Spatial Photomultiplier Array Counting and
Integrating Readout Chip} (\acrshort{SPACIROC}), which has been designed especially for JEM-EUSO by the $\Omega$mega microelectronics group of \acrshort{CNRS}. This ASIC includes
both a photon counting circuit and a charge integration read out.
In the current design, each EC is powered by an individual high voltage power supply.
JEM-EUSO's placement on the ISS results in a strict power limitation of less than 1 kW during operation, and this precludes the use of a resistive power-supply for the MAPMTs. 
In addition, lightning, meteors, TLE, or man made light (cities) can be expected to give photon fluxes up to 10$^{6}$ times higher than the background level, which is estimated to be 500 photons/m$^{2}$ sr ns or $\sim600~$kHz/pixel. 
As studying these phenomena is a science goal of JEM-EUSO, the dynamic range of the instrument must also span these 6 orders of magnitude, and, as the MAPMTs must be protected from large currents, the switches must operate in 2 to 5 $\mu$s. 
The chosen solution is a \gls{CW-HVPS} with an integrated switch system. The CW-HVPS has been designed especially for JEM-EUSO and is discussed in more detail in chapter~\ref{CHAPTER:CWHVPStests}.
 
\subsubsection{The M64 Photomultiplier Tube}

\begin{figure}
\centering
\includegraphics[angle=270,width=0.8\textwidth]{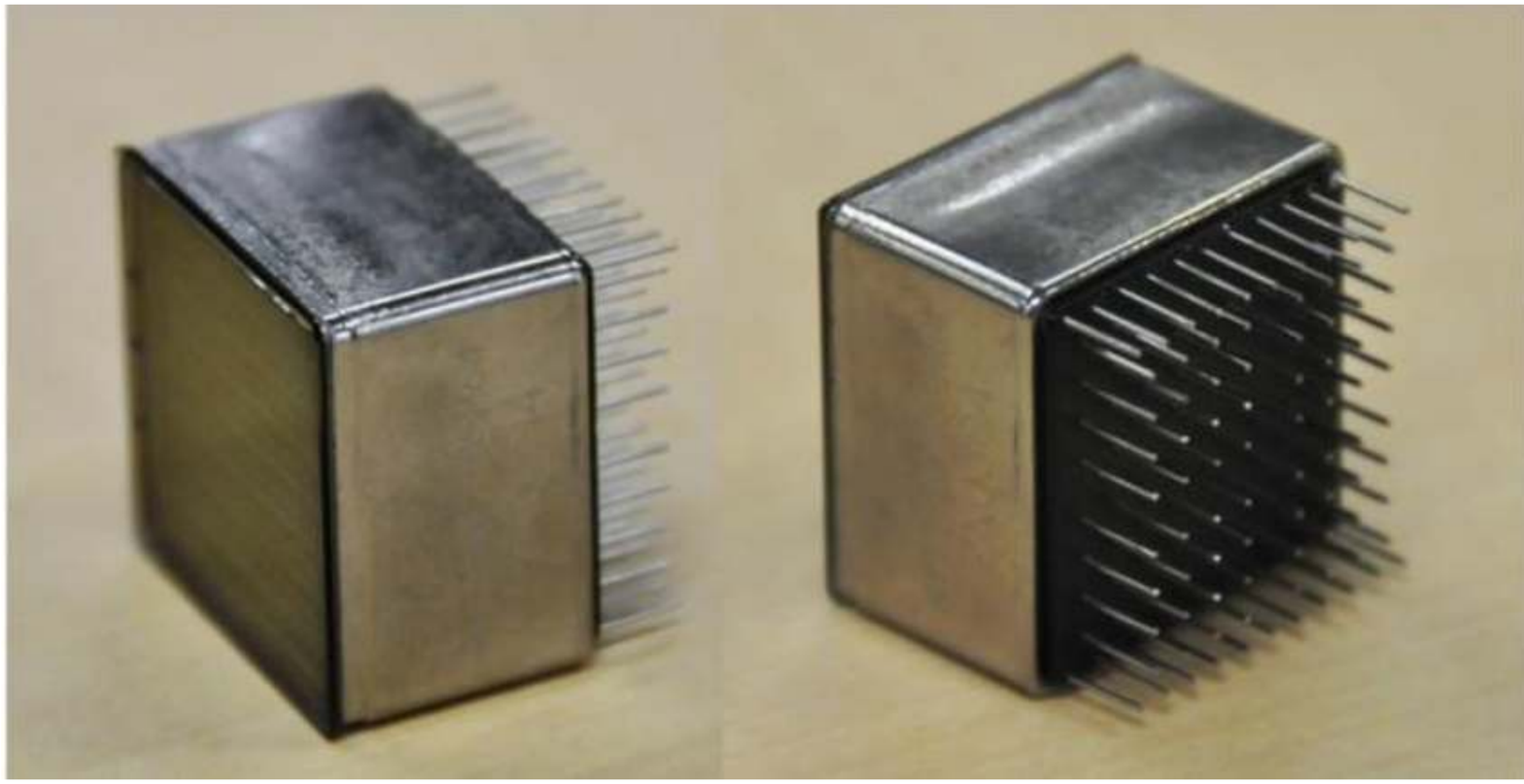}
\caption[The Hamamatsu M64 MAPMT]{\label{fig:JEM-EUSOM64-PMT} Two photographs of the Hamamatsu R11265-M64 multi-anode PMT, taken from two different angles. On the left side, the photocathode of the M64 can be seen, along with the 
pixel structure. On the right, the M64 is turned to show the anode pins. The 11 pins on the top and bottom are the cathode and dynode pins (some pins are repeated).}
\end{figure}

The primary considerations for the choice of photodetector in JEM-EUSO are speed, pixel density, detection efficiency, and a low background count rate. 
Due to this, a multianode vacuum photomultiplier tube is the preferred photodetector. 
For JEM-EUSO the chosen photodetector is the Hamamatsu R11265-M64 multi-anode PMT, which is shown in \fig\ref{fig:JEM-EUSOM64-PMT}. 
This PMT has been developed for JEM-EUSO as a collaboration between Hamamatsu Photonics and RIKEN.

The M64 has 64 individual pixels, each 2.88 mm square, and an ultra bi-alkali photocathode with a quantum efficiency of 35-45\% for light in the 290 to 430 nm wavelength range. 
The MAPMT amplifies photoelectrons by means
of a stack of 12 metal channel dynodes, with typical gains of $10^{6}$ at a cathode voltage of 900 V.

\begin{table}
 \begin{center}
\begin{tabulary}{1.0\textwidth}{LL}
\toprule
\multicolumn{2}{l}{Hamamatsu R11265-M64 Multi-anode Photomultiplier}\\
 \cmidrule(l){1-2}
Specification & Value\\
\cmidrule(l){1-2}
Physical Dimensions & $26.2~$mm$~\times 26.2~$mm$~\times 20.5~$mm \\
Mass & 27.3 g\\

Photocathode Material & Super Bi-alkali\\
Spectral Range & 185 nm to 650 nm\\
Quantum Efficiency & $> 35\%$\\

Pixels & 64 ($8\times8$) \\
Pixel Side & 2.88 mm \\
Pixel Pitch & 2.88 mm \\
Sensitive Area &  23.04 mm$~\times23.04~$mm\\

Number of Dynodes & 12 \\
Maximum Supply Voltage & 1100 V\\
Gain & $10^{6}$ at 900 V\\

Dark Current & 0.4 nA\\
Anode pulse Rise-Time & $\sim 1~$ns\\
Anode Gain Uniformity  & 1:3\\
Pixel Cross-Talk & $\sim 1\%$\\
Operating Temperature & $-10^{\circ}$ to $30^{\circ}~$C\\
MAPMT Anode Capacitance & $\sim2.8~$pF \\
Magnetic Field Sensitivity & $\sim0.1$ Relative Gain variation at 2 G\\
\bottomrule
\end{tabulary}
\caption[M64 Characteristics]{
\label{tab:M64MAPMTCharacteristics}
A table showing some characterisitics of the Hamamatsu R11265-M64 multi-anode photomultiplier tube.
}
 \end{center}
\end{table}

A summary of the characteristics of the M64 can be found in table~\ref{tab:M64MAPMTCharacteristics}. 
The rate of photons from an EAS on the focal surface is extremely low, and so the most beneficial read-out strategy is to use photon counting. This means counting 
the individual anode pulses coming from the collection and multiplication of single photoelectrons.  
At the same time, the observation of other atmospheric phenomena, such as TLE, implies a photon flux several orders of magnitude larger. 
In this situation, the arrival rate of photons is such that anode pulses from single photoelectrons will overlap, giving large pulses of light or a DC level, depending on the time profile of the phenomena.

To accommodate these two response modes, the frontend electronics contain both a single photoelectron counting part (a preamplifier and discriminator) and a charge integration part.
The response of photomultiplier tubes in single photoelectron mode will be introduced in detail in chapter~\ref{CHAPTER:PMT}, and the M64 will be studied in detail in the main body of this thesis.

\subsubsection{Elementary Cells}
\begin{figure}
\centering
 \subfigure[Diagram of the EC]{\label{fig:JEM-EUSOM-EC:diagram} \includegraphics[angle=270,width=0.8\textwidth]{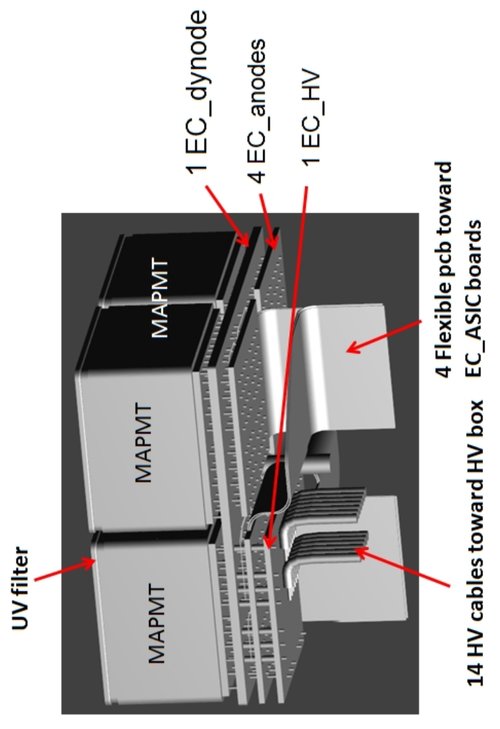}}
\subfigure[Photograph of an EC]{\label{fig:JEM-EUSOM-EC:photo}\includegraphics[angle=270,width=0.9\textwidth]{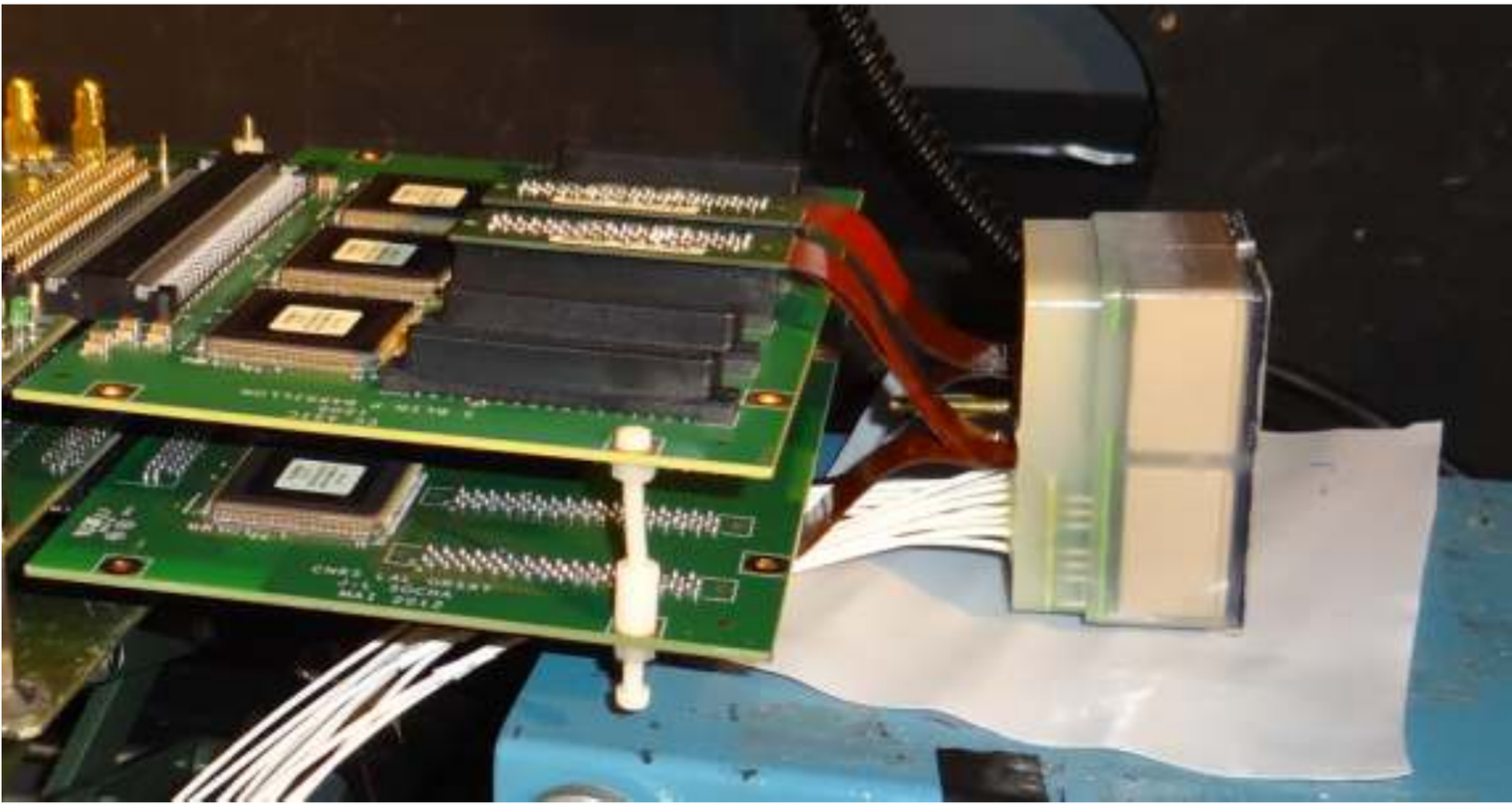}}
\caption[Photographs of ECs]{\label{fig:JEM-EUSOM-EC} \fig\ref{fig:JEM-EUSOM-EC:diagram} shows a diagram of the EC design for EUSO-Balloon. Each element of the EC is labeled.
The EC is made of 4 MAPMT, each with an attached BG3 filter (Labeled UV filter).
The EC\_HV board takes the high voltages from the power supply and distributes them to each element (dynode) of the MAPMT through the EC\_Dynode board. 
The EC\_Anode board connects each of the  anode pins to 1 of 4 flat ribbon cables, which take the anode signals to the read-out ASIC. (Labeled EC\_ASIC). 
The photograph in \fig\ref{fig:JEM-EUSOM-EC:photo} shows an assembled and potted EC connected to two ASIC boards in the black box at APC.}
\end{figure}

Within the focal surface, the photomultipliers are grouped into squares of four, called Elementary Cells. 
A diagram of an EC is shown in \fig\ref{fig:JEM-EUSOM-EC:diagram}.
Each M64 has a Schott BG3 filter glued on the 
photocathode. Several figures showing details of the EC are shown in \fig\ref{fig:JEM-EUSOM-EC}.
The EC is the largest flat area of the focal surface, as the PDM are curved to match the radius of curvature of the focal surface. 
For EUSO-Balloon, the ECs are potted, and this is also foreseen for JEM-EUSO. 

A photograph of an assembled and potted EC is shown in \fig\ref{fig:JEM-EUSOM-EC:photo}. 
Each EC is individually powered, and, due to the high density of pins (64 anode plus 14 high voltage pins per M64 in an area of $\sim69~$mm$^{2}$), the design of the EC boards is ``tight''. 
This fact complicates the mapping between pixels and the output (flat ribbon) cables which lead to the front end electronics. 
For JEM-EUSO itself another design is in development which would place the ASICs directly on the back of the MAPMTs. This would reduce the weight by dispensing with the ASIC board, and also improve
the MAPMT signal quality. 

\subsubsection{Photodetection Modules}

\begin{figure}
\centering
 \subfigure[]{\label{fig:JEM-EUSO-PDM:TA}\includegraphics[angle=270,width=0.48\textwidth]{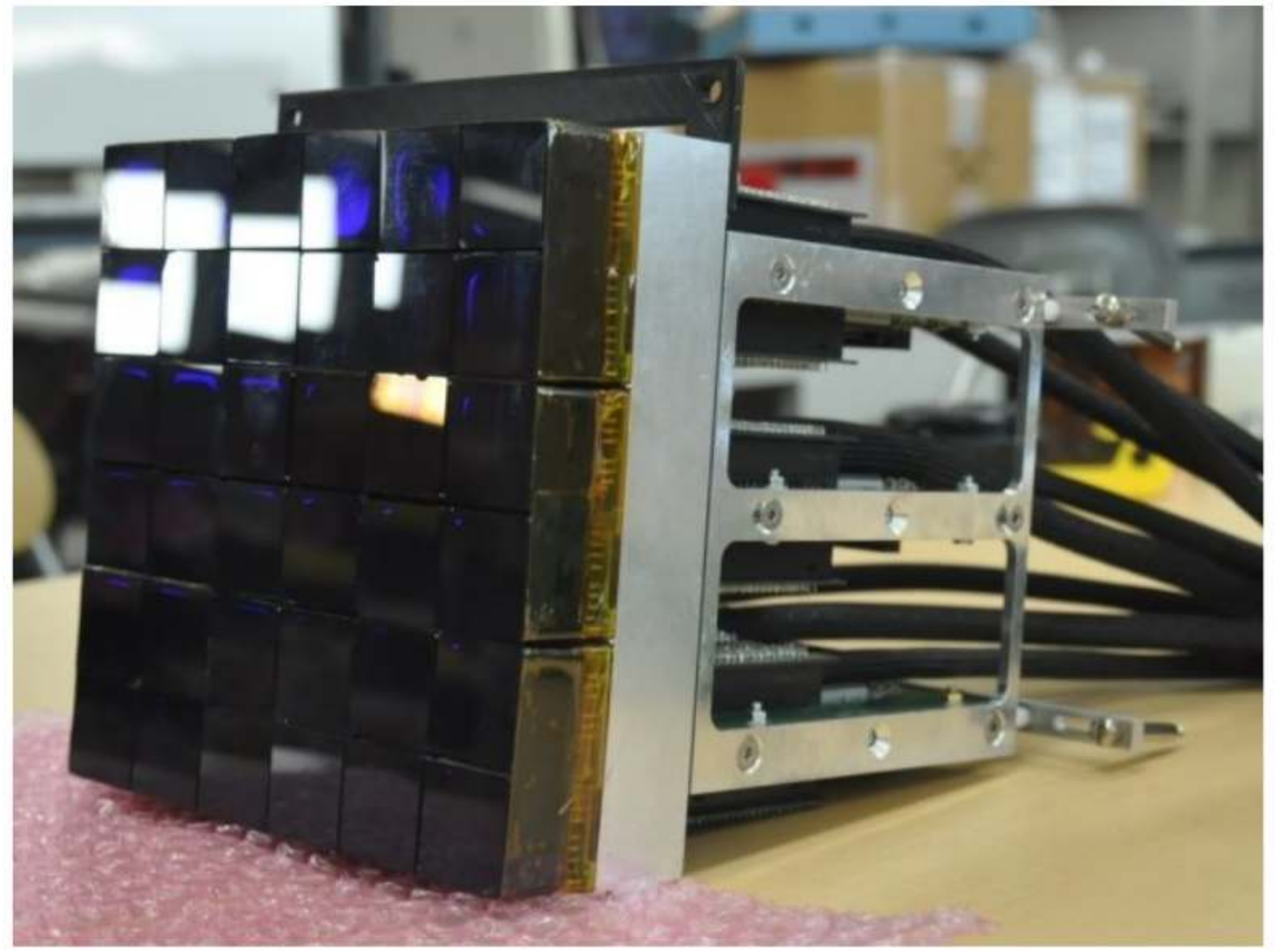}}
 \subfigure[]{\label{fig:JEM-EUSO-PDM:Balloon}\includegraphics[angle=270,width=0.48\textwidth]{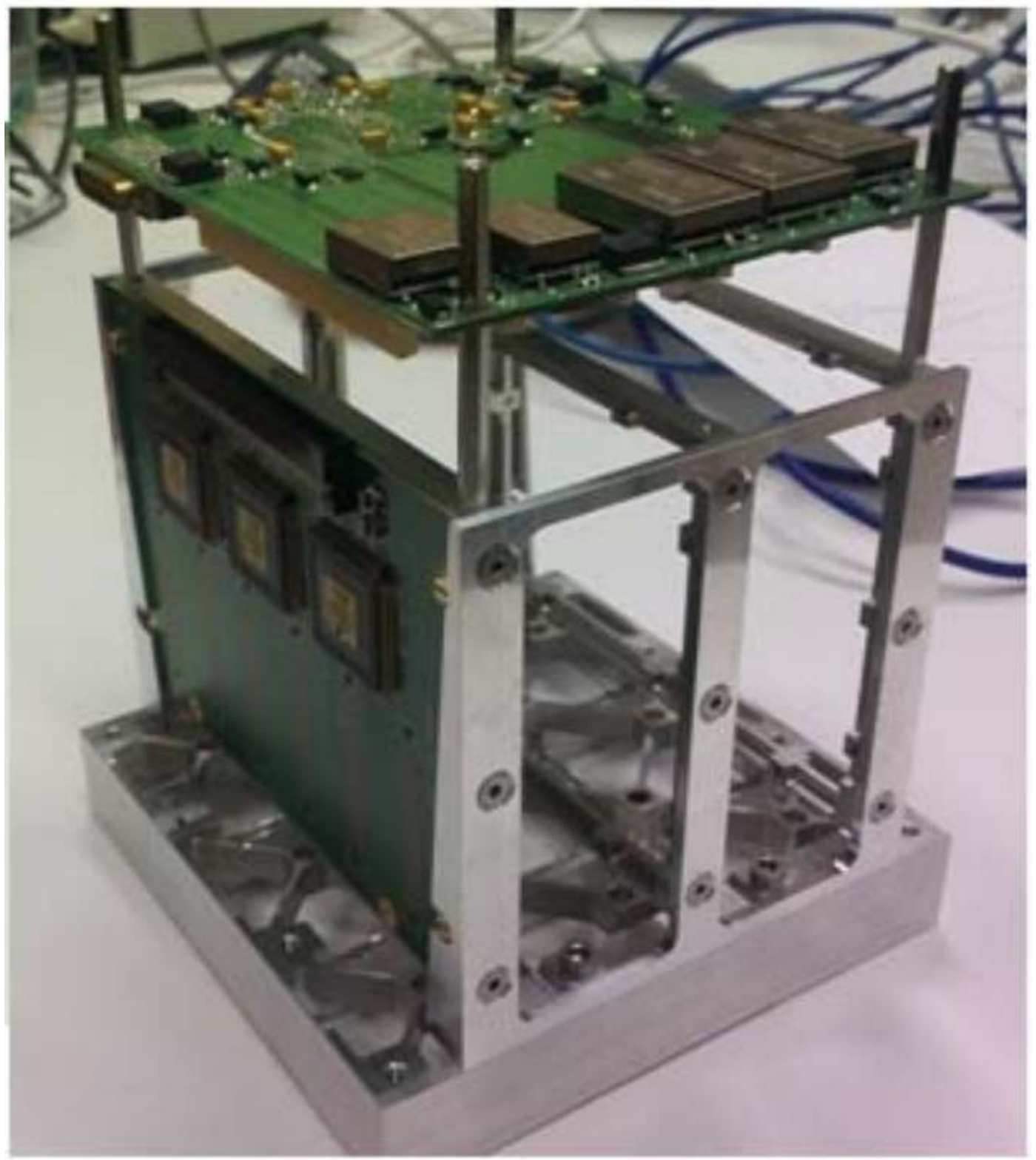}}\\
 \subfigure[]{\label{fig:JEM-EUSO-PDM:ASIC}\includegraphics[angle=0,width=0.48\textwidth]{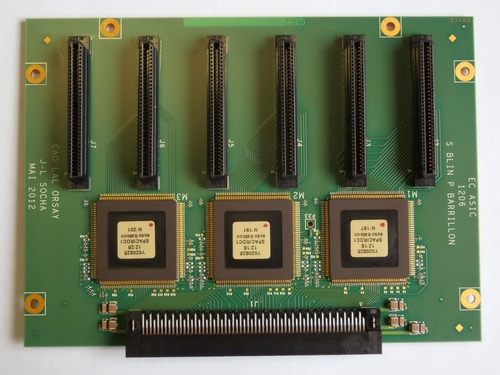}}
\caption[Photographs of PDM]{\label{fig:JEM-EUSO-PDM} Photographs of the assembled EUSO-TA PDM (in \fig\ref{fig:JEM-EUSO-PDM:TA}) and the mechanics of the EUSO-Balloon PDM (\fig\ref{fig:JEM-EUSO-PDM:Balloon}). 
Unlike in EUSO-Balloon, the EC for EUSO-TA are not potted. The large cables in \fig\ref{fig:JEM-EUSO-PDM:TA} lead to the high voltage power supply. In \fig\ref{fig:JEM-EUSO-PDM:Balloon}, both one ASIC board (the vertical \acrshort{PCB})
and the PDM board (the horizontal PCB at the end of the PDM) can be seen. \fig\ref{fig:JEM-EUSO-PDM:ASIC} shows a photograph of the ASIC board, which connects to six MAPMT. 
}
\end{figure}

The PDM structure, with its read-out electronics, is the smallest independent subunit of the focal surface. The PDM is made up of 9 ECs, i.e.\ 36 M64 photomultipliers,  and their read-out electronics.
The electronics consist of one ASIC per photomultiplier, one CW-HVPS per EC, and a PDM board which manages the data from the entire PDM. 
The ASIC board can be seen in \fig\ref{fig:JEM-EUSOM-EC:photo} and Figs.~\ref{fig:JEM-EUSO-PDM:Balloon} and \fig\ref{fig:JEM-EUSO-PDM:ASIC}.
The PDM also includes the supporting mechanics. 
Photographs of the PDM mechanical structure and the integrated EUSO-TA PDM are shown in \fig\ref{fig:JEM-EUSO-PDM}.
 
\subsection{Focal Surface Electronics}

\begin{figure}[h]
\centering
\includegraphics[angle=270,width=1.0\textwidth]{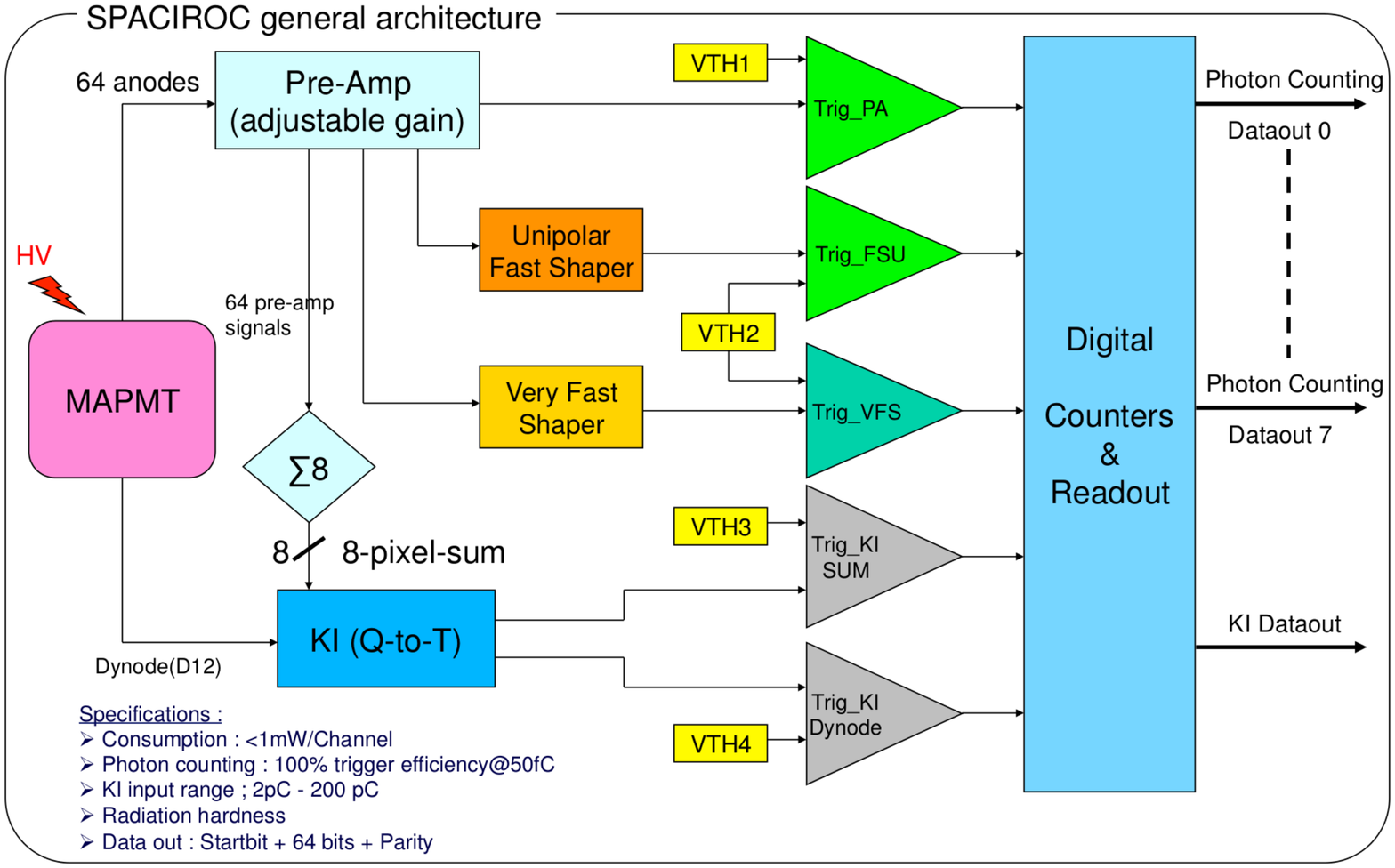}
\caption[]{\label{fig:JEM-SPACIROC} A diagram of the general architecture of the SPACIROC ASIC. Each of the 64 pixels of the MAPMT go through a integrating preamplifier with an adjustable gain. The signals are then sent through
a unipolar fast shaper and on to discriminator (Trig\_FSU). The number of counts over threshold is returned each GTU for all 64 pixels. At the same time, the SPACIROC includes a
time over threshold charge integrator (KI) which takes the sum of 8 pixels. This read-out mode is used for phenomena which give a large illumination. A few properties of the ASIC are listed in the bottom-left corner of the figure. }
\end{figure}

The task of the frontend electronics is to digitize the output signals of the multi-anode photomultipliers 
in successive time slices. In JEM-EUSO the read out is by so-called \glspl{GTU} of $2.5~\mu$s.
The length of a GTU was chosen for several reasons. Firstly, from \fig\ref{fig:PhotonArrival} is can be seen that an EAS develops completely in around 200$~\mu$s,
and so the read-out should be on the microsecond scale. A higher read-out rate, however, implies a larger frontend electronics power consumption, and a greater difficulty in design.
At the same time, the size of single JEM-EUSO pixel projected on the ground is $\sim 0.5~$km. This makes one time slice of $2.5~\mu$s, as a length in $z$ at the speed of light, roughly comparable to
the size in $x$ and $y$ of a pixel on ground.

The first level of the frontend electronics are the read-out ASICs of each M64 PMT. The ASIC contains both an analog part which handles the 64 anode signals from the phototube, and a
digital part which returns photon counting and charge integration data. The structure of the ASIC is shown in \fig\ref{fig:JEM-SPACIROC}.
At the end of each GTU, the ASIC of each MAPMT returns the number of pulses over a set threshold in each of the 64 pixels into a ring memory to wait for trigger assertion. 
For large signals, the charge in a grouping of 8 pixels is returned using a time over threshold method.

The counting and charge integration output of each ASIC are then passed on to the PDM electronics board. The PDMs are grouped into units of eight, each of which are controlled by a \gls{CCB}.
The PDM board implements a complex trigger in order to reduce the overall data rate. This trigger operates on a two level scheme.

The first level trigger is implemented by the PDM board, and is designed to reject the majority of background fluctuations by requiring a locally persistent signal over a predefined threshold during the course of several 
GTUs. In this \gls{PTT}, the pixels are grouped into 3 by 3 boxes. 
The trigger is issued if the activity in any pixel within the box is greater than the threshold $n_{\text{thr}}^{\text{pix}}$, and the total number of counts in the entire box is greater than a second threshold value $n_{\text{thr}}^{\text{box}}$.
For an average background of $\sim 1.1~$counts per GTU per pixel, $n_{\text{thr}}^{\text{pix}}$ is around 2 and $n_{\text{thr}}^{\text{box}} \approx 32$.

If the PTT is issued, then the photon counting data in the ring buffer of the PDM is requested by the CCB, which implements the second level trigger.
The second level trigger is known as the \gls{LTT}. The LTT is designed to follow the movement of the EAS spot within the PDM over a predefined time window in order to distinguish the pattern of an EAS from
the background. The photon count along each predefined line through the PDM is integrated across GTUs, and, if the integration along any direction exceeds a preset threshold, then the LTT is issued. 
The lines through the PDM are set to cover the entire phase space of possible track directions. 

\begin{figure}[h]
\centering
\includegraphics[angle=270,width=0.8\textwidth]{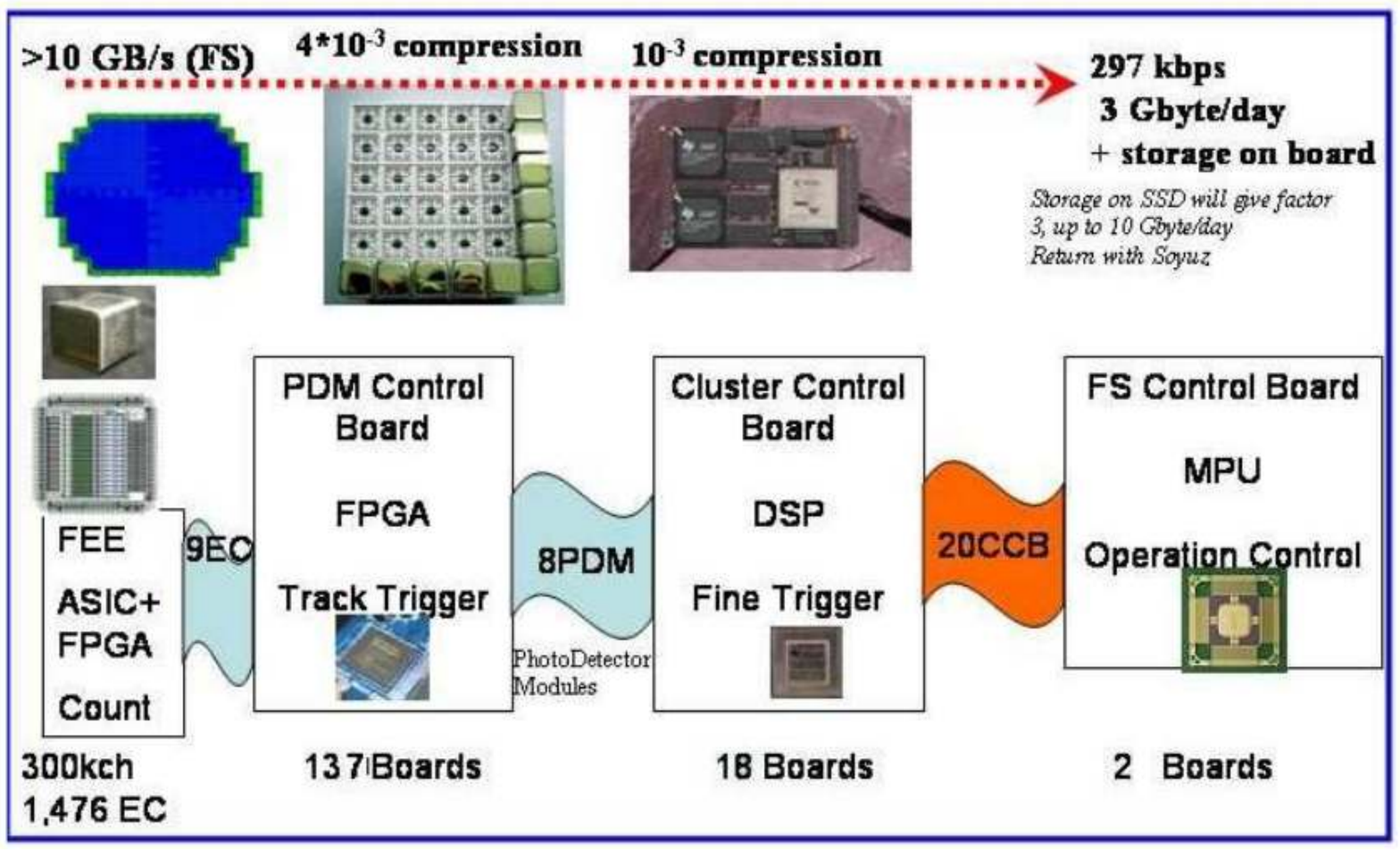}
\caption[]{\label{fig:JEM-DataHandling} A diagram of the data handling in JEM-EUSO. Because it is a space mission, the data budget available is extremely limited, requiring a large reduction in
data rate between the raw focal surface read out and event data. There are two triggers, one at the PDM level, and one at the CCB level, as described in the text. There are a total of 20 CCB, each controlling at most 8 PDMs.
Using this trigger scheme the expected data rate is reduced from 10 GB/s to 297 kb/s.}
\end{figure}

If the LTT trigger is issued, then the data from the entire cluster of 8 PDMs is saved, along with data from the atmospheric monitoring system. 
The threshold of the LTT is dependent on the background rate, and is tuned to reduce the event rate (including fake triggers) to $\sim 0.1~$Hz over the entire focal surface.
This scheme is necessary to 
\begin{inparaenum}[i\upshape)]
\item extract real EAS events from the background and
\item reduce the overall data rate to fit within the limited budget.                             
\end{inparaenum}
An overview of the data processing scheme is shown in \fig\ref{fig:JEM-DataHandling}. The raw data raw of more than 10 GB/s is reduced at each level down to a final data rate of $297~$kb/s for storage or down-link via TDRS.

The CCB also interfaces the PDMs with the overall JEM-EUSO housekeeping system, which issues general commands to each subsystem. 
Other data handling electronics include the \gls{MDP}, \gls{TCU}, \gls{IDAQ}, and a Clock \& Time Synchronization Board.
All of these electronics systems must be radiation hard in order to operate correctly in a space environment for 5 years and must run on a limited power budget.

\subsection{On-board Calibration System}

It is expected that the photodetection properties of the focal surface will show some change over the course of the mission, and so an on-board calibration
system has been foreseen. This on-board system will be used to perform a relative calibration with respect to the precise absolute pre-flight calibration.
It is also planned to use external light sources such as reflected moonlight. This is in addition to the on-ground lasers and Xenon flashers of the Global Light System. 

\begin{figure}[h]
\centering
\includegraphics[angle=0,width=0.9\textwidth]{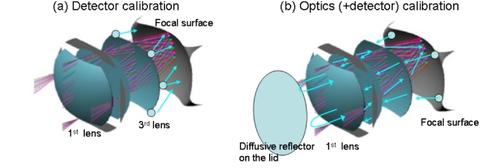}
\caption[Diagram of the On-Board Calibration]{\label{fig:OnboardCalibration} A diagram of the two on-board calibration modes. In (a) the focal surface is illuminated directly 
by light sources placed around the edge of the last lens. This gives the relative shift in gain and photon counting efficiency of the entire focal surface. 
In (b), the focal surface is illuminated indirectly by a second set of light sources around the edge of the focal surface. This gives the combined response of the focal surface and the optics throughput.}
\end{figure}

The on-board calibration light source is essentially a miniature version of the calibration setup which will be discussed in chapter~\ref{CHAPTER:PMT}.
This is a small integrating sphere equipped with a UV LED and a NIST photodiode to monitor the light intensity. Several of these light sources will be placed behind the rear lens in such a manner as to illuminate the focal surface in a roughly uniform way.
This is shown in \fig\ref{fig:OnboardCalibration}. By setting the intensity of the light sources low enough, the single photoelectron response of the focal surface can be measured in order to find any shifts in gain or efficiency.
If a large change in gain is found, then the photon counting thresholds are updated accordingly or the high voltage supply to the MAPMTs is adjusted. 

At the same time, another set of light sources can be placed around the edge of the focal surface to illuminate the rear lens. This light will transit through the optics and reflect off of the diffuse surface of the lid.
In this way the convolution of the optical and the focal surface response can be obtained. The two measurements together give the change in the focal surface response and optics throughput with time. There is, however,
no way to compensate for the loss of transparency. The loss of transmittance on the first lens, due to atomic oxygen exposure and proton bombardment, is wavelength dependent and is also strongly affected by the orientation of the 
surface of the first lens relative to the ram direction (the direction of flight). The transmittance reduction is larger for lower wavelengths, and is likely to be between a few percent at 400 nm
up to 10\% at 300 nm over the course of a five-year mission for a PMMA lens, if a protective coating of Si0$_{2}$ is used\footnote{As a very rough estimate, see \cite{PurpleBook} and \cite{AtomicOxygenDamage}}.

In the event of a large change in the efficiency of the focal surface, the possibility of a in-flight absolute calibration using reflected moonlight or the GLS is foreseen. These methods would give a precision on the order of 20\%.
\subsection{Atmospheric Monitoring System}

The final component of JEM-EUSO is the Atmospheric Monitoring System (AMS). 
AMS provides information on the optical properties of the atmosphere and 
the distribution of clouds and aerosol layers in the main telescope's FoV.
This information is vital for properly reconstructing the energy of an observed EAS, as was discussed for ground arrays in the last chapters. 

An importance difference for JEM-EUSO, however, is that the ability to view EAS in cloudy conditions depends less on the presence of the clouds, but rather on the height
of the cloud tops. 
In some cases the reconstruction of an EAS is aided by the reflection of Cherenkov light off the cloud, as long as the shower maximum is above the cloud top.
This means that determining the cloud-top height with a precision of around 1 km or better is mandatory.
At the same time, monitoring the cloud coverage is important in estimating the effective observation time of JEM-EUSO, increasing the accuracy of the UHECR flux measurement.
The AMS system includes an infrared (IR) camera and a Light Detection And Ranging (LIDAR) system. Slow data (with a ground speed of 7 km/s) from the focal surface itself, on the background rate for example, is 
also useful for atmospheric monitoring.

\subsubsection{IR Camera}

The detector of IR camera is a UL04171 infrared opto-electronic device consisting of a $\mu$bolometer Focal Plane Array. 
The total camera weight and power consumption are 11 kg and $\leq 15~$W, including the germanium optics, electronics, and built-in calibration system.
The working temperature of the detector is around $30^{\circ}$ C, and the operational wavelength range of the IR-camera is between a wavelength of 10 to 12 $\mu$m.
This corresponds to a temperature range of 220 to 320 degrees Kelvin, which is the annual 
cloud temperature variation plus a $20^{\circ}$ K margin. The expected temperature accuracy is $3^{\circ}~$K, translating into a accuracy of 500 m on the cloud-top height. 
The camera FoV is matched to that of the main telescope, and the expected spatial resolution is $0.1^{\circ}$.
The IR-Camera will take images every 17 s, during which time the ISS moves 1/4 of the total on-ground FoV.  

\subsubsection{LIDAR}

The goal of the LIDAR is to provide measurements of the extinction and scattering properties of the atmosphere along the path of EAS development and between the EAS and JEM-EUSO. 
Measurements by LIDAR are complementary to those taken by
the infrared camera. While the infrared camera provides knowledge on the overall distribution of
optically thick clouds in the telescope FoV, the LIDAR reveals optically thin clouds (e.g.\ high-altitude cirrus
clouds) and aerosol layers, while also measuring the optical depth and scattering properties of the clouds.

The LIDAR is composed of a transmission system and a receiver. 
In JEM-EUSO, the transmitter is a steered Nd:YAG laser, while the main telescope acts as the receiver.
This gives the LIDAR
a wide field of view, which is quite different from typical LIDARs, where the receiver has a FoV on the order of only a few arcminutes. 
The UV optics of JEM-EUSO will require that the LIDAR operates at the 355 nm third harmonic of the Nd:YAG laser.
The steering of the LASER is done using \gls{MEM} mirrors, with a response time on the order of a millisecond.
The basic operational scheme is to implement the pointing and shooting of the LIDAR directly into the EAS trigger, 
so that the region of the recently observed EAS is probed within $\sim 300~$ms. The expected rate of EAS triggers is about 0.1 Hz, so that 
a LIDAR repetition rate of 1 Hz would give time to take measurements in several predefined directions around the location of the observed EAS.

\section{Performance and Exposure of JEM-EUSO}

\begin{table}
 \begin{center}
\begin{tabulary}{1.0\textwidth}{LLL}
\toprule
Parameter & Value & Note(s)\\
\cmidrule(l){1-3}
\emph{Optics} & & \\ 
Optical Aperture & 4.5 m$^{2}$ & Baseline\\
Ensquared Collection Efficiency & 35\% (15\%) & For $\lambda = 350~$nm\\
Ensquared Energy & 86\% (80\%) & For $\lambda = 350~$nm\\
Optical Bandwidth & 290-430 nm & \\
Field of View & 0.85 sr & \\
Observational Area & $1.4~10^{5}~$km$^{2}$\\
 & & \\
\emph{FS Detector and Electronics} & &  \\ 
Number of Pixels & $3.2~10^{5}$ & \\
Spatial Angular Resolution & $0.074^{\circ}$ & \\
Pixel Size at Ground & 0.51 km (0.61 km) & For $H_{o}=400~$km\\
PMT efficiency & $\sim32 \%$ & For $\lambda = 350~$nm \\
Cross Talk & $< 2\%$ & \\
Transmittance of UV filter (BG3) & 97\% & For $\lambda = 350~$nm \\
Sampling Time & $2.5~\mu$s & \\
\end{tabulary}
\caption[JEM-EUSO Characteristics]{
\label{tab:JEMEUSOCharacteristics}
The characteristics of the JEM-EUSO telescope design. Values apply at the center of the FoV, while values in parenthesis apply at the edge of the FoV.
The ensquared collection efficiency is the ratio of the number of photons focused within a pixel area to the number incident on the entrance aperture of the optics.
Similarly, the ensquared energy is the ratio of photons focused within the area of a pixel to those reaching the focal surface.
}
 \end{center}
\end{table}

\begin{figure}
\centering
\includegraphics[angle=270,width=0.9\textwidth]{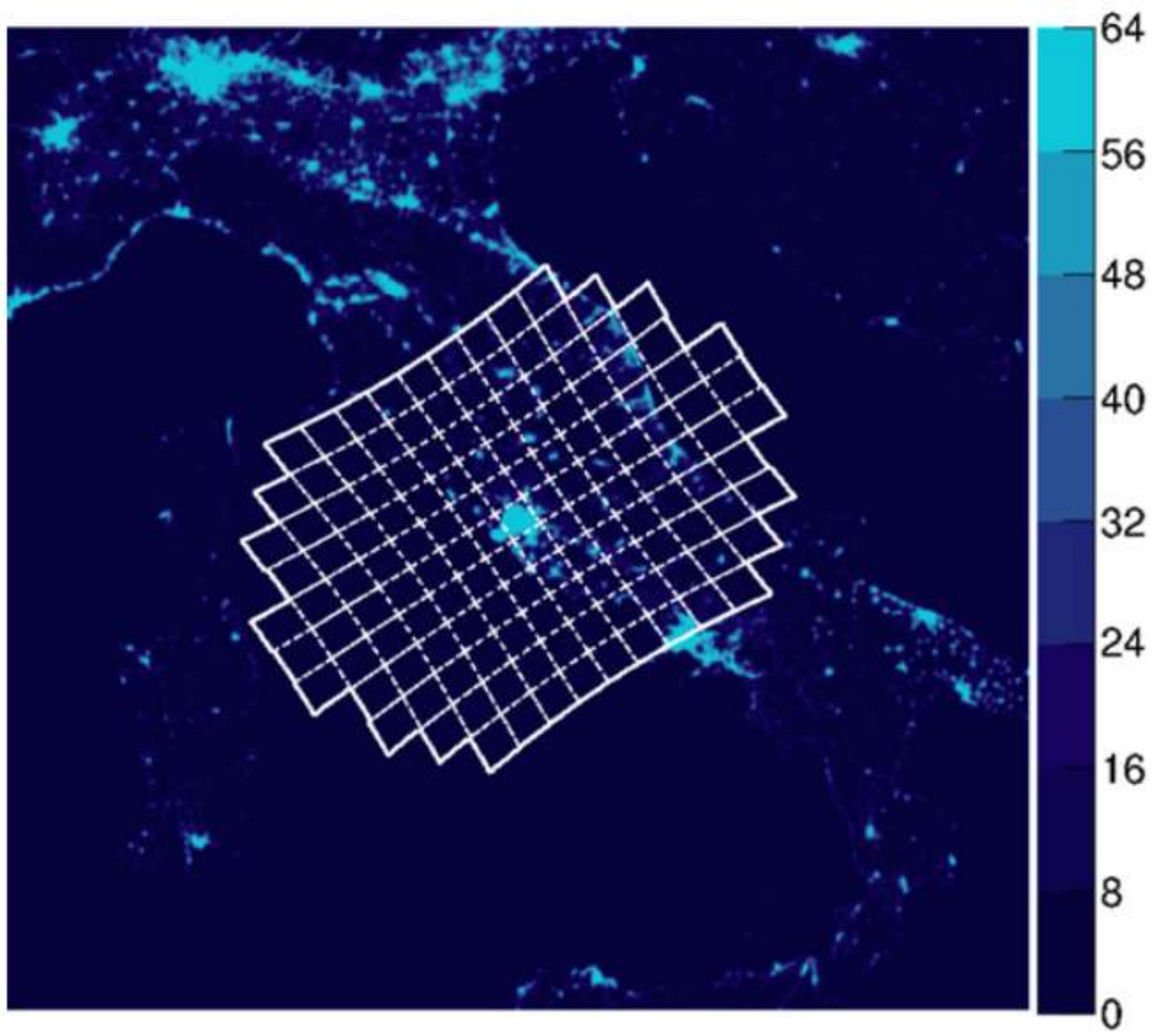}
\caption[JEM-EUSO Focal Surface Mapped on Earth]{\label{fig:FSmappedtoEarth}  The JEM-EUSO focal surface mapped over a \gls{DMSPac} image of the 
average light intensity in the visible band (here over central Italy). The Scale denotes DMSP units \cite{DMSP, AdamsJr201376}. }
\end{figure}

Having presented the design of the JEM-EUSO instrument, a short discussion of the effects which will impact JEM-EUSO's performance and exposure will now be presented. 
This discussion follows the results presented by the JEM-EUSO collaboration in \cite{AdamsJr201376}.
The area of the Earth's surface observed by JEM-EUSO is determined by the projection of the FoV of the optics and the area of the focal surface.
As shown in table \ref{tab:JEMEUSOCharacteristics}, the baseline aperture of the telescope is $4.5~$m$^{2}$. 
The field of view of the telescope can be estimated by using ray tracing, giving a FoV of $\Omega_{\text{FoV}}\sim 0.85~$sr.
A mapping of JEM-EUSO's focal surface on the Earth, done by ray tracing, is shown in \fig\ref{fig:FSmappedtoEarth}. The background of the figure shows the 
distribution of visible light taken from the \gls{DMSPac} \cite{DMSP}. The solid lines indicate the limits of the focal surface, while the dashed lines indicate the 
individual PDMs within the focal surface. 

In the baseline design of JEM-EUSO, the telescope is cut on two sides in order to fit into the HTV transport vehicle, with the result that the telescope will have a major and a minor axis.
The dimensions of the field of view are $64^{\circ}$ on the major axis and $45^{\circ}$ on the minor axis, giving a projected length on the Earth's surface of $500~$km and $330~$km for a height of $H_{o} = 400~$km.
The size of the observation area is a function of $H_{o}$:
\begin{equation}
 S_{\text{obs}}(\text{km}^{2}) = \Omega_{\text{FoV}}H^{2}_{o} = 1.4~10^{5}\left(\frac{H_{o}}{400[\text{km}]}\right)^{2}
\end{equation}
The wide FoV of JEM-EUSO allows a measurement of the entire EAS development from the first detection to the fade out or impact of the EAS on the Earth. 
This fact is useful for EAS at large zenith angles and gamma-ray or neutrino EAS.

The ability of JEM-EUSO to detect EAS within its FoV is influenced primarily by the UV background light.
The observational duty cycle of JEM-EUSO is given by
\begin{equation}
 \eta (< I_{\text{bkg}}^{\text{thr}}) = \eta_{\text{night}} \int_{0}^{I_{\text{bkg}}^{\text{thr}}} p(I_{\text{bkg}}) dI_{\text{bkg}}
\end{equation}
where $I_{\text{bkg}}$ is the intensity of the diffuse background light, $I_{\text{bkg}}^{\text{thr}}$ is a given threshold value, and  $p(I_{\text{bkg}})$ is the probability 
density function of $I_{\text{bkg}}$. The quantity $\eta_{\text{night}}$ is the fraction of time which is night, defined as the absence of the Sun in the visible sky. At $H_{o}=400~$km 
this requires a Solar Zenith angle greater than $109^{\circ}$, in which case $\eta_{\text{night}} = 34 \%$.

Background sources during night include terrestrial sources, such as night-glow, TLE, and localized light sources such as cities. In addition, outside sources of light, such as
reflected starlight and moonlight must be considered part of the background. Of these, moonlight is the largest variable component which would affect the entire focal surface at once.
The moonlight contribution will depend on the Moon phase angle $\beta_{\text{M}}$ and zenith angle $\theta_{\text{M}}$. The background intensity can be parametrized as
\begin{equation}
 I_{\text{BG}} = I_{\text{M}}(\theta_{\text{M}}, \beta_{\text{M}}) + I_{o},
\end{equation}
where $I_{\text{M}}$ is the intensity of moonlight. $I_{o}$ is taken as  constant value of $\simeq 500~$photons/m$^{2}$ sr ns \cite{Garipov2005400,Barbier2005439, Catalano:2002sc}. 
Using measured data to account for the moon position and brightness, the fraction of time  $\eta (< I_{\text{bkg}}^{\text{thr}})$ during which $I_{\text{BG}}$ is below
$I_{\text{bkg}}^{\text{thr}} = 1500~$photons/ m$^{2}$ sr ns was found to be $\eta_{0}=20\%$.
This results in an average background of 550 photons/m$^{2}$ sr ns. 
The threshold value of $1500~$photons/m$^{2}$ sr ns is an arbitrary reference value, which gives an increase in signal-to-noise ratio of $\sim \sqrt{3}$ compared to 500 photons/m$^{2}$ sr ns.
At this signal-to-noise ratio, the observation of EAS would still be possible, but with a slightly higher threshold.

The operational duty cycle is further reduced by non-diffuse background, such as lightning, flashes, auroras, and cities. 
The reduction in duty cycle due to lightning was estimated using data from the Tatiana satellite \cite{Garipov2005400}.
Making the (conservative) assumption that the entire instrument is blind if any such event is inside the FoV, and that each light source remains inside the field of view for $\sim70~$s 
(the maximum time for a source to cross the major axis of the telescope) it was found that the duty cycle is reduced by $\approx 2\%$.
A similar estimate for aurorae found a reduction on the level of 1\%, assuming maximum solar activity.

The effect from city light was estimated on the PDM level using DMSP data. The assumption was that no EAS observation could occur if a single pixel within the PDM detected a light intensity 
of greater than 3 times the average level (a DMSP unit of 7). The orbital path of the ISS results in the ocean accounting for 72\% of the total area observed by JEM-EUSO.
This gives an inefficiency of about 7\% from man-made light. It should be noted, however, that the DMSP data is in the visible band, and the expectation is that cities are brighter in visible wavelengths than in the UV.
Tatiana (which, however, had no focusing optics), for example, measured an intensity over cities such as Houston or Mexico City of 2-3 times that over the ocean \cite{Garipov2005400}.  
The combination of all the non-diffuse backgrounds gives a coverage loss of $f_{\text{loc}} = 10\%$. 

\begin{figure}
\centering
\includegraphics[width=0.8\textwidth]{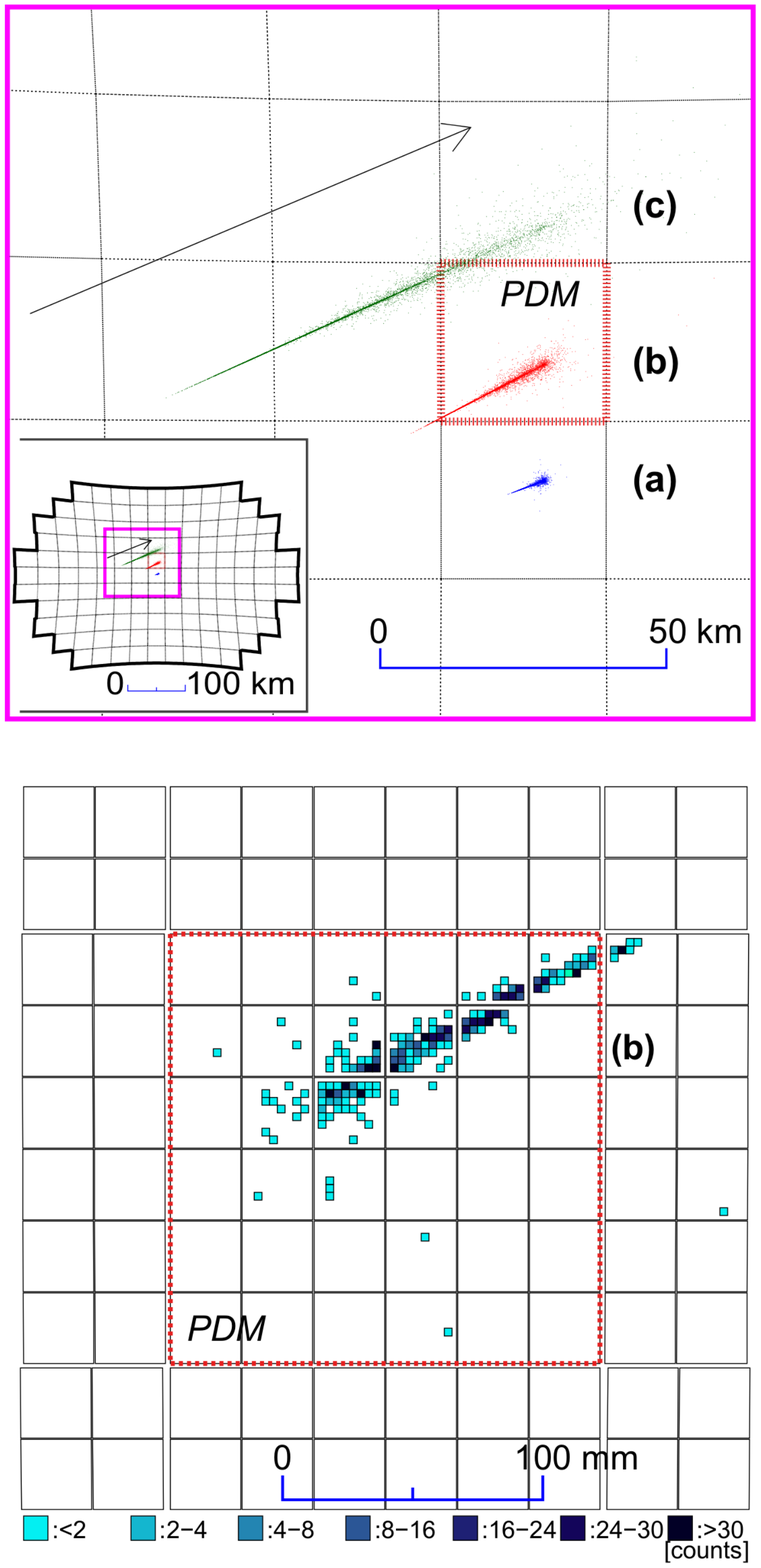}
\caption[Simulated EAS Tracks in JEM-EUSO]{\label{fig:JEM-EUSO:lightTracks} Two figures showing simulated EAS tracks on the JEM-EUSO focal surface. The top panel shows the 
projected tracks on the Earth's surface for EASs with $E_{o}= 10^{20}~$eV and zenith angles of (a)$~\theta = 30^{\circ}$, (b)$~\theta = 60^{\circ}$, and (c)$~\theta = 75^{\circ}$.
The dashed curves show the FoV of different PDMs, while the location of the plot within the FoV is shown in the sub-panel by the solid lines.
The bottom panel shows a blow-up of (b) with the integrated counts in each pixel shown in discrete scale. The red dashed lines in both plots indicate the same region (rotated by 180 degrees).}
\end{figure}

A final effect on the total exposure of JEM-EUSO comes from clouds. The possibility of reconstructing an EAS in the presence of clouds depends on the 
geometry of the shower and the height of the top of the offending cloud. In some instances, the presence of an optically thick cloud can enhance the Cherenkov reflection, and thus aid the reconstruction of the EAS as long
as the height of the cloud is well-known. It is for this reason that JEM-EUSO includes a LIDAR and IR camera system. 
A detailed study of the distribution of clouds was performed using data from numerous satellites, such as those of the International Satellite Cloud Climatology Project \cite{ISCCP} and the 
Climatic Atlas of Clouds Over Land \cite{CACOLO}, and the impact of clouds was studied in combination with the event trigger, which was discussed in the instrument section.

\begin{figure}
\centering
\includegraphics[angle=270,width=0.8\textwidth]{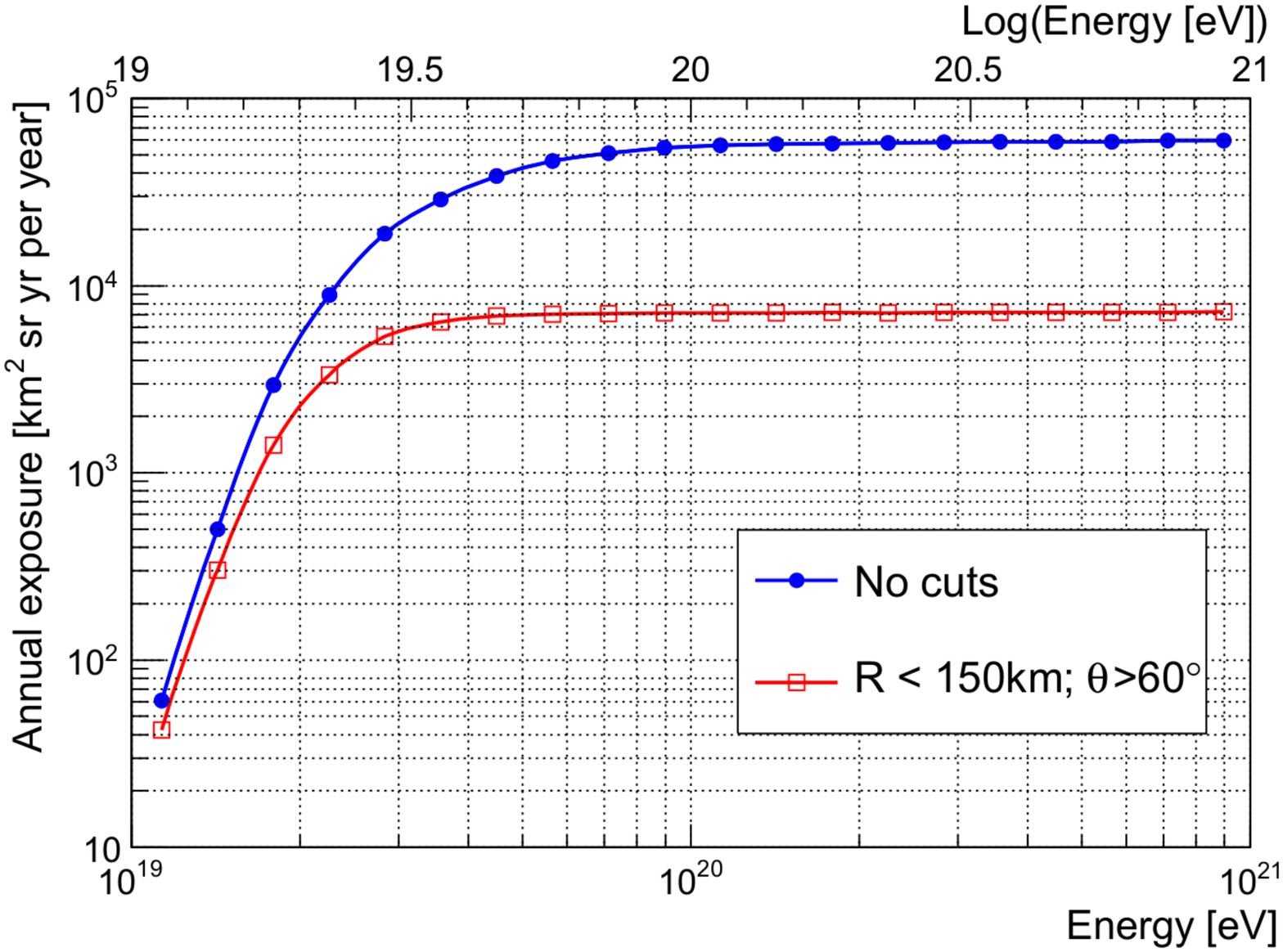}
\caption[Annual Exposure of JEM-EUSO]{\label{fig:JEMEUSOAnnualExposure_frompaper}  The Annual Exposure of JEM-EUSO, calculated accounting for background, cloud cover, and general duty cycle (see text and \cite{AdamsJr201376}).
The exposure is shown as a function of energy for two extreme cases 
\begin{inparaenum}[(a\upshape)]
 \item entire observation area (filled blue circles), and
\item with cuts on distance (on ground from the center of the FoV) $R < 150~$km and zenith angle $\theta > 60^{\circ}$ (open red squares). 
\end{inparaenum}
In the later case, the annual exposure is well-controlled down to lower energies, while the full aperture gives nearly an order of magnitude increase at the highest energies. }
\end{figure}

The three main factors which define the trigger efficiency for a given night-glow background level and set of atmospheric conditions 
are the optics response, the zenith angle of the EAS, and the distance effect. A series of simulated EAS in the JEM-EUSO telescope are shown in \fig\ref{fig:JEM-EUSO:lightTracks}.
From these simulations it was found that the effective exposure $\kappa_{\text{c}}$ is 72\%. The annual exposure of the instrument is then given by
\begin{equation}
 (\text{Annual Exposure}) = A(E) \kappa_{\text{c}}\eta_{o}[1-f_{\text{loc}}(1~\text{yr})] 
\end{equation}
Where $A(E)$ is the geometrical aperture which includes the triggering efficiency and thus depends on event energy. Using the estimated values of the observation inefficiencies the annual exposure is approximately
\begin{equation}
 (\text{Annual Exposure}) = 0.13\cdot A(E) (1 \text{yr})
\end{equation}
The annual exposure of JEM-EUSO as a function of energy is shown in \fig\ref{fig:JEMEUSOAnnualExposure_frompaper}, and is expected to be 9 times 
that of Auger at energies around $10^{20}~$eV. JEM-EUSO should reach full efficiency at $3~10^{19}~$eV for a restricted sub-set of events (representing $\sim1/8$ of the aperture), and at $E_{0} \geq 7~10^{19}~$eV
for the full aperture.

\begin{figure}
\centering
\includegraphics[angle=270,width=0.8\textwidth]{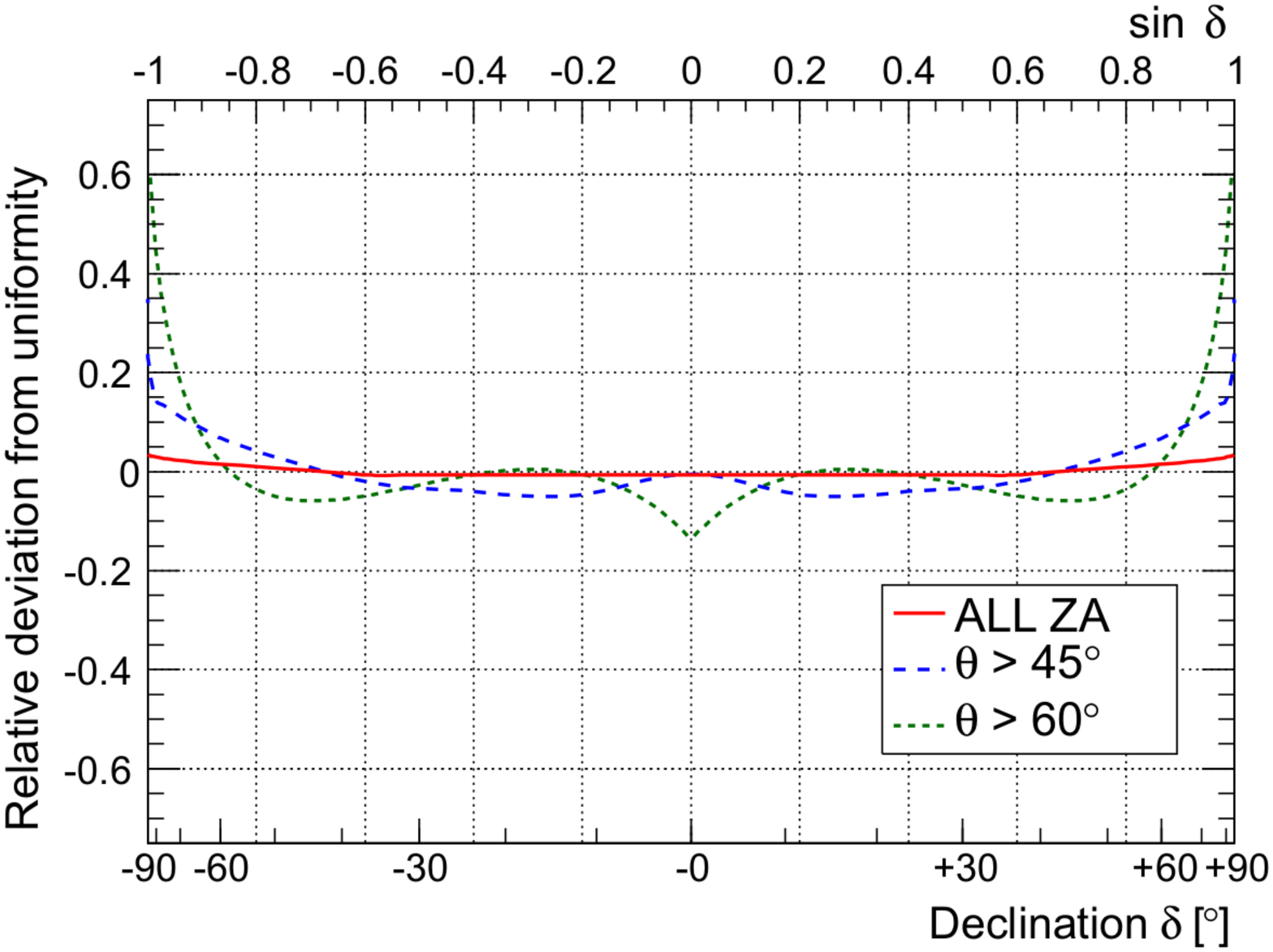}
\caption[JEM-EUSO Exposure by Declination]{\label{fig:JEMEUSOExposureByDeclination_frompaper}  A plot of the exposure of JEM-EUSO as a function of declination. The exposure is showed for three different cases corresponding to
one with no cut on minimum zenith angle $\theta_{\text{c}}$ (red solid line), $\theta_{\text{c}} > 45^{\circ}$ (blue dashed line), and $\theta_{\text{c}} > 60^{\circ}$ (green dotted line). The vertical axis indicates the 
deviation from uniform exposure.}
\end{figure}

At the same time, the exposure distribution was estimated and is shown in \fig\ref{fig:JEMEUSOExposureByDeclination_frompaper}. The exposure distribution is basically flat in right ascension.
The non uniformity in declination can be affected by local or seasonal variation in background or cloud cover. If these are ignored, the actual exposure is a simple function of the applied zenith angle cuts. 
For strict cuts minor excesses in exposure and deficits in exposure arise near the Celestial poles and Equator (respectively) as the ISS has slightly longer residence time at higher latitudes. 

\section{The JEM-EUSO Pathfinders}

\subsection{EUSO-TA}
\begin{figure}
\centering
\includegraphics[angle=270,width=0.8\textwidth]{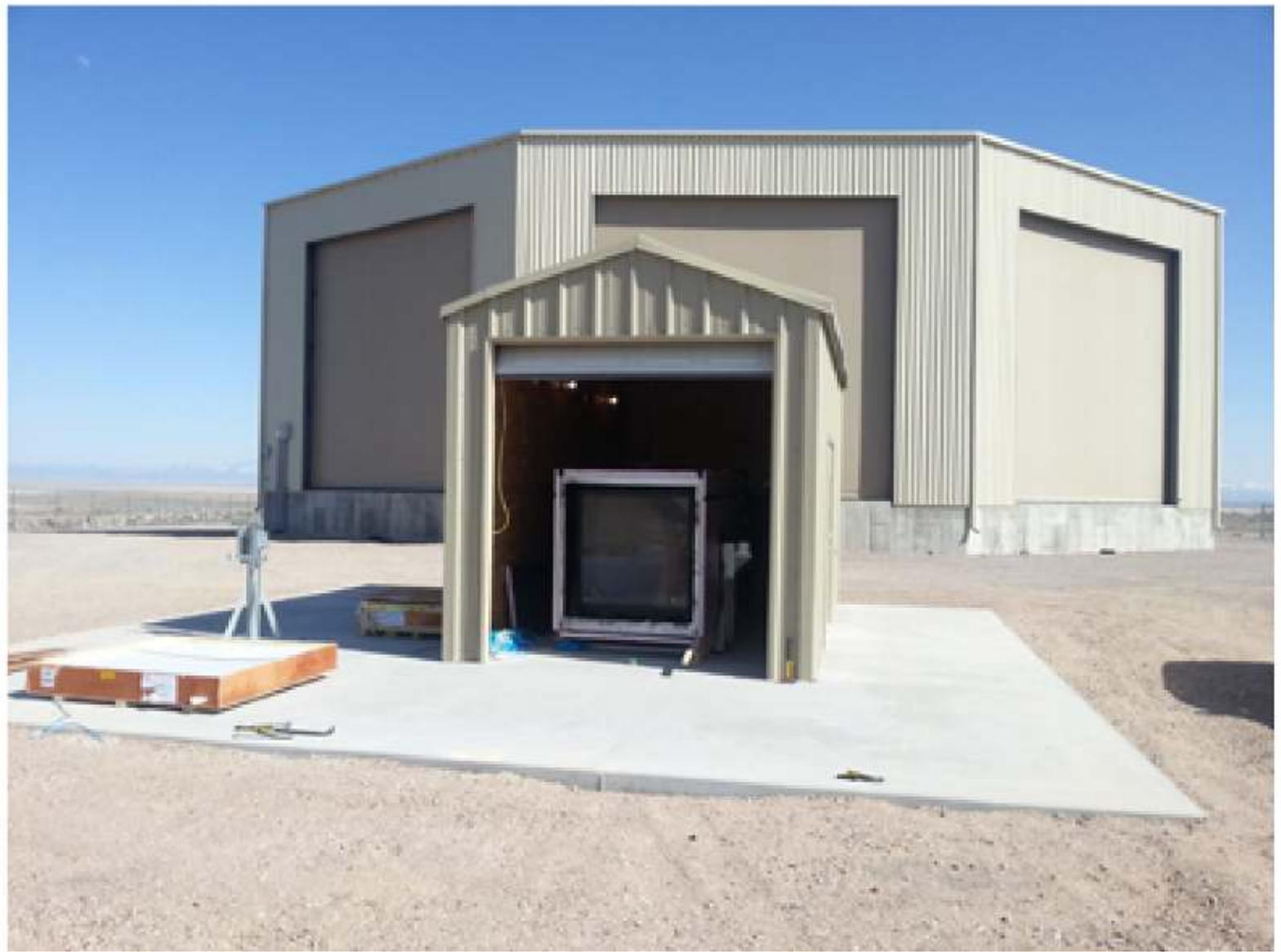}
\caption[A Photograph of EUSO-TA on-site]{\label{fig:JEM-EUSOTA-OnSite} A photograph showing the shed housing EUSO-TA at the Telescope Array Black Rock Mesa site.
The Telescope Array fluorescence detector can be seen in the background. The EUSO-TA optics (two Fresnel lens of 1 m$^{2}$) have already been installed and can be seen inside the shed.
}
\end{figure}

EUSO-TA is a ground-based prototype of JEM-EUSO which will be installed at the Black Rock Mesa site of Telescope Array. 
The detector is a single JEM-EUSO PDM with two 1 meter square Fresnel lenses, as there is no need for the middle diffractive lens.
This 2 lens system gives a field of view of $8^{\circ}\times8^{\circ}$, and a point-spread function of 8 mm, much less than the 
Telescope Array fluorescence detector.
The frontend electronics and other design aspects are the same as in JEM-EUSO.

The future home of EUSO-TA is shown in \fig\ref{fig:JEM-EUSOTA-OnSite}. From this location directly in front of the Telescope Array fluorescence detector, EUSO-TA will view the 
same EAS and the Electron Light Source (ELS) and Central Laser Facility (CLF)-induced events with a finer granularity than the Telescope Array fluorescence detector. 
For EAS detection a trigger will be given by the TA detector. The aim of the project is to cross calibrate the
response of the EUSO telescope with the TA fluorescence detector in the presence of a shower of known
intensity and distribution, and to analyze the distribution of particles inside the EAS. The optics of EUSO-TA have already been installed at Black Rock Mesa, and the PDM is undergoing testing.

\subsection{EUSO-Balloon}
\begin{figure}
\centering
\includegraphics[angle=0,width=0.8\textwidth]{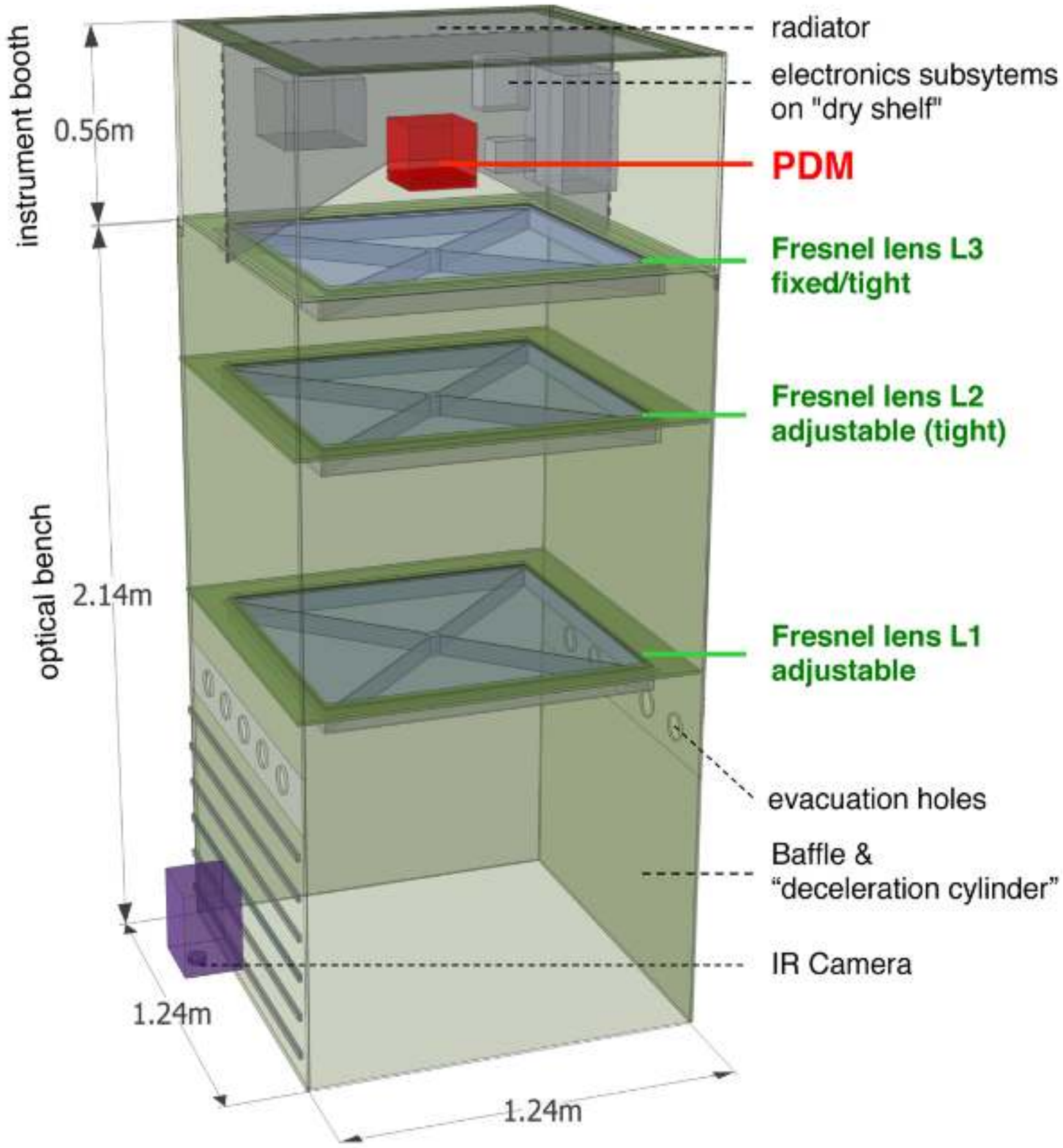}
\caption[Overview of EUSO-Balloon]{\label{fig:JEM-EUSOBalloon-Overview} An overview of the EUSO-Balloon instrument. The PDM is a copy of the current JEM-EUSO design. The PDM and electronics
are located on a dry shelf behind the third Fresnel lens. This metal shelf also acts as a radiator towards the top of the gondola to maintain the internal temperature of the instrument (not trivial at 3 mbar). The optics
are a system of three Fresnel lens, adapted from the JEM-EUSO design.  }
\end{figure}

EUSO-Balloon is a JEM-EUSO pathfinder mission led by the French part of the JEM-EUSO collaboration and organized as a mission of the French Space Agency CNES.
The EUSO-Balloon detector is composed of a single JEM-EUSO PDM with optics launched in a balloon-borne gondola.
A diagram of the instrument is shown in \fig\ref{fig:JEM-EUSOBalloon-Overview}. The balloon gondola is a rectangular booth, and
the optics are an adaptation of the baseline JEM-EUSO three Fresnel lens design with a point-spread function of about 4 mm. 
An open baffle in front of the first lens acts as a deceleration cylinder for possible water landings.
The PDM and electronics are placed in a water tight section behind the third lens for protection during water landings. 
The PDM electronics (ASIC, CW-HVPS, and PDM board) are the JEM-EUSO designs, while the data processing and housekeeping systems are adaptations.  EUSO-Balloon
also includes an IR camera for cloud tracking, which is mounted in its own compartment at the front of the gondola.

The philosophy in designing and building EUSO-Balloon is to follow as closely as possible the actual hardware design and requirements of JEM-EUSO.
In some cases the design requirements for EUSO-Balloon are more strict than in JEM-EUSO.
One example is high voltage isolation, which is more difficult at an altitude of 40 km, where the pressure is $\sim 3~$mbar, than in the vacuum of near-Earth orbit.

The objectives of the EUSO-Balloon mission are to:
\begin{inparaenum}[a\upshape)]
 \item act as a technology demonstrator for JEM-EUSO,
 \item perform a study of data acquisition and an analysis of the UV background with a focusing instrument, and
\item  possibly view a few EAS from above (as a bonus objective).
\end{inparaenum}
As noted above, the driving idea behind the EUSO-Balloon design is for it to be a full scale end-to-end test of the key components of JEM-EUSO. 
This includes a flight test of the CWHVPS and switch system, the frontend electronics, and the triggering algorithms.

\begin{table}
 \begin{center}
\begin{tabulary}{1.0\textwidth}{LLL}
\toprule
  &  JEM-EUSO & EUSO-Balloon\\
\cmidrule(l){1-3}
Number of PDMs & $\sim137$ & 1 \\
Flight Altitude & 400 km & 40 km \\
Diameter of Optics & 2.5 m & 1 m \\
Field of View/PDM &  $3.8^{\circ}$ & $12^{\circ}$\\
PDM Size at Ground & 28.2 km & 8.4 km \\
Field of View/pixel & $0.08^{\circ}$ & $0.25^{\circ}$ \\
Pixel Size at Ground & 580 m & 175 m\\
Signal w.r.t.\ JEM-EUSO & 1 & 17.6 \\
BG w.r.t.\ JEM-EUSO & 1 & 0.9-1.8\\
S/$\sqrt{\text{N}}$ w.r.t.\ JEM-EUSO & 1 & 10-20\\
Threshold Energy & $3~10^{19}~$eV & $1.5-3~10^{18}~$eV\\
\bottomrule
\end{tabulary}
\caption[EUSO-Balloon Characteristics]{
\label{tab:EUSOBalloonCharacteristics}
The characteristics of EUSO-Balloon, compared to JEM-EUSO. 
The field of view of EUSO-Balloon, and therefore the pixel size, has been tuned to give a background measurement comparable to JEM-EUSO. 
}
 \end{center}
\end{table}

The characteristics of EUSO-Balloon compared to JEM-EUSO are shown in table~\ref{tab:EUSOBalloonCharacteristics}.
EUSO-Balloon would be the first measurement of the UV background from above with focusing optics and a high spatial resolution. 
As measuring a representative background for JEM-EUSO is a primary goal of the balloon project, the EUSO-Balloon FoV, and hence the projected pixel size on ground
have been selected so that the effective background rate is equivalent. 
In order to test trigger algorithms and function, a series of xenon-flashes and LASER-induced events originating from airplanes are planned during the first flights.

Although it is not a primary goal of the Balloon mission, simulation studies have shown that 0.2-0.3 events above $E_{o} = 2~10^{18}~$eV can be expected during a night-flight of 10 hours. 
The uncertainty in the estimation assumes also the presence of a moderate cloud fraction. This makes it probable that EUSO-Balloon will view one or more UHECR events during later long-duration flights.
The EUSO-Balloon PDM is currently being integrated at APC, and the first flight is foreseen to be in the fall of 2014 in Timmins, Canada.

\section{Conclusion}

The JEM-EUSO mission proposes to deliver an order of magnitude increase in exposure at the highest UHECR energies. 
This would allow an exploration of phenomena in UHECR astrophysics which would be highly complementary to the current and next generation ground observatories.
At the same time, JEM-EUSO would continuously monitor the Earth's atmosphere in UV, allowing numerous studies in atmospheric science.
The JEM-EUSO project is progressing rapidly thanks to the momentum provided by the EUSO-TA and EUSO-Balloon pathfinders, and 
both EUSO-Balloon and EUSO-TA are expected to begin taking data in 2014.	  
     \printbibliography[heading=subbibliography]
    \end{refsection}

\part{Photodetection Aspects of JEM-EUSO}
    Imagining an experiment and making it a reality are two very different tasks, especially in the case of a large space-based experiment such as JEM-EUSO. 
After setting the stage by presenting a overview of ultra-high energy cosmic ray physics, current experiments in the field, and the EUSO experiments in Part.~\ref{PART1}, 
I will now discuss some of the work that I have been involved in to make JEM-EUSO a reality.
Within the JEM-EUSO collaboration, the APC group provides expertise in photodetection at extremely low light levels -- i.e.\ photomultiplier tube operation, characterization, and calibration at the single photoelectron level.
The operation of PMTs in single photoelectron mode requires a very low background testing environment, a specially designed black box.
This gives my group an important role as a test-bench, and I have been directly involved in the testing of all JEM-EUSO hardware which interfaces with the photomultiplier tubes. Such testing is particularly important
given the particularities of designing a space-based experiment. 

Chapter~\ref{CHAPTER:PMT} is an introduction to the characteristics of photomultiplier tubes, how they operate, and how they can be calibrated. Within this chapter, section \ref{sec:INTRO:PMT} 
gives a overview of how photomultiplier tubes function and the parameters that characterize them. Section \ref{subsec:PhotoDetection} describes the use of photomultiplier tubes as photodetectors, focusing
on single photon counting, and section \ref{sec:PMTCalibration} discusses experimental techniques to measure the absolute single photon detection efficiency of photomultiplier tubes with an uncertainty of a few percent. 

After this introduction to photomultiplier tubes, chapter \ref{CHAPTER:AIRFLUOR} considers an application of an absolutely calibrated photomultiplier to the measurement of the nitrogen fluorescence yield in air. 
The aim in this chapter is to show both the conceptual design of a precision experiment where knowledge of the absolute detection efficiency is crucial and to use the calibration of the photomultiplier as a detailed example of the technique 
presented in section \ref{sec:PMTCalibration}.  
Section \ref{sec:A New Absolute Measurement of the Fluorescence Yield} details the design of a new air fluorescence yield measurement which aims to make an absolute measurement of the yield in all relevant conditions of air temperature, pressure, and humidity. 
In this setup both the absolute yield and the absolute integrated spectrum will be measured simultaneously for each point in the parameter space.
With these sections as the background, section \ref{sec:Calibration of Photomultiplier Tubes for the Air Fluorescence Measurements at LAL} discusses the absolute calibration of the photomultiplier tubes to be used in the measurement.

Chapter \ref{CHAPTER:CWHVPStests} moves to the topic of the high voltage power supply and switch design used in JEM-EUSO.
JEM-EUSO's placement on the ISS results in a strict power limitation of less than 1 kW during operation, and this precludes the use of a resistive power-supply for the PMTs. 
In addition, lightning, meteors, transient luminous events (TLE), or man made light (cities) can be expected to give photon fluxes up to 10$^{6}$ times higher than the background level, which is estimated to be 600 kHz/pixel. 
As studying these phenomena is a science goal of JEM-EUSO, the dynamic range of the instrument must span these 6 orders of magnitude, and, as the PMTs must be protected from large currents, the switches must operate in 2 to 5 $\mu$s. These factors require 
a high voltage power supply designed especially for JEM-EUSO. 
As will be shown in chapter~\ref{CHAPTER:PMT} however, the high voltage power supply is an integral part of the photomultiplier tube in terms its performance. Due to this, I have worked closely with the designers of the high voltage system for JEM-EUSO. 
The first part of the chapter gives an introduction to the high-voltage system requirements and design. 
Sections~\ref{sec:Cockcroft-WaltonandSwitchDesign} presents the tests done to provide design feedback for the high voltage power supply and section~\ref{sec:Test of the CW-HVPS and Switch Prototype} 
reports the measurement done to check the final performance of the JEM-EUSO high voltage system, which will also be used in both EUSO-balloon and EUSO-TA.

Chapter \ref{CHAPTER:PMT Sorting} discusses the building of a complete test bench capable of handling multi-anode photomultiplier tubes (MAPMTs). 
This was motivated by the need to measure both the gain and efficiency of 64 pixels of the Hamamatsu M64 MAPMT which will be used in JEM-EUSO.
Because of the linearity properties of the SPACIROC-ASIC and variation in both gain and efficiency from one M64 MAPMT to the next, it is necessary to sort each photomultiplier to be used according to gain. 
In the case of JEM-EUSO this means measuring the gain and efficiency of 64 pixels for more than 5,000 photomultiplier tubes.
On a (slightly) smaller scale, there are a total of 36 MAPMT in the EUSO-balloon. Section \ref{sec:The Sorting Setup} discusses the requirements and construction of a data acquisition system which 
allows taking the spectra for 64 pixels simultaneously with the statistics needed to reach 1\% precision, and later sections in this chapter present the hardware, software, and characterization aspects of this work.

In chapter~\ref{CHAPTER:EUSOBalloonMeasurements} this calibration system is used to perform a first absolute calibration of the entire focal surface of EUSO-balloon at the EC level. 
Within this chapter, section~\ref{sec:PixelScanningResults} shows measurements of the pixel width and the dead-space between photomultiplier tubes within an EC, again using the same test system developed in chapter~\ref{CHAPTER:PMT Sorting}.
The last chapter discusses the extension of these measurements to a final absolute calibration of the EUSO-Balloon PDM, and concludes this part of the thesis.

     \begin{refsection}
  \chapter{Photomultiplier Tubes and Their Calibration}
    \label{CHAPTER:PMT}
     \section{Introduction to Photomultiplier Tubes}
\label{sec:INTRO:PMT}
\begin{figure}[h!]
 \includegraphics[angle=270,width=1.0\textwidth]{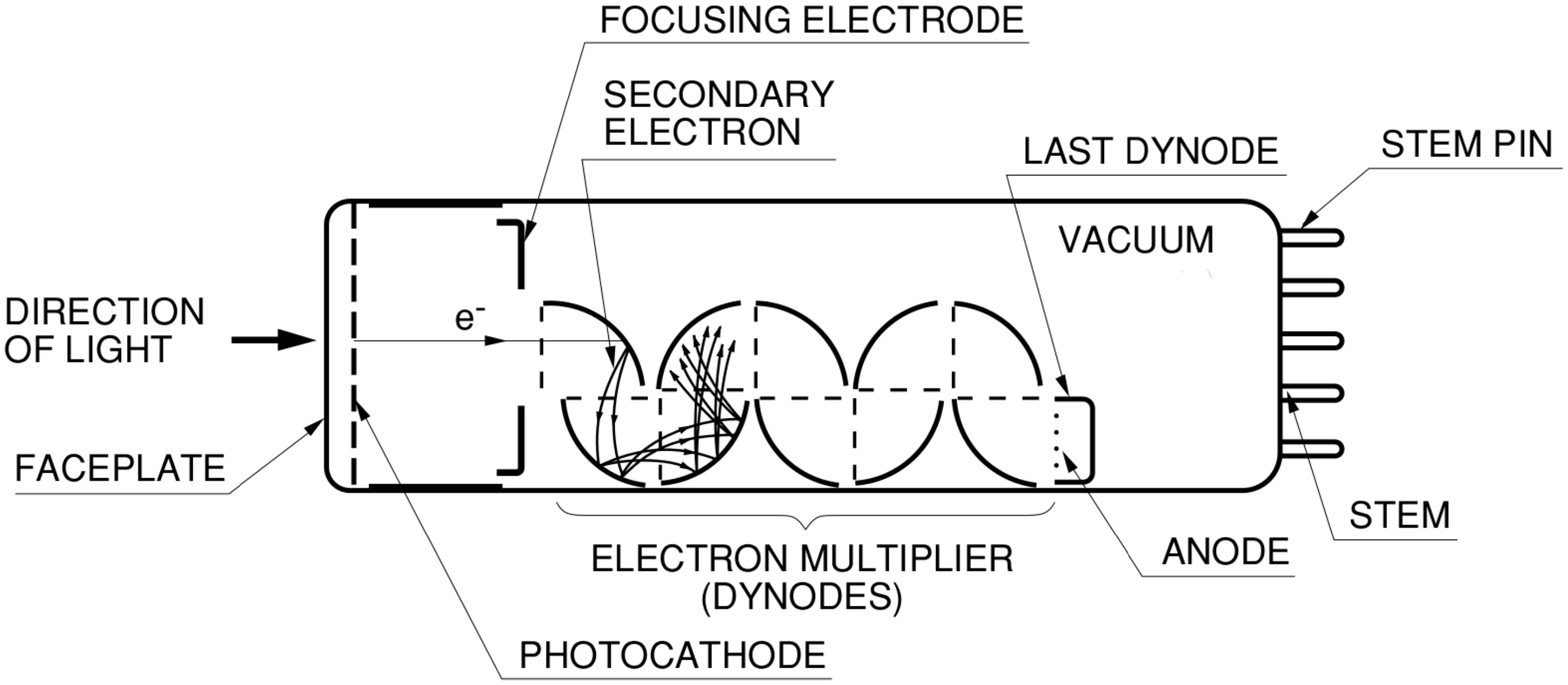}
\caption[Schematic Diagram of a Photomultiplier Tube]{A schematic diagram of a photomultiplier tube (PMT), here one with a semi-transparent window and box configuration dynodes. Light enters the vacuum tube through the front window (left-hand side) and impinges on 
the photocathode, where it liberates electrons into the vacuum through the photoelectric effect. These electrons are collected by focusing electrodes into the electron multiplier stage, where they are multiplied by secondary emission.
The resulting electron shower is collected at the anode. Figure taken from ref.~\cite{HamamatsuManual3a}.}
\label{fig:PMTDiagram}
\end{figure}

A \gls{PMT} is a vacuum tube which converts incident photons into an electric current. The intensity of the current produced is proportional to the 
luminous power deposited on the sensitive surface of the PMT.
A very generalized schematic of a PMT is shown in \fig \ref{fig:PMTDiagram}. The basic elements of a PMT are a faceplate (input window), a photocathode, focusing electrodes, an electron multiplier, and an anode.
These elements are sealed in an evacuated glass tube.

When light strikes the metal or semiconductor photocathode, electrons are emitted into the vacuum through the photoelectric effect. An electron
emitted by the photocathode due to an incident photon is known as a \gls{pe}.  
Once one or more photoelectrons are emitted by the photocathode, they are collected by the focusing electrodes into the electron multiplier. 

In the multiplication stage of the PMT, 
the number of electrons is increased in order to achieve a detectable current. The multiplication takes place using secondary emission through a series of electrodes held at a certain voltage, known as dynodes.
The voltage difference between dynodes, which are in vacuum, accelerates the emitted electrons, providing energy to fuel secondary emission at the next dynode.
Each primary electron absorbed by the dynode imparts energy to electrons in the dynode material through ionization, x-ray generation, and excitation of electrons between valence bands. These energized electrons diffuse through 
the dynode material, and those reaching the surface with enough energy escape into the vacuum towards the next dynode. 

The resulting electron shower arrives at the anode, where it is measured as a current. Photomultiplier tubes thus operate as 
a current source with close to zero impedance. The electron shower has a width in time of several nanoseconds to several tens of nanoseconds, depending on the design of the 
PMT.
The rise-time of this shower gives the time-resolution of the PMT response, and any slower phenomenon will be 
reproduced well by the current output from the anode of the PMT. 

The properties of a PMT design in terms of spectral sensitivity, signal-to-noise ratio, response delay and pulse width, sensitivity to magnetic fields, etc. are determined by numerous factors, such as the
photocathode and input window material, the layout and number of dynodes, the dynode material, the size of the PMT, and so on. All of these characteristics must be chosen according to the desired application, if it is indeed the case that a PMT
is the best tool for the task. Each stage of the PMT can be characterized by one or more efficiencies which together give the primary response characteristics of the PMT.
These are the:



\subsubsection{Quantum Efficiency, $\epsilon_{q}$}
\begin{quote}
The quantum efficiency $\epsilon_{q}$ characterizes the efficiency of converting photons into electrons at the photocathode. 
It is defined by the ratio 
\begin{equation}
\label{eq:QuantumEff}
 \epsilon_{q}=N_{\text{pe}}/N_{\text{photon}},
\end{equation}
and is a property of the photocathode material and thickness. The photocathode thickness is a compromise between the probability for a photon to interact, which increases with thickness, and the 
energy loss of the liberated photoelectron as it exits the material into the vacuum, which should be as low as possible.

The spectral sensitivity of the PMT is determined by 
the variation of $\epsilon_{q}$ with wavelength and the optical properties of the PMT entrance window.
As the photocathode is not perfectly uniform, the quantum efficiency depends on the 
position of the incident photon on it. This effect is greater for PMTs with a single large photocathode. The amount of variation also depends on the manufacturing process. Older PMTs, 
where the evaporation of the photocathode is controlled by hand, show a greater variation than those in which the cathode is deposited in an automated process, for example. 

The incidence angle of the oncoming photon also factors into the quantum efficiency. Some photons may pass through the semitransparent photocathode. These photons can
be reflected back to photocathode from the interior of the PMT or hit photoemissive surfaces on the inside of the PMT. Both the transmission and reflection are dependent on the angle of incidence. 
The photocathode thickness is also a factor here, as a thicker layer reduces the amount of light transmitted.  
 \end{quote} 

\subsubsection{Collection Efficiency, $\epsilon_{\text{coll}}$}
\begin{quote}
The collection efficiency $\epsilon_{\text{coll}}$ is the efficiency of collecting created photoelectrons into the multiplication stage of the PMT.
It is the probability that photoelectrons from the photocathode will land on the effective area of the first dynode, and can be defined by the ratio
\begin{equation}
\label{eq:CollectionEff}
  \epsilon_{\text{coll}}=N_{\text{d1}}/N_{\text{pe}}
\end{equation}
The collection efficiency depends on
the electrostatic field between the photocathode and the first dynode, as this affects the trajectories taken by photoelectrons.
\end{quote} 

\subsubsection{Electron Multiplication}
\begin{quote}
The electron multiplication stage of the PMT is characterized by the secondary emission ratio of each dynode $\delta_{i}$ and the collection efficiency $\rho_{i}$ of the following inter-dynode space. 
The efficiency $\rho_{i}$ is the ratio of
electrons emitted into the vacuum following dynode $i$ to the number created in the dynode -- i.e., it characterizes the collection of electrons into
the space between dynode $i$ and dynode $i+1$ such that the electrons have trajectories leading towards the sensitive area of the next ($i+1$) dynode. In practice, the collection
efficiencies of each inter-dynode space are very close to one.
The secondary emission ratio $\delta_{i}$ is the average number of electrons emitted by dynode $i$ for each incident electron.
The gain of dynode $i$ is defined by $g_{i} = \rho_{i}\delta_{i}$.\end{quote} 

The \textit{single photoelectron} gain $\mu$ is defined as the number of electrons at the anode for each photoelectron collected into the multiplier, or in terms of the gain of each stage in an n-stage PMT as
\begin{equation}
\label{eq:SinglePhotonGain}
 \mu= \prod^{n}_{i=1} \rho_{i}\delta_{i} = \prod^{n}_{i=1} g_{i}
\end{equation}
Both $\rho_{i}$ and $\delta_{i}$ increase with the applied voltage \cite{PhotonisManual}, and, if the collection efficiency of each stage tends towards 100\%, then the 
single photoelectron gain can be written as a function of the total applied voltage:
\begin{equation}
\label{eq:PMTGain}
 \mu =  \prod^{n}_{i=1} g_{i} =  \prod^{n}_{i=1} k_{i}V_{i}^{\alpha} =  \prod^{n}_{i=1} k_{i} f_{i}^{\alpha}V_{i}^{\alpha} = K V^{n\alpha}
\end{equation}
using the fact that each $V_{i}$ is a fraction $f_{i}$ of the total supply voltage. Due to the nature of secondary emission, there is a large fluctuation in the number of electrons 
from shower to shower. 
The gain of the \textit{photomultiplier tube}, often measured as the ratio of the anode current to the cathode current, $G = I_{a}/I_{k}$, is given by the product of the single photoelectron
gain and the collection efficiency $\epsilon_{\text{coll}}$ of the first dynode.

The \textit{photomultiplier tube efficiency} is the product of the quantum and collection efficiencies (when the multiplication collection efficiencies $\equiv1$),
\begin{equation}
 \label{eq:DetectionEff}
  \epsilon(\lambda, V_{0})= \epsilon_{q}(\lambda)\cdot\epsilon_{\text{coll}}(V_{0})
\end{equation}
where $\lambda$ is the wavelength of the incident light and $V_{0}$ is the voltage difference between the cathode and the first dynode. The efficiency is the proportionality between
number of photons incident on the photocathode of the PMT and the number of photoelectrons which make their way into the multiplication stage of the PMT.
This efficiency is dependent both on
the wavelength, incidence angle, and location on the photocathode of the incident photons through $\epsilon_{q}$, and on the high voltage in the PMT through $\epsilon_{\text{coll}}$.
Both the single photoelectron gain and efficiency are therefore sensitive to variations in the PMT high-voltage power supply, which include
voltage drift and ripple, changes with temperature, input regulation fluctuations, and load regulation issues \cite{HamamatsuManual3a}.

In addition to the gain and detection efficiency, there are other properties of PMTs which are useful to mention here: sensitivity to magnetic fields, dark current, and afterpulses.
PMTs are sensitive to ambient magnetic fields as these affect the trajectories of both photoelectrons emitted from the photocathode and electron showers in the multiplication stage. 
This means that the collection efficiency and gain of the PMT depend on the strength of the magnetic field and on the PMT's orientation within it. For long cylindrical PMT, for example, even rotating the PMT
about its $Z$-axis in the Earth's magnetic field can have a noticeable affect on detection efficiency. This is less of an issue for more compact PMT designs.

The current present in a PMT when it is not illuminated is known as dark current.
This dark current increases with PMT supply voltage, but the rate of increase is not constant with voltage, reflecting the different components of which it is composed: thermionic emission from the 
photocathode and dynodes, leakage current between elements of the PMT, field emission current, ionization current from residual gases in the vacuum tube, and noise from cosmic rays or radiation from isotopes 
in the PMT glass envelope \cite{HamamatsuManual3a,PhotonisManual}. The thermionic emission is strongly dependent on the photocathode material and temperature and is the largest single dark current component which contributes pulses in the 
signal range. Photocathodes are by definition made of materials with a low work function\footnote{The work function is the minimum energy needed to liberate an electron from a solid material into the vacuum immediately outside the surface of the material}, 
and as the cathode's sensitivity is extended to longer and longer wavelengths the rate of thermionic dark current increases exponentially. PMTs operating in the IR thus have the highest rate of dark current, while those working in
the UV show  the lowest rates \cite{HamamatsuManual3a}.  

In addition to being a function spectral sensitivity of the photocathode and the magnitude of the high voltage, the dark current is also affected by the polarity of the voltage repartition. 
In the more usual positive polarity, with the photocathode at a large negative voltage and the anode at ground, the 
voltage difference across the glass of the photocathode, which is in contain with the air, the scintillation medium, the PMT housing, etc.\ is large. This leads to numerous small discharges which create noise that can be picked up at the anode. 
This can result in a factor of 100 increase in dark current. A negative polarity, on the other hand, with the  cathode at ground and the anode at a large positive voltage, gives a much lower dark current.
The anode signal must be passed through a capacitor, however, which limits the signal bandwidth.

Afterpulses are extra pulses which are time correlated with true anode pulses. These afterpulses come from two main sources: ionization of residual gases inside the PMT, and luminous reactions.
These two afterpulse sources can be distinguished by the delay of the afterpulse from the signal pulse when the signal pulse is marked by an external trigger. 
Residual gas ionization afterpulses are due to the excitation and photoemission of either trace gases left in the PMT after the vacuum tube is evacuated,
gases outgassed from electrodes, or helium which has migrated through the glass window of the PMT. Typically this type of afterpulse occurs more than a microsecond after the signal pulse, depending on the ion species \cite{PhotonisManual}.
Luminous reaction afterpulses are caused by photon emission when electrodes are bombarded by electrons. These photons from the electrodes can make their way back to the 
photocathode, causing the emission of a photoelectron, typically 20--100 ns after the signal pulse \cite{PhotonisManual}. Such afterpulses are generally much lower in amplitude than the original signal pulse and 
so they can neglected when working in single photoelectron mode.

    \section{Photon Detection}
\label{subsec:PhotoDetection}
The operational use of PMTs can be broken up into two broad categories: 
\begin{inparaenum}[i\upshape)]
 \item the observation of pulsed light, and
 \item the observation of continuous light.
\end{inparaenum}

For pulsed light, some number of photons arrive at the photocathode in a single well-defined pulse.
This is the case in a scintillator detector, for example, where a number of photons proportional to the energy 
deposited by the particles passing through the scintillator arrive through a light guide to the photocathode of the PMT in a short time. 
The charge delivered to the anode of the PMT is proportional to the number of photons in the pulse.

For continuous light, photons arrive at the photocathode of the PMT randomly at some average rate. If the rate of photons is high enough, the anode signal is a measurable current which is proportional to the 
photon rate. The change in the current gives the variation of the light incident on the PMT.   
 
\begin{figure}[h!]
\includegraphics[angle=270,width=0.90\textwidth]{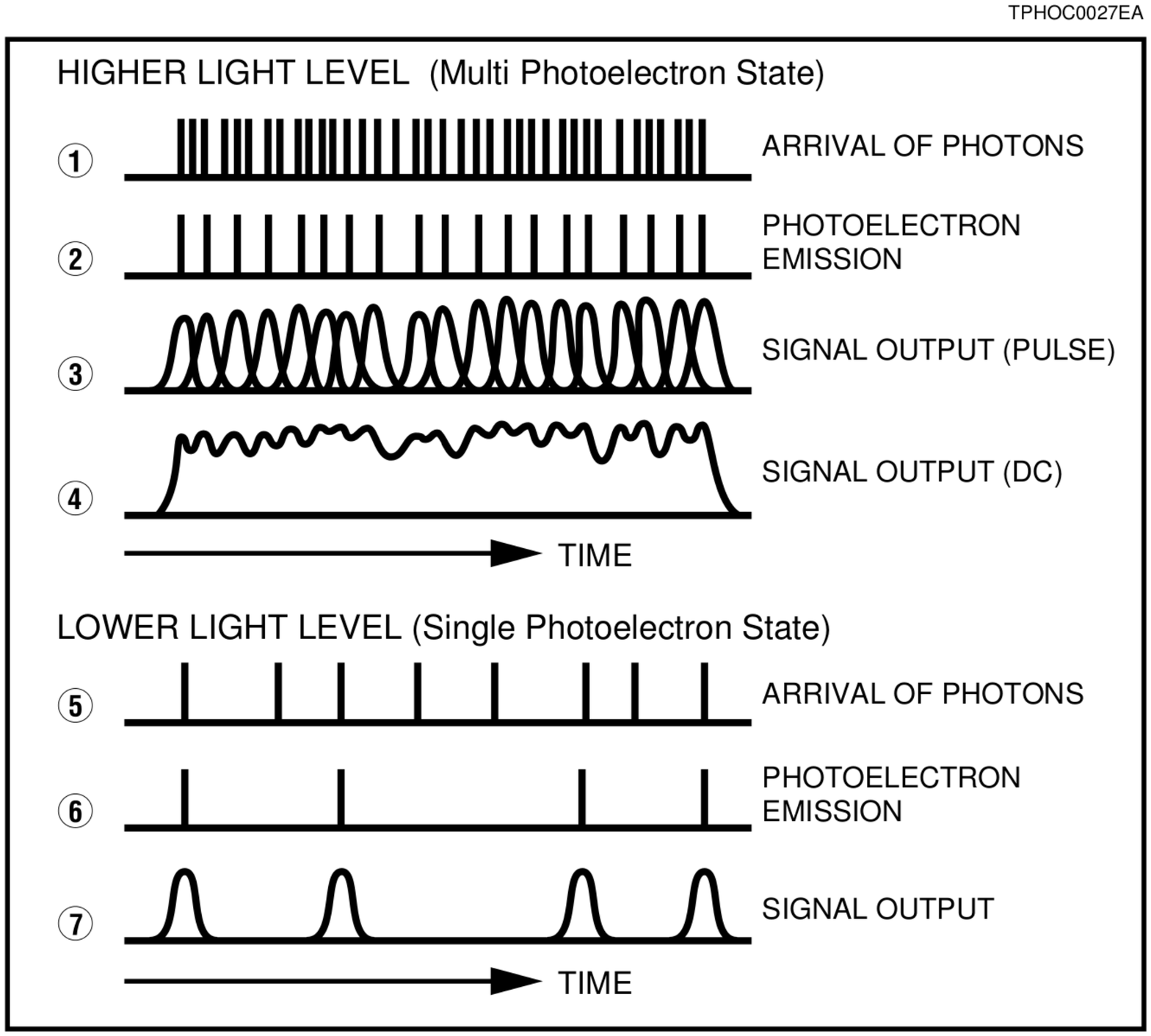}
\caption[Diagram of Photomultiplier Operation Modes]{\label{fig:LightLevelDiagram} A diagram taken from ref.~\cite{HamSPECounting98} showing the analog and single photoelectron operation of a PMT.
The top part of the figure shows the response of a PMT when the rate of arriving photons is high (1), leading to a high rate of photoelectrons (2). 
Because of the finite time-width of the electron shower from each collected photoelectron, the signal pulses on the PMT overlap (3) in a situation which is known as pile-up. The usable readout 
signal is a DC current from the superposition of the individual pulses (4) which is proportional to the incident light. For this reason this state is often called analog mode. 
The bottom half of the figure shows the response of a PMT when the number of arriving photons is low (5). The number of emitted photoelectrons is lower still (6), and so the output pulses from each photoelectron are separated.
In this case the count rate corresponding to the rate of photoelectron emission can be measured (7), and the PMT is said to be operating in digital or single photoelectron counting mode. }
\end{figure}

A particular case of both pulsed light, when the amplitude of each pulse is small, and continuous light, when the average rate is low, is \textit{single photoelectron counting}.
In single photoelectron counting, photons arrive at the photocathode at a rate such that only one or zero photoelectrons are produced at a time, and so the anode pulse from each collected photoelectron can be separated, as is shown at the bottom of \fig\ref{fig:LightLevelDiagram}.
In this case the anode signal can be sent into a discriminator circuit with a threshold set to reject noise, and each count from the discriminator then corresponds to the 
arrival and conversion of at least one photon at the photocathode. 

As the rate of single photoelectrons increases, the rate of anode pulses also increases until individual anode pulses begin to overlap. 
This is known as \textit{pile-up}, and in this situation the count rate begins to saturate, as shown in the upper part of \fig\ref{fig:LightLevelDiagram}.
The onset of pile-up is determined by the time resolution of the PMT, i.e.\ the single photoelectron pulse width, and the time resolution of the read-out electronics. If the time resolution of 
the read-out electronics is the limiting factor, then the range in which the single photoelectron counting is linear can be extended up to that of the 
PMT by introducing a correction to the measured count rate $n_{\text{measured}}$ given by:
\begin{equation}
\label{eq:countratecorrection}
 n_{\text{true}} = \frac{n_{\text{measured}}}{1-n_{\text{measured}}t_{\text{res}}}
\end{equation}
where $t_{\text{res}}$ is the time resolution of the read-out electronics \cite{HamSPECounting98}. This correction gives the true count rate with an error of around 1\%.

As pile-up increases beyond the time-resolution of the PMT, resolving single pulses is no longer possible and the corrected count rate is no longer accurate. In this case the anode output is a direct current including dark current and shot-noise fluctuations. 
This situation is often referred to as \textit{analog} mode, and here the PMT gives an output current which is proportional to the incident light. The constant of proportionality
is the product of the single photoelectron gain and PMT efficiency. 
If the current at the anode increases too far, then the response of the PMT is no longer linear due to space charge effects in the electrodes \cite{Tubs064}.
The linearity characteristics of the cathode and anode are dependent only on the current through them if the supply voltage is constant \cite{HamamatsuManual3a}.

\subsection{Single Photoelectron Counting}
  \label{subsec:Single-Photon Counting}
\begin{figure}[h]
\centering
\includegraphics[width=0.90\textwidth]{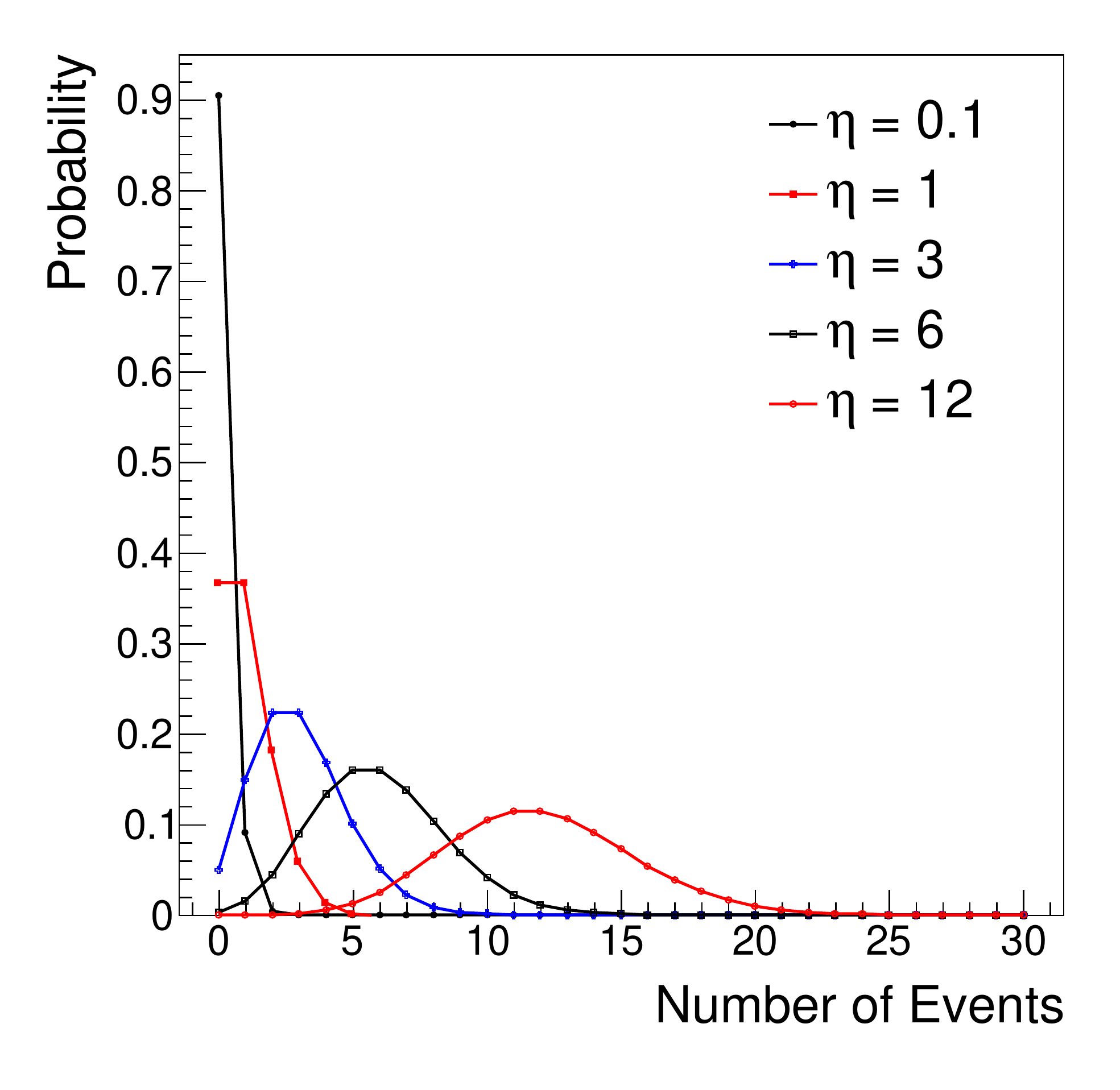}
\caption[The Poisson Distribution]{ \label{fig:poisson} The Poisson distribution, given by \eq\ref{eq:poisson}), plotted for increasing values of $\eta$. The discrete distribution gives the probability of observing $n$
occurrences in a fixed interval of time with an expected number $\eta$. For low values of $\eta$, the distribution is highly asymmetric. As the expected number of occurrences increases, the probability
tends towards a normal distribution.}    
\end{figure}

Single photoelectron counting has advantages over analog measurement in terms of signal-to-noise ratio \cite{HamamatsuManual3a}. 
In particular, current offset from electronics and low-amplitude pulses originating in the electron multiplier are eliminated, and the signal-to-noise ratio improves with the square-root of the total counting time \cite{PhotonisManual}.
If the PMT efficiency $\epsilon$ and the threshold to separate photon pulses from noise are known, then the number of single photoelectron counts is a measurement of the number of single photons incident on the PMT.

To operate in single photoelectron counting mode, the probability that more than two
photoelectrons are emitted from the photocathode within the time-resolution of the measurement must be negligible. Here the time-resolution is either the PMT single photoelectron pulse width if the pulses are 
counted with a discriminator or the integration time (gate) if the anode charge is integrated in a defined window.
The number of photoelectrons per pulse is given by the product of the average number of photons per pulse, the photocathode quantum efficiency, and the collection efficiency: $n_{\text{pe}} = n_{\gamma}\epsilon$.

The creation and collection of any given photoelectron is independent of the same for other photoelectrons and can be assumed to occur at some constant average rate.
Let us assume that the probability of one photoelectron being collected in a short interval of time $\Delta t$ is $\eta \Delta t$, with $\eta$ constant if $\Delta t$ is short enough that the 
probability to collect two photoelectrons is negligible. The probability to collect $n$ photoelectrons in a time interval $t + \Delta t$ is then $P_{n}\left(t+ \Delta t\right)$. 
If $n > 0$, then $P_{n}\left(t+ \Delta t\right)$ is the sum of two mutually exclusive events 
\begin{equation}
\label{eq:PoissonA}
 P_{n}\left(t + \Delta t\right) = P_{n}\left(t\right)P_{0}\left(\Delta t\right)+P_{n-1}\left(t \right)P_{1}\left(\Delta t\right)
\end{equation}
By assumption, the probability of one photoelectron in $\Delta t$ is $\eta \Delta t$, and so the probability of no photoelectrons in $\Delta t$ is $1-\eta \Delta t$.
Substituting this in \eq\ref{eq:PoissonA}) gives:
\begin{equation}
  P_{n}\left(t + \Delta t\right) = P_{n}\left(t\right)\left(1-\eta \Delta t\right) + P_{n-1}\left(t\right)\eta \Delta t
\end{equation}
which can be rearranged into
\begin{equation}
 \frac{P_{n}\left(t + \Delta t\right) - P_{n}\left(t\right)}{\Delta t} = \eta P_{n-1}\left(t\right) - \eta P_{n}\left(t\right)
\end{equation}
Taking the limit $\Delta t \rightarrow 0$ gives the differential equation
\begin{equation}
\label{eq:PoissonB} 
\frac{dP_{n}\left(t\right)}{dt} = \eta P_{n-1}\left(t \right) -\eta P_{n}\left(t\right) 
\end{equation}
Integrating \eq\ref{eq:PoissonB}) for $n=0$ gives $P_{0} = e^{-\eta t}$. The equation can then be solved recursively, and setting $t=1$ gives us the Poisson distribution\footnote{This derivation of the Poisson distribution is taken from M. Boas \cite{BoasTextBook}}:
\begin{equation}
\label{eq:poisson}
p(n;\eta)= \frac{\eta^{n}}{n!}e^{-\eta}.
\end{equation}
 The Poisson distribution gives the probability to find a given number of occurrences $n$ in a given length of time if the occurrences are independent and happen an average rate per unit time $\eta$.
The Poisson distribution is asymmetric, and at low values of $\eta$ the probability of having a larger number of photoelectrons quickly decreases. 
For large values of $\eta$ the Poisson distribution tends towards a normal distribution, as shown in \fig\ref{fig:poisson}.  

For each single photoelectron entering the electron multiplier, the number of electrons on the anode is the single photoelectron gain $\mu$.
There is a large fluctuation in the number of electrons in each anode pulse, due to the secondary emission process in the electron multiplier. This
is dominated by statistical fluctuations in secondary emission at the first dynode. The charge spectrum can be obtained by integrating the total charge received in a defined time window. Alternatively,
the output pulses can be sent through an integrating preamplifier, and the amplitude distribution can be taken with a multi-channel pulse height analyzer. 
The charge or pulse-height spectrum taken at a extremely low light level, such that the response of the PMT to a single photon can be measured with high accuracy, is known as a \textit{single photoelectron spectrum}.  

\subsection{The Single Photoelectron Spectrum}
\label{subsec:The Single Photoelectron Spectrum}

\begin{figure}[ht]
\centering
\includegraphics[angle=0,width=0.9\textwidth]{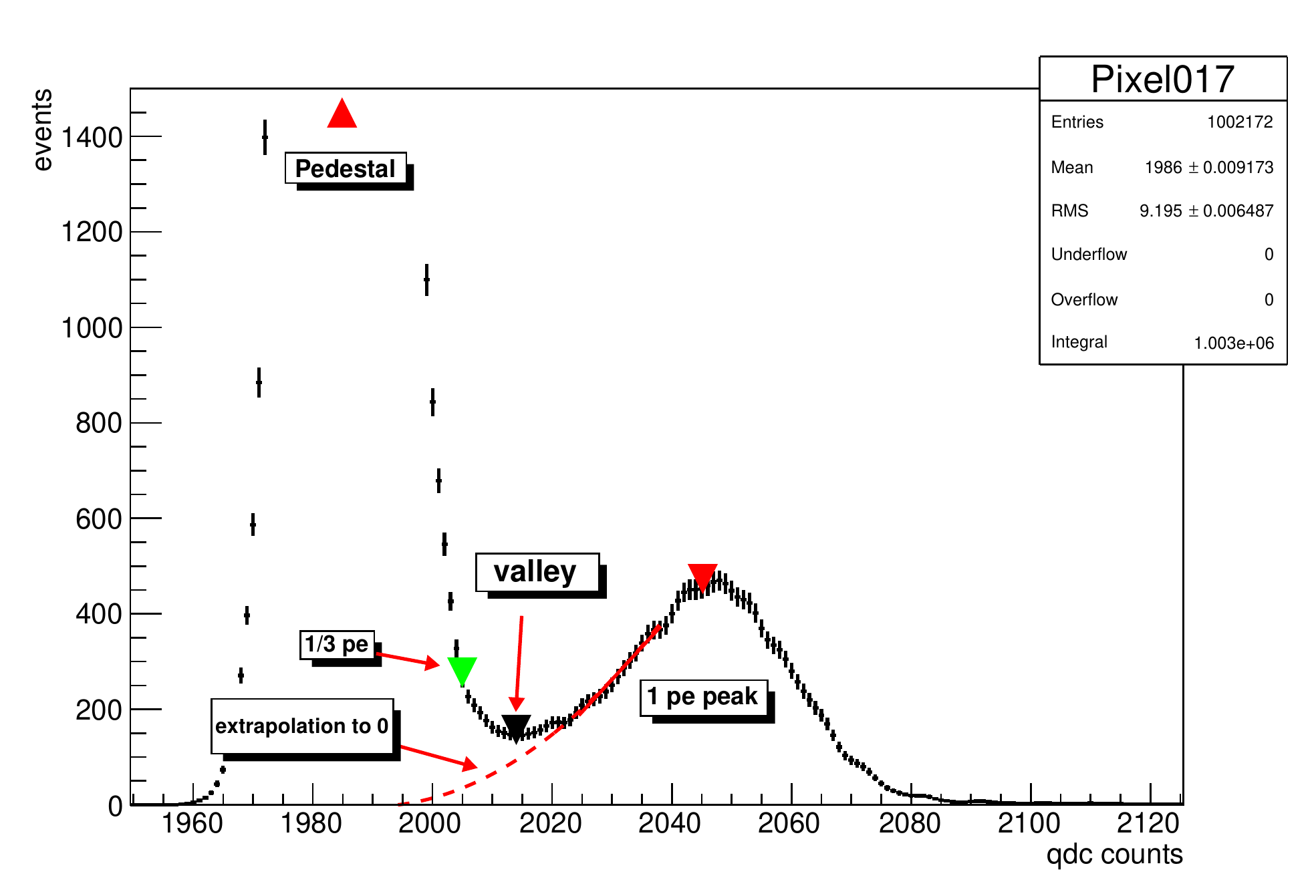}
\caption[Example of a Single Photoelectron Spectrum]{ \label{fig:ExampleSPEspectra}  An example of a \gls{spe} spectrum which I measured using the setup which will be 
presented in chapter~\ref{CHAPTER:PMT Sorting}. The spectrum is shown as a histogram of the number of pulses 
with a given charge, in counts, returned by a charge-to-digital converter. The first peak is the pedestal, corresponding to pulses in which no pe are collected. The second peak, on the right, corresponds to
pulses in which one pe is collected. The two pe peak is not visible, as the ratio of one pe events to zero pe events is $\sim1\%$, and so the number of two pe
events is negligible ($<0.5\%$ the number of one pe) as a consequence of the Poisson statistics. The gain $\mu$ is the difference between the means of the zero pe and one pe peaks (shown by red markers), while
the efficiency $\epsilon$ of the PMT is proportional to the surface of the one pe peak. The black marker shows the valley between the pedestal and one pe peak, and the green marker shows the charge of 1/3 of a pe.
The dashed red line is an extrapolation of the one pe peak from the valley to the mean of the pedestal. }    
\end{figure}

An example of a single photoelectron spectrum which was taken\footnote{Using a Hamamatsu M64 multi-anode PMT, and the setup which will be presented in chapter~\ref{CHAPTER:PMT Sorting}} 
by integrating the charge during a fixed gate (here each gate is called an event) is shown in \fig\ref{fig:ExampleSPEspectra}. 
On the left hand side of the spectrum is the peak corresponding to events during which no photoelectron was collected into the electron
multiplier of the PMT. This peak is often called the pedestal. In theory, the pedestal will be centered on zero charge with noise fluctuations, but in practice the pedestal has a non-zero mean due to low-charge dark current, such as leakage current,
and to the readout electronics offset \cite{Bellamy1994468}. The surface area of the pedestal gives the number of events in which no photoelectron was collected.

The peak on the right is the single photoelectron peak. Several key pieces of information about the PMT can be extracted from the spectrum:
\begin{itemize}
  \item The mean charge of the single photoelectron peak minus the mean charge of the pedestal gives the single photoelectron gain $\mu$ in coulombs.
  The width of the single photoelectron peak is due to the inherent fluctuation in the secondary emission at the first dynode, convolved with the width of the pedestal.
  The fluctuation in gain is dominated by the fluctuation in the secondary emission at the first dynode, which follows Poisson statistics. 
  If, for example, the secondary emission ratio of the first dynode is 3, then the single photoelectron peak resolution will be $\sigma=1/\sqrt{3}$ 
  \item The peak-to-valley ratio can be used as a figure of merit for both the spectrum and the PMT itself.
  The greater the peak-to-valley ratio, the better is the resolution of the PMT. 
  The pulse-height resolution of the PMT, defined as the ratio of the \acrshort{FWHM} of the single photoelectron peak to the height of the peak, increases with peak-to-valley ratio. 
  \item The surface area of the single photoelectron peak gives the number of one photoelectron events. 
 \end{itemize}
If the single photoelectron gain and peak-valley ratio are high enough then it may be possible to resolve a third peak corresponding to two photoelectron events at twice the charge of the single photoelectron peak. 
From \eq\ref{eq:poisson}), the ratio of the number of two photoelectron events to single photoelectron events can be expressed in terms of the ratio of the number of pedestal events to single photoelectron events as
\begin{equation}
\label{eq:SPEcondition}
\frac{p(2;\eta)}{p(1;\eta)} = \frac{\eta^{2}}{2!}\frac{1!}{\eta} = \frac{\eta}{2} = \frac{1}{2}\frac{p(1;\eta)}{p(0;\eta)}
\end{equation}
This allows the contamination of two photoelectron events in the single photon electron spectrum to be estimated: if the ratio of pedestal events to one photoelectron events is less that 1\%, then the ratio of two photoelectron events
to one photoelectron events is less than 0.5\%.

In addition to measuring single photoelectron spectra of  PMTs by taking a charge or pulse height spectrum, the equivalent information can be obtained in a counting experiment. To do this the anode signal of the PMT
is sent through an integrating preamplifier and then through a discriminator circuit. The discriminator gives a output pulse whenever it receives an input pulse with a voltage over a set threshold. 
The plot of the count rate vs threshold is known as an S-curve.
As the single photoelectron spectrum is the charge distribution of the PMT signal, and the S-curve is simply its cumulative distribution
function (in volts, rather than charge, because of the integrating preamplifier), and so the single photoelectron spectrum can be recovered by derivation. 

\begin{figure}[h]
\centering
\includegraphics[angle=0,width=0.9\textwidth]{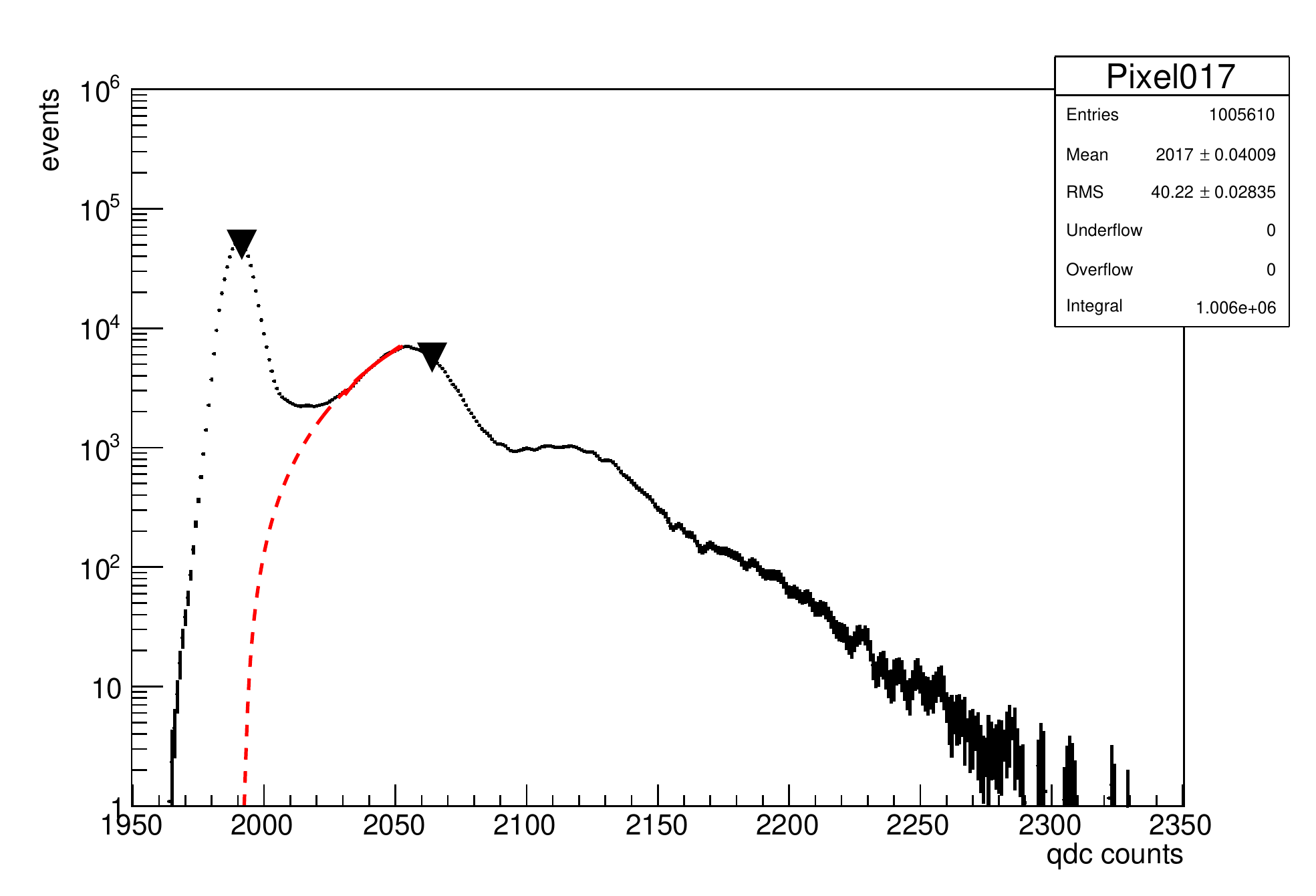}
\caption[Single Photoelectron Spectrum in Logarithmic Scale]{ \label{fig:ExampleSPEspectra_logwith2pe}  
A logarithmic plot of a pseudo-single photoelectron (pe) spectrum for the same pixel of the same PMT as shown in \fig\ref{fig:ExampleSPEspectra}. The two spectra are not identical, as in this spectrum the 
ratio of one pe to pedestal events is approximately 3\%, leading to a larger number of two pe events.
Because of this, and the good peak-to-valley ratio of this PMT, the two pe peak can be resolved at twice the gain of the one pe peak. The black markers show the arithmetic means of the pedestal, calculated between zero charge and 
the valley (between the pedestal and one pe), and of the one pe peak, calculated between the valley and the histogram maximum. 
Notice how the larger number of two pe counts has pulled the one pe mean higher compared to that in the true single pe spectrum (\fig\ref{fig:ExampleSPEspectra}), where the number of 
two pe counts is negligible. The dashed red line shows an extrapolation of the one pe peak to zero.}    
\end{figure}
   
\subsection{Larger Light Pulses}

\begin{figure}[ht!]
\includegraphics[angle=0,width=1\textwidth]{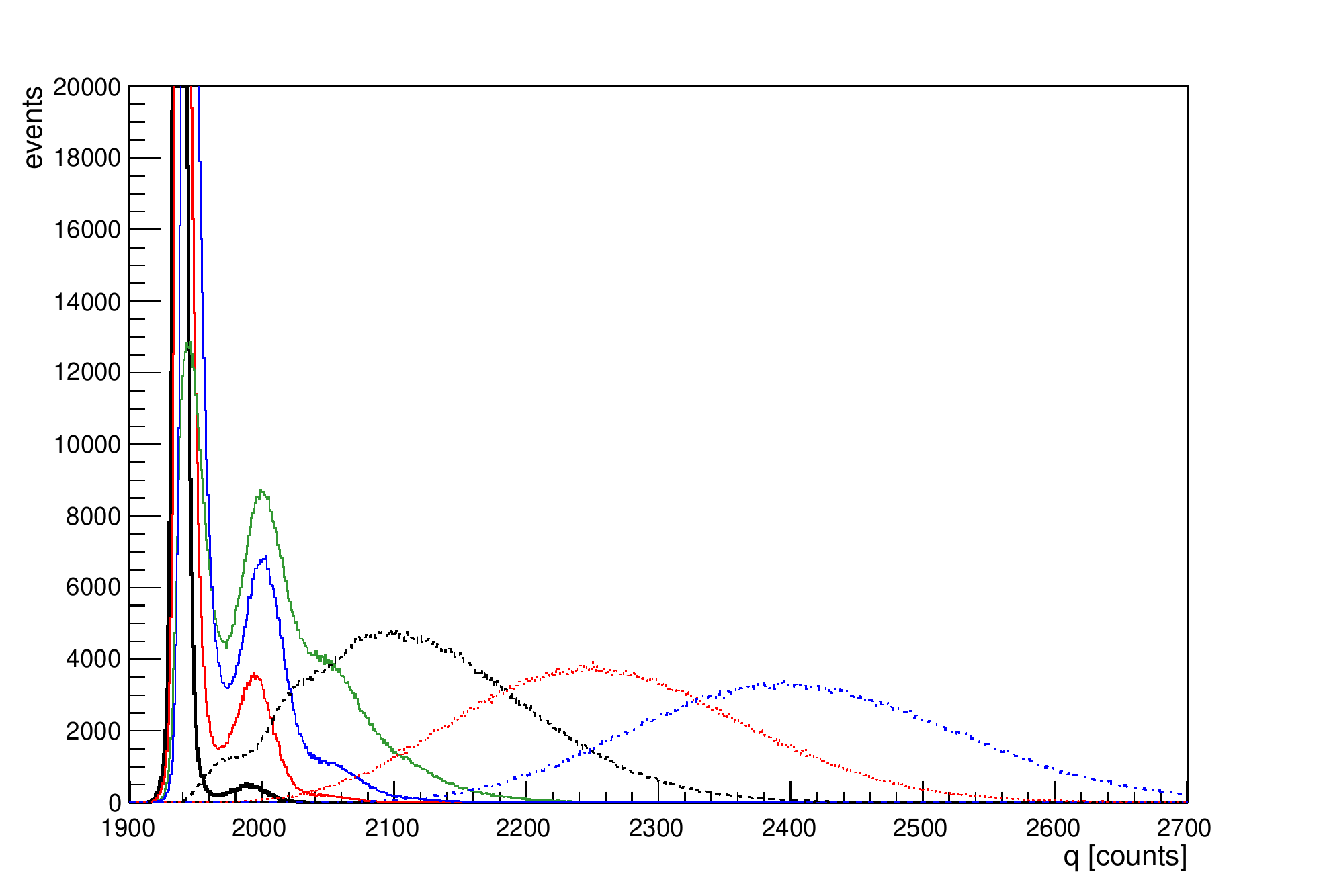}
\caption[Example PMT Spectrum with Increasingly Large Light Pulses]{
\label{fig:LargeLightPulseSpectra} PMT charge spectra overlaid on one another to demonstrate the effect of the Poisson statistics on the PMT response. 
The solid-black curve on the left is a single photoelectron (pe) spectrum, as shown in \fig\ref{fig:ExampleSPEspectra}. 
The solid-\textcolor{red}{red} and solid-\textcolor{blue}{blue}, curves show the response of the PMT as the light pulse amplitude increases but is still low enough that the probability of having zero pe in an event dominates. 
In the solid-\textcolor{green}{green} curve, $\sim50\%$ of the total events are in the pedestal. 
In the dashed-black curve, the pedestal can still be barely seen, but zero pe events are increasingly rare, and the number of photons incident on the 
photocathode is high enough that most events give 1 or more pe.
In the  dashed-\textcolor{red}{red} and dashed-\textcolor{blue}{blue} curves in the middle and far right of figure the pedestal can no longer be seen, 
and the spectrum is a Gaussian peak with a mean given by the product of the average number of pe per event and the single pe gain.
}
\end{figure}

The spectrum of the PMT is a convolution of 
the individual photoelectron peaks, and the relative occurrence of a given number of photoelectrons per pulse is related to the average number of photons per pulse by the Poisson statistics.
As the number of photons per light pulse (or the rate of pulses in an integration window) increases, a transition from the single photoelectron counting mode to the same response as in analog mode can be seen.  
This is demonstrated in \fig\ref{fig:LargeLightPulseSpectra}, which shows examples of charge spectra that I measured for increasingly large light pulses 
using the same pixel of a multi-anode PMT as for the single photoelectron spectrum of \fig\ref{fig:ExampleSPEspectra}.

The solid-black curve on the left is a single photoelectron spectrum of this pixel, as shown in \fig\ref{fig:LargeLightPulseSpectra}. The single photoelectron spectrum is dominated by the pedestal, but
the one photoelectron peak can clearly be seen, along with a tail from pulses which give two photoelectrons. 
The solid-\textcolor{red}{red} and solid-\textcolor{blue}{blue} spectra show the response of the PMT as the number of photons per pulse increase, but are still few enough that the probability of having no photoelectrons in an event is high. 
The proportion of 1, 2, and higher photoelectron events is greater than in the single photoelectron spectrum, 
following the Poisson distribution for increasing $\eta$ (compare to \fig\ref{fig:poisson}).

In the solid-\textcolor{green}{green} spectrum of \fig\ref{fig:LargeLightPulseSpectra}, approximately 50\% of the total events are in the pedestal. 
If the number of photons per pulse is increased even further, then nearly every event gives 1 or more photoelectrons. In the dashed-black curve, there are some zero photoelectron events. These are increasingly rare, however, and the number of photons incident on the 
photocathode is high enough that most events give 1 or more photoelectrons.

A still further increase in the number of photons per pulse gives the dashed-\textcolor{red}{red} and dashed-\textcolor{blue}{blue} spectra in the middle and on the right of \fig \ref{fig:LargeLightPulseSpectra}.
The pedestal can no longer be seen in these spectra, and the spectrum is a Gaussian peak with a mean given by the product of the average number of photoelectrons per event and the single photoelectron gain.

\begin{figure}[ht]
\includegraphics[angle=270,width=1\textwidth]{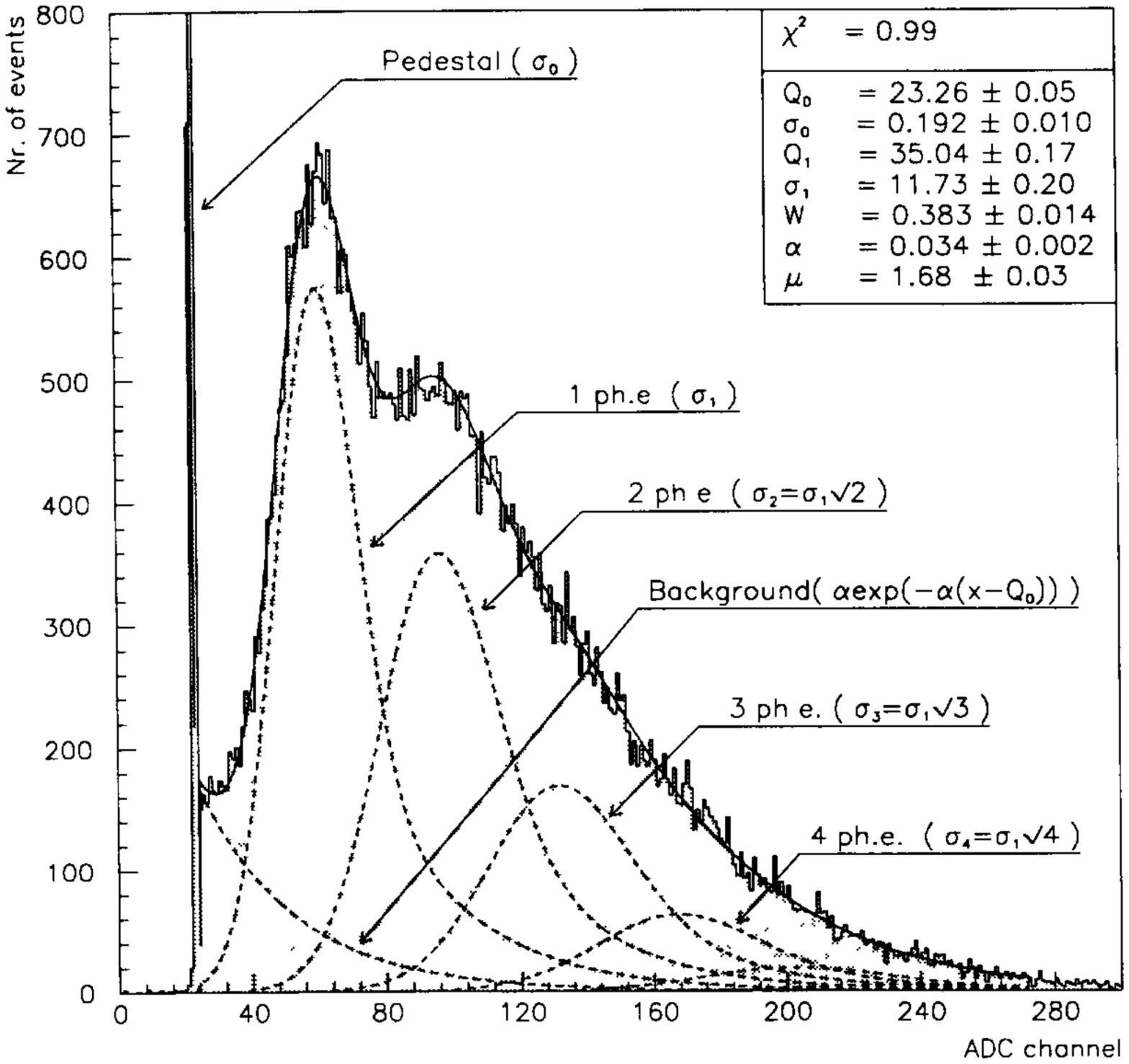}
\caption[Example of PMT Spectrum Deconvolution]{Plot from ref.~\cite{Bellamy1994468} showing the deconvolution of a PMT spectrum. The spectrum in this plot 
is similar to that shown by the solid-\textcolor{green}{green} curve of \fig\ref{fig:LargeLightPulseSpectra}. The peaks from 0, 1, 2, etc.\ photoelectron peaks which compose the total spectra are shown. An exponential component corresponding
to the thermionic dark current is also included in the fit. The parameters $Q_{0}$ and $\sigma_{0}$ are the pedestal mean and width, and $Q_{1}$ and $\sigma_{1}$ are the one photoelectron mean and width. The mean and width of higher photoelectron peaks
are given by $Q_{n} = nQ_{1}$ and $\sigma_{n}=\sigma_{1}\sqrt{n}$, respectively. $W$ and $\alpha$ are the background parameters, with $W$ being the background event probability. The relative heights 
of each peak are determined by the Poisson distribution, here $\mu$ is the Poisson parameter -- i.e., the average number of photoelectrons per event.}
\label{fig:SpectrumDeconvolution}
\end{figure}

If the gain and peak-to-valley ratio of the PMT were high enough compared to the resolution of the charge measurement, these spectra would be resolved into multiple peaks each corresponding to a certain number of photoelectrons.
In the case of a PMT such as the one used in this spectrum, where the individual photoelectron peaks cannot be resolved, they can be recovered by a deconvolution if assumptions about the nature of the PMT response are made \cite{Bellamy1994468,Tokar:1999xza}.
An example of spectrum deconvolution is shown in \fig \ref{fig:SpectrumDeconvolution}, which shows both the total spectrum and the peaks which correspond to a given number of photoelectrons per pulse. Such a fit is complex and 
very sensitive to both the simplifying assumptions made in the PMT-response model and to the initial choice of fit parameters. An analysis of this type is not necessary in JEM-EUSO, as the observation of UHECR showers will be by single photoelectron counting.

    \section{Photomultiplier Calibration}
\label{sec:PMTCalibration}
In many applications the absolute number of photons incident on the detector is needed with a low uncertainty. 
In JEM-EUSO, for example, the observation of UHECR extensive air showers using the air fluorescence technique provides a calorimetric measurement of the energy deposited in the atmosphere by the electromagnetic part of the shower. The number of photons
emitted per eV of deposited energy is known as the fluorescence yield. Reconstructing the energy of the primary cosmic ray thus requires knowledge of the absolute number of photons 
received by the detector, and therefore information on the  efficiency $\epsilon_{i}$ of each multi-anode PMT pixel in the JEM-EUSO focal surface. To meet the science requirements of JEM-EUSO, our goal is to know each $\epsilon_{i}$
with an overall accuracy of 5\% or better.  

The \textit{relative} quantum efficiency of the cathode as a function of wavelength is generally provided by PMT manufacturers with an error on the order of a few percent. 
The absolute PMT efficiency is more difficult, and is generally not given by the manufacturer to better that 20\%. 
Even in physics experiments which have undertaken a dedicated measurement, the precision of the techniques they have used was no better than 10--20\%. 
For a high energy particle physics experiment such as JEM-EUSO, this uncertainty is much too high, and a method for measuring the absolute efficiency with an uncertainty of a few percent is needed.  

The PMT efficiency is the product $\epsilon(\lambda) = \epsilon_{q}(\lambda)\cdot\epsilon_{\text{coll}}$, and so knowledge of $\epsilon_{q}$ can be translated into $\epsilon$ by measuring the collection efficiency.
The collection efficiency can be measured using the ratio of the PMT gain, \eq\ref{eq:PMTGain}), 
over the single photoelectron gain, \eq\ref{eq:SinglePhotonGain}), but this method is difficult and imprecise for several reasons.

The first difficulty is measuring the cathode current $I_{k}$. Even if the count rate is as high as 1 MHz, $I_{k}$ will only be a few nA, which is difficult to measure precisely. The work-around to this is to 
measure $I_{k}$ at very high light level so that the current is on the order of a pA.  At a gain of $10^{7}$ such an amount of light would damage or destroy the anode, as the anode current $I_{a}$ must typically be kept below $\approx 100~\mu$A. 
The response of the PMT is also not linear at very large anode currents.
The PMT should thus be operated at a gain of 
$\sim10^{4}$--$10^{5}$. The single photoelectron gain must be higher than $10^{6}$ to be measurable, however. This means that the PMT gain must be extrapolated 2--3 orders of magnitude.
 
One way around this problem is to operate the PMT in diode mode, placing only the photocathode and first few dynodes at high voltage. This way $I_{k}$ can be measured at a very high light level without saturating the anode. 
After $I_{k}$ is measured, the light flux is attenuated 
several orders of magnitude using neutral density filters with a known attenuation coefficient $\alpha$. The PMT is then put at full gain so that the anode current can be measured assuming a cathode current of $I_{k}/\alpha$. 

Care must be taken here that there are no fluctuations in the emission of the source between the two measurements. 
Assuming that the anode current is attenuated as $I_{k}/\alpha$ also ignores the dark current component of $I_{k}$. 
Many elements of the dark current, such as the thermionic emission or ohmic leakage current, are not reduced by the attenuation of the incident light. For low $I_{k}$ this can be a significant source of error.
There is also a large source of error in the knowledge of the attenuation factor $\alpha$, which is often not known to better than 10\% due to reflection and diffusion effects.

Another technique is to directly measure the efficiency of the PMT as the ratio of the number of detected signals to the number of incident photons. 
If we measure the single photoelectron spectrum of the PMT, the number of single photoelectron events is given by the surface of the one photoelectron peak. The contamination of two photoelectron events is determined by the 
Poisson statistics and is less than 1\% if the ratio of one photoelectron events to pedestal events is less than 2\%, as shown in \eq\ref{eq:SPEcondition}). 
In this case, the experimental difficulty is to determine the absolute number of photons incident on the photocathode.
This can be determined by using a light source of known emission characteristics, which means either i) comparing the PMT to a calibrated source or ii) comparing the PMT to another calibrated detector.

\subsection{Comparison to a Calibrated Source}
A calibrated light source is any source which gives a known number of photons per second per steradian. These may be a calibrated lamp, a laser, synchrotron radiation, or Cherenkov emission.  
If the power spectrum $dP^{3}/dS d\lambda d\Omega$ of the source is known with high accuracy, then the emissive surface, flux, and solid angle can be accounted for in the measurement.
This is a delicate task experimentally, and the variation of the flux with temperature and time must also be taken into account.

If the solid angle subtended by the emission of the source is small, such as for a laser, the flux is generally high in the emission region compared to the operational range of the PMT and must be attenuated by multiple orders of magnitude. 
For sources in which the emission subtends a large solid angle, such as many calibrated lamps, the flux may require less attenuation, but then a precise knowledge of the spatial variation is critical.
The uncertainty involved in attenuating light sources in a controlled way can reach nearly 20\% and so the precision on the measurement of the efficiency using this type of method is 
limited \cite{tubs091}. The uniformity of the source is also a problem in either case as
attenuation filters can create lobes and other variations in intensity which, if unaccounted for, can be a source of systematic error. 

Using Cherenkov or synchrotron radiation is subject to errors in calculating the Cherenkov yield, determining the effective aperture, and controlling the viewing distance. For radioactive Cherenkov sources there are also 
systematic uncertainties due to photons trapped in the source by total internal reflection. There can also be an effective reduction in the number of photons incident on the PMT due to photon coincidences when using Cherenkov emission \cite{Biller:1999ik}.
The chance of photon coincidences is increased because the distribution of Cherenkov light from each charged particle emitted is focused in a cone around the direction of the charged particle.

\subsection{Comparison to a Calibrated Detector}
Comparison to a calibrated detector effectively creates a continuously calibrated source, and so eliminates the problem of intensity variations with temperature and time. The spatial variation of 
the source must be still accounted for, however.

Any absolutely calibrated detector, such a NIST\footnote{National Institute of Standards and Technology, USA} photodiode, NPL\footnote{National Physics Laboratory, UK} photocell, or CNAM\footnote{Conservatoire National des Arts et Métiers, France} photodiode, can be used as the reference detector. The same flux must be viewed simultaneously by both the PMT and the photodiode.
This presents the most obvious problem with this method: the gain of a PMT is on the order of $10^{6}$ or higher, while the gain of the photodiode is 1 or less.

The PMT calibration of \gls{SNO} attempted to overcome this by placing the reference photodiode at a position closer to the source than the PMT \cite{Biller:1999ik}. 
The problem is then shifted to the inference of the total flux at the PMT from the measured flux per steradian and knowledge of solid angle subtended by the PMT. This method also fails to 
account for the non-uniformity of the light source, which would have a dependence in angular shape of $\cos\theta$ in their case (as they used a Lambertian source). 
Comparison to a calibrated detector thus represents a clear improvement over the use of a calibrated source, but it also presents difficulty in the need to match the gain of the reference detector to that of the PMT and in 
measuring the luminosity distribution of the source.

    \section{A Precise Method for Measuring Absolute Efficiency}
\label{sec:Measuring Absolute Detection Efficiency}

As we have seen there are two ways, philosophically speaking, to measure the absolute detection efficiency of a PMT: a) send a known light, or b) compare to a known detector.
Having looked at the experimental difficulties in both cases, the question remains: how can we measure the critical parameters of the PMT, the single photoelectron gain and PMT efficiency, to a high accuracy?

To answer this problem, Lefeuvre et al.\ \cite{LefeuvreThesis, Lefeuvre:2007jq} (G. Lefeuvre being my predecessor at APC) developed the previous setup of ref.~\cite{Biller:1999ik} to its logical conclusion. They proposed an absolute
calibration by comparing the response of the PMT \emph{directly} to an absolutely calibrated photodiode. Here directly is emphasized because although previous methods used a comparison with a known detector, the indirect nature of 
their comparison was a source of systematic uncertainties. The calibration technique developed by Lefeuvre et al.\ is novel enough that it led to a patent.

In this direct comparison method the PMT to be calibrated and the reference photodiode view the same light in real time, and their gains are matched by attenuating the light by a factor of $\sim 10^{6}$--$10^{7}$ in a stable and repeatable way.
To do this, an integrating sphere is used as a stable and well-characterized splitter. The attenuation is itself measured using a second absolutely calibrated photodiode. This second photodiode directly replaces the PMT so that it is at the same 
distance and uses the same opening of the sphere. In this way the intensity and spatial uniformity are the same in both cases.
The basic hardware ingredients of this method will be briefly presented in sections \ref{subsec:The Testing Enviroment: A Black Box}, \ref{subsec:Integrating Spheres}, and \ref{subsec:Reference Detector: NIST Photodiode}, 
and then the total procedure will be explained in section \ref{subsec:Absolute Calibration Setup}.

\subsection{The Testing Environment: A Black Box}
\label{subsec:The Testing Enviroment: A Black Box}

The most essential element of the calibration setup, regardless of technique, is the measurement environment. 
Working with PMTs requires that the immediate setup be placed inside a black box to control the exposure of the high-gain PMT to light. This is to both ensure that the PMT is not damaged by over exposure and that the signal-to-background ratio of the measurement is good.
Several pictures of our black box are shown in Figs.~\ref{pic:BlackBox:1} and \ref{pic:BlackBox:2} in the appendix.
Keeping the level of background light as low as possible and ensuring the safety of the PMTs each require their own consideration.

An estimate of the maximum allowed background light can be made by considering that operating in single photoelectron mode requires that less than 1\% of the measurement cycles give a photoelectron, as per the Poisson statistics.
As an example, say that we wish to test a PMT with a 20 cm$^{2}$ photocathode at the single photoelectron level using a discriminator on the anode signal of the PMT, giving a measured count rate. We illuminate the PMT with a constant light, so that photons arrive at the photocathode randomly and with some average rate. 
If the double-pulse resolution of the discriminator is 40 ns, then a light level which gives a count rate of $\approx$ 25 kHz, would be at single photoelectron mode, i.e., there would be less than 1\% missed counts due to having two photoelectrons emitted in the same 40 ns.
If the efficiency of the PMT is 25\%, then a signal-to-background ratio of would 100 require that the flux of photons in the black box be less than 50 per second per cm$^{2}$, integrated in the entire spectral sensitivity of the PMT. The allowed background rate
scales with the size of the PMT, a single larger photocathode requires a lower background flux.

To eliminate as much background light as possible, the black box must be optically sealed in a similar manner to a film camera.
Although putting the setup inside a closed box seems simple, it is far from trivial. The choice of material is important; a wooden box is the simplest to work with and to make light tight. 
The box has to be constructed using lap joints or a similar technique, so that there is no possibly of small gaps at corners or at joints between sides which would give a straight path for photons to enter the box.
There cannot be any holes in the box, and so every cable which goes through the black box must pass through a connector. The connector design has to be one that is light tight, which is not necessary true for many common connector types. For example, ribbon cables and \gls{USB} connectors are both difficult in this sense, and special
connectors for each had to be found or even fabricated. The entry door to the black box is sealed using a camera baffle, which assures that the door is light tight every time it is closed, unlike a rubber seal. 

Inside the box, every surface is painted matte black to minimize reflections. Obviously, there cannot be any lights inside. This is often a problem with electronics, and, as an example, the LEDs on several test read-out boards had to be cut off because their designers did not
provide any way to turn them off.

For the safety of the PMTs there is also a protection from human forgetfulness built into the black box. Imagine that a physicist, busy with his measurements, might forget to turn off the high voltage to the PMT before opening the box once in a hundred times.
This would result in a destroyed PMT, at a cost of several thousand euros, once a week$\ldots$
To avoid this, the high voltage connectors are placed so that the door of the black of box cannot be opened while the high voltage is connected. A photograph of this essential mechanism is shown in \fig\ref{pic:BlackBox:LockandCables} in the appendix. 
Once the proper testing environment is available, the other necessary aspects of the calibration setup can be considered.

\subsection{The Splitter: An Integrating Sphere}
\label{subsec:Integrating Spheres}

\begin{figure}
\centering
 \subfigure[The Exchange Factor]{\label{fig:ISdiagrams:ISdifferentialElements} \includegraphics[angle=270,width=0.8\textwidth]{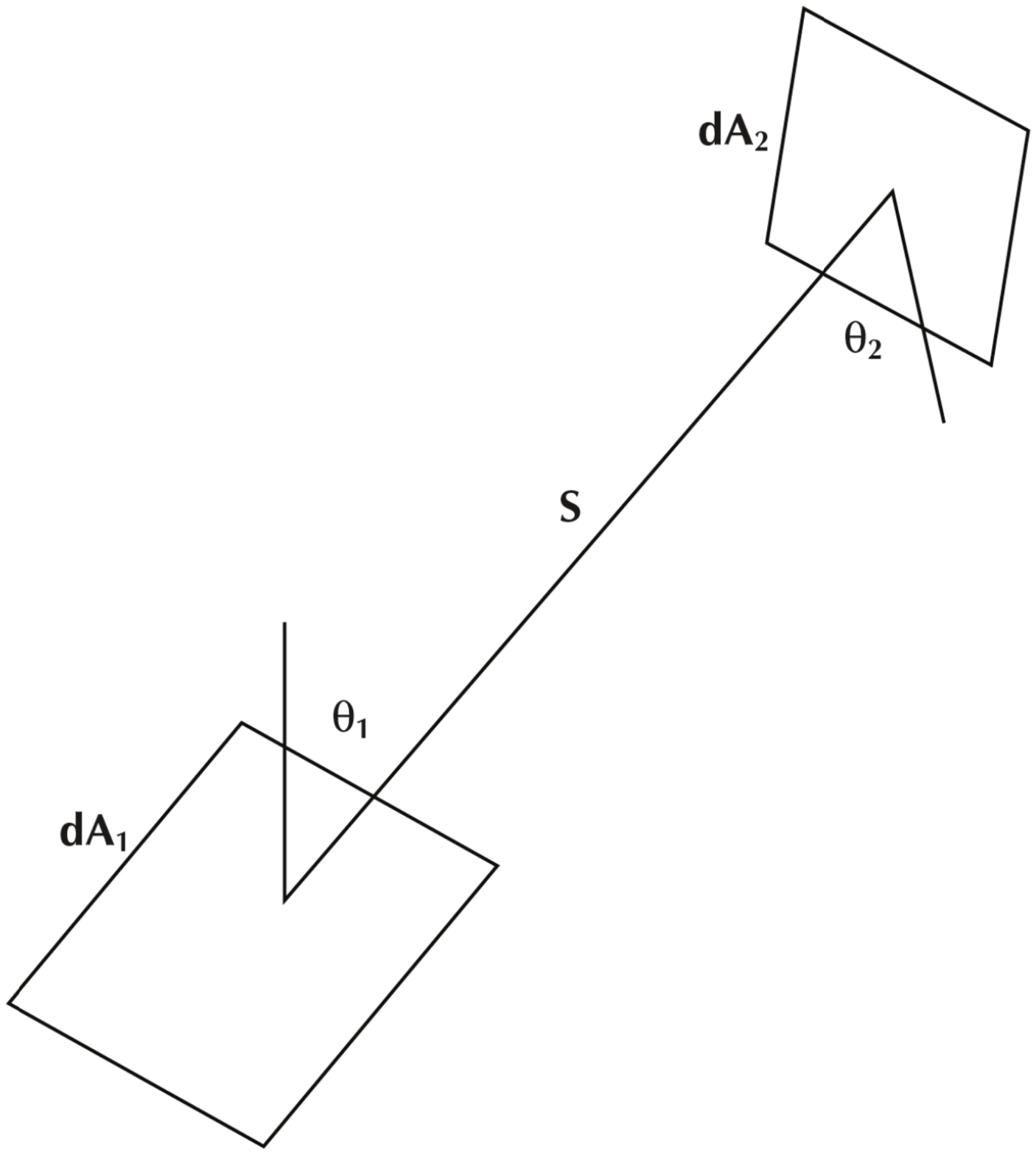}}
  \subfigure[Diagram of the Flux]{\label{fig:ISdiagrams:ISfluxdiagram} \includegraphics[angle=270,width=0.8\textwidth]{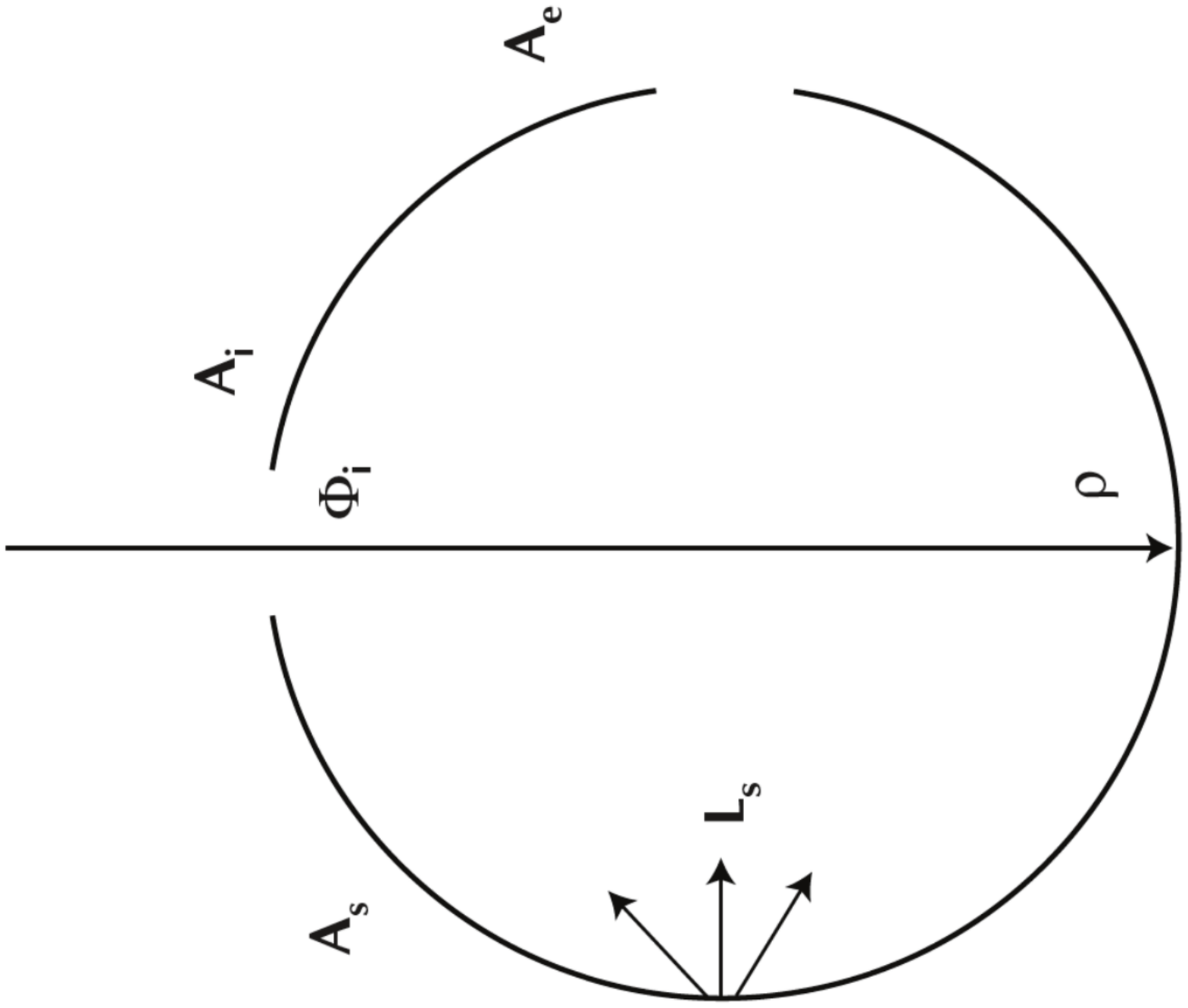}}
\caption[Diagrams for Deriving Integrating Sphere Properties]{ \label{fig:ISdiagrams} A collection of diagrams, taken from refs.~\cite{LabSphere01,LabSphere02}, to demonstrate the properties of integrating spheres. \fig\ref{fig:ISdiagrams:ISdifferentialElements} shows the radiative exchange between
two differential surface elements $dA_{1}$ and $dA_{2}$. The two elements are at a distance $S$ from one another and are oriented at angles $\theta_{1}$ and $\theta_{2}$ normal to each surface. In this situation, the 
exchange factor $dF$, the fraction of energy leaving surface 1 and arriving at surface 2, is given by \eq\ref{eq:ISExchangeFactor}). The second diagram, \fig\ref{fig:ISdiagrams:ISfluxdiagram} shows an integrating sphere
with a total internal surface area $A_{\text{s}}$ and two ports with areas of $A_{\text{i}}$ and $A_{\text{e}}$. The reflectance of the interior walls of the sphere is $\rho$, and there is an input flux $\Phi_{\text{i}}$ entering the sphere through
the top port. In this case, the radiance inside the sphere $L_{\text{s}}$ is given by \eq\ref{eq:TotalRadiance}). }
 \end{figure}

An integrating sphere is a hollow sphere with a diffusive material coating the inside surface.
The basic function of an integrating sphere is to spatially integrate radiant flux, and
the general properties of an integrating sphere can be derived by considering the radiation exchange between two differential elements of a diffuse surface, as shown in \fig\ref{fig:ISdiagrams:ISdifferentialElements}.
The exchange factor $dF$, the fraction of energy leaving surface element $dA_{1}$ and arriving at $dA_{2}$, is given by
\begin{equation}
  dF = \frac{\cos\theta_{1}\cos\theta_{2}}{\pi S^{2}}dA_{2}
\end{equation}
where the angles $\theta_{1}$ and $\theta_{2}$ are measured normal to each surface. 
If we have two differential elements $dA_{1}$ and $dA_{2}$ on the inside of a sphere with a diffuse surface, then the distance $S$ is equal to $2R\cos\theta_{1} = 2R\cos\theta_{2}$.
The exchange factor is then
\begin{equation}
\label{eq:ISExchangeFactor}
 dF = \frac{dA_{2}}{4\pi R^{2}} = \frac{dA_{2}}{A_{\text{s}}}
\end{equation}
and the fraction of the radiant flux received by a finite surface $A_{2}$ is simply the fraction of the total sphere surface area it represents.
This result is important as it is independent of the viewing angle between the surfaces, the distance between them, and the size of the emitting part of the surface \cite{LabSphere01, LabSphere02}.

\subsubsection{Radiance}
The radiance $L$, defined as the flux density per steradian, is a useful description of the light emanating from an integrating sphere. For a diffuse surface the radiance is given by
\begin{equation}
 L = \frac{\Phi_{\text{i}}\rho}{\pi A_{\text{s}}}
\end{equation}
where $\rho$ is the reflectance of the surface and $\Phi_{\text{i}}$ is the input flux. 

\fig\ref{fig:ISdiagrams:ISdifferentialElements} shows an integrating sphere with an input port of area $A_{\text{i}}$, an exit port of area $A_{\text{o}}$, and an input flux of $\Phi_{\text{i}}$. In such a situation,
the input flux is diffused by the initial reflection, and the flux incident on the total internal surface of the sphere is
\begin{equation}
\label{eq:ISTotalFlux1refl}
 \Phi_{\text{total},1} = \Phi_{\text{i}}\rho\left(\frac{A_{\text{s}} - A_{\text{i}} - A_{\text{o}}}{A_{\text{s}}}\right) =  \Phi_{\text{i}}\rho \frac{A_{\text{eff}}}{A_{\text{s}}}
\end{equation}
where the quantity $A_{\text{eff}}$ is the effective surface of the sphere.

By the same logic, the contribution to the flux from a second reflection is $\Phi_{\text{total},2} = \Phi_{\text{i}}\rho^{2}\left(A_{\text{eff}}/A_{\text{s}}\right)^{2}$.
After $n$ reflections the total flux inside the sphere is
\begin{equation}
  \Phi_{\text{total},n} = \Phi_{\text{i}}\sum^{n}_{i=1}\rho^{i}\left(\frac{A_{\text{eff}}}{A_{\text{s}}}\right)^{i}
\end{equation}
and, as the number of reflections goes to infinity, this becomes
\begin{equation}
  \Phi_{\text{total}} = \Phi_{\text{i}} \frac{\rho A_{\text{eff}}}{A_{\text{s}} -\rho A_{\text{eff}}}
\end{equation}
The final radiance is then
\begin{equation}
\label{eq:TotalRadiance}
 L = \frac{\Phi_{\text{total}}\rho}{\pi A_{\text{s}}} = \frac{\Phi_{\text{i}}}{\pi A_{\text{s}}} \frac{\rho A_{\text{eff}}}{A_{\text{s}} -\rho A_{\text{eff}}}
\end{equation}
An exact analysis of the distribution of radiance inside an integrating sphere depends on the distribution of incident flux, the geometrical details of the actual sphere design, 
and the reflectance distribution function of the sphere coating and all surfaces of each device
mounted at a port opening or inside the integrating sphere \cite{LabSphere01}.

The effective surface area of the sphere $A_{\text{eff}}$ generalizes to any number of openings as $A_{\text{eff}}= A_{\text{s}} - \sum_{1}^{m}A_{i}$ 
with a corresponding decrease in the spatial integration of the radiant flux as $\rho$ and $A_{\text{eff}}$ decrease. 
A general rule is that $A_{\text{eff}}$ should be more than 95\% in order to maintain a high reflected flux -- i.e., the sum of all the ports on the integrating sphere should be less than 5\% of the total surface area.

\subsubsection{Time Response}
Aside from its effect on the spatial distribution of light, an integrating sphere also has a temporal effect.
For rapidly varying light signals, such as short pulses, the output signal can be noticeably distorted by the ``pulse stretching'' caused by multiple diffuse reflections. 
The shape of the output signal is the convolution of the input signal with the impulse
response of the integrating sphere, which is of the form $e^{-t/\tau}$.
The time constant $\tau$ can be calculated as 
\begin{equation}
\label{eq:IStimeconstant}
 \tau = - \frac{2}{3}\frac{d_{\text{sphere}}}{c \ln \bar{\rho}}
\end{equation}
where $d_{\text{sphere}}$ is the diameter of the sphere and $\bar{\rho}$ is the average reflectance of the inner walls of the sphere. 
Typical integrating sphere time constants range from several nanoseconds to a few tens of nanoseconds.

\subsubsection{Spatial Uniformity}
\begin{figure}[p]
\centering
 \subfigure[The Irradiance]{\label{fig:ISirradiance:diagram} \includegraphics[angle=270,width=0.90\textwidth]{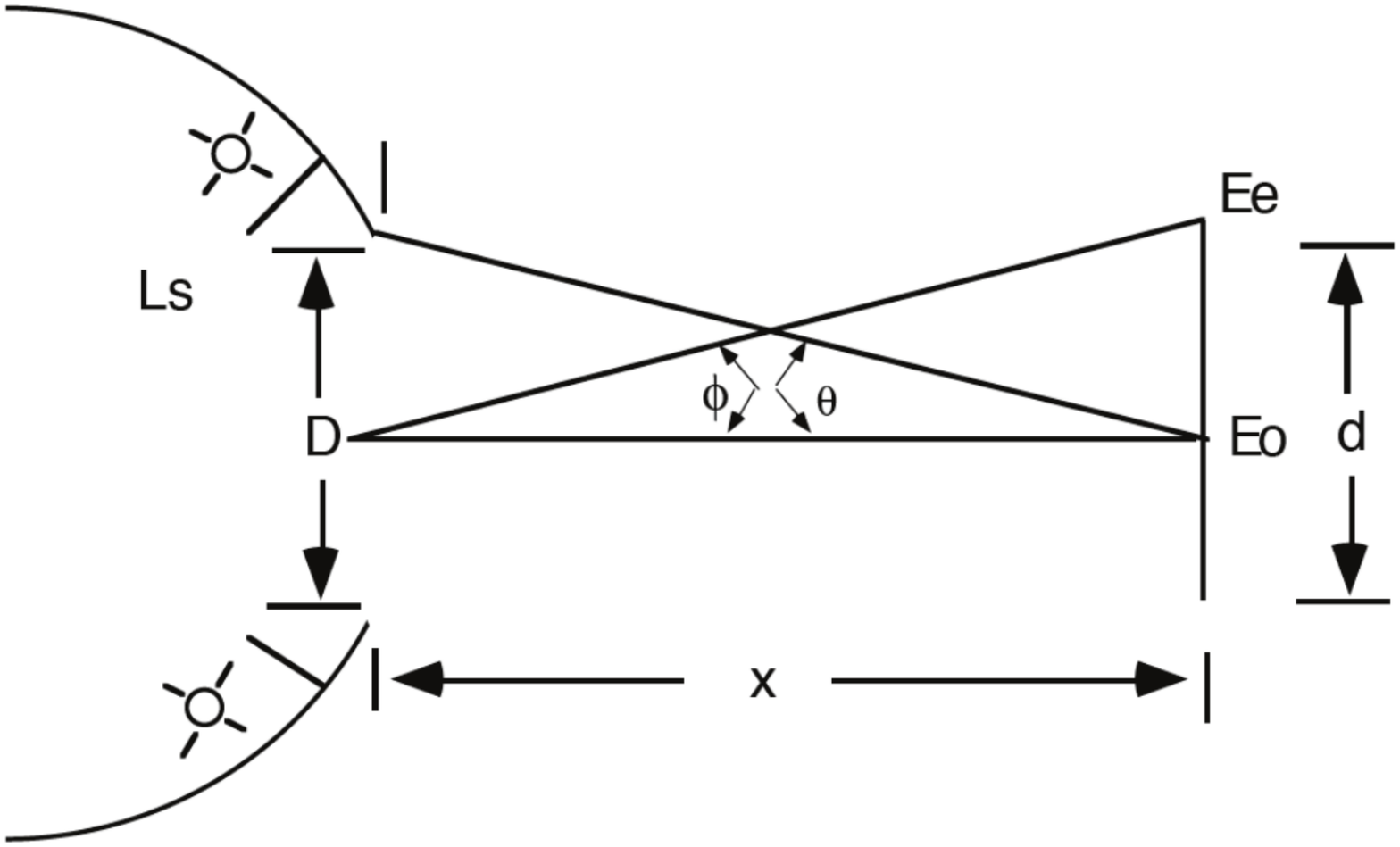}}\\
  \subfigure[A Plot of the Uniformity]{\label{fig:ISirradiance:Plot} \includegraphics[angle=270,width=1.0\textwidth]{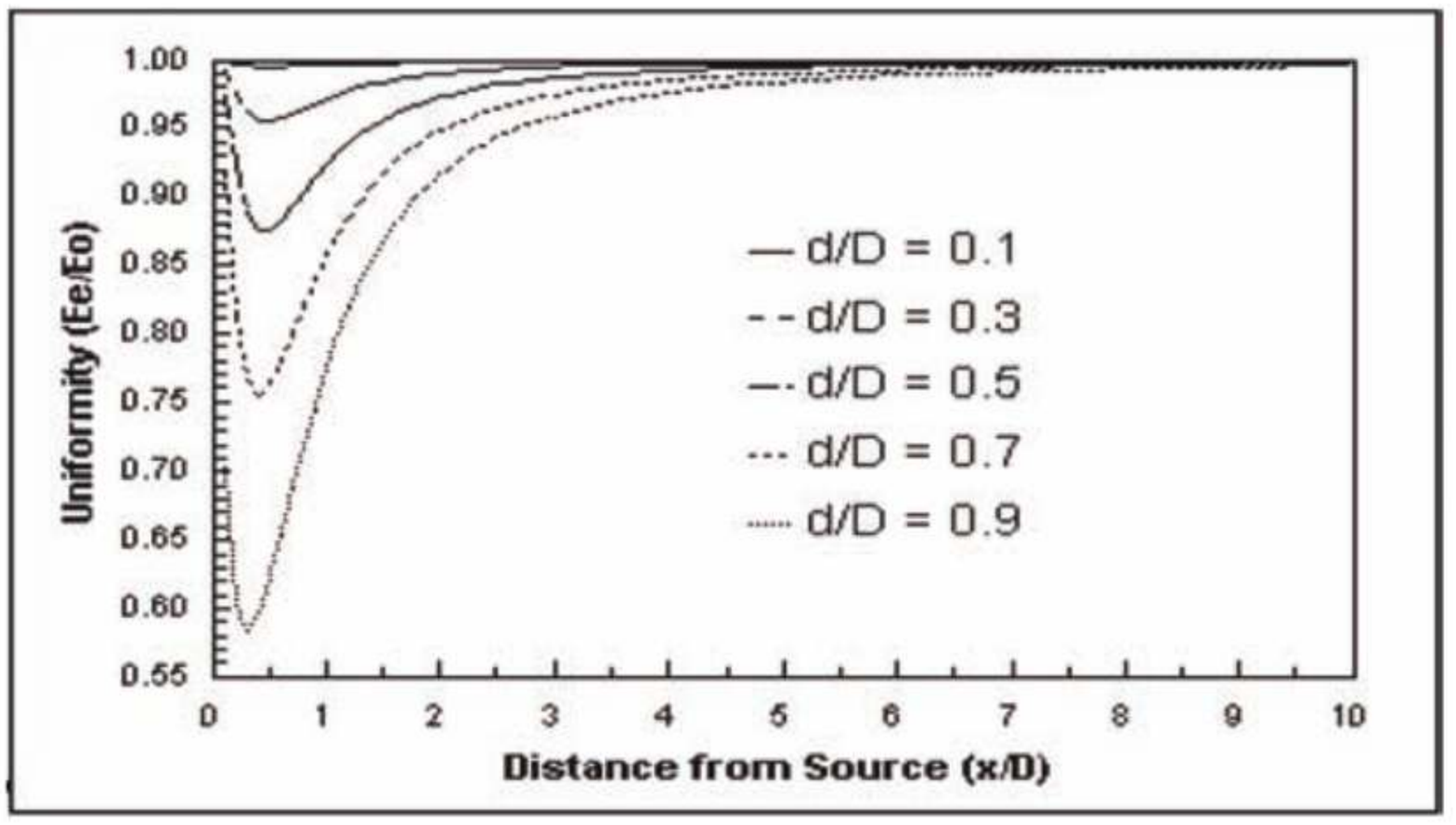}}
\caption[Integrating Sphere Spatial Uniformity]{ \label{fig:ISirradiance} Two figures concerning the spatial uniformity of irradiance from an integrating sphere taken from ref.~\cite{LabSphere02}.
The first, \fig\ref{fig:ISirradiance:diagram} shows an object with a diameter $d$ at a distance $x$ from the port of an integrating sphere with a radiance of $L_{\text{s}}$. The Uniformity of the irradiance is defined
as the ratio of the irradiance of the edge of the object $E_{\text{e}}$ to that on-axis $E_{0}$. As shown in \fig\ref{fig:ISirradiance:Plot}, the uniformity $E_{\text{e}}/E_{0}$ is near 1 at the plane of the port, and then decreases
sharply as the object is moved away. As the distance, given as the ratio of the total distance $x$ to the port diameter $D$, continues to increase, the uniformity recovers. The uniformity is shown for several values of $d/D$, the ratio of
the object diameter $d$ to the port diameter.}
 \end{figure}

Another useful property of integrating spheres which will later be used is their ability to act as source of uniform irradiance. 
Consider the setup shown in \fig \ref{fig:ISirradiance:diagram}, in which there is an integrating sphere with a port of diameter $D$ acting as a source with a given radiance $L_{\text{sphere}}$. 
The irradiance on the axis of the illuminated object is then $E_{0}=\pi L_{\text{sphere}} \sin^{2}\theta$, where $\theta$ is the angular radius of the integrating sphere port as seen by the illuminated object.
As a Lambertian source, the luminous surface of the integrating sphere port has a diffuse radiance which is independent of viewing angle \cite{LabSphere02}.
However, there will still be a variation in the irradiance across a plane at a finite distance from the port of the sphere, 
and the uniformity of the irradiance across an object can be expressed as the ratio of the irradiance on axis to that at the edge of the object.

The behavior of the irradiance uniformity is shown in \fig \ref{fig:ISirradiance:Plot} as a function of the distance $x$ between the integrating sphere port and the object, expressed as the ratio of the distance over the port diameter $x/D$. 
The irradiance uniformity curve is given for various object sizes, also expressed as a ratio $d/D$ of the object diameter $d$ to the port diameter $D$. 
As can be seen in the figure, the irradiance is nearly 100\% uniform directly at the plane of the port. It decreases as the object is moved away from the port, before recovering at large distances. 

It is useful to note from this plot that 
an object of approximately the same size as the port of the integrating sphere can be illuminated with an irradiance that is uniform to 1\% across its entire surface, if it is placed at a distance from 
the sphere port of at least 10 times the port diameter \cite{LabSphere02}. 
The uniformity can also be calculated as
\begin{equation}
\label{eq:Uniformity}
 \frac{E_{\text{e}}}{E_{0}} =  \frac{1}{2\sin^{2}\theta}\left(1-\frac{1 + \left(x/r\right)^{2} - \left(R/r\right)^{2}}{\sqrt{\left[1 + \left(x/r\right)^{2} + \left(R/r\right)^{2}\right]^{2} -4\left(R/r\right)^{2}}}\right) 
\end{equation}
where $x$ is the distance between the object to the port, $R$ is the radius of the port, $r$ is radius of the illuminated object, and $\theta$ is one half the angular size of the port, as shown in \fig\ref{fig:ISirradiance:diagram}.

\subsubsection{Our Integrating Sphere}
The sphere which used in our setup is manufactured by labsphere and is shown in \fig\ref{pic:NISTfigures:ISpicture} (and in \fig\ref{pic:IntegratingSphere:1} in the appendix). 
The integrating sphere has an internal diameter of 10.16 cm (4 in) and three ports, each located 90$^{0}$ from one another. The largest port has a diameter of 3.81 cm (1.50 in), and the two smaller ports both have 
a diameter of 2.54 cm (1.00 in), where the dimensions given by labsphere in inches. The interior of the sphere is coated with a proprietary Spectralon$^{\text{\textregistered}}$ material which has a diffuse reflectivity $\rho$ of 
$\approx0.97$ for 400 nm light.  Using \eq\ref{eq:IStimeconstant}), the time response of our integrating sphere has a time constant $\tau$ of about 7 ns.

\subsection{The Reference Detector: A NIST Photodiode}
\label{subsec:Reference Detector: NIST Photodiode}

\begin{figure}[t]
\centering
 \subfigure[The Integrating Sphere]{\label{pic:NISTfigures:ISpicture} \includegraphics[angle=270,width=0.42\textwidth]{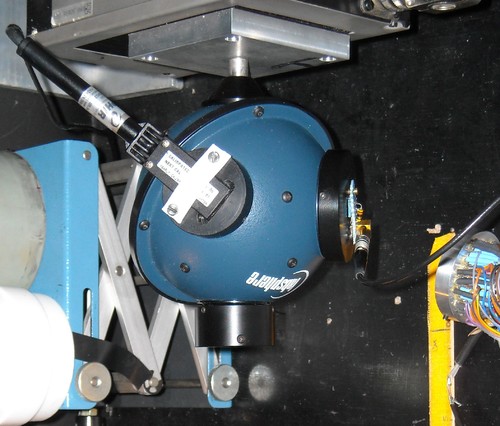}}
 \subfigure[A Diagram of the NIST Photodiode]{\label{fig:NISTfigures:diagram} \includegraphics[angle=270,width=0.52\textwidth]{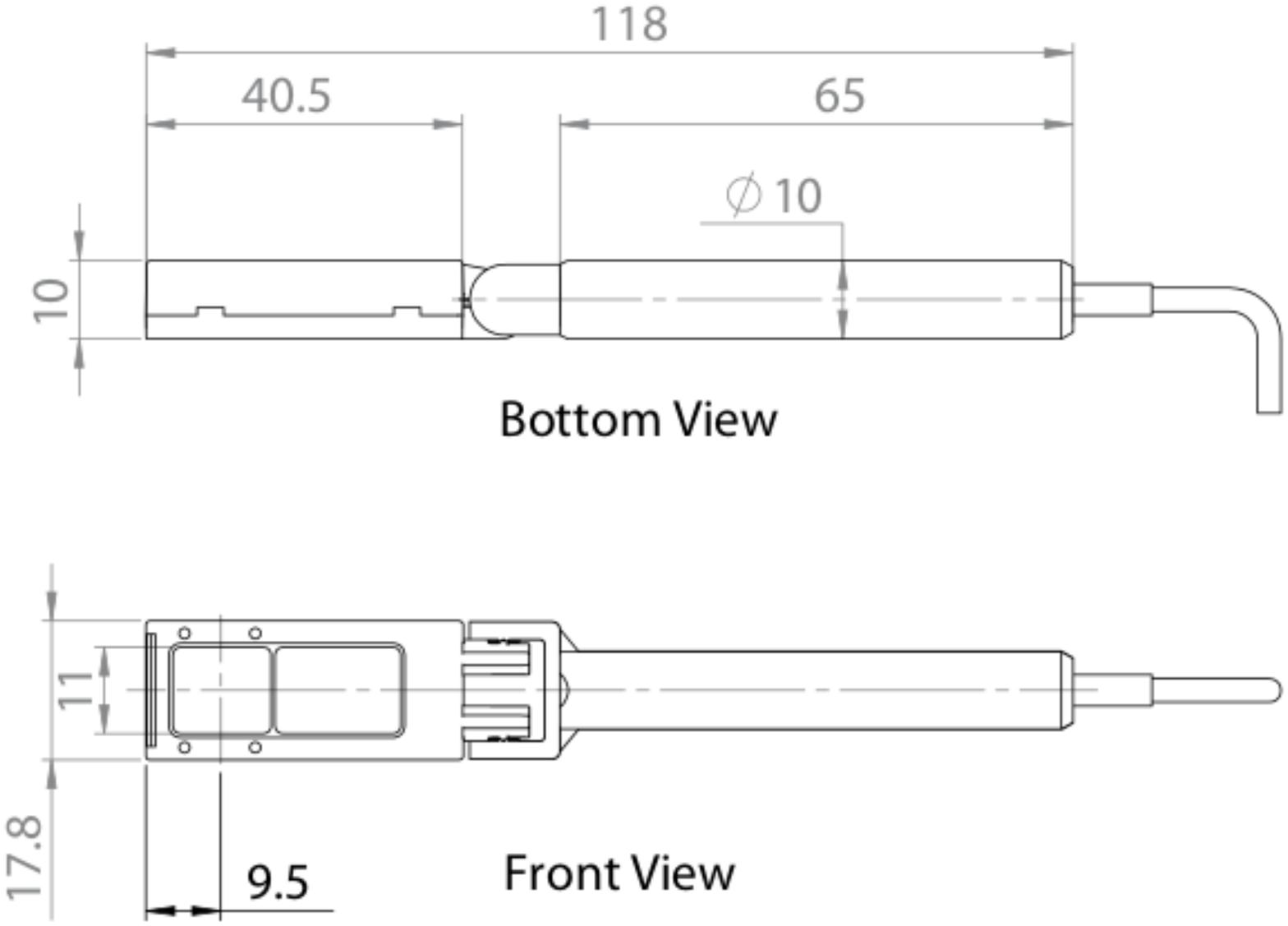}}
\caption[Diagram of the NIST Photodiode and Picture of the Integrating Sphere Assembly]{ \label{fig:NISTfigures} 
\fig\ref{fig:NISTfigures:diagram} shows a diagram of the NIST photodiode with dimensions, taken from the Ophir datasheet \cite{OphirDataSheet}. 
As can be seen, the sensitive area is about approximately 11 mm$^{2}$. \fig\ref{fig:NISTfigures} shows a photograph of our integrating sphere, as it is setup inside our black box.
On the top port, directly facing the camera, is the NIST photodiode. 
The LED is mounted on the lower port. The exit port is on the left, with an attached collimator. 
The sphere itself is mounted on an X-Y stage, allowing a precise placement of the light spot on the PMT photocathode.
}
 \end{figure}

The reference detector used in our setup is a silicon photodiode. 
Silicon photodiodes consist of either a p-n junction or a PIN diode. When an incident photon with enough energy strikes the diode it creates an electron-hole pair. The p-n structure causes the electron to move towards the cathode, producing a photocurrent. 
Photodiodes can be operated as un-polarized devices, and in this configuration they are be stable over almost 10 decades of incident power.
Just as with a PMT, the primary parameter defining a photodiode is the quantum efficiency. The quantum efficiency of a typical photodiode is in the range of 80\% for wavelengths in the 800 to 900 nm band \cite{PhotodiodeTheory}. 

\subsubsection{Photodiode Characteristics}
In addition to the quantum efficiency, photodiodes can be described by several other characteristics.
The responsivity $R_{\lambda}$ of the photodiode is the ratio of amperes of photocurrent to watts of incident illumination. $R_{\lambda}$ is related to the quantum efficiency 
of the photodiode and the wavelength of the incident light as:
\begin{equation}
 R_{\lambda} = \frac{\epsilon_{\text{quantum}} \lambda e }{h c} = \frac{\epsilon_{\text{quantum}}(\%) \lambda(\mu m)}{124}
\end{equation}
As can be seen from this relation, the responsivity increases with wavelength.  
 
A photodiode can also be characterized by its response time and minimum detectable power. The time response is the quadratic sum of the charge collection time and the RC time constant of the photodiode 
(from the series and load resistance and the junction and stray capacitance). A typical signal rise-time for a photodiode operating in unbiased mode is of the order of $0.5~\mu$s, depending on the operating wavelength \cite{PhotodiodeTheory}. 
Such a rise-time is appropriate for our application, as we wish to measure the average power received from an light source which is pulsed at something like 10 kHz.
The minimum detectable power is the minimum incident power on the photodiode needed to generate a photocurrent equal to the total noise current. This 
is defined as the noise equivalent power, or NEP. The NEP is given by the ratio of the RMS noise current to the responsivity of the photodiode.  

\subsubsection{Our Photodiode}
We use two model PD300-UV silicon photodiodes produced by Ophir. A diagram of these photodiodes with dimensions is shown in \fig\ref{fig:NISTfigures:diagram}. 
The overall sensitive area of this model of photodiode is $\approx 11 \times 11$ mm, and it is sensitive in a wavelength range from 200 nm to 1100 nm.
The photodiode's operational power range is from 20 pW to 3 mW with a resolution of 0.001 nW, and the output noise level is on the order of $\pm$ 1 pW. 
The temperature dependance of the photodiode response is very low, less than  0.1\%  per $^{\circ}$C between the short wavelength end of its spectral sensitivity and 800 nm.

By far the most important characteristic for our application is the wavelength-dependent absolute efficiency of the photodiode.
This calibration is traceable to the \gls{NIST} with an uncertainty of 1.5\% in the wavelength range $270$--$950$ nm.
The photodiode is readout by a specially designed Laserstar power meter, which is essentially a pico-ammeter which applies the absolute calibration curve of the photodiode. Once the wavelength of the source is selected on
the ammeter it gives a direct reading of the power incident on the photodiode. The precision on the current measurement using the LaserStar is 0.5\% \cite{LaserStar}.
 
\subsection{The Absolute Calibration Setup} 
\label{subsec:Absolute Calibration Setup}

In order to match the gain of the NIST photodiode to that of the PMT it is necessary to determine the needed attenuation. The light source is one or more \glspl{LED} which are placed on the larger, 3.81 cm, port of the integrating sphere.
The LEDs are pulsed in coincidence with the integration gate of the charge measurement.

The photodiode is placed on one of the 2 smaller integrating sphere ports. The aperture leading to the photodiode is 9 mm to limit reflections from the material surrounding it.
The PMT is illuminated through the 2nd small port, with a diaphragm to reduce the effective size of the port, and thus the flux the PMT receives.
From Eqs.~(\ref{eq:ISExchangeFactor}) and (\ref{eq:ISTotalFlux1refl}), the flux received (in the first reflection) by a given surface area of the sphere is the ratio of that area to the total inner surface of the sphere.
The flux on the photodiode is then
\begin{equation}
 \Phi_{\text{NIST}} = \Phi_{\text{i}}\rho \frac{A_{\text{NIST}}}{A_{\text{s}}} = \Phi_{\text{sphere}} \frac{A_{\text{NIST}}}{A_{\text{eff}}}.
\end{equation}
and the flux seen by the PMT, assuming that a diaphragm with a radius $r$ is placed on the port, is
\begin{equation}
 \Phi_{\text{PMT}} = \Phi_{\text{sphere}} \frac{\pi r^{2}}{A_{\text{eff}}}.
\end{equation}
so that the ratio of the flux on the PMT to that on the photodiode is
\begin{equation}
 \frac{\Phi_{\text{PMT}}}{\Phi_{\text{NIST}}} = \frac{\pi r^{2}}{A_{\text{NIST}}} \approx \frac{r^{2}}{ (0.45~\text{cm})^{2}}
\end{equation}

The noise level of the photodiode is $\sim1~$pW, which I will use as an estimate of the minimum background. From this, there must be around 0.1 nW on the photodiode to have a good signal-to-noise ratio. 
If the light source is a LED with a wavelength of 398 nm, and we assume that the photodiode has an efficiency of 80\%, this would correspond to $\approx 2.5~10^{8}$ photons per second incident on the exposed surface of the photodiode. 

From the Poisson statistics, the number of two photoelectron events in the single photoelectron spectrum is less than 1\% of the number of one photoelectron events if $\sim99\%$ of all events give no photoelectron 
(see section \ref{subsec:Single-Photon Counting}).
If the LED is pulsed at a rate of $\sim1$ kHz, and the PMT detection efficiency is around 25\%, this corresponds to $\approx40$ photons per second incident on the PMT. 

Using these estimates, the reduction in flux between the PMT and the photodiode must be $\sim10^{7}$. It is useful to note that as the LED pulse rate increases, so does the integrated power on the photodiode. 
One way to reduce the attenuation needed, while keeping both a good signal-to-noise ratio on the photodiode and the same number of photons per pulse on the PMT, is to increase the overall pulse rate of the LED. 

To achieve a reduction of $\sim10^{8}$ using our above assumptions, the radius of the diaphragm at the PMT must be $\approx2~10^{-4}$ cm. Thus the setup cannot be built using a single integrating
sphere with only a diaphragm at the PMT port. The solution of Lefeuvre et al.\ was to add a second integrating sphere. For multi-anode PMT calibration this it not ideal, as an integrating sphere is a Lambertian source. This means that 
the illumination from the port of an integrating sphere will follow a $\cos^{4}\theta$ dependence, even if a diaphragm is used to reduce the effective radius of the port.
To illuminate the PMT in a more uniform way, and restrict this illumination to a small area of the photocathode, we add a collimator rather than a second sphere. 
The exact dimensions of the collimator vary depending on the circumstance, but typical values are an entrance diameter of 1 mm, a length of 30 mm, and an exit diameter of between 1 and 0.3 mm.
The attenuation due to this collimator is not calculated, but measured directly in the second step of the calibration procedure.

\begin{figure}[p]
\centering
\includegraphics[angle=0,width=0.75\textwidth]{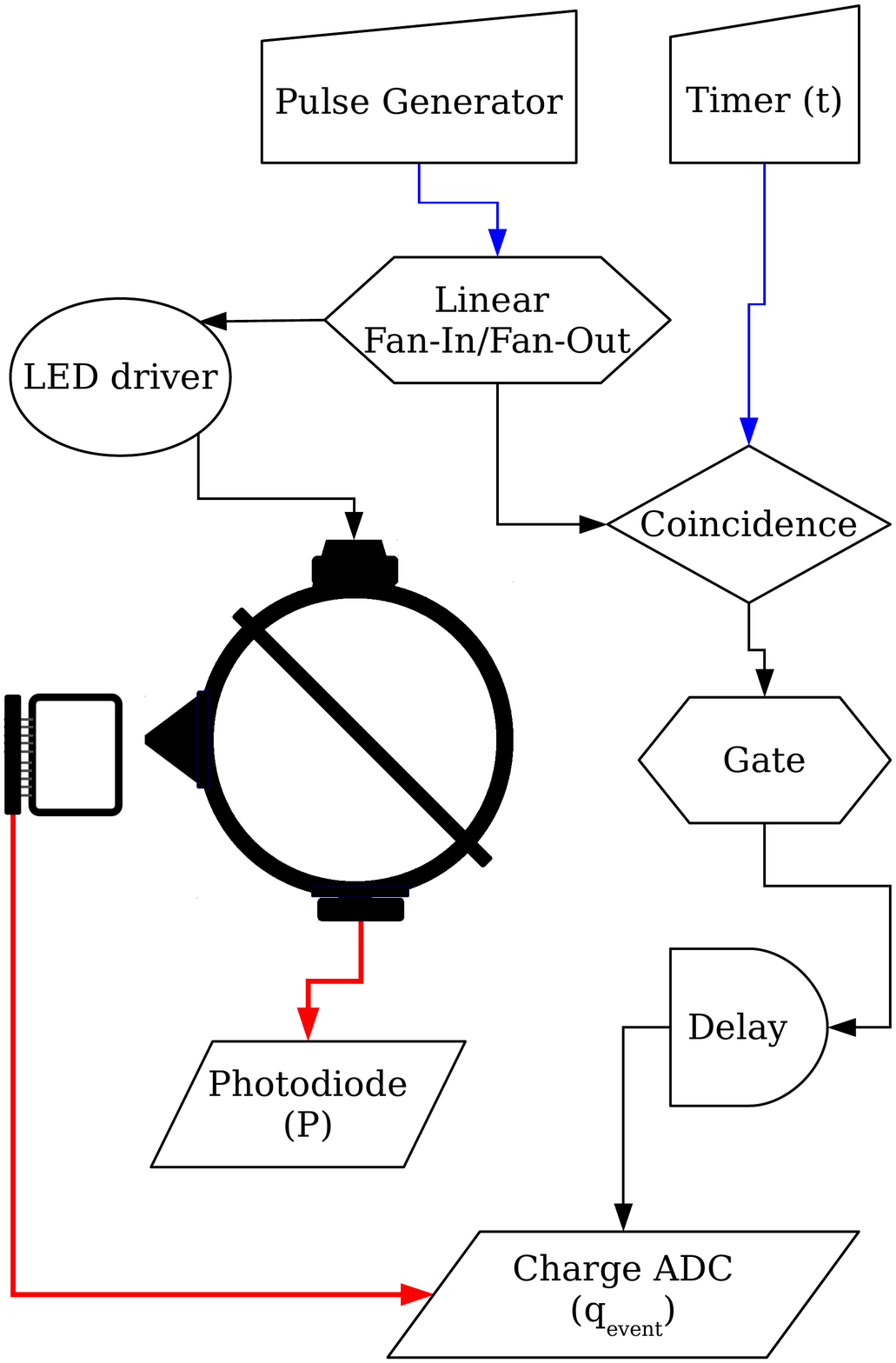} 
\caption[Signal Logic of the Calibration Setup]{ \label{fig:CalibrationSetup:Logic} The general signal logic of the calibration setup (e.g., implemented with NIM modules). 
This diagram is to illustrate the method; details for our actual setup, are omitted for clarity and will be shown in later chapters. 
Starting from the top of the figure:  A short pulse of the desired rate is created using a pulse generator and is sent to
a fan-in/fan-out logic module where is it copied. One copy of the pulse is sent to a LED driver which pulses the LED. The second copy can be put in coincidence with a timer module to control the total time $\tau$ of the measurement. The coincidence signal
(or the second copy itself) is used to generate a gate.  This gate is used as the integration window for some type of charge-to-digital conversion electronics. 
For each cycle, i.e., an event, the anode charge from the PMT is integrated in this time window and a measured charge is returned. The power $P$ received by the NIST photodiode is recorded simultaneously with the single photoelectron spectrum.}
 \end{figure}

\begin{figure}[p]
\centering
 \subfigure[Taking the PMT Spectrum]{\label{fig:CalibrationSetup:PMT} \includegraphics[angle=0,width=0.75\textwidth]{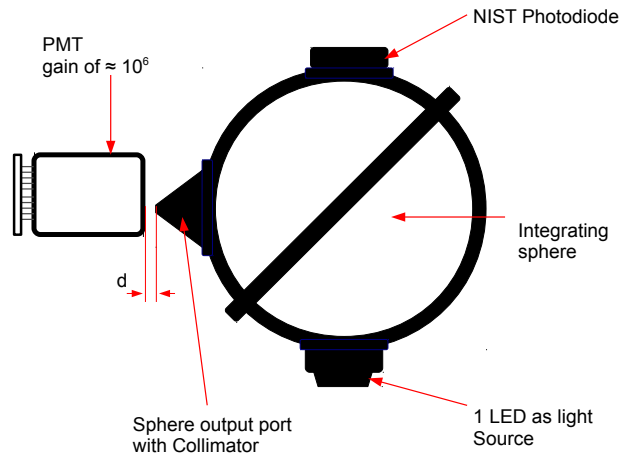}}
  \subfigure[Calibrating the Attenuation]{\label{fig:CalibrationSetup:NIST} \includegraphics[angle=0,width=0.75\textwidth]{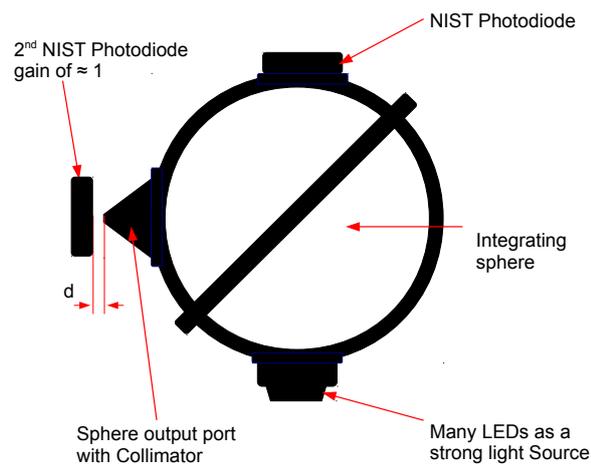}}
\caption[The Calibration Procedure]{ \label{fig:CalibrationSetup} Two diagrams showing the overall procedure for measuring the PMT efficiency. \fig\ref{fig:CalibrationSetup:PMT} shows the measurement of 
the PMT single photoelectron spectrum. The PMT is placed at a distance $d$ from the exit of the collimator, and is illuminated using a single pulsed LED as the light source with the integrating sphere plus collimator as a splitter. 
The NIST photodiode is attached to the 3rd port of the integrating sphere and measures the power. The attenuation of the integrating sphere and collimator assembly is calibrated in the second step, as shown in \fig\ref{fig:CalibrationSetup:NIST}. Here 
the PMT is directly replaced with a second NIST photodiode at the same distance $d$. The single LED is also replaced with a collection of LEDs, so that the power received on the 2nd photodiode is high enough to give a good measurement. The ratio of the 
power on the 2nd photodiode to that on the 1st gives the attenuation $\alpha$.}
 \end{figure}

The absolute measurement of the detection efficiency of the PMT proceeds as follows: 

\begin{itemize}
 \item The first measurement is shown in \fig\ref{fig:CalibrationSetup:PMT}. A generalized version of the signal logic of the setup is shown in \fig\ref{fig:CalibrationSetup:Logic}. 
The PMT is illuminated by a pulsed LED through the combination of the integrating sphere and collimator. A single photoelectron spectrum is taken using a charge to analog convertor, which integrates the 
total charge received during the gate (a fixed time window). The gate is generated in coincidence with the LED pulse, and both the gate length and delay are adjusted, so that the anode pulses from the PMT are contained within the gate.
Then the light level is reduced until the number of events (gates) which give one photoelectron is about 1\% the number of events which give no photoelectron.
We then know from the Poisson statistics that the contamination of two photoelectron events in the spectrum is less than 1\% the number of single photoelectron events.  

We then take a single photoelectron spectrum with enough total events to give the needed statistical uncertainty on the number of one photoelectron counts $N_{\text{pe}}$. At a rate of one photoelectron per 100 events, this means one million events to reach a
$\mathcal{O}\left(1\%\right)$ statistical uncertainty. The total time $\tau$ over which the spectrum is taken is also measured, for example by putting a timer in coincidence with the pulse generator to form the gate. 

The power $P$ received by the NIST photodiode attached to the integrating sphere is recorded simultaneously.

The resulting single photoelectron spectrum (see \fig\ref{fig:ExampleSPEspectra} as an example) is analyzed to determine $N_{\text{pe}}$, which is threshold dependent. $N_{\text{pe}}$ should be determined either
for a chosen working threshold, for example 1/3 of the mean single photoelectron charge $\mu$, or extrapolated to the zero of the pedestal.

\item The PMT is then replaced by a second NIST photodiode, as shown in \fig\ref{fig:CalibrationSetup:NIST}. The photodiode is placed at the same distance from the exit of the collimator as the PMT was previously.
The single LED is also replaced with a collection of LEDs, so that the power on the second photodiode is high enough to have a good signal-to-noise ratio.
Here we use the fact that the photodiode is linear across 10 decades of power.
The ratio of the power measured by the photodiode on the sphere and the second photodiode gives the attenuation:
\begin{equation}
 \label{eq:AttenuationFactor}
 \alpha = \frac{P_{\text{PMT}}}{P_{\text{sphere}}}
\end{equation}
where $P_{\text{sphere}}$ is the power measured by the photodiode on the sphere and $P_{\text{PMT}}$ is the power measured by the photodiode replacing the PMT.
\end{itemize}

From these measurements, the efficiency of the PMT is given by the equation
\begin{equation}
 \label{eq:EfficiencyCalculation}
 \epsilon = \frac{N_{\text{pe}}}{P\alpha\tau}\frac{hc}{\lambda}
\end{equation}
 where:
\begin{itemize}
 \item $N_{\text{pe}}$ is the number of single photoelectron events in the spectrum
 \item $P$ is the average photodiode power measured while taking the spectrum
 \item $\tau$ is the time of the measurement
 \item $h$ is the Planck constant
 \item $c$ is the speed of light in vacuum
 \item $\alpha$ is the measured attenuated factor
\end{itemize}

The overall uncertainty on the measured PMT efficiency can be broken into a systematic and statistical part:
\begin{equation}
 \label{eq:EfficiencyStatPlusSys}
 \left|\frac{\delta \epsilon}{\epsilon}\right| \leq \left|\frac{\delta \epsilon}{\epsilon}\right|_{\text{stat}} + \left|\frac{\delta \epsilon}{\epsilon}\right|_{\text{sys}}
\end{equation}

The systematic uncertainty is dominated by the uncertainty on the power measured by the reference photodiode and
by the calibration of the attenuation. There is a further contribution, however, from the systematic underestimation of the true $N_{\text{pe}}$ due to the contamination of two photoelectron
events in the spectrum. This is less than $1\%$ if the single photoelectron spectrum is taken under the proper conditions ($\geq98\%$ of all events in the pedestal).
The photodiode used to measure the power while taking the spectrum $P$ and the reference power on the sphere during the attenuation calibration are the same. 
The systematic uncertainties on $P$ and $P_{\text{sphere}}$ are thus completely correlated and will cancel out in the quantity $P\alpha$. 
The systematic error is thus given by: 
\begin{equation}
 \left(\frac{\delta \epsilon}{\epsilon}\right)^{2}_{\text{sys}}= \left(\frac{\delta P_{\text{PMT}}}{P_{\text{PMT}}}\right)^{2}+ \left(\leq1\%\right)^{2}
\end{equation}
The two sources of systematic error have been added in quadrature since they are independent and random relative to one another.

The statistical uncertainty is dominated by the statistical uncertainty of $N_{\text{pe}}$, which can easily be brought below $\mathcal{O}\left(1\%\right)$.
The uncertainty on $h$ and $\lambda$ are completely negligible. The error on $\tau$ is also negligible in our case, but depending on the experimental setup -- i.e., the measurement rate, this may not necessarily be true.
The statistical uncertainty on the measurement of the attenuation is negligible, as the mean value can be determined to arbitrary statistical precision, but
the statistical uncertainty on the measurement of the mean power while taking the spectrum may not be negligible, depending on the sampling rate and the $\sigma$ of sample distribution.
There is also a contribution to the statistical uncertainty on $P$ from the read-out of the current from the photodiode. For the LaserStar pico-ammeter this is 0.5\%.
All of these uncertainties are independent and random in nature and so we add them in quadrature:
\begin{equation}
 \left(\frac{\delta \epsilon}{\epsilon}\right)^{2}_{\text{stat}}= \left(\frac{\delta P}{P}\right)_{\text{stat}}^{2}+ \left(\frac{1}{\sqrt{N_{\text{pe}}}}\right)^{2}
\end{equation}

From this the total error on the PMT efficiency is:
\begin{equation}
 \label{eq:TotalGeneral}
 \left|\frac{\delta \epsilon}{\epsilon}\right| \leq \left|\left(\frac{\delta P}{P}\right)_{\text{stat}}^{2}+ \left(\frac{1}{\sqrt{N_{\text{pe}}}}\right)^{2}\right|^{1/2}_{\text{stat}}
 + \left| \left(\frac{\delta P_{\text{PMT}}}{P_{\text{PMT}}}\right)^{2}+ \left(\sim0.5\%\right)^{2}\right|^{1/2}_{\text{sys}}
\end{equation}

If we use the systematic error of our NIST photodiode (1.5\%), take a single photoelectron spectrum such that the statistical error on $N_{\text{pe}}$ is negligible and the number of two photoelectrons is 0.5\% the number of one photoelectrons, and assume that the statistical error of $P$ is that of the LaserStar current read-out,
then the total uncertainty on the PMT efficiency using this technique is:
\begin{equation}
 \label{eq:Total}
 \left|\frac{\delta \epsilon}{\epsilon}\right|
 \leq \left(0.5\%\right)_{\text{stat}} + \left(1.6\%\right)_{\text{sys}}
\end{equation}
these systematic and statistical errors are independent and summation in quadrature gives a estimate for the total uncertainty of
\begin{equation}
 \frac{\delta \epsilon}{\epsilon} = \pm 1.7\%
\end{equation}

     \printbibliography[heading=subbibliography]
     \end{refsection}

     \begin{refsection}
  \chapter{Measurement of the Air Fluorescence Yield}
   \label{CHAPTER:AIRFLUOR}
   
The physics of air fluorescence was discussed in detail in chapter~\ref{sec:INTROtoAFY} of the introduction. While this topic is interesting in its own right, 
the physics of air fluorescence is also critical for the detection of UHECR extensive air showers using the air fluorescence technique, as presented in section~\ref{sec:AirFluorEAS}. 
The air fluorescence yield, which gives the number of fluorescence photons emitted per unit of deposited energy, is a non-negligible calibration parameter
for estimating the energy of an EAS from the photon flux observed in a fluorescence telescope.

At the same time, as shown in the second half of chapter~\ref{sec:INTROtoAFY}, the measurement of the absolute fluorescence yield with a high precision is experimentally challenging,
and there are still gaps in our understanding of air fluorescence. The uncertainty on the fluorescence yield is a major source of systematic uncertainty on
the determination of the UHECR primary energy with fluorescence telescopes.
The best precision on the absolute yield can be achieved by measuring it directly, in air, for all combinations of temperature, pressure, and humidity.
For this reason, we propose a measurement of the absolute spectrum and integrated yield in all conditions, using the same setup, with an uncertainty of 5\% or less. 
As previously mentioned, this measurement would be strongly complementary to both the 
average value of Rosado et al.~\cite{RosadoICRC0377} and the absolute yield measurement of AirFly \cite{Ave:2012ifa}. 
The latter point is particularly true, as our absolute 
calibration will use the techniques presented in chapter~\ref{CHAPTER:PMT}, which represents an entirely different approach compared to that of AirFly.

Section~\ref{sec:A New Absolute Measurement of the Fluorescence Yield} will present the proposed measurement setup and some preliminary laboratory work which has been done.
The absolute calibration of two PMTs for this measurement is shown in section~\ref{sec:Calibration of Photomultiplier Tubes for the Air Fluorescence Measurements at LAL}.
This absolute calibration is performed using the PMT efficiency measurement technique presented in section~\ref{sec:Measuring Absolute Detection Efficiency}, and
serves as a detailed example of the application of this technique.

\section{A New Absolute Measurement of the\\ Fluorescence Yield}
\label{sec:A New Absolute Measurement of the Fluorescence Yield}
The basic philosophy of this fluorescence yield measurement is to simultaneously perform an absolute measurement of both the total yield and 
the resolved spectrum, and to do so for every point of interest in the kinematic phase space of air.
This requires building a fluorescence bench which is completely characterized in terms of detecting fluorescence emission and
in which there is complete control over the temperature, pressure, and gas mixture.

\subsection{Measurement Setup}

\begin{figure}[p]
  \centering
  \includegraphics[width=1.0\textwidth]{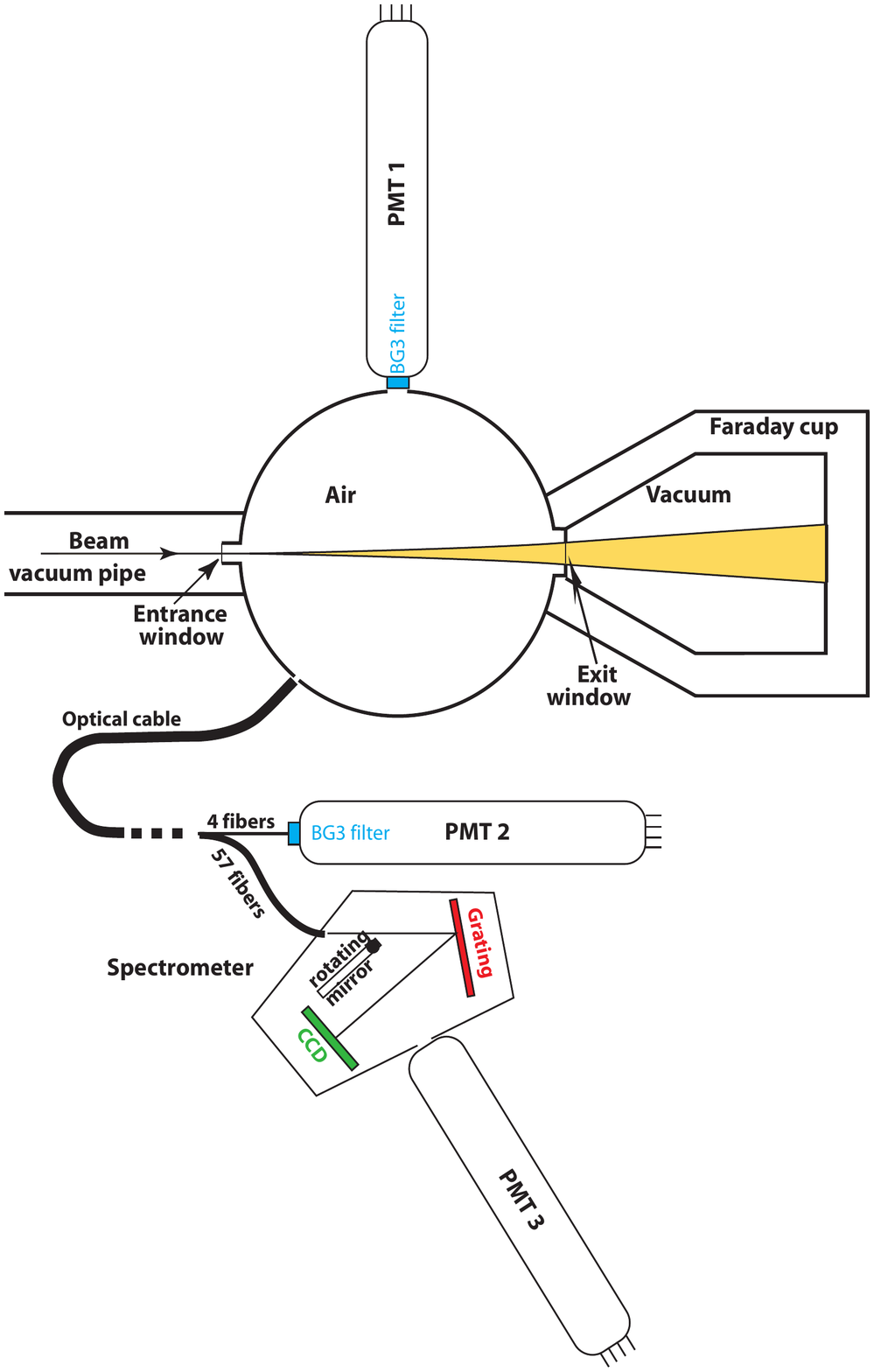}
 \caption[Design of Our Air Fluorescence Bench]{A diagram showing the design of our air fluorescence bench, taken from one of our previous publications \cite{AirFlICRC13}. An electron beam is shown coming from the right side. It enters into the 
integrating sphere, which is filled with air at a controlled temperature, pressure, and humidity. The light in the sphere is measured by one PMT, with an attached BG3 filter, placed directly on one port of the sphere. A bundle of 61 optical fibers leaves from
another port of the sphere towards a grating spectrometer and a second PMT. The light in the grating spectrometer is measured with a LN$_{2}$-cooled CCD. The CCD is calibrated using a 3rd PMT, which is also shown.
Every attenuation coefficient is measured during the calibration of the bench, which is shown in \fig\ref{fig:AFYmeasurementCalibration}.
After passing through the integrating sphere, the total charge of the electron bunch is measured using an evacuated Faraday cup.}
  \label{fig:AFYmeasurementDesign}
 \end{figure}

A diagram of the proposed measurement setup is shown in \fig\ref{fig:AFYmeasurementDesign}. 
An electron beam is shown entering from the left of the diagram. The source of primary electrons is an electron accelerator with an energy in the MeV range. Originally we planned to use the 
PHIL linear electron accelerator at LAL\footnote{Laboratoire de l'Acc\'el\'erateur Lin\'eaire, Univ Paris-Sud, CNRS/IN2P3, Orsay, France }.
This may not, however, be possible due both to programmatic issues with PHIL, and the fact that it may not be the best solution for our measurement, due to the large transverse size of their beam.
For the sake of example, I will assume an electron beam with the general properties of PHIL \cite{PHILweb, brossard:in2p3-00643766}. 
This is a pulsed beam with a bunch rate of 5 Hz, bunch width of 8 ps, and a bunch charge of 100 pC. I assume a primary electron energy in the range of 3-5 MeV.

As seen on the diagram, the beam passes from the vacuum of the beam pipe through an entrance window
into the target volume. The target volume is an integrating sphere. By using an integrating sphere, we benefit from the fact that the collection of fluorescence light emitted inside the sphere is independent of the location of the emission. 
After traversing the air volume, the electron beam exits from the sphere back into vacuum through the exit window shown on the right side of the diagram. 
The integrating sphere and its coupling mechanics were custom made at LAL.

The electron beam disperses in the air of the target volume, by an amount depending on the air pressure and the distance traveled. The exit window must be large enough to contain the dispersed beam.
After exiting the integrating sphere, the electrons are collected into a Faraday cup which gives a precise measurement of the total charge, and thus the total number of electrons which pass through the sphere.

At the top of the diagram is a PMT, labeled PMT-1,  with a BG3 filter on its photocathode. 
PMT-1 is attached to one port of the integrating sphere with a collimator to reduce the number of photons it receives per bunch.
This PMT acts like a NIST photodiode; the total charge on PMT-1 is integrated for each electron bunch received, monitoring the overall variation of light in the integrating sphere.

A bundle of 61 quartz-silica fibers, labeled ``optical cable'' in the diagram, 
is attached to another port of the sphere. Each individual fiber in the bundle has a diameter of 0.1 mm and a numerical aperture of 0.22. 
Out of these 61 fibers, 57 go to a grating spectrometer. These fibers arrive at the spectrometer as a single, vertical, mono-fiber layer with a height of 5.7 mm and a thickness of 0.1 mm. 
The entrance slit of the grating spectrometer is closed to 0.1 mm to match the mono-fiber width. This limits the angular dispersion on the grating, and constrains the wavelength resolution of the spectrometer to 0.1 nm.
The grating of the spectrometer has 600 grooves per nm, spanning a total of 100 nm. 

The light detection in the spectrometer is by a LN$_{2}$-cooled \gls{CCD} with 1024 horizontal and 256 vertical pixels. This allows a measurement of the entire spectrum between 
300 and 400 nm at once, and, because of the LN$_{2}$-cooling, the CCD has a background rate on the order of 1 electron per pixel per hour. 
This low background rate allows a very precise measurement of the number of incident photons with a high signal-to-noise ratio (better than $10^{4}$, see below). 

The maximum number of counts per CCD pixel is approximately $3~10^{5}$. This sets the upper limit on the statistics which can be collected in any single pixel per run, so as to not saturate the pixel.
The mono-fiber is placed within a rectangular window of 100 pixels vertically.
In order to reach a statistical uncertainty of better than 1\% on the number of photons detected in each spectral line, there must be at least $10^{4}$ counts in the CCD for each line.
Each spectral line will cover about 2 pixels, and, as a rough estimate from \fig\ref{fig:FYSpectrum}, there is a factor of 250 in magnitude between the 337 nm line and the minor spectral lines.  
A total of $10^{4}$ counts in each of the minor spectral lines would thus give $2.5~10^{6}$ counts in the 337 line.
For the 337 nm line, given the beam, the sphere size, the spectrometer and CCD characteristics, 
the acquisition time for such statistics can be estimated to be about 10 minutes at a pressure of 1 atm. 

The other 4 fibers from the bundle lead to a second PMT, labeled PMT-2 in the diagram. PMT-2 is also equipped with a BG3 filter.
Not shown is an additional collimator between the bundle of 4 fibers and the PMT, to adjust the number of incident photons per bunch to an appropriate level,
and 2 other small ports on the integrating sphere for gas control.
In later measurements a Dewar flask will surround the integrating sphere, so that the temperature, pressure, and humidity inside the sphere can be completely controlled.
The ranges for the measurement are planned to be from 1 atm to 0.1 atm in pressure, and from room temperature to -60$^\circ$C.
At each temperature and pressure, the humidity will be set by introducing a known fraction of water vapor, from complete saturation to 1\% of the saturation level.
The integrating sphere is made from machined Teflon to give good diffusion properties.
Also shown in \fig\ref{fig:AFYmeasurementDesign} is a third PMT attached to the spectrometer. This PMT is used in the calibration of the CCD, which is discussed in the next section.

 
\subsection{Calibration and Other Considerations}

\begin{figure}[p]
  \centering
  \includegraphics[width=1.0\textwidth]{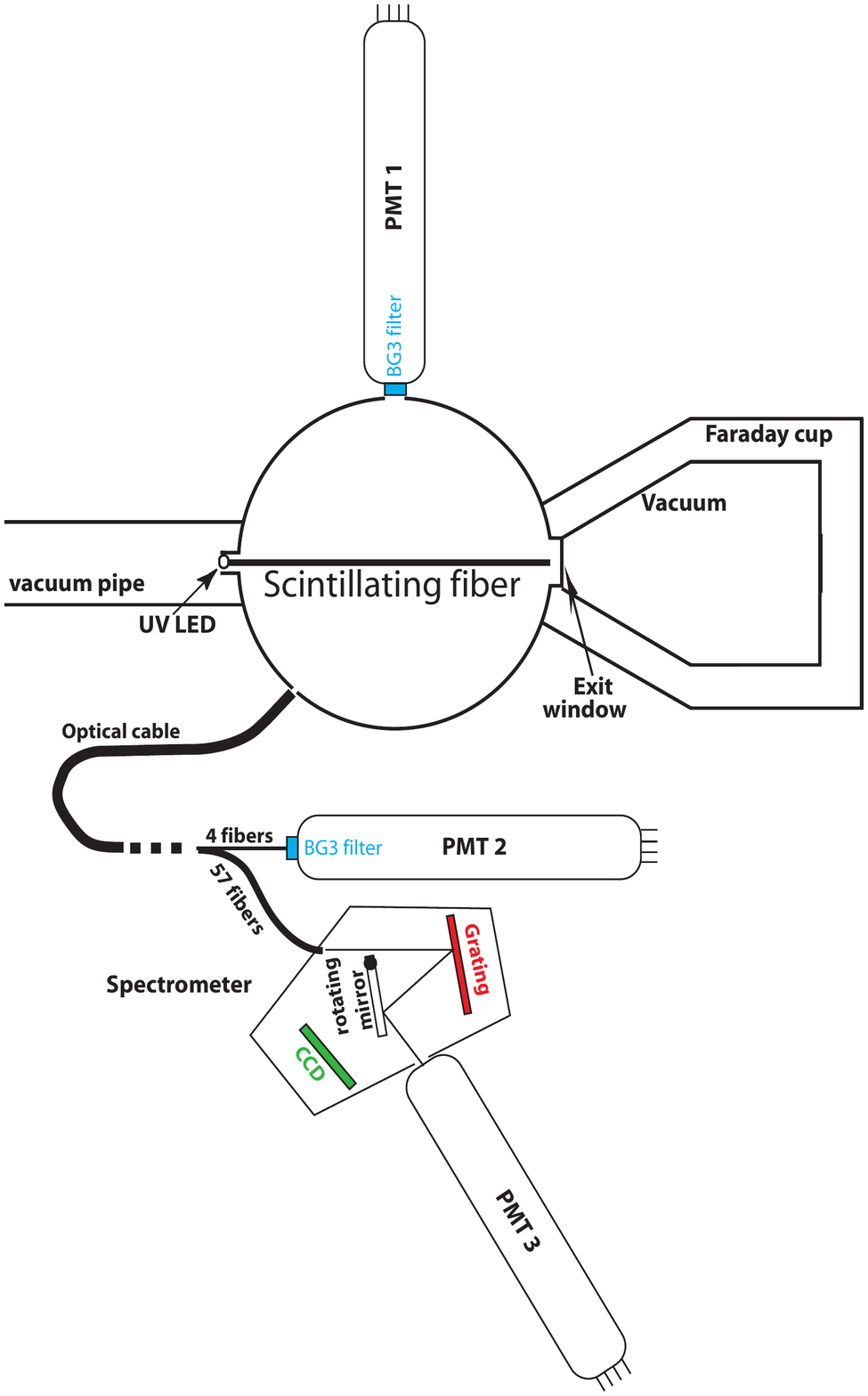}
  \caption[Calibration of Our Measurement Bench]
{A diagram showing the calibration of the air fluorescence bench, again taken from ref.~\cite{AirFlICRC13}. For the calibration, the fluorescence emission is replaced with a LED coupled to 
a scintillating fiber. Each PMT can also be replaced with a photodiode to measure the attenuation of each component. The CCD is calibrated by comparison to PMT-3 by redirecting the light 
using a rotating mirror inside the grating spectrometer.}
  \label{fig:AFYmeasurementCalibration}
 \end{figure}


The absolute integrated yield is proportional to the number of photoelectrons measured by PMT-2, divided by the total number of incident electrons. 
Similarly, the spectrally-resolved yield is proportional to the number of electrons counted by the CCD, divided by the total number of incident electrons. 
The constant of proportionally of both measurements is determined by a complete calibration of the setup.
A diagram of this calibration is shown in \fig\ref{fig:AFYmeasurementCalibration}.

Each of the PMTs used for the measurement are themselves calibrated with an uncertainty of less than $2\%$ using the technique discussed in section~\ref{sec:Measuring Absolute Detection Efficiency}. 
The absolute calibration of two PMTs was done at APC\footnote{Laboratoire Astroparticule et Cosmologie, Paris 7 Universit\'{e}} and is shown in the next section as a detailed first example of the application of this calibration technique.
The calibrated PMTs are used as a reference detector for the absolute calibration of the CCD, and the measurement of the transmission of each component of the setup. 

For the measurement of the transmission, the fluorescence emission (i.e., the electron beam) is replaced by a monochromatic UV LED. The LED is placed outside the integrating sphere and is coupled to a 1 mm scintillating fiber to mimic the emission profile of 
the fluorescence from the electron beam. Both PMT-1 and the fiber bundle can be replaced by a NIST photodiode to accurately measure the ratio of light collected by their respective integrating sphere ports. The fiber bundle is then placed back on the sphere 
and the ratio of the amount of light exiting the bundles of 57 and 4 fibers to that at the PMT-1 port is measured. This gives the transmission of the two branches of the fiber bundle. 
Several different wavelengths of LED will be used to measure the spectral dependence of the transmission. This spectral dependence can be due, for example, to the variation of reflectivity of the inside of the integrating sphere with wavelength.   

The transmission of the spectrometer and the efficiency of the CCD are measured by comparison to PMT-3, which is located on a second exit port of the spectrometer. The light from 
the grating, which is usually incident on the CCD, can be redirected to PMT-3 by rotating a mirror inside the spectrometer. The reflectivity of this mirror is 99\% and is known with high accuracy. 
Like the other 2 PMTs, the absolute efficiency of PMT-3 is measured in single photoelectron counting mode with an uncertainty of less than $2\%$. By scanning the emission profile of the scintillating fiber,
we can compare the yield of the CCD to the yield of the absolutely calibrated PMT-3. This comparison takes into account all the effects from sphere reflection, numerical aperture and transmission of the fibers, and entrance slit aperture.  

Once all the transmission coefficients and efficiencies are known, the final ingredient is the determination of the actual local energy deposit; 
i.e.\ the correction for energy carried away by secondary electrons which escape the viewing region (in the case of an integrating sphere by impacting on the walls of the sphere).
This energy loss and its importance in a precision measurement of the fluorescence yield was discussed in sections~\ref{sec:Physics of Air Fluorescence} and~\ref{sec:ExperimentalMeasurementsofAirFluorescence}.
For a given integrating sphere radius $R$ and gas pressure $P$, the difference between the energy loss, given by the Bethe-Bloch formula, and the local energy deposit is well defined and can be 
determined from simulations. The error on this correction is on the order of 5-10\% \cite{Rosado:2011yx}. 
This correction for the local energy deposit increases as the thickness $P\cdot R$ decreases, but can be estimated to be up to 20\%. 
This gives an uncertainty on the local energy deposit on the order of 2\% after correction.

In addition, the minimum size of the integrating sphere is limited by the requirement that the sum of all the ports of the sphere cannot be more than 5\% of the total surface in order to retain the radiance properties of the sphere (which was discussed in
section~\ref{subsec:Integrating Spheres}).
As the thickness $P\cdot R$ increases, the dispersion of the electron beam passing through the integrated sphere increases. The exit window of the integrating sphere must have a diameter of 8$\sigma_{\text{beam}}$ to contain the entire electron beam. 
There is a conflict, then, between increasing the radius of the sphere, which allows a larger exit port, and the dispersion of the beam, which increases as the electrons must travel farther in air.
This problem can be reduced by having a well focused beam at the sphere entrance and by having an optimal choice of entrance window. The choice of window is itself not simple, as the window thickness and 
material both affect the dispersion of the beam traversing the window, and the window must be able to hold the vacuum of the beam pipe. The overall dispersion in air can also be reduced by increasing the energy of the electron beam, and the 
dispersion is no longer a large issue for electrons with energies above $\sim 10$ MeV.
A series of GEANT4 simulations of the detection volume have been performed to study the beam dispersion and energy deposit for a variety of beam energies, and entrance window materials and thicknesses.
This work will not be detailed here, however, as there are too many other things to discuss.

The lower bound for the radius of the integrating sphere is set by these two considerations, once the lowest working pressure, and what constitutes a reasonable energy containment are determined. 
If the lowest measurement pressure is 0.1 atm, which corresponds to $\approx10$ km altitude, then 
a sphere of radius 20 cm will contain approximately 85\% of the total energy deposited by primary electrons with an energy of 4 MeV \cite{Rosado:2011yx}.  
The contained energy is close to 91\% for the same radius at 1 atm. 
This size of integrating sphere is a reasonable compromise between the total energy containment and the size of the integrating sphere, which must be surrounded by a Dewar flask for temperature and pressure control.
Two spheres will be constructed however, a smaller one of 6 cm diameter for test purposes, and a larger 20 cm sphere for final measurements.   

For a 4 MeV beam, the highest energy delta is $\approx2$ MeV, which would have a range of around 8 m in air at 1 atm. An integrating sphere with a radius of 8 m would be impractical, and the only way to eliminate the need for a
correction to the deposited energy is to act on the other component of the thickness $P\cdot R$, the pressure. By decreasing the primary electron energy and increasing the pressure in the sphere, we can come to a situation in 
which the entire beam will be stopped within the sphere. In this case, the local energy deposit and the energy loss will be the same. 
The number of electrons per bunch must then, however, be estimated from the known characteristics of the accelerator.

Considering all these factors, the total uncertainty of our measurement is expected to be 5\% for both the total integrated yield and each spectral line of the resolved measurement. This uncertainty would apply to every combination of 
temperature, pressure, and humidity used. The key to this low total uncertainty is the absolute calibration of the PMTs used in the measurement, and this calibration is presented in the next section.

   \section{Calibration of Photomultiplier Tubes for Our Air Fluorescence Measurements}
\label{sec:Calibration of Photomultiplier Tubes for the Air Fluorescence Measurements at LAL}

\begin{figure}[h]
\centering
 \subfigure[Diagram of the XP2020Q PMT]{\label{fig:XP2020Q:Diagram}\includegraphics[width=0.40\textwidth]{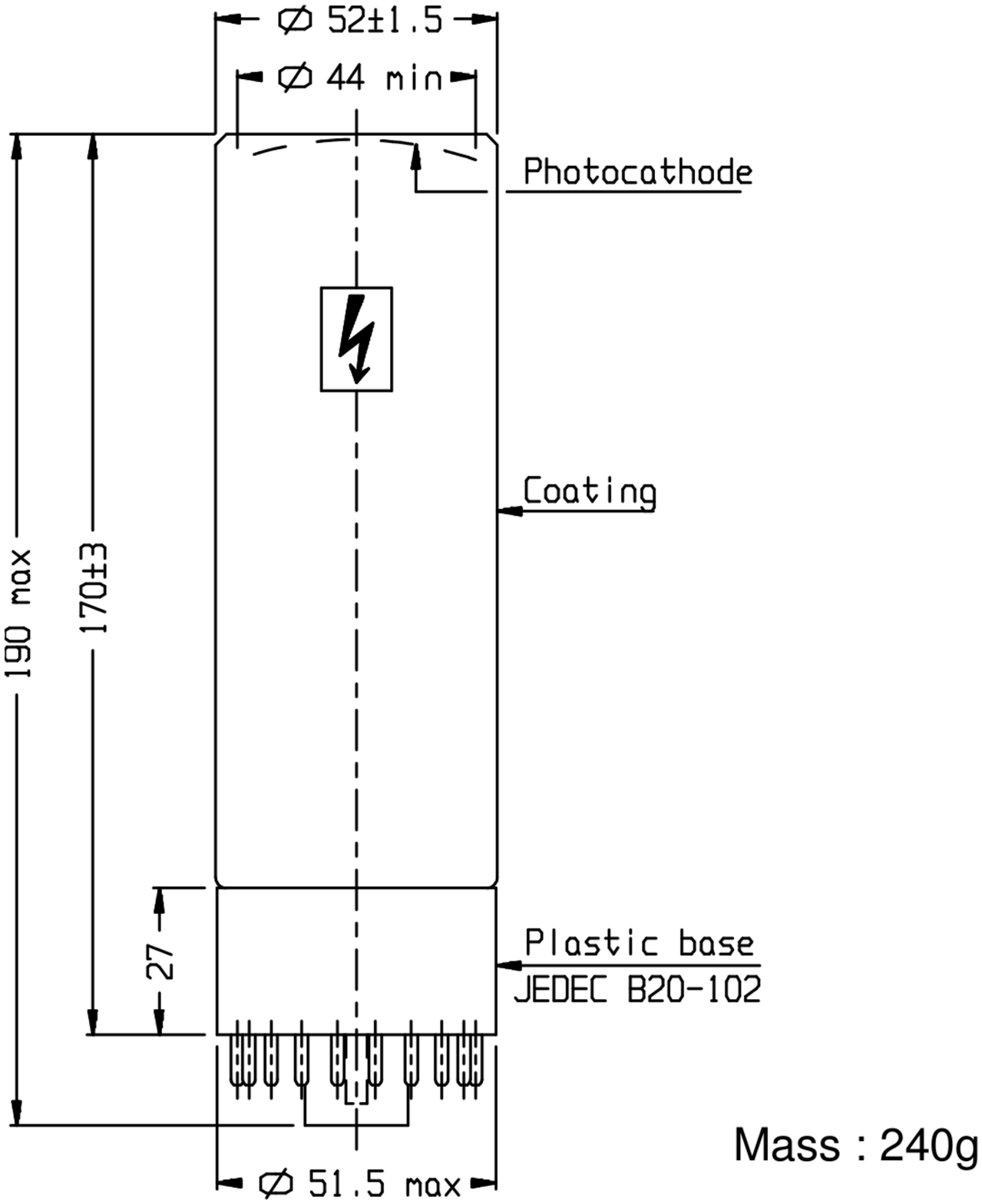}}
 \subfigure[XP2020Q Spectral Sensitivity]{\label{fig:XP2020Q:spectralsens}\includegraphics[width=0.40\textwidth]{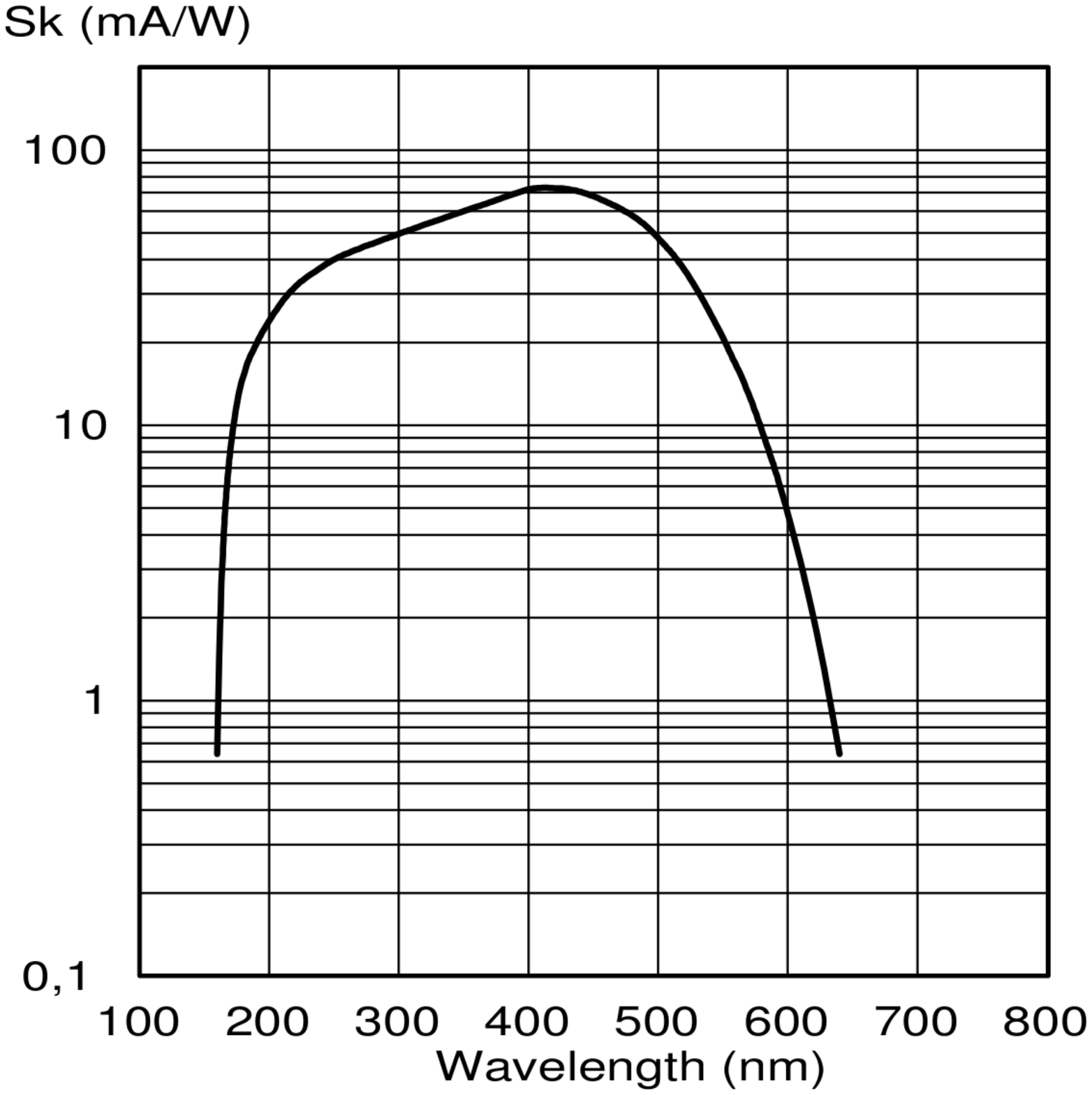}}
\caption[Diagram of the XP2020Q PMT and its Spectral Sensitivity]{ \label{fig:XP2020Q} The spectral response of the XP2020Q PMT as given by its data sheet \cite{XP2020Qdatasheet}. Figure \ref{fig:XP2020Q:spectralsens}
gives the photocathode response in mA per Watt as a function of wavelength, while figure \ref{fig:XP2020Q:Diagram} shows a diagram of the XP2020Q PMT with dimensions.}
\end{figure}

The PMTs which are used for the air fluorescence measurement presented in the last section are Photonis model XP2020Q PMTs. The Q designates 
a PMT model with a quartz-silica entrance window, which gives the PMT an improved sensitivity in the UV. A diagram of 
this type of PMT with dimensions is shown in \fig\ref{fig:XP2020Q:Diagram}.  
The XP2020Q type PMT is cylindrical with a diameter of 51 mm. At the front of the PMT is a fused silica window with a bi-alkali photocathode. 
The spectral sensitivity of the bi-alkali photocathode is shown in \fig\ref{fig:XP2020Q:spectralsens}.
This photocathode has a spectral range of 150-650 nm with a maximum sensitivity at 420 nm. The refractive index of the window is 1.48 for light of 420 nm wavelength.

The electron multiplier part of the PMT is a linearly focused structure with 12 stages.
This gives the XP2020Q a typical single photoelectron gain on the order of $10^{7}$ using supply voltages around 2300 V.

Both PMTs are powered by a resistive base which divides the supply voltage according the repartition recommended by Photonis. There is a critical modification to this base, however: the polarity of both PMTs have been inverted.
The photocathode is at ground, and the anode is at the (positive) supply voltage.
When the PMTs are mounted on the fluorescence bench, their photocathode will be in contact with the integrating sphere, the spectrometer, or the fiber mounting. In the usual negative polarity, discharges 
can occur between the photocathode, which is at high voltage, and the mounting, which is at ground, inside the silica window.
Lefeuvre et al.\ found that these discharges created a large background count rate \cite{LefeuvreThesis}.
The background rate was reduced to a lower level once they inverted the polarity of the voltage divider. 
A typical ``good'' XP2020Q has a background count rate of $\sim 300~$Hz in negative polarity and $\sim3~$Hz in positive polarity.
In order to connect the anode, which is now at high voltage, the anode signal is sent through a
capacitor integrated into the PMT base. 
The capacitance must be high enough that there is no signal loss for high counting rates, but in any cases the anode output signal will be insensitive to analog light (when the anode signal is a voltage level rather than resolvable single pulses).
This is not a problem for the two PMTs calibrated, as they will be used in single photoelectron counting mode in our air fluorescence measurement.

The laboratory work for the calibration was done together with my colleagues from the air fluorescence measurement (P.\ Gorodetzky and D.\ Monnier Ragaigne).
I was responsible for creating the data acquisition setup and the detailed analysis of our results.
Two XP2020Q PMTs were calibrated in total.
The first, with serial number 16361, has an attached BG3 filter, and this one will be referred to as ``PMT-2''. The BG3 filter is glued onto the 
PMT window using Epotec N 301-2 glue, which has the same refractive index as both the filter and the glass.
The second XP2020Q, with serial number 42175, does not have an attached filter, and I will refer to this PMT as ``PMT-3''. 
Their respective names refer to their presumed role in the air fluorescence measurement setup, as discussed in section~\ref{sec:A New Absolute Measurement of the Fluorescence Yield} and shown in \fig\ref{fig:AFYmeasurementDesign}.
The third PMT for the fluorescence bench, PMT-1 in Figs.~\ref{fig:AFYmeasurementDesign} and \ref{fig:AFYmeasurementCalibration}, has not yet been calibrated.

\subsection{Setup}
The calibration of these two PMTs was done in single photoelectron mode. An overview of this technique was given in section~\ref{sec:Measuring Absolute Detection Efficiency}; 
here its application will be discussed in detail. In order to take the single photoelectron spectrum of the two PMTs we used a VME \gls{QDC} module. This QDC was a CAEN model V792-N
QDC with 16 input channels in LEMO 00 format. The VME crate was linked to a PC running Linux using a CAEN V1718 USB-to-VME bridge. For each channel of the QDC, the charge input during a time window defined by a gate signal
is converted to a voltage level by charge-to-amplitude conversion electronics, multiplexed, and then converted by two fast 12-bit \gls{ADC} modules. The input range of this QDC is from 0 to 400 pC with a resolution of 100 fC per count.

The software controlling the data acquisition was a heavily modified version of the 
``Multi Instance Data Acquisition System--Prospectus'' software \cite{MIDASUK}. The data acquisition system
which was used for this calibration measurement took several months to put together and is distinct from the one which was later built. 
This later system uses the Midas data acquisition framework, and the data acquisition software and hardware will be discussed in more depth in chapter~\ref{CHAPTER:PMT Sorting}. 

This QDC has a 32 event buffer, which allowed it to be readout by a continuous block transfer.
With this combination of software and QDC hardware we where able to take data for one QDC channel at a sustained rate of around 20 kHz. Although the system was fast, the QDC itself had several disadvantages. The first is the relatively 
low charge resolution, which required us to use an amplifier with a gain of 10 to take a good single photoelectron spectra even for a PMT with a gain of $10^{7}$. 
The second issue is that the QDC is very sensitive to positive voltage levels. This causes problems with using the amplifier and with pulse overshoot or ringing, and so 
extreme care had to be taken when setting up any measurement. 

The signal logic of the setup was as shown in \fig\ref{fig:CalibrationSetup:Logic}. A pulse generator was used to create NIM pulses with a width of 20 ns at a rate of 20 kHz. These pulses where sent through a fan-in/fan-out module to 
copy them. One copy of the pulse was sent to a delay unit and then to a discriminator, which was used to generate a gate signal with an adjustable width. The other copy was sent to a LED driver with an adjustable amplitude.

Using this setup, the light source was pulsed in coincidence with the charge integration gate, which eliminates the importance of the dead-time of the data acquisition system.
The timing of the single photoelectron pulses from the PMT within the integration gate was adjusted using an oscilloscope. 

The adjustable amplitude of the LED driver is used to tune the number of single photoelectron events in the spectra. 
The ratio of the number of single photoelectron events to the number of pedestal events was checked by taking spectra and by reducing the LED pulse 
height until no more than 1.5\% of the events where outside the pedestal. This ensures that the contamination of two photoelectron events is less than 0.75\% in every spectrum we took afterward. 

\begin{figure}[t]
\centering
\includegraphics[width=0.7\textwidth]{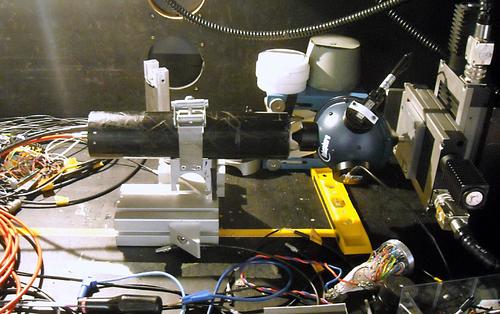}
\caption[Photograph of the PMT Calibration Setup]{ 
\label{pic:AirFlPMTbeingCalibrated:picture} A photograph of the one of the XP2020Q PMTs in our black box. 
The integrating sphere with NIST photodiode, LED, and collimator can be seen on the 
right side of the picture. The X-Y movement on which the integrating sphere is attached can also be seen.
The PMT is housed inside the black plastic cylinder in the center of the photo. 
}
 \end{figure}

The source LED for the PMT spectra was a single LED with a wavelength of 378 nm. 
The PMT was illuminated by the LED through an integrating sphere, with an additional collimator, as shown in \fig\ref{fig:CalibrationSetup:PMT}. 
The collimator which was used had an exit pinhole with a diameter of 0.3 mm, and the photocathode of each PMT was 1.5 mm from the collimator exit. 
Using a collimator has the advantage that the light is nearly parallel, so that any small change in the distance to the photocathode (or NIST photodiode) is negligible. 
The collimator can be seen in \fig\ref{pic:ThreeCollimators}, where it is the middle collimator. This distance was preserved using a thickness gauge each time the 
assembly was moved. The integrating sphere is mounted on an X-Y movement to allow scanning the photocathode of the PMT with an accuracy on the relative position of better than 0.1 mm.
A photograph of one of the XP2020Q PMTs in the black box is shown in \fig\ref{pic:AirFlPMTbeingCalibrated:picture}.

The power received by the NIST photodiode on the integrating sphere was read at a rate of 15 Hz during each absolute measurement. For shorter relative measurements, which were done first, the photodiode was read only as an average value
during the middle of the run. This was due to an update in the computer software, which was finished only while the data was being taken.
   
\subsection{Measurements}
As was shown in section \ref{sec:INTRO:PMT}, the PMT efficiency is affected by the voltage repartition, external magnetic fields, and the location of the incident photons on the photocathode.
To take into account all these factors we measure:
\begin{itemize}
\item The gain as a function of the supply voltage 
\item The absolute efficiency at one position and orientation to make the relative mapping an absolute one
\item A map of the relative PMT efficiency as a function of the position of the beam spot on the photocathode
\item The change in efficiency as the PMT is rotated around its long axis relative to the Earth's magnetic field
\end{itemize}
 
\subsubsection{Gain}
\begin{figure}[t]
\centering
\includegraphics[width=0.95\textwidth]{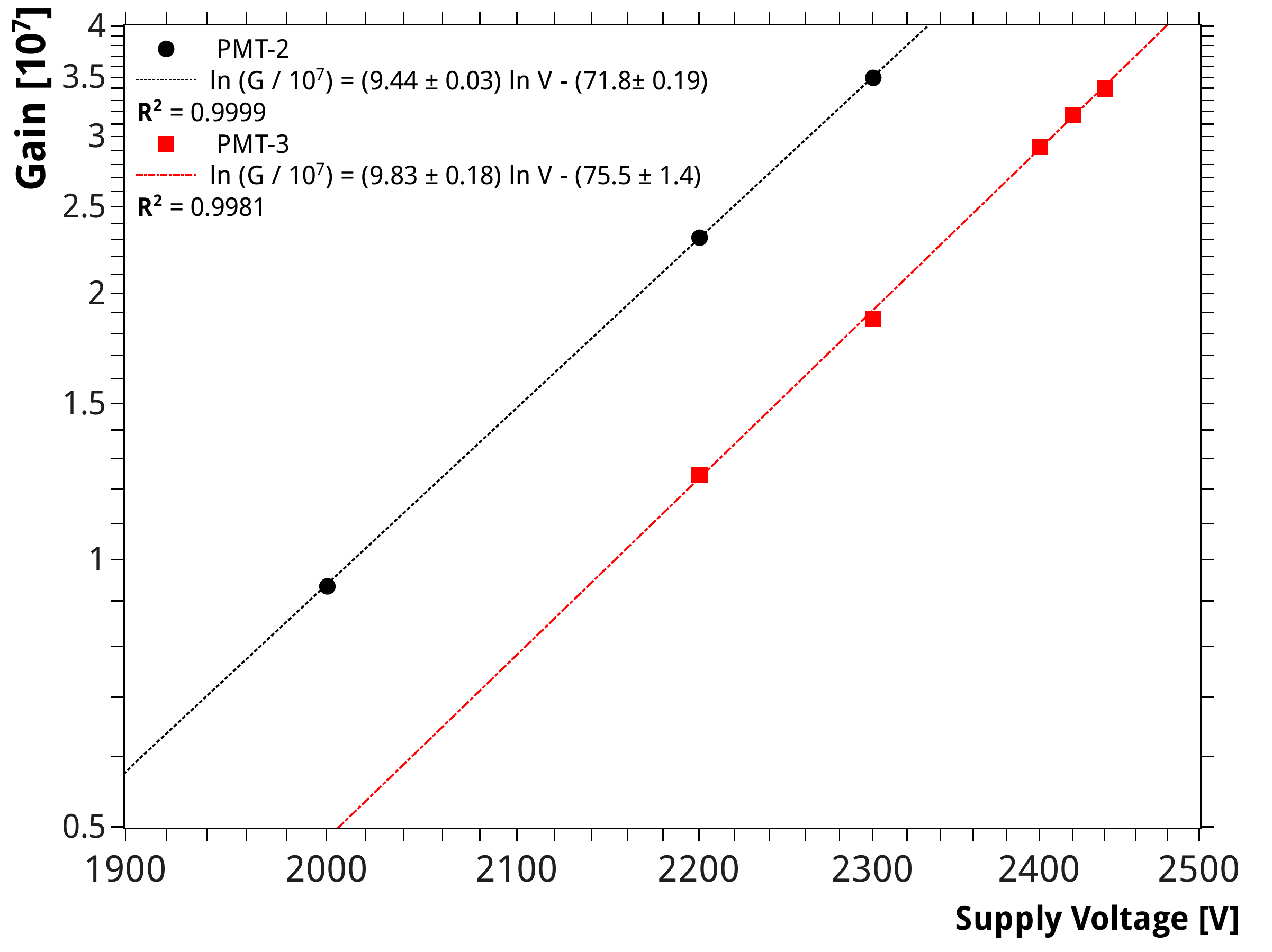}
\caption[Gain Curves for the 2 Calibrated PMTs]{ \label{fig:XP2020Q:gain} The gain $\mu$  of the two PMTs as a function of the supply voltage $V_{o}$. As can be seen, the gain of PMT-3 is a factor of 1.87 lower than the gain of PMT-2. 
The slope of the gain curve is similar for both PMTs, which is typical as both have the same internal structure. The red and black dashed lines show a linear fit to the natural logarithm of the data. 
The $R^{2}$ value of the fit is compatible with a linear relationship between $\ln(\mu)$ and $\ln(V_{o})$ in both cases. }
 \end{figure}

Using the setup just described, we took numerous single photoelectron spectra for both PMTs in order to study the four points listed above.
We started by measuring the dependance of the gain on the supply voltage, and took spectra for a range of voltages between 2000 V and 2500 V. 
In our measurements the gain is defined using the mean of the single photoelectron peak, as opposed to the peak maximum.
Both gain curves are shown plotted in $\log$-$\log$ in \fig\ref{fig:XP2020Q:gain}. 
The dependence of the gain on the supply voltage, given by \eq\ref{eq:PMTGain}), is such that the gain should be
proportional to the supply voltage to the power $n\alpha$, where $n$ is the number of stages in the PMT. The gain curves should thus be a straight line in $\log$-$\log$. In \fig\ref{fig:XP2020Q:gain}, the markers show 
the measured data, and the dashed lines show a linear fit to the natural logarithm of the data. For both PMTs the fit is in good agreement with \eq\ref{eq:PMTGain}), as shown by the $R^{2}$ values.
However, the error on the extrapolation is quite large, due to the low number of measurements and the logarithmic dependence on the voltage. 
The slope of both gain curves is similar, which is natural, as both PMTs have the same design. The gain of PMT-2 is a factor of 1.87 higher than the gain of PMT-3 at the same supply voltage however. This level of variation
between individual PMTs is typical.

\begin{table}
\begin{center}
\begin{tabulary}{1.0\textwidth}{LCCCCCCC}
 \toprule
\multicolumn{2}{l}{Voltage}  & 2000 & 2200 & 2300 & 2400 & 2420 & 2440 \\
\cmidrule(l){2-8}
\multirow{2}{*}{Gain} & PMT-2 & 0.936 & 2.31 & 3.50 &  -- & -- & --\\
& PMT-3 & -- & 1.25 & 1.87 & 2.93 & 3.18 & 3.40  \\
\bottomrule
\end{tabulary}
\caption[PMT Calibration Gain Results]{ \label{tab:TableOfXP2020QGains} The measured gain of both PMTs as a function of supply voltage. All values are given in units of $10^{7}$. These values are plotted in \fig\ref{fig:XP2020Q:gain}. 
As can be seen both here and in the plot, the gain of PMT-3 is almost a factor of 2 lower than that of PMT-2 at the same supply voltage. We wished to have a similar working gain for both PMTs. For PMT-2 we worked at a gain $3.5~10^{7}$, which
was possible at a supply voltage of 2300 V. A similar gain for PMT-3 required a supply voltage of 2440 V.}
 \end{center}
\end{table}

Each measured single photoelectron gain, as plotted in \fig\ref{fig:XP2020Q:gain}, is given in Table~\ref{tab:TableOfXP2020QGains}.
Based on the measured gain of PMT-2 we chose a working supply voltage of 2300 V, where this PMT has a gain of $3.5~10^{7}$ .
Both PMTs should be operated such that they have a similar single photoelectron gain.
The lower gain of PMT-3 required a supply voltage of 2440 V, where that PMT had a single photoelectron gain of $3.40~10^{7}$.

The error on the relative gain is very low. 
As long as the response of the QDC is linear in the charge range used for the measurement, the uncertainty on the relative gain comes only from the statistical error on the mean of the pedestal and single photoelectron peak. 
This uncertainty is on the order of 1\% or less. 
The uncertainty on the absolute gain is larger, as this includes the systematic uncertainty on the response of the QDC. For these measurements we had only the hardware specifications for the conversion 
slope of the QDC. This value is determined simply by the range of the QDC and the number of return bits (i.e., the number of charge bins). The error on the conversion slope of the QDC could be as high as 5\%. 

\subsubsection{Absolute Efficiency}
\label{subsubsec:AFYPMT:Absolute Efficiency}
\begin{figure}[ht]
\centering
\includegraphics[width=12cm]{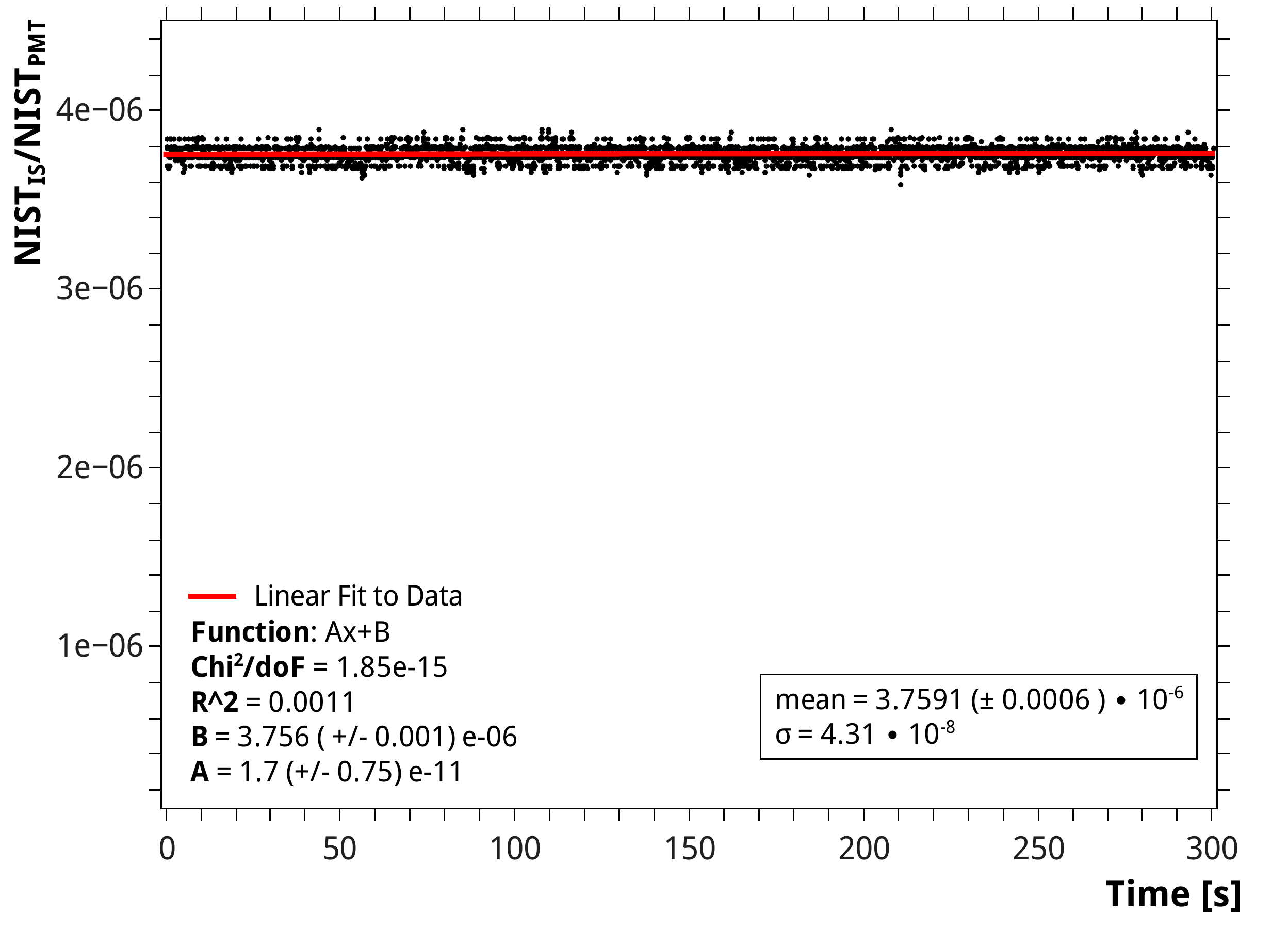}
\caption[Calibration of the Attenuation]{ \label{fig:NISTratioXP2020Q} Measurement of the attenuation in power between the photodiode on the integrating sphere and the PMT photocathode.
The ratio is taken between the first NIST photodiode placed on the integrating sphere, and a second NIST photodiode which replaces the 
PMT. Data was taken over 300 seconds at a sample rate of 15 Hz, giving a mean of $\alpha = 3.7591~(\pm~0.0006)~10^{-6}$, with the standard error quoted. 
The red line shows a linear fit to the data, which was done to check the stability of the ratio during the measurement. }
\end{figure}

The first information needed to calculate the absolute efficiency is the attenuation of the integrating sphere and collimator. 
This is measured by replacing the PMT with a second NIST photodiode, as shown in \fig\ref{subsec:Absolute Calibration Setup} and discussed in section \ref{fig:CalibrationSetup}. The LED on the sphere is 
also replaced with a stronger one so that the illumination on the second photodiode is high enough to give a good signal-to-noise ratio on the measured incident power.
Like the photodiode on the sphere, the second photodiode is also sampled at a rate of 15 Hz. The power incident on both photodiodes was measured for a period of 300 seconds to give a large number of samples for the calculation of the mean ratio. 
The ratio of each sample is shown in \fig\ref{fig:NISTratioXP2020Q} plotted versus time. The attenuation $\alpha$ is taken as the mean value of the ratio and is $3.7591~(~\pm 0.0006 )~10^{-6}$. The quoted error is the standard
error (calculated as $\sigma/\sqrt{n-1}$, where $n$ is the number of samples), and does not include the systematic error of either photodiode. The systematic uncertainty of the photodiode on the sphere cancels in the calculation of 
the efficiency, while that of the second photodiode remains in the result.

\begin{figure}[h]
\centering
  \includegraphics[width=10cm]{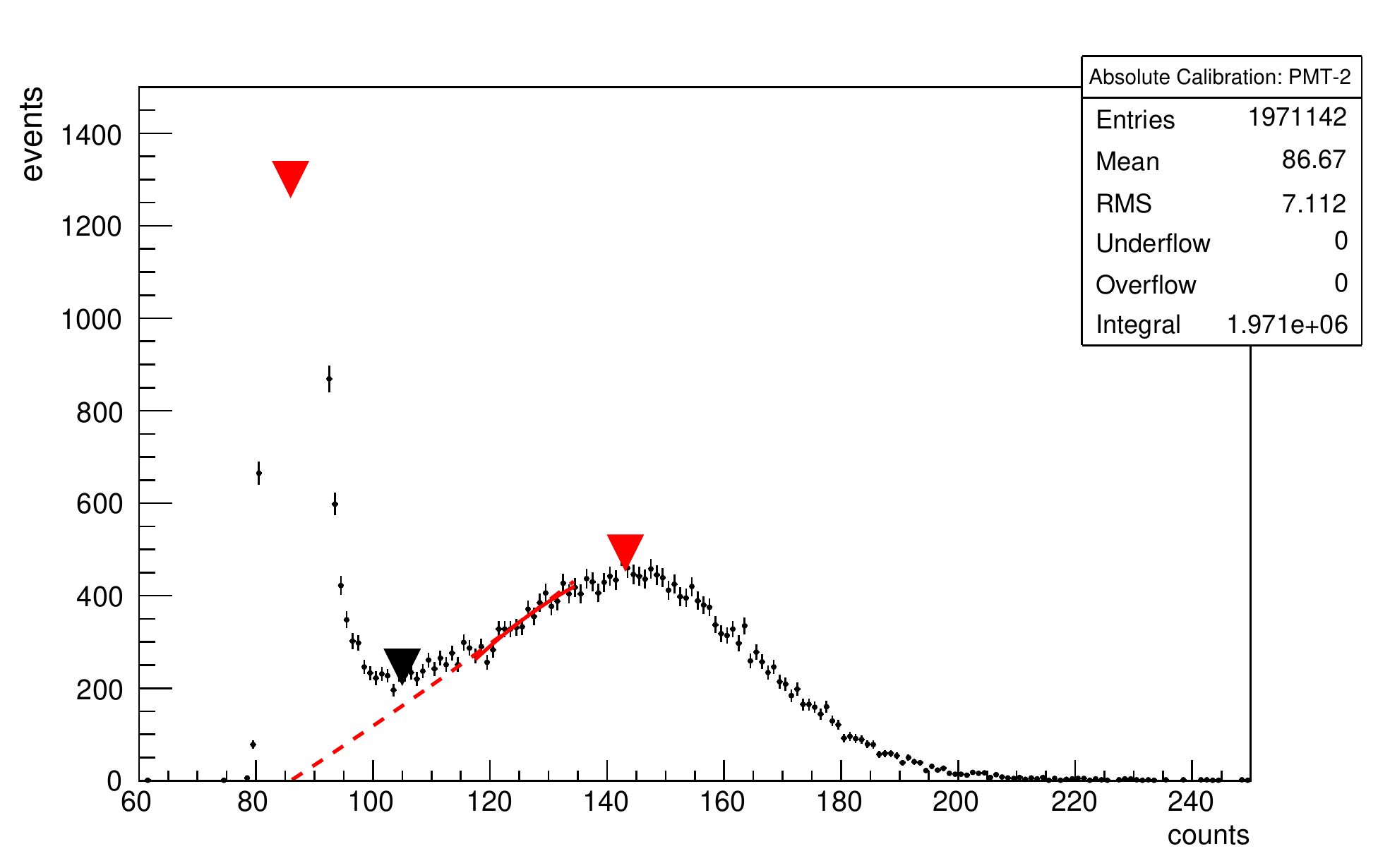}
\caption[Single Photoelectron Spectrum of ``PMT-2'']{ \label{fig:XP2020Qspectrum:16361} The single photoelectron (spe) spectrum for the absolute calibration of the PMT-2. The pedestal can be seen on the left with its peak (far out of the 
plot range) at 85.89 counts. The Spe peak, on the right, is located at 142.0 counts, giving a gain of 56.11 pC or $3.50~10^{7}$. The mean value of each peak is shown by a red marker.
Using this gain, the $1/3$ photoelectron threshold is at 104 counts, shown by the black marker. The number of events above
$1/3$ of a photoelectron is 25736. The dashed red line shows a polynomial extrapolation of the Spe peak to the pedestal mean, used to find the extrapolated number of Spe events.}
 \end{figure}

 \begin{figure}[h]
\centering
 \includegraphics[width=10cm]{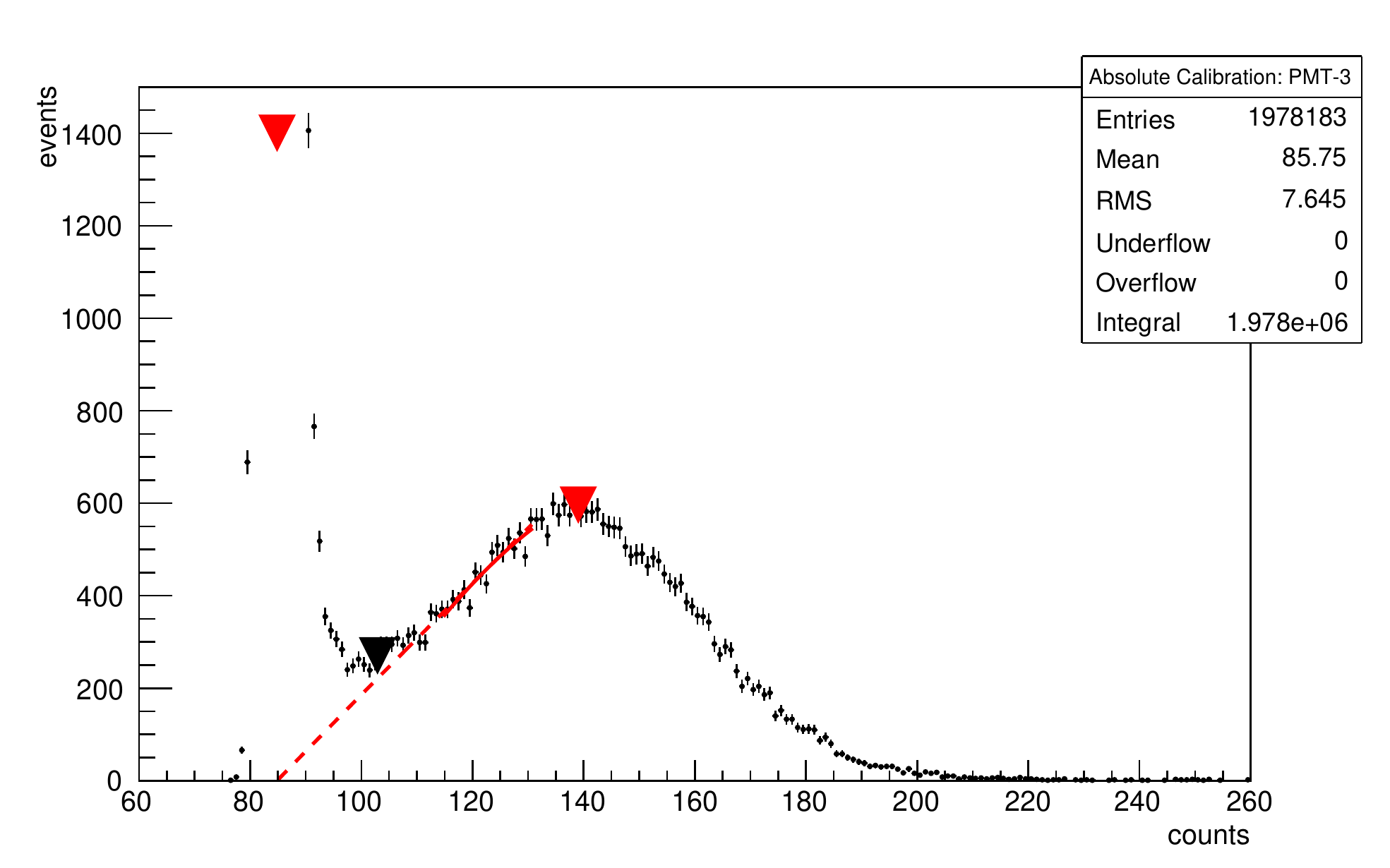}
\caption[Single Photoelectron Spectrum of ``PMT-3'']{ \label{fig:XP2020Qspectrum:42175} The single photoelectron (Spe) spectrum for the absolute calibration of the PMT-3. The pedestal can be seen on the left with its peak (far out of the 
plot range) at 84.83 counts. The Spe peak, on the right, is located at 139.3 counts, giving a gain of 54.48 pC or $3.40~10^{7}$. The mean value of each peak is shown by a red marker.
Using this gain, the $1/3$ photoelectron threshold is at 103 counts, shown by the black marker. The number of events above
$1/3$ of a photoelectron is 32095. The dashed red line shows a polynomial extrapolation of the Spe peak to the pedestal mean, used to find the extrapolated number of Spe events.}
 \end{figure}

After finding a good supply voltage for each PMT, we began working towards measuring the efficiency of each PMT. 
The single photoelectron spectra taken for the absolute efficiency measurement are shown in Figs.~\ref{fig:XP2020Qspectrum:16361} and \ref{fig:XP2020Qspectrum:42175}, for PMT-2 and PMT-3, respectively.
The absolute efficiency of PMT-2 was measured at a supply voltage of 2300 V, while the efficiency of PMT-3 was measured at a supply voltage of 2440 V.
Each of these spectra where taken in a run of 100 seconds, at 20 kHz, giving a total of 2 million events in each spectrum. As around 1\% of the total events are single photoelectrons this gives us around 20 thousand single photoelectron events 
per spectrum, keeping the statistical error at less than 1\%. During both runs the power incident on the NIST was recorded at a rate of 15 Hz to track variations during the run. The mean value of the power was used for the calculation of the 
efficiency.

Looking at the spectrum of PMT-2, \fig\ref{fig:XP2020Qspectrum:16361}, the pedestal can be seen on the left with its peak far out of the 
plot range. The mean of the pedestal is determined by taking the weighted mean of all bins between 0 counts and the valley between the pedestal and the 
single photoelectron spectrum. The calculated mean of the pedestal is 84.83 counts, marked by the first red triangle. The mean of the single photoelectron peak, on the right side of the spectrum, is at 142.0 counts. 
It is marked by the second red triangle. 
The mean of the single photoelectron peak is determined in the same way as the pedestal, taking all the bins between the valley and tail of the spectrum. 
The difference between the mean of the pedestal and the single photoelectron peak, 56.11, gives the gain in units of QDC counts. Conversion of this relative measurement into an absolute gain depends on the properties of the QDC.   

The efficiency of the PMT is proportional to the surface area of the single photoelectron peak, i.e., the number of single photoelectron events. A determination of the number of single photoelectron events in the 
spectrum is threshold dependent, as is the useful efficiency. For this reason, it is normal to measure the efficiency at either the threshold which will be used, or at a threshold which is well defined. The
threshold is taken at $1/3$ of a photoelectron, which is a typically used value. This is a threshold in charge, or voltage if an integrating preamplifier is used, which corresponds to $1/3$ of the single photoelectron gain. 
For PMT-2 the single photoelectron gain is 56.11 counts, and so $1/3$ of a photoelectron plus the pedestal is 104.6 counts. This threshold is shown by the 
black marker in \fig\ref{fig:XP2020Qspectrum:16361}. The data is binned by the QDC, and so this means taking the number of events at 105 counts and above.

In addition to the 1/3 photoelectron efficiency,
we also determine a threshold independent efficiency by extrapolating the single photoelectron peak to the pedestal mean. This extrapolation is shown in \fig\ref{fig:XP2020Qspectrum:16361}
by the dashed red line. The extrapolation is a 4th order polynomial which is found by fitting the backside of the single photoelectron peak (shown by the solid red line). A correction to the 
number of single photoelectron events is determined by integrating the polynomial extrapolation between the pedestal mean and the threshold at $1/3$ of a photoelectron. These events are added to the 
number found above the threshold. Generally, this correction is on the order of 5-10\% with a probable error of 5-10\%.

All told, the resulting number of single photoelectron events in the spectrum shown in \fig\ref{fig:XP2020Qspectrum:16361} is $25736\pm160$ events above $1/3$ of a photoelectron and $1295$ events from the extrapolation. 
The analysis of the spectrum for PMT-3, shown in \fig\ref{fig:XP2020Qspectrum:42175}, is performed in the same way, giving $32095\pm179$ events above $1/3$ of a photoelectron and $1703$ events from the extrapolation.
Using the time of the run $\tau$, the power measured by the NIST photodiode $P$, the attenuation between the photodiode and the PMT $\alpha$, and the wavelength of the LED $\lambda$, the absolute efficiency is given by \eq\ref{eq:EfficiencyCalculation}). 
The resulting efficiencies, extrapolated to the pedestal mean, are  $0.2024$ for PMT-2 and $0.2121$ for PMT-3, both at $\lambda = 378$ nm. A summary of these results is given in table~\ref{tab:TableOfXP2020QEff}.

The uncertainty on the efficiency is given by \eq\ref{eq:Total}). The largest contribution to the uncertainty is the 1.5\% systematic error from the calibration of the NIST photodiode. In the spectra of both PMTs the ratio of 
one photoelectron event to pedestal events is less than 1.2\%, so the contamination of two photoelectron events is less than 0.6\% the number of one photoelectron events. This percentage is added to the systematic uncertainty, as it represents 
a systematic under-counting of two as ones. 
For the extrapolated efficiency, the statistical error on the number of one photoelectron events takes into account the estimated 10\% uncertainty on the number of events found by the extrapolation to the pedestal mean. The 0.5\% statistical
error on the measurement of the current from the NIST photodiode is also included.
All of these uncertainties are added in quadrature, as it is reasonable to assume that they are all independent and random relative to one another. This gives a total uncertainty of $\pm~1.9\%$ for the extrapolated efficiency
and $\pm~1.8\%$ for the efficiency above 1/3 of a photoelectron.

\begin{table}
\begin{center}
\begin{tabulary}{1.0\textwidth}{LCCCCC}
 \toprule
PMT & P & $N_{\text{1/3pe}}$ & $N_{\text{corr}}$ & $\epsilon_{\text{1/3}}$ & $\epsilon_{\text{ex}}$\\
\cmidrule(l){2-6}
2 (\#16361) & $1.867~10^{-10}$ & $25736$ & $1295$& $0.1927\pm 1.8\%$ & $0.2024\pm1.9\%$ \\
3 (\#42175) & $2.227~10^{-10}$ & $32095$ & $1703$& $0.2014\pm1.8\%$ & $0.2121\pm1.9\%$\\
\bottomrule
\end{tabulary}
\caption[PMT Calibration Efficiency Results]{
\label{tab:TableOfXP2020QEff}
A table of results for the absolute efficiency of PMT-2 and PMT-3. These results are taken from 
the spectra shown in Figs.~\ref{fig:XP2020Qspectrum:16361} and \ref{fig:XP2020Qspectrum:42175}.
$N_{\text{1/3pe}}$ gives the number of events above 1/3 of a photoelectron (pe). This is taken as the number of single pe events.
$N_{\text{corr}}$ is the number of events found by an extrapolation of the single pe peak from the 1/3 pe threshold to
the pedestal mean. $\epsilon_{\text{1/3}}$ is the efficiency at threshold of 1/3 of a pe and $\epsilon_{\text{ex}}$ is the efficiency extrapolated to the pedestal mean.
}
 \end{center}
\end{table}

\begin{figure}[h!]
\centering
 \subfigure[PMT-2]{\label{fig:ThresholdVariation:16361} \includegraphics[width=0.70\textwidth]{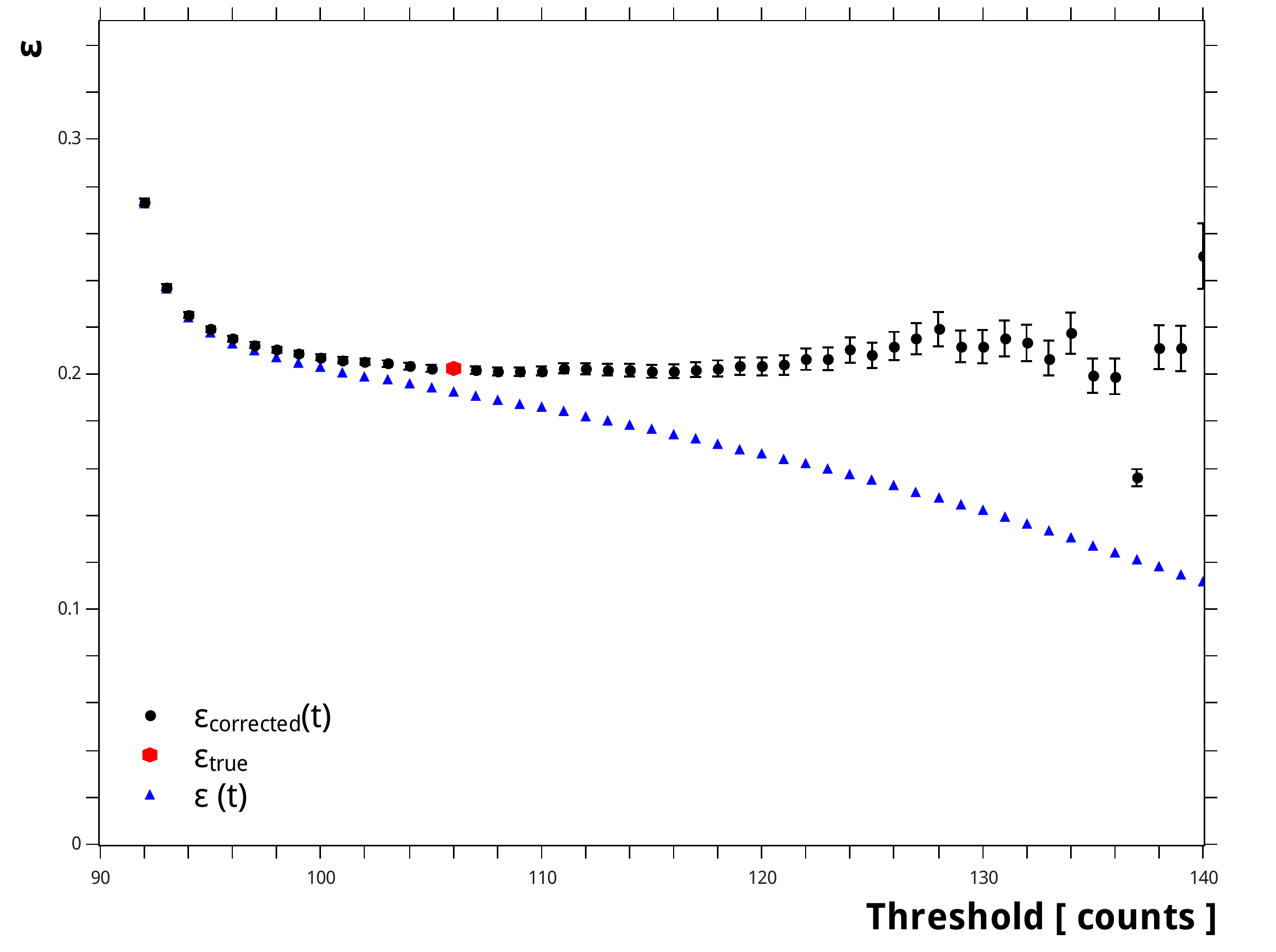}}\\
  \subfigure[PMT-3]{\label{fig:ThresholdVariation:42175} \includegraphics[width=0.70\textwidth]{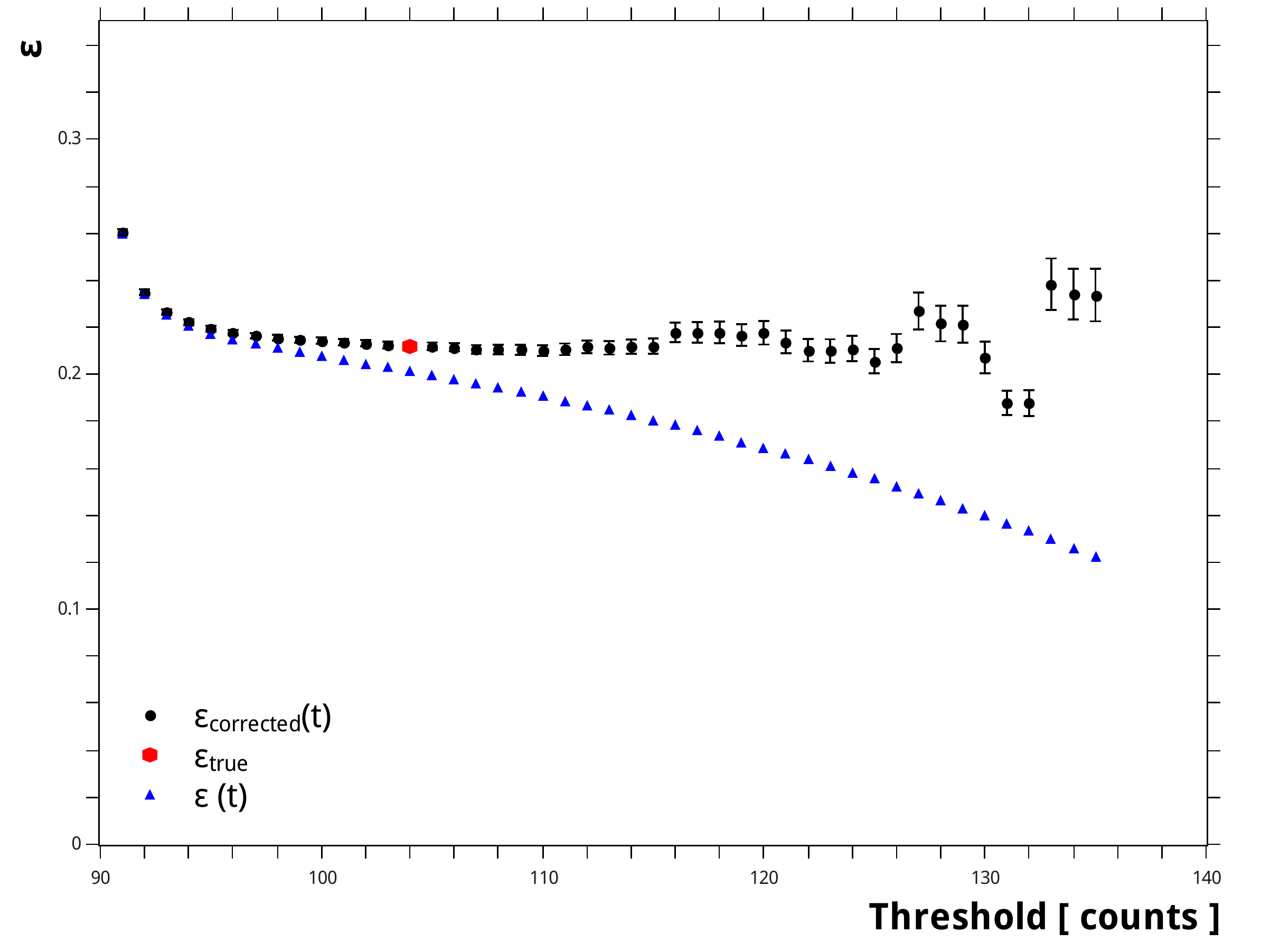}}
\caption[Dependence of the Efficiency on Threshold]{ 
\label{fig:ThresholdVariation} Two plots showing the effect of threshold choice on the efficiency. The black circles show the efficiency extrapolated to the pedestal mean. 
This efficiency has a plateau around the valley, where the sum of events above the threshold and the events from the extrapolation is the same despite the threshold.
The red hexagon shows the actual reported efficiency (see text), using a threshold at 1/3 of a photoelectron. The blue triangles show the efficiency taking only counts above the threshold (i.e.\ no extrapolation), 
and are basically the S-curves (within this small threshold range) of the single photoelectron spectra.
}
 \end{figure}

To study what effect the choice of threshold has on the efficiency result, the efficiency is plotted versus the chosen threshold, which is shown in \fig\ref{fig:ThresholdVariation:16361} for PMT-2 and \fig\ref{fig:ThresholdVariation:42175} for PMT-3.
In both plots the black circles show the efficiency (ordinates) calculated using the 
number of single photoelectron events above the chosen threshold (abscissa) plus the events found for an extrapolation of the single photoelectron peak from that threshold to the pedestal mean. The black error bars show the 
statistical error, which increases as the threshold moves higher. This is natural, as the number of actual events decreases, and the number of events from the extrapolation, with an assumed error of 10\%, increases. 
The apparent efficiency increases sharply as the threshold moves into the pedestal, as more and more noise is counted as single photoelectron events. 
The efficiency calculated using a threshold at $1/3$ of a photoelectron and an extrapolation to zero is shown by the red hexagonal marker. 
There is a plateau around this value where the extrapolated efficiency is not highly sensitive to the choice of threshold. 

This is in contrast to the efficiency without correction, i.e., using only the events above the current 
threshold, shown in Figs.~\ref{fig:ThresholdVariation:16361} and \ref{fig:ThresholdVariation:42175} by the blue triangles. 
These two curves are essentially S-curves, the cumulative distribution function of each spectrum, and give the efficiency as 
a function of threshold. 
This efficiency naturally decreases rapidly as a higher threshold is used,
and does not show a strong plateau around the valley (at least not in this case). If it is the S-curve itself which is measured, as in
 a counts over threshold setup, then the single photoelectron spectrum, and thus the position of the 
valley, can be recovered by derivation.

\subsubsection{Efficiency Variation Over the Photocathode Surface}
\begin{figure}[h!]
\centering
 \subfigure[PMT2]{\label{fig:surfacescan:16361} \includegraphics[width=0.48\textwidth]{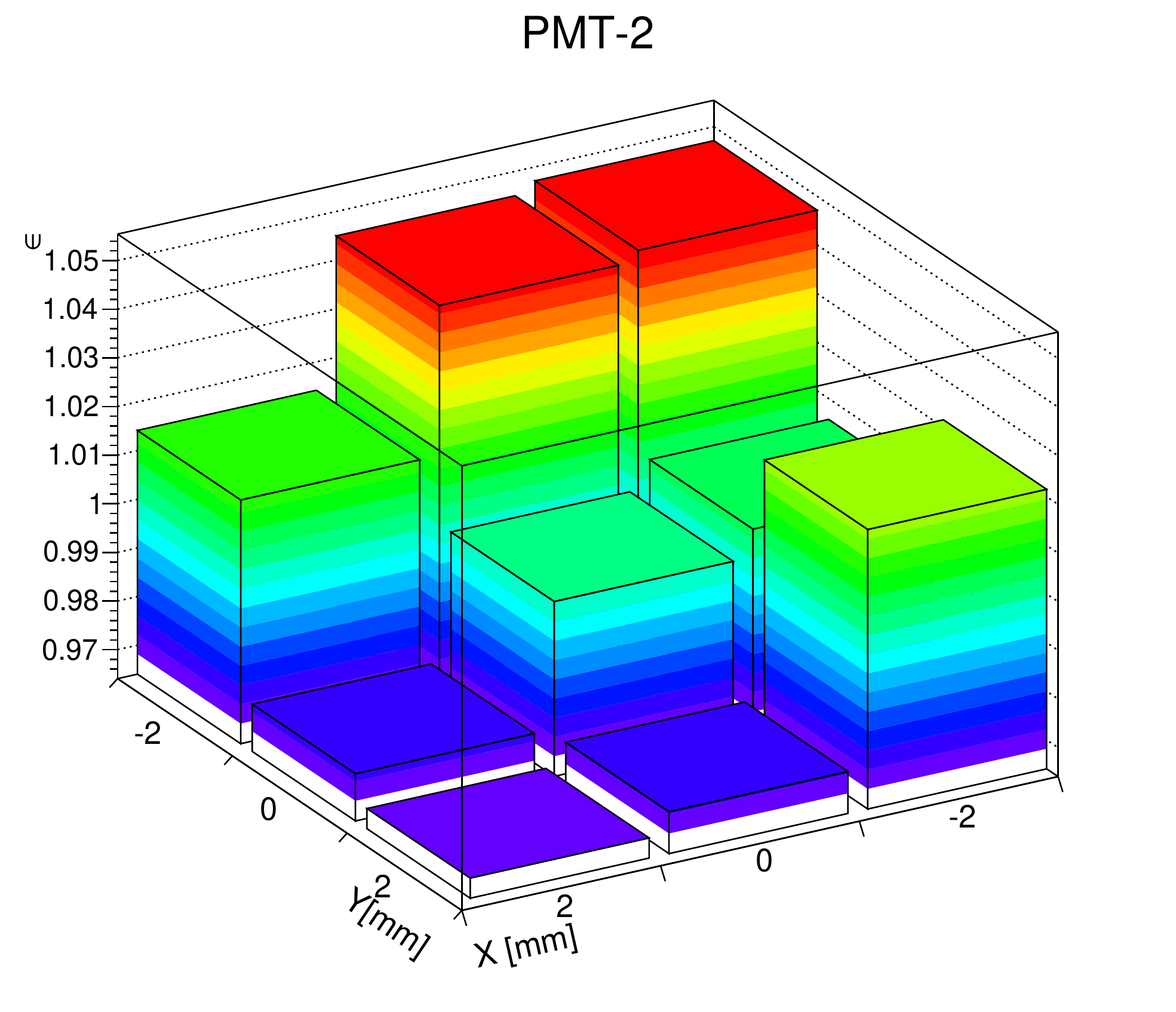}}
  \subfigure[PMT3]{\label{fig:surfacescan:42175} \includegraphics[width=0.48\textwidth]{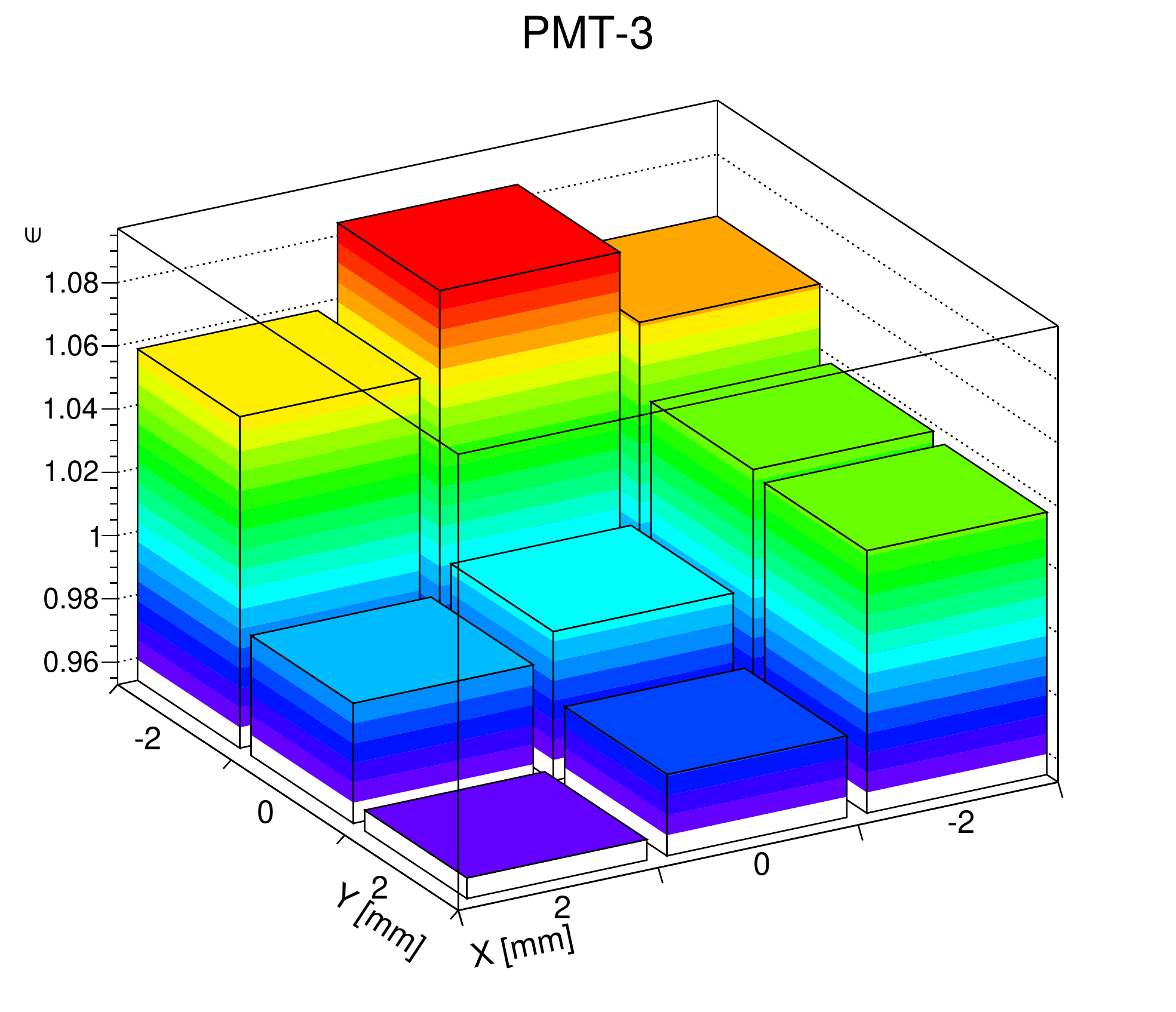}}\\
  \subfigure[Previous Scan of PMT2]{\label{fig:surfacescan:Gwen:PM2} \includegraphics[angle=270,width=0.48\textwidth]{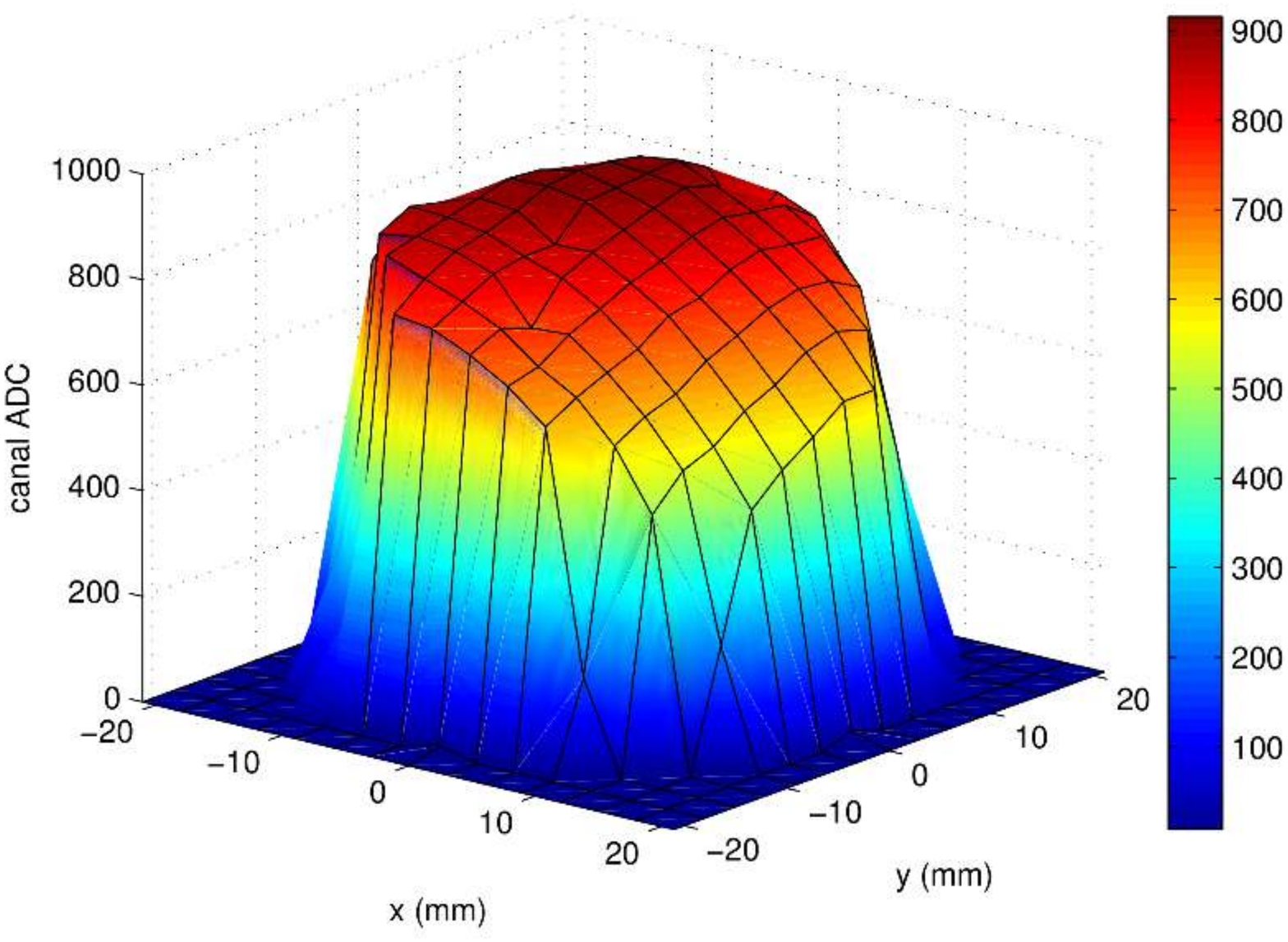}}
  \subfigure[Previous Scan of PMT3]{\label{fig:surfacescan:Gwen:PM3} \includegraphics[angle=270,width=0.48\textwidth]{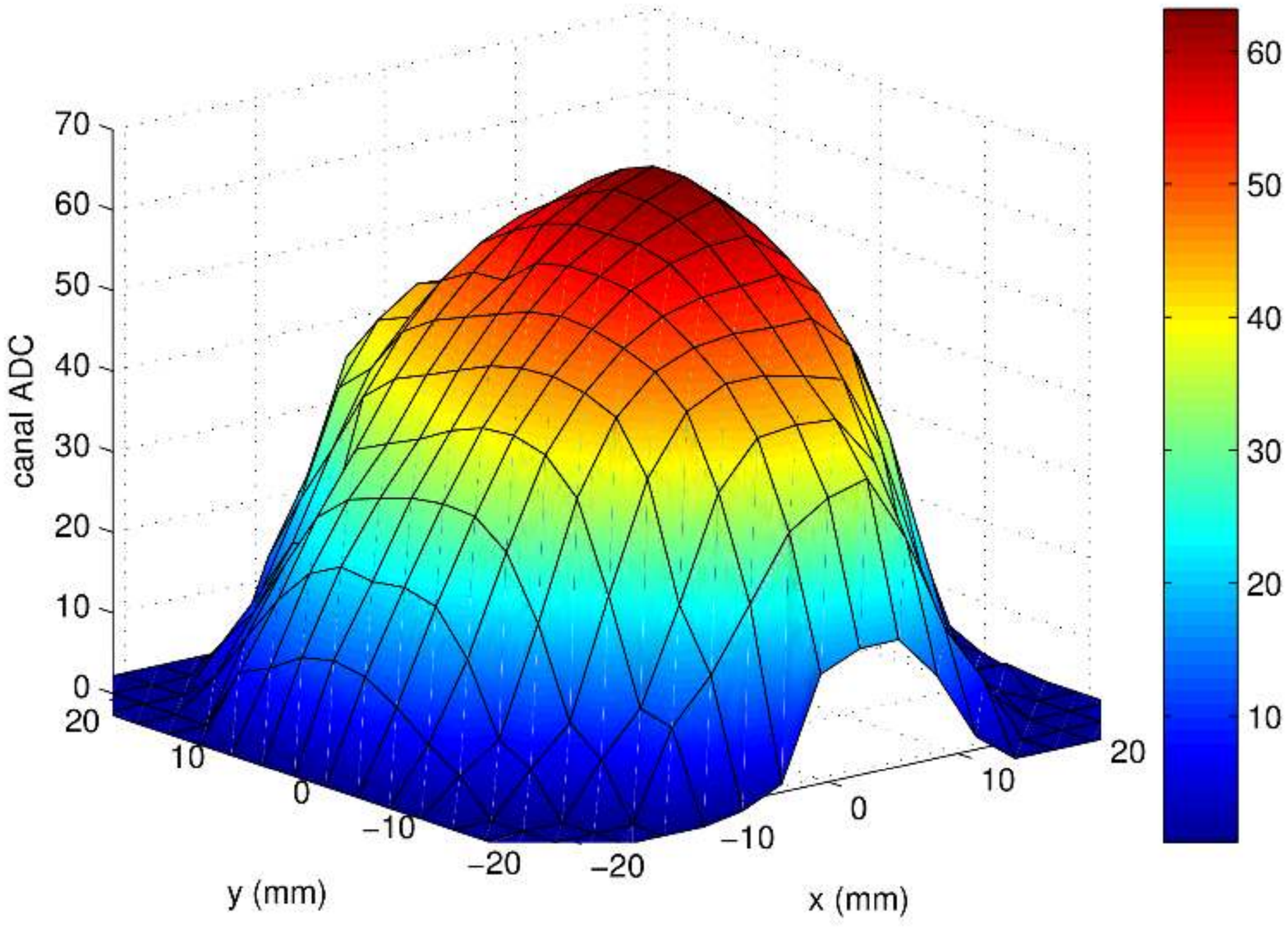}}
\caption[Scan the Photocathode Surface]{
\label{fig:surfacescan} Surface Scan of each PMT in a rectangle of $\pm$ 2 mm around the center (the point where the absolute efficiency was measured) of the photocathode.
The average variation in the efficiency is $1.2\%$ per mm for PMT-2 and $1.9\%$ per mm for PMT-3. 
In the fluorescence bench the incident light spot must be in the same position on the photocathode as the absolute measurement, and must be smaller in size than the scale of the variation of the efficiency with position.
Figs.\ref{fig:surfacescan:Gwen:PM2} and \ref{fig:surfacescan:Gwen:PM3} show previous scans of these PMTs from \cite{LefeuvreThesis} for comparison. Note that the previous scans were over a much larger area of the photocathode, whereas
the scans done here where only around the ``central'' point.
}
 \end{figure}

The absolute efficiency of both PMTs was measured at one point on the photocathode, at the apparent center. 
The PMT was placed inside a mu-metal tube which was slightly larger than the PMT itself, and the alignment of the PMT was done by allowing it to settle into the bottom of the tube. 
After taking the absolute spectrum at the center, we scanned in a rectangle of 2 mm around this central point. 
At each point we took a single photoelectron spectrum and determined the efficiency of PMT, to check the variation of the efficiency with the position of the incident light on the photocathode.

\fig\ref{fig:surfacescan} shows two plots, one for each PMT, of the efficiency measured at each position, relative to the absolute efficiency measured at the center.
Both plots have been rotated so that every bin is visible. The X and Y coordinates are defined with the positive Y direction upwards and the positive X direction to the right, when facing the photocathode from the front. This was with the 
PMT rotated upside down (such that the high voltage and anode BNC connectors of the base where on the bottom). 

Both PMTs show the same trend, with the efficiency being greater towards the bottom of the region scanned. 
It could be that the change in efficiency is dominated by a variation in the collection efficiency, and so we are seeing an effect from the orientation of the focusing electrode. 
It could also be the case that we are slightly ($\leq1~$mm) off of the true center of the photocathode due to the 
way the PMTs were laying in their housing. It is, in either case, not important as the efficiency must in any case be measured with the same light-spot size and position as will be used in the fluorescence bench.

The measurement is sensitive to this difference, as the difference is larger than the $\sim1\%$ statistical error on the efficiency. This level of variation is expected, and
is due on one hand to the change in the quantum efficiency across the photocathode surface, and on the other, to the variation of the collection efficiency with the position of the liberated photoelectron
(see section \ref{sec:INTRO:PMT} for more discussion on this topic).
Both of these effects are more obvious on a PMT with a single relatively large photocathode, such as the XP2020Q.
This means that the surface used on the photocathode is an important consideration and when the PMT is used for a precise measurement:
\begin{inparaenum}[i\upshape)]
 \item the incident light spot should be restricted to a small area of the PMT where the efficiency is effectively constant, and
 \item the light spot should be placed in the same spot on the photocathode as where the absolute efficiency was measured.
\end{inparaenum}
This is also one reason why a large decrease in uncertainty can be gained by switching from a single large PMT to
multiple smaller PMTs in many applications.


\subsubsection{Effect of the Earth's Magnetic field}
\begin{figure}[h!]
\centering
\includegraphics[width=0.70\textwidth]{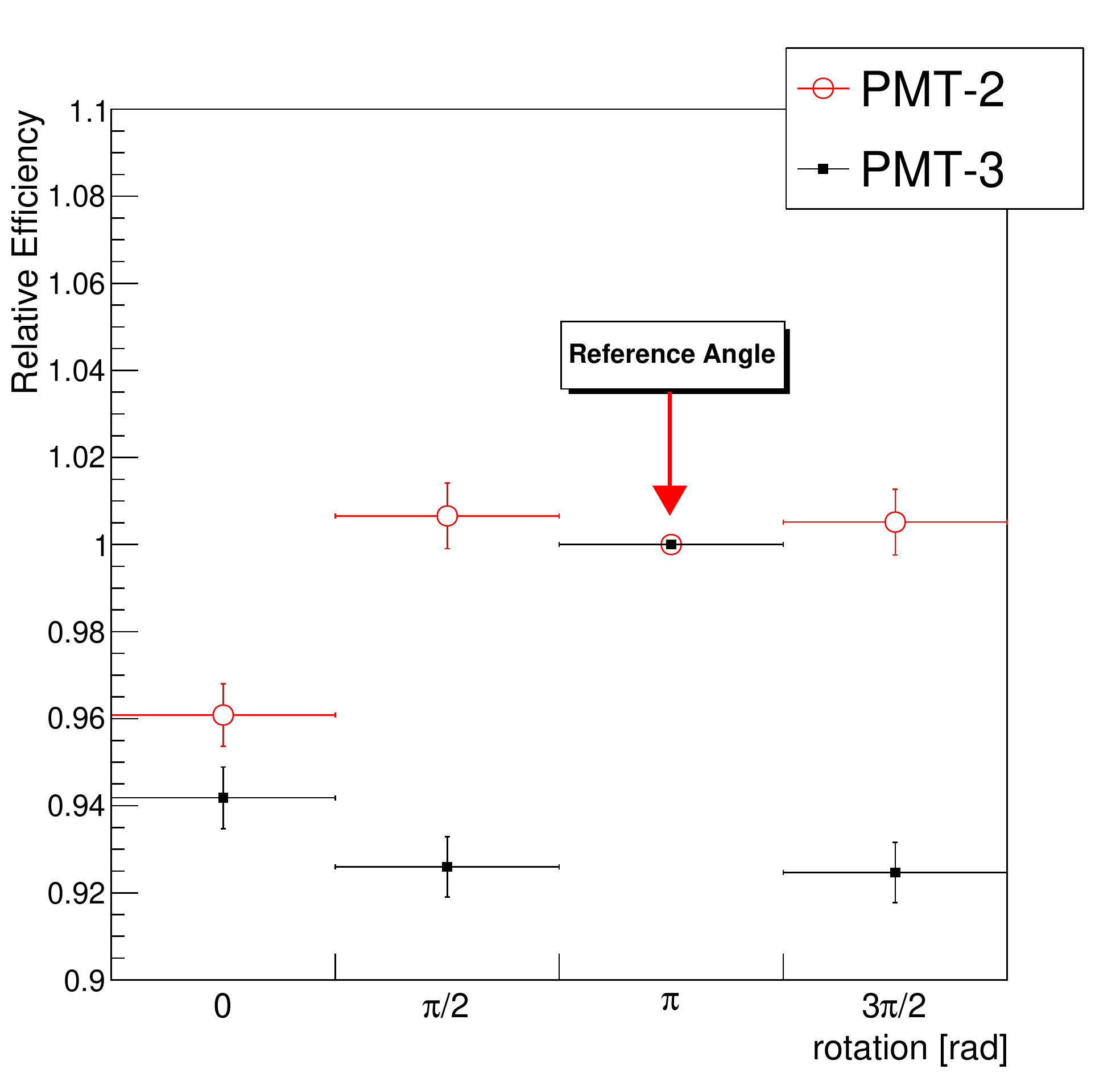}
\caption[Effect of Magnetic Fields on the  Efficiency]
{\label{fig:magneticscan}  The change in efficiency caused by rotating the PMT around its z-axis. 
The efficiency is relative to the value at a rotation of $\pi$, the orientation where the absolute efficiency was measured (defined as the SHV and BNC connectors of the PMT base pointing down). A clear
difference can be seen with the orientation in the Earth's magnetic field. The same rotation must be preserved between the calibration and the use of each PMT in the fluorescence bench. }
 \end{figure}

The efficiency of a PMT is also sensitive to ambient magnetic fields. Magnetic fields affect the trajectories of photoelectrons emitted from the photocathode and the electron showers in the multiplication stage, meaning that 
the collection efficiency and gain of the PMT depend on both the strength and orientation of the magnetic field. To guard against stray magnetic fields in the laboratory, both PMTs are housed inside a mu-metal tube. 

They are still sensitive, however to their orientation in the Earth's magnetic field, at the level of a few percent. To check this effect, we rotated both XP2020Qs around their $Z$-axis and took a single photoelectron spectra at steps of 
$\pi$/2. The rotation was clockwise, defined relative to the orientation with the BNC and SHV connectors on the PMT base on top.

\fig\ref{fig:magneticscan} shows the result of these measurements. The red line shows the efficiency of PMT-2 and the black line shows the efficiency of PMT-3. The efficiency is given as a value relative to the measured efficiency at a rotation of 
$\pi$ radians. A clear difference can be seen with the orientation, particularly between 0 radians and the other orientations. These where some of the first measurements we did, and all the later
 measurements where done with the PMTs rotated by $\pi$, i.e., with the connectors facing towards the bottom. The important point is that the orientation in which the efficiency is measured must be conserved when setting up the air fluorescence measurement, otherwise the small 
change on the order of 1-2\% would be a source of systematic error.

\section{Conclusion}

A accurate and complete knowledge of air fluorescence is essential for the observation of UHECR extensive air shower using the fluorescence technique. The physics of air fluorescence is itself an interesting and rich field which is experimentally challenging.
We have proposed a new measurement of the air fluorescence which aims to measure both the absolute yield and the absolute spectrum with high precision and in every set of atmospheric conditions relevant to UHECR physics. 

The accuracy of this proposed setup relies on an absolute calibration of the photomultiplier tubes used in the bench with an uncertainty of less than 2\%. The first measurement of two XP2020Q PMTs for this setup was shown in the last section.
It has been shown how the measurement of the absolute efficiency of the PMT
takes into account all the variations in efficiency due to the location of the incident light on the photocathode, the orientation of the PMTs in the Earth's magnetic field, and the supply voltage of the PMT.

In order for this measurement of the efficiency to be valid, the PMT must calibrated in the same conditions under which it will be used. This means that the supply voltage must be the same, the single photoelectron threshold must be the same, and the 
numerical aperture of the collimator (or fiber) must be the same. To this end the final efficiency of all the PMTs used in our air fluorescence measurement will be checked during the calibration of the bench. Notice, in particular, that the calibration
phase of the fluorescence bench, shown in \fig\ref{fig:AFYmeasurementCalibration}, is essentially the same as the PMT calibration setup.
Once the transmission of the fiber bundle is 
characterized, PMT-2 can be re-calibrated with the scintillating fiber by placing a photodiode on the integrating sphere
in the place of PMT-1. In this case the efficiency of PMT-2 would be known using the same electronics (threshold), spatial pattern of illumination, etc.\ as will be present in the final fluorescence measurement. 

In effect, our proposed fluorescence bench is a PMT calibration setup in which 
the calibration light source is later replaced by air fluorescence emission.   
This eliminates any systematic errors associated with either trying to recreate the exact conditions used in a separate calibration measurement, or extrapolating the efficiency to a difference set of conditions, and it is this which allows us to transform the 
accuracy of the PMT calibration into an accurate and absolute measurement of the air fluorescence yield in all relevant atmospheric conditions.
  
    \printbibliography[heading=subbibliography]
     \end{refsection}

    \begin{refsection}
  \chapter{The High Voltage Power Supply and Switches in EUSO}
   \label{CHAPTER:CWHVPStests}

%
%
%
%

   
Because both the single photoelectron gain and efficiency of a photomultiplier tube are strongly dependent on the voltage supplied to the dynodes,
a high voltage power supply and PMT should be characterized together. 
The gain of a PMT is extremely high, and the gain is a logarithmic function of the voltage. Due to this, the stability of the PMT gain 
is strongly affected by the stability of the high voltage. Generally speaking, if the gain must be maintained within 1 percent, then the high voltage must be maintained within better than 0.1 percent.
Each of the PMT's dynodes must also be maintained at the proper voltage relative to one another for the PMT to function optimality in terms of collection efficiency, gain, and, in the case of multianode PMTs, focusing. 

Typically a given PMT has a recommended voltage repartition which is tuned to
give a certain set of properties. 
For the XP2020Qs calibrated in section \ref{sec:Calibration of Photomultiplier Tubes for the Air Fluorescence Measurements at LAL}, for example, there is a recommended voltage repartition
tuned for maximum gain, and one tuned for better linearity.
In order to implement these voltage repartitions, it is possible to power a PMT by using a separate power supply for each dynode voltage required. A more convenient approach is to build a circuit which generates the required voltages from one input 
voltage. 
Most often the desired voltage repartition is implemented using a resistive voltage divider. For this type of PMT power supply, the highest voltage, the cathode voltage 
(or the anode voltage if a positive polarity is used as was the case in section \ref{sec:Calibration of Photomultiplier Tubes for the Air Fluorescence Measurements at LAL}) is input into the 
divider, which then gives each intermediate voltage by using the voltage drop across a resistor. 
These circuits are simple to construct and have the advantage that the voltage division can be arbitrarily selected to optimize the properties of the PMT.
There is some subtlety in their design, however, because no real circuit can maintain the exact voltage on each dynode without some feedback effect from the total current through the divider \cite{Tubs069}.    

The great disadvantage of a resistive divider, however, is the power dissipation in the resistors and the fact that the current in the divider is in parallel to the
current in the PMT. For JEM-EUSO, this disadvantage is a deal-breaker, as JEM-EUSO is a space mission with a limited power budget.
We can estimate the power consumption of a resistive divider by considering an estimated background count rate of 1 photoelectron per pixel per gate time unit ($1~\text{GTU} = 2.5~\mu$s), averaged over the JEM-EUSO focal surface.
If each PMT in the focal surface is operating at a gain of $1~10^{6}$, then the current drawn by the entire focal surface of 315 kPixels will be $\approx20.2~$mA, and
there should be a factor of 100 more current flowing in the resistive divider to avoid loading the high voltage power supply in the event of an increase in light. As the M64 PMT needs a cathode voltage of 
900~V to have gain near $1~10^{6}$, a resistive divider would require a power of $\approx1.8~$kW. The power budget for the entire JEM-EUSO focal surface is $\sim 500$ W.
This immediately shows that this approach is unworkable, and a different design philosophy is needed. In addition, this power is dissipated as heat, which must then be removed from the Focal Surface. 
Cooling systems are complicated in a space environment, add weight, and may possibly need additional power.

Our solution is to use a Cockcroft-Walton voltage multiplier circuit to power the PMTs in JEM-EUSO, EUSO-Balloon, and EUSO-TA. 
The Cockcroft-Walton circuit is so-named for J.\ Cockcroft, and E.\ Walton, who used this circuit to power the accelerator on which they performed their experiments with atomic disintegration \cite{Cockcroft01071932}.
The Cockcroft-Walton circuit generates a higher voltage from an AC or pulsed DC input voltage using a ladder of capacitors and diodes. 
The voltage output at any stage of an ideal Cockcroft-Walton voltage multiplier is twice the peak input voltage times the number of stages, and voltages can be taken at multiple stages of 
the ladder. 

Like all high voltage power supplies, real Cockcroft-Walton circuits show voltage sag and ripple. In Cockcroft-Walton circuits these increase as the number of stages increases. 
The voltage ripple increases with output current, and so Cockcroft-Waltons are most ideal for applications which require a low current.
In addition to accelerators, high voltage power supplies based on the Cockcroft-Walton have previously been used to power PMTs in many instances (e.g.~\cite{CockCroftPMTBase, CockcroftFEU84, CockcroftKM3Net}).   
The M64 is well suited to the use of a Cockcroft-Walton power supply as the voltage repartition recommended by Hamamatsu has an equal step between the majority of the dynodes. 

In addition to the basic power supply concerns, there is also the question of dynamic range of the PMT. 
Although the primary goal of JEM-EUSO is the observation of extensive air showers created by UHECR, there are numerous other phenomena which can be studied by a Earth-viewing UV telescope. 
These phenomena range from Transient Luminous Events (TLE), meteoroids and meteors, and anthropomorphic phenomena as discussed in \ref{sec:INTROtoJEMEUSO}. 
The dynamic range of the readout electronics is limited, however, and the anode of a PMT suffers radiation damage due to electron bombardment. This can affect the gain of the PMTs if the anode receives too high of an electron flux. 
For the M64, Hamamatsu recommends that the sum of the current through all 64 pixels be less than $100~\mu$A to avoid permanent damage. This is a typical anode current limit for PMTs. 
The dynamic range between the other phenomena  of interest and extensive air showers requires a factor of nearly $10^{6}$ difference in gain. 
In order to study the phenomena across this range of light levels, and to protect the PMTs from damage due to overexposure at high gain, there must be a system to rapidly 
lower the number of electrons reaching the anode. This gain reduction must be fast to keep the energy received by the anode over its operational lifetime below the level which would damage it.
This gives a required switching time on the order of a few microseconds.

The single photoelectron gain of a PMT is determined by the repartition of voltages across its dynodes, and the dependence of the gain on the voltage is given by \eq\ref{eq:SinglePhotonGain}).
The magnitude of the gain varies logarithmically with the cathode voltage, $V_{\text{c}}$ when the overall voltage repartition is scaled. 
That is to say that $\mu=K V^{n\alpha}$ (\eq\ref{eq:PMTGain})) holds when all the voltages are lowered together proportionally so that
the voltage on each dynode has the same ratio at the new cathode voltage as at the old.
Switching the gain in this way means that the voltage on 14 separate elements must be changed at once. This is slow due to the capacitance of the PMT, and is complicated due to the need to have 14 switch circuits in parallel if we do not used 
a voltage divider.

There is another solution if we recognize that we really want to change the PMT gain and not necessarily the single photoelectron gain.
The current through the anode is proportional to the PMT gain, which is the product of the single photoelectron gain and the collection efficiency. The collection efficiency depends on the voltage difference between the photocathode and the first dynode. 
The other solution to reduce the current received by the anode of the PMT, then, is to change only the voltage on the photocathode, reducing the collection efficiency. 

In this case, the single photoelectron gain of the PMT will be roughly the same at each switch step, but the number of photoelectrons which will be collected into the multiplication stage of the PMT will be greatly reduced.
In this later scenario there is only one element to be switched, which simplifies the design of the switching circuit. 
The design of the Cockcroft Walton high voltage power supply (CW-HVPS) for JEM-EUSO and the associated switch, built as an integrated part of the power supply, have been created within the JEM-EUSO collaboration by J. Szabelski et al.~\cite{Adams:2012tt}.
Since we have the equipment and expertise to test PMTs at APC, however, we were responsible for defining the requirements, giving feedback for design decisions, and testing the final design of the CW-HVPS.

\section{High Voltage Power Supply and Switch Design}
\label{sec:Cockcroft-WaltonandSwitchDesign}
During the course of the CW-HVPS design we did periodic test measurements using prototype boards.
These tests where done to check the characteristics of the M64-PMT using the CW-HVPS, and to give feedback for the improvement of the CW-HVPS.
One important decision which had to be made early in the development of the CW-HVPS was the exact voltage repartition to be used in the CW-HVPS design. 

The voltage distribution recommended by Hamamatsu for good gain and focusing characteristics is intended for use in a resistive divider. In a resistive divider,
the voltages on each dynode can be adjusted easily by choosing the value of each resistor in the circuit, and so the voltage drop between any two elements can be arbitrarily selected.
In contrast to this, the voltage step between levels in a Cockcroft-Walton circuit is the same for each step in the chain. 
Any voltage difference in the repartition which is not an integer multiple of the step voltage will either require a second power supply or the addition of a resistive divider between two stages of the Cockcroft–Walton circuit. 
To minimize the power consumption, and to simplify the design of the power-supply as much as possible, it is beneficial to avoid any resistive dividing element whenever possible.
This means modifying the voltage repartition to suit a CW circuit, as long as the modification does not sacrifice the characteristics of the PMT.

At the early stages when we where doing these measurements, the baseline PMT to used in JEM-EUSO had been changed from the M36 multianode PMT to the M64. The M64 had the same sensitive surface, but 64 pixels compared to the 36 pixels of the M36.
There is an additional internal difference, however, in the structure of the electrodes. The M36, like many PMTs, has a focusing electrode between the photocathode and the first dynode. In the M64, however, this element
was removed and replaced with an electrode between the 12th dynode and the anode. This element is called a ``guard-ring'' in the Hamamatsu literature. This change may have been motivated by the fact that 
the distance between the photocathode and the first dynode in the M64 is reduced compared to the M36, so that the focusing element was not as necessary.

In the typical case, the voltage on the focusing electrode affects the collection efficiency of the PMT and, in multianode PMTs, the focusing of photoelectrons from one pixel towards the first dynode of that same pixel. 
The guard-ring of the M64, on the other hand, is intended to provide an electrostatic focusing at the anode and reduce the spill-over of an electron shower from one pixel onto its neighboring pixels.
All the previous measurements where done with the M36, however, so that the actual effect of the guard-ring in the M64 was not clear.  
Therefore, one of the first measurements which I did was to determine the effect of the guard-ring voltage, $V_{\text{gr}}$, 
on the characteristics of the M64 PMT by measuring the gain, focusing, and dynode currents. These measurements were done using two different voltage repartitions, the standard
Hamamatsu voltage repartition, and a repartition modified to better suit the CW-HVPS.  Both of these voltage repartitions are shown in table \ref{tab:M64MAPMTvoltageRepartions}.

\subsection{Study of the Focusing on the Anode by the Guard-Ring}
\label{subsec:Focusing}

To study the focusing properties of the guard-ring element in the M64, each electrostatic element of the PMT was first placed at the proper voltage
using a CAEN model A1532 high voltage power supply with the photocathode at 1000 V.
A lab power supply was used for the Grid, because the CAEN HVPS has a high impedance under 100 V, and therefore cannot polarize properly at low voltages. 
Due to this, the PMT will not function properly if the CAEN HVPS is used to power the guard-ring.  

The PMT was placed in our typical calibration setup, as described in section \ref{sec:Measuring Absolute Detection Efficiency}.
The light source was a single LED with a wavelength of 378 nm connected to an integrating sphere. 
A collimator with an entrance diameter of 1 mm, an exit diameter (towards the PMT) of 0.3 mm, and a length of 23.5 mm was attached to the exit port of the sphere.
The PMT was positioned directly at the aperture of the collimator, and the LED was pulsed at a rate of 1 kHz in coincidence with the charge integration gate of the QDC. 
This rate was due to the limitations of the data acquisition system we used at the time, which was a single Lecroy 2249 CAMAC QDC controlled through Labview. To 
overcome the low charge resolution of this QDC we used an amplifier with a factor of ten gain.
The pulse amplitude sent to the LED was adjusted so that $99\%$ of the pulses would produce no photoelectron in the PMT. 
The remaining $1\%$ of the pulses are single photoelectron pulses, with a less than 1\% contamination of two photoelectron events (see section \ref{subsec:The Single Photoelectron Spectrum} for a discussion of ``single photoelectron mode'').

In the first measurements we had difficulty obtaining a good single photoelectron spectra due to the relatively low gain of our test PMT, which was a very early example of the M64. To overcome this, we scaled the cathode 
voltage to 1100 volts to increase the gain. 
Using this scaled voltage distribution it was possible to obtain a usable single photoelectron spectra with a $\times 10$ fast amplifier. 

The focusing properties of the guard-ring were then studied by looking at the number of events seen in a $3\times3$ block of pixels surrounding pixel 10, numbered as defined on the Hamamatsu specifications of the M64 (which is shown in \ref{fig:M64pindiagram}).
The collimator was first positioned by eye at the cross between the upper right four pixels in our $3\times 3$ block. The light spot was then centered on pixel 10 by equalizing the number of single photon events in the 4 pixels using the X-Y movement 
which held the integrating sphere (this procedure is described in more detail in section~\ref{sec:PixelCenetering}).
This was done to an accuracy of better than $\approx0.1~$mm. This position was then taken as the origin. 
The Hamamatsu specifications for the M64 give a pixel width and height of 2.88 mm, which has been verified to be accurate in previous measurements \cite{GoroPrivate}, and so
the beam was moved 1.44 mm vertically and 1.44 mm horizontally to place it in the center of pixel 10.
 
%
%

\begin{table}
 \begin{center}
\begin{tabulary}{1.0\textwidth}{LCCC}
\toprule
Element & \multicolumn{3}{c}{Repartition}\\
 \cmidrule(l){2-4}
 &  Hamamatsu &  Low-Power & CW-HVPS\\
\cmidrule(l){1-4}
K & 2.3 & 2.9 & 2.7\\
D1 & 1.2 & 1 & 1.3\\
D2 & 1 & 1 & 1\\
D3 & 1 & 1 & 1\\
D4 & 1 & 1 & 1\\
D5 & 1 & 1 & 1\\
D6 & 1 & 1 & 1\\
D7 & 1 & 1 & 1\\
D8 & 1 & 1 & 1\\
D9 & 1 & 1 & 1\\
D10 & 1 & 1 & 1\\
D11 & 1 & 1 & 1\\
D12 & 1 & 0.67 & 0.67\\\
GR & 0.5 & 0.33 & 0.33\\
Total & 15 & 14.9 & 15\\
\bottomrule
\end{tabulary}
\caption[M64 Voltage Repartions]{
\label{tab:M64MAPMTvoltageRepartions}
A table showing the voltage repartitions used with the Hamamatsu M64 PMT. The ``Hamamatsu'' repartition is the normal voltage repartition 
given by Hamamatsu on the M64 data sheet. The ``low-power'' repartition is a modified voltage repartition more suited to a Cockcroft-Walton voltage multiplier which was tested in 
section~\ref{sec:Cockcroft-WaltonandSwitchDesign}. The ``CW-HVPS'' repartition is the actual repartition used by the prototype CW-HVPS, which was tested in section~\ref{sec:Test of the CW-HVPS and Switch Prototype}. This CW-HVPS design is 
used in EUSO-Balloon and EUSO-TA. 
}
 \end{center}
\end{table}

\begin{figure}[t]
\begin{center}
\subfigure[Single Photoelectron Gain]{\includegraphics[width=0.48\textwidth]{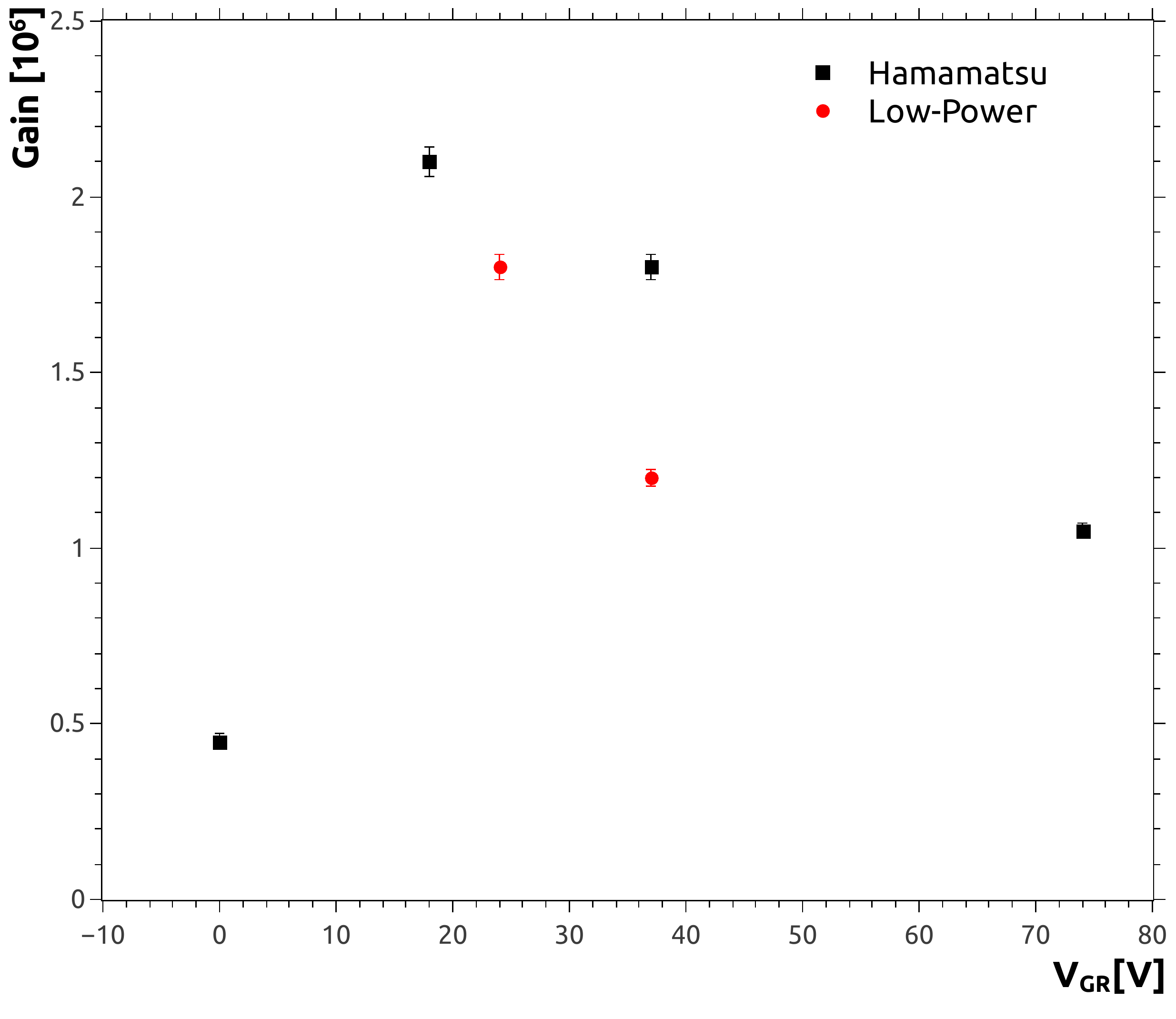}\label{fig:gain}}
\subfigure[Focusing]{\includegraphics[width=0.48\textwidth]{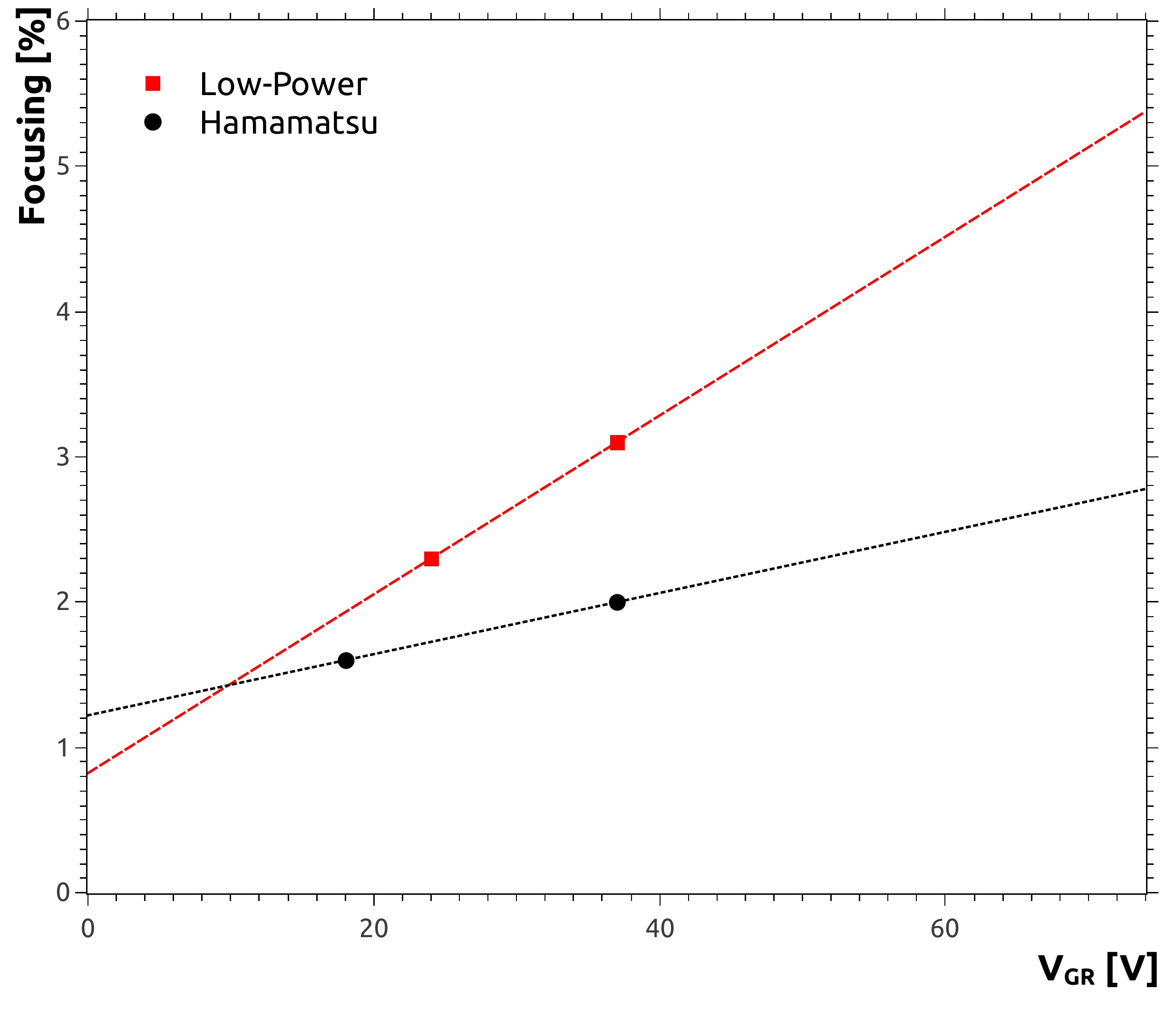}\label{fig:focus}}
\caption[Gain and Focusing Results]{Results for gain, \ref{fig:gain}, and focusing, \ref{fig:focus}, as a function of grid voltage $V_{\text{GR}}$ for the M64 PMT using the Hamamatsu and the
 ``Low-Power'' voltage distributions (see table \ref{tab:M64MAPMTvoltageRepartions}). The gain versus $V_{\text{GR}}$ is lower overall for the Low-Power distribution. The overall focusing is also better for the Hamamatsu
voltage distribution, but the difference is very small. 
}
\label{fig:focusinggain}
\end{center}
\end{figure}

The total number of single photo-electron events for each pixel were then counted in a run of 1 million events. This measurement was repeated
for several different guard-ring voltages, keeping all the other voltages the same. The focusing is calculated as the percentage of single photoelectron counts in the surrounding pixels relative to the number in 
the central pixel.
The measured change in the gain and focusing with changing guard-ring voltage $V_{\text{GR}}$ are plotted in figure \ref{fig:focusinggain}. 
The focusing and gain of the PMT are highest for a $V_{\text{GR}}$ voltage of 18 V (24 V for the Cockcroft-Walton distribution), and reduce slightly for $V_{\text{GR}}=37\,$V. At more extreme values of the guard-ring voltage, such as those approaching 
ground (the anode voltage) or the voltage of dynode twelve (74 V in this case), the gain of the PMT decreased to the point that it was not possible to see a single photoelectron peak in the central pixel.  

\subsection{Dynode Currents and Counting Rate}

We then studied the behavior of the current on individual dynodes using a different voltage repartition in which
the voltage step between the last dynode and the anode was changed to be the same ratio as between the other dynodes. 
This was motivated by a desire to simplify the design of the Cockcroft-Walton power supply.


The single photoelectron counting rate was measured by sending the anode signal of the PMT through an amplifier and then into a discriminator. 
The discriminator gives a NIM logic signal whenever the input is over a threshold voltage, where the threshold is chosen to correspond to the valley between the pedestal and the single photoelectron peak in the single photoelectron spectrum. 
The output of the discriminator was sent into a scaler which counted the number of logic signals, giving the rate of photoelectrons. If the efficiency of the PMT is known, then the count rate gives the rate of photons incident on the PMT 
(see section \ref{subsec:PhotoDetection}). 
 
For the count rate measurement, the single LED was replaced with a collection of $\sim100$ high intensity LEDs with a wavelength of 398 nm wavelength. These LEDs where connected in parallel with a buffer resistance to an external DC power supply. The
amount of light given by the LEDs at a particular input voltage was measured using the NIST photodiode attached to the top port of the integrating sphere.
The PMT was positioned approximately 30 cm from the exit port of our integrating sphere, to which a 3 mm aperture was attached. 
We then measured the counting rate and dynode currents while we varied the light level from zero (LED off) to the level at which the CAEN power-supply reached over-current. 
  
\begin{figure}
\begin{center}
 \includegraphics[width=1.0\textwidth]{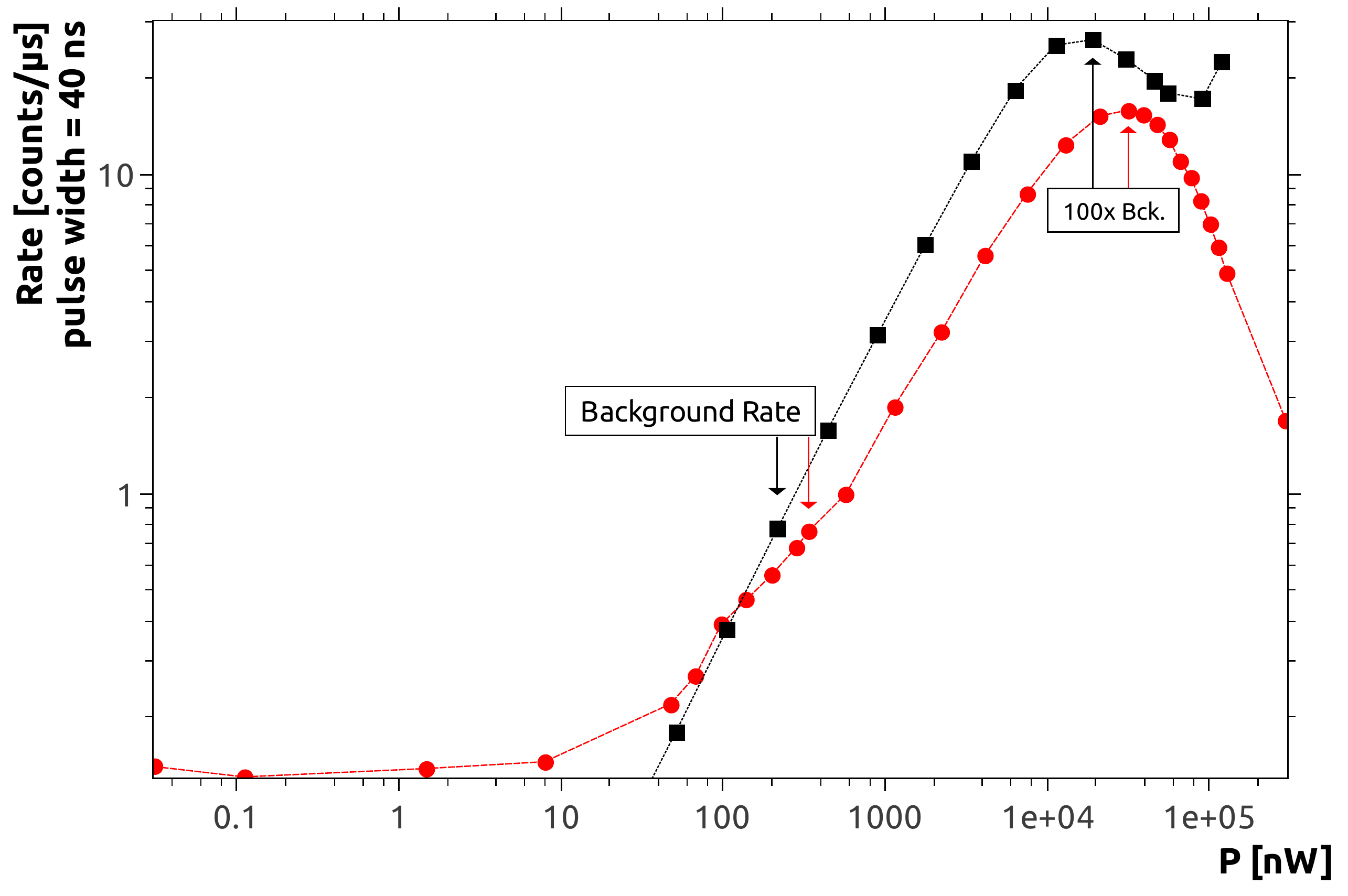}
\caption[Counting Rate]{The counting rate (per $\mu$s) for a pulse width of 40 ns plotted against NIST reading for the M64 PMT.  The rate using the Hamamatsu voltage distribution
 with the cathode at 1000 volts is shown by the black squares, and the rate using the Low-Power distribution is shown by the red circles. The label ``Background Rate'' corresponds to the JEM-EUSO background rate of $\sim600~$kHz/pixel.
The turning point in each curve is due to the onset of pile-up.
}
\label{fig:countrate}
\end{center}
\end{figure}

The measured count rate is shown in \fig\ref{fig:countrate} for both the Low-Power repartition and the Hamamatsu voltage repartition as a function of the measured incident power. 
The response of the PMT in the two cases is consistent, with the rate curve shifted higher for the Low-Power distribution. 
This is due to the lower gain of the PMT in that configuration, which causes a larger number of photoelectrons to be below the threshold of the discriminator.
The label ``background rate'' in the plot refers to working estimate of the background rate in JEM-EUSO of 0.62 MHz per pixel (500 photons/m$^{2}$ ns sr \cite{AdamsJr201376}). In both cases, the counting rate saturates at around 100 times the background rate due to pile-up.

\begin{figure}
\begin{center}
 \includegraphics[width=1.0\textwidth]{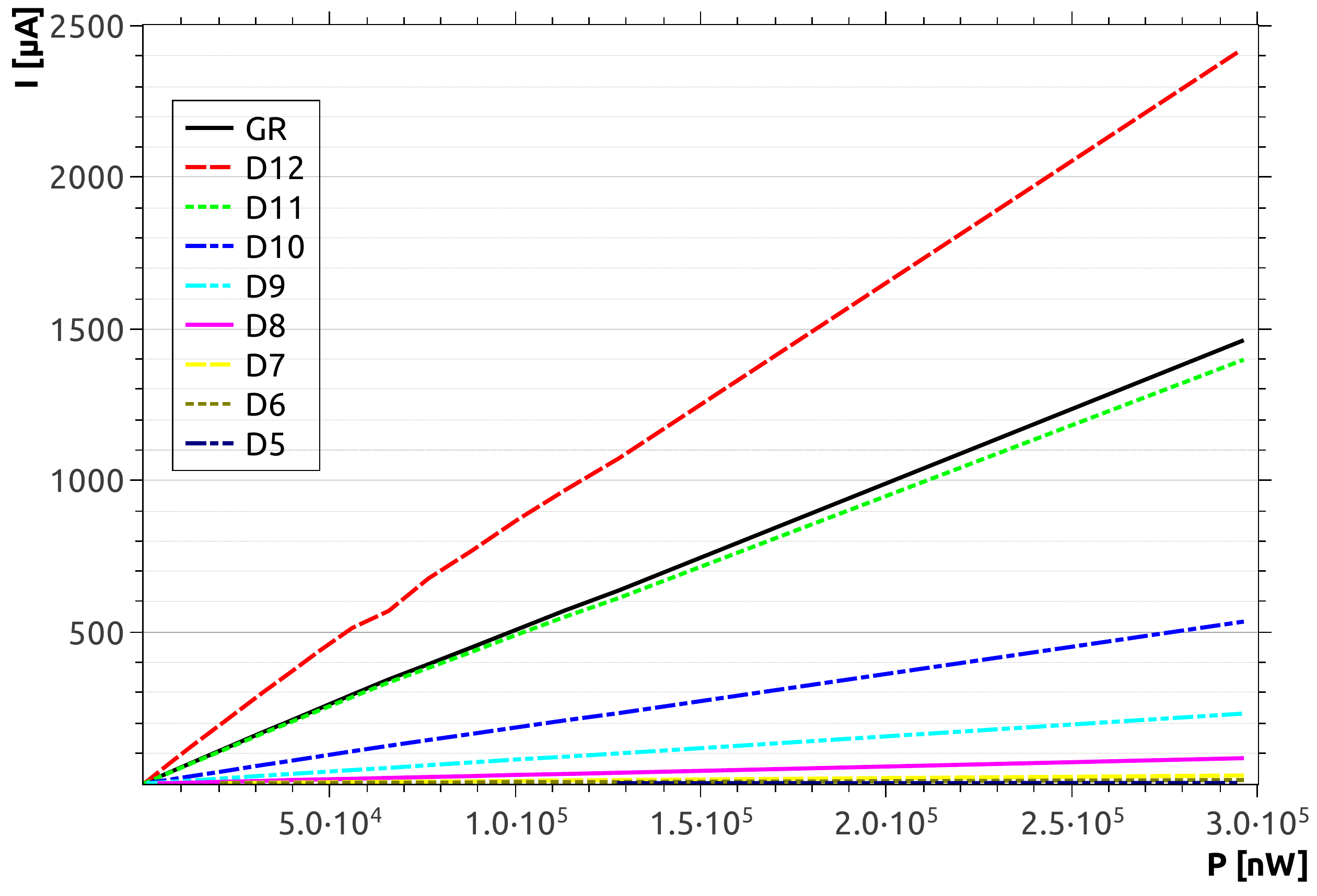}
\caption[Dynode Currents]{Measurement of dynode current versus power incident on the NIST photodiode (i.e., the light level) for the M64 PMT using the ``Low-Power'' voltage distribution.
}
\label{fig:dynodecurrent}
\end{center}
\end{figure}

The current drawn by each element of the PMT during the count rate measurement using the CW repartition is shown in \fig\ref{fig:dynodecurrent}.
The primary conclusion we drew from these measurements is that the guard-ring element present in the M64 behaves differently from the normal focusing electrode.
The guard-ring does provide electrostatic focusing on the anode pixels, much as the more typical grid electrode focuses on the first dynode. However, it can clearly be seen from the current drawn by the guard-ring that it also behaves in some way like a dynode.

\begin{figure}
 \includegraphics[width=1.0\textwidth]{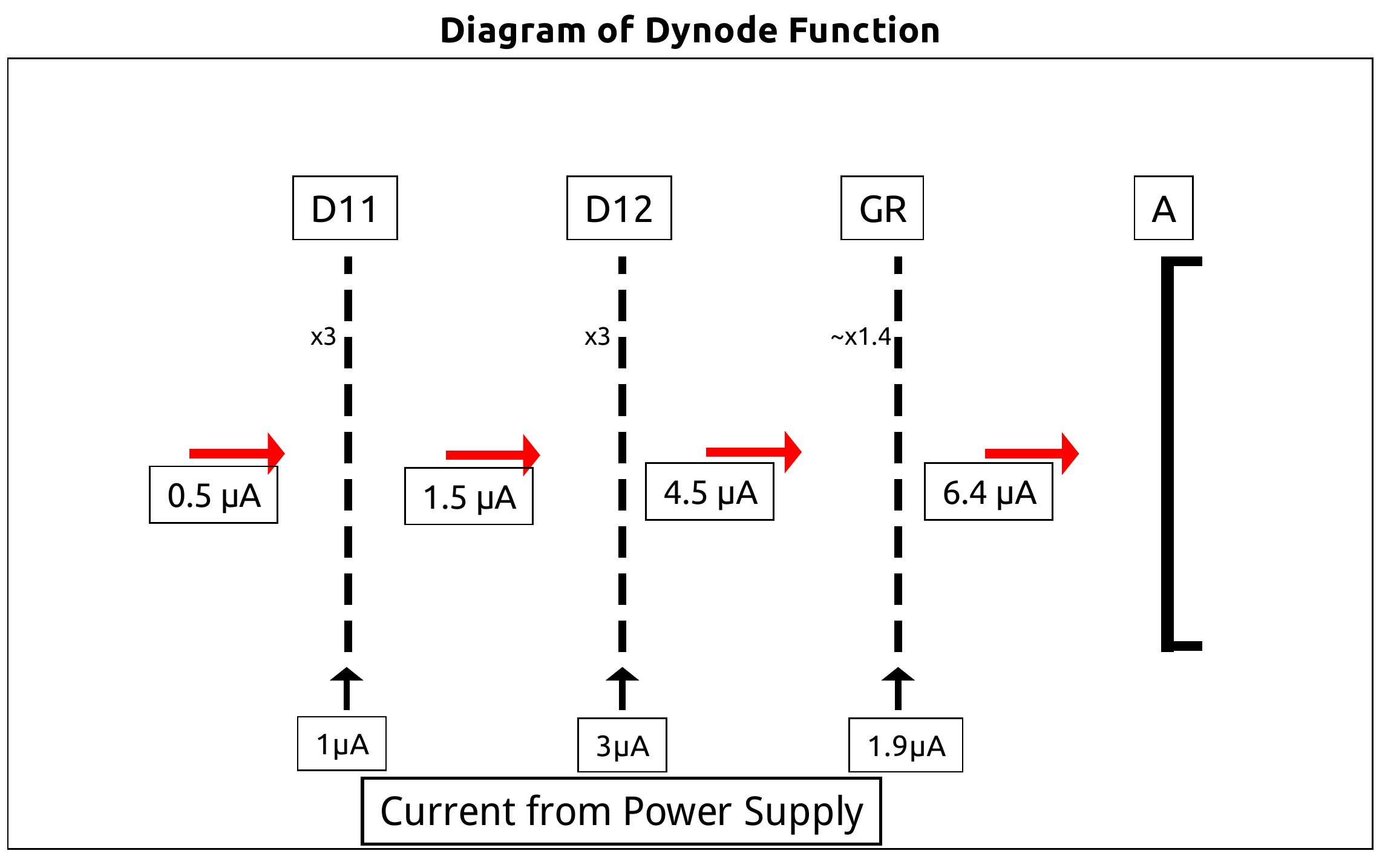}
\caption[Diagram of the Multiplication of Photoelectrons]{Schematic illustrating the multiplication of photo-electrons in the last three stages of the M64. The currents listed at the bottom are the measured currents draw by each element at the JEM-EUSO background 
level (600 kHz/pixel) using the ``low-power'' voltage distribution. The electron currents through the center of the diagram are based on the measured currents supplied by the
power supply.}
\label{fig:multiplicationdiagram}
\end{figure}

This can be understood in the following manner: if each arbitrary dynode has a secondary emission ratio of 3, then 3 electrons exit on average for every electron incident on a dynode. A total of $2/3$ of
the current through that dynode is drawn from the power supply, and $1/3$ of the current is drawn from the preceding dynode. This is illustrated in figure \ref{fig:multiplicationdiagram}. 
It is clear from measurements of $I_{\text{GR}}$ that the guard-ring provides some level of multiplication, therefore acting as a dynode, but at a lesser level than an actual dynode 
(which is natural as the voltage difference between GR and the anode is 1/3 the typical step). 
This is also supported by the fact that the value of $V_{\text{GR}}$ effects the gain of the PMT to the extent that letting $V_{\text{GR}}$ go to either $V_{\text{anode}}$ or $V_{\text{D12}}$ effectively kills the single photoelectron gain of the PMT.

A comparison of the two voltage distributions studied shows that the measured gain and focusing for the M64 are slightly lower in the repartition optimized for the Cockcroft-Walton.
This difference is negligible, however, compared to the variation in gain between PMTs, which can be a factor of 2 or more. In contrast, designing a simple and efficient power-supply is of great importance.
These results where used to determined the actual voltage repartition used in the Cockcroft-Walton high voltage power supply board for EUSO-balloon, which is shown in table~\ref{tab:M64MAPMTvoltageRepartions}.    

\section{Test of the CW-HVPS and Switch Prototype}
\label{sec:Test of the CW-HVPS and Switch Prototype}
\begin{figure}
\centering
\subfigure[The CW-HVPS Prototype]{\label{pic:CWHVPSphotos:CWHVPS} \includegraphics[width=0.7\textwidth]{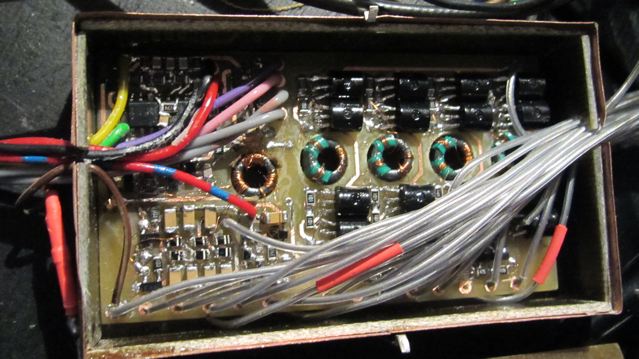}}\\
\subfigure[The Switch Controller]{\label{pic:CWHVPSphotos:Controller} \includegraphics[width=0.7\textwidth]{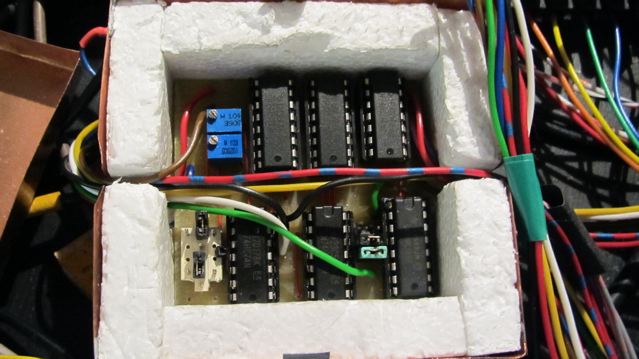}}
\caption[Photographs of the Protoype CW-HVPS]{ Two photographs of the CW-HVPS prototype boards. \fig\ref{pic:CWHVPSphotos:CWHVPS} shows the HVPS prototype board, the 14 gray wires going to the right are the cathode and dynode outputs.
\fig\ref{pic:CWHVPSphotos:Controller} shows the switch controller, which provides the control logic to the switches that are integrated into the CW-HVPS board.
\label{pic:CWHVPSphotos}}
\end{figure}

\begin{figure}[p]
\includegraphics[angle=90,width=1.0\textwidth]{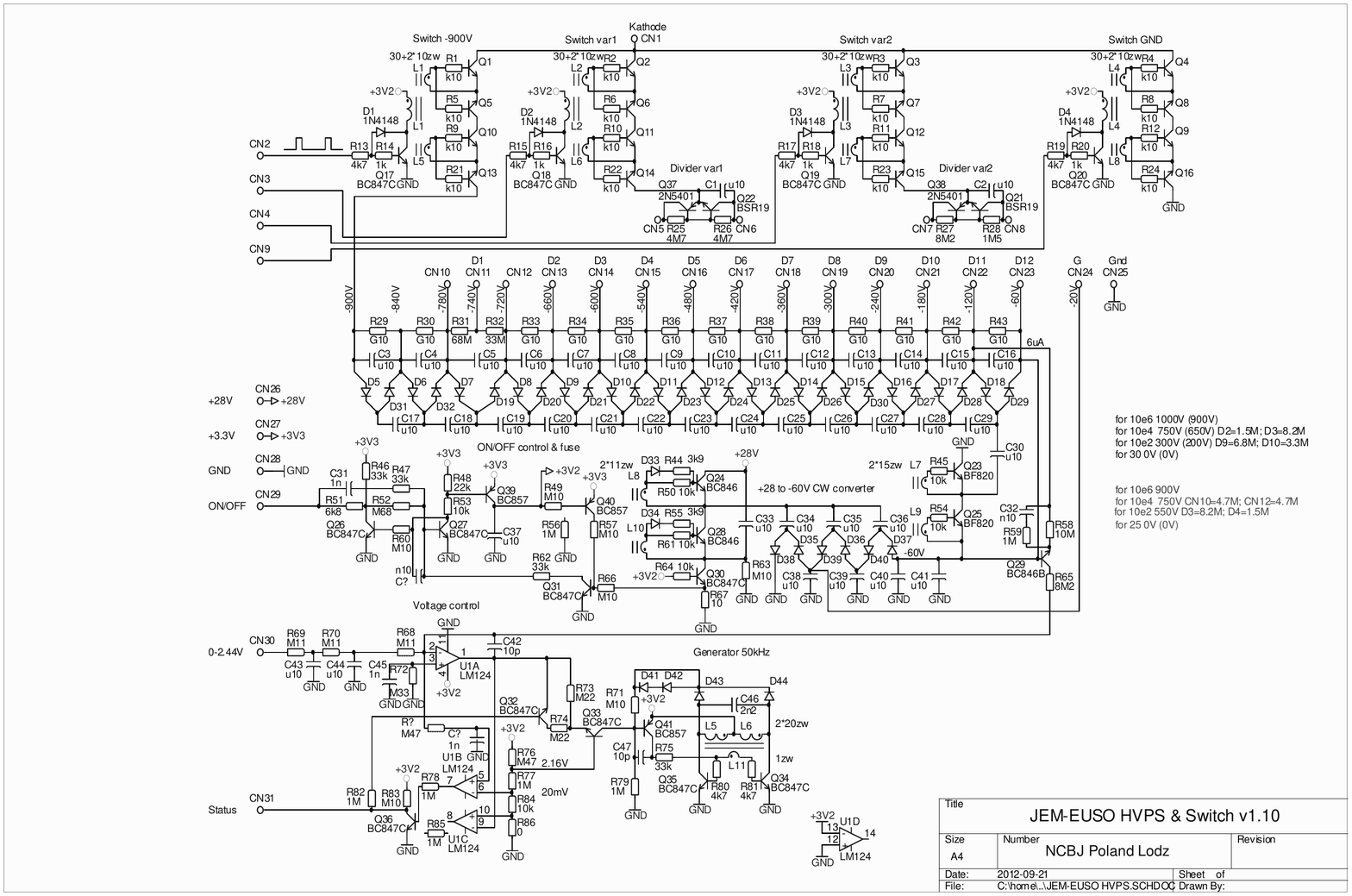}
\caption[Schematic of the EUSO-Balloon CW-HVPS]{\label{fig:CWHVPS-schematic} 
The electrical schematic of CW-HVPS and switch board, as it exists in EUSO-Balloon. The CW-HVPS design and the schematic shown are the work of J. Szabelski et al.~\cite{Adams:2012tt} }
\end{figure}

A prototype of the Cockcroft-Walton high voltage power supply (CW-HVPS) and the switches was later tested, before the printed circuit-boards were to be made for EUSO-Balloon. A picture of the prototype 
board is shown in \fig\ref{pic:CWHVPSphotos}, and photographs of the EUSO-Balloon CW-HVPS are shown in \fig\ref{pic:EUSOBalloonCWHVPS}.
The schematic of the combined CW-HVPS and switch circuit is shown in \fig\ref{fig:CWHVPS-schematic}. The complexities of the design will not be discussed here for brevity.
The CW-HVPS board takes an input of +28 volts, which is used for the multiplication, and +3.3 volts, which is used for the control logic and clock. An output regulation voltage
between 0 and $\simeq2.44~$V can be set on the board by changing a potentiometer. This regulation voltage scales the voltage step size, and thus the overall scale of the CW-HVPS output voltage. 

The switches are integrated directly into the CW-HVPS circuit. In the prototype board the switches where commanded by a second control board, which is shown in \fig\ref{pic:CWHVPSphotos:Controller}. 
The switches allow a change between four voltage levels:
\begin{inparaenum}[i\upshape)]
\item   full cathode voltage, 
\item two different intermediate voltages,  which are each taken from a resistive bridge between any two stages of the Cockcroft-Walton, and
\item the power supply ground.                
\end{inparaenum}
The high voltage must be isolated from the switch control, which allows two possible switch techniques: an optical switch, or a transformer-based switch. 
Optical switching systems are too slow for this application, typically operating in several microseconds, and so a faster transformer switch is used in the CW-HVPS.
In JEM-EUSO, the switching logic is controlled by the light level measured by the PMTs. This is currently done, in EUSO-Balloon and EUSO-TA, with the charge integration read out, but in the future will instead use the dynode 12 current. 

These tests where the first study of the reduction in gain caused by changing only the
cathode voltage of the M64 MAPMT. Changing only the cathode voltage while leaving the dynode voltages unchanged affects primarily the
collection efficiency of the PMT and only slightly reduces the actual gain.
Therefore the current through the PMT is reduced not by reducing the multiplication of photoelectrons, but by reducing the number of photoelectrons which
are collected.

\begin{figure}
\begin{center}
 \includegraphics[width=1.0\textwidth]{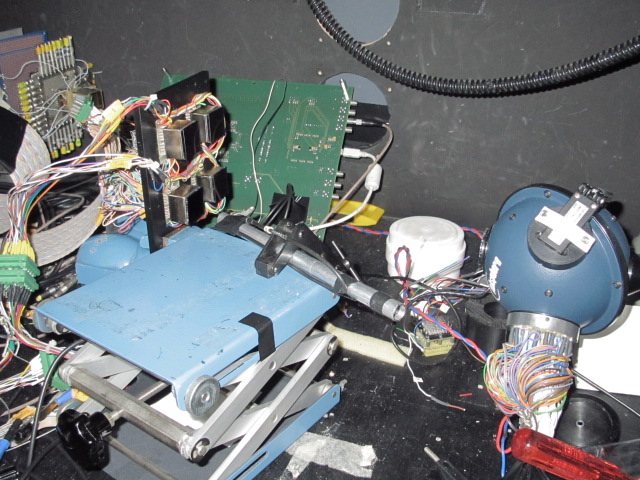}
\caption[A Picture of the Test EC]{ The test EC set up in our black box. The background illumination is provided by the collection of LEDs which are attached to the integrating sphere. The 2nd collimated LED is in the center of the photograph, mounted on the 
blue collapsing stage with a clamp. This LED is aimed at the center of PMT-3, which is in the bottom left corner of the EC. 
The large ribbon cable behind the EC carries the anode signals to the C205 QDC. All the unconnected anodes of the EC are grounded through 50 $\Omega$ resistors.}
\label{pic:testECsetup}
\end{center}
\end{figure}

These measurements were also the first in which the CW-HVPS was tested with a complete elementary cell (EC) of four M64 PMTs.
The four PMTs where each mounted on individual printed circuit boards with 64 anode pins on the backside and lines for the 
14 voltage supplies. Each of these four printed circuit boards where mounted with epoxy onto a large black PVC board with soldered connections in parallel for each of the cathodes, dynodes, and guard-rings. The actual test EC is shown in figure
\ref{pic:testECsetup}. The cables out to the high voltage supply are arranged along the top and left side. All of the anodes which are not used are grounded through a 50 $\Omega$ resistor. In our tests we looked at pixel 21 (Hamamatsu numbering,
as defined in figure \ref{fig:M64pindiagram}) of each of the four PMTs.  

Initially, we used 4 M64 PMTs which we had for testing to build the EC. A large oscillation was found on all four PMT anode signals at a frequency of about 10 Hz. Several days were spent completely rebuilding and testing the EC, 
before it was determined that several of these PMTs had been damaged during previous tests at very high illumination. Due to this, all four of the old PMTs were replaced with ones which where intended for EUSO-balloon, and so all the
PMTs used here include a attached BG3 filter. This was a calculated risk, and so we were careful in our tests, so as to not damage or degrade any of the PMTs by over-illumination.

Once the EC was working, the first thing which was done was to measure the single photoelectron gain of each PMT. 
For this we used a single pulsed 378 nm LED, with the EC at a distance from the integrating sphere exit such that the illumination over 
all 4 PMTs was roughly uniform. We then lowered the LED amplitude so that the number of single photoelectron events was less than 10\% of the number events with zero photoelectrons.
Here the number of two photoelectron events are is no longer negligible compared to the number of 1 photoelectron events. This is dangerous if we were measuring the efficiency. 
However, as long as we are below 10\% the affect on the measurement of the gain is negligible, and we worked here specifically to improve statistics when measuring the 
gain. 

The data acquisition system at the time used CERN prototypes of the C205 CAMAC QDC. These QDC modules measure 32 input charges of up to 900 pC during a gate of between 100 ns and 5~$\mu$s using a charge-to-voltage 
convertor. After conversion, each output voltage is simultaneously amplified by a 1$\times$ and a 7.5$\times$ amplifier, and then digitized into two 12-bit words by two parallel 12-bit ADCs. 
This allows taking measurements on a range of 0-120 pC with 7.5$\times$ amplification and then to directly continue onto a range of 0-900 pC. 
The nominal charge resolution of the C205 
is 33 fC per count for the lower range.
The ability to use a gate width of 2.5~$\mu$s was particularly convenient for tests using higher illumination, as this allows directly measuring the response per JEM-EUSO gate time unit ($1~\text{GTU} = 2.5~\mu$s).
The overall story of the DAQ system is told in section \ref{sec:INTROtoSorting}.

\begin{figure}
 \includegraphics[width=1.0\textwidth]{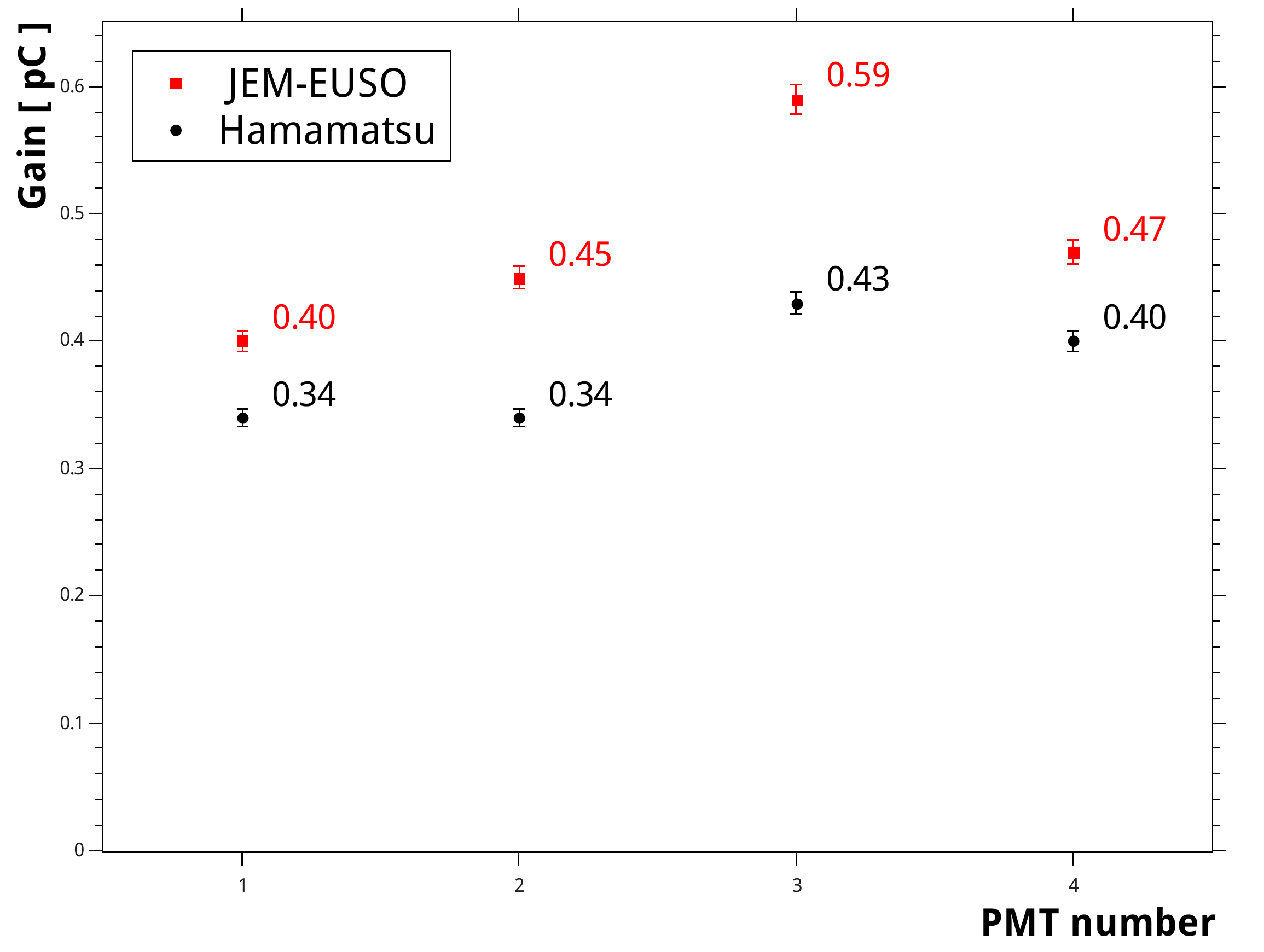}
\caption[Measured Single Photoelectron Gains of the PMT in the EC]{The measured single photoelectron gain for each PMT in the EC. All measurements where taken using the CAEN power supply and with a cathode voltage of 1000 volts. 
The black points are the gains measured using the Hamamatsu voltage distribution, while the 
red points are the gains measured using the voltage distribution of the EUSO-Balloon Cockcroft-Walton power supply.
} 
\label{fig:EC_Gains_VC1000_CAENPS}
\end{figure}

The PMTs were first powered with our high capacity CAEN power supply, using the Hamamatsu specified voltage distribution with a cathode voltage of 1100 volts. The cathode voltage was then reduced (scaling all voltages accordingly) to 1000 volts, 
where all later tests were done. After measuring the gain at 1000 V using the Hamamatsu voltage division, the voltage repartition was changed to that of the Cockcroft-Walton power supply and the gains where again measured at a supply voltage of 1000 V.
Both voltage distributions are shown in table \ref{tab:M64MAPMTvoltageRepartions}. 
The measured single photoelectron gains are plotted in fig \ref{fig:EC_Gains_VC1000_CAENPS}. It can be seen
that the gain is higher for all four PMTs using the voltage distribution of the CW-HVPS in this measurement.   

We then switched from the single pulsed LED to a collection of one hundred 398 nm wavelength LEDs driven by a DC power supply.
This collection directly replaced the single LED at the same port of the sphere. 
The total charge received by each PMT anode was measured using the same C205 QDC. The gate was a NIM pulse with a width of 1 GTU. 
The current drawn by dynodes twelve through nine was measured using the current monitor of the CAEN power supply.    

The pedestal for all four PMTs was measured by taking data with the LEDs off.
The voltage of the LEDs was then adjusted so that the illumination on the EC was approximately the estimated JEM-EUSO background rate of 0.62 MHz per pixel, or 1.6 photoelectrons per GTU.
Previous measurements showed that this corresponded roughly to an illumination, as measured by the NIST photodiode on the sphere, of approximately 0.209~$\mu$W.

\begin{figure}
\includegraphics[width=1.0\textwidth]{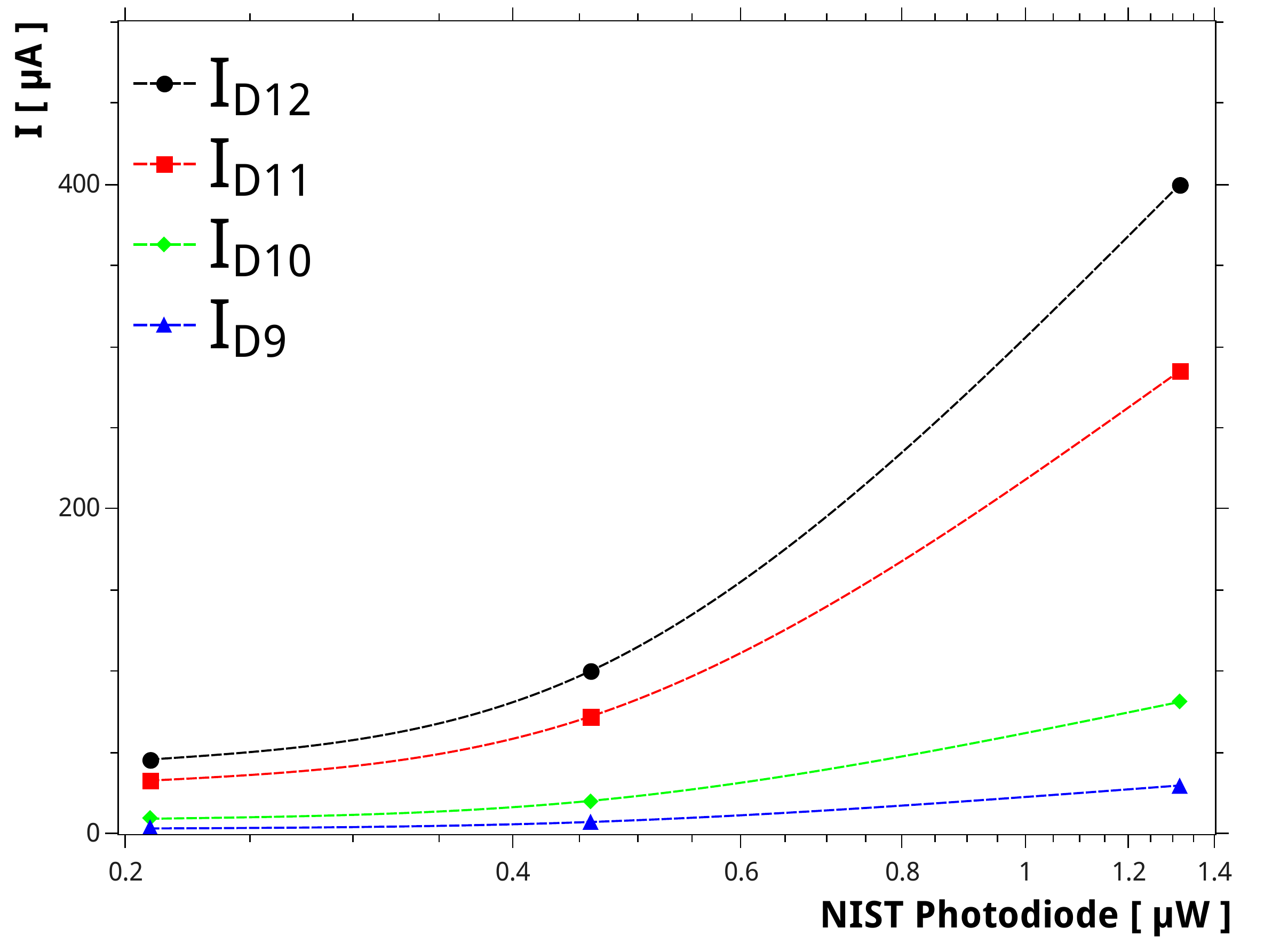}
\caption[Dynode Current Vs. Illumination]{ \label{fig:HVPStest_CAEN_currents} The measured dynode currents as a function of the light level, measured as the power incident on the NIST photodiode mounted on the integrating sphere. Here the 
current is that drawn by the EC of four PMTs, so that the limit of $100~\mu$A per PMT was respected.} 
\end{figure}

\begin{figure}
\includegraphics[width=1.0\textwidth]{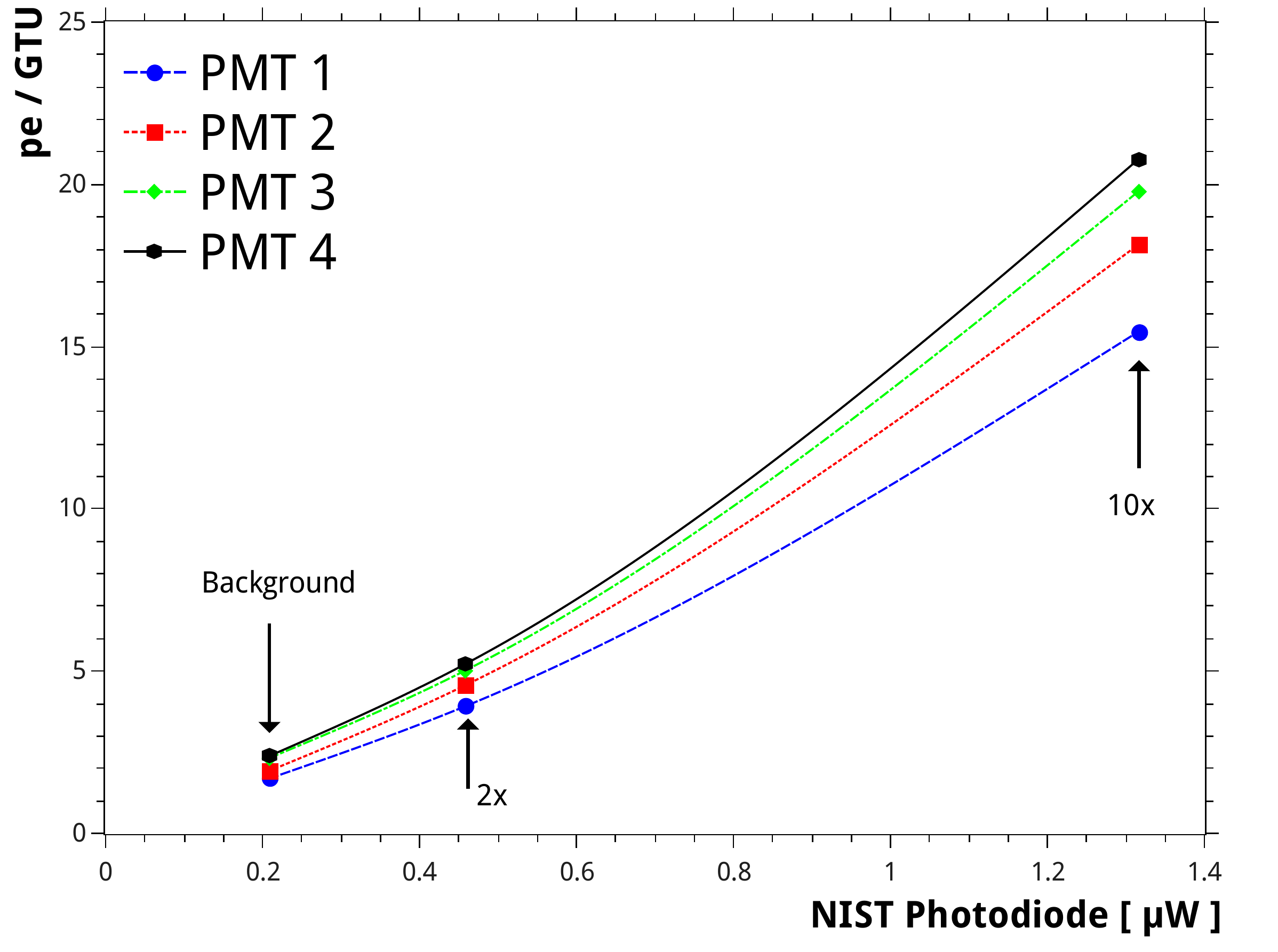}
\caption[Counting Rate Vs. Illumination]{\label{fig:HVPStest_CAEN_pe}  The rate of photoelectrons per $2.5~\mu$s (1 GTU) measured using the QDC. This measurement was taken at the same time as the dynode current measure shown in 
\fig\ref{fig:HVPStest_CAEN_currents}. The incident light corresponds to the JEM-EUSO background rate of $\sim~600~$kHz, twice the background rate, and ten times the background rate. } 
\end{figure}

At this illumination the current on dynode twelve was measured to be 45.5~$\mu$A, or $\sim$12~$\mu$A per PMT. The light was then increased by a factor of 2, and 
then by a overall factor of 100 to verify the response of the EC. 
A plot of the dynode currents drawn by the EC is shown in \fig\ref{fig:HVPStest_CAEN_currents}. 
The number of photoelectrons per GTU measured at the same time are shown for each PMT in \fig\ref{fig:HVPStest_CAEN_pe}. The rate of photoelectrons per GTU was determined by measuring the integrated charge over $2.5~\mu$s, subtracting the 
pedestal charge, and dividing by the measured single photoelectron gain.  As can be seen in the figure, the rate at the background power of $0.209~\mu$W was in fact $\approx2.4~$pe per GTU.
 
We then added a second light source illuminating a few pixels of the third PMT.
This light source was a single collimated LED, aimed at a small area of one PMT. The second LED can be seen in photograph of the EC unit in \fig\ref{pic:testECsetup}. This is meant to mimic high illumination, as in JEM-EUSO the expectation is that
there will be a large signal confined to a small area of a one PMT, while the EC as a whole is illuminated at or near the background level. 

After setting up the second LED, the response at the maximum current of 100~$\mu$A per PMT was tested. 
If the estimated background rate is 0.62 MHz per pixel ($\approx 1.6~$pe/GTU), and the single photoelectron
gain of the pixel is $1~10^{6}= 160~$fC, then the background current on the anodes of one M64 PMT is
\begin{equation}
\left( \frac{0.62~\text{MHz}}{\text{pixel}}\right)\left(  64~\text{pixels}\right)\left(  160~\text{fC}\right) = 6.3~\mu\text{A}
\end{equation}
or $0.1~\mu$A per pixel. If the total current limit on the anodes is $100~\mu$A, then this limit corresponds to 15 times the background. Thought of another way, the background of $6.3~\mu$A leaves up to $93~\mu$A for higher signals. From this, the number of pixels
which can be at 100 times the background at once is then
\begin{equation}
93~\mu\text{A} \left(\frac{\text{pixel}}{0.1~\mu\text{A}}\right)\frac{1}{100} = 9.3~\text{pixels} 
\end{equation}

To account for the actual gain of the PMTs in our test EC we used the measured current on dynode twelve at the nominal background, which was $45.5~\mu$A for the four PMTs together or $\approx12~\mu$A per PMT (our working ``background'' rate was higher than
the JEM-EUSO background by a factor of $\sim2$). The second LED illuminates only one
PMT, and so a maximum of $88~\mu$A is allowed from the extra illumination, or $133~\mu$A maximum on dynode twelve for the entire EC.
The voltage was therefore increased on the second LED until $I_{\text{D12}}$ = 133 $\mu$A. As expected, the number of photoelectrons on PMT
1, 2, and 4 stayed near the background level, while a count rate of 62 pe per GTU was measured on the highly illuminated third PMT. 

We then stopped there, and switched to using the prototype Cockcroft-Walton high voltage power supply.
The voltage scale of the CW-HVPS is determined by a regulator voltage $V_{\text{reg}}$, which in the prototype board is adjusted using a potentiometer.
The CW-HVPS was set up so that both the current drawn by dynode twelve and the regulator voltage could be measured. 
As a starting point, the regulator voltage was adjusted until 67 volts was measured on the dynode twelve output of the CW-HVPS. This corresponded to a regulator voltage of 2.20 volts and should give a cathode voltage of 1000 volts, assuming that 
the voltage repartition is correct.
Using the same DC LED light source and 2.5~$\mu$s gate as before, we returned to the same background conditions by adjusting the light level until there was again a current on dynode twelve of 45.5~$\mu$A. The number of
pe per GTU was consistent with the same measurements taken using the CAEN power supply. 

\begin{figure}[t]
\includegraphics[angle=0,width=1.0\textwidth]{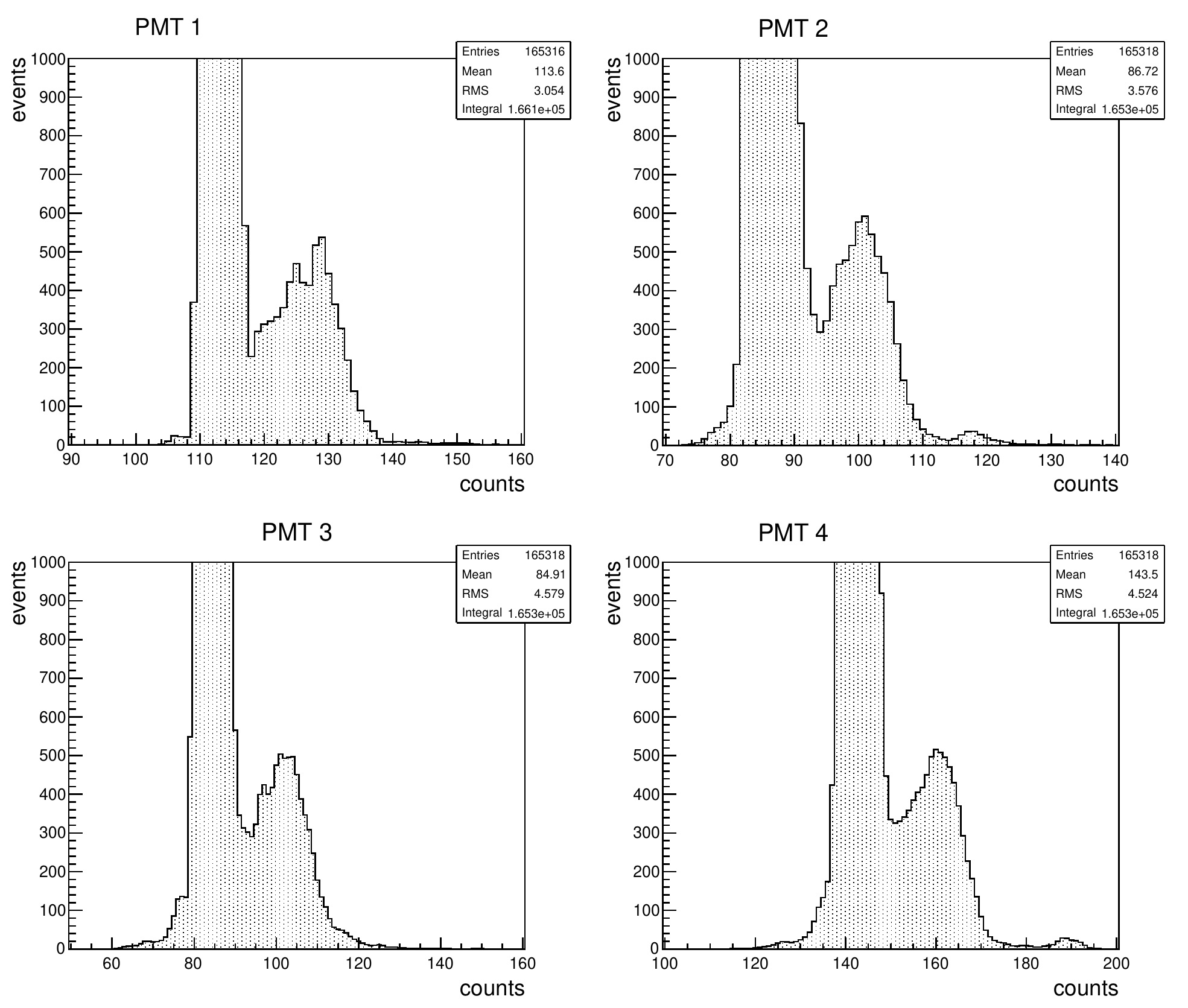}
\caption[Single Photoelectron Spectra of the EC with the CW-HVPS]{ \label{fig:SpeSpectraforCWtests} The measured single photoelectron spectra for the four PMTs in the test EC. These spectra were taken with the EC powered by the CW-HVPS 
at a cathode voltage of 1000 V ($V_{\text{reg}} = 2.20$). }
 \end{figure} 

Having validated that the CW-HVPS functioned in the most basic case, we switched back to the single pulsed LED light source. 
The single photoelectron spectra of all four PMT where checked using a LED pulse width of 20 ns and a gate width of 100 ns. All four spectra are shown in \fig\ref{fig:SpeSpectraforCWtests}. The single photoelectron gains of the measured pixel in
four PMTs were found to 0.42 pC, 0.45 pC, 0.54 pC, and 0.53 pC for PMT-1, PMT-2, PMT-3, and PMT-4 of the test EC.  

We then switched back to DC illumination using the collection of LEDs, and  a few pixels of PMT-3 were again illuminated separately using a second collimated LED. 
With the second LED off, the voltage was set on the background LEDs so that the total current drawn by the EC was 
45.5~$\mu$A, which corresponds to the approximate background rate.
The second LED was then turned on and set so that the total current on dynode twelve was 136~$\mu$A, i.e. $\sim$100~$\mu$A on PMT 3 and $\sim$12~$\mu$A on the other three PMTs. 

\begin{figure}
 \includegraphics[width=1.0\textwidth]{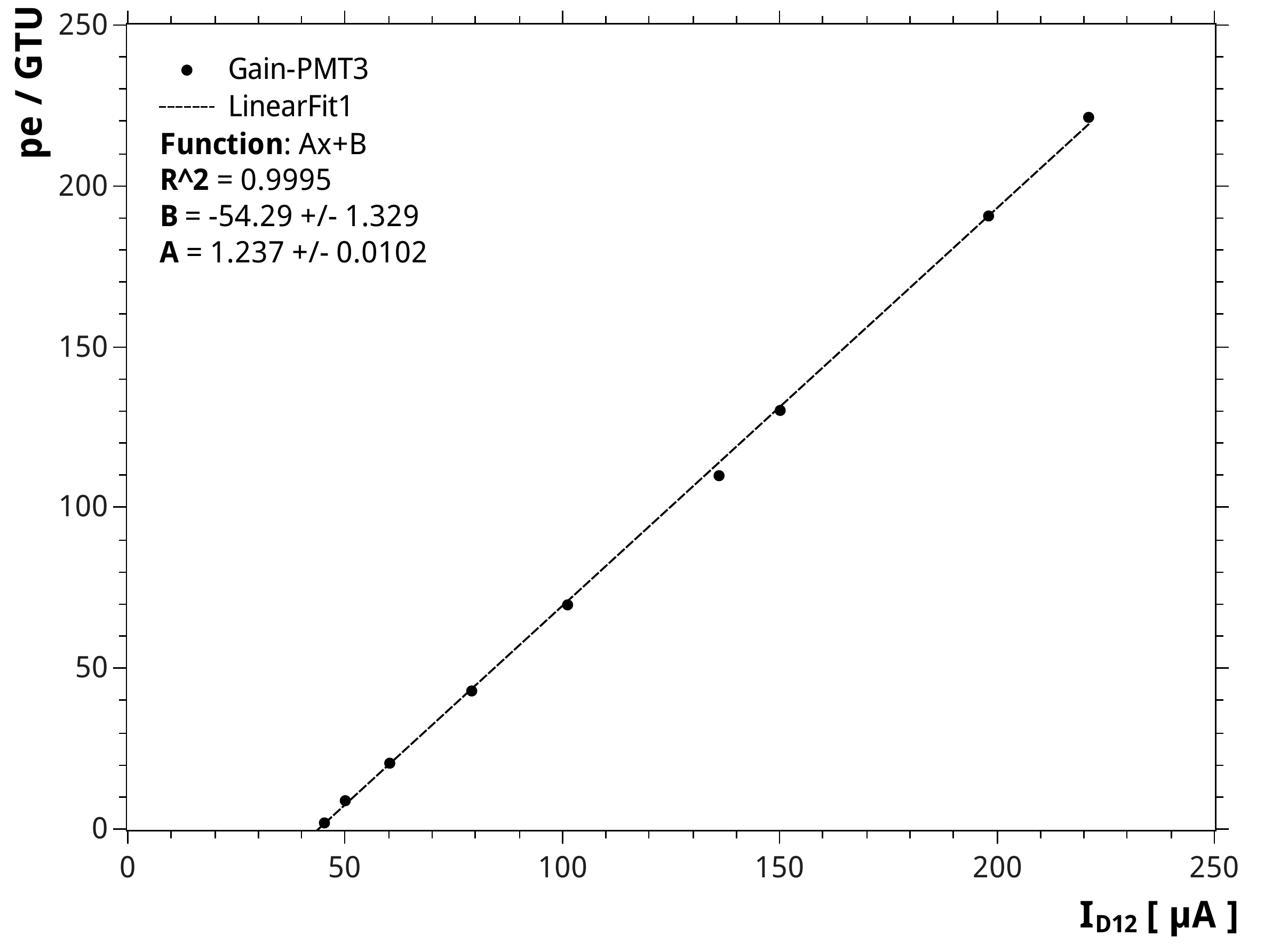}
\caption[Counting Rate Vs. Dynode Current]{The counting rate versus the  measured dynode current with the EC powered by the CW-HVPS. The count rate is given in photoelectrons per $2.5~\mu$s (GTU). The dashed line is a linear least-squares fit 
to the measured data. As can be seen from the fit, the linearity of the current with the counting rate is very good from 1 pe/GTU up to and beyond 200 pe/GTU. This shows the quality of the CW-HVPS.
} 
\label{fig:pePerGTU_CWHVPS}
\end{figure}

The charge measured per $2.5~\mu$s on PMT-1, PMT-2, and PMT-3  was equal to 1.7, 1.9, and 2.2 photoelectrons per GTU, consistent with the background illumination. The charge, in QDC counts, measured per $2.5~\mu$s on PMT-3 was equivalent
to 110 pe per GTU. 
With this as a starting point, the number of photoelectrons per GTU on PMT-3 was measured as a function of $I_{\text{D12}}$. The result of this measurement is shown in \fig\ref{fig:pePerGTU_CWHVPS}, and as can be seen,
the current on dynode twelve is very linear with the photoelectron rate. 

During this measurement the current drawn by the $+28$ and $+3.3$ volt power supplies to the CW-HVPS board were also measured. The $+3.3$ volts is the supply for the 
control of the Cockcroft-Walton circuit and the $+28$ volts provides the power to the PMT. 
With the EC illuminated at the background level, 0.211~$\mu$W on the NIST photodiode, a current of 1.6 mA was measured on the $+28$ volt supply
and a current of 7.6 mA was measured on the 3.3 volt supply.
This gives a power consumption of $0.045 + 0.025 = 70~$mW per EC at twice the JEM-EUSO background illumination.  

The voltage output of the CW-HVPS as a function of the regulator voltage $V_{\text{reg}}$ was also calibrated. 
The cathode voltage from the CW-HVPS was measured as a voltage difference relative to the output of the CAEN high-voltage power supply.
This was done, because of the difficulty of matching the impedance of the CW-HVPS to the laboratory voltmeter. 
In order to work around this, the calibration was performed by setting a voltage on the CAEN HVPS and then adjusting the regulator voltage on the CW-HVPS until the voltage difference was zero. 
The curve was measured starting at $V_{\text{cathode}} \approx 700$, and the two voltages were increased together so that the difference was always within the range of our laboratory voltmeter.
The output voltage of the CW-HVPS at the current value of $V_{\text{reg}}$ was then given by the voltage monitor of the CAEN HVPS.

\begin{figure}
 \includegraphics[width=1.0\textwidth]{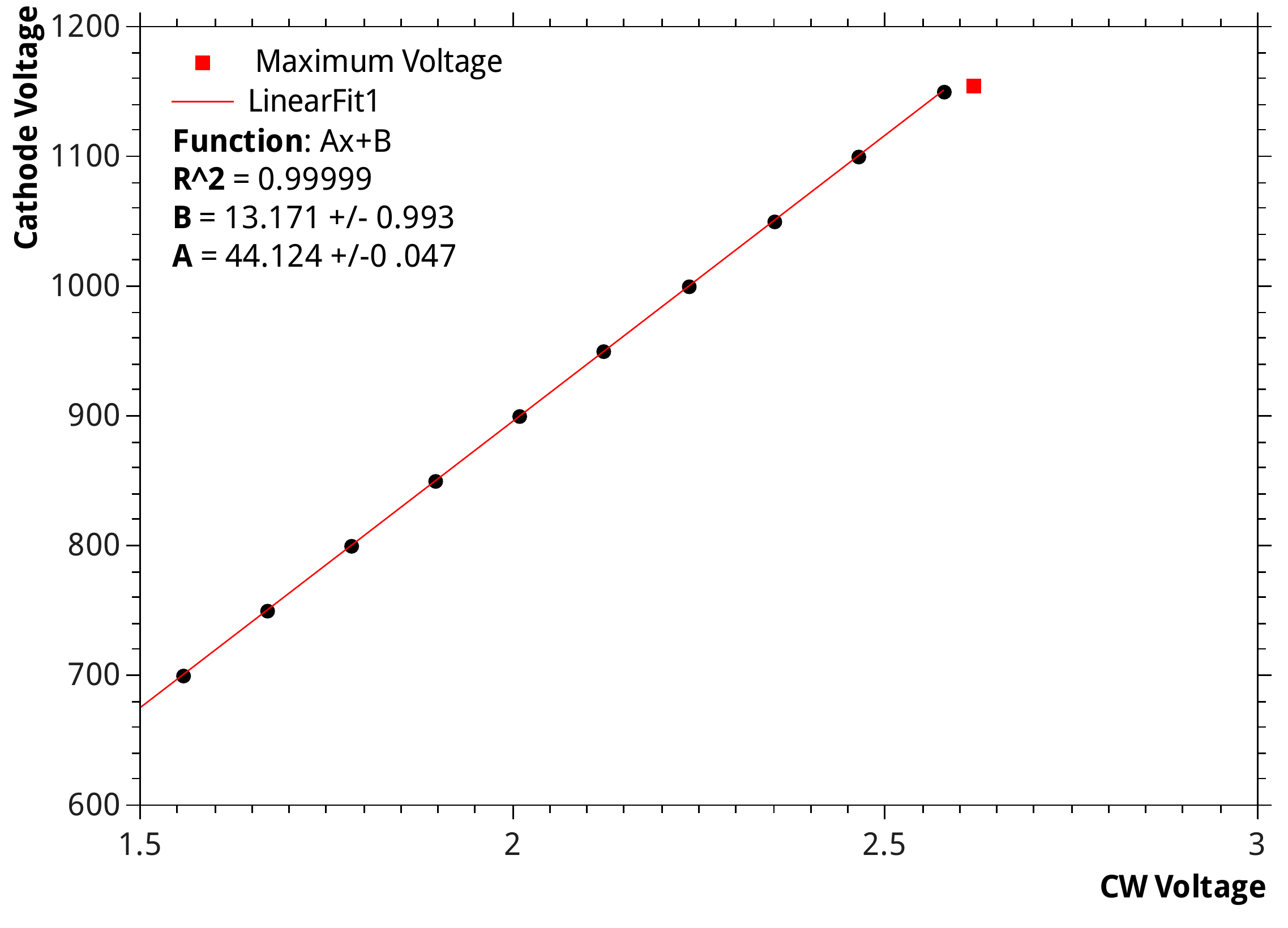}
\caption[Calibration of the CW-HVPS Output Voltage]{The calibration of the output voltage of the CW-HVPS versus its regulator voltage $V_{\text{reg}}$. The output voltage was measured as a voltage difference 
between the CW-HVPS cathode voltage and a voltage supplied by the CAEN HVPS. The maximum voltage output of th CW-HVPS is shown by the red square. Below this maximum voltage the output of the CW-HVPS is linear with $V_{\text{reg}}$.
} 
\label{fig:DirectVoltageCalibration}
\end{figure}

The measured curve is shown in \fig\ref{fig:DirectVoltageCalibration}. The resolution of the CAEN HVPS voltage monitor is 0.5 V, so the cathode voltage at each $V_{\text{reg}}$ was measured 
with an uncertainty of less than 1\%.  The measurement was done up to $V_{\text{reg}} = 2.58$ V, slightly past the maximum designed $V_{\text{reg}}$ of 2.44 V.
The voltage output of the CW-HVPS is completely linear within this range. The minimum measured cathode voltage was 700 V at $V_{\text{reg}} = 1.56$ V and the maximum was 1150 at $V_{\text{reg}} = 2.58$ V.
The red square in the plot shows the point at which the CAEN and CW output could no longer be equalized, here we measured a cathode voltage of 1155 for $V_{\text{reg}} = 2.62$.

\section{Tests of the Switches and ``Gain'' Reduction}

After measuring voltage output of the Cockcroft-Walton HVPS, we began a test of the switches. To do this, the same test-EC was powered using the CW-HVPS with its integrated switches. The regulation
voltage on the CW-HVPS was set to 2.01 V to give a cathode voltage of 900 V.
As before, all four PMTs were illuminated with a constant background illumination, and one area of the PMT3 was additionally illuminated with a second LED.
The control of the switches is by TTL levels and allows applying one of four possible voltages to the photocathode:
\begin{enumerate}[i\upshape)]
 \item The full cathode voltage, depending on $V_{\text{reg}}$, but nominally 900 V (Gain $\simeq1~10^{6}$)
 \item A voltage from a resistive bridge around D1, 739 V in this case (Gain $\simeq1~10^{4}$)
 \item A voltage from a resistive bridge between D8 and D9, here 250 V (Gain $\simeq1~10^{2}$)
 \item 0 V (Gain $\simeq10$)
\end{enumerate}
Each of the resistive bridges has a potentiometer which allows the exact output voltage, and thus the gain reduction, to adjusted.
For this test, the TTL signal was a square pulse such that the switch on the 900 V supply went from ``on'' to ``off'' at intervals of 1.2 seconds, while
the switch on the 739 V went from ``off'' to ``on'' at the same moment. 

\begin{figure}
\begin{center}
\subfigure[900 V to 739 V (200 ms/division)]{\label{fig:SwitchDown900to739:FullView} \includegraphics[width=0.48\textwidth]{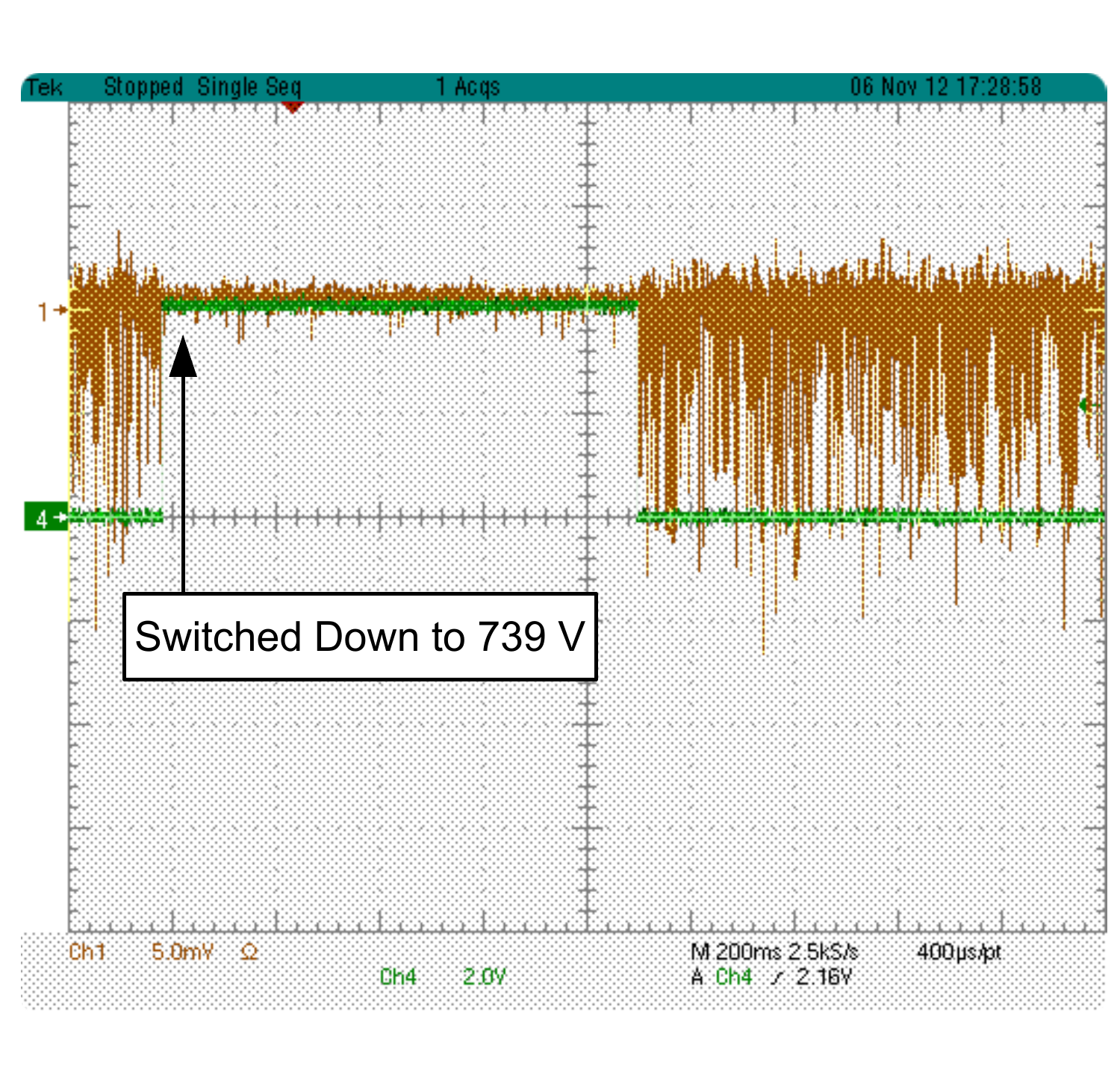}}
\subfigure[900 V to 739 V ($2.0~\mu$s/division)]{\label{fig:SwitchDown900to739:Zoomed} \includegraphics[width=0.48\textwidth]{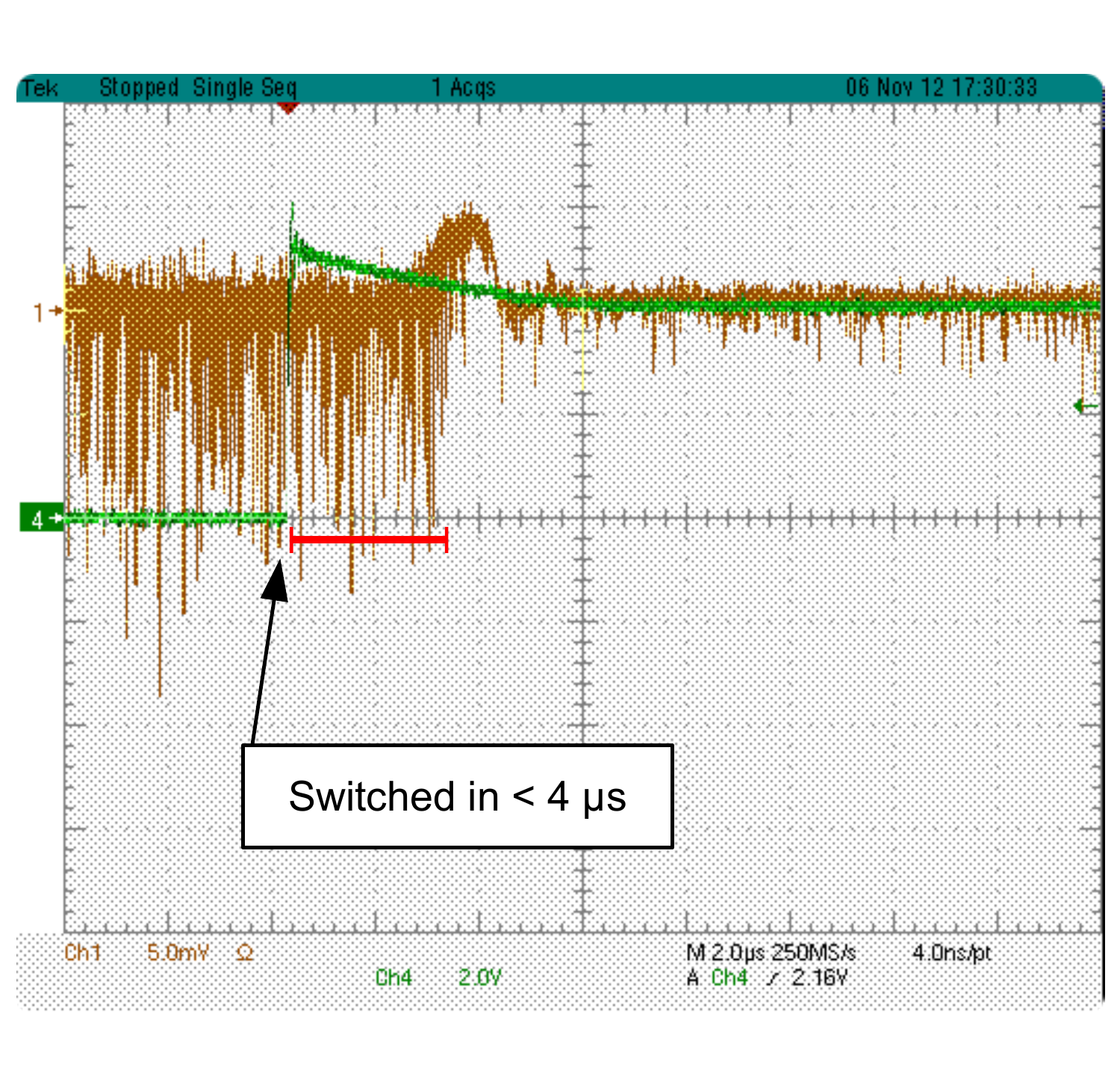}}
\subfigure[739 V to 250 V ($2.0~\mu$s/division)]{\label{fig:SwitchDown739to250}\includegraphics[width=0.5\textwidth]{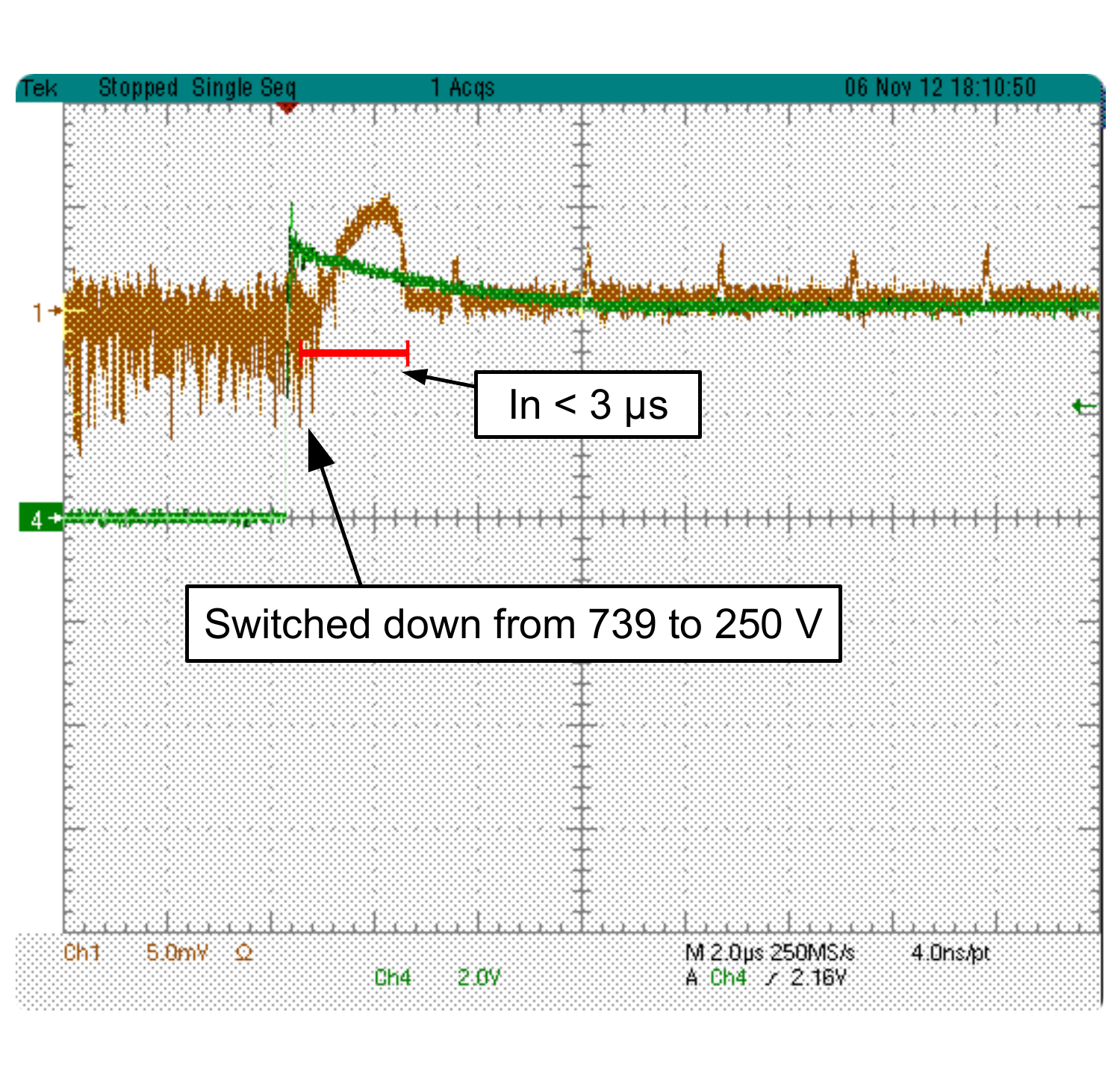}}
\caption[Switching Down from 900 V to 739 V to 250 V]{\label{fig:SwitchDown900to739} Oscilloscope screen captures of the switch operation. The switch command is the green line. 
The switch period is 1.2 seconds, and the high TTL level sets the cathode voltage down to 739 V from 900 V. The anode signal of pixel 21 of PMT-3 of the test EC is shown by the brown line.
The reduction in PMT gain can be clearly seen.
As can be seen in \fig\ref{fig:SwitchDown900to739:Zoomed} the gain is lowered in less than $4~\mu$s. 
\fig\ref{fig:SwitchDown739to250} shows switching from 739 V to 250 V, in which case the gain is lowered in less than $3~\mu$s.
}
\end{center}
\end{figure}

The timing and function of the switches was checked by looking at the anode signal of pixel 21 of PMT-3 on an oscilloscope. The first test was switching the cathode voltage down from 900 V to 739 V. 
The oscilloscope screen is shown in \fig\ref{fig:SwitchDown900to739}. The green line is the TTL command to the switches, and the golden-brown line is the anode signal from PMT-3. 
The switch functions properly and the reduction in cathode voltage also reduces the current on the anode, as can be seen in
\fig\ref{fig:SwitchDown900to739:FullView}. The same oscilloscope view with a time division of $2.0~\mu$s is shown in \fig\ref{fig:SwitchDown900to739:Zoomed}. The reduction in PMT gain occurs in less than $4~\mu$s.
The same result can be seen in \fig\ref{fig:SwitchDown739to250} for switching the cathode voltage from 739 V to 250 V, and in this case the gain is reduced in less than $3~\mu$s. 

\begin{figure}
\begin{center}
\subfigure[From 250 to 739 V (10 ms/division)]{\label{fig:SwitchUpIntoBigLight:250to739} \includegraphics[width=0.48\textwidth]{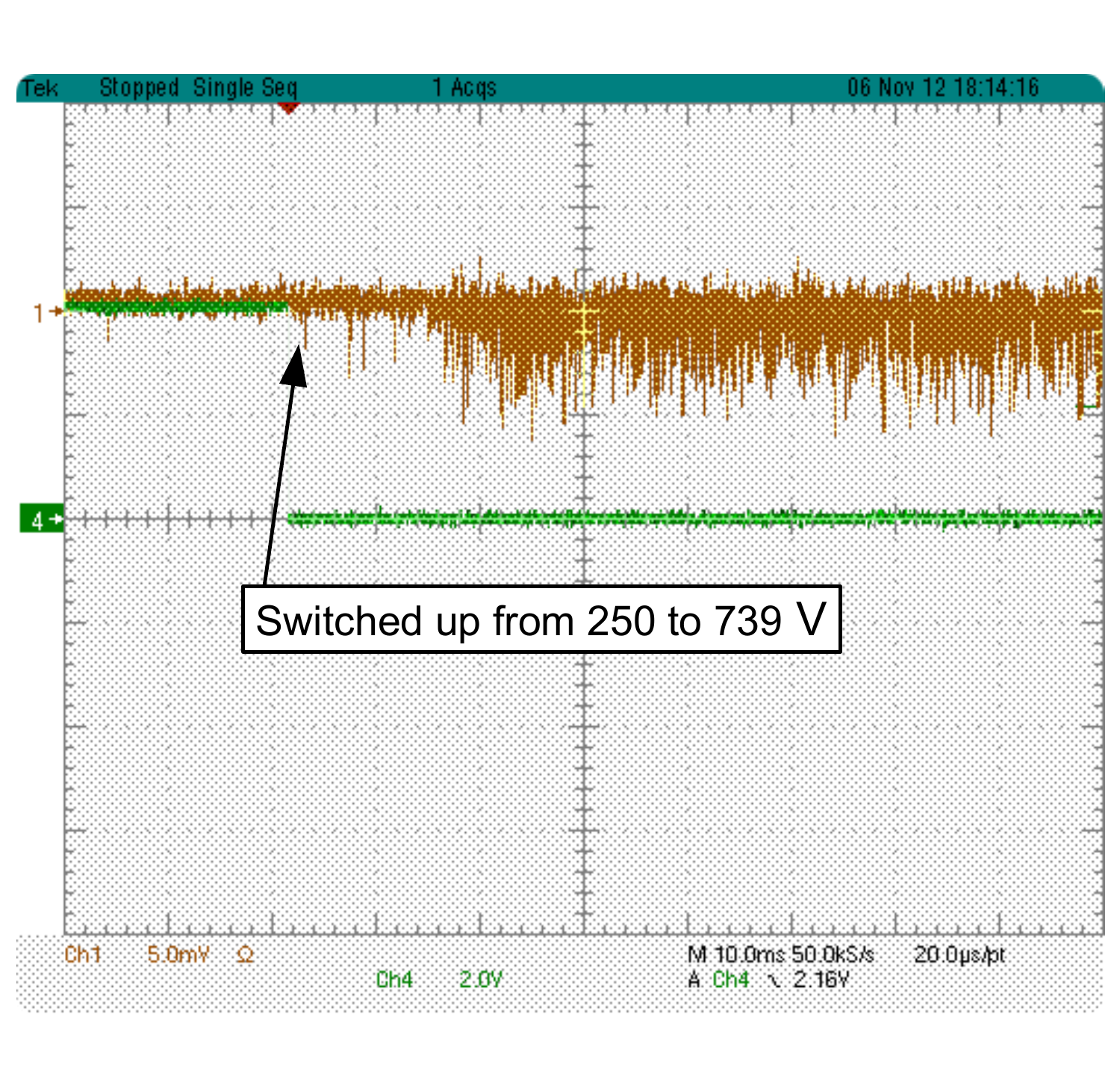}}
\subfigure[From 739 to 900 V ($200~\mu$s/division)]{\label{fig:SwitchUpIntoBigLight:739to900} \includegraphics[width=0.48\textwidth]{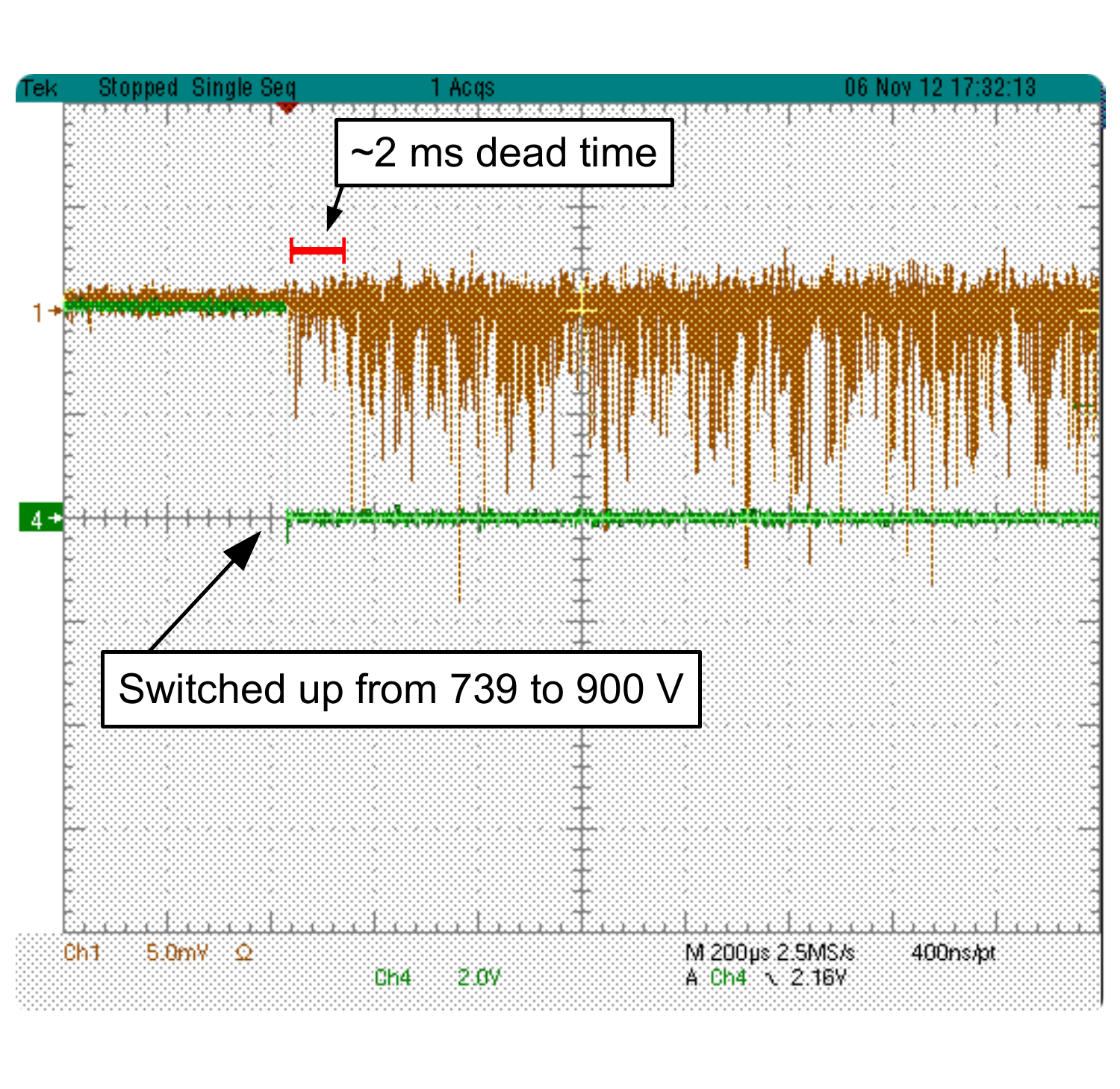}}
\caption[Switching Up]{\label{fig:SwitchUpIntoBigLight} Oscilloscope screen captures of the switch operation (as in \fig\ref{fig:SwitchDown900to739}), here showing the switching down. 
The return to high gain at a cathode voltage of 900 V is shown in \fig\ref{fig:SwitchUpIntoBigLight:739to900}. In this case the gain recovery takes $\sim2~$ms. This is only when switching into a load, however, as explained in the text.
The actual expected recovery time when switching back up after a high illumination has decreased to $\simeq3~\mu$s, as shown in \fig\ref{fig:SwitchUBckOnly:739to900}}
\end{center}
\end{figure}

\begin{figure}
\begin{center}
\includegraphics[width=1.0\textwidth]{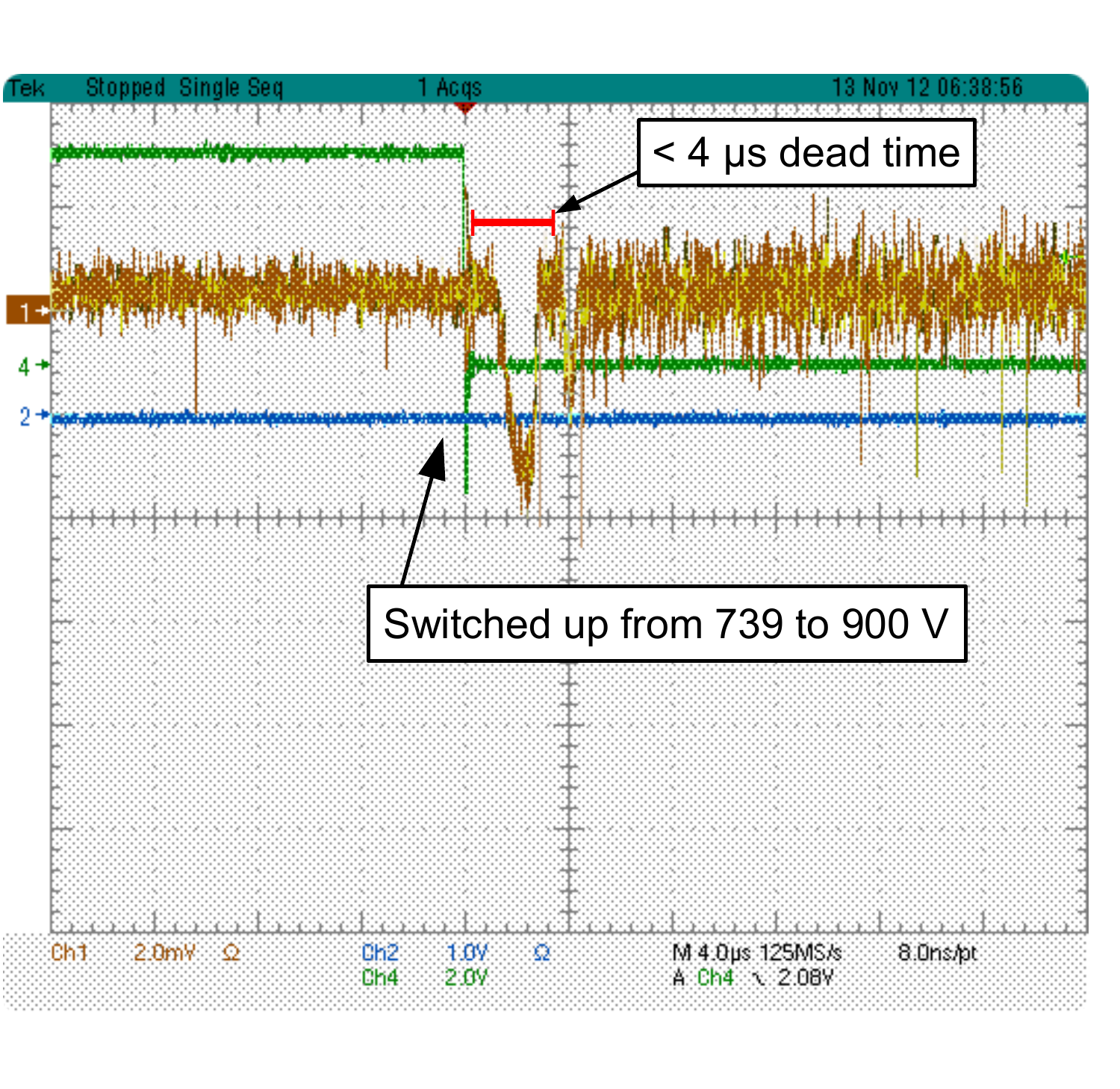}
\caption[Switching Up with No Load]{\label{fig:SwitchUBckOnly:739to900} Switching up from 739 V to 900 V. Here the second LED is synchronized with the switch command so that the cathode voltage is switched when the LED is off. The background
LEDs are always on. In this case, the CW-HVPS
is not switching into a high load, and so the gain recovers in less than $4~\mu$s. This situation more closely resembles the actual operation of the switches, as opposed to switching up while the high illumination is still present.}
\end{center}
\end{figure}

Switching up \emph{appeared} to be rather slow, as is shown in \fig\ref{fig:SwitchUpIntoBigLight}.  Here switching from 739 V on the photocathode to 900 V required $\simeq2$ ms. We realized, however, that this could be due to the 
fact that the CW-HVPS is switching into the load created by the high illumination. To test this, the second LED was synchronized with the switch command, so that the LED was on during the switch down to 739 and then powered off before the switch back up to 900 V.
This is a better approximation of what would occur in JEM-EUSO, where the cathode voltage would be switched down in response to a large signal in a few pixels, and then switched up as the signal settles back to the background level.
The function of the switch with this setup is shown in \fig\ref{fig:SwitchUBckOnly:739to900}. Here the gain recovers in less than $4~\mu$s.

\subsection{``Gain'' Reduction}
After testing the basic function of the integrated switches, the reduction in effective gain caused by lowering the cathode voltage was studied.
These results were used to tune the intermediate voltage levels of the four switches so that overall reduction in PMT gain was a factor of $\sim10^{6}$.

For the first measurement, the test EC was powered by the CAEN HVPS using the voltage distribution of the CW-HVPS at 900 V. 
Using the CAEN HVPS allowed us to set the cathode at an arbitrary voltage.
The pedestals of all four PMT in the EC where taken, and then the EC was again illuminated at the background level using the collection of LEDs.
A few pixels of PMT-3 were again illuminated additionally using a second LED.

The charge received on the anode of pixel 21 of PMT-3 during a gate of $2.5~\mu$s (1 GTU), minus the pedestal charge, was then measured as a function of cathode voltage. 
This charge is proportional to the PMT gain, that is the product of the single photoelectron gain and the collection efficiency.
The illumination was increased several times as the cathode voltage was reduced, so that the anode signal could still be seen in the QDC as the PMT gain decreased. Each time 
the light level was changed the response was remeasured at the same voltage in order to account for the change in the number of photons incident on the pixel per GTU.

\begin{figure}[ht]
\begin{center}
\includegraphics[width=1.0\textwidth]{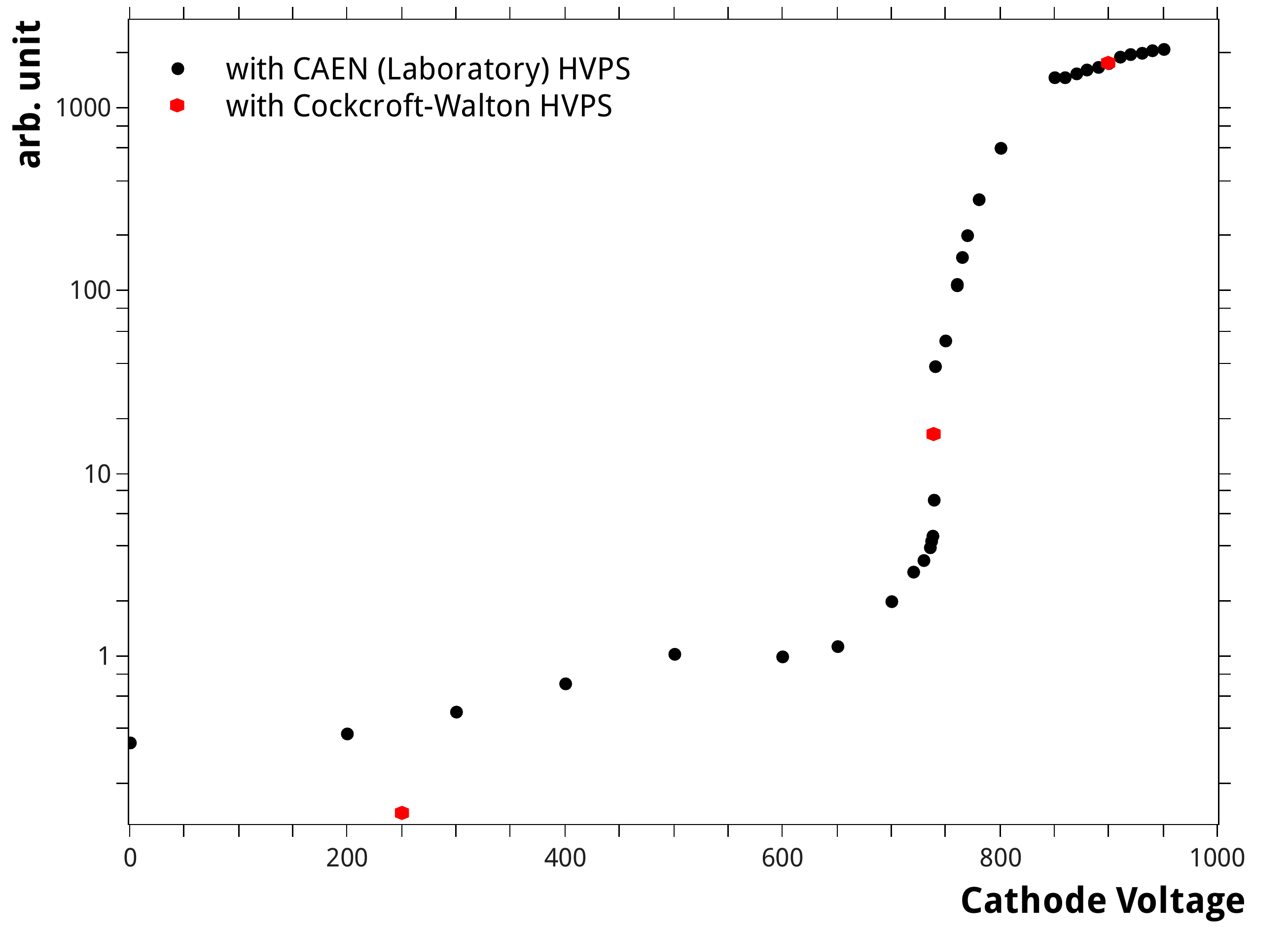}
\caption[Gain Reduction by Switching Cathode Voltage]{\label{fig:GainReductionCurve} A plot of the reduction in PMT gain caused by changing only the cathode voltage. The cathode voltage is shown in abscissa, and the 
number of QDC counts minus the pedestal is shown in ordinates. The response in the QDC is proportional to the product of the single photoelectron gain and the number of photoelectrons, and thus to the PMT gain at a constant illumination.
The full curve is taken using the CAEN HVPS, while the red hexagons show points measured using the CW-HVPS. The difference between the two measurements is probably due to the high impedance of the CAEN-HVPS.}
\end{center}
\end{figure}

The response was measured for a cathode voltage between 900 V and 0 V (cathode grounded), and is shown in \fig\ref{fig:GainReductionCurve}. 
Below 200 V the CAEN HVPS has difficulty polarizing properly, and so no measurements were made between 200 V and 0 V.
The dependence of the PMT gain on the cathode voltage is rapid in the range between 700 and 800 volts, and this makes it difficult to adjust the switched cathode voltage for an exact reduction of a factor of 100.
The dynamic range shown by the plot is a factor of $5.2~10^{3}$ between 900 and 0 V. 
The dynamic range at a given gain is factor of $\sim 200$, as shown by \fig\ref{fig:pePerGTU_CWHVPS}, for a total dynamic
range of more than $1~10^{6}$.

The same measurement was then done using the CW-HVPS itself. In this case the response was measured in the same manner as before, now at a cathode voltage given by the four switch levels: 
 900 V, 739 V, 250 V and 0 V.
These measurements are shown on \fig\ref{fig:GainReductionCurve} by
the red hexagons. The difference in illumination was corrected by normalizing the measured charge at 900 V to the previous value measured using the CAEN HVPS. This correction was less than 2\%. 
The reduction in PMT gain when using the CW-HVPS is similar to that seen when using the CAEN HVPS, but not exactly the same, and the difference is much greater at lower voltages. 
This is most likely due to the trouble which the CAEN HVPS has polarizing at lower voltages.
With the CW-HVPS, the gain reduction was measured to be factor of 107 between 900 V and 739 V, a factor of 118 between 739 V and 250 V, and a further factor of 1.25 between 250 V and 0 V. 
This gives a range in gain of $1.6~10^{4}$, for a total dynamic range of $\simeq3~10^{6}$.

\subsection{Uniformity of the Gain Reduction}

Just after testing the switches, an update to the data acquisition system was finished which allowed us for the first time to read out all 64 pixels of one M64 PMT. 
This was immediately put to use to study the uniformity of the gain reduction across the surface of the photocathode. 
Changing the photocathode voltage relative to the voltage of the first dynode reduces the total gain of the PMT by destroying the collection efficiency. 
The change in the collection efficiency is a function of the electrostatics between the photocathode and the first stage, which are not necessarily the same throughout the PMT. This is 
especially true in a rectangular design such as the M64, and because of this it is likely that the reduction in gain will not be equal across pixels.
This uniformity was studied using the same setup as for study of the gain reduction, except that every pixel of PMT-3 of the test EC was now read out by the QDC with an integration gate of $2.5~\mu$s.

\begin{figure}
\begin{center}
\subfigure[2nd LED On]{\label{fig:BigLightOnSomePixels:FullV} \includegraphics[width=0.48\textwidth]{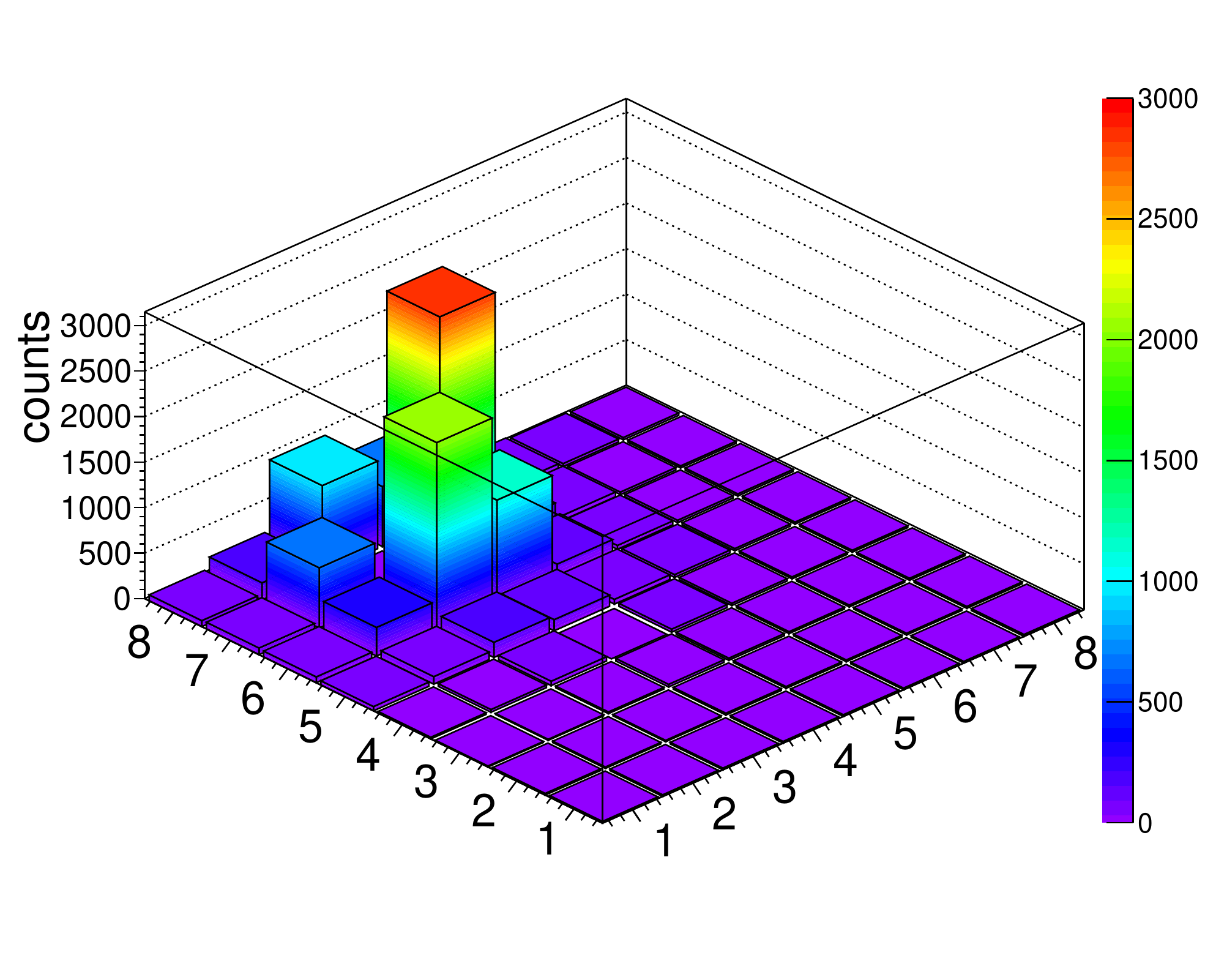}}
\subfigure[2nd LED Off]{\label{fig:BigLightOnSomePixels:LEDoff} \includegraphics[width=0.48\textwidth]{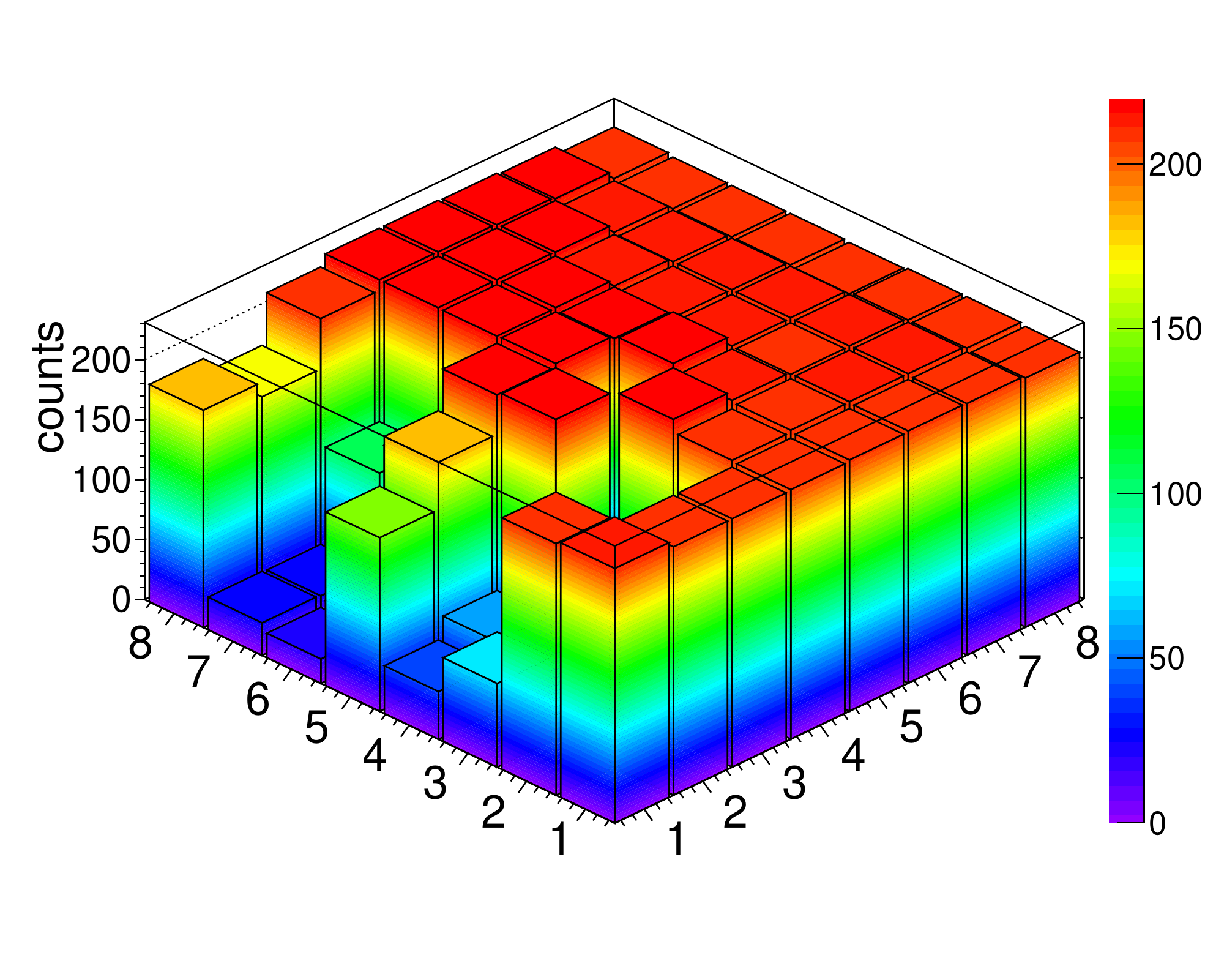}}
\caption[Full Visualization of the Illumination on One PMT]{\label{fig:BigLightOnSomePixels} A histogram of the mean number of QDC counts returned for each of the 64 pixels of PMT-3 of the test EC. 
Shown in \fig\ref{fig:BigLightOnSomePixels:FullV} is the response with the PMT illuminated by the both the background LEDs and the second collimated LED (as discussed in the text). The high illumination is confined to only a few pixels.
\fig\ref{fig:BigLightOnSomePixels:LEDoff} shows the response with only the background LEDs, and the shadow from the collimator of the second LED can clearly be seen.}
\end{center}
\end{figure}

The pedestal of every channel was taken with the LEDs off in a first run.
We then took one run with the second collimated LED illuminating its area of the PMT, just to see way that the second LED had been illuminating the PMT in the previous measurements. 
A map of the mean counts returned by the QDC for each pixel with the 2nd LED on is shown in \fig\ref{fig:BigLightOnSomePixels:FullV}.
The high illumination in a few pixels can be clearly seen. Another run was then taken with the second LED off, and the background LEDs on,
shown in \fig\ref{fig:BigLightOnSomePixels:LEDoff}. Here is was clear that the second LED, with its collimator, was casting a shadow on the PMT.
In this run we also found that pixel 28 was not connected, and it was removed from all the further analysis.

The collimator was moved out of the way before the next measurements, in which a run was taken with an illumination of $\sim 15$ times the background. 
For each of the 64 pixels, the pedestal was subtracted to give the charge $q_{i}$ (in QDC counts) per GTU in each pixel $i$. 
The average $\bar{q}$ over all 64 pixels was found, and an equalization factor $\sigma_{i}$  was calculated for each pixel, 
so that $q_{i}\sigma_{i} = \bar{q}$. We then looked to see if each $\sigma_{i}$ was constant while switching the cathode voltage, that is to say if the gain of each pixel relative to the others remained the same at each voltage. 
After equalizing all the pixels, we took one run each with the cathode switched down to 739 V, 250 V, and 0 V. 

\begin{figure}
\begin{center}
\subfigure[Equalized at 900 V]{\label{fig:15xBck:Equalized} \includegraphics[width=0.48\textwidth]{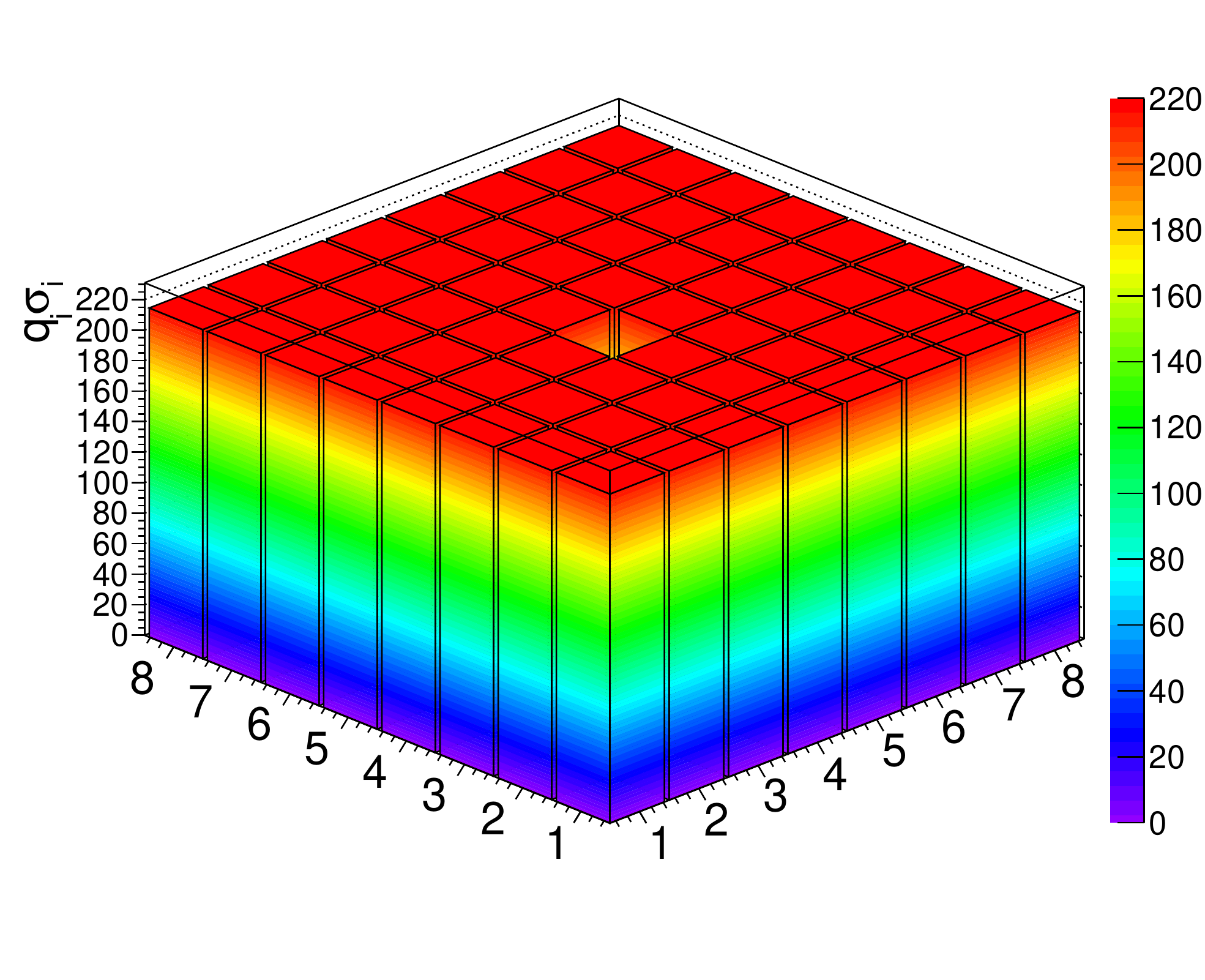}}
\subfigure[Switched to 739 V]{\label{fig:15xBck:D1V} \includegraphics[width=0.48\textwidth]{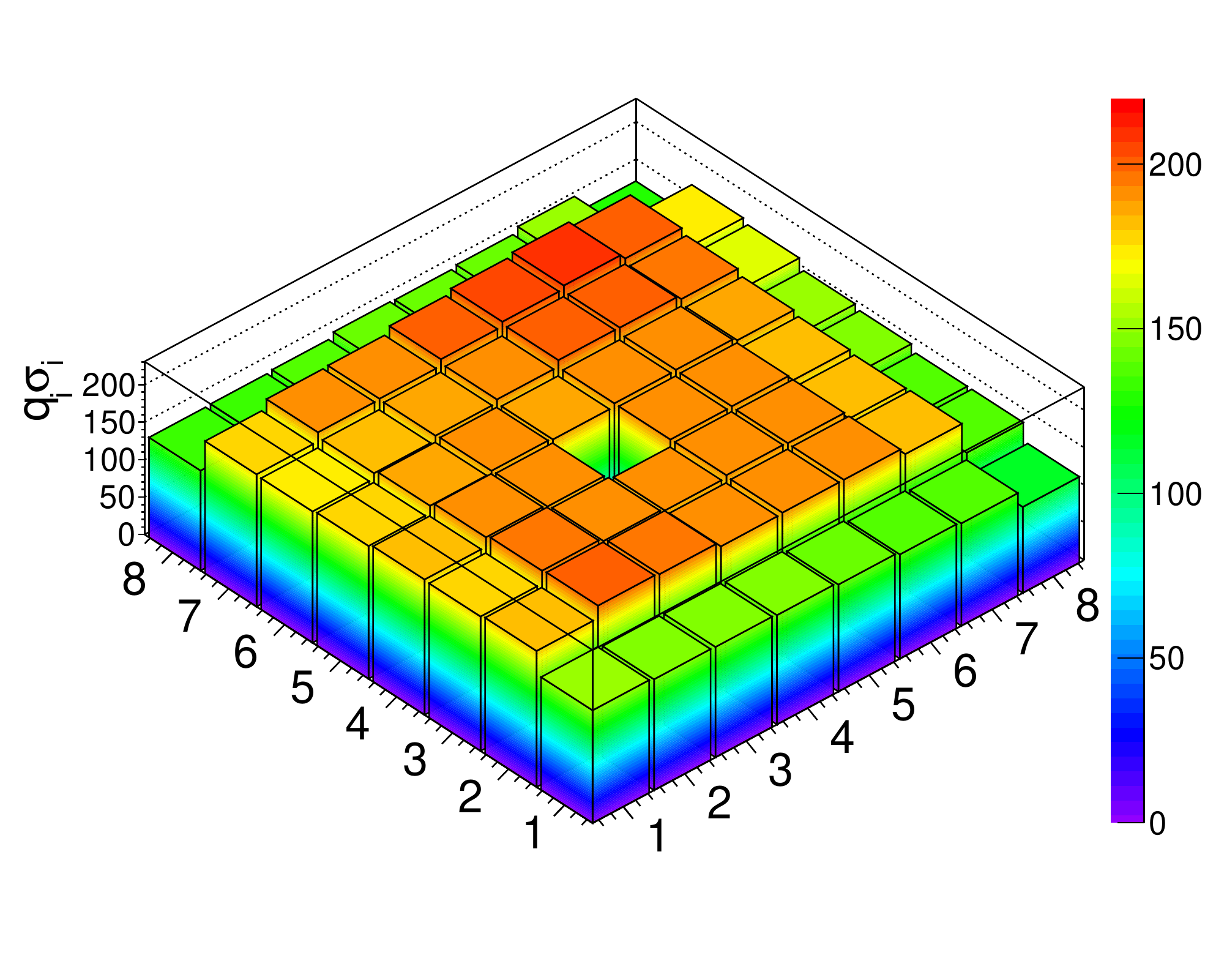}}
\subfigure[Switched to 250 V]{\label{fig:15xBck:D9V} \includegraphics[width=0.48\textwidth]{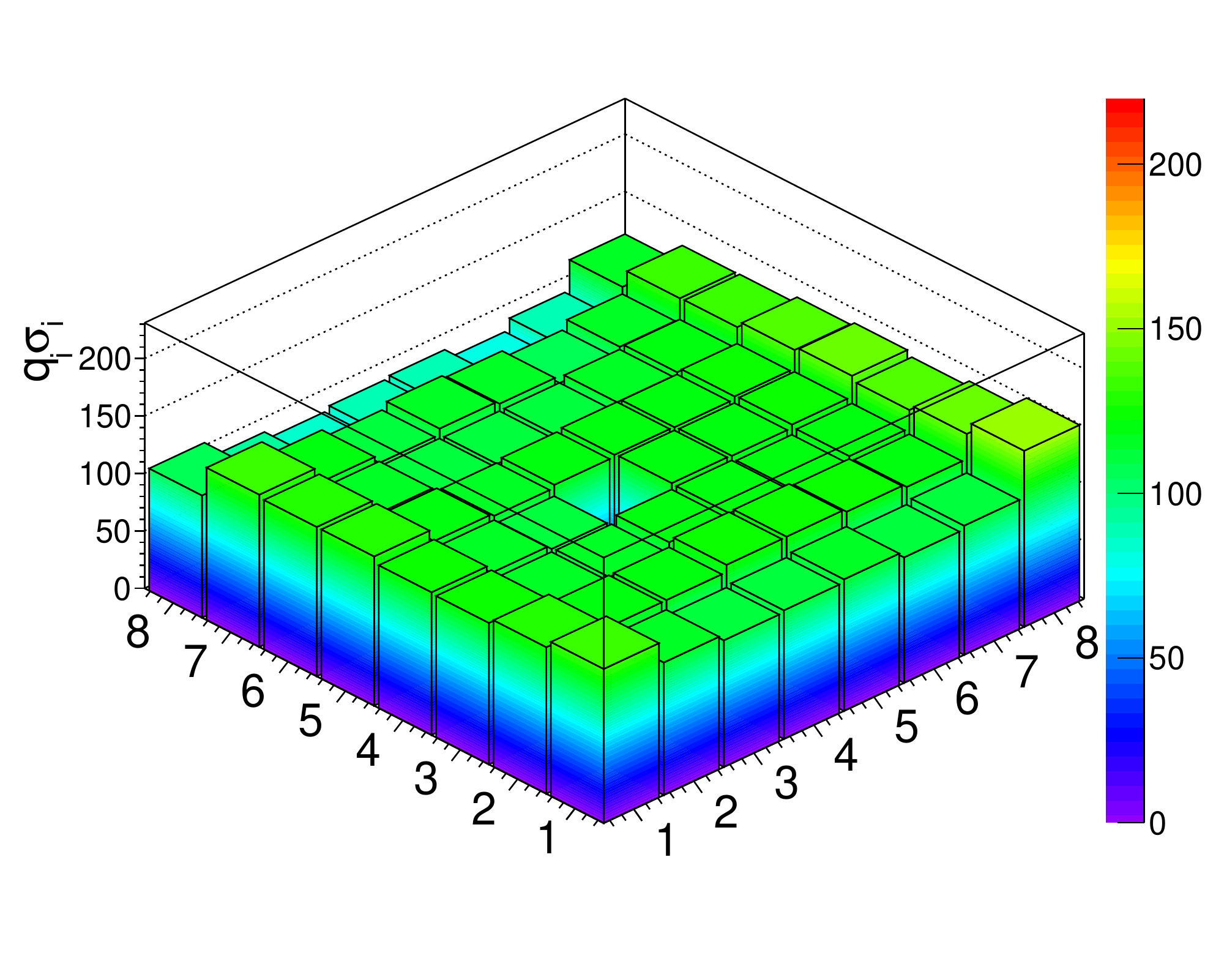}}
\subfigure[Switched to 0 V]{\label{fig:15xBck:0V} \includegraphics[width=0.48\textwidth]{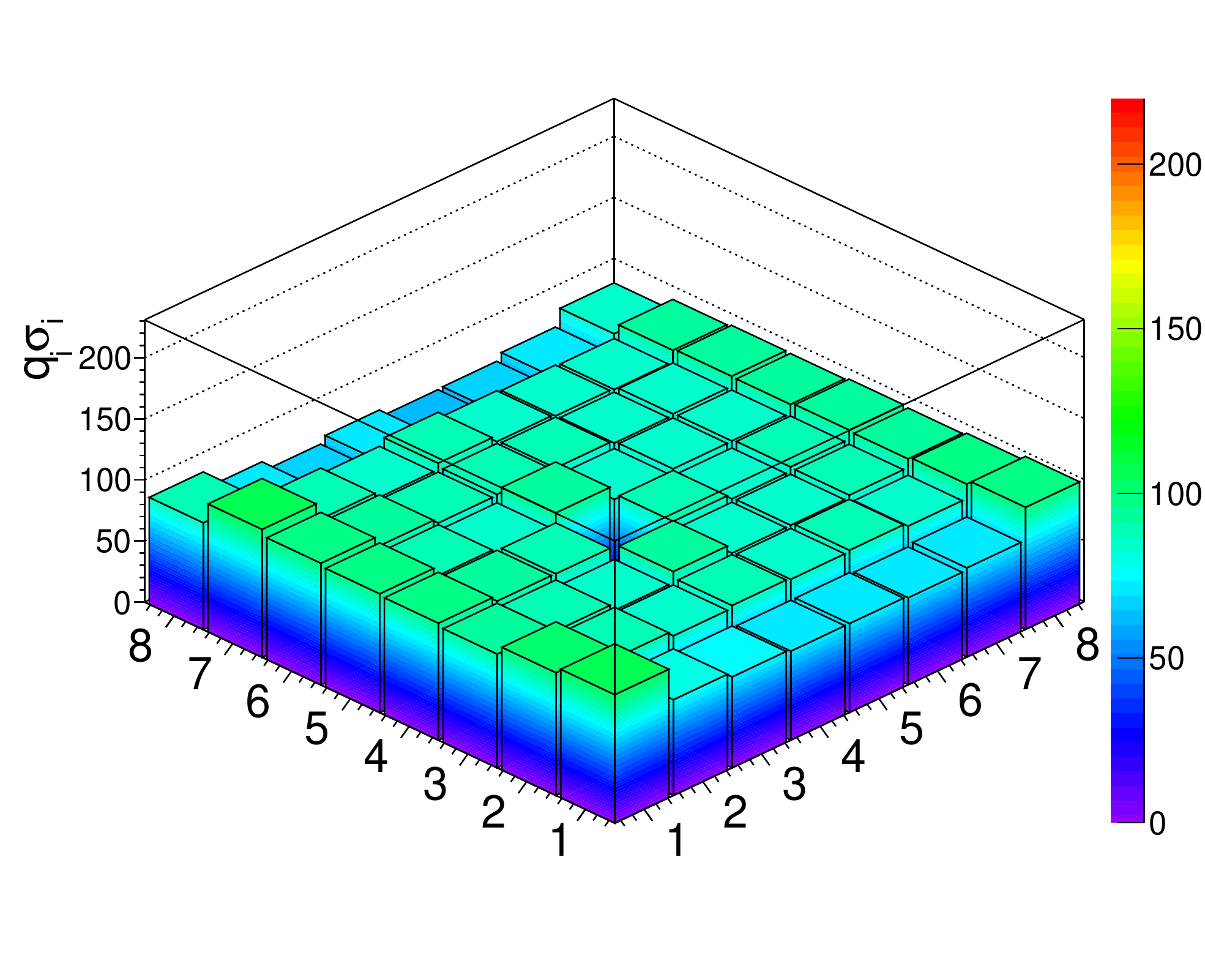}}
\caption[Uniformity of the Gain Reduction]{ \label{fig:15xBck}  Four plots showing the uniformity of the gain reduction, as discussed in the text. 
In each plot PMT-3 of the test EC is illuminated at $\sim15$ times the background rate. The $X$ and $Y$ coordinates of each plot are the pixel row and column, with pixel 1 at (1,1). 
The ordinates are the QDC counts minus the pedestal times the equalization factor $\sigma_{i}$.
\fig\ref{fig:15xBck:Equalized} shows the the gains of all 63 (pixel 28 is disconnected) pixels equalized to their average at a cathode voltage of 900 V.  
\fig\ref{fig:15xBck:D1V} shows the response of each pixel multiplied by the same $\sigma_{i}$ with the cathode voltage reduced to 739 V, while Figs.~\ref{fig:15xBck:D1V} and~\ref{fig:15xBck:0V} show the same at a 
cathode voltage of 250 V and 0 V respectively. It can be seen that the gain of each pixel changes relative to the others at each voltage step.
Both the color and axis scale are the same in all four plots, but the illumination was increased as the gain was reduced (which means that the counts in each plot are not proportional to the gain reduction).
}
\end{center}
\end{figure}

Each of these runs is shown in \fig\ref{fig:15xBck}.
Fig.~\ref{fig:15xBck:Equalized} shows the equalized $q_{i}\sigma_{i}$ at a cathode voltage of 900 V. 
The next plot, \fig\ref{fig:15xBck:D1V} shows the response of each pixel, multiplied by the same factor of $\sigma_{i}$, at a cathode voltage of 739 V. 
The reduction in PMT gain in the central pixels is relatively uniform, but does show some difference. At the edge of the PMT, however, the pixels show a larger decrease in gain. 
Fig.~\ref{fig:15xBck:D9V} shows the next step, down to a cathode voltage of 250 V. The difference in gain reduction in the central pixels is still more uniform than in the edge pixels.
The pixels on the right and left edge of the PMT have now reduced less in gain than the central pixels, while the pixels along the top and bottom edge have reduced more. This trend continues when switching down 
to 0 V. These results clearly show the complicated electrostatics of this PMT, which should be accounted for on a PMT-by-PMT basis.

\section{Conclusions}
 
Our conclusion from these tests on the Cockcroft-Walton HVPS prototype was that it not only worked as intended, but better.
The single photoelectron gain of the M64 multianode PMT (MAPMT) using the CW-HVPS was as good or better than with the normal Hamamatsu voltage repartition.
The CW-HVPS was found to be very quiet, and did not induce any unwanted noise on the anode signals of the test EC. In addition, the power supply 
does not saturate at 100 times the background illumination and shows a very good linearly.  

The primary concern of power consumption is also met: our test EC used 70 mW at the background illumination. As each EC will have one HVPS, this gives a power consumption of 
630 mW for an entire photodetection module of 9 EC. This can be compared to 13 W for a resistive voltage divider, as calculated in the introduction to this chapter.

The switching circuit of the CW-HVPS was also found to work well, and it is capable of switching the cathode voltage down in 1 GTU or $2.5~\mu$s. Recovering back up to a higher cathode voltage 
is equally fast when not switching into a load.
The four switch levels give an overall dynamics of
\begin{enumerate}[i\upshape)]
 \item 1-200 photoelectrons per GTU at 900 V
 \item 200 to $2~10^{4}$ photoelectrons per GTU at 739 V
 \item $2~10^{4}$ to $2~10^{6}$  photoelectrons per GTU at 250 V
 \item $2~10^{6}$ to $\sim3~10^{6}$  photoelectrons per GTU at 0 V
\end{enumerate}
This Dynamic range allows JEM-EUSO to explore the full range of UV phenomena, including extensive air showers, meteoroids and meteors, lighting or other atmospheric emission, and human-made light.


    \printbibliography[heading=subbibliography]
     \end{refsection}
   \begin{refsection}
  \chapter{A Setup for Photomultiplier Tube Calibration and Sorting for EUSO} 
   \label{CHAPTER:PMT Sorting}
  \section{Introduction: Why Sorting?}
\label{sec:INTROtoSorting}

I come now to the largest work in which I have participated, the realization of a setup for sorting and calibrating photomultiplier tubes for JEM-EUSO and EUSO-Balloon. 
This setup is motivated by two needs. The first is the general need to have an experimental setup which allows working, in single photoelectron
mode, with every channel of a multi-anode PMT simultaneously. This is a basic requirement.
The second necessity is the requirement to sort the PMTs which will be used.

The need to sort PMTs for JEM-EUSO is due to their grouping into Elementary Cells (ECs), and the characteristics of the read-out electronics.
Each EC of four PMTs is powered by its own Cockcroft-Walton high-voltage power supply (HVPS) which has been designed to meet the power consumption requirements of the JEM-EUSO mission.
The design and testing of this power supply was discussed in chapter~\ref{CHAPTER:CWHVPStests}.
The gain of the four PMTs within each EC can be adjusted together by regulating the HVPS output.
The gain of each pixel within a single multi-anode PMT (MAPMT) can be modified individually using the preamplifier of the \textit{Spatial Photomultiplier Array Counting and Integrating Read-Out Chip}
(SPACIROC-I), the PMT read-out Application-Specific Integrated Circuit (ASIC) which has been designed for JEM-EUSO \cite{AhmmadASICex}. An overview of the SPACIROC-I was given in the introduction to JEM-EUSO in section \ref{sec:INTROtoJEMEUSO}.
The SPACIROC-I preamplifier has a range of up to a factor of two. JEM-EUSO will use SPACIROC-III, an improved version of SPACIROC-I and II.

At a given high voltage, each MAPMT can be characterized by the average single photoelectron gain across pixels, the standard deviation of the gain, and the minimum and maximum gain within the PMT. 
There is a factor of $\approx$ 4 variation in this average single photoelectron gain between individual M64 PMTs.
These differences in gain come from small manufacturing variations in the multiplication stage of the PMT, where a relatively small
change in electrostatics between dynodes or a small difference in the semiconductor coating of the dynodes can have a large impact on the gain of each pixel. 

Within each PMT, there is a further variation of $\approx$ 25\% in gain from pixel to pixel. This variation is due mainly to the change in electrostatics with location on the photocathode, 
and is larger for a rectangular PMT like the M64.
The gain variation above is a rough estimate based on the general properties of PMTs. In the next chapter measured results will be used to check this estimate.

\begin{figure}
  \centering
  \includegraphics[angle=0, width=1.0\textwidth]{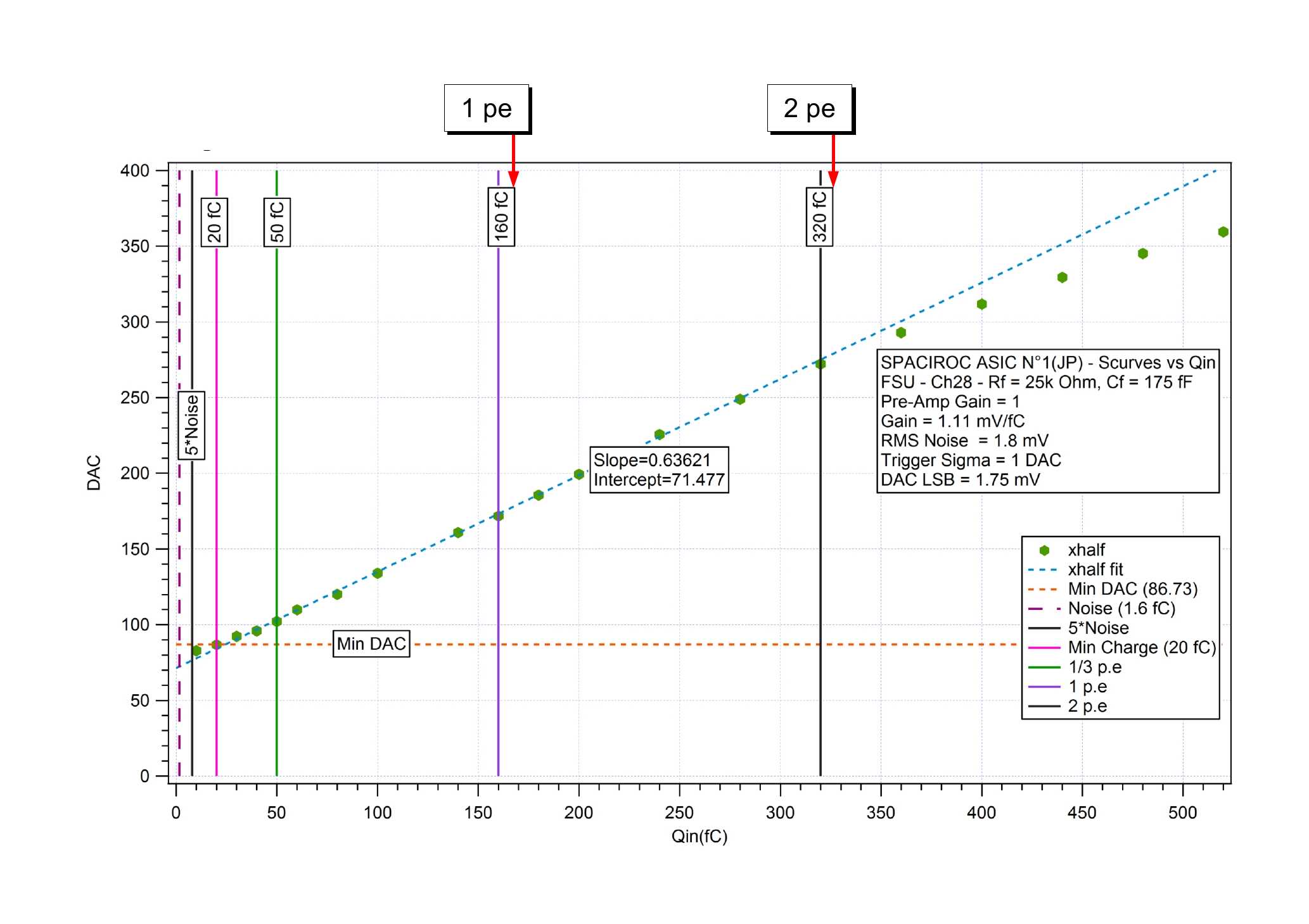}
  \caption[The Linearity of the SPACIROC FSU]{\label{fig:FSU_Linearity_Salleh} A plot showing the linearity of the Fast Shaper Unit (FSU) of the SPACIROC ASIC\cite{SallehThesis}.
The DAC (threshold) value which gives a 50\% triggering efficiency is shown versus the charge input with a pulse generator. Vertical lines are shown at 160 fC (1 photoelectron at a gain of $10^{6}$)
and 320 fC (2 pe). As can be clearly seen, the triggering efficiency saturates above 2 pe. The pulse width in this measurement, however, was twice the width of an actual M64 pulse. 
As the saturation of the FSU is in voltage, rather than charge, the FSU is expected to in fact saturate above 
1 pe when used with the M64 PMT.}
 \end{figure}

The characteristics of the SPACIROC have been measured using both test pulses and an MAPMT. I participated in the latter measurements using an MAPMT, which were done in our black box \cite{SallehThesis}. 
The most important consideration in the present context is the linearity of the integrating preamplifier and pulse shaper, and the photon counting linearity of the ASIC.
The measured photon counting linearity is easily understood by considering the double-pulse resolution of the counting 
discriminator. This will be mentioned again during the discussion of the future measurement of the efficiency of EUSO-balloon ECs using the ASIC in chapter~\ref{CHAPTER:FutureMeasurementsAndDiscussion}.

The linearity of the integrating preamplifier and the \gls{FSU} is important due to its interaction with the gain of the PMT. The integrating preamplifier gives an output pulse which is proportional to the charge of the input pulse. 
The SPACIROC is a high-performance design which pushes the limits of microelectronic
circuit performance, and to achieve this performance the dynamic range of the preamplifier is limited to a small input charge range.
The proportionally of the output pulse height to the input charge thus saturates above a certain input charge, and
the response of the FSU measured by injecting a test pulse into the ASIC is shown in \fig\ref{fig:FSU_Linearity_Salleh}.

As can be seen, the preamplifier is linear up to a charge of 320 fC, or two photoelectrons at a gain of $1~10^{6}$.
One point to note, however, is that the test pulse which was used had a full-width half maximum of 2 ns. This is in fact double the width of a typical single photon pulse from the M64 PMT. As the 
FSU saturates in pulse height, we expect that the FSU will actually begin to saturate above 160 fC, or only 1 pe.

Since a factor of four difference in average gain between PMTs at the same supply voltage is not atypical,
within an EC one PMT could be at a gain of $1~10^{6}$, where the preamplifier is linear, while another PMT could have an average gain of up to $4~10^{6}$, where the FSU is saturated. 
This causes two problems: one being that the saturation of the preamplifier makes it difficult to measure the gains and equalize them, 
and the other being that the difference in gain between the two PMTs is beyond the ability of the preamplifier to compensate.

If all the PMTs within the EC have a similar average gain, then they can all be brought to a gain of around $1~10^{6}$ by adjusting the high voltage. The smaller variation of the 
individual pixels then allows each pixel to brought to the same gain using the gain adjustment of the preamplifier. The PMTs must therefore be sorted in order to group them into gain bins
so that each EC can be build from PMTs with nearly the same gain at the same high voltage. 

There is also the need to test each PMT before it is assembled into an EC. This is particularly important in light of the fact that all four PMTs within an EC share a common HVPS, and so an electrostatic fault in 
any one PMT can make the entire EC unusable. 
As an example, it was found during the sorting for EUSO-Balloon that in a small number (less than 1 in 10) of M64 PMTs one of the first dynodes draws a large current regardless of the incident light.
This is due a to a low resistance between this dynode and the cathode, on the order of $\simeq10$ M$\Omega$ rather than several G$\Omega$.
These PMTs still function for light detection (as evidenced by the fact that they passed Hamamatsu testing), but the large current drawn by the dynode is a potential problem, and these PMTs can not be used.

From these considerations, sorting a sufficiently large number, in principle every PMT to be used in JEM-EUSO, is necessary before any EC can be constructed. 
Because the single photoelectron gain and the PMT efficiency can be measured together by taking a single photoelectron spectrum, the sorting is also a first calibration.

The scale of the sorting should be mentioned: for JEM-EUSO, sorting all of the PMTs requires measuring the gain and efficiency of 64 pixels per PMT for some 10,000 photomultiplier tubes. Each pixel can be though of as an individual PMT with its own single 
photoelectron gain and efficiency.
On a smaller scale, there were a total of 40 PMTs to sort for EUSO-Balloon. This serves as model and test run for future sorting during the construction of JEM-EUSO. 

In the end, the PMTs for EUSO-Balloon were not sorted according to our gain results. 
Data was taken for every PMT, but the QDC which was used did not have the needed resolution to make a reliable determination of the gain of the majority of pixels. There was also not enough time to get the new data acquisition system 
working before the PMTs had to be handed over to be integrated into EC units, and so in the end the PMT gain data given by Hamamatsu was used. This gain is the PMT gain, measured as the ratio of the 
photocathode current to the anode current, and so is in fact the product of the single photoelectron gain and the collection efficiency. This means that sorting the PMTs according to the Hamamatsu data is not correct, and this can be 
confirmed in latter measurements of the completed EC units, where a relatively large difference in the single photoelectron gain between PMTs in the same EC can be seen.

The sorting measurement was not in vain, however, because several defective PMTs were found, as mentioned before. These PMTs were sent back to Hamamatsu and exchanged. 
The overall system which was built for the sorting is also directly useful for any PMT characterization measurement.
The sorting setup was used
to measure the efficiency of each pixel in the completed  EUSO-Balloon EC units. These results will be presented in the next chapter, after discussing the development of the sorting bench.
All told, the sorting campaign includes: 
\begin{itemize}
 \item the set up of a \gls{DAQ} system, both hardware and software,
 \item taking single photoelectron spectra for each MAPMT, and
 \item analyzing these spectra to determine the gain and efficiency of each pixel.
\end{itemize}


\section{The Development of the Sorting Data Acquisition System}
\label{sec:The Sorting Setup}
The first thing needed for PMT sorting and calibration on a large scale is the correct measurement setup and a Data Acquisition System (DAQ) to control it.
The basic technique which we use is the same method which was presented in section \ref{sec:Measuring Absolute Detection Efficiency}, and which was used to calibrate two 
PMTs for our air fluorescence measurement in section \ref{sec:Calibration of Photomultiplier Tubes for the Air Fluorescence Measurements at LAL}.
This technique uses a LED light source, with an integrating sphere and a collimator acting as a stable splitter, and a photodiode acting as a calibrated reference detector. The general aspects of this setup where shown in \fig\ref{fig:CalibrationSetup}.

The main consideration here is how to adapt this technique to calibrating a large number of PMTs in a reasonably short time, and
one of the main challenges in working with the M64 is to scale the data acquisition to handle 64 anodes at once. This requires both a large amount of hardware work, to put 64 cables through the black box, etc., and a large amount of software work to run the DAQ hardware and collect data in an efficient way.
The data acquisition hardware and software went through several iterations over the course of two years. The goal from the beginning was to arrive at a setup for the sorting, 
but the arc of the developed was dominated by the needs of the measurements which were being done at the time, the hardware available, and the time which it took to get new things running.
The development of the sorting system will be explained here in narrative form.

The requirements on the calibration setup are given by both the needs of the sorting, and the typical parameters of the M64 MAPMT.
In order to be useful for sorting and MAPMT calibration the data acquisition system must:
\begin{itemize}
  \item implement our calibration technique or the equivalent,
  \item handle at least 64 channels simultaneously,
  \item operate at a data rate such that a spectrum can be taken in a few minutes,
  \item have a good enough charge resolution to reliably separate the single photoelectron peak from the pedestal at a gain as low as $1~10^{6}$,
  \item give a measurement of the anode pulse charge, and therefore the gain with a systematic uncertainty of $\sim1\%$, and
  \item be able to control other hardware, such as the readout of the photodiode, X-Y movements, etc., in order to perform more complicated measurements.
\end{itemize}
In addition to these online requirements the system must all handle and store the resulting data. Particularly, every spectrum needs to be analyzed to reliably extract the gain and the number of
single photoelectron events, and each spectrum must be assigned to the correct pixel. 

Our approach to these points was also influenced by the status of JEM-EUSO and EUSO-balloon. Namely, we did not have a large amount of funds available to
buy completely new hardware. We do have, however, a large library of \gls{CAMAC} and \gls{NIM} modules, and some amount of \gls{VME} equipment, and, so, we therefore went about 
using this legacy hardware to build a fast system in a similar manner as done in ref.~\cite{LegacyCAMAC}.  

\subsection{The Data Acquisition Hardware}
\label{sec:DAQ}

There are in principle two 
possible readout methods to take the single photoelectron spectrum. The first is to measure it directly by taking a charge spectrum (or pulse height spectrum if an integrating amplifier is used). Another method is to 
send the anode signal through an integrating amplifier and then to a discriminator, which gives a number of counts above a set threshold. A scan across thresholds then gives a plot known as an S-curve, which is
the cumulative distribution function of the single photoelectron spectrum. The latter method is used by the ASIC frontend electronics of JEM-EUSO, whereas taking a charge spectrum is more typical in a laboratory 
setting. 

The advantage of discriminator readout is that it is typically faster than charge measurement. The charge measurement however, has the benefit that it 
measures directly the number of electrons on the anode, which, keeping in mind that a PMT is a nearly perfect current source, is the measurement which returns the most information. 
For the  currently existing calibration bench  a bank of charge-to-digital conversion modules are used to take charge spectra. 
This allows our measurements to act as a cross-check of measurements using the ASIC.
In the future, however, using a well characterized ASIC may be the better solution for the laboratory calibration bench. 

The required charge resolution of the readout can be determined from the lowest gain that must be measured.
The M64 PMT has a typical gain of $\simeq 1~10^{6}$ at a cathode voltage of 900 V. To be sure that the worst pixels of each PMT can be measured, the spectra are taken at a cathode voltage of 1100 V, where the gain is factor $\approx 7$ higher. 
Considering the variation of the gain between MAPMT, the worst PMT which can be expected might have a gain of around  $1~10^{6}$ at 1100 V. 
In this worst case, the separation between the pedestal and single photoelectron peak is 160 fC. 
The resolution of the PMT readout must be high enough to divide this charge into enough bins that the two peaks can be reliably separated.

The fact that we must measure 64 spectra in parallel (one for each pixel) makes the use of amplifiers difficult, and, more importantly, expensive.
It happened, that CAEN produces a newer model of QDC with a better charge resolution, and   
because of this, we decided to rely on the resolution of the QDC to resolve the single photoelectron peak and the pedestal.
If at least 8 charge bins are demanded between the pedestal and the mean of the single photoelectron peak at a gain of $10^{6}$, then the largest QDC resolution which achieves this is $\sim20$ fC.

Similarly, the data rate which is needed can be estimated from the desired statistical precision and the length of time required to take a spectrum. A $1 \%$ statistical error requires $10^{4}$ signal events in each spectrum.
Working in single photoelectron mode requires a signal-to-background event ratio of no more than $1\%$, in order to keep the contamination of two photoelectron events below 0.5\%. 
This means that at least $10^{6}$ events are needed per spectrum.
If a spectrum is to be taken in a few minutes, then this requires a DAQ with an acquisition rate of better than 1 kHz over all 64 channels.

The need for a data acquisition rate of several kHz can be easily satisfied by CAMAC or VME hardware. As mentioned,  
the first system which we used had one LeCroy 2249 Charge-to-Digital convertor (QDC) module. This CAMAC module is a Wilkinson-type QDC with 12 input channels in LEMO 00 format. This QDC has a charge resolution of 250 fC and a conversion time of 60 $\mu$s.
The charge resolution requires a times ten amplifier to work with PMTs that have a gain of $1~10^{6}$. For early measurements this was not a problem,
as measurements were done with a block of at most 9 pixels at a time (cf. section \ref{sec:Cockcroft-WaltonandSwitchDesign}), and we have 16 fast amplifiers available. 

A large limitation of this system was the control software.
The data acquisition software was written in LabView, and was extremely slow when more than 100 thousand events where taken. This was due to the algorithm of the program and the (Windows) PC on which it was run. Unfortunately, it was not 
trivial to port this existing software to a new PC, because the CAMAC crate was controlled through a National Instruments VME to PCI bridge. This bridge requires proprietary drivers, of which there was no version for 64-bit operating systems.
Rather than attempt to develop hardware drivers or downgrade the new PC for the sake of preserving an old system, it was opted to move to Linux, use some VME hardware which was available, and move the data acquisition software to the C programing language.
 
The VME hardware which we had at the time was a single CAEN V792N QDC and a CAEN V1718 USB-to-VME bridge. 
The V792N QDC operates by using a charge to amplitude conversion. The signals are then multiplexed and converted into digital numbers by two fast 12-bit ADCs.
The N model of this QDC has 16 LEMO-type inputs with a charge resolution of 100 fC and a conversion time of 2.8 $\mu$s over the full number of inputs.
To control this system, I developed a heavily modified version of the Prospectus branch of the \textit{Multi Instance Data Acquisition System} (\acrshort{MIDAS-UK}) \cite{MIDASUK}. 

This system was able to work at a rate of 
up to 30 kHz for one input channel by using a continuous block transfer from the event buffer of the V792N, and this was used for a several measurements (cf. section \ref{sec:Calibration of Photomultiplier Tubes for the Air Fluorescence Measurements at LAL}).
This QDC was not ideal, however. The first difficulty was that, as this QDC has a resolution of 100 fC, an amplifier was still needed to take good spectra in many cases. 64 amplifiers would 
not only be expensive (costing again as much as the QDC hardware itself), but also potentially self-defeating because of the noise introduced by the amplifiers.
Using amplifiers is also complicated by the fact that this QDC is very sensitive to positive voltage levels, 
and any spike around +25 mV could potentially saturate its respective QDC channel. For one or two inputs the level of the amplifier could be adjusted to suit this limit, but positive afterpulses from the PMT could still be a problem. What's more,
the gate timing of the QDC was not completely understood. Considering the cost of the module, it was decided not to invest in buying three more to build a complete 64-channel system.

At this point work towards a full 64-channel data acquisition system was begun for the first time. 
It was decided to use the many LeCroy 2249 QDCs which we had laying around the lab. This venerable QDC is very robust and simple to operate compared to the V792N.
A 64 channel data acquisition system was put together using five LeCroy 2249 QDCs, under the assumption that it might be possible to 
acquire 64 amplifiers. To control the CAMAC crate, the CAMAC controller was again interfaced to VME using a CBD 8210 CAMAC branch driver.
The V1718 USB-to-VME bridge was dispensed with, as it is only efficient when using block transfers. In its place, a Motorola VME processor was used. 
This processor board directly runs the acquisition program, and is connected to a backend PC by Ethernet.
At the same time, I developed a fully customized software system based on the \textit{Maximum Integrated Data Acquisition System} (\acrshort{MIDAS}) framework \cite{MIDAS}. 

Shortly after this, two C205 QDCs were inherited from the Opal experiment by way of a colleague at CERN\footnote{D. Laza-Lazic, a former graduate student of P. Gorodetzky}.
This QDC is a CERN design housed in a CAMAC module with 32 input channels in a $32+32$ pin flat connector format. 
The C205 converts input charge to a voltage level. This 
voltage level is amplified by a $1\times$-gain amplifier and a $7.5\times$-gain amplifier, and then digitized in two parallel 12-bit analog to digital convertors. This gives two ranges with a charge resolution of 
$\approx33$ fC and $\approx250$ fC respectively, and a conversion time of 1.6 ms for all 32 channels. 

This system was used to do tests of the high voltage and switches (cf. chapter~\ref{CHAPTER:CWHVPStests}) and the sorting of the 36 PMTs for EUSO-Balloon. Several episodes occurred with these modules. 
In one instance, one of the C205 modules was connected to the CAMAC crate
using extension mechanics so that the pedestal trim potentiometers could be adjusted. Unfortunately, this extender was loose, which allowed the contacts of the module to wiggle in the crate and short. This killed the module. 
Luckily, however, it was possible to hunt down the problem, a destroyed fuse, and repair it. 

In the end there where several major disadvantages which we found with the C205. The first difficultly is that the charge resolution of the module is not ideal, and the flat cable is rather noisy. 
Also, the C205 operates as a ``quasi-differential'' QDC and so gives strange behavior if used with a negative-polarity integrating amplifier. 
Furthermore it is not possible to debug individual input channels because of the flat cable. This resulted in many spectra which where unusable due either to noise or to a low gain in that pixel. 
In addition, the readout of the combined 64 channels was limited to a rate of 500 Hz.
This was because of 
\begin{inparaenum}[i\upshape)]
\item  the QDC dead time, and 
\item  the fact the converted charge values are read from a FIFO which stores the events for both charge ranges in sequence.                     
\end{inparaenum}
To read the values for the low range, you must first read the value for the 
high range, meaning that in each event 64 read operations are required per 32 channels. Because each module handles 32 input channels, the conversion time is higher than in a 12 or 16 channel QDC. Additionally, many channels of these QDCs have a very high
differential non linearity, which is not possible to correct in this module. The only solution is to re-bin the data or to increase the bin error bars.
To work around this, two runs  were taken per PMT; the cables were switched between the two runs, so that 
the pixels which where formerly in a bad QDC channel could be measured.

Once these problems with the C205 became obvious, and there were funds to consider buying either amplifiers or new QDCs, we began to look at the existing options.  
It was found that there is a stark difference between the 
simplicity and robustness of the ``outdated'' LeCroy 2249 compared to the ``modern'' VME QDCs. In particular, the LeCroy 2249 has very low integral and differential non-linearity compared to many later QDC designs, and is very simple to use.
It is a general trend that, while VME QDCs are faster, the best available ones have a conversion resolution in the range of 100 fC per count.
The high transfer rate of digital VME modules, and the fact that they are not shielded, makes VME less suited to precision charge measurements than CAMAC.

The hardware solution which was found is the newer C1205 CAMAC QDC. The C1205 is a Wilkinson-type QDC with 3 independent charge ranges.
The lowest of these is 0 to 80 pC with a 12 bit resolution, giving a theoretical conversion of 21 fC per QDC count. 
Each C1205 has 16 channels with inputs in LEMO 00 format, meaning that 4 modules are needed per PMT. The LEMO 00 format of the QDC inputs is an advantage in terms of signal quality and the ease with which a single pixel can manipulated.
The fact that each input channel can be viewed separately on an oscilloscope allowed us to systematically debug each of the 64 pixels. 

However, handling 64 LEMO 00 cables is not so easy. We had to not only put 64 feed-through connectors through our black box, but also 
make special connectors which allowed us to connect the anode pins of the PMT to the bundle of 64 LEMO 00 cables.
The hardware interface between the CAMAC crate and the control PC for the four C1205 is the same as for the C205 QDC. 
The C1205 is a very modern design which is controlled by an internal FPGA. This required a large change to the previously simple CAMAC front end routine. 

During the course of development, other measurement hardware was included in the DAQ. This included a precision X-Y movement for moving the integrating sphere and the read-out of the NIST photodiode.
Measuring the power incident on the photodiode by computer was particularly important, as it gave a better precision on the average power.
Similarly, interfacing the X-Y movement directly to the DAQ allowed us to quickly perform more complicated measurements, such as centering the light beam with high precision on a given pixel of an MAPMT. 

\subsection{The Data Acquisition Software}
\label{subsec:DAQsoftware}

\begin{figure}
  \centering
  \includegraphics[width=1.0\textwidth]{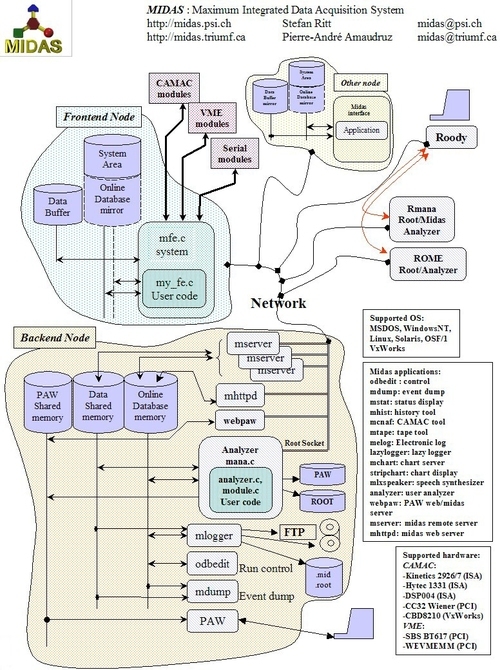}
  \caption[The MIDAS Software Framework]{\label{fig:MIDASsystem}  A diagram showing the overall structure of the MIDAS data-acquisition framework, taken from \cite{MIDAS}. The overall system is divided into numerous sub-processes which 
handle specific tasks. A full guide to MIDAS can be found at ref.~\cite{MIDASwebsite} and a guide to the implementation in this system is given in section~\ref{Chapter:DAQuserguide} of the appendix.}
 \end{figure}

The data acquisition software to run the hardware described in the last chapter was developed in a long winding road, starting from
the original LabView program which was running a single LeCroy 2249 QDC. This system was inherently slower than it could be because the data acquisition
loop was written for one channel and then copied 9 times. This limited the rate to around 1 kHz for nine channels. A more important limitation was caused by 
the memory handling of the program, with the result that the system would grind to a halt when taking spectra with statistics higher than $\sim100$ kevents. 
Due to this, it was decided to develop a new software system. 

This was motivated by a move to a new PC, and a switch to a VME QDC read out through a USB-to-VME bridge. To run this hardware configuration a heavily modified version
of the Prospectus branch of the \textit{Multi Instance Data Acquisition System} (MIDAS-UK) was used \cite{MIDASUK}. The modifications needed to get this software running took several months.
The benefit of this system was that it made good use of the VME QDC's multiple event buffer. This allowed the QDC to be read by block transfer, which gave a very high acquisition rate of up to 32 kHz when 
using only one channel of the QDC. 

The MIDAS-UK system was not very modular, and was very difficult to modify or extend. The ability to add other hardware was very limited, for example, and would require extensive modifications to the 
core code. This became a problem later when switching to a different combination of QDC and VME read out was required. 
Once work towards a system with 64 QDC channels was begun, the primary advantage of our older MIDAS-UK acquisition system, the block transfer, disappeared, as the older CAMAC QDCs do not have an internal event buffer.

At this point, a data acquisition program was written especially for this new setup in C/C$++$ using the MIDAS data acquisition framework \cite{MIDAS}.
In MIDAS the DAQ software is divided into front-end programs which collect data, and back-end programs which handle background processes such as run control, data analysis, and storage. 
A diagram of the structure of MIDAS is shown in \fig\ref{fig:MIDASsystem}.
All of the data acquisition processes can be distributed across various computers, connected by a standard network. This allows for an extremely flexible system. 

The event data collected 
by each frontend program is arranged into banks, and each analysis program running on the backend is split into modules which act independently. 
Each module can access one or more frontend event banks, perform calculations on them, and then pass the result into a new bank which can in turn be used as 
an input for another module in the analysis routine. The data in each bank, both raw and calculated, are logged to disk in an event by event format.
The analysis routine includes the CERN ROOT libraries, which are used for histogramming. Generated histograms can be viewed both online and offline using the Roody histogram viewer \cite{ROODY}.

MIDAS also allows multiple frontend and analysis programs to be connected together. 
This allows data to be collected from multiple sources. In addition to the hardware which is 
directly supported by MIDAS, adding new support is simply a matter of including the needed C or C++ libraries in the frontend code. This allows a wide range of hardware, such as USB, Serial, and \gls{GPIB} to be interfaced with the system 
using readily available function libraries.
Although in this case all the software has been developed on Linux, the entire system is portable and capable of running on any operating system. 

All the data for a given experimental configuration in MIDAS is stored in a central \gls{ODB}. 
The ODB contains experiment run parameters, information from the logging channel, parameters for frontend programs and analyzers, slow control values, status and performance data, and any other information defined by the user.
Every variable within the ODB can be accessed from within any MIDAS frontend or backend process. This allows each process to have access to information generated by the other processes.
MIDAS includes a run sequencer program, which can access the ODB information. This allows a complex series of measurements to be taken, including using feedback from event data.

Within the frontend code which runs the CAMAC crate, the CAMAC commands are generated by a standard CAMAC crate controller. The crate controller is driven by a CBD 8210 CAMAC branch driver.
The CBD 8210 is a VME module which interfaces to the CAMAC branch. The DCAMLIB library written by D. Kryn \cite{DCAMLIB} was used to control the CBD 8210.
The DCAMLIB directly accesses the CBD-8210 in user state, and gives a set of functions which are implementations of the ESONE CAMAC functions \cite{ESONECAMAC}. 
By including this library in the frontend, the CAMAC crate can be controlled by calling the appropriate function with the correct CAMAC address. This, combined with 
the relative simplicity of most CAMAC hardware, makes programing a complicated data acquisition system reasonably straight forward.

The VME part of the acquisition system is controlled by a Motorola VME processor board in the VME crate. The MIDAS frontend code can be compiled and run directly on the processor board. The frontend program communicates by Ethernet with 
the backend processes which histogram, analyze, and store the data, and
as the VME processor board has direct access to the VME back-plate, including VME hardware in the system is also easily doable.

The newer C1205 QDC was a source of problems, however. This CAMAC QDC is a very new design which uses an internal FPGA to control numerous on-board options. This increases the complexity of the 
control program, as various registers have to be set at power on. A more important difference, though, is the fact that this QDC gives 24-bit data words, rather than the 12-bit words which are typical of older CAMAC hardware, such as the LeCroy 2249.
The CB 8210 can address the VME back-plate only with 16-bit cycles, and so must perform 2 VME accesses sequentially to access each 24-bit data word. 

When the programing of the frontend for the C1205 was started, the data words which were received by the analyzer did not make sense. 
Each 24-bit word from the C1205 encodes information about the QDC channel, range, event number, and converted value in different bit ranges, depending on the 
word type. These words seemed to be scrambled such that the header information was correct, but the actual returned counts did not correspond to the bits given on the CAMAC data-way display which was used for debugging.
It took almost a month of stripping the CAMAC crate down and testing it component by component until the problem was found. 

The problem was related to the 24-bit read cycle through the CB 8210. When the DCAMLIB was written,
there was not any CAMAC hardware available which gave 24-bit words, and so this functionality could not be tested. The Library itself was correct, but it was found, after managing to hunt down the manual of the CB 8210, that the specification 
given there for the read order of the two 12 bit sequences in the 24-bit access was not correct.

In this last section, an attempt was made to give an idea of the work involved in building the data acquisition system, including the intermediate steps, roads not taken, and dead ends. 
In depth information on MIDAS itself can be found at ref.~\cite{MIDASwebsite}.
The next section will summarize the DAQ setup as it currently exists.  After that, the calibration of the QDCs will be presented. 
Actual PMT measurements for EUSO-balloon which were done using this DAQ will be discussed in chapter~\ref{CHAPTER:PMT Sorting}.

\section{The Current Data Acquisition System}

\begin{figure}[p]
  \centering
  \includegraphics[angle=90,width=1.0\textwidth]{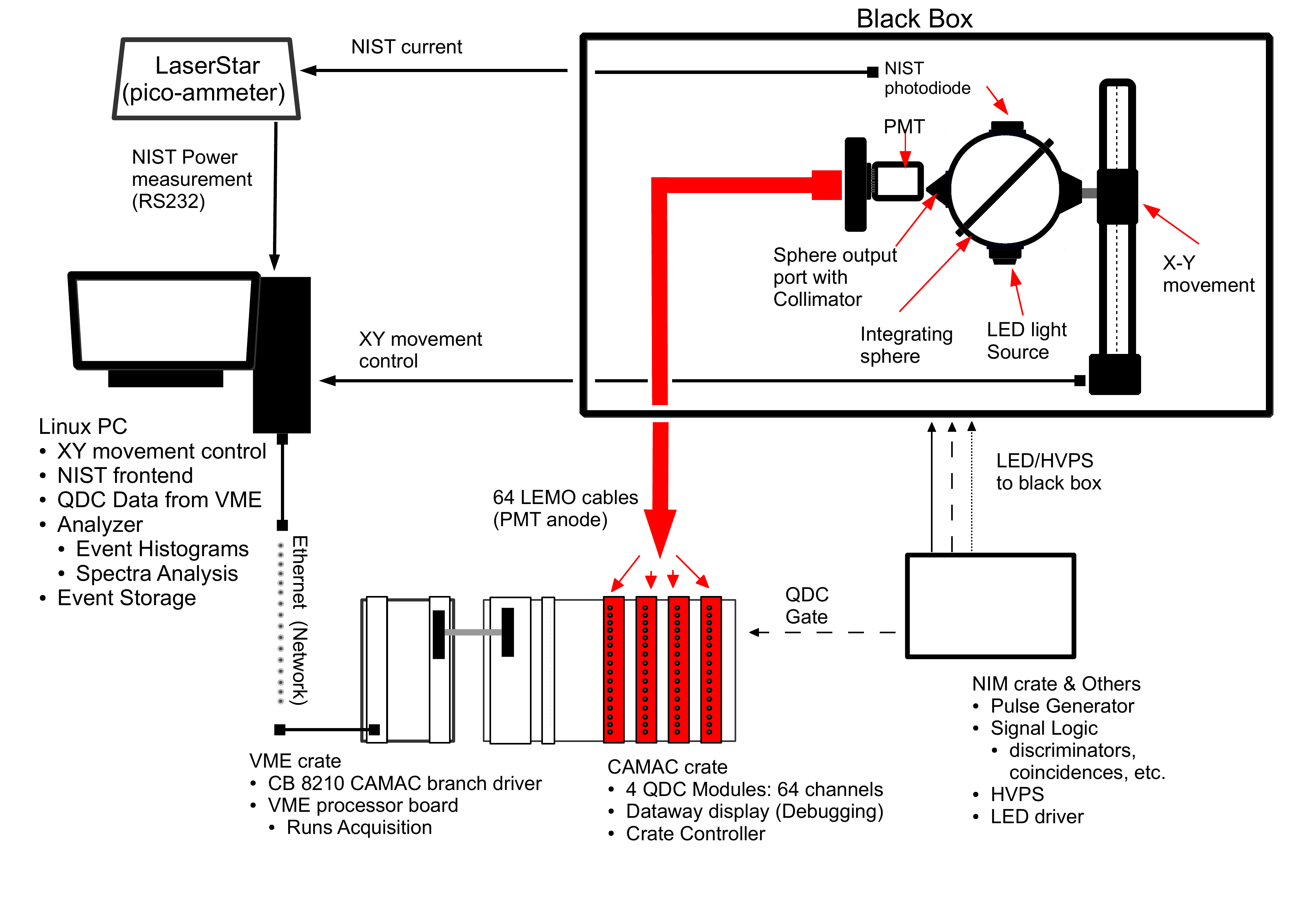}
  \caption[The PMT Sorting Setup]{\label{fig:DAQdiagram} A diagram of the entire PMT sorting setup, as described in the text. 
The PMT is placed inside a black-box and is illuminated using the output of an integrating sphere, as described in section~\ref{sec:Measuring Absolute Detection Efficiency}. The absolute light output of the sphere
is monitored by a NIST photodiode. The NIST photodiode is read out by the LaserStar, which is simply a pico-ammeter which applies the absolute calibration curve of the photodiode and returns the incident power.
The exit port of the integrating sphere can be equipped with a collimator to illuminate a single MAPMT pixel at a time, and the integrating sphere is mounted on a X-Y movement to allow scanning the photocathode of the PMT. 
The anode signals of the PMT are measured using a bank of CAMAC QDCs, controlled by the DAQ software described in section~\ref{subsec:DAQsoftware}.}
 \end{figure}

An overall diagram of the current DAQ system is shown in \fig\ref{fig:DAQdiagram}, and a user guide to the current DAQ system is included in the section \ref{Chapter:DAQuserguide} of the appendix. 
This guide is a general end-user manual for running the system, and gives detailed information 
about the frontend and analysis routines. Photographs of various parts of the DAQ can be found in section \ref{appendix:chapter:Photographs} of the appendix.

After the long-winded discussion in the last section, this section will summarize the hardware of the DAQ for clarity. The core of the DAQ is a standard CAMAC crate containing 4 CAEN C1205 charge-to-digital conversion modules.
These modules have a charge resolution on the order of 20 fC per count. The CAMAC crate is controlled by a SEN CC 2089 ``A2'' crate controller, and the CAMAC branch is driven by a CBD 8210 branch driver, which interfaces the CAMAC branch to a VME 
crate. The VME crate is controlled by a Motorola MVME 3100 VME processor board running Debian Linux on 32-bit Power PC architecture.

The frontend program which controls the VME and CAMAC crates runs directly on the VME processor board, which is connected to the backend PC by Ethernet.
This setup allows the acquisition to proceed at a rate of 2 kHz for all 64 channels in parallel with zero dead time. In this configuration, the rate is limited by the CAMAC signal definitions. 
As the C1205 is compatible with the FAST-CAMAC standard, a upgrade of the CAMAC crate controller to a FAST-CAMAC compatible model would increase the readout rate by a factor of 2-10 \cite{FASTCAMACa, FASTCAMACb}.

\label{sec:TheDAQ}

In addition to the frontend driving the QDC hardware, several other frontend programs for other hardware are run on the backend PC. The first program implements the slow-control of two Zaber T-LSM200
movements mounted together, orthogonal to each other. This X-Y movement holds the integrating sphere (light source), and is controlled through RS-232. 
The LaserStar power meter which reads the NIST photodiode is controlled, again through RS-232, by a third frontend routine. The photodiode is read periodically at a rate of 10 Hz, and this event information is
stored event by event along with the QDC data and the position of the X-Y movement.

\begin{figure}[p]
  \centering
  \includegraphics[angle=90, width=1.0\textwidth]{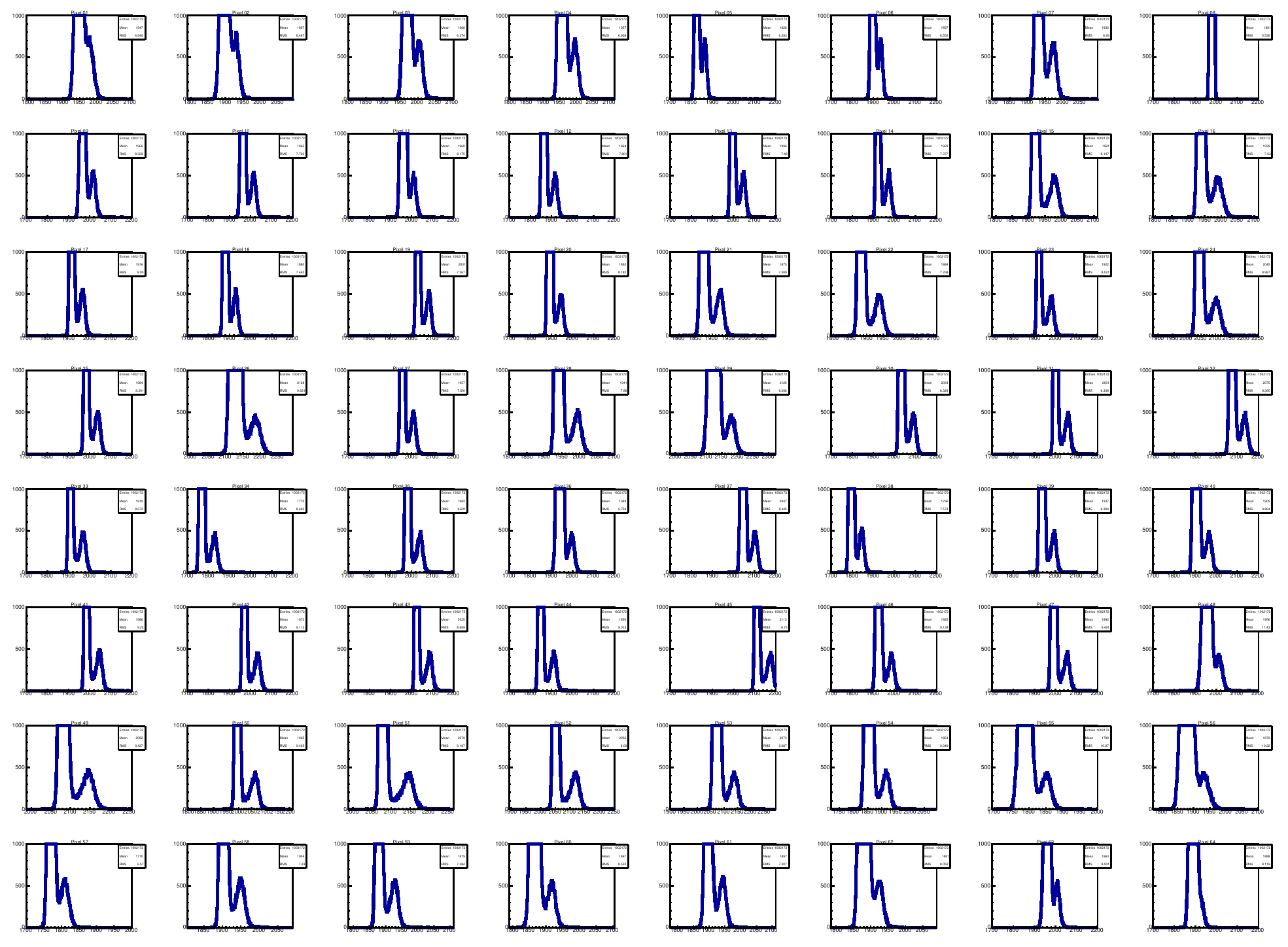}
  \caption[Spectra for One M64 MAPMT]{64 spectra taken for one M64 MAPMT (here PMT-D of EC 109) in a single run with uniform illumination, before analysis. The single photoelectron peak can clearly be seen for the 
majority of pixels.}
  \label{fig:64Spectra}
 \end{figure}

The analysis routine is made up of 4 modules. Each module accesses the raw data words from the QDC, decodes them, and performs a certain analysis task.
The C1205 QDC includes three independent charge ranges. In the most general operation mode, the charge range which the returned count value is on is encoded in the event header.  
The first analysis module defines one histogram for each QDC channel for each charge range, and bins the incoming data in these histograms. 
This gives the most general possible view of the incoming data (i.e., one set of histograms for each charge range, listed by module and channel number). An example of 64 spectra measured during one run are shown in figure \ref{fig:64Spectra}

\begin{figure}[p]
  \centering
  \includegraphics[angle=90,width=1.0\textwidth]{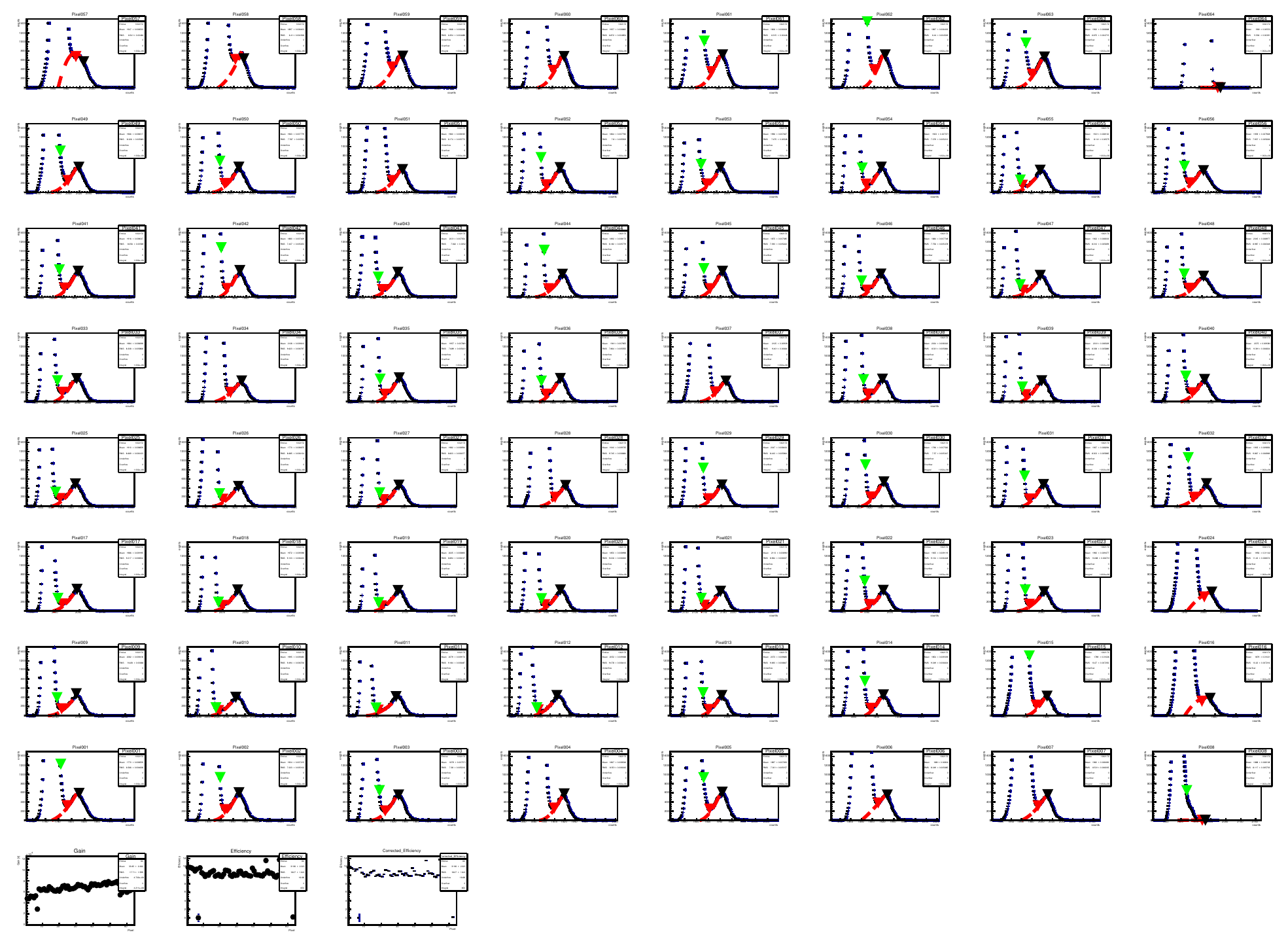}
  \caption[64 Spectra Analyzed]{The 64 spectra of figure \ref{fig:64Spectra}, after analysis by the routine discussed in section~\ref{sec:TheDAQ} and chapter~\ref{Chapter:DAQuserguide}.
 For each pixel, the location of the 1 pe peak mean and the valley have been found and marked. The red line 
shows an extrapolation of the single photoelectron peak below the valley. A close-up of an analyzed spectrum can be seen in detail in \fig\ref{fig:ExampleSPEspectra}.}
  \label{fig:64SpectraAnalyzed}
 \end{figure}

The second analysis module performs a specific analysis for single photoelectron spectra. This routine assigns each QDC channel a pixel number according to a pixel map stored in the ODB. Since the map is not hard-coded, it is easy to load 
a new configuration from a file. This is useful, for example, for measuring each of the four PMTs within an EC, where the mapping between pixels and connector pins is not the same for each PMT (as each PMT is rotated by $\pi/2$ relative to its neighbors). At the end of the run, each spectrum is 
analyzed to extract the number of single photoelectron events and the mean of the single photoelectron peak.
The 64 spectra in figure \ref{fig:64Spectra} can be seen after analysis in figure \ref{fig:64SpectraAnalyzed}. For each pixel, the valley, shown by the red marker, has been found and the mean of the
1 pe peak has been determined, shown by the black marker. The line in each spectrum is an extrapolation of the single photoelectron peak from the valley to the mean of the pedestal. The first example of a single photoelectron spectrum
shown in section~\ref{subsec:PhotoDetection}, \fig\ref{fig:ExampleSPEspectra}, was analyzed by the same routine. 
Details about the analysis procedure can be found in chapter~\ref{Chapter:DAQuserguide} of the appendix. 

The gain and efficiency are calculated from the results of the analysis. The gain calculation includes the conversion 
between QDC counts and charge which is held in the ODB. These values were measured for each QDC channel, and this measurement is described in the next section. 
The efficiency calculation uses the mean value of the power incident on the NIST photodiode, returned by a third analysis routine which is fed data 
from the photodiode read-out. 

A fourth and fifth analysis routine were written which interface the results of the single photoelectron spectrum analysis with the control of the X-Y movement. 
These analysis routines allow, for example, centering with high precision on a given pixel using the 
response of the PMT itself. This centering algorithm in will be discussed in section~\ref{sec:PixelCenetering}. 
 
\subsection{QDC Characterization}
\label{sec:QDCCharacterization}
\begin{figure}
  \centering
  \includegraphics[angle=270,width=1.0\textwidth]{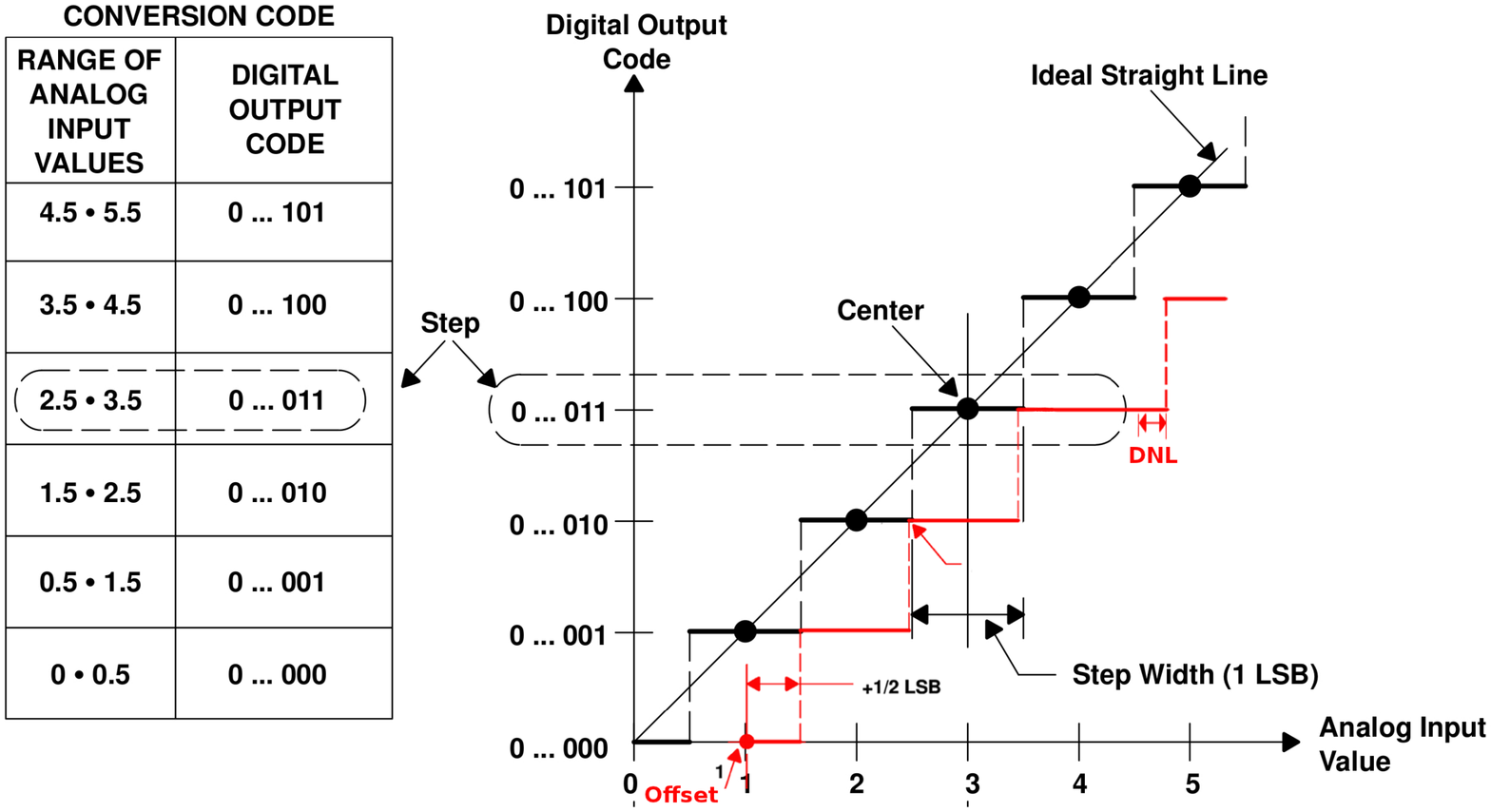}
  \caption[Diagram of an ADC Transfer Function]{\label{fig:ADCtransfer} An example of an ADC transfer function, based on a figure taken from \cite{ADCerror1}. The ideal transfer function is 
shown in black. The bin center of each return code lies on the ideal linear response of the ADC (shown by the black line). In the ideal case, the step width (width of each bin) is exactly 1 least significant bit (LSB).
The ideal response goes through the origin, giving a return code of 0 for a input of amplitude 0.  
The red line, on the other hand, shows a more realistic transfer function. This real transfer function does not go through the origin, because the return code of 0
corresponds to an input amplitude of 1, or in other words the ADC has an offset error of 1 LSB. In addition, not all the bins of the red transfer function are equal in width. This
variation in bin width is known as differential non-linearity (DNL).}
 \end{figure}

To extract the absolute gain from the measured spectra, the conversion from QDC counts to Coulombs must be known with an accuracy on the order of 1\%. 
The conversion slope given by the manufacturer does not include the linearity properties of the QDC, and the quoted inter-channel uniformity is 2\%.
Due to this it is necessary to verify the linearity of each QDC channel, and to determine the response of each channel relative to one another (if not the absolute response).

A QDC is simply an analog-to-digital convertor (ADC) which is sensitive to the charge of the input rather 
than amplitude. Any ADC transforms an input signal into a digital number according to some transfer function. A diagram of a typical ADC transfer function 
is shown in \fig\ref{fig:ADCtransfer}. An ideal ADC transfer function represents a certain range of analog inputs uniquely with a finite number of return values 

The two basic characteristics which define an ADC are its \gls{FSR} and \gls{LSB}. The FSR is the simply the total range of analog inputs over which the ADC operates. 
The LSB is the width of one return bin and so is defined by the FSR of the ADC and the total number of return codes available. 
The least significant bit is so called because it is the difference in input which in theory flips the least significant bit of the ADC return code.
In an $n$-bit ADC, the number of return codes is $2^{\text{n}}$, which define $2^{\text{n}}$ steps. The first and last step of the 
ADC are, however, only one half of a full step width and so the FSR is in fact divided into $2^{\text{n}}-1$ codes. 
The LSB is therefore given by:
\begin{equation}
 \text{LSB} = \frac{\text{FSR}}{\left(2^{\text{n}}-1\right)}
\end{equation}
The C1205 model QDC is a 12 bit ADC with a FSR of 80 pC on the lowest charge range, giving a theoretical LSB of 19.54 fC. The charge resolution given in the manual for the same range is 21 fC.  

Aside from the quantization error on any given input, which is $\pm1/2$ LSB, there are four distinct types of error possible in an QDC transfer function \cite{ADCerror1,ADCerror2}.
These are offset error, gain error, differential non-linearity, and integral non-linearity. 
The offset error is defined as the mid-step value of the analog input when the digital return code is zero. An offset error which is negative
implies that an analog input of zero does not give a digital return code of zero, and so the ADC has an inherent pedestal.
The gain error is the difference in slope between the actual and ideal transfer function, measured once the offset error is corrected. Both the gain and offset error
can easily be corrected and are present even when relationship between the digital output and the analog input is linear. 

The \gls{DNL}, on the other hand, is the variation of the width a of bin step from the ideal value of one LSB. If the DNL is greater than 1 LSB, then there may be missing
codes in the ADC output. It is not possible to directly correct DNL, but it can be handled by using the sliding scale technique \cite{SlidingScale}. This technique consists of adding a randomly selected, but known level to the analog signal. 
This way the same analog value is converted in different regions of the ADC conversion range each time it occurs, with the overall effect of averaging out the width of each bin of the ADC.
The sliding scale is built into many modern QDC designs, including the C1205.

The total deviation of the transfer function from a straight line is known as \gls{INL}. The INL at a given bin is the sum of the DNL
from zero up to that bin. The exact magnitude of the INL depends on the straight line to which the transfer function is compared. The easiest definition is the end-point INL, which is defined as the difference between the 
end points of the ADC transfer function once both the gain and the offset error have been corrected. The best-straight-line INL is defined relative to a best-fit straight line which minimizes the deviation of
the INL. The best-straight-line INL is in principle a better representation of the actual transfer function of the ADC, and contains information about the offset and gain error \cite{ADCerror2}.

\begin{figure}
  \centering
  \includegraphics[width=0.85\textwidth]{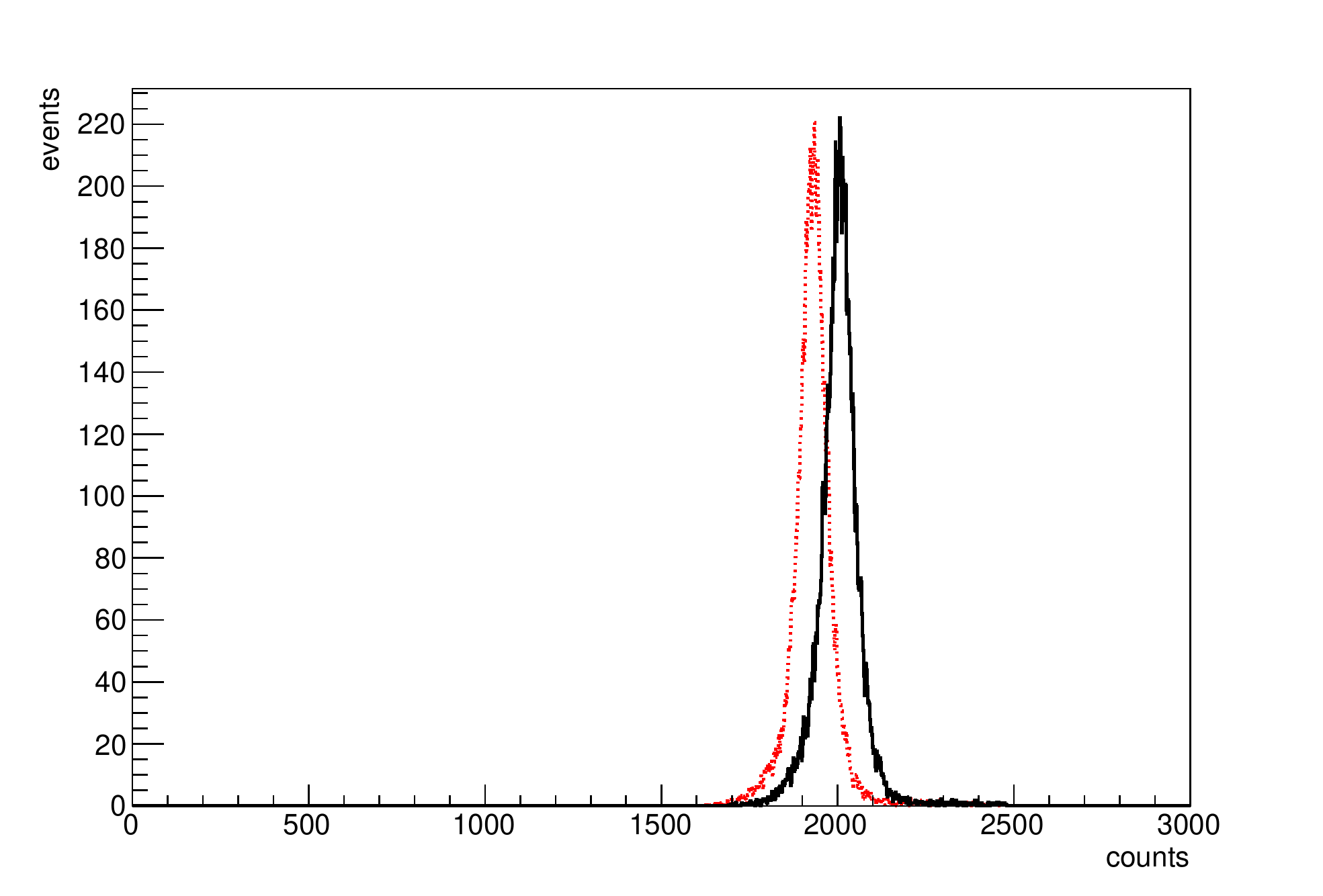}
  \caption[One QDC Response Measurement]{\label{fig:QDCMod1_Ch16:OnePoint} An example of a pair of spectra taken during the calibration of the QDC. This set is from QDC module 1, channel 16, with 
an integration gate width of 10 ns. The pedestal measurement is shown in red, and the measurement with the input current on is shown in 
black. A total of 20 kevents were taken for each spectrum.}
 \end{figure}

The response curve of each channel of each of the four C1205 QDCs in the DAQ system was measured by inputting a known test charge and checking the return code of the QDC.
This was done by using a NIM level from a CAMAC status register module with a resistance in series to give an input current on the order of $0.15~\mu$A.
The exact voltage of the NIM level was not important, as the actual current through the input was measured with a high-accuracy Keithley 6514 system electrometer (operating as a pico-ammeter). The fact that
the voltage level was controllable through CAMAC, was however, essential as will be shown below.

The integration gate for the QDC was created using a Agilent 81101A digital pulse generator. The actual magnitude of the test charge
was varied by changing the gate width. For each gate width, two spectra were taken, one with a level of -800 mV (NIM on) and one at 0 mV (NIM off) to take the QDC pedestal. Taking the pedestal is essential because the pedestal (offset error) of 
the C1205 increases with the gate width at a rate of approximately 0.2 counts per ns.

At least 20 thousand QDC events were taken in each spectrum, so that the statistical error on the mean counts returned by the QDC was less than 1\%. 
A histogram of one pair of measured spectra is shown in \fig\ref{fig:QDCMod1_Ch16:OnePoint}.
The pedestal is shown in red, and the QDC response with the NIM signal on is shown in black.
The response of the QDC across its full range was taken by varying the gate width from 10 ns (the minimum allowed gate 
width) to 490 ns (500 ns is the maximum). 
For each measured point, the largest uncertainty on the test charge comes from the gate width. This is due to the 1-2\% error on the pulse width setting of 
the pulse generator and a lack on information on the true integration time of the QDC -- i.e., it was not known a priori if there was any gate opening and closing offset. 
Compared to this, the error on the measured current was negligible. 

\begin{figure}
  \centering
  \includegraphics[width=0.85\textwidth]{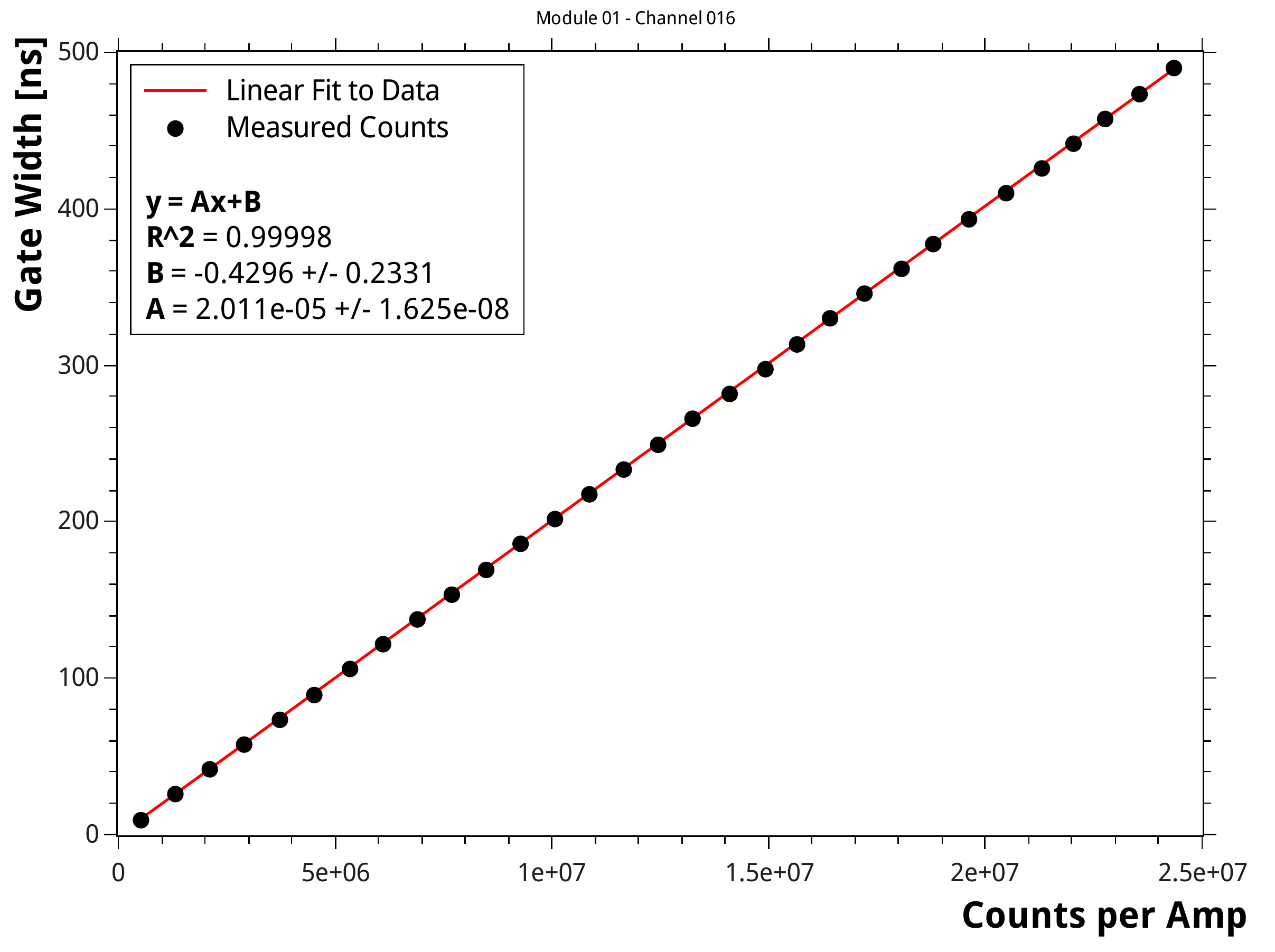}
  \caption[QDC Module 1 Channel 16 Response Curve]{\label{fig:QDCMod1_Ch16:TotalCurve} The response of channel 16 of QDC module 1 (C1205$\# 88$) measured across the range of the QDC. Each point is taken by 
passing a measured current through the QDC channel for a given gate width, as described in the text. The charge resolution is determined by a least-squares fit to the measure data, shown by the red line.
The linearity of the QDC response is very good, and the correlation coefficient $R^{2}$ is very close to 1. The best-straight-line INL is on the order of 4 LSB.
 }
 \end{figure}

The resulting response curve for one channel of one module is shown in figure \ref{fig:QDCMod1_Ch16:TotalCurve}, plotted with the gate width in ordinates and the ratio of the QDC counts to the measured current in abscissa.
The conversion slope was determined by performing a linear least-squares fit of the curve.
For every channel, the agreement of the QDC response with a linear assumption was very good, with $R^{2}$ typically equal to 1 within 0.1 part per thousand. 
The offset error was found to be very small, and, as the single photoelectron gain of a PMT is the difference between the single photoelectron peak and the pedestal, any offset will have no effect on the gain measurement.  
The best-straight-line INL was calculated by taking the sum of the absolute value of the difference between the measured curve and the least-squares fit. This INL was found to be on the order of 80 fC, or 4 LSB.

\begin{figure}
  \centering
  \includegraphics[width=0.85\textwidth]{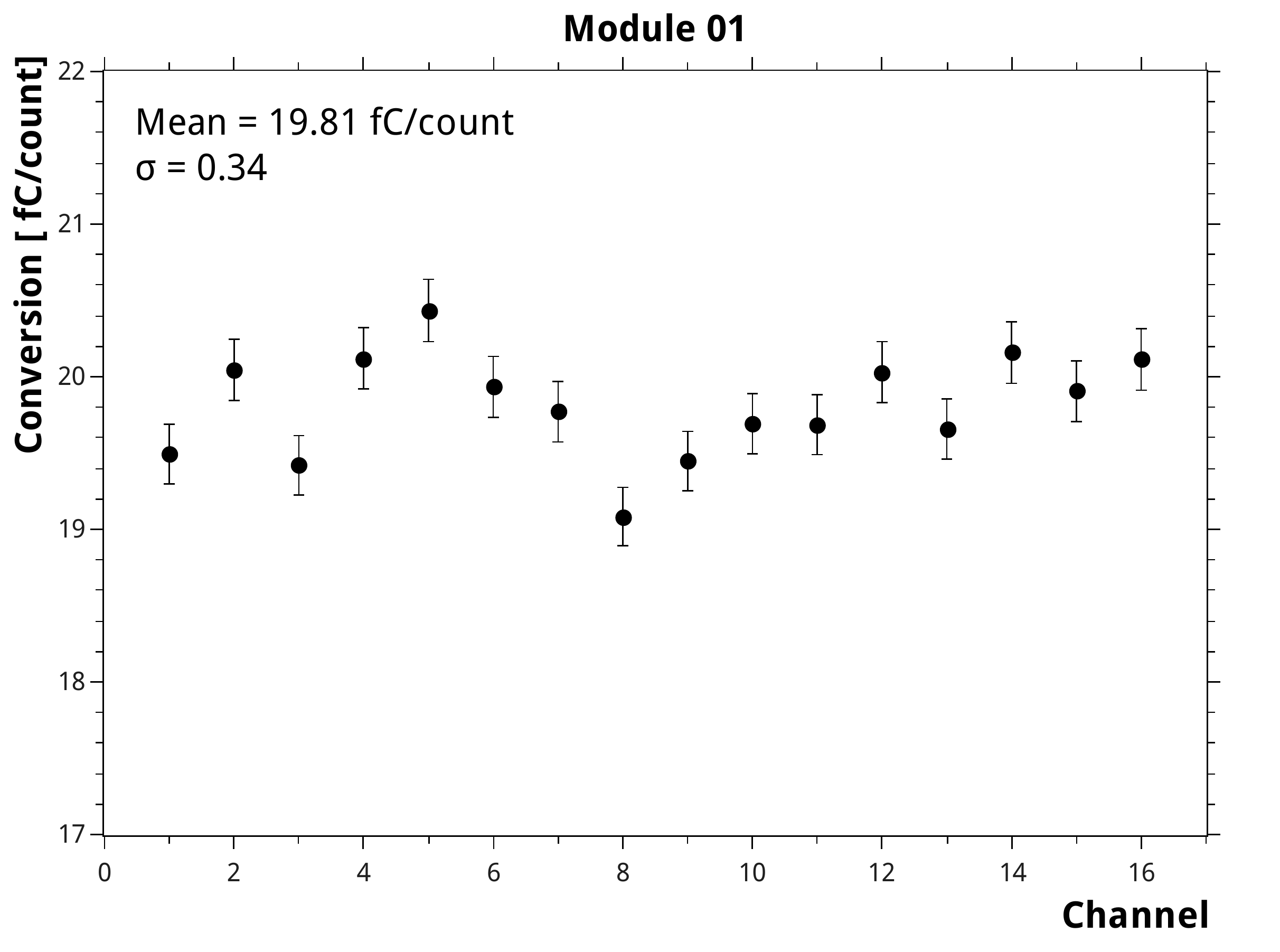}
  \caption[QDC Module 1 Charge Resolutions]{\label{fig:QDCMod1_overview} A plot showing the results for the charge resolution for all 16 channels of QDC 1 (C1205$~\# 88$). Each QDC channel is measured independently, as described in the text.
The abscissa is the channel number within the module (numbered from 1 to 16), and the ordinates
are the resolution in fC per count returned. }
 \end{figure}

The measurement of the absolute PMT gain depends on converting the difference between the pedestal and single photoelectron peak, in QDC counts, into a charge. To do this, the slope of the transfer 
function must be known, and the uncertainty on this slope translates directly into an uncertainty on the gain.
In order to reach better than $1~\%$ accuracy on the slope, at least 30 data points per curve where needed, making more than 3840 measurements to complete a full characterization of all 64 QDC channels.
To made this feasible, the readout of the  pico-ammeter and the gate width setting of the pulse generator where
controlled by the DAQ through GPIB. The input current was switched on and off through CAMAC, and the measurement of each response curve was scripted using the DAQ run control. A plot of the measured conversion constant for all 16 channels
of one QDC module can be seen in figure \ref{fig:QDCMod1_overview}.
The abscissa gives the channel number within the module, and the ordinates are the measured conversion slopes in units of fC per QDC count. 
The same plots for the three other QDCs can be seen in Figs.~\ref{fig:QDCMod2_overview}, \ref{fig:QDCMod3_overview}, and \ref{fig:QDCMod4_overview} in the appendix.

The least squares fit should in principle account for the error in the gate width, and the estimate of $\sigma_{y}$ from the fit result is on the order of 0.5 ns.
The error on the value of the slope calculated by the linear fit is negligible, but this is far too optimistic. For one it does not account for the statistical error on the number of counts at each measured point. With 20 thousand events per spectra 
the uncertainty on the difference between the pedestal and the number of QDC counts returned with the test input on is on the order of 1\% and this is used as an estimate of the uncertainty on the conversion slope.

\begin{table}
 \begin{center}
\begin{tabulary}{1.0\textwidth}{LCCCC}
\toprule
 & \multicolumn{4}{c}{Module} \\
\cmidrule(l){2-5}
Channel & 01 & 02 & 03 & 04\\
\cmidrule(l){1-5}
1 & 19.49 & 19.98 & 19.59 & 18.57\\
2 & 20.05 & 20.24 & 20.28 & 19.29\\
3 & 19.42 & 19.27 & 19.35 & 18.98\\
4 & 20.12 & 19.70 & 19.34 & 19.12\\
5 & 20.43 & 19.95 & 19.21 & 18.89\\
6 & 19.93 & 19.51 & 20.71 & 20.10\\
7 & 19.77 & 20.04 & 19.86 & 19.95\\
8 & 19.08 & 19.33 & 19.62 & 19.92\\
9 & 19.45 & 19.71 & 19.21 & 20.79\\
10 & 19.69 & 18.85 & 19.44 & 20.11\\
11 & 19.69 & 20.00 & 19.33 & 20.00\\
12 & 20.03 & 19.82 & 19.69 & 20.08\\
13 & 19.66 & 18.61 & 19.04 & 20.12\\
14 & 20.16 & 19.39 & 19.41 & 20.58\\
15 & 19.91 & 19.63 & 19.47 & 19.59\\
16 & 20.11 & 19.16 & 19.42 & 20.39\\
\bottomrule
\end{tabulary}
\caption[QDC Conversion Constants]{ \label{tab:QDCconversionConstants} Measured results for the slope of the QDC transfer function. A value is given
for each channel of each of the four QDC modules. Each conversion constant has an estimated error of 1\%.}
 \end{center}
\end{table}

The thermal stability  of the C1205 QDC is given in the manual as +3 counts/C maximum. For the measurement of the QDC properties 
and all subsequent measurements using the QDC, the fan-cooled CAMAC crate was turned on before hand so that the electronics would reach a stable operating temperature. The temperature of the laboratory itself was also noted. 
As the photodetection laboratory at APC is climate controlled, the temperature was never seen to vary more than a few tenths of a degree from the typical temperature of $24^{\circ}$ C.

\section{Conclusion}
The DAQ described in the last sections is a complete implementation of the absolute calibration technique described in section~\ref{sec:Measuring Absolute Detection Efficiency} for 64 pixel MAPMTs. 
This implementation includes not only CAMAC QDCs to read the anode signals of the 
MAPMT, but also a read-out of one or more photodiodes, and the control of a precision X-Y movement. This allows complete flexibility to perform complex measurements, such as scanning a MAPMT photocathode pixel by pixel.
In addition, a complete analysis of all 64 channels is performed at the end every run to reliably extract results from the measured single photoelectron spectra. All data are saved event-by-event, which makes complex analyses possible,
such as searching for coincidences between events in different pixels.

The software system which controls the DAQ is both flexible and powerful, and this allows new hardware to be added quickly. This point, in fact makes the present system a powerful \emph{general} tool for the photodetection laboratory, as any combination 
of CAMAC and VME hardware can be implemented in the time it takes to read the manual. The flexibility of the system was used to include control of other laboratory hardware, such as electrometers and pulse generators, to
create an automated calibration setup for the C1205 QDCs. All 64 QDC channels were characterized using this setup, as discussed in the last section.
Several software tools were created as part of the DAQ to simplify tasks such as centering on a given MAPMT pixel, or measuring the response of the PMT as a function of the position of the incident light. 
The next chapter will present several measurements done using the DAQ and its associated tools. 

    \printbibliography[heading=subbibliography]
     \end{refsection}
   
   \begin{refsection}
    \chapter{Characterization of EUSO-Balloon Elementary Cells} 
   \label{CHAPTER:EUSOBalloonMeasurements}

The Data Acquisition system (DAQ) described in last chapter was used to do a preliminary measurement of the gain and efficiency of each pixel of the assembled and potted EUSO-Balloon elementary cells (EC).
The first goal of these measurements was to check that each pixel functioned properly, and that the analog signals from each pin of the EC assembly could be seen.
A true measurement of the absolute efficiency was performed for each completed EC unit.
The measurements themselves are presented in sections~\ref{sec:Pixel-by-Pixel Illumination: Measurement of the Absolute Efficiency} and \ref{sec:Uniform Illumination: Measurement of the Relative Efficiency}, and
the results are discussed in section~\ref{sec:Efficiency and Gain Results}.
These results where used to estimate the typical spread in PMT gain and efficiency, which is presented in section~\ref{subsec:Statistics on the Single Photoelectron Gain and the Efficiency}. 
After measuring the efficiency of every pixel of each EC, further measurements were done to characterize the response within individual pixels by scanning the photocathode with a small light spot.
This work is discussed in section~\ref{sec:PixelScanningResults}.

The SPACIROC Application Specific Integrated Circuit (ASIC) for EUSO-Balloon is designed for use with MAPMTs with a single photoelectron gain of $1~10^{6}$, which equals an input charge of 160 fC for a single photoelectron. 
The ASIC sets a photon counting threshold using a 10-bit digital-to-analog convertor, such that one DAC count is $\approx 1~$fC, whereas the QDC resolution is $\approx 20~$fC per count with no preamplifier.
Because of this, the QDC-based DAQ is clearly not as sensitive as the ASIC, and so the spectra had to be taken at a cathode voltage of 1100 V. The voltages were supplied independently to each element using a CAEN high voltage power supply, and a voltage repartition 
identical to that of the Cockcroft Walton HVPS of EUSO-Balloon was used. The advantage of using the CAEN HVPS was that the current on the individual elements could be automatically measured to check for problems on the shared high voltage cabling of 
the four MAPMT in the EC.

One such problem was found involving the central screw of the EC unit. Pictures of the EC can be seen in Figs.~\ref{pic:ECUniformIllumination:Closeup} and \ref{pic:ECUniformIllumination:Overview}, and in chapter~\ref{CHAPTER:JEMEUSO}. 
The central screw is intended to hold the EC in the PDM mechanical structure and is made of metal. There is only a small distance between this screw and an extra pin of the M64, which is connected to the cathode. This pin is supposed to be cut and the 
potting should isolate it from the screw. During the first tests of the EC units, a large noise was found on the MAPMT anodes, and a large current ($\sim 60~\mu$A at a cathode voltage of 800 V) was drawn by the cathode when the screw was grounded. This would imply that there are discharges occurring between the 
EC screw and the cut cathode pin. To overcome this problem, in all the following measurements the screw was isolated from the mechanics holding the EC in the black box and was set to the cathode voltage.
The PDM mechanics of EUSO-Balloon were also redesigned so that EC screw is isolated from ground, and the aluminum frame for the PDM units was replaced with a Delrin frame. 
 
After the initial tests and debugging using the first assembled EC we received, a complete measurement of each EC unit was begun.
The efficiency was measured using a comparison to a NIST photodiode, as described in section~\ref{sec:Measuring Absolute Detection Efficiency}. 
In order to give an absolute efficiency, however, the number of photons $N_{\text{photon}}$ which touch the photocathode must be known with high precision. 
If an entire PMT is illuminated at once, then it is difficult to know $N_{\text{photon}}$ for each pixel with a precision at the level of 1\%. 
This is the same general consideration regarding calibration techniques as was discussed in sections~\ref{sec:PMTCalibration}, and
can be solved by using a collimator so that the light spot is restricted to an area of the photocathode which is small enough to be replaced by a NIST photodiode, $\approx11~$mm square. 
For the M64 MAPMT, a single pixel is 2.88 mm square, and the pinhole of the collimator used is 0.3 mm.

Unfortunately, this is not the whole story, as the time required to take a single spectrum at the 1\% statistical level is about 10 minutes. Scanning each pixel independently would then require 16 days. 
To overcome this challenge, the efficiency measurement of each PMT is done in two steps:
\begin{enumerate}
 \item The absolute efficiency is measured for several pixels of each MAPMT by illuminating each with a light spot such that only one pixel is illuminated at a time, thus creating a set of ``NIST'' pixels.
 \item Each MAPMT is illuminated uniformly to measure the efficiency of each pixel relative to the reference pixel(s).
\end{enumerate}
The gain of each pixel is taken from the measurement using uniform illumination, and
as the DAQ is able to read out 64 channels at once, each EC is measured PMT by PMT.

\section{Pixel-by-Pixel Illumination: Measurement of the Absolute Efficiency}
\label{sec:Pixel-by-Pixel Illumination: Measurement of the Absolute Efficiency}
The signal logic of the setup was as presented in section~\ref{subsec:Absolute Calibration Setup}, and was 
shown in \fig\ref{fig:CalibrationSetup:Logic}.
A pulse generator was used to create NIM pulses with a width of 20 ns at a rate of 2 kHz. 
These pulses are sent through a fan-in/fan-out module to copy them. One copy of
the pulse is sent to a delay unit and then to a discriminator, which was used
to generate a gate signal. After studying the dependence of the pedestal width on the gate width and the timing of the single photoelectron pulses in the gate, the optimum gate width for the C1205 was found to be $\sim 40~$ns.

The second copy of the pulse was sent to
a LED driver with an adjustable amplitude, so that light source is pulsed in coincidence with the charge
integration gate. The timing within the integration gate of the single photoelectron pulses from the PMT was adjusted using an oscilloscope.
For the 64 channel DAQ, the gate is daisy chained through a series of discriminators to create four gates, one for each QDC module, with the same delay relative to the LED pulse.
This daisy chain can be seen in the picture of the NIM crate in \fig\ref{fig:NIMmodules_CloseUp}.

The adjustable amplitude of the LED driver was then used to tune the number of
single photoelectron events in the spectra. The ratio of the number of single
photoelectron events to the number of pedestal events was checked by taking
spectra and by reducing the LED pulse height until no more than 1.5\% of the
events where outside the pedestal. This ensures that the contamination of two
photoelectron events is less than 0.75\% in all the spectrum we took afterward.

The source LED for the PMT spectra is the same single LED (wavelength 378 nm) as was used for previous efficiency measurements in section~\ref{sec:Calibration of Photomultiplier Tubes for the Air Fluorescence Measurements at LAL}. 
The PMT was illuminated by the LED through an integrating sphere, with an additional collimator.
The collimator had to be changed, however, compared to previous measurements. The background of the photodiode with the LED off is on the order of 5 pW, and so a power of 500 pW was needed to give a proper signal-to-noise ratio. 
At the same time, the number of photons incident on the MAPMT must be low enough to stay in single photoelectron mode.

In order to do this, the attenuation of the collimator had to be increased, and this difference is due to the change in the data acquisition system. 
As described in the last sections, the older data acquisition system operated a single channel of a VME QDC at a rate of 20 kHz.
Because the NIST photodiode effectively integrates the power received in a short time, the power measured by the photodiode is proportional to the acquisition rate when using a pulsed LED.

This means that for the same number of photons at the photocathode per pulse, the power measured by the reference photodiode will be a factor of 10 lower compared to previous measurements. 
As a consequence, the attenuation of the collimator must be increased to get the same signal to noise ratio on the reference photodiode. All the collimators used are shown in \fig\ref{pic:ThreeCollimators} in the appendix, 
where the collimator used for the pixel-by-pixel measurement is the right-most.
This collimator has an exit pinhole with a diameter of 0.03 mm, an entrance diameter of 1 mm, and a length of 64 mm.

\begin{figure}
\centering
\includegraphics[width=1.0\textwidth]{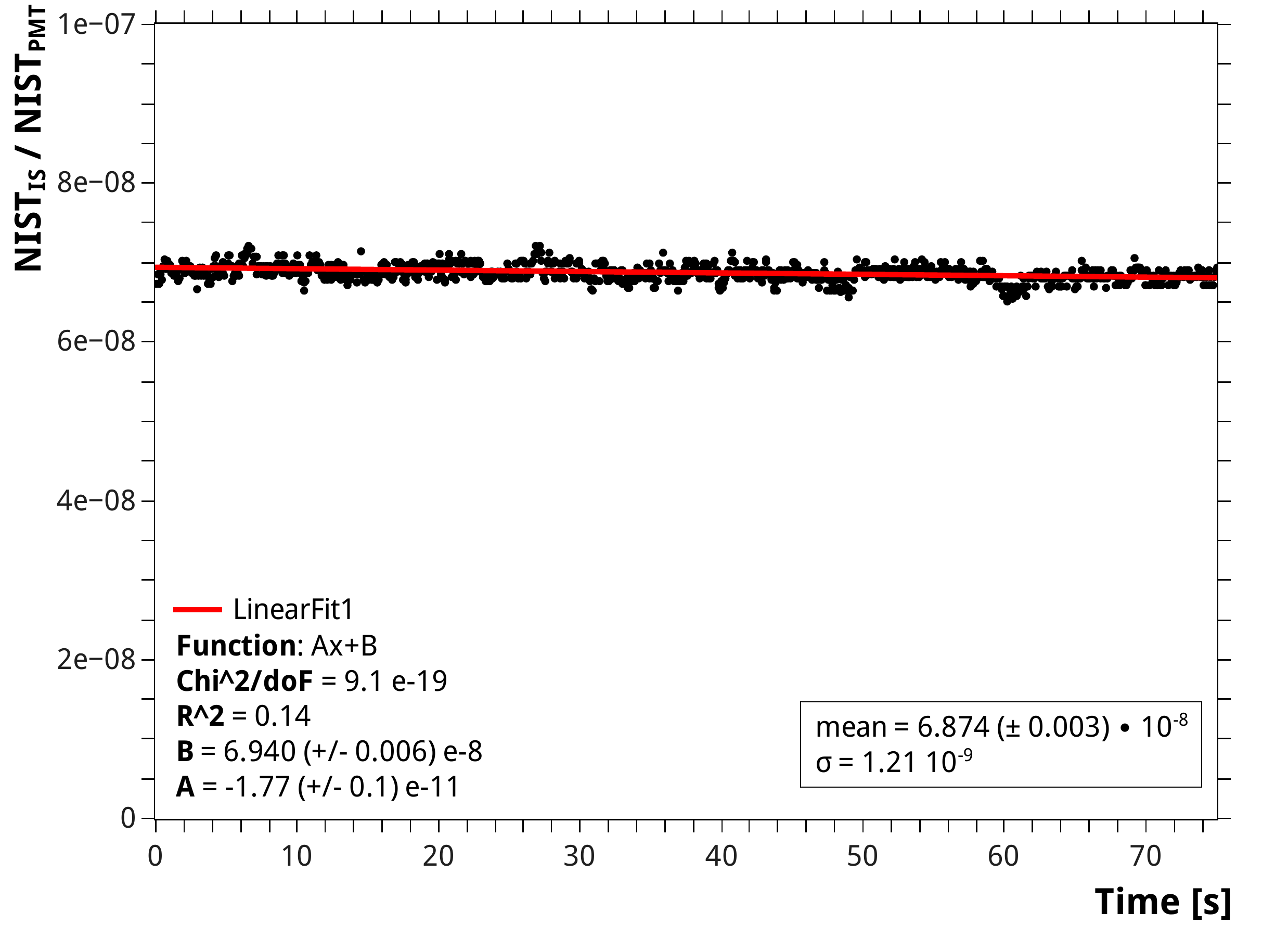}
\caption[Calibration of the Attenuation]{ \label{fig:NISTratioEC} Measurement of the attenuation in power between the photodiode on the integrating sphere and the PMT photocathode.
The ratio is taken between the first NIST photodiode placed on the integrating sphere and a second NIST photodiode which replaces the 
PMT. Data was taken over 70 seconds at a sample rate of 15 Hz, giving a mean of $\alpha = 6.874~(\pm~0.003)~10^{-8}$, with the standard error quoted. 
The red line shows a linear fit to the data, which was done to check the stability of the ratio during the measurement. }
\end{figure}

The attenuation between the NIST photodiode on the sphere and the photocathode of the PMT was calibrated by replacing the PMT with a second NIST photodiode, 
as shown in \fig\ref{fig:CalibrationSetup} and discussed in sections~\ref{subsec:Absolute Calibration Setup} and \ref{sec:Calibration of Photomultiplier Tubes for the Air Fluorescence Measurements at LAL}.
The single pulse light was also replaced by a collection of 398 nm wavelength LEDs. The ratio of the power
measured on the two photodiodes is shown in \fig\ref{fig:NISTratioEC}. As the attenuation is almost 2 orders of magnitude higher, the current through the LED had to be much higher 
to get a large enough incident power on the second photodiode (the one replacing the PMT). Due to this, the ratio was measured for only $\approx 1$ minute (rather than the $5$ minutes in the previous case, \fig\ref{fig:NISTratioXP2020Q}),
to avoid over heating the LEDs.

The measured attenuation ratio is $\alpha = 6.874~(\pm~0.003)~10^{-8}$, where the quoted uncertainty is the standard error. This result can be compared to 
$\alpha = 3.7591~(\pm~0.0006)~10^{-6}$, as measured during the calibration of two PMTs in section~\ref{sec:Calibration of Photomultiplier Tubes for the Air Fluorescence Measurements at LAL}. The difference is a larger than
the simple factor of ten from the change in acquisition rate because the LED pulse height and width were also changed (hence the amount of light per pulse). 

\begin{figure}
\centering
\includegraphics[width=1.0\textwidth]{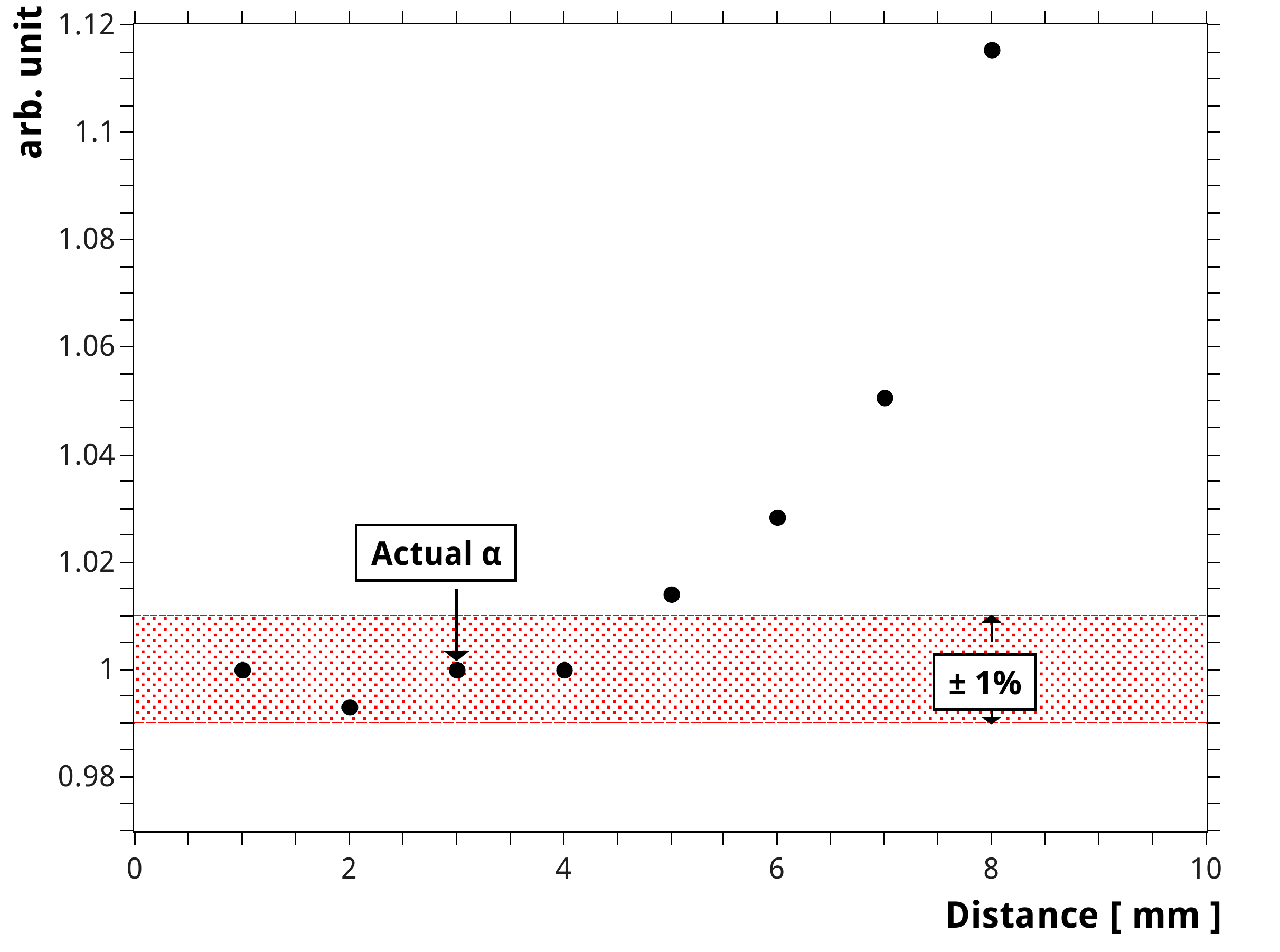}
\caption[The Attenuation as a Function of Distance]{\label{fig:NISTratioWdistances} The change in the attenuation $\alpha$ with distance, expressed relative to the measurement of the attenuation in \fig\ref{fig:NISTratioEC}.
The distance between the photocathode and the collimator in \fig\ref{fig:NISTratioEC} was 3 mm. Each EC is expected to be within less than 0.5 mm of the original distance.
As can be seen in the plot, $\alpha$ is within $\pm 1\%$ of the reference value between 2 to 4 mm, and so no significant change in $\alpha$ is expected for each measurement.}
\end{figure}

The actual distance $d$ between the photocathodes of the PMT and the collimator exit (and where $\alpha$ was measured) was 3.0 mm.
This distance was preserved across EC units by mounting the EC on a hand-operated movement. The mounting of the EC in the black box can be seen 
in Figs.~\ref{pic:ECUniformIllumination:Closeup} and \ref{pic:ECUniformIllumination:Overview} in the appendix. Mounting the EC in this way allowed it to be moved away from the collimator while it was being connected and cleaned, 
and then be placed back at the original distance reliably. 
To check what effect a variation on $d$ would have on the attenuation $\alpha$, the ratio of the two NIST photodiodes was measured as a function of distance. This is shown in \fig\ref{fig:NISTratioWdistances}. As can be seen, $\alpha$
varies less than 1\% between a distance of 2 to 4 mm; a variation in distance which is much larger than the precision of the placement of the EC unit away from the collimator.

After calibrating the attenuation of the integrating sphere and the collimator, the measurement of the absolute efficiency was started.
To illuminate each pixel one by one requires that the light spot is well centered the pixel to be measured. 
The EC unit is not mounted on a optical table or the like, and so the position of each pixel is not well-known relative to the X-Y movement. 
This is especially true when the EC is moved, such as when the next one to be tested is put in the black box.

\subsection{Pixel Centering}
\label{sec:PixelCenetering}
The position of the X-Y movement relative to the pixels of the MAPMT is determined using the single photoelectron response of the MAPMT itself. 
First, a single photoelectron spectrum is taken for four adjacent test pixels in order to find the valley between the single photoelectron peak
and the pedestal. The valley position is used as a threshold for single photoelectron counting. To begin the centering process, the light spot is placed roughly at the cross between the four pixels. 
The number of single photoelectron counts in each pixel is proportional to the surface of the light spot on that pixel. 

\begin{figure}
  \centering
  \includegraphics[width=0.90\textwidth]{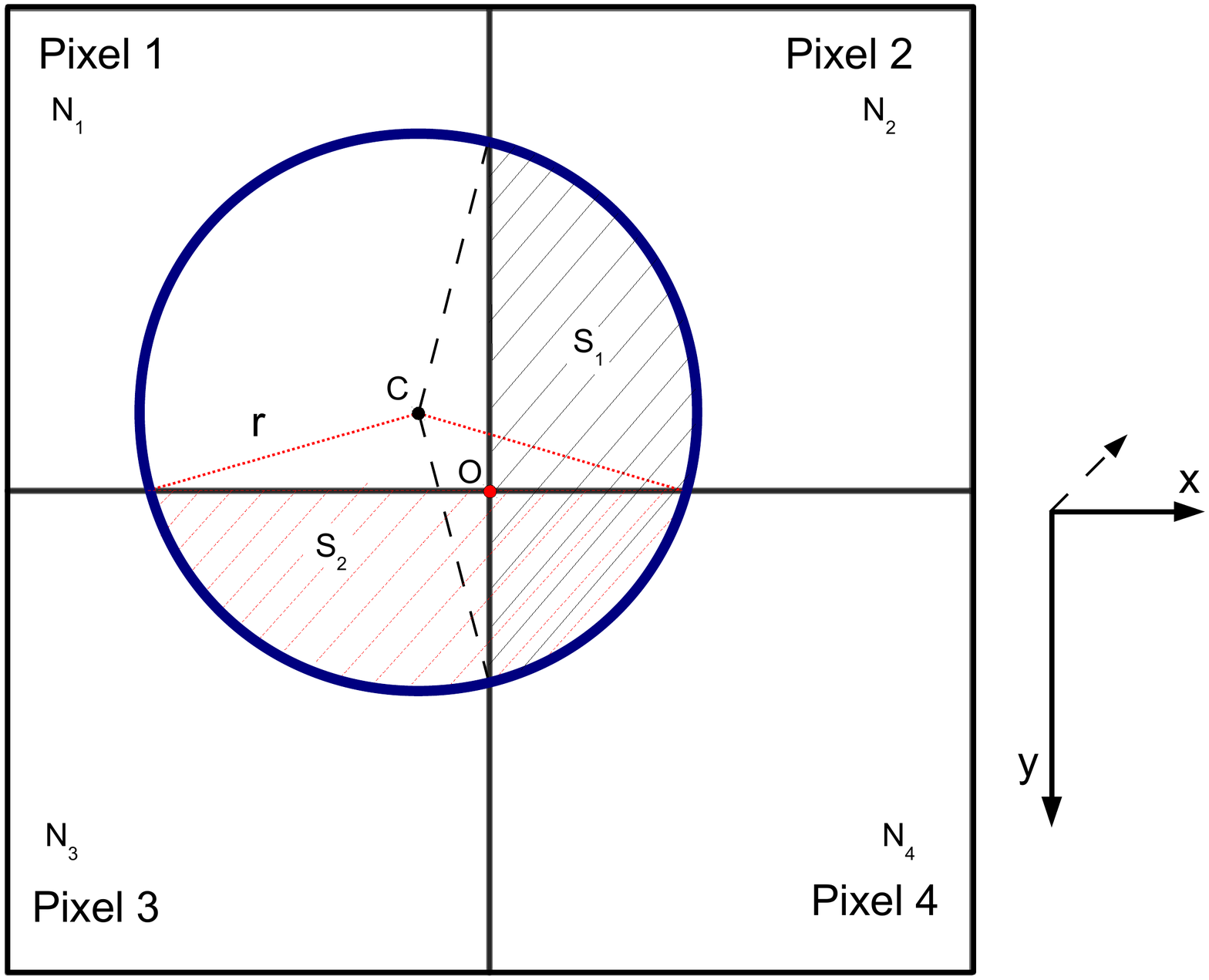}
  \caption[Diagram of Pixel Centering]{\label{fig:CenteringDiagram} A diagram of the pixel centering. Four pixels are shown with the incident light spot represented by the blue circle. The center of the light spot is at some position $C$ and 
it is the cross between the four pixels $O$ which must be found. The number of photoelectron counts $N_{i}$ in each pixel is proportional to the surface of the light spot in that pixel. The horizontal distance between
$C$ and $O$ is given by the surface area $S_{1}$, and the vertical distance is given by $S_{2}$. The actual coordinate system of the DAQ is shown in the bottom right corner of the figure, defined facing the photocathode of the PMT.}
 \end{figure}

This situation is shown in \fig\ref{fig:CenteringDiagram}. In this figure a light spot of radius $r$ is incident on four pixels. The center $C$ of the light spot is some distance in $X$ and $Y$ from the cross between the four pixels. 
The axes of the X-Y movement are defined within the DAQ such that the positive $X$ direction is towards the right and the positive $Y$ direction is towards the bottom, when facing the photocathode of the MAPMT.
 In the vertical direction, the sum of the number of single photoelectron counts 
in pixels 3 and 4 is proportional to the surface $S_{2}$, while in the horizontal direction, the sum of the number of counts in pixels 2 and 4 is proportional to $S_{1}$.
Neglecting any difference in efficiency between the four pixels, the correction in the vertical and horizontal directions are then proportional to:
\begin{equation}
 \Delta X =\frac{\left(N_{1} + N_{3}\right) - \left(N_{2} + N_{4}\right)}{\left(N_{1} + N_{2} + N_{3} + N_{4}\right)}
\end{equation}
and
\begin{equation}
 \Delta Y = \frac{\left(N_{1} + N_{2}\right) - \left(N_{3} + N_{4}\right)}{\left(N_{1} + N_{2} + N_{3} + N_{4}\right)}
\end{equation} 
One approach, if the radius of the light spot is known, is to calculate the distance of the spot center C from the cross between the four pixels geometrically. 
A second approach, which was used in this case, is to  multiply $\Delta X$ and $\Delta Y$ by an assumed length scale to give a correction to the position. The actual position of the 
light spot will converge towards the cross over several iterations. 

In the DAQ, the centering is handled automatically by an analysis routine which returns the number of counts in each pixel and calculates a correction to the X-Y movement position. 
This centering routine selects the 4 central pixels of the PMT, and the run sequencer of the DAQ is
used to script a series of runs with decreasing length scales, so that the only starting requirements are the photoelectron counting thresholds and that the initial position of the light spot be within one of the four pixels.
The centering loop continues until $\Delta X$ and $\Delta Y$ are compatible with zero within the statistical error. 

In practice, the statistics per run was such that the uncertainty on the number of counts in each pixel was $\approx2.5\%$. The two movements have a micro-step resolution of 
0.047625 $\mu$m with one step, the difference of the least significant bit in the X-Y movement position setting, equal to 64 micro-steps, or 0.003048 mm. 
The light spot size is approximately 0.3 mm in diameter. This gives a centering at the cross between the four pixels to within $\pm$ 0.005 mm on each axis.
The Hamamatsu specifications give a pixel width of 2.88 mm for the M64 MAPMT. 
This pixel width has been verified in previous measurements \cite{GoroPrivate}, but a further measurement of the width will be presented in section~\ref{sec:PixelScanningResults}.
Once the cross between the pixels is found, this is taken as the origin, and the light spot is moved to the center of a chosen pixel by moving $(n+0.5)\times2.88~$mm, where n is the number of pixels which must be crossed to reach the destination.

\begin{figure}
\centering
  \subfigure[The Hamamatsu Pixel Definitions]{\label{fig:M64pindiagram} \includegraphics[angle=0,width=0.44\textwidth]{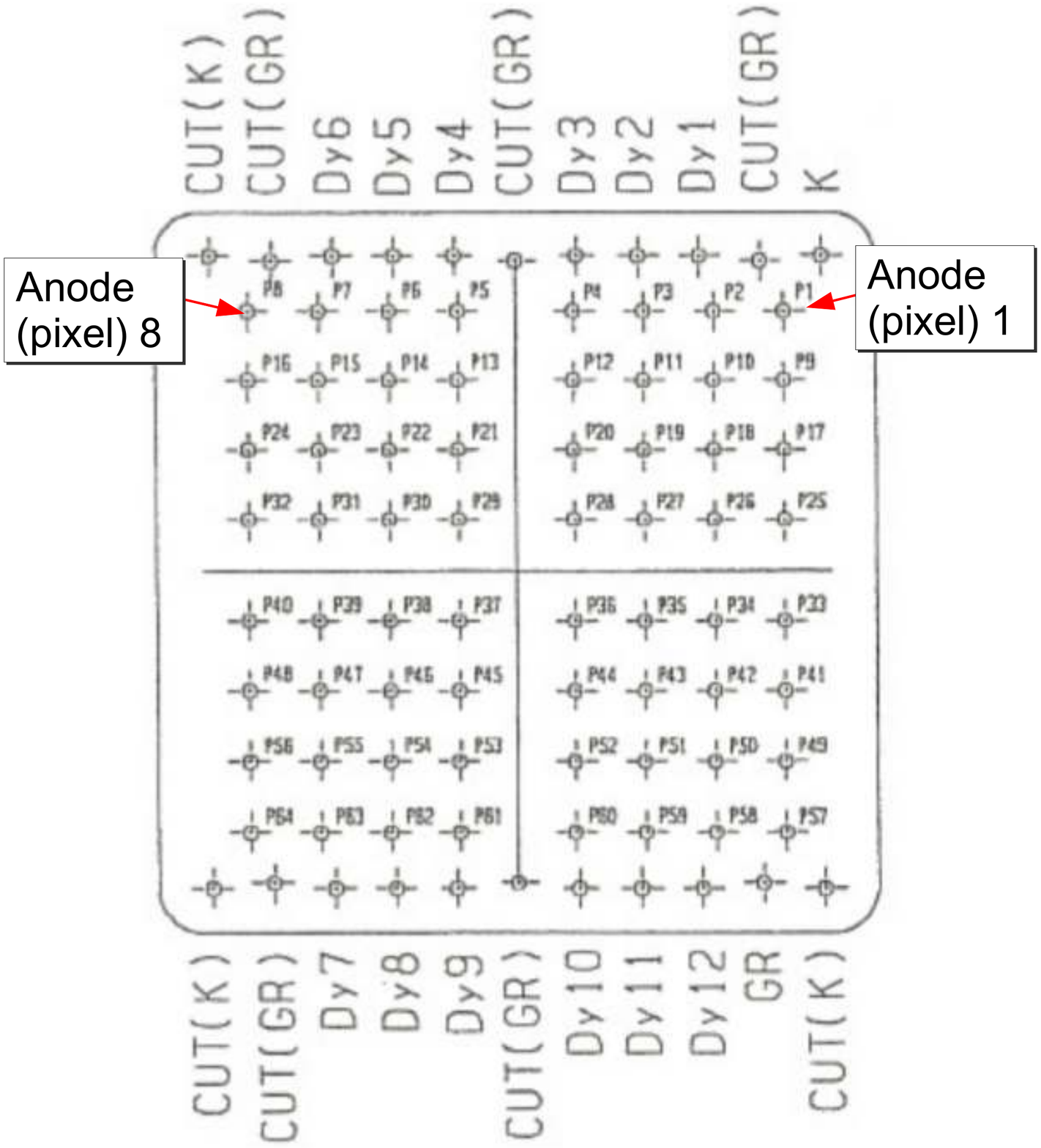}}
 \subfigure[The layout of the EUSO-Balloon PDM]{\label{fig:PDMlayout} \includegraphics[angle=0,width=0.50\textwidth]{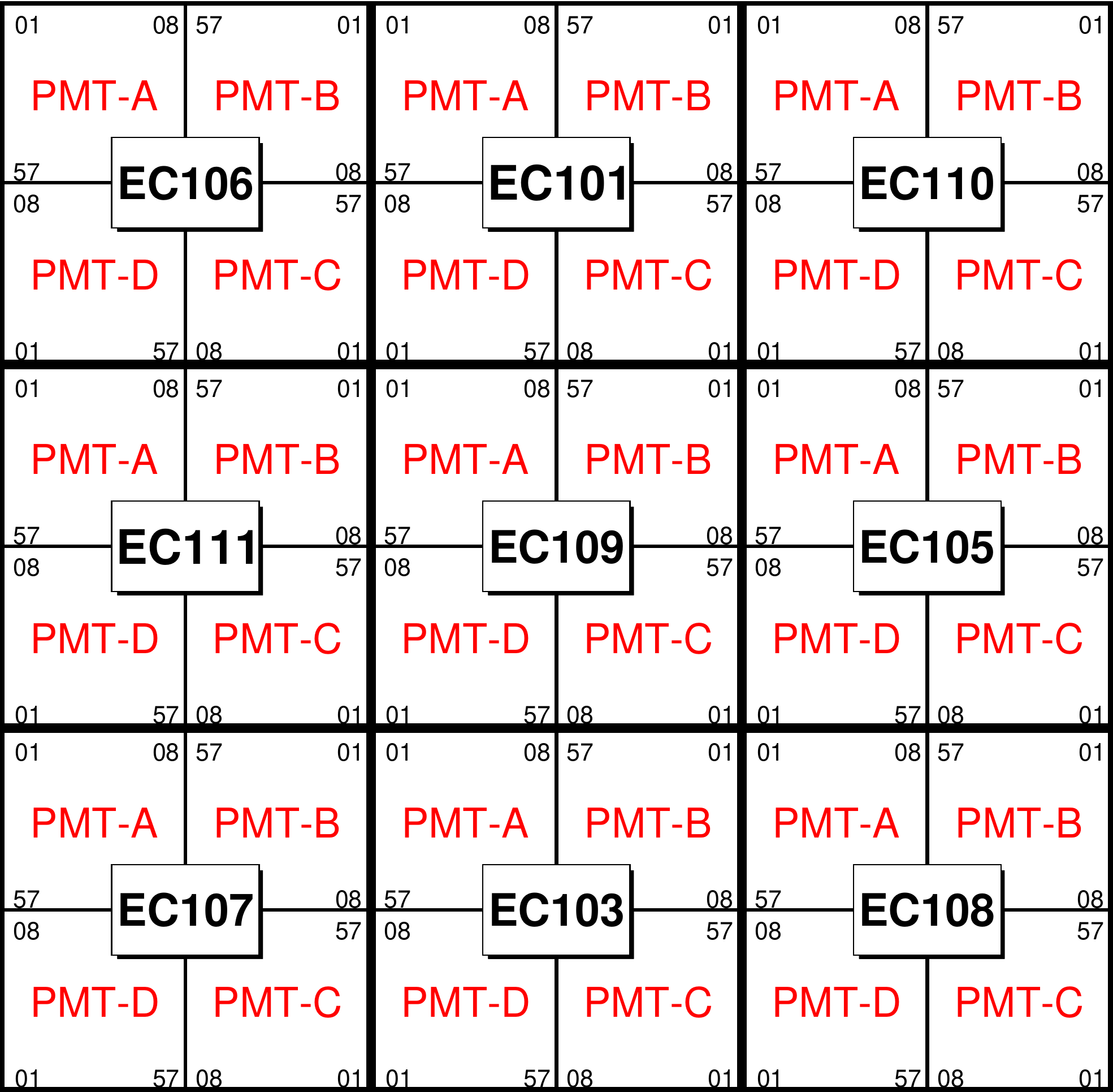}}
\caption[PDM Pixel Mapping]{ \label{fig:Mapdiagrams} Two diagrams which show the definition of the pixel numbers within a PMT and the layout of these pixels within the EUSO-Balloon PDM. 
 \fig\ref{fig:M64pindiagram} shows the Hamamatsu definition of pixels in the M64 PMT, which is used here. In this diagram the PMT is viewed from the back, showing the anode pins. 
Pixel 1 is in the top-right corner next to the cathode pin (labeled K), and pixel 8 is in the top-left corner. \fig\ref{fig:PDMlayout} shows the layout of the EC units within the EUSO-Balloon PDM.
Here the PDM is viewed from the front, facing the photocathodes. Each EC is labeled with a number, and within each EC the PMTs are labeled alphabetically, clockwise, starting from the top-left corner. 
Each PMT is rotated $90^{\circ}$ relative to its neighbor, so that pixel 1 is always on the outside corner. 
This complicated mapping is due to the very compact design of the EC unit.}
\end{figure}

A diagram of the pixel numbering used by Hamamatsu is shown in \fig\ref{fig:M64pindiagram}, and this numbering scheme is used in the DAQ. 
The layout of the Hamamatsu pixels within each EC is shown in \fig\ref{fig:PDMlayout}. Each MAPMT within the 
the EC is rotated $90^{\circ}$ relative to its neighbor, and there are two different flat anode pin cables for PMTs A and D and PMTs B and C respectively. This is due to the compact design of the EC unit (cf.\ \fig\ref{fig:JEM-EUSOM-EC}). 
and results in two different mappings between the QDC channels and the M64 pixels. 

The mapping was the first thing which was checked. Naturally, if the either the assignment of pixel numbers to QDC channels is incorrect or the rotation of the pixels relative to 
the $X$ and $Y$ coordinates of the movement is off, then the centering routine will not converge. It will, for example, calculate a correction to a left which is in fact right, or an up which is in fact down. 
Indeed, in the first runs the centering routing did not converge, because the mapping table was not correctly understood.

Once the mapping was corrected, it was possible to center the light spot, and a series of pixels where checked for each of the four PMTs within the EC. 
It was confirmed from this check that the mapping between Hamamatsu pixels and EC anode connectors is correct.

\subsection{Absolute Efficiency Measurement}
\label{subsec:AbsoluteEC}
\begin{figure}
  \centering
  \includegraphics[width=0.90\textwidth]{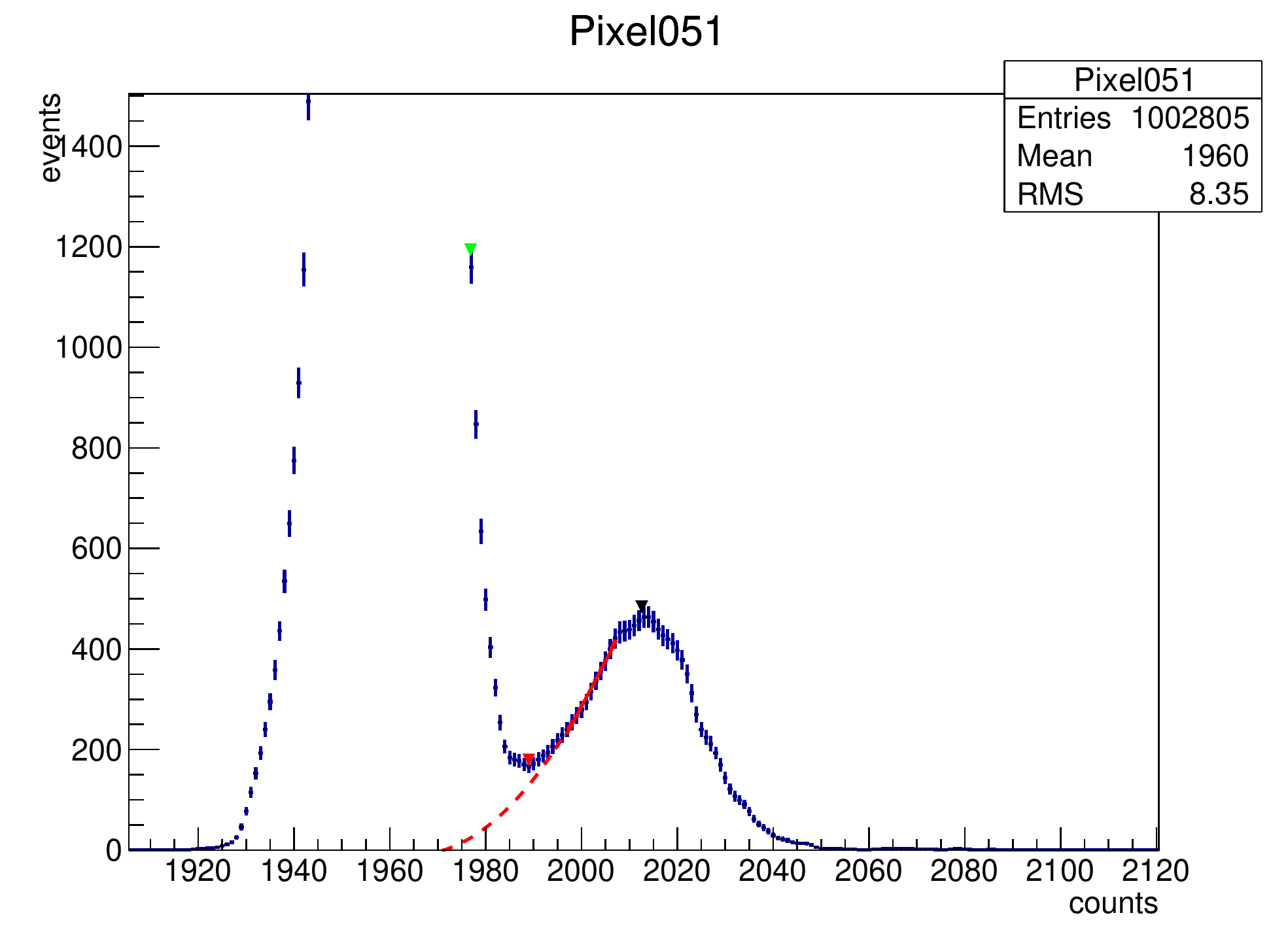}
  \caption[Absolute Spectrum for EC 109, PMT-D]{\label{fig:ReferencePixelSpectrum} The single photoelectron (pe) spectrum taken for pixel 51 of PMT-D of EC 109. This spectrum gives the absolute efficiency, as the pixel was individually illuminated
and so the number of incident photons is given by the power measured by the reference photodiode. 
The pedestal mean is at 1959 counts, and the mean of the single pe peak (shown by the black marker) is at 2012 counts, giving a gain $\mu = 6.59~10^{6}$. 
The number of single pe events above the valley (shown by the red marker) is $N_{\text{pe}}=14214 \pm 119$. The extrapolation of the single pe peak from the valley to the mean of the pedestal (shown by the dashed-red line) gives 724 events.
The resulting efficiency for this pixel is 0.263 above the valley and 0.276 extrapolated, with an uncertainty of 2\% in both cases.
}
 \end{figure}

To measure the absolute efficiency, one EC was mounted in the black box with the photocathode at a distance of 3 mm from the exit of the collimator. The EC anode cable of one MAPMT was connected to the
DAQ. After the box was closed and the HVPS was turned on, the current on each element was checked to make sure that there were no faults with the high voltage of the EC. The correct pixel map for the current PMT was
loaded into the online database of the DAQ, and the automatic centering script was started.

Once the absolute center of the PMT was found, a run script was used which sequentially took spectra for several pixels in both the center and the edge of the MAPMT. Typically these where pixels 17, 22, 33, and 51.
As each pixel was illuminated
individually, the number of photons delivered to that pixel is proportional to the product of the power measured by the reference photodiode (on the integrating sphere) and the attenuation ratio. 
The single photoelectron spectrum analysis routine of the DAQ finds the number of single photoelectron events for each pixel $i$ and calculates the absolute efficiency $\epsilon_{i}$ from \eq\ref{eq:EfficiencyCalculation}).  
This analysis routine is discussed in more detail in chapter~\ref{Chapter:DAQuserguide} of the appendix.

The reference spectrum for pixel 51 of EC 109 PMT-D is shown in \fig\ref{fig:ReferencePixelSpectrum}. The number of photoelectron counts is determined both above the valley (shown by the red marker), and extrapolated to the pedestal mean (the extrapolation is
shown by the red line). 
Unlike in section~\ref{sec:Calibration of Photomultiplier Tubes for the Air Fluorescence Measurements at LAL}, it was not possible to define a efficiency using 1/3 of a photoelectron as the threshold (1/3 pe is shown by the green marker).
The charge corresponding to 1/3 of a photoelectron was generally inside the pedestal, because of the combination of the pedestal width, QDC resolution, and the gain of the M64 PMT (mostly due to the gain). 

The absolute measurement was repeated for each PMT in each of the eight available EC units.
A full scan of every pixel of EC 108, PMT-C was also taken. This pixel-by-pixel scan took 9.5 hours and was used as a cross check. 

\section{Uniform Illumination: Measurement of the Relative Efficiency}
\label{sec:Uniform Illumination: Measurement of the Relative Efficiency}
After one or more absolute spectra were taken for each PMT in the EC, the entire mounting was moved away from the integrating sphere, and the collimator was removed to illuminate the entire PMT at once.
The distance between the photocathode of the MAPMT and the exit port of the integrating sphere
was slightly further than 35 cm. 
In this setup the ratio of single photoelectron counts was again kept below 1.5\% to ensure that the contamination of two photoelectron events was negligible.
To do this, a diaphragm was added to the port of the integrating sphere to reduce the light reaching the photocathode (by reducing the area of the integrating sphere port, cf. \eq\ref{eq:ISExchangeFactor})).
This diaphragm has an opening of 3 mm and is shown in \fig\ref{pic:ThreeCollimators}.

For the illumination of an object using an integrating sphere, the uniformity of the illumination across the object can then be calculated from \eq\ref{eq:Uniformity}). Considering the dimensions of the diaphragm and the PMT, and the distance between then, the ratio of the illumination at 
the edge of one PMT to that at the center of the same PMT is 99.7\%.
Because the collimator is removed, the attenuation between the reference photodiode on the sphere and the PMT is also much lower. While this 
does not matter for the measurement of the absolute efficiency, keeping the PMT in single photoelectron mode required that power on the NIST was only on the order of 30 pW. 
This is not a problem in this measurement, however, as the relative efficiency is determined for all 64 pixels together in one run. The number of photons incident on each pixel is equal within the uniformity of the illumination,
and the absolute number of photons reaching the photocathode is not important as long as the light is low enough that contamination of two photoelectron events in the spectrum is also negligible.

For each PMT, several runs of 1 million events where taken, each run giving 64 spectra, as shown in \fig\ref{fig:64Spectra}. From the resulting single photoelectron spectra, the gain and the number of single photoelectron
counts is extracted by the DAQ analysis routine, as was shown in \fig\ref{fig:64SpectraAnalyzed}. 
Just as for the absolute, pixel-by-pixel spectra, the number of single photoelectron events above the valley (the surface of single photoelectron peak above the threshold), 
the position of the valley, and the number of events from the extrapolation of the single photoelectron peak to the pedestal mean are recorded. 
The relative efficiency of each pixel $\eta_{i}$ is given by the ratio of the number of single photoelectron events $N_{i}$ in each pixel to 
the number of single photoelectron counts in the reference ``NIST'' pixel. The absolute efficiency then comes from the pixel-by-pixel measurement of the efficiency of the ``NIST'' Pixel. 

\section{Efficiency and Gain Results}
\label{sec:Efficiency and Gain Results}
From the measured efficiency of the reference pixel in each PMT and the relative efficiency measurement, the absolute efficiency $\epsilon_{i}$ of each pixel is given by
\begin{equation}
\label{eq:EffMapCalculation}
 \epsilon_{i} = \eta_{i}\epsilon_{\text{ref}}
\end{equation}
In each PMT, the reference pixel with the cleanest spectra was chosen. For three PMTs: PMT-B and PMT-C of EC 102, and PMT-A of EC 103; it was not possible to find a good reference pixel. Due to this, no
results are reported for these PMTs. This does not mean that these PMTs are useless, however, as the ASIC is much more sensitive than the QDCs used in the DAQ. 

For each PMT with a good reference pixel, the extrapolated efficiency
is reported, as this is the only value which is not threshold dependent and is a better value for comparison between pixels and PMTs.
Bad spectra were removed from the uniform measurement by placing requirements (cuts) on the peak-to-valley ratio and the percentage of single photoelectron events which come from the extrapolation (as opposed to above the valley). 
The peak-to-valley ratio
is a figure of merit for the single photoelectron spectrum. 
Similarly, the number of 
single photodetection events from the extrapolation is a function of the separation between the two peaks. The extrapolation has a probable error on the order of 10\%, and so the 
uncertainty on the extrapolated efficiency increases as the percentage of events from the correction increases. The peak-to-valley ratio and the number of events from the extrapolation are correlated.


The exact cuts used were tuned by looking at all 64 pixels of EC 109 PMT-D and determining both the minimum cuts which removed all bad spectra and the maximum cuts which removed no good spectra.
By eye is was determined that 6 of the 64 spectra for this PMT should be rejected, with one additional border-line spectrum. 
This is achieved by a large range of cut choices, and it was decided to use the minimum cuts which rejected the bad spectra, but not the border-line spectrum. 
This was a minimum peak-to-valley ratio of 1.1 and a maximum of 36\% of the total single photoelectron events from the extrapolation. 
Cross-checking these cuts with the spectra from other PMTs showed that this choice is robust, giving a similar rejection of bad spectra for other PMTs.

\begin{figure}
  \centering
  \includegraphics[angle=0, width=1.0\textwidth]{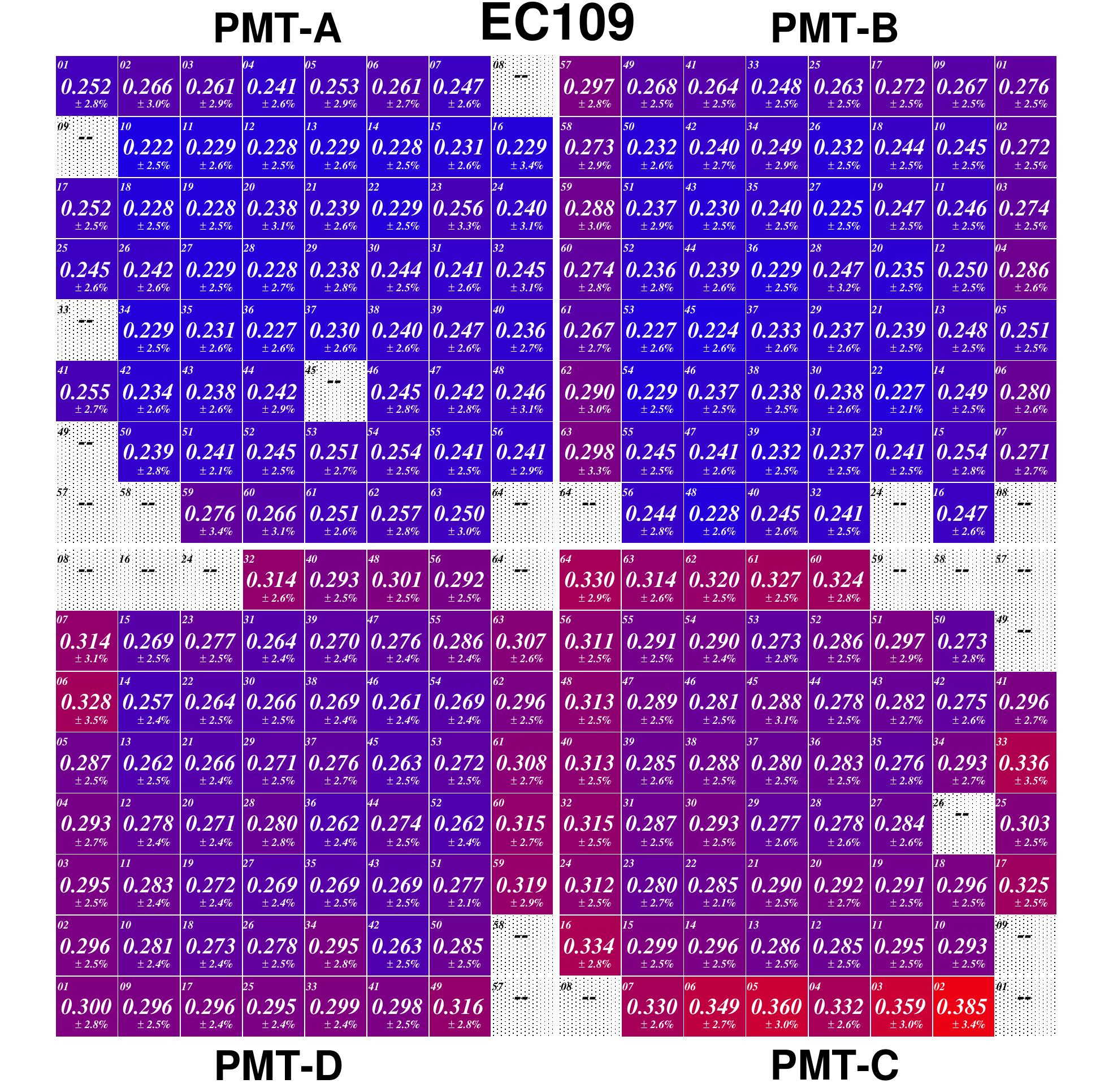}
   \caption[A Map of the Efficiency of EC 109]{\label{fig:EC109_EffPlot} A map of the extrapolated efficiency of each pixel in EC 109 measured at a supply voltage of 1100 V.
Each pixel is represented by a square with the Hamamatsu pixel number given in the upper-left corner.
The central number is the efficiency of the pixel, and the relative uncertainty is given at the bottom of the each square. 
The color of each square is proportional to the efficiency.
Blank squares indicate pixels which gave spectra that did not pass the cut in peak-to-valley ratio and $N_{\text{ex}}/N_{\text{spe}}$, as described in the text. For 
those pixels no result is reported.}
 \end{figure}
 
\begin{figure}
  \centering
  \includegraphics[angle=0, width=1.0\textwidth]{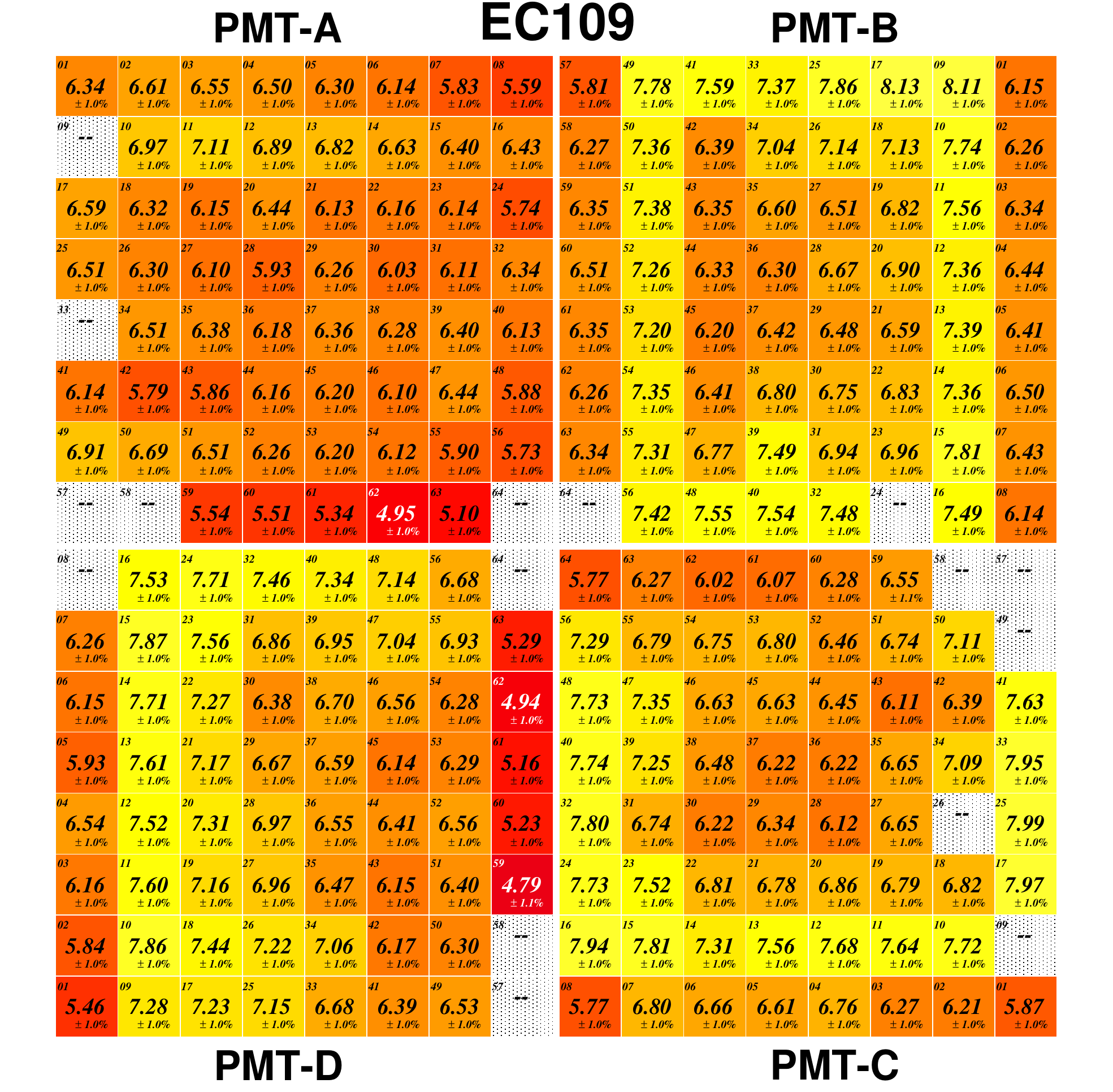}
   \caption[A Map of the Gain of EC 109]{ \label{fig:EC109_GainPlot} A map of the gain of each pixel in EC 109 measured at a supply voltage of 1100 V.
Each pixel is represented by a square with the Hamamatsu pixel number given in the upper-left corner.
The central number is the single photoelectron gain of the pixel, with the relative uncertainty given at the bottom of each square. 
The color of each square is proportional to the gain.
Blank squares indicate pixels with a spectra in which the peak-to-valley ratio was less than 1.10.
}
 \end{figure}

\begin{figure}[p]
\centering
  \includegraphics[angle=0, width=1.0\textwidth]{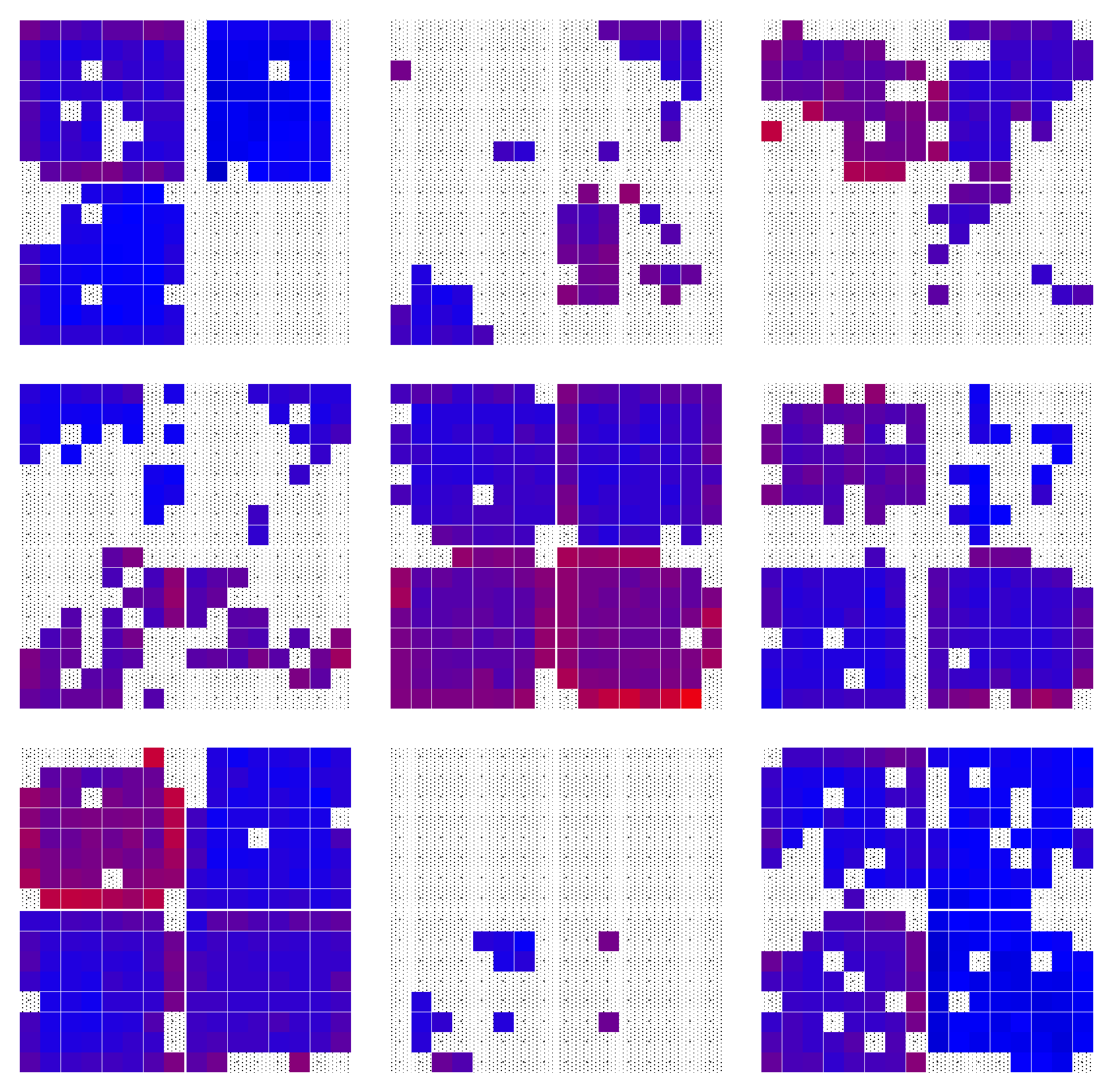}
   \caption[A Map of the Efficiency of the EUSO-Balloon Photodetection Module]{ \label{fig:PDMeff_v1} A map of the efficiency for the entire EUSO-Balloon PDM. 
The layout of the PDM can be seen in \fig\ref{fig:PDMlayout}. 
The color scale in this plot is the same as in \fig\ref{fig:EC109_EffPlot}.
Grayed-out squares indicate pixels for which no result is given.
No result is given for PMT-C of EC 106 (top-left), PMT-D of EC 110 (top-right), and  PMT-A and PMT-B of EC 103 (bottom-middle) as these PMTs did not 
give any spectra which passed the quality cuts. It is emphasized again that this is only because the QDC is not as sensitive as the ASIC.}
 \end{figure}

\begin{figure}[p]
\centering
  \includegraphics[angle=0, width=1.0\textwidth]{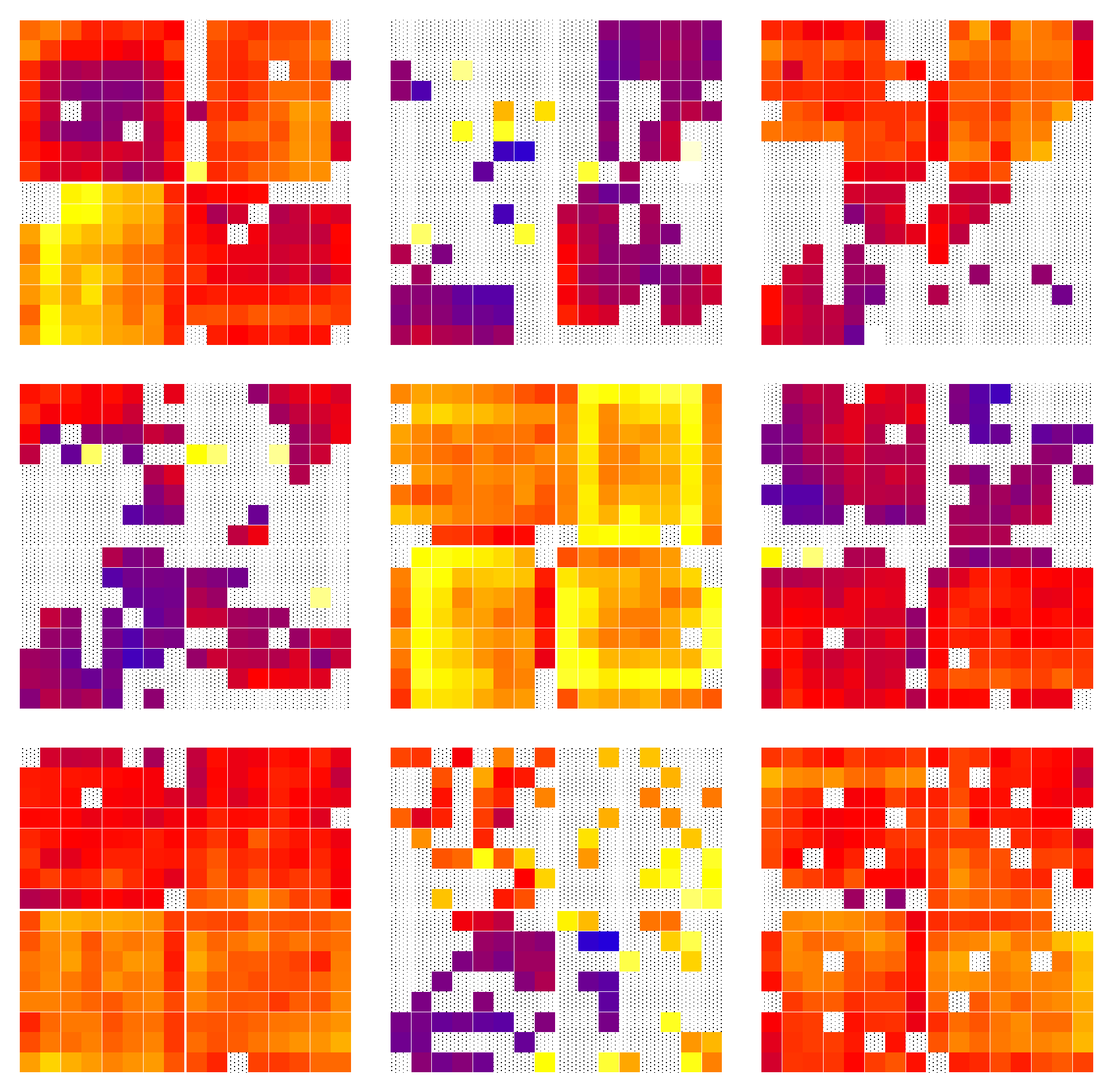}
   \caption[A Map of the Gain of the EUSO-Balloon Photodetection Module]{ \label{fig:PDMgain_v1} A map of the gain for the entire EUSO-Balloon PDM. 
The color scale in this plot is the same as in \fig\ref{fig:EC109_GainPlot}.
The layout of the PDM can be seen in \fig\ref{fig:PDMlayout}.
Grayed-out squares indicate pixels for which the peak-to-valley ratio of the single photoelectron peak was less than 1.10.}
 \end{figure}

The resulting map\footnote{These plots where created with the author's own program, using ROOT.} of the efficiency of each pixel of EC 109 PMT-D is shown in \fig\ref{fig:EC109_EffPlot}, 
and a similar map of the gain for the same PMT can be seen in \fig\ref{fig:EC109_GainPlot}. 
The efficiencies $\epsilon_{i}$ of every PMT in EC 109 can be found in the tables in chapter~\ref{appendix:chapter:Figures} of the Appendix.
Finally, a map of the efficiency of gain of the entire PDM, with the EC units laid out according to \fig\ref{fig:PDMlayout} can be seen in Figs.~\ref{fig:PDMeff_v1} and \ref{fig:PDMgain_v1}. 

The final uncertainty on each $\epsilon_{i}$ is
\begin{equation}
 \label{eq:TotalUncertaintyECeff}
 \left|\frac{\delta \epsilon_{i}}{\epsilon_{i}}\right| \leq \left|\frac{\delta \epsilon_{\text{ref}}}{\epsilon_{\text{ref}}}\right| + \left|\frac{\delta \eta_{i}}{\eta_{i}}\right|
\end{equation}
where the error on the absolute efficiency of the reference pixel is given by \eq\ref{eq:Total}). 
As in section~\ref{subsubsec:AFYPMT:Absolute Efficiency}, the largest contribution to the uncertainty is the 1.5\% systematic error from the calibration of the NIST photodiode. In every spectrum taken, the ratio of 
one photoelectron events to pedestal events is 1.5\% or less, so that the contamination of two photoelectron events is less than 0.75\% of the number of one photoelectron events. This percentage is added to the systematic uncertainty, as it represents 
a systematic under-counting of two as ones. 

For the extrapolated efficiency, the statistical error on the number of one photoelectron events takes into account the estimated 10\% uncertainty on the number of events found by the extrapolation to the pedestal mean. The 0.5\% statistical
error on the measurement of the current from the NIST photodiode is also included.
All of these uncertainties are added in quadrature, giving a total uncertainty on the order of 2.0\% for $\epsilon_{\text{ref}}$. The exact value depends on the percentage of single photoelectron events which come from the extrapolation
in each spectrum and therefore on the correlation between the gain and the number of counts above the valley for each pixel. 

The uncertainty on each relative efficiency $\eta_{i}$ is purely statistical. The error from the contamination of two photoelectron events cancels in the calculation of $\eta_{i}$. The statistical error on the number of single photoelectron
events, the uncertainty from the extrapolated correction, and the uncertainty on the uniformity of the illumination remains. 
The calculated value of the uniformity at the edge of the PMT is 99.7\%, so in practice the illumination was assumed to be completely uniform with an uncertainty of 1\%.
This gives an uncertainty on the relative efficiency on the order of 1.3\%, again depending on the spectrum.

The uncertainty on the relative and the reference efficiencies are added in quadrature, and so the total uncertainty on the calculated absolute efficiency is on the order of 2.5--3\%.
Even though the efficiency of the PMT and the single photoelectron gain can be measured separately in a single photoelectron spectrum, that is not to say that the two parameters are independent. 
Obviously, the useful efficiency, which is proportional to the number of counts above the valley, will be higher if the single photoelectron gain is larger. Here the efficiency has been extrapolated
below the valley to the pedestal mean, in order to better compare one pixel or PMT to another. 
In principle, the extrapolated efficiency should be independent of the gain of the pixel, as long as the gain is high enough to give a good single photoelectron peak so that the extrapolation itself is good.

In practice, however, the extrapolation is determined by fitting the backside of the single photoelectron peak with a 4th-order polynomial, and the estimated uncertainty on the resulting number of single photoelectron counts
between the valley and the pedestal mean is $\sim10\%$.
This means that the extrapolated efficiency is still sensitive to the gain, as a lower gain will result in larger percentage of events from the extrapolation and thus a large uncertainty on the 
efficiency. This was accounted for in the calculation of the total uncertainty given above, but, as the gain decreases, the extrapolation itself is also likely to be less precise.
It is therefore very probable that the uncertainty on the extrapolation, and therefore on the extrapolated efficiency is asymmetric.
It is more likely, for example, that the extrapolation  overestimates the number of single photoelectron events below the valley, rather than underestimating this number. 
No attempt has been made here to quantify this, but it should be kept in mind when looking at the results.

There are also two general points which must be taken into account in the measurement of the efficiency of individual pixels of MAPMT: 
\begin{inparaenum}[i\upshape)]
\item  the averaging of the efficiency over a pixel, and
\item the escape of single photoelectrons from one pixel to its neighboring pixels.                  
\end{inparaenum}
Just as in a PMT with a single large photocathode, the measured efficiency is the average over the area of the photocathode on which the light spot is incident.
For high-precision single photoelectron counting applications is it mandatory to restrict the incident light to
an area of the photocathode small enough that the variation in efficiency is less than the desired uncertainty, and to measure the efficiency at the exact same region of the photocathode which will be used.
In the case of JEM-EUSO, the focal surface of the telescope is created from MAPMT pixels, and so the light is distributed across each entire pixel. This implies that the actual working efficiency of 
each pixel is closer to the one measured with the pixels uniformly illuminated. 

As discussed at the beginning of the chapter, however, the absolute efficiency can only be measured if the number of photons incident on the pixel is well-known.
The simplest way to do this is to use a collimator or optical fiber so that the light is completely contained inside the pixel being measured. In this case however, the measured efficiency
will  not include the response of the pixel to photons incident near the edges of the pixel, and so it will be slightly different than the efficiency which would be measured with uniform illumination (if the number of incident photons
was well known in both cases). In the measurement of the efficiency presented here, the relative efficiency was measured using uniform illumination, and so the efficiency of each pixel relative to one another
takes into account the entire surface of each pixel. This therefore also automatically accounts for differences in the surface area of each pixel. 

The absolute efficiency of the reference pixel was measured using a small light spot, however, and so this absolute scale does not take into account the entire surface of the reference pixel. 
The difference between the efficiency of a small spot within a pixel and the efficiency averaged over the entire pixel can be accounted for by mapping the reference pixel, as will be done in the section~\ref{sec:PixelScanningResults}.

The second effect which must be considered is the escape of photoelectrons from one pixel to its neighboring pixels. This is simply that with one pixel illuminated at a time, an emitted photoelectron can fail to be collected in that pixel
and instead be collected in one of the adjacent pixels. This photoelectron must be accounted for, because it does result in a count on some anode, but it is not included in the spectrum of the illuminated pixel. 
P. Gorodetzky et al.~\cite{GoroPrivate} measured this by looking at a block of nine pixels with only the central pixel illuminated. 
They did a correlation analysis to eliminate events that appeared simultaneously in two or more pixels (cross talk events) and found that a correction of 4\% should be included to
 account for the photoelectrons which were collected in the neighboring pixels.

The relative efficiency, automatically includes this correction, as each pixel loses counts to its neighbors while receiving counts as well. The correction is therefore a correction to the absolute scale of the efficiency in each PMT. The 
results shown in Figs.~\ref{fig:EC109_EffPlot} and~\ref{fig:PDMeff_v1} and in the tables in section~\ref{app:sec:EfficiencyTables} \emph{do not} include this correction, which can be applied by adding 4\% to all the reported values.

\subsection{Statistics on the Single Photoelectron Gain and the Efficiency}
\label{subsec:Statistics on the Single Photoelectron Gain and the Efficiency}

\begin{figure}[p]
\centering
 \includegraphics[angle=0,width=1.0\textwidth]{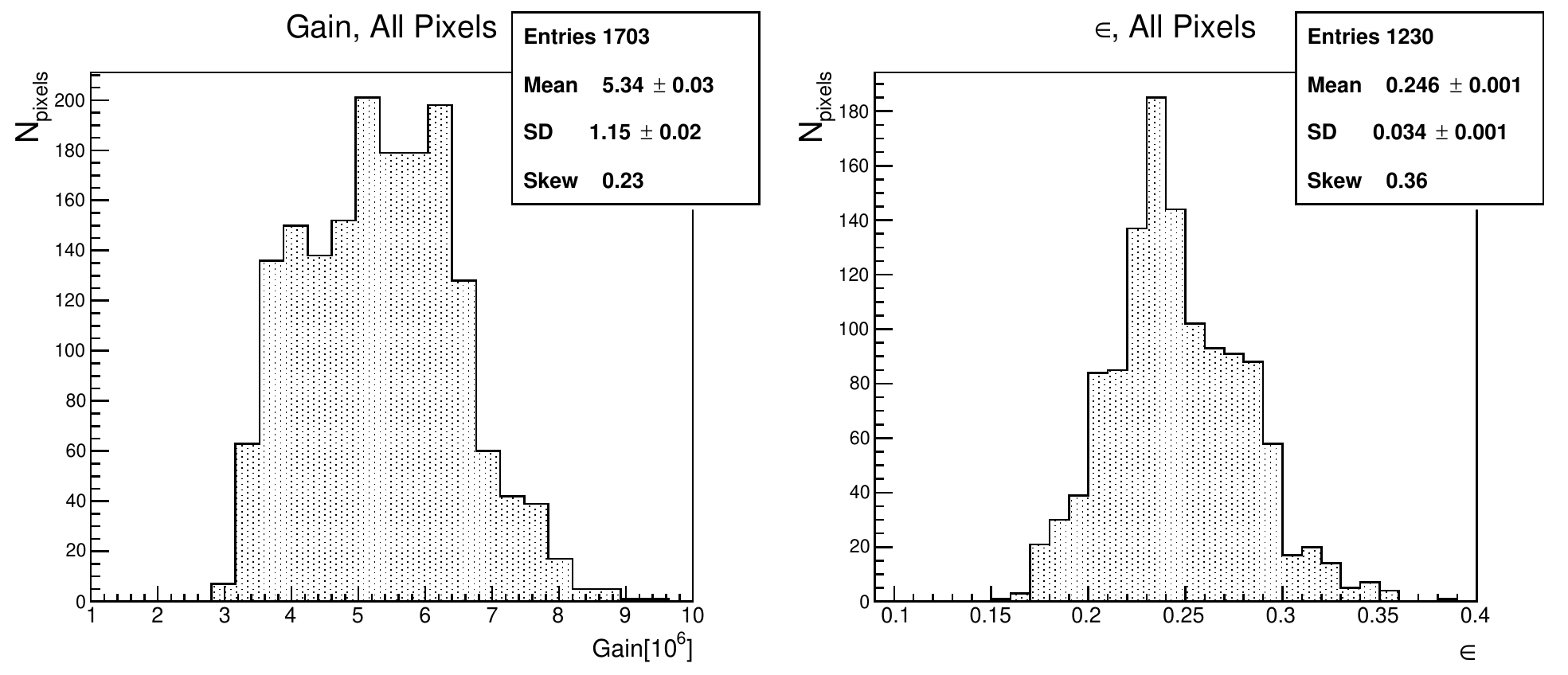}
\caption[Statistics of All Measured Pixels]{\label{fig:AllpixelsStatistics} Histograms of the gain and efficiency across all measured pixels. The gain includes only those pixels with spectra which had a peak-to-valley ratio of 1.1 or better. The efficiency
histogram includes only pixels for which the spectra passed the cuts discussed in the text. See the text for a discussion of these two histograms.}
 \end{figure}

These results, for over 1000 pixels in single photoelectron mode, allow us to estimate the spread of the efficiency $\epsilon$ and single photoelectron gain $\mu$ within EUSO-Balloon and in future M64 MAPMTs. For the gain, the histograms include 
every pixel which gave a spectrum with a peak-to-valley ratio higher than $1.10$, while the efficiency histograms include every result which had a peak-to-valley ratio greater than 1.1 and $N_{\text{ex}}/N_{\text{total}}$ less than 0.36.
A histogram of the $\mu$ and $\epsilon$ of all measured pixels is shown in \fig\ref{fig:AllpixelsStatistics}.
The average gain of all measurable pixels was $5.34~10^{6}$ and their average efficiency (extrapolated) was 0.246. 

Here it can be seen that the standard deviation of $\mu$ across all pixels is 22\% of the average, while 
the standard deviation of $\epsilon$ is 14\% of the average value. Both distributions have a positive skewness, indicating that the distributions have a larger tail towards positive values than negative ones.
 This points out the obvious bias in this data set: if the gain of a pixel is too low, then no result can be obtained for that pixel, and so the distribution will be truncated on the lower side by the quality cuts.
For this reason the estimate of the spread of these values should be considered as a lower limit.
 
\begin{figure}
  \centering
  \includegraphics[angle=0, width=1.0\textwidth]{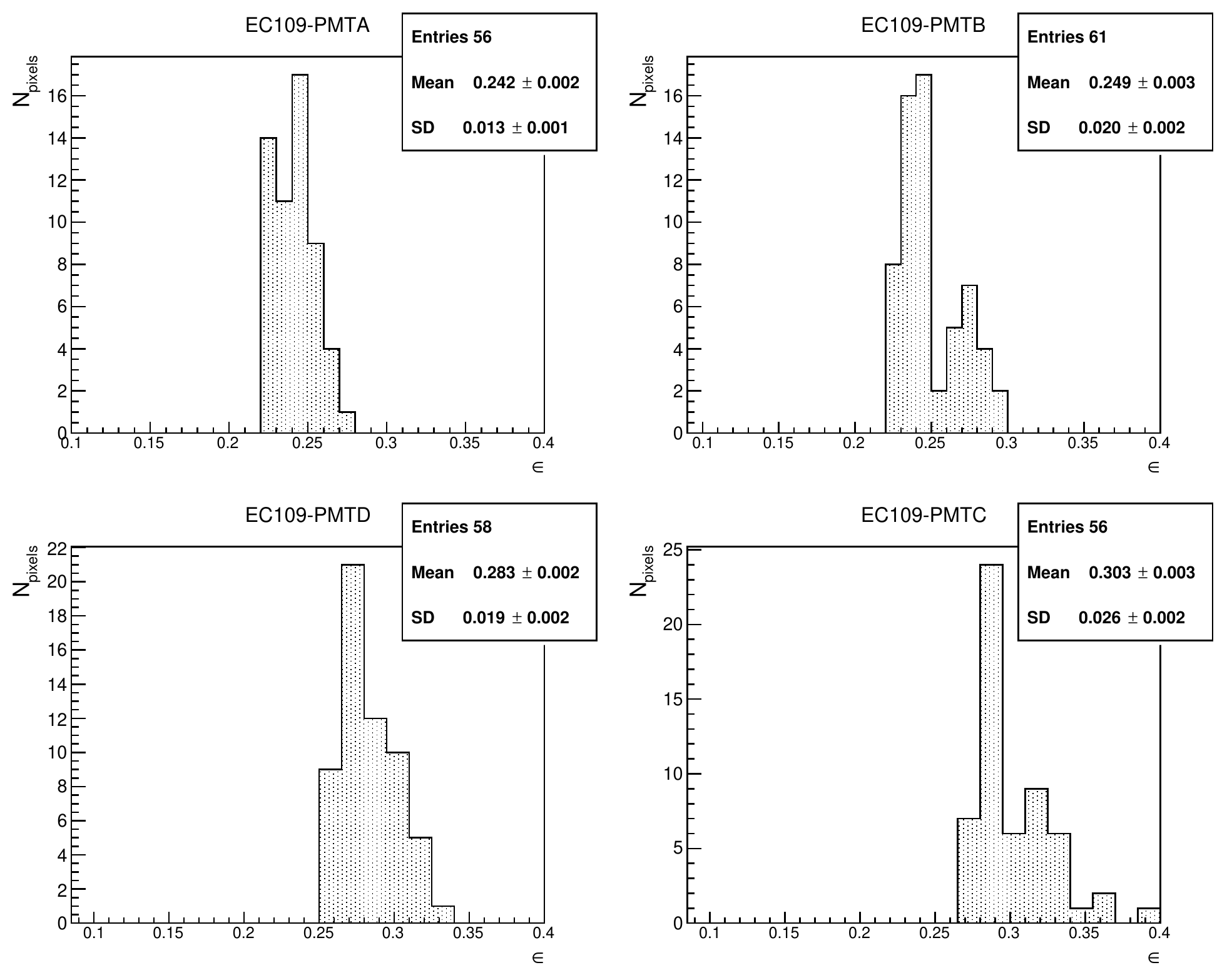}
   \caption[Histograms of the Efficiency in EC 109]{\label{fig:EC109-eff_histo} Four histograms of the efficiency across pixels, one for each PMT in EC 109. Only pixels which passed a cut in peak-to-valley ratio and $N_{\text{ex}}/N_{\text{total}}$
are included, as described in the text.}
 \end{figure}

\begin{figure}
  \centering
  \includegraphics[angle=0, width=0.95\textwidth]{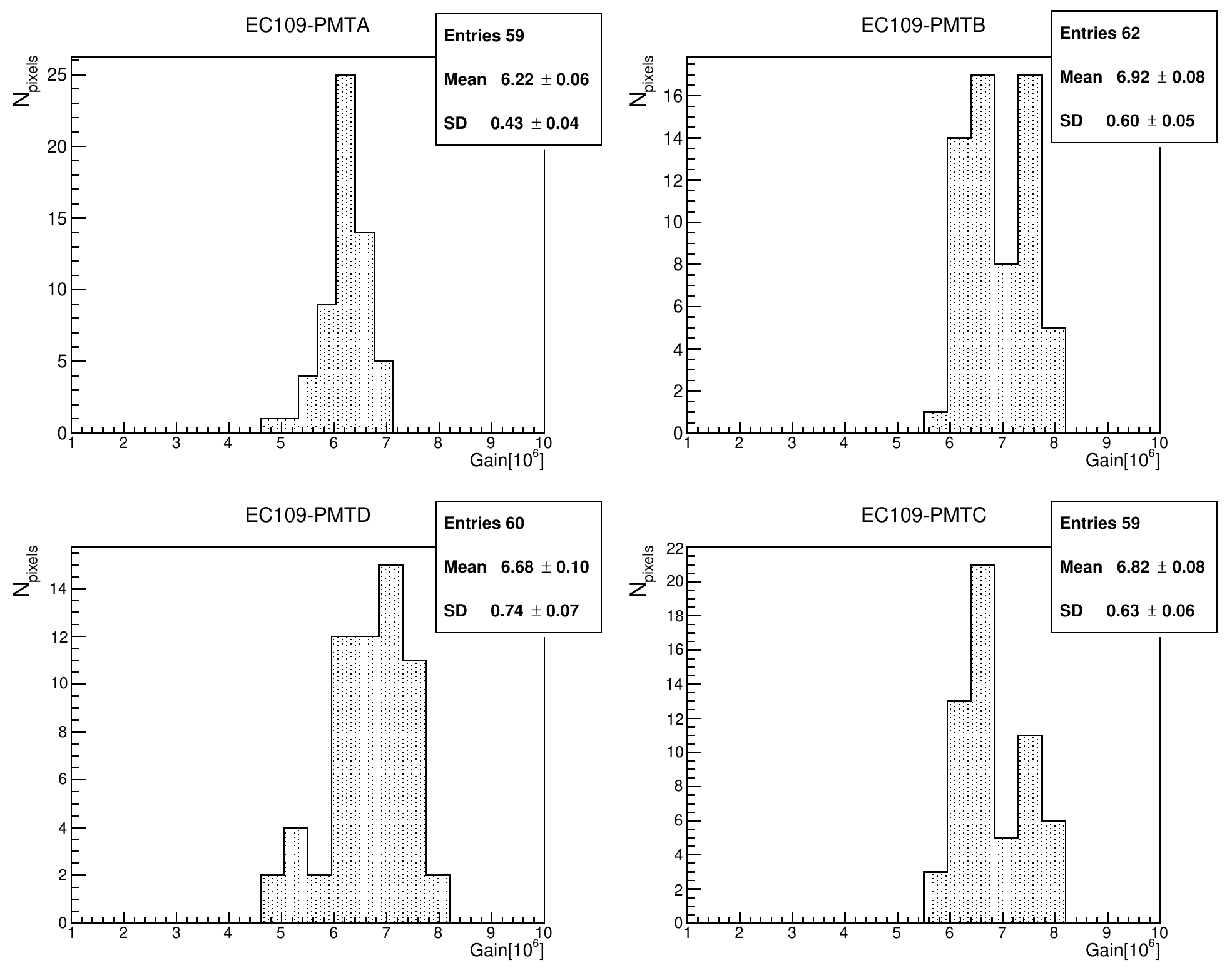}
   \caption[Histograms of the Gain in EC 109]{  \label{fig:EC109-gain_histo} Four histograms of the gain across pixels, one for each PMT in EC 109. Only pixels which passed a cut in peak-to-valley ratio 
are included.}
 \end{figure}

Within each PMT, the mean and the standard deviation of the gain and efficiency can be calculated across pixels. 
These statistics within a PMT are more important, as they relate directly to the sorting of the PMTs into EC units. Namely, the typical spread of the gain within 
a PMT sets a limit on how close together the average gain of the 4 PMTs must be.
Such results are shown in Figs.~\ref{fig:EC109-eff_histo} and~\ref{fig:EC109-gain_histo} for the efficiency $\epsilon$ and gain $\mu$ of the four PMTs in EC 109. The average gains of these four PMTs are reasonably close together, being
6.22, 6.92, 6.82, and $6.68~10^{6}$ for PMT A, B, C, and D respectively. In the case of the efficiency, PMTs D and C show a slightly higher average efficiency than PMTs A and B. 
The standard deviation of the gain within each PMT is on the order of 7-11\%, while the standard deviation of the efficiency is on the order of 5-8\%.
The results for this EC unit are relatively complete, having very few rejected pixels, and so the average and standard deviation of the gain and efficiency should be robust.
There is still a slight bias, however, from the fact that the extrapolated efficiency will have a larger $N_{\text{ex}}/N_{\text{total}}$ in those pixels with a lower gain, and so the uncertainty on
the efficiency result in those pixels will be larger.   

\begin{figure}[p]
\centering
  \subfigure[Mean]{\label{fig:MetaStats:Mean}\includegraphics[angle=0,width=1.0\textwidth]{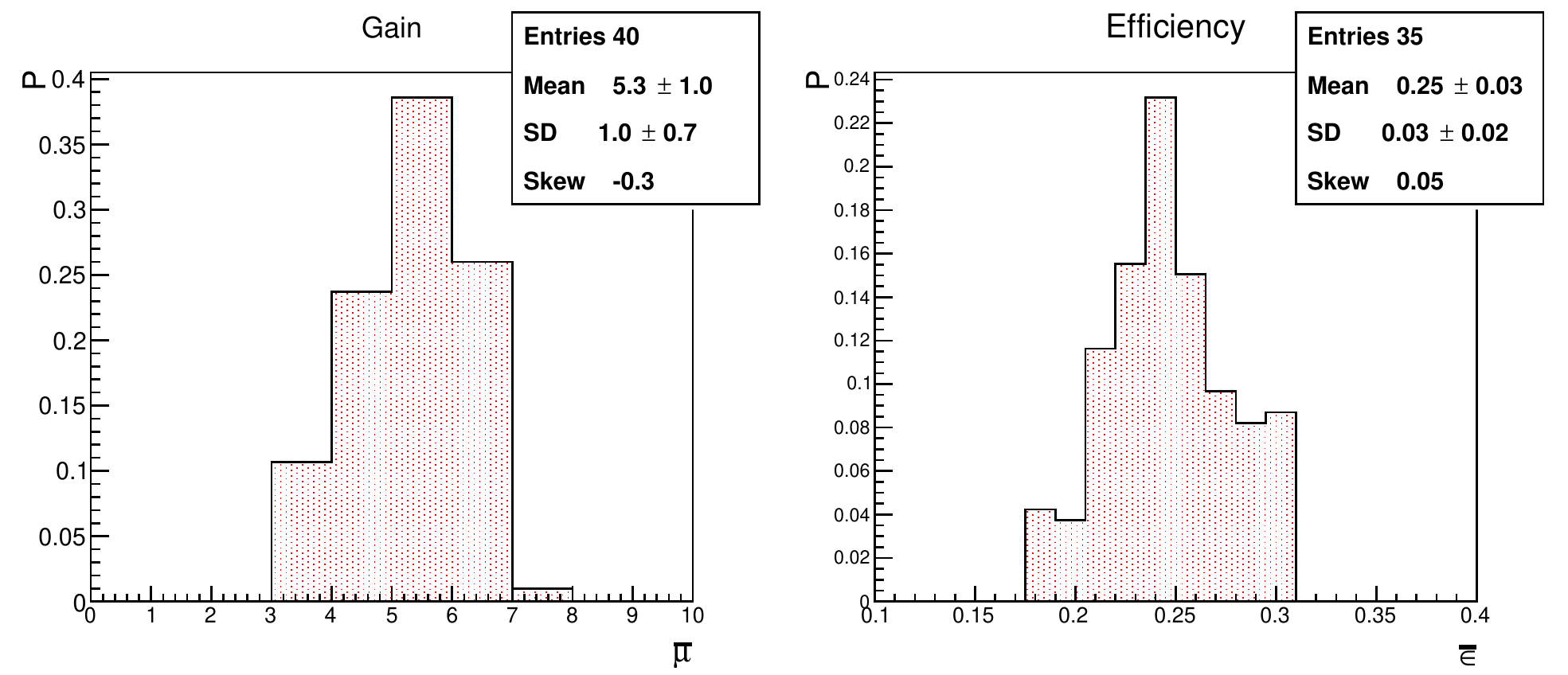}}\\
  \subfigure[Standard Deviation]{\label{fig:MetaStats:Sigma}\includegraphics[angle=0,width=1.0\textwidth]{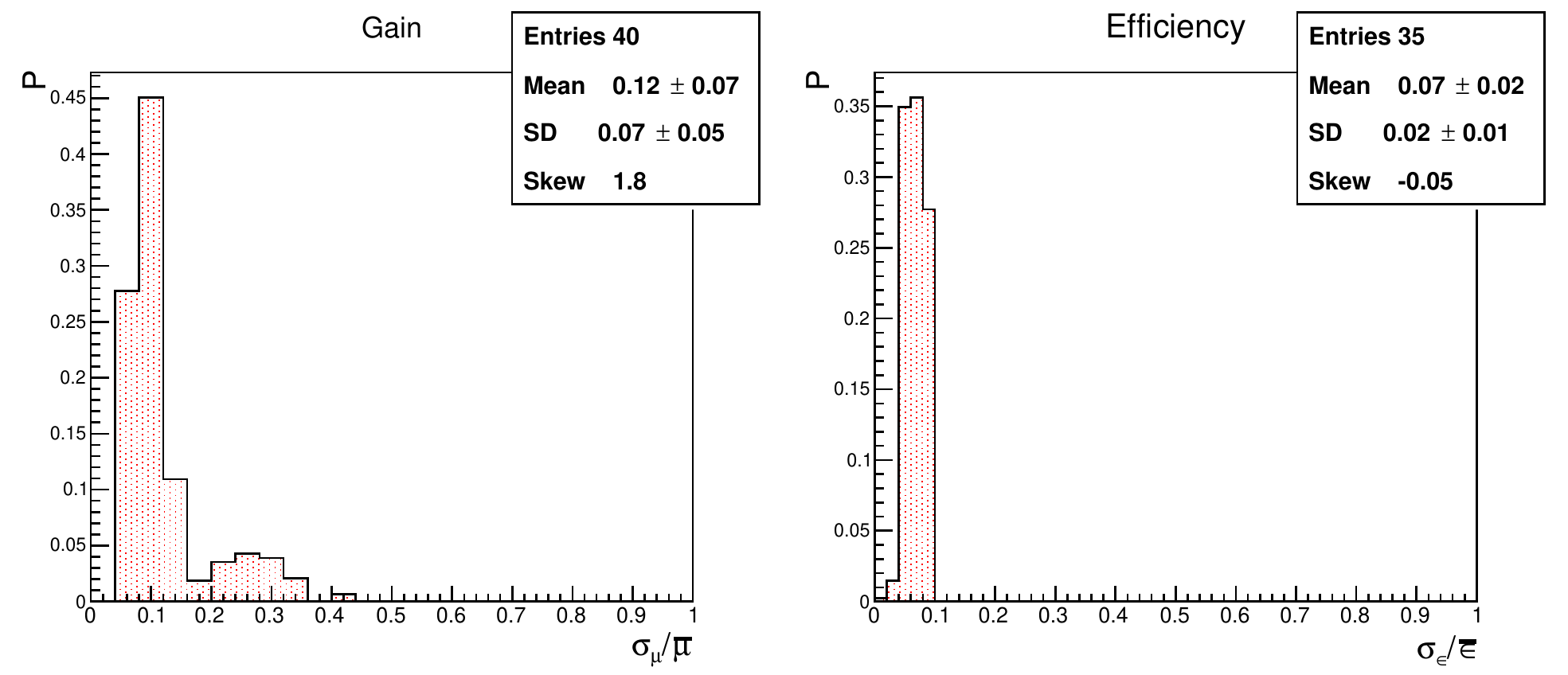}}
\caption[Histograms of the Mean and Standard Deviation Across PMTs]{ \label{fig:MetaStats} Histograms of the mean, \fig\ref{fig:MetaStats:Mean}, and standard deviation, \fig\ref{fig:MetaStats:Sigma}, of the gain and efficiency across all measured PMTs. 
The standard deviation is expressed as a percentage of the mean. It can be seen that the typical standard deviation of the gain within a PMT is 12\% of the mean value and the typical spread of the efficiency within a PMT is 7\% of the mean value.
}
 \end{figure}

The average gain and efficiency of all 40 PMTs are histogrammed in \fig\ref{fig:MetaStats:Mean}. Here, the average result from each PMT was histogrammed with a weight given by the number of 
good pixels (spectra which were not rejected) in the PMT. The average gain across PMTs is $5.3~10^{6}$ with $\sigma_{\bar{\mu}}=1.0$ (the standard deviation divided by the average) or 19\%. The average efficiency across
PMTs is 0.25 with $\sigma_{\bar{\epsilon}}=0.03$ or 12\%. This gives an indication of the typical gain and efficiency which can be expected on a PMT-by-PMT basis (at, of course, 1100 V).
As before, these results are biased towards higher gains, and so are most likely overestimating the average gain and underestimating the standard deviation. This bias is less for the efficiency, as
the efficiency of each pixel is not completely correlated with the gain of the pixel. 

\fig\ref{fig:MetaStats:Sigma} shows similar histograms of the standard deviation of the efficiency and gain, expressed as a percentage of the average. 
It can be seen in the two histograms that the typical standard deviation of the extrapolated efficiency within a PMT is on the order of $7~(\pm 2) \%$,
 and $12~(\pm 7)\%$ for the gain. 
The distribution of  $\sigma_{\mu}/\bar{\mu}$ has large positive skewness. This is most likely from the cut of the low-gain tail, which affects this parameter twice, 
once in the calculation of $\sigma_{\mu}$ and again in an overestimation of $\bar{\mu}$. This would imply that this estimate of the mean of
$\sigma_{\mu}/\bar{\mu}$ is probably low. As before, $\sigma_{\epsilon}/\bar{\epsilon}$ is less sensitive to this bias, and so the estimate for this parameter is likely more robust.
 
\section{Pixel Scanning Results}
\label{sec:PixelScanningResults}
Beyond measuring the efficiency of each pixel, the power of the DAQ can be leveraged to further characterize the M64 PMT. This includes the measurement of the effective 
pixel width, and a characterization of the response of the pixels at the edge of the PMT. 
To perform such measurements, a further analysis routine was written based on the centering algorithm. This routine simply returns the number of single photoelectron events above 
a defined threshold for a series of requested pixels and prints this number to a list. The number of single photoelectrons counts is normalized according to the power received
by the NIST photodiode on the sphere in order to account for fluctuations in the output of the LEDs within and between runs. 
Using the run sequencer of the DAQ, a series of positions on the photocathode can be scanned to give the 
response of each pixel with the light spot at that point.

\begin{figure}
\centering
\includegraphics[angle=0,width=1.0\textwidth]{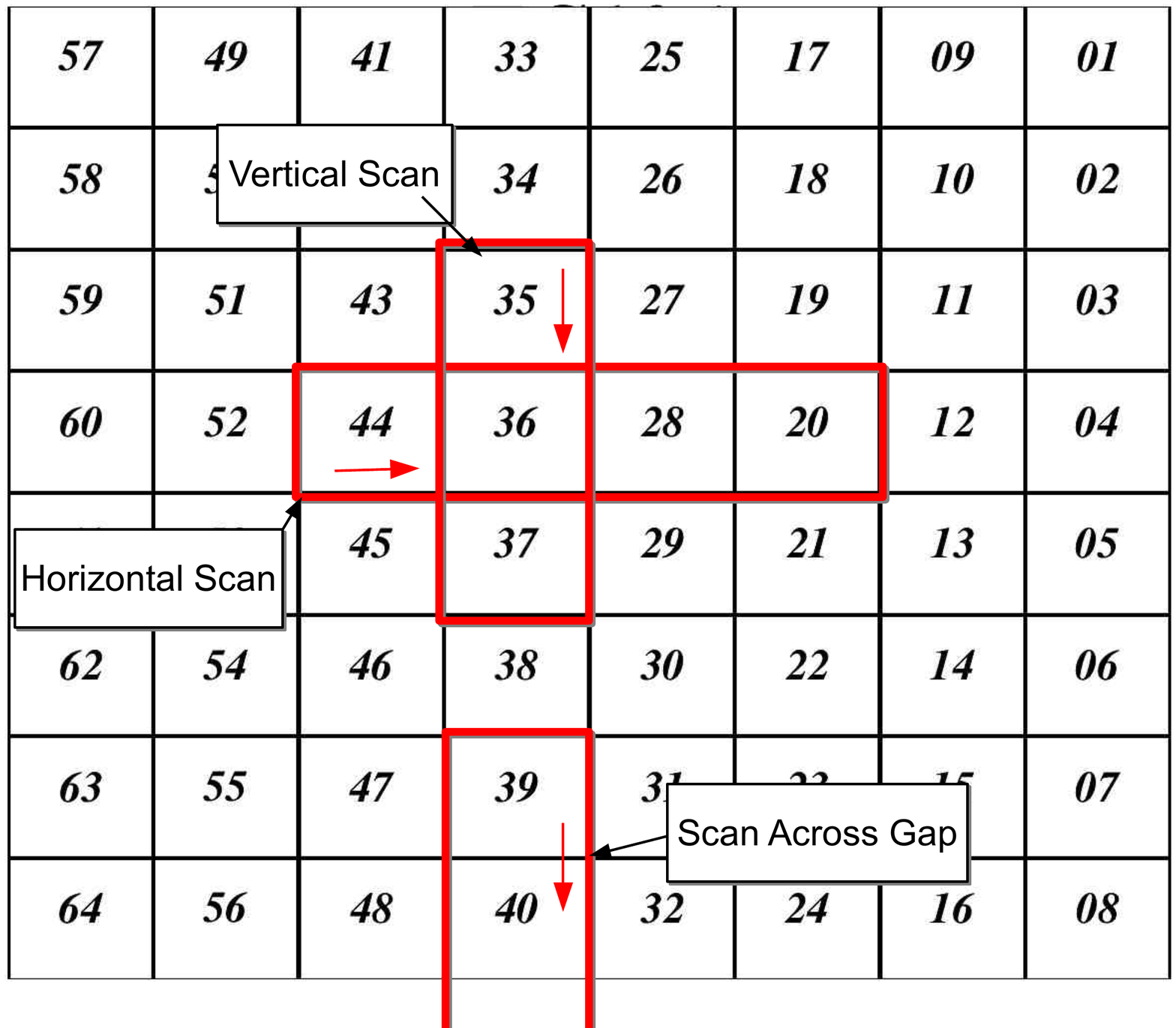}
\caption[Diagram of Pixels Scanned]{ \label{fig:PixelScan:Diagram} A diagram showing the orientation of the PMT and the pixels scanned. The pixel numbers are according to the Hamamatsu definition (shown in \fig\ref{fig:M64pindiagram}).
The photocathode is seen from the front in this diagram. With this setup, three independent scans where performed: i) a vertical scan starting in pixel 35, ii) a horizontal scan starting in pixel 44, and iii) a scan across the gap between PMTs in the EC,
starting in pixel 39 of PMT-A and ending in pixel 61 of PMT-B.
}
\end{figure}

The same setup as for the absolute measurement (section \ref{subsec:AbsoluteEC}) was used. The exit from the collimator had a diameter 0.03 mm,
giving a light spot of approximately the same size at a distance of only a few millimeters.
After first centering the X-Y movement, a horizontal scan across pixels 44, 36, 28, and 20  of EC 108 PMT-A was performed. 
The EC was placed in the black box with PMT-C on the bottom-left corner, this puts PMT-A in the top-left corner, with pixel 1 in the top-left corner.
A diagram of the pixel orientation, with scanned pixels marked, is shown in \fig\ref{fig:PixelScan:Diagram}.

A single photoelectron spectrum was first taken for each pixel to find the valley, which was used as the threshold for single photoelectron counting.
For each position, the number of single photoelectron counts, normalized to the power received by the NIST photodiode on the sphere, was returned after a run of 100 thousand events. 
The total scan was done in a series of 44 steps, with a movement of $\Delta X =0.144~$ mm per step. 

\begin{figure}
\centering
\subfigure[Horizontal]{\label{fig:PixelScan:Horizontal} \includegraphics[angle=0,width=0.7\textwidth]{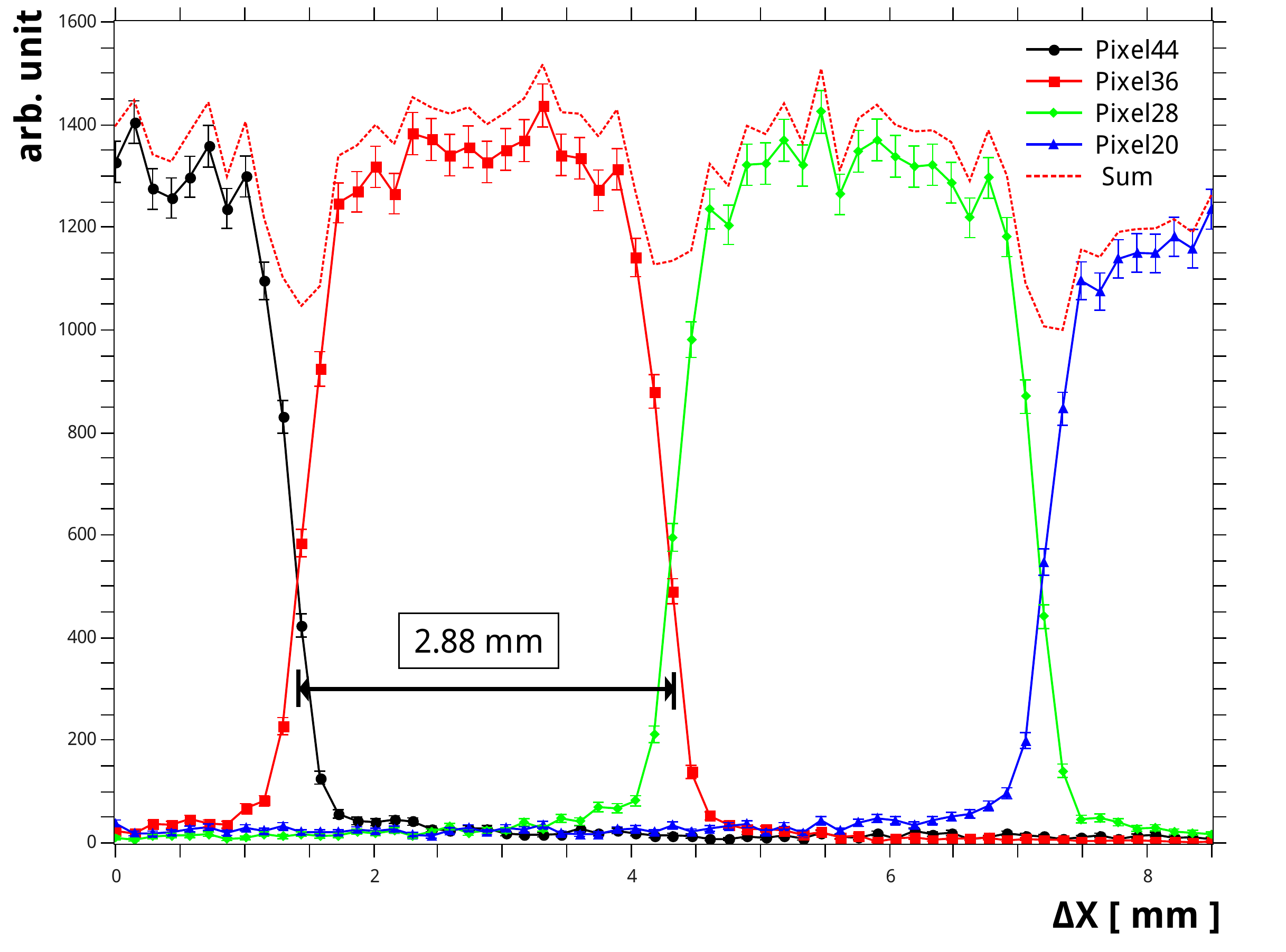}}
\subfigure[Vertical]{\label{fig:PixelScan:Vertical} \includegraphics[angle=0,width=0.7\textwidth]{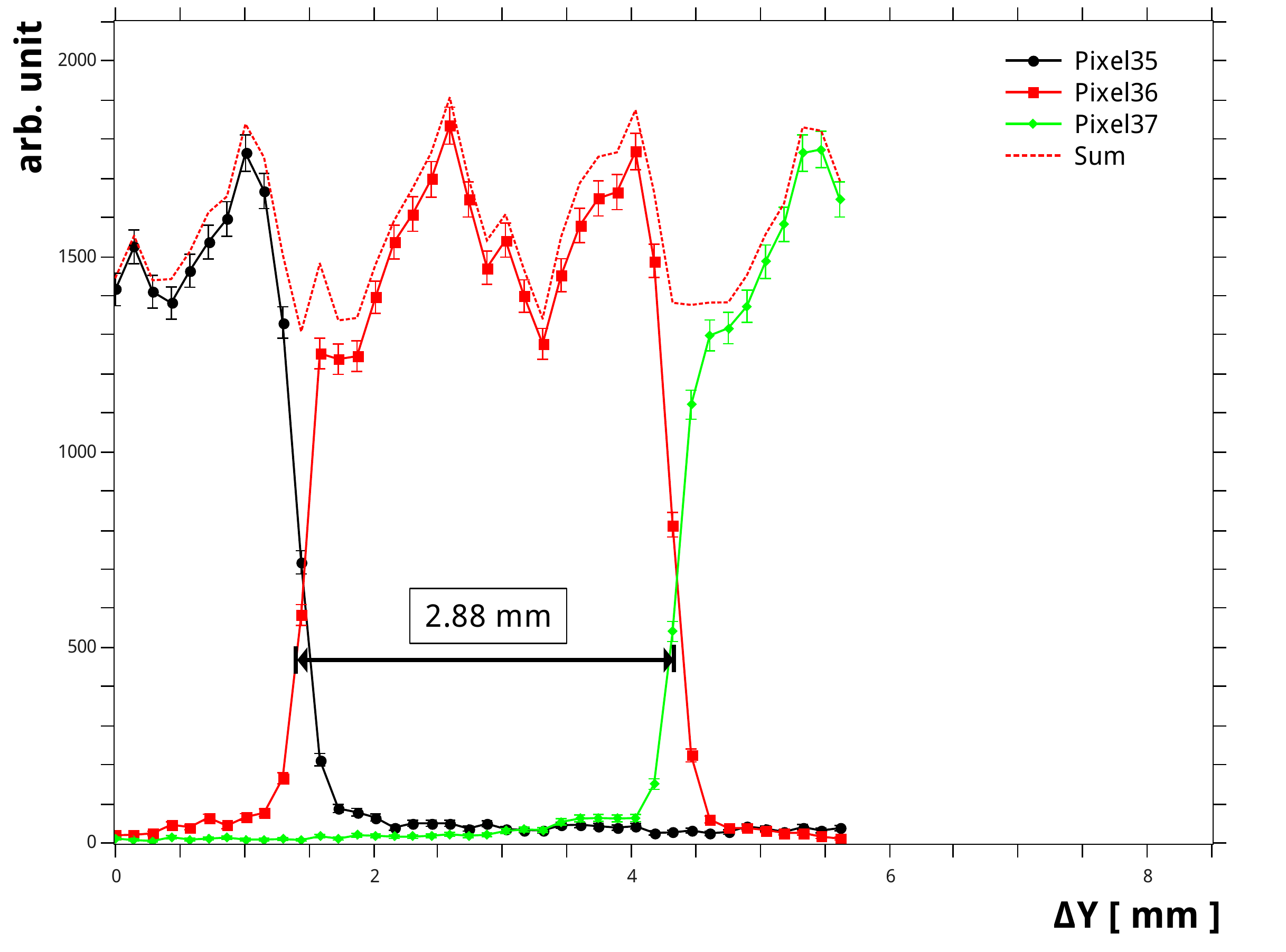}}
\caption[Scan of a Central Pixel]{ \label{fig:PixelScan} 
The results for horizontal and vertical scans through several central pixels. The parameters in the scan are discussed in the text and shown in \fig\ref{fig:PixelScan:Diagram}. 
The dashed red line shows the sum of counts in all measured pixels. The width of each pixel is consistent with the value of 2.88 mm given by Hamamatsu. An area of lower 
efficiency can be seen between pixels, where the total number of single photoelectron events is $\approx 20\%$ lower than in the center of the pixel. A structural difference can 
also be seen within each pixel in the horizontal vs. the vertical scans 
}
\end{figure}

The response of each pixel at each position during the horizontal scan is shown in \fig\ref{fig:PixelScan:Horizontal}. A similar scan was also done
vertically across pixels 37, 36, and 35 with 40 points total and $\Delta Y =0.144~$mm per step. The result of this scan is shown in \fig\ref{fig:PixelScan:Vertical}. 
The width of each pixel is consistent with 2.88 mm, which is indicated by the black arrows.  
In both plots the sum of all pixels is shown by the 
dashed red line. As can be seen, there is a small area between pixels where the sum of the counts in each pixel is lower due to a decrease in efficiency.

This loss can be estimated by comparing the average sum inside the pixel to the value at the gap between pixels. In the horizontal scan the 
average sum between $x=1.73~$mm and $x=4.18~$mm is 1411 photoelectrons. At the minimum between pixel 44 and pixel 36, at $x=1.44~$mm, the sum in all pixels is
only 1046 photoelectrons. This gives a loss of $26(\pm3)\%$.  
Similarly, the percentage of photoelectrons lost between pixels 37 and 36 in the vertical scan is $21(\pm3\%)$.  

A difference in the response within the pixel can also be seen when scanning vertically versus horizontally. In the horizontal the response of the pixel is more or less flat, within the statistical uncertainty.
In the vertical direction, the number of photoelectrons in the pixel shows a saw-tooth structure. That the structure is not an artifact in one pixel
is clear as it can be seen in every pixel which is scanned vertically (keeping in mind the relative rotation of PMTs within an EC). This difference is due to the dynode structure. 

\begin{figure}[p]
\centering
\subfigure[Gap Between PMTs in an EC]{\label{fig:GapScan:EC}\includegraphics[angle=0,width=1.0\textwidth]{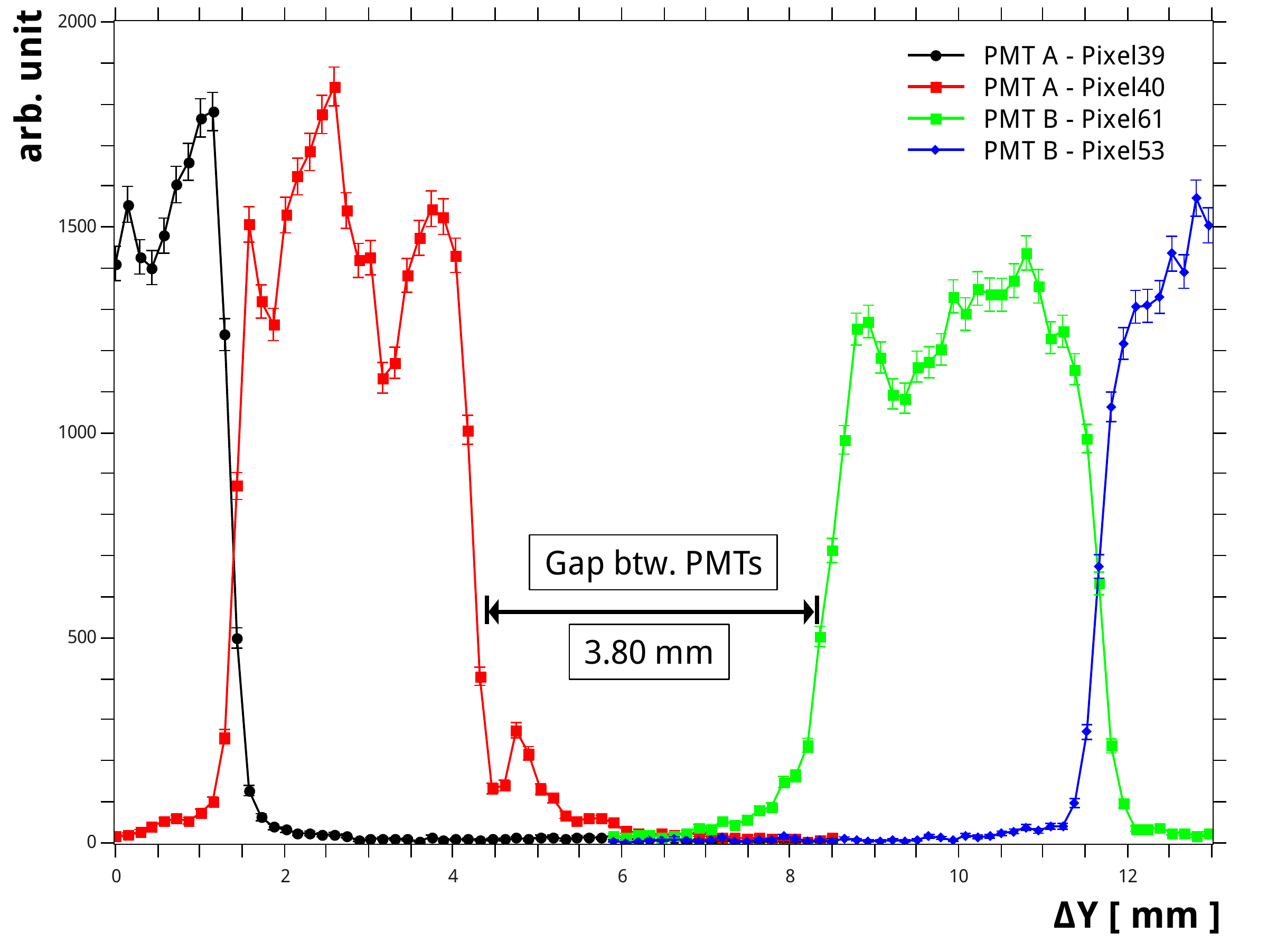}}
\subfigure[Edge of Filter with no Potting]{\label{fig:GapScan:FilterEdgeNoPotting}\includegraphics[angle=0,width=1.0\textwidth]{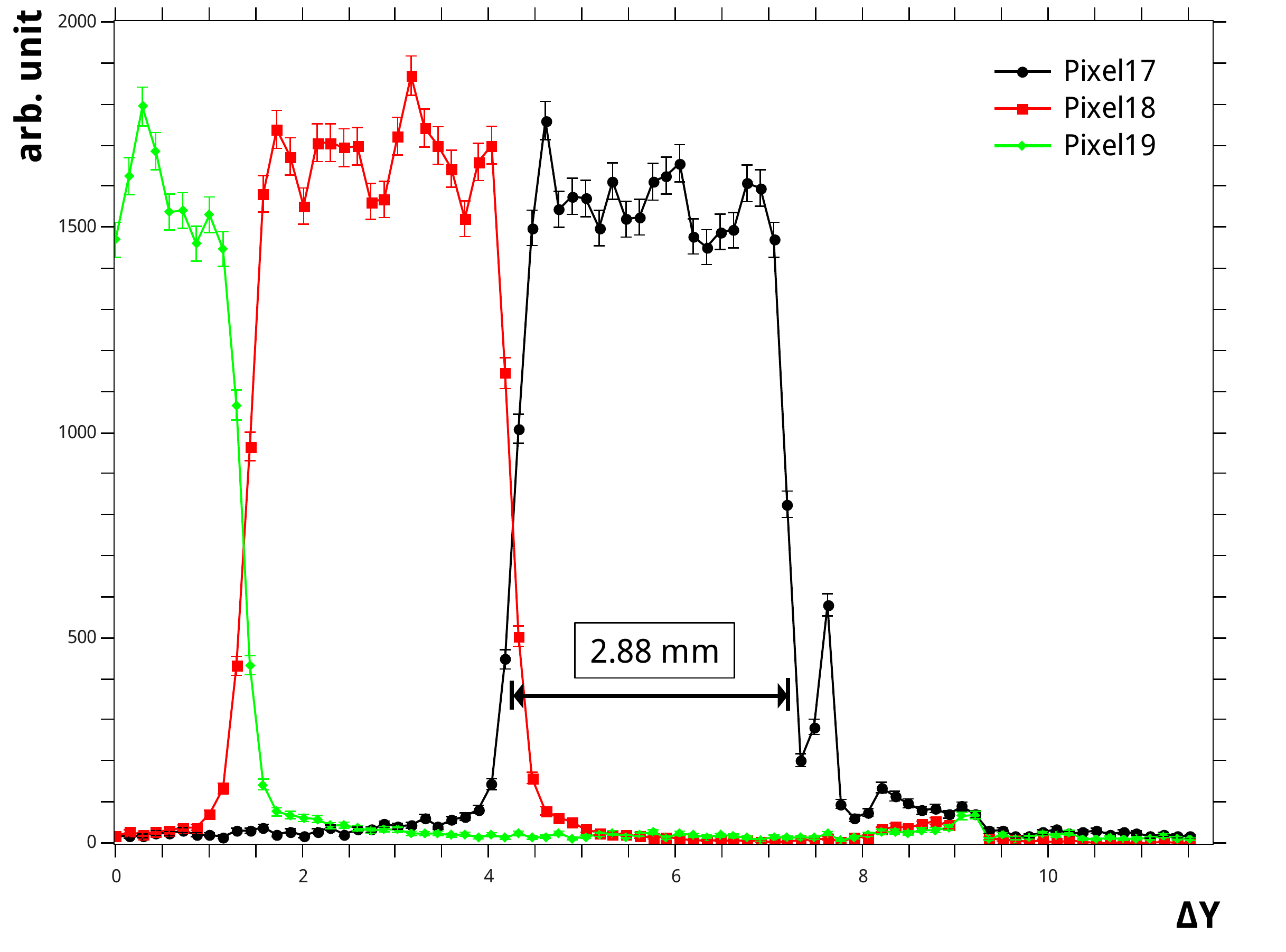}}
\caption[Scans of the Edge of a PMT]{ \label{fig:GapScan} \fig\ref{fig:GapScan:EC} shows a Scan across the gap between PMTs in an EC, as discussed in the text. The width of the edge pixel is $\approx 2.88~$mm, which implies that 
the Winston cone structure does not function. \fig\ref{fig:GapScan:FilterEdgeNoPotting} Shows a similar scan, with a PMT which is not potted, showing that the failure of the Winston cone is not due to the 
refractive index of the potting.}
\end{figure}

The gap between two PMTs in the EC was also done to assess the efficiency of the ``Winston Cone'' of the BG3 filter. The BG3 glued to the MAPMT photocathode has a trapezoidal cross section
with the large base facing outward, and the smaller base facing the photocathode. This is intended to collect photons incident at the edge onto the photocathode by internal reflection and so enlarge
the sensitive area of the PMT, with the goal of reducing the dead space between PMTs.

This measurement was complicated by the fact that the EC anode cable had to be switched to the next PMT in the middle of the run, but this was possible with some care. 
The step size was again $0.144~$mm, and the scan was done vertically across pixels 39 and 40 of PMT-A, continuing through pixels 61, 53, and 45 of PMT-B. The result of the scan is shown in
\fig\ref{fig:GapScan:EC}. This is a small drop and then recovery in the number of photoelectrons in pixel 40 at the edge of the PMT.
The gap between the two PMTs is $3.80~$mm, and the width of the edge pixel is effectively $2.88~$mm. This implies that the Winston cone does not increase the effective area of the 
edge pixel. 

At first we thought that this could be due to the refractive index of the potting, as the potting comes up to the backside of the filter. To check this, the scan was repeated using a single M64 MAPMT with an attached BG3 filter, and no potting. 
The PMT was oriented so that pixel 1 was in the top-left corner (when facing the photocathode) and a block of nine pixels in the upper left corner were read. A single photoelectron spectrum was taken for each pixel to find the valley.
After this, a scan was then done starting from pixel 19 through pixels 18 and 17 to the outside edge of the PMT. The $\Delta X$ is this scan was the same as in the previous cases. 
This scan is shown in \fig\ref{fig:GapScan:FilterEdgeNoPotting}, and
it can seen that the effective width of the pixel is still equivalent to $2.88~$mm. The same structure can also be seen at the edge of the PMT. 

We later realized that the fact that the Winston cone does not work is obvious. The edge of the filter
is at an angle of $45^{\circ}$ which means that photons incident at the edge of the PMT, normal to the photocathode, will not be reflected into the edge pixels. This means that the shape of the Winston cone must be reconsidered in the future.

\begin{figure}[p]
\centering
\subfigure[Horizontal]{\label{fig:CornerScan:Horizontal} \includegraphics[angle=0,width=0.6\textwidth]{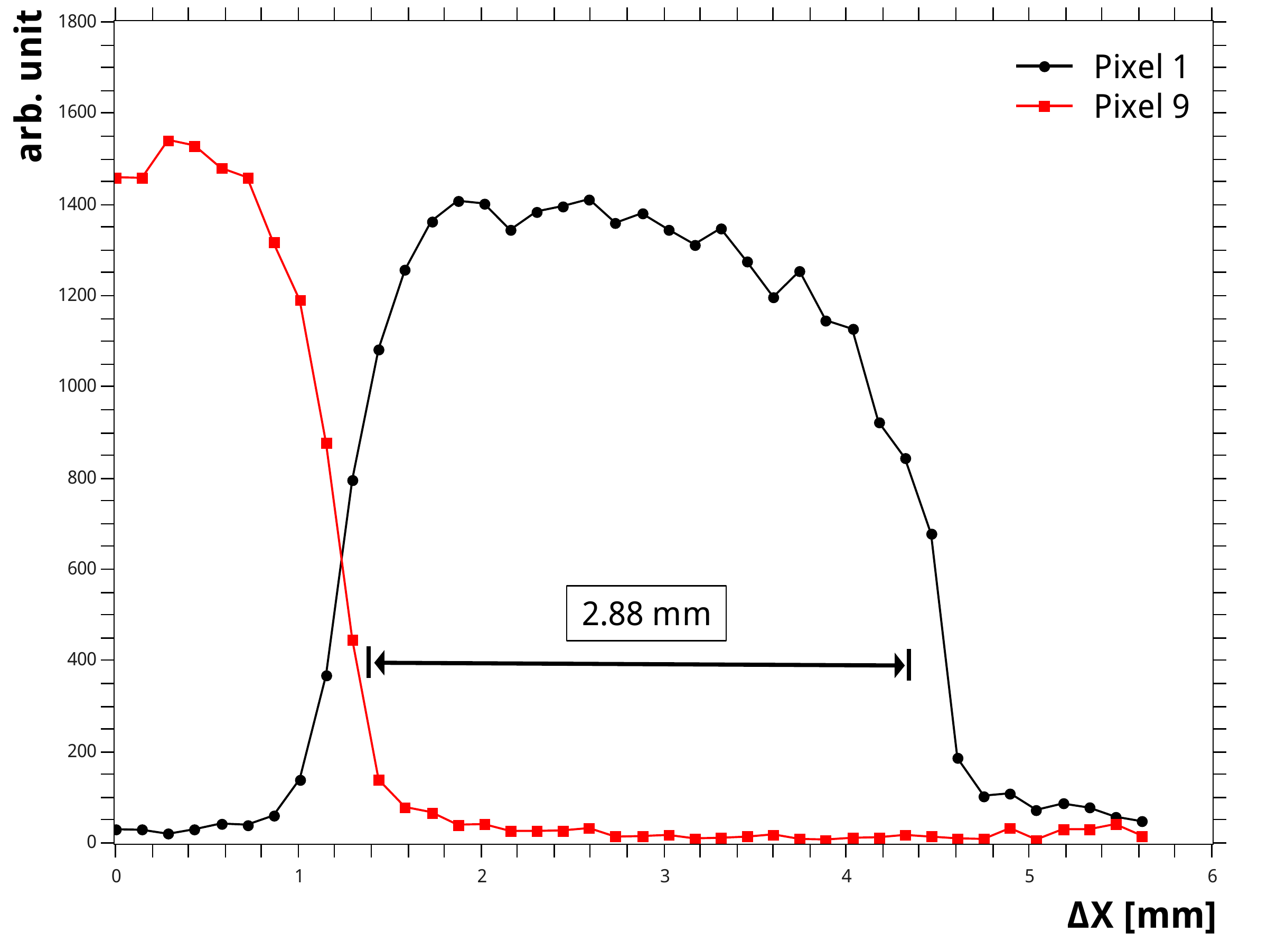}}\\
\subfigure[Vertical]{\label{fig:CornerScan:Vertical} \includegraphics[angle=0,width=0.6\textwidth]{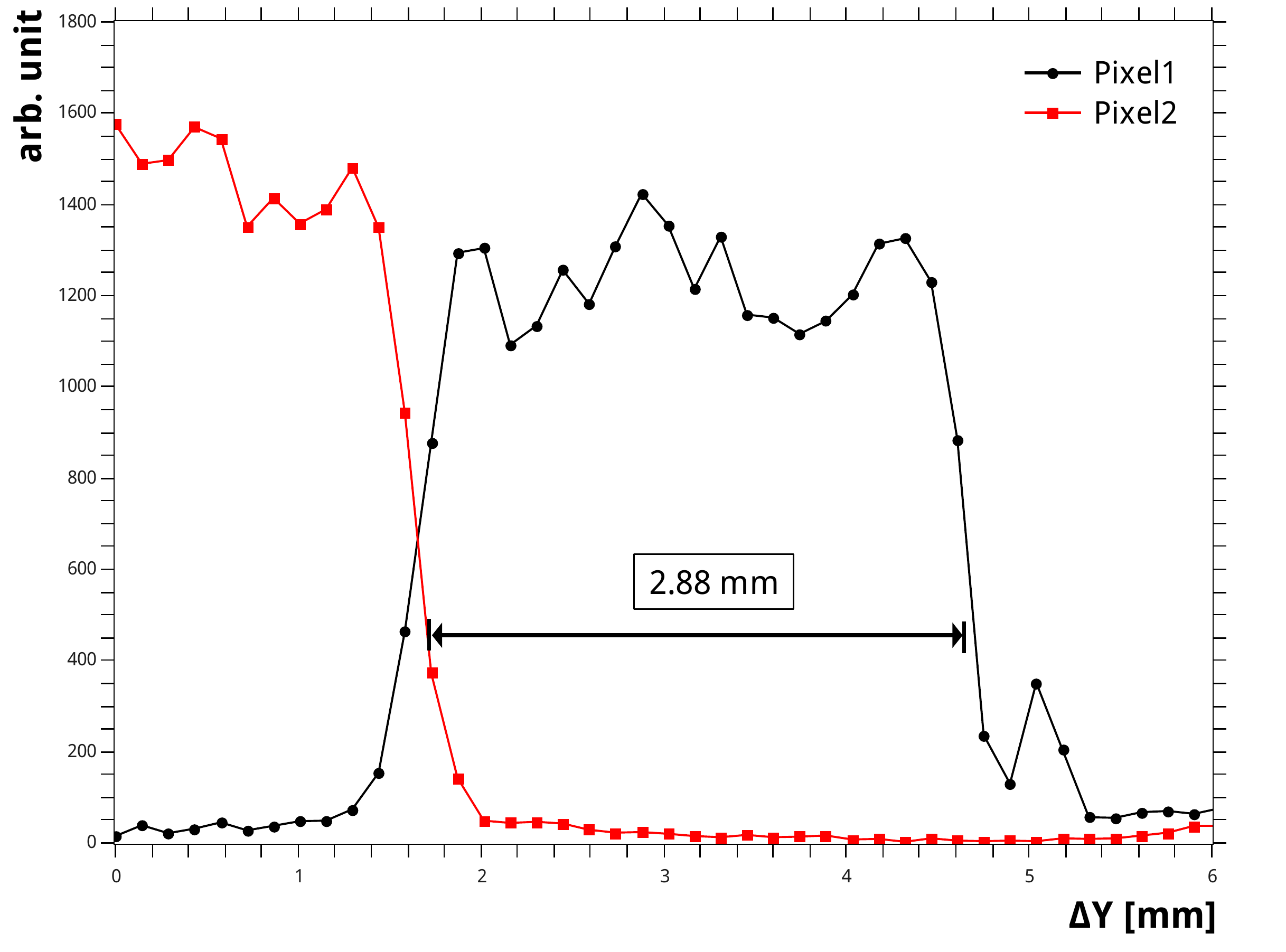}}
\caption[Scans of a Corner Pixel]{ \label{fig:CornerScan} Two scans of the corner of the PMT (pixel 1) in the horizontal and vertical directions. The effective size of this pixel is slightly larger than 2.88 mm. 
While a gentle roll-off the response can be seen in the horizontal direction, the pixel shows the same internal structure as the central pixels in the vertical direction. 
}
\end{figure}
After the scan of the edge, another two scans were performed through the corner pixel (pixel 1). One scan was done horizontally through pixel 1 and a second scan was performed vertically. 
In this case, the $X$-direction is towards the left edge of the PMT, and the $Y$-direction is towards the top edge.
Both of these scans are shown in \fig\ref{fig:CornerScan}.
From these results, the corner pixel appears to be sightly larger than the central pixels. The same difference in the pixel response between the vertical and horizontal scans can be seen as in the previous scans, including the 
spike in the response at the edge of the PMT in the vertical direction. 
In the x direction, however, the corner pixel shows a gentle drop-off in response.

\begin{figure}[p]
\centering
\subfigure[Response in Pixel 1]{\label{fig:2Dscan:Pixel1} \includegraphics[angle=0,width=0.8\textwidth]{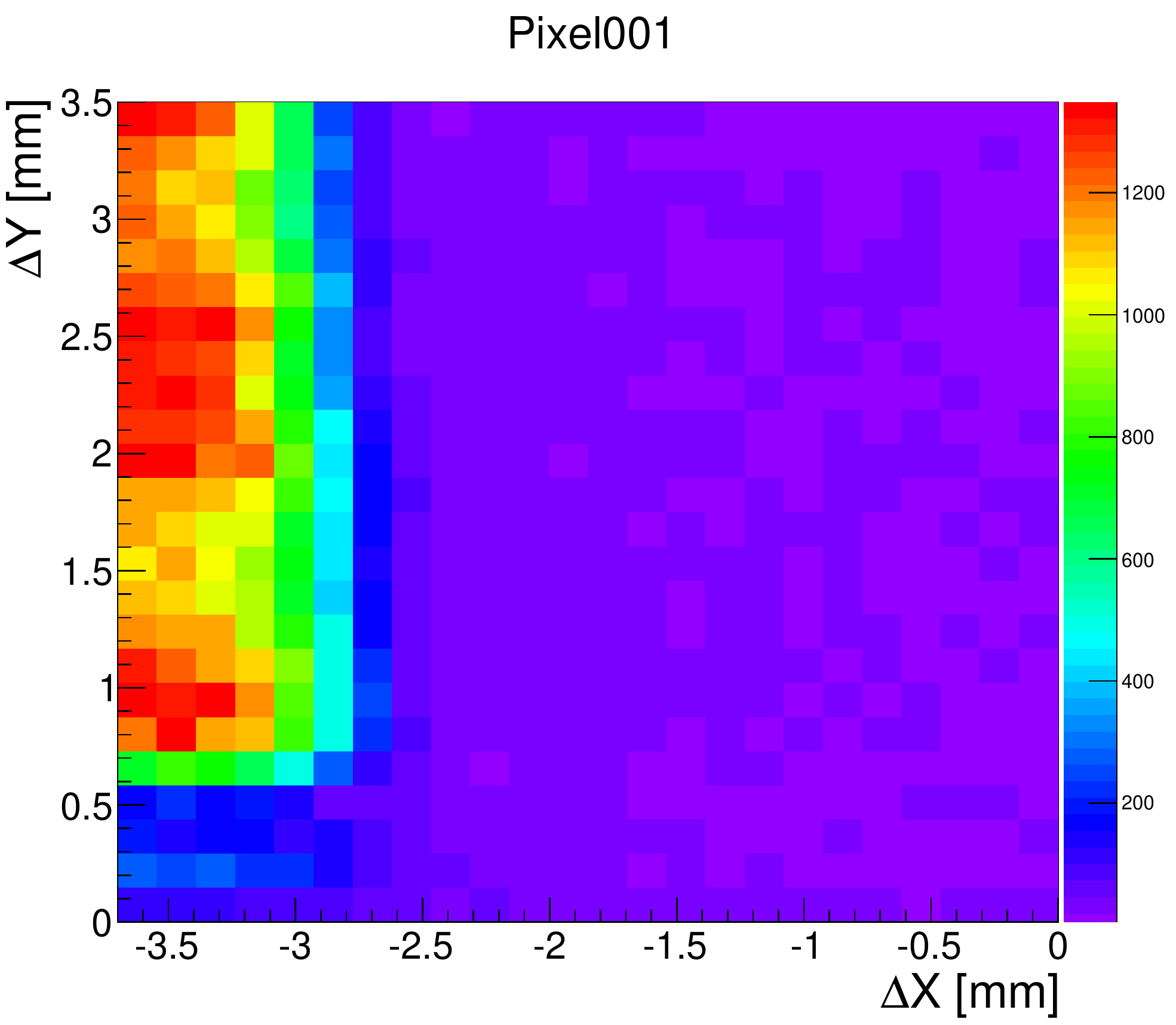}}
\subfigure[Response in Pixel 9]{\label{fig:2Dscan:Pixel9} \includegraphics[angle=0,width=0.8\textwidth]{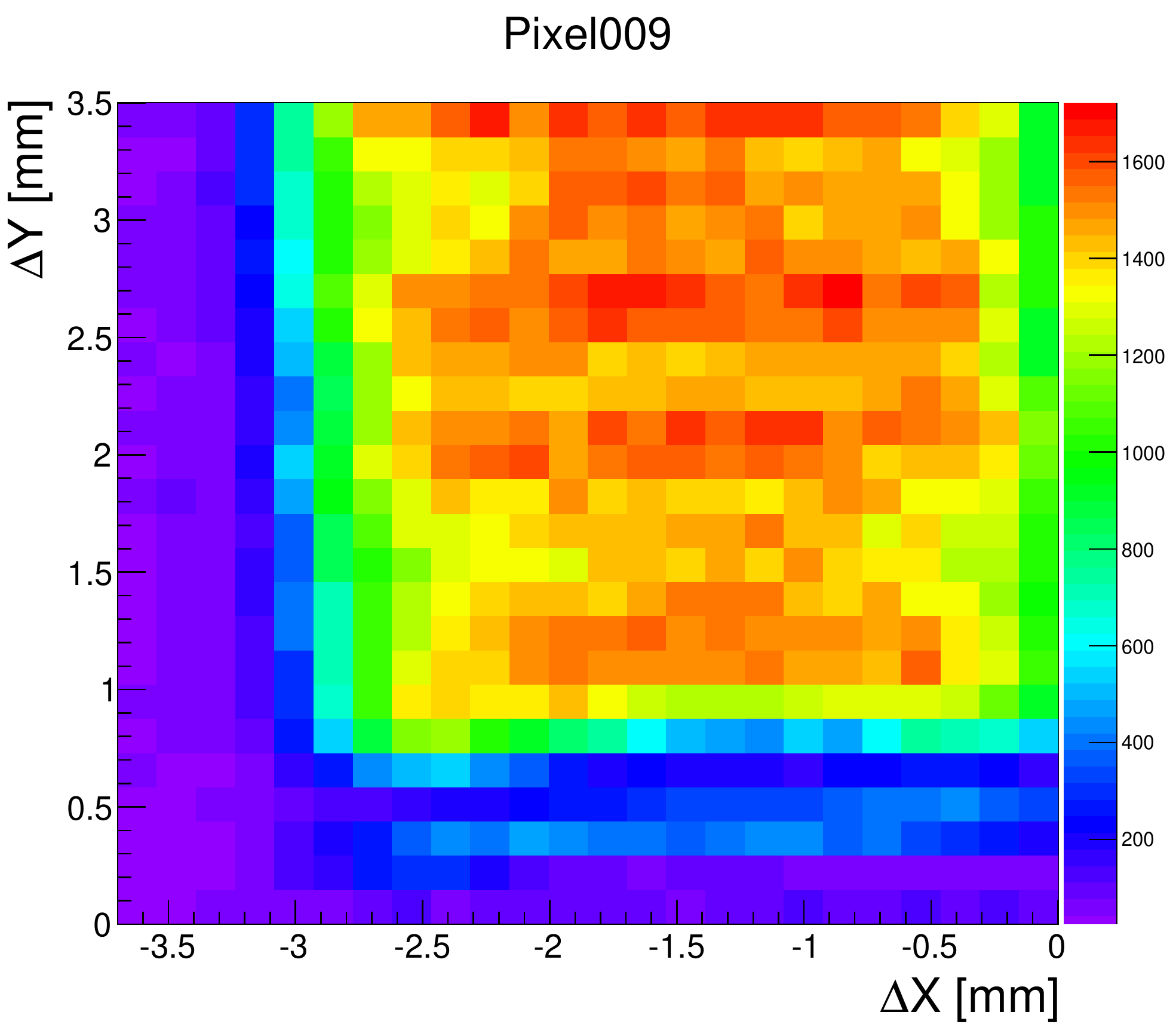}}
\caption[2D Scans of a Pixel]{ \label{fig:2Dscan} Two dimensional scans of the area between pixel 1 and pixel 9 of one M64. 
Each figure shows the number of counts in the respective pixel, as a function of the position of the light spot.
625 points were taken over a region of $3.5~$mm in $X$ and $Y$. The internal structure of
the pixels (horizontal bands of higher response in the center) can be clearly seen. 
The overlap between pixels can also be seen, as well as the roll-off of the response at the lower edge of pixel 9.}
\end{figure}

Finally, a 2 dimensional scan was done at in the area of pixels 1, 2, 9, and 10. This scan was done in 625 points $\left(25\times25\right)$ with a $\Delta X$ and $\Delta Y$ of 0.15 mm. The scan
thus covered about 14.06 mm$^{2}$ of the photocathode. The response of pixel 1 is shown in \fig\ref{fig:2Dscan:Pixel1}, and pixel 9 is shown in \fig\ref{fig:2Dscan:Pixel9}. 
These plots clearly show the structure inside the pixels, the red lines of higher response, which were previously seen in the vertical scan along the $Y$-axis. The overlapping area between the 
two pixels can also be clearly seen. 

\section{Conclusion}

The pixel scans just presented were the last characterization measurements done on the elementary cells of EUSO-Balloon. 
These results, combined with the measurement of the absolute efficiency of much of the EUSO-Balloon PDM, greatly improves our understanding of the photodetection response of the PDM. 
This knowledge is also important for refining the PDM hardware for JEM-EUSO. Unfortunately, the measured absolute efficiency is not directly applicable to EUSO-Balloon, due to the fact that it was measured
at 1100 V. Working at 1100 V was necessary due to the lower sensitivity of the QDCs compared to the ASIC. As the collection efficiency depends on the electrostatics in a complex manner, it not possible to 
extrapolate these results to a lower gain with precision. These results do, however, give us preliminary information for a complete characterization of the entire EUSO-Balloon PDM using the ASIC and PDM board read-out chain. 
 
At the present, eleven ECs have been assembled and potted, and each of these ECs are included in the results given here.
Of these EC 104, had an electrical fault and was disassembled to recover the PMTs. Each of the four PMTs where quickly tested by powering them and looking for single photoelectron pulses on several anodes with an oscilloscope. 
During this test, three of the four recovered PMTs were found to be working. These three PMTs where shipped to RIKEN to have a new BG3 filter glued on, as the depotting removed the original filters. 

These 3 PMTs were received back at APC along with 4 new PMTs and the gain was measured for all seven using the QDC DAQ system. It was found that another one of the 3 recovered PMTs, which had before shown only a very low gain, was in fact dead.
The six working PMTs where combined with 2 spare PMTs according to the sorting results, and were assembled into the tenth EC unit and one spare EC.

In the mean time, the ASIC is being debugged and completely characterized. When this work is complete, the PDM of EUSO-Balloon will undergo integration at APC, and, once the PDM is operating and debugged, the efficiency of every 
pixel will be measured a second time using the ASICs themselves. This will be the true, absolute, calibration of EUSO-Balloon. A test of the PDM trigger using a moving light spot is also planned. 
These future measurements and an overall outlook will be presented in the next chapter. 

    \printbibliography[heading=subbibliography]
     \end{refsection}

     \begin{refsection}
    \chapter{Future Measurements and Conclusion} 
   \label{CHAPTER:FutureMeasurementsAndDiscussion}

As ever, there is still more work to be done.
After the characterization of the EUSO-Balloon ECs presented in the last chapter, the assembled EC units are waiting to be integrated to form a complete Photodetection Module (PDM).
The data processing and frontend electronics are being tested at APC by their responsible groups.
Once the PDM is assembled, there is a full range of performance measurements which must be done. 
These include
\begin{inparaenum}[i\upshape)]
 \item a final measurement of the absolute efficiency, and
 \item a test of the response of the PDM,
\end{inparaenum}
in addition to general hardware functionality tests.

\section{Absolute Calibration of the PDM}
The first measurement is the absolute calibration of the PDM using the SPACIROC ASIC. 
This must be done for several reasons:
\begin{itemize}
 \item The sensitivity of the ASIC is much higher, meaning that the working voltage of the EC units should be $\simeq 900~$V, as opposed to the 1100 V used during the measurements with the QDCs. This change in working voltage will 
 have an effect on the efficiency.
 \item The measurement of the efficiency using the ASIC will give the absolute efficiency as a function of the photoelectron counting threshold.
\end{itemize}
The need to characterize the PDM with its read out electronics is the same as the need to characterize the ECs after sorting the PMTs: There could be, \textit{a priori}, a difference
in efficiency at the level of a few percent between the ASIC and QDC, just as the potting and shared high voltage power supply can affect the gain and efficiency of PMTs which have been combined into an EC.

Ignoring this possible difference, and the fact that the QDC is less sensitive than the ASIC, it would be possible to use the results presented in chapter~\ref{CHAPTER:EUSOBalloonMeasurements} as the absolute reference calibration.
A major error which would be introduced by doing this is the ill-definition of the single photoelectron counting threshold. Any threshold is arbitrary as long as it is high enough to reject noise (i.e.\ above the valley in the
single photoelectron peak). The threshold in QDC counts could be translated into a charge threshold and then compared to the ASIC threshold in volts,
but it is difficult to believe that this could be done with negligible uncertainty. If the uncertainty on the threshold voltage is 5\%, 
then the resulting uncertainty on the absolute efficiency could be quite large (cf.\ \fig\ref{fig:ThresholdVariation}).

In addition, the ASIC sets the photoelectron counting threshold using an internal 10 bit \gls{DAC}. The absolute output of
this DAC is linear and is stable in time for a given setting, but varies with the voltage supplied to the ASIC. Due to this, the absolute threshold for a given DAC setting varies 
systematically with the position of the ASIC on the ASIC board (as each ASIC is daisy chained). Thus even a direct conversion between a charge threshold and a voltage threshold would not completely account for the ASIC response.
 
The ASIC incorporates an integrating preamplifier and a discriminator in order to count the number of single photoelectrons arriving at the anode of each M64 pixel. 
The response of each PMT is therefore measured as a number of counts over a given number of GTU (time intervals of 2.5 $\mu$s) for a given choice of threshold. 
The full set of measurements, one at each threshold, gives an S-curve. As discussed in chapter~\ref{CHAPTER:PMT}, the S-curve is the integral of the single photoelectron spectrum, and so the 
valley between the noise (the pedestal) and the single photoelectron peak can be found by taking the derivative of the S-curve. 
The measurement of the efficiency is then a simple counting experiment, with the efficiency at a given threshold $\epsilon(\tau)$ given by 
\begin{equation}
\label{eq:ThresholdDepEff}
 \epsilon(\tau) = \frac{N_{\text{pe}}(\tau)}{N_{\gamma}}
\end{equation}
where $\tau$ is the threshold in arbitrary units (DAC counts) and $N_{\gamma}$ is the number of incident photons.
$N_{\gamma}$ is known using our ``standard'' calibration procedure. 
The error on \eq\ref{eq:ThresholdDepEff}) is the combination of the statistical error on the number of counts and the uncertainty on $N_{\gamma}$. 
There are then two interrelated problems which must be addressed:
\begin{enumerate}[i\upshape)]
 \item the dead time of the ASIC-PDM board read-out chain, and
 \item the length of time that a measurement of the efficiency for the full range of thresholds will take. 
\end{enumerate}
The second item relates to both the dead time, the statistics required, and the requirement of being in single photoelectron mode. This last item begs the question of ``what does single photoelectron mode mean?''
As will be shown, this point will have an important impact on our ability to perform this measurement.

The whole point of what has so far been called single photoelectron mode is the application of the Poisson statistics to single photoelectron spectra so that the number of two photoelectrons in the spectra is negligible.
For the efficiency this is important because the read-out is not sensitive to multiple 
photoelectrons in a short period of time, and so counts will be lost if there are a significant occurrence of two photoelectrons, biasing the efficiency. 

This is obvious for a QDC, where two photoelectrons within the same integration gate will give twice the charge. The problem in this case is one of resolving the 
one photoelectron and two photoelectron peaks, which can be done in some cases, depending on the properties of the photodetector. 
For a discriminator-based measurement, this is more dangerous. With a single discriminator the only information obtained is that the pulse is larger than a set threshold, but it is not 
known how much larger. Two photoelectrons at nearly the same moment thus give only one count. The ability of a discriminator to resolve two pulses close together in time is known
as the double-pulse resolution, given as the minimum separation in time needed to resolve the two pulses.

If the efficiency is measured using a discriminator, as in the ASIC, and the PMT is illuminated with a single, short, LED pulse ($\sim 20~$ns or so) per GTU then any time the LED pulse creates two photoelectrons the second photoelectron
is likely to be lost. That this loss is negligible can be ensured by invoking the Poisson distribution, with the result that 1\% of GTUs can give a photoelectron, i.e.\ an average rate of 0.01 pe/GTU.
Since one GTU is $2.5~\mu$s, this gives a count rate of 4 kHz. 
As the S-curve is the integral of the single photoelectron spectrum, the statistics in the bin which corresponds to the valley must be at least 10,000 counts, so that the statistical uncertainty
on the efficiency at that threshold is 1\%. With a rate of 0.01 pe/GTU, we must count for 1 million GTU or 2.5 seconds.
If a scan over 200 different thresholds is taken, then the measurement of an S-curve will take 500 seconds. This is approximately the same amount of time as needed to take a spectrum with the QDC.

This measurement time assumes a continuous and completely efficient read-out of the ASIC, which is unfortunately not true.
For laboratory tests using the ASIC, we currently have several ``test-boards'' which incorporate a FPGA to control the ASIC. One test-board can handle one EC-ASIC board of 6 ASICs.

These boards were designed for engineering tests of the ASIC, and are controlled by LabView software with the communication between the board and the PC being through USB.
Because of this, and the manner in which the firmware of the board is programmed, there is an enormous dead time (more than 90\%). 
This dead time has been estimated from the fact that a full, high-statistics, S-curve takes upwards of 4 hours. 
The dead time in this case is also very unstable, as the data write from the FPGA has to wait for the USB write cycle.

An acquisition time of 4 hours is a problem because multiple runs are needed for a full characterization, and because is it only possible to read out one EC at a time. A full characterization of 
the entire PDM will therefore take more than 36 hours.
It is better in this case to use the full EUSO-Balloon read-out chain of ASICs and PDM board.
However, even the full PDM board and data processing chain of EUSO-Balloon are not designed for a continuous read out of all channels. 
The system is designed to continuously sample the rate and then return 128 consecutive GTUs on a trigger, according to the JEM-EUSO trigger scheme discussed in chapter~\ref{CHAPTER:JEMEUSO}. 
In EUSO-Balloon, the full read-out of all 2304 channels over 128 GTUs creates a pause of approximately 20 ms.
This gives a read-out of 128 GTU per cycle at a cycle rate of $\sim 48~$Hz. Keeping in mind that the rate of photoelectrons should be 0.01 pe/GTU, a single set of S-Curves for the entire PDM would take $\sim9~$hours.

Which returns to the important question of what is single photoelectron mode. If, instead of a LED pulsed in coincidence with the gate, a continuous (DC) LED is used, the situation is very different. 
In DC illumination, the photons will arrive at a randomly at the photocathode with some average rate. One mistake which could be made (and indeed was) is to assume that this rate must be so low as 0.01 pe/GTU, which is not the case.
The length of the GTU is completely unimportant when using DC illumination.
The important thing is instead the probability of having two photoelectrons so close together that the discriminator misses one,
and the time scale which matters in this case is the double-pulse resolution of the photon counting circuit (the combination of the preamplifier, pulse shaper, and discriminator).

Using the Poisson distribution, \eq\ref{eq:poisson}), the probability of two photoelectrons in a time $\delta t$ is a function of the average number $\eta$ per $\delta t$:
\begin{equation}
 P(2,\eta) = \frac{\eta^{2}}{2!}e^{-\eta}
\end{equation}
The condition for a clean measurement, with no significant loss of photoelectrons is that $P(2) < 0.01 P(2)$ or
\begin{equation}
\label{eq:SpeCondiiton2}
 \frac{P(2)}{P(1)} = \frac{\eta}{2} < 0.01
\end{equation}
which gives $\eta < 0.02$. For a measurement with the ASIC, the time-interval $\delta t$ in which to avoid having two photoelectrons is the double-pulse resolution $t_{\text{dpr}}$, as a second photoelectron separated from the first by 
more than  $t_{\text{dpr}}$ would simply give a second count. Remembering that $\eta$ is the average rate per $\delta t$, the number of counts $n$ over a total counting time $\Delta t$ must be less than
\begin{equation}
 n < 0.02 \frac{\Delta t}{t_{\text{dpr}}} 
\end{equation}
The ASIC double-pulse resolution $\tau_{\text{dpr}}$ is $30~$ns \cite{SallehThesis}, and, so the counting rate can be up to 1.6 pe/GTU.
This is an increase of more than 100 compared the use of a pulsed LED, and this decreases the total time needed to take a S-curve for the entire PDM from 9 hours to 4 minutes.

\begin{figure}[h]
  \centering
  \includegraphics[angle=270, width=0.90\textwidth]{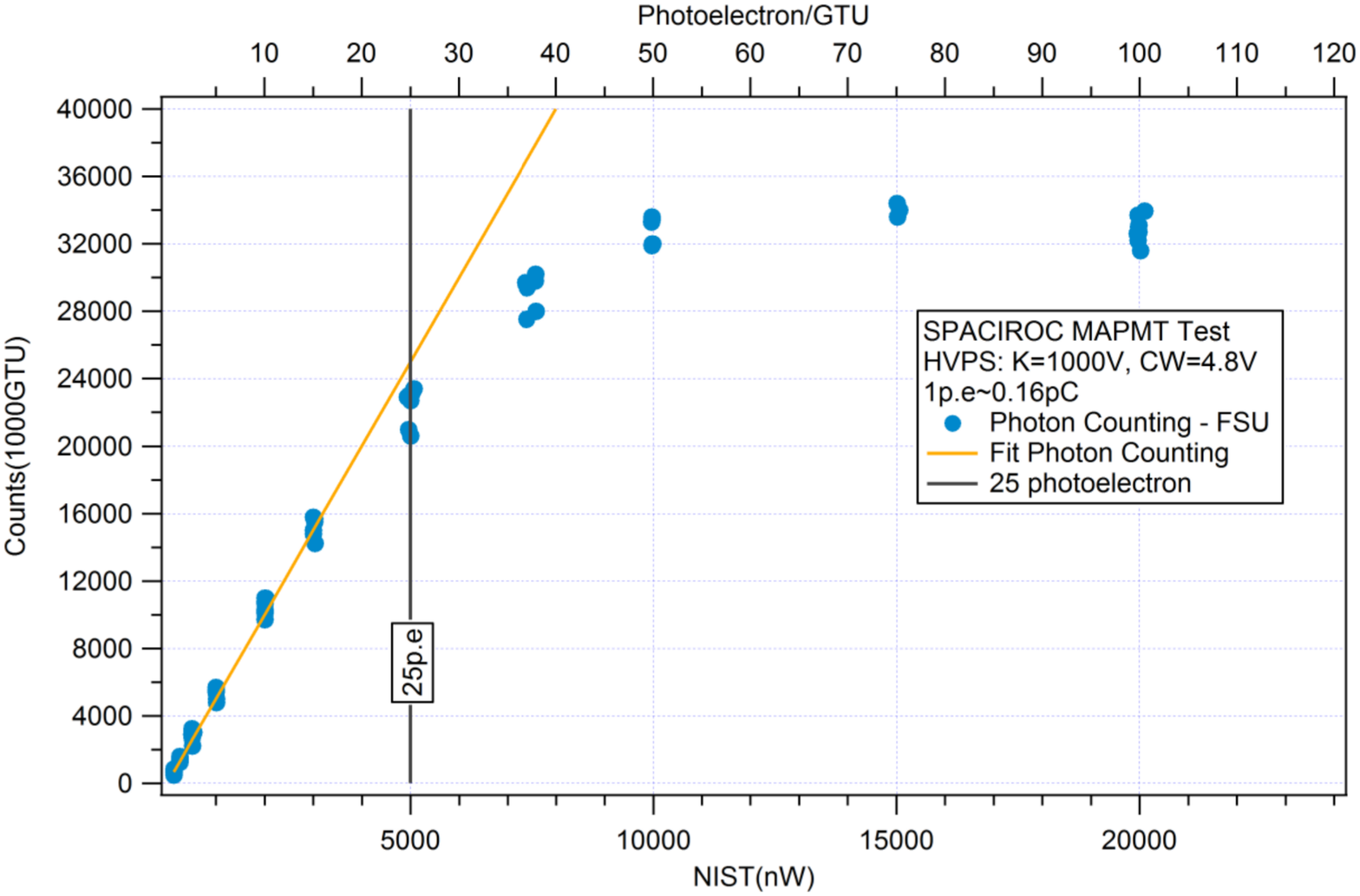}
   \caption[Photon Counting Linearity in the SPACIROC ASIC]{ \label{fig:PhotonCountingLinearity_Salleh} A plot of the photon counting linearity of the SPACIROC ASIC taken from \cite{SallehThesis}. 
 This measurement was done used an M64 PMT powered at 1000 V, giving a gain of $1~10^{6}$ or $\sim 0.16$ pC.
The count rate in photoelectrons per GTU (2.5 $\mu s$) is plotted versus the power on the NIST photodiode. 
As can be seen, the count rate begins to saturate around 25 photoelectrons per GTU. }
 \end{figure}

This result can be cross checked by looking at the photon counting linearity using the ASIC, as shown in \fig\ref{fig:PhotonCountingLinearity_Salleh}.
In this figure, the count rate per GTU is shown versus the incident power measured by the NIST photodiode. For lower amounts of light, the count rate is linear with the power, as expected. 
As the rate increases however, it is no longer proportional to the incident power due to the pile up of photoelectrons within the double-pulse resolution.
Looking at the previous calculation, a rate of 1.6 pe/GTU corresponds to a ratio of 1\% two photoelectrons to one photoelectrons within one interval $t_{\text{dpr}}$. 

From \eq\ref{eq:SpeCondiiton2}), the ratio is proportional to the rate. It the measured number of counts increases to 20 pe/GTU, then the rate is approximately 12.5 times higher, and the
ratio of two photoelectrons to one photoelectrons can be estimated to be 12.5\%. 
This means that there were in fact $\approx 2.5$ photoelectrons which were lost because two photoelectrons were collected in the same 30 ns. A similar analysis for three and four photoelectrons within the double-pulse resolution shows that 
true photoelectron rate is $\sim 25~$pe/GTU. This is consistent with the saturation seen in \fig\ref{fig:PhotonCountingLinearity_Salleh}, which is already more than 10\% at a measured rate of 20 pe/GTU.
Looking more closely however, it can be seen that a measured count rate of $\approx 21~$pe/GTU corresponds to $\approx 24.5~$pe/GTU from the linear fit. 
This implies that the 30 ns double-pulse resolution is a slight over estimate (perhaps not accounting for the chance that some pulses separated by less than 30 ns might still be counted separately).

If the measured count rate is corrected using \eq\ref{eq:countratecorrection}), with the measured ASIC double-pulse resolution $\tau_{\text{dpr}}=30~$ns \cite{SallehThesis}, then we get
\begin{equation}
  n_{\text{true}} = \frac{20~\text{pe/GTU}}{1-(20~\text{pe/GTU})(30~\text{ns})} = \frac{8~\text{MHz}}{ 1 - 0.24} = 25~\text{pe/GTU}
\end{equation}
This shows the importance of knowing the double-pulse resolution of the ASIC with high accuracy and applying a correction to the measured rate, especially for higher count rates which would occur inside a shower track in JEM-EUSO.

This important digression aside, at a rate of 1.6 pe/GTU the number of photoelectrons $N_{\text{pe}}$ can be obtained for the entire range of thresholds in a few minutes.
The final component of the calibration is then the number of incident photons $N_{\gamma}$, and here the dead time is again important.
Using our calibration technique, $N_{\gamma}$ is known from the power measured by the reference NIST photodiode and the time of the 
measurement. If there is some  dead time $\delta$ in the read out, expressed as a fraction, then the true time of the measurement is $t_{\text{true}} = t_{\text{total}}\delta$
Putting this into the calculation of the efficiency, \eq\ref{eq:EfficiencyCalculation}) simply gives
\begin{equation}
\label{eq:EffwithDeadTime}
  \epsilon(\tau) = \frac{N_{\text{pe}}(\tau)}{N_{\gamma}\delta}
\end{equation}
and the uncertainty on the dead time directly adds to the uncertainty on the efficiency.

The number of GTU read out before writing the data to buffer is directly controlled by the read out software.
As the number of bytes written to the buffer after each set of GTUs is always the same when using the full PDM read out chain (but not the USB test card), the write time is also stable, but not precisely known.
The true percentage of time the ASIC is active (counting) can be measured by comparing the absolute run time, measured with a timer or the CPU clock, with the number of GTUs collected.
The total number of GTUs read is known precisely, and, so, the uncertainty on the dead time is given by the uncertainty on the run time. 
If the run time is long enough, the dead time can be found with arbitrary statistical uncertainty. 
Creating a histogram of the write time of each cycle and repeating the full measurement several times will give an estimate of the stability of the dead time.

A second source of dead time is created within a GTU by the so-called ``val-event''. There is noise on the PMT anode signals within the ASIC due to the start bit of the GTU, and, due to this noise, the leading $\sim 200~$ns of each GTU
is blocked. The length of the val-event can be easily be estimated with an uncertainty of 10\% or less, and as 200 ns is less than 8\% of the total GTU, this would give a less than 1\% uncertainty on this source of dead time.

Once the dead time is known, then the efficiency for each pixel in the PDM can be measured using \eq\ref{eq:EffwithDeadTime}). An important benefit of measuring the efficiency in this way is that not only will the absolute 
efficiency within the read-out 128 GTUs be known (both with and without the val-event), thanks to the measurement of the dead time, but also the \emph{effective} efficiency for a given incident photon flux.
As was the case in the previous measurements in chapter~\ref{CHAPTER:EUSOBalloonMeasurements}
it will take too long to measure every pixel independently, and so the absolute efficiency will be taken for a subset of ``NIST'' pixels. The efficiency of every pixel
in the PDM will then be measured relative to these reference pixels under uniform illumination, as was previously done.
At the same time, the need to center on a certain subset of pixels for the ``NIST pixel'' measurement will allow a check of the PDM pixel mapping.

\section{PDM Trigger Tests}
After the final calibration of the PDM, the last set of tests envisioned at APC is a test of the PDM's ability to recognize a pattern and trigger properly.
In order to test the track recognition, a moving point of light is needed. 
Tests in the past have used a laser or other light source and a rotating mirror to generate a moving track.
This is difficult to control and not capable of creating more complex patterns without a substantial increase in hardware complexity.

\begin{figure}[h]
  \centering
  \includegraphics[angle=0, width=0.90\textwidth]{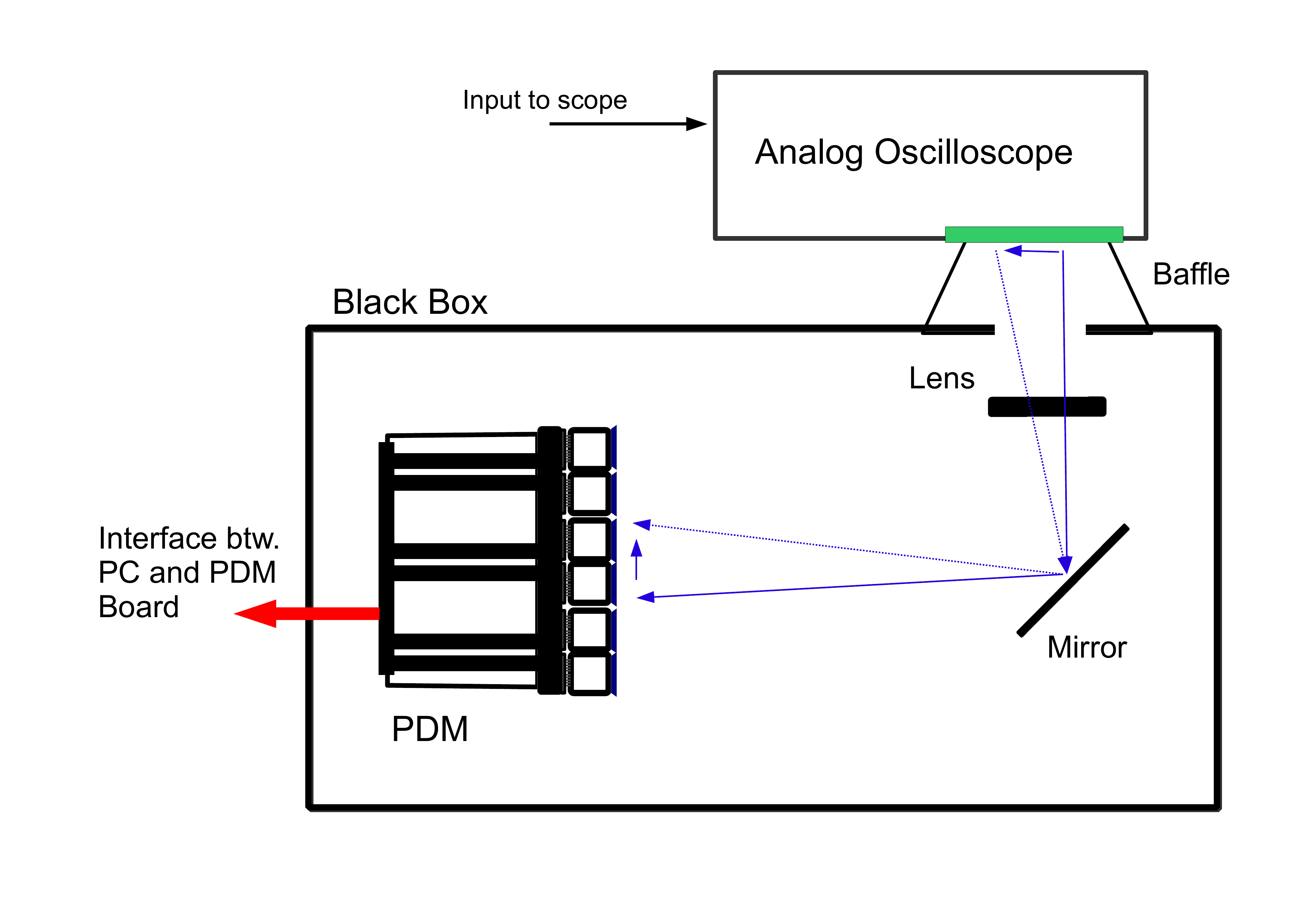}
   \caption[Light-Track Test of EUSO-Balloon PDM]{ \label{fig:TriggerTest} A diagram of the proposed trigger test for the EUSO-Balloon PDM. The spot from an analog oscilloscope (which has been measured to
give a signal to background ratio of better than 100 in the near UV) is sent into the box black. A set of baffles is used to ensure a light tight seal around the oscilloscope screen (so the black box stays black).
The light spot is sent through a lens and then incident on a mirror which directs it towards the PDM. By setting the oscilloscope in X-Y mode, a track of light can be sent across the PDM, mimicking a EAS.
Any variety of complex patterns can be easily generated, allowing a complete test of the PDM read-out chain.  }
 \end{figure}

One solution is to use an analog oscilloscope in X-Y mode. One such oscilloscope has been tested at APC using a spectrometer. It was seen that the phosphor screen gives a good intensity in the UV, with
a signal-to-noise ratio of over 100 in a lit room for a wavelength of $\sim 370~$nm. The dot of the oscilloscope would be used to simulate an EAS track moving across the PDM
as shown in \fig\ref{fig:TriggerTest}. The light from the oscilloscope will enter through a baffle designed to maintain the blackness of the black box, and will be focused through and a mirror onto the PDM.
The spot will be moved on the PDM surface using the X-Y inputs of the oscilloscope, which will allow many different light patterns to be generated. 
The speed of the spot on the PDM surface can be matched to that of an EAS by adjusting the combination of the true spot speed on the oscilloscope screen and the lengths between the mirror and the PDM and the mirror and oscilloscope. 
This test, along with the reconstruction of the track image, will be one of the final photodetection tests of EUSO-Balloon. 
 
\section{Conclusion}
This part of the thesis has presented several bodies of work related to the photodetection aspects of JEM-EUSO and EUSO-Balloon. This work is centered around a deep understanding of photomultiplier behavior, including techniques for gain calibration
and precision measurements of the absolute photon counting efficiency, which were presented in chapter~\ref{CHAPTER:PMT}. While 
chapter~\ref{CHAPTER:PMT} was not original work, it is a direct inheritance from the photodetection group at APC and lays the foundation for the later chapters.

These techniques were first applied to a proposed measurement of the air fluorescence yield in chapter~\ref{CHAPTER:AIRFLUOR}, which in the context of the
UHECR fluorescence detection technique is a bridge between instrument response and the physical observables. The actual fluorescence yield measurements are ongoing, but the first complete calibration of 
two PMTs was presented in this chapter. Further simulation work for the fluorescence measurement was also done, but was not discussed in this thesis.
The presentation of the PMT efficiency measurement in this chapter was intended to show, in detail, how the calibration techniques of chapter~\ref{CHAPTER:PMT} can be applied in practice.

Chapter~\ref{CHAPTER:CWHVPStests} moved to the testing and validation of a high voltage power supply and switch system for JEM-EUSO, which was another body of original work. A complete set of tests were presented, showing the response of a test EC across
its full dynamic range. This high voltage system will be used in EUSO-Balloon, and these tests were part of the successful CNES phase B review of the EUSO-Balloon instrument.

After the high voltage power supply tests, chapter~\ref{CHAPTER:PMT Sorting} discussed the creation of a PMT sorting system for JEM-EUSO. This setup was motivated by the foreseen requirement to test and sort more than 5000 multi-anode PMTs
for JEM-EUSO. A detailed discussion of the data acquisition system was presented, but it should be stressed that this setup is a continuing area of research and development for the future. 

In addition to the sorting, the data acquisition setup presented in chapter~\ref{CHAPTER:PMT Sorting} was used for a first characterization of the elementary cells of EUSO-Balloon. 
A complete measurement of the absolute efficiency was done for 256 pixels for each of eight elementary cells with a total uncertainty of $\sim 3\%$. This work was presented in chapter~\ref{CHAPTER:EUSOBalloonMeasurements}. 
These measurements were limited by the sensitivity of the setup available, but, at the same time, were only made possible by the work which went into creating the data acquisition system.
 
Finally, this chapter discussed several future measurements related to the final characterization of the EUSO-Balloon PDM. As is always the case with research and development, these future measurements are made possible by 
building on a large amount of past work. After the flight of EUSO-Balloon, the lessons learned here will have to be applied to the even greater challenge of JEM-EUSO. This will include working towards the true goal
of going from a photodetection calibration to a true physics calibration and complete understanding of the JEM-EUSO instrument, which it is hoped will shed light on the mysterious sources of UHECRs.
    \printbibliography[heading=subbibliography]
     \end{refsection}

\part{The Ultra-High Energy Cosmic Ray Sky}

This part of the thesis presents theoretical work done in the area of UHECR phenomenology. 
This work represents the quantification of relatively simple ideas, motivated by questions in the area of UHECR composition and anisotropy.
Unlike the experimental part of this work, a dedicated introductory chapter will not be given. The reader is instead referred to Part I (remember way back when?).

Chapter~\ref{CHAPTER:UHECR Source Maximum Energy and the UHECR Spectrum} studies a generic class of models for UHECRs, in which the sources accelerate protons and nuclei with a power-law spectrum having the same index, 
but with different values for the maximum proton energies, distributed according to a power-law. It will be shown that, for energies sufficiently lower than the maximum proton energy, 
such models are equivalent to single-type source models, with a larger effective power law index and a heavier composition at the source. 
The resulting enhancement of the abundance of nuclei will be calculated resulting in typical values of a factor 2--10 for Fe nuclei.
 At the highest energies, the heavy nuclei enhancement ratios become larger, and the granularity of the sources must also be taken into account. 
The conclusion from this work is that the effect of a distribution of maximum energies among sources \emph{must} be considered when one tries to understand both the energy spectrum and the composition of UHECRs, as measured on Earth.

In chapter \ref{CHAPTER:UHECR Source Statistics}, the statistics of UHECR sources are studied quantitatively as a function of energy for a range of models
compatible with the current data; varying source composition, injection spectrum, source density and luminosity distribution. 
This is motivated by the idea that the Greisen-Zatsepin-Kuzmin (GZK) effect
will result in a drastic reduction of the number of sources contributing to the observed flux above
$\sim 60$~EeV.
Various realizations of the source distribution are also explored, and it is found that, in typical cases, 
the brightest source in the sky contributes more than one-fifth of the total flux above 80~EeV and about one-third of the total flux at 100~EeV. 
It is further shown that typically between two and five sources contribute more than half of the UHECR flux at 100~EeV. 
With such low source numbers, the isolation of the few brightest sources in the sky may be possible 
for experiments collecting sufficient statistics at the highest energies, even in the event of relatively large particle deflections.
This work is noteworthy (in the context of a thesis) as it has been adopted as part of the JEM-EUSO science case.

    \begin{refsection}
  \chapter[UHECR Source Maximum Energy and Spectrum]{UHECR Source Maximum Energy and the UHECR Spectrum}
   \label{CHAPTER:UHECR Source Maximum Energy and the UHECR Spectrum}

A key aspect of ultra high energy cosmic ray (UHECR) phenomenology is the nuclear composition at their sources.
The presence of nuclei in addition to protons in UHECRs, a mixed composition, implies a harder source spectrum, along with different UHECR rigidities and thus a change in magnetic deflection. 
The spectrum of UHECRs at their sources is often modeled using a power-law of the form $E^{-x}$. The spectral index $x$ needed to fit UHECR observations is typically $\simeq 2.2$-$2.3$ in mixed composition models, 
as opposed to $x\simeq 2.6$-$2.7$ in the case of a pure proton composition, depending on the cosmological evolution of the source power \cite{Allard+08}. 
A mixed composition also implies that the energy at which the extra-galactic component of the total UHECR flux becomes larger than
the Galactic component should be somewhere around the so-called ankle -- i.e., $\sim 3$-$5~10^{18}$~eV, while this transition is found at a lower energy in pure proton models \cite{BerGri88,Allard+05,Allard+07,Aloisio+07}.

It goes without saying, that any accurate description of the UHECR data should reproduce both the measured spectrum \emph{and} composition in a consistent way.
While the measurement of the UHECR composition at Earth remains a difficult observational challenge, important progress has been made in the recent years.
Notable results include those of the Pierre Auger Observatory \cite{AugerCompo09,AugerXmax10}, which has provided hints that the composition becomes heavier and heavier above $\sim 10^{19}$~eV. 
In contrast to this, other experiments such as HiRes \cite{Abbasi+04} and Telescope Array \cite{Kawai+08,Thomson+10} have shown results which are compatible with a pure proton scenario.

Propagation effects are known to modify the composition of UHECRs, as the energetic nuclei are photo-dissociated in interactions with background photons.
 Horizon analyses show that Fe nuclei and protons with energies above $\sim 6~10^{19}$~eV can propagate over roughly the same distance without losing a significant
 fraction of their total energy, while intermediate mass nuclei are suppressed at shorter distances \cite{Harari+06,Allard+08}. 
As a result, UHECRs at these high energies, should be dominated by either protons, Fe (or sub-Fe) nuclei, or a combination of the two. 
The Auger results on UHECR 
composition can therefore be understood if the proton component is cut at the source at a relatively low energy, around $10^{19}$~eV, and Fe nuclei are accelerated 
up to higher energies, eventually dominating the overall spectrum \cite{1984ARA&A..22..425H}. This would be natural, for instance, in a scenario where each nuclear species reaches a 
maximum energy at the source which is proportional to its charge, $Z$.

Although it has been shown that it is indeed possible to fit the UHECR energy spectrum in such a scenario \cite{Allard+08}, this requires a source composition which is richer in Fe nuclei than would be expected from a simple extrapolation of the low energy cosmic ray source composition.
In principle, this can occur if the source environment is richer in Fe nuclei than the typical interstellar medium, or if the UHECR are accelerated by a mechanism which somehow discriminates
nuclei on the basis of mass. 

One simplifying assumption which is often made is that all sources of UHECR cosmic rays accelerate nuclei with a spectrum which cuts off at the same maximum energy $E_{\text{max}}$.
Although this assumption seems innocuous, it will be shown here that this assumption is not toothless. 
We propose another mechanism to produce a heavier \emph{effective} composition at the sources, independent 
of the acceleration model. The assumptions made are that the maximum energy reached by the particles accelerated in a given source is not a universal quantity, and that the distribution of
sources with respect to their maximum energy follows a power-law. Under these two assumptions, the contribution of all sources is equivalent to a scenario where identical sources 
not only inject UHECRs with a softer spectrum, as already shown by \cite{KacSem06}, but also with a heavier source composition. 
The Fe-to-proton enhancement ratio $\eta$, which quantifies this effect, is computed as a function of the source parameters.

\section[Source $E_{\text{max}}$ Distributions and the UHECR Source Spectra]{Source $E_{\text{max}}$ Distributions and Resulting UHECR Source Spectra}
\label{sec:compositionenhancement:source}

For most proposed cosmic ray sources the maximum energy which can be reached is limited by the ability of the source to contain particles in the acceleration region. The containment of particles
is determined by their rigidity, defined as $R \equiv p/q$. In
the relativistic limit which is applicable to UHECR acceleration, the rigidity is proportional to $E/Z$, where $Z$ is the charge of the accelerated nuclei. 
Unless other mechanisms come into play to limit the energy of specific nuclei (such as photo-dissociation processes), the maximum energy, $E_{\text{max}}^{(i)}$, of nuclei of type $i$ at the source is simply proportional to their charge, $Z_{i}$:
\begin{equation}
E_{\text{max}}^{(i)} = Z_{i}\times E_{\text{max}}^{(\text{p})}
\label{eq:emaxrelation}
\end{equation}
where $E_{\text{max}}^{(\text{p})}$ is the proton maximum energy.

For simplicity, UHECR models usually assume that all cosmic ray sources are identical, having the same spectrum extending up to the same energy. 
It is clear, however, that the maximum energy will differ among sources depending on their individual properties, notably their size, magnetic field strength, intrinsic power, age, etc. 
In order to avoid introducing extra free parameters, the simplest assumption (beyond assuming an identical $E_{\text{max}}$) is to assume a power-law distribution for the number of sources as a function of $E_{\text{max}}^{(p)}$.
This is the same assumption as used in ref.~\cite{KacSem06}.

The number of sources which are able to accelerate UHECR to a given maximum energy is
\begin{equation}
\label{eq:numbersources}
n_{\text{sources}}(E_{\text{max}}^{(p)})= n_{0}\left(\dfrac{E_{\text{max}}^{(p)}}{E_{0}}\right)^{-\beta} H\left(E_{\text{sup}}-E_{\text{max}}^{(p)}\right),
\end{equation}
where $E_{\text{sup}}$ has been introduced as the highest possible proton energy in any source and $H(x)$ is the Heaviside step function. 
The parameter $E_{0}$ is an arbitrary reference energy, which has been introduced here simply to clarify the dimensionality of the various quantities.
The parameter $\beta$ is expected to be a positive number, as the number of sources able to accelerate nuclei to high energies should presumably decrease with increasing energy.
$E_{\text{sup}}$ serves not only to avoid a divergence in the total number of sources, but also physically represents the limit of particle acceleration mechanisms if such a limit exists. 

The individual sources are assumed to each produce a power-law spectrum of UHECRs with the same spectral index, $x$. The number of nuclei of type $i$ injected per second and per unit energy by a given source is then:
\begin{equation}
 \label{eq:injection}
Q_{i}(E) \equiv \dfrac{\text{d}^{2}N}{\text{d}E\,\text{d}t}=Q_{0}\,\alpha_{i}\left(\dfrac{E}{E_{0}}\right)^{-x} H(E_{\text{max}}^{(i)}-E),
\end{equation}
where $\alpha_{i}$ is the abundance of nuclear species $i$, with $\sum \alpha_{i} = 1$.

Assuming that the sources are homogeneously distributed over space, the total number of cosmic rays injected per unit time, per unit energy, and per unit volume is then given by:
\begin{equation}
q_{i}(E) = \int_{0}^{\infty} Q_{i}(E)\times n_{\text{sources}}(E_{\text{max}}^{(p)})\,\text{d}E_{\text{max}}^{(p)}.
\end{equation}

Replacing from Eqs.~(\ref{eq:numbersources}) and~(\ref{eq:injection}), one gets:
\begin{equation}
q_{i}(E) = Q_{0}n_{0}\alpha_{i}\left(\frac{E}{E_{0}}\right)^{-x}\,\int_{0}^{E_{\text{sup}}} \left(\frac{E_{\text{max}}^{(p)}}{E_{0}}\right)^{-\beta}H(Z_{i}E_{\text{max}}^{(p)} - E)\,\text{d}E_{\text{max}}^{(p)},
\end{equation}
 which integrates (for $\beta \neq 1$) into:
\begin{equation}
q_{i}(E) = \frac{Q_{0}n_{0}E_{0}}{\beta-1}\,\alpha_{i}Z_{i}^{\beta-1}\left(\frac{E}{E_{0}}\right)^{-x-\beta+1}\left[1 - \left(\frac{E}{Z_{i}\,E_{\text{sup}}}\right)^{\beta-1}\right].
\label{eq:effectiveInjection}
\end{equation}

The situation $\beta = 1$ bears note. In that case there would be the same number of sources in each decade of $E_{\text{max}}$, leading to a logarithmic divergence in the integral if $E_{\text{sup}}$ tends to infinity. 
This is not physical, and in any case it would be most natural to expect $\beta$ to be larger than 1.

\section{The Effective Spectrum and Composition in the limit $E\ll E_{\text{sup}}$}

For values of $E_{\text{sup}}$ much larger than $E$ (and $\beta > 1$), \eq\ref{eq:effectiveInjection}) is equivalent to a single source power-law distribution given by:
\begin{equation}
q_{i}(E) = q_{0}\,\alpha_{i}Z_{i}^{\beta-1}\left(\frac{E}{E_{0}}\right)^{-x-\beta+1}.
\end{equation}

This effectively shows that the simple model described here is equivalent to the usual ``universal source model'' with an effective source spectral index
\begin{equation}
x_{\text{eff}} = x + \beta - 1,
\label{eq:xEff}
\end{equation}
and with a modified source composition corresponding to the effective nuclear abundances:
\begin{equation}
\alpha_{i,\text{eff}} = \alpha_{i}\times Z_{i}^{\beta-1}.
\label{eq:alphaEff}
\end{equation}

The effect of a distribution of $E_{\text{max}}$ values among sources is thus a modification of both the energy spectral index \emph{and} the abundances of nuclei. This 
modification is correlated through the parameter $\beta$, showing that these two aspects of UHECR phenomenology are not independent.

\begin{figure}[ht!]
\begin{center}
 \includegraphics[width=0.93\textwidth]{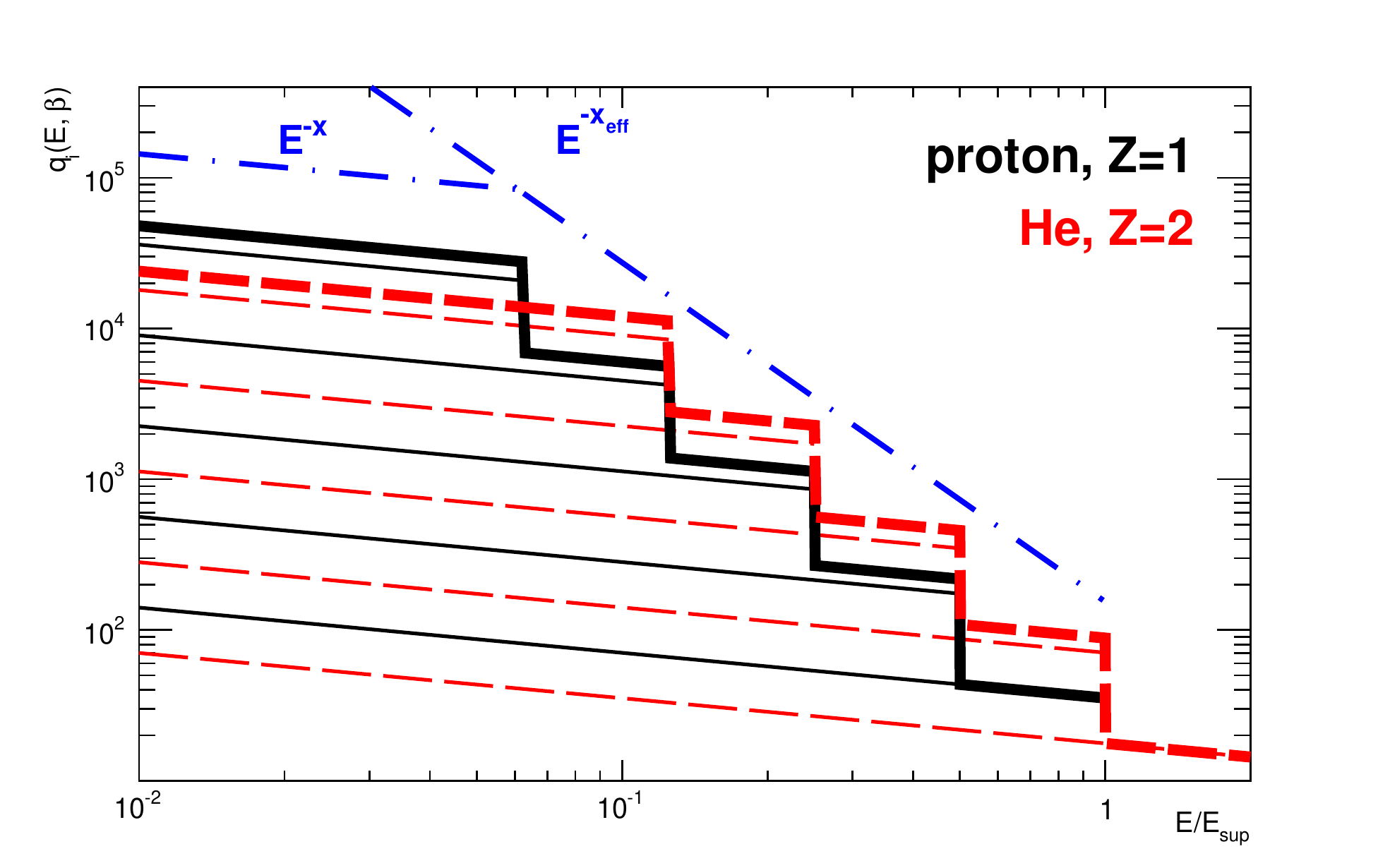}
\caption[$E_{\text{max}}$ Distribution Illustration]{An illustration of the effect of an $E_{\text{max}}$ distribution on the ratio of different nuclei in UHECRs. 
Here, we consider an artificial distribution of sources with 5 different maximum energies, as explained in the text. 
The solid lines are the proton spectra for each source, while the dashed lines denote the helium spectra, extending to twice higher energies. 
The total effective spectrum for each species (before propagation) is given by the bold lines. 
In this example, even though the abundance of protons at the source is higher, in the UHECR region the spectra of the heavier nuclei are enhanced due to the distribution of the number of sources with respect to $E_{\text{max}}$.}
\label{fig:distributions}
\end{center}
\end{figure}

This behavior is easily understood by considering the simple example of, say, 5 discrete sources, as illustrated in \fig\ref{fig:distributions}. 
In this plot, the plain lines show an identical proton injection spectrum, extending up to a maximum energy that is different for each source. 
The dashed lines show the corresponding helium injection spectra, with an assumed abundance ratio of $\alpha_{\text{He}}/\alpha_{\text{p}} = 0.5$, and a maximum energy twice as large for each source, as per \eq\ref{eq:emaxrelation}). 
The bold lines are then simply the sum of all 5 contributions, showing an enhancement of the relative helium abundance. 
In this illustrative example, while the helium nuclei are less abundant than protons in each individual source, they dominate the overall flux injected at high energy.

\fig\ref{fig:distributions} also shows how the original source spectrum steepens as fewer and fewer sources contribute to the flux at higher and higher energies. 
Below the $E_{\text{max}}$ of the least energetic source, the sum spectrum obviously has the same form as the individual source spectra.
Above that energy, and up to $\sim E_{\text{sup}}$, that is the $E_{\text{max}}$ of the most energetic source, the resulting spectrum is a new power-law with the larger spectral index, $x_{\text{eff}}$. 
The resulting spectra in \fig\ref{fig:distributions} is shown by the dashed line fit.

In an actual astrophysical context, extensive propagation studies have determined the effective spectral index needed to reproduce the observed UHECR energy spectrum for a given assumed source composition. 
In the case of pure proton sources, the composition effect demonstrated above is irrelevant, but for a mixed composition scenario \eq\ref{eq:alphaEff}) 
shows that the effective composition to be used in the models is systematically richer in heavy nuclei than the individual source composition, as soon as $\beta > 1$. 
Given \eq\ref{eq:xEff}), this condition is equivalent to saying that the effective source spectrum is steeper than the intrinsic source spectra ($x_{\text{eff}} > x$). 
For instance, if the source spectral index is $x = 2.0$, as would be expected from standard diffusive shock acceleration, 
the effective source spectrum needed to reproduce the data in the case of a mixed composition model, namely $x_{\text{eff}} \simeq 2.3$, requires a source distribution index $\beta \simeq 1.3$. 
This in turn implies that the effective Fe nuclei abundance is larger than the Fe abundance in individual sources by a factor $\eta_{\text{Fe}} = Z_{\text{Fe}}^{\beta - 1} \sim 2.7$.

\begin{figure}
\begin{center}
  \includegraphics[angle=90,width=0.9\textwidth]{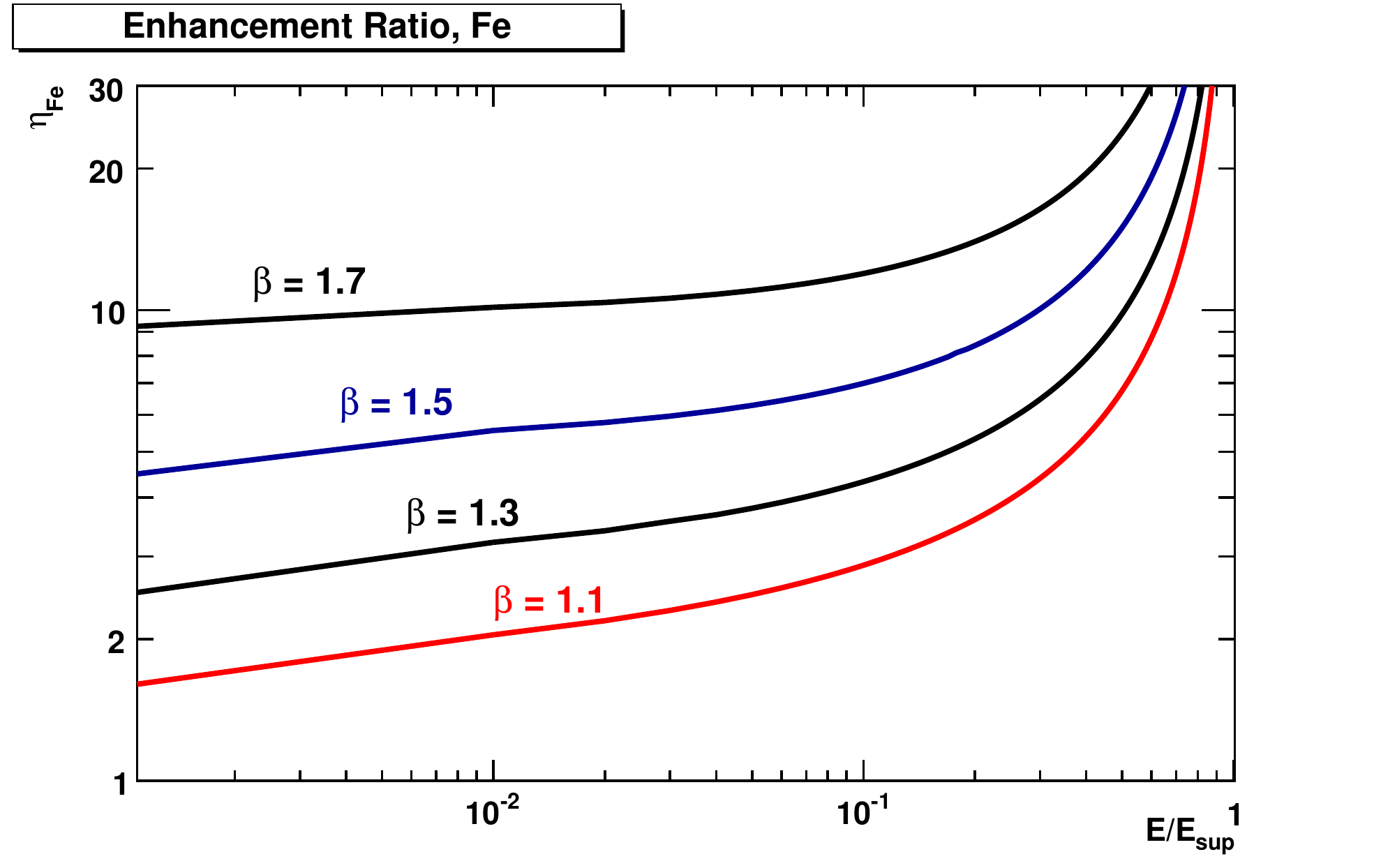}
\caption[The Enhancement Ratio $\eta_{\text{Fe}}$]{The enhancement ratio of Fe nuclei, $\eta_{\text{Fe}}$, \eq\ref{eq:enhancementratio}), as a function of $E/E_{\text{sup}}$, for several values of $\beta$, the slope of the $E_{\text{max}}$ distribution.}
\label{fig:ratioplot}
\end{center}
\end{figure}

\section{The Abundance Enhancement Ratio at the Highest Energies}

At the very highest energies, when E becomes closer to $E_{\text{sup}}$, the last factor in \eq\ref{eq:effectiveInjection}) is no longer equal to $\sim 1$. The effective spectrum is then no longer a power-law, and the effective composition becomes dependent on energy.
We can define the enhancement ratio of the abundance of the various nuclei relative to protons, $\eta_{i}$, as
\begin{equation}
\label{eq:enhancementratio}
 \eta_{i}(E,\beta)\equiv \dfrac{q_{i}(E)/q_{\text{p}}(E)}{\alpha_{i}/\alpha_{\text{p}}} = \frac{Z_{i}^{\beta-1} - (E/E_{\text{sup}})^{\beta-1}}{1 - (E/E_{\text{sup}})^{\beta-1}}\,,
\end{equation}
where $q_{i}(E)$ was taken from \eq\ref{eq:effectiveInjection}).

\begin{figure}
\begin{center}
\includegraphics[angle=90,width=0.87\textwidth]{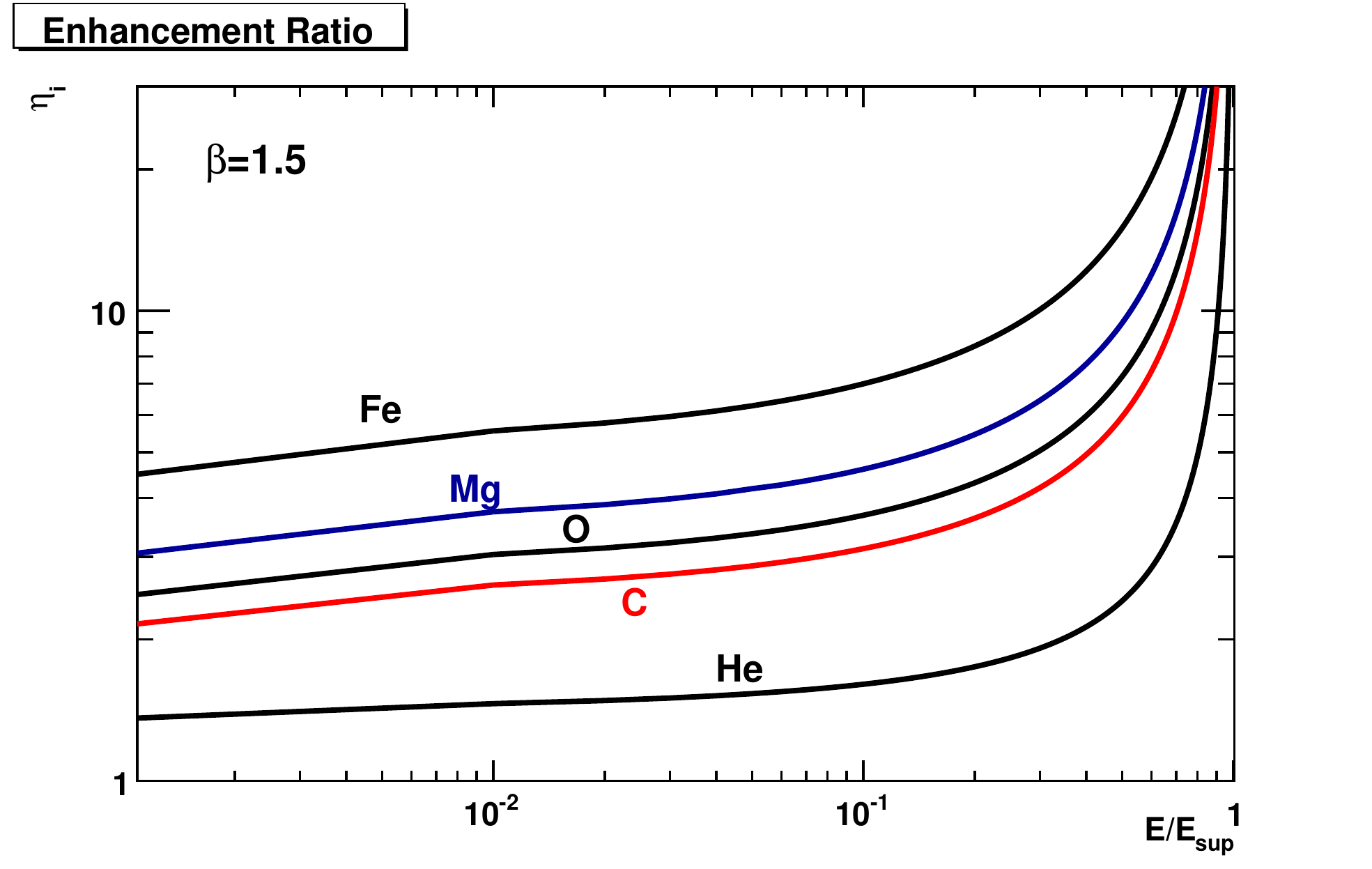}
\caption[The Enhancement Ratio $\eta_{i}$] {\label{fig:ratioplotNuclei} The enhancement ratio, $\eta_{i}$, for various nuclei, as indicated, as a function of $E/E_{\text{sup}}$.}
\end{center}
\end{figure}

As an illustration, the behavior of the Fe enhancement ratio, $\eta_{\text{Fe}}$, as a function of $E/E_{\text{sup}}$ is plotted in \fig\ref{fig:ratioplot} for several values of $\beta$. 
The low energy limit is given by $\eta_{i} = Z_{i}^{\beta-1}$ (see \eq\ref{eq:alphaEff})). 
The enhancement ratio increases with energy. Likewise, the enhancement ratio is larger for heavier nuclei, as shown in \fig\ref{fig:ratioplotNuclei} for $\beta = 1.5$.

At the highest energies, close to $E_{\text{sup}}$, the values shown on the plot should no longer be taken seriously, since in practice the granularity of the sources would have to be taken into account. 
The features of the cosmic ray energy spectrum and composition in this energy range are determined by the properties of the few highest energy sources, which are subject to ``cosmic variance''. 
That is to say that the observed configuration of sources is one realization of the many possible configurations of Universe. 
It is obvious, however, that at energies above the proton maximum energy, $E_{\text{sup}}$, the UHECRs should be completely devoid of protons, and thus the propagated UHECRs will be dominated by Fe and sub-Fe nuclei~\cite{Allard+08}.

\section{Conclusion}

In sum, a distribution of UHECR sources as a function of the maximum energy to which they can accelerate particles changes both the effective source spectrum and the effective source composition.
Although the results shown here correspond to a specific set of assumptions, they show that these effects are significant, and 
thus should be kept in mind when studying the phenomenology of UHECRs within the simplified framework of identical source models.

In particular, a UHECR composition which is heavier than the composition at the sources can be obtained very naturally under the astrophysically sensible assumption that the number of sources able 
to accelerate particles up to a given energy decreases sufficiently rapidly with that energy ($\beta > 1$). 
This ``abundance enhancement effect'' depends on the charge of the nuclei, and is less pronounced for lighter nuclei. 
The tendency is an enhancement of the contribution of Fe nuclei with respect to intermediate mass nuclei, which are also more affected by propagation effects due to their shorter horizon scales~\cite{Allard+08,Harari+06}.

It is well-known that a heavier composition is needed to fit the most recent UHECR data, within a model where a low value of $E_{\text{sup}}$ is invoked to explain the apparent transition towards a heavy-nuclei-dominated component above $10^{19}$~eV.
These models naturally account for a relatively sharp cut-off of the proton-dominated UHECRs, in the energy range where the sources cease to accelerate protons. 
If, however, the heavier components are not sufficiently abundant in this energy range, this proton cut-off would appear in the energy spectrum as a visible feature. 
The consistency of these type of models therefore requires a higher abundance of heavy nuclei than what would be expected for the actual source composition.

As has been shown here, an enhancement ratio of the order of 3 for Fe nuclei is natural if the $E_{\text{max}}$ distribution function has a power law index $\sim 1.3$. 
This value of $\beta$ is needed to go from an intrinsic source spectrum of $E^{-2.0}$ to an effective source spectrum of $E^{-2.3}$,
as is typically needed to fit the observed spectrum using a mixed composition scenario~\cite{Allard+05,Allard+07b}. 
Larger enhancement ratios are possible if the $E_{\text{max}}$ distribution function is steeper, which in turn implies that the individual source spectra are harder. 
A source spectrum of $E^{-1.8}$, for example, would need a value of $\beta = 1.5$ to mimic a single-type source distribution with a spectrum of $E^{-2.3}$.
This would result in an enhancement ratio of $\sim 5$ for Fe nuclei.

An important note, however, is that if the value of $E_{\text{sup}}$ is low, UHECR experiments are in fact detecting cosmic rays just below and above $E_{\text{sup}}$. 
The composition effects would then be expected to be \emph{larger} than discussed above (as apparent in Figs.~\ref{fig:ratioplot} and~\ref{fig:ratioplotNuclei}). 
In this transition range, across $E_{\text{sup}}$, the analytical treatment used here is no longer relevant, and the local distribution of sources should be taken into account. 
For the same reason, a simple description of the overall spectrum in terms of power laws may not be possible for the highest energy cosmic rays. 
Even in the framework of the above continuous model, \eq\ref{eq:effectiveInjection}) is no longer a power-law in this energy range, and it is not obvious that it is possible to define an effective power-law index as in \eq\ref{eq:xEff}).


Although the $E_{\text{max}}$ distribution function introduces an extra free parameter in the general phenomenology of UHECRs, it is one which can be related to the astrophysical parameters of a given source population.
More specifically, the value of $\beta$ can in principle be derived from physical considerations related to the acceleration model,  within a given UHECR source model. 
The introduction of this parameter therefore enriches the possibility to understand the UHECR phenomenon in an astrophysical context. 
In other words, this extra parameter allows us to relate the parameters of UHECR phenomenology, derived from fits of the data within  single-identical-source models, 
to more meaningful astrophysical parameters related to the actual acceleration mechanism and acceleration environment present in the sources of UHECR, whatever they may be.


%
%
%
%
%
%
%
%
%
%
%
%
%
%

      \printbibliography[heading=subbibliography]
       \end{refsection}
 
    \begin{refsection}
  \chapter{UHECR Source Statistics} 
    \label{CHAPTER:UHECR Source Statistics}
    

The continued absence of a clear signal of anisotropy or correlation with some classes of astrophysical objects 
has raised doubts about the utility of continuing to search for the sources of the highest energy particles in the universe.
Because it is in line with some indications of heavy nuclei among the UHECR statistics,
it is legitimate to question the ability of both present and future detectors to identify sources and to study their properties through the particle channel.

On the other hand, even though, due to large particle charges or strong magnetic fields, it may be the case that cosmic rays are highly deflected at ultra-high-energy, 
individual sources could still be isolated in the sky if the UHECR flux is dominated by the contribution of a limited number of sources. 
In fact, the reduction of the horizon associated with the GZK effect implies that fewer and fewer sources contribute to the flux at higher and higher energy. 
In this regard, the GZK effect can turn into a useful phenomenon, provided that the low statistics which come with it can be overcome in future experiments.

This question will be addressed here quantitatively by studying the contribution of individual sources to the overall flux. 
To this end, several astrophysical scenarios with different source densities, source spectra, and compositions are simulated.
In addition, both the ``cosmic variance'' associated with different possible realizations and the effect of a distribution of intrinsic source luminosities are explored. 
Because they depend strongly on the assumptions made regarding magnetic fields and source composition, the study of particle deflections is omitted for the present, and  
no attempt is made to draw realistic sky maps. Rather, the focus is put on the number of contributing sources as a function of energy. 
This is in contrast to some earlier studies of source statistics which were developed in the context of UHECR clustering and multiplet analyses (such as in refs.~\cite{BlaDeMar04,Harari+04,Younk09}).

\section{The method}
\label{sec:method}

As the energy of cosmic-ray particles increases, their propagation length decreases due to their interaction with the photon background. The result of this is that the contribution of far-away sources is attenuated with increasing energy, 
and so fewer and fewer sources are visible in the UHECR sky with increasing energy. This phenomenon is known as the GZK effect~\cite{Greisen1966,ZatKuz66}. 
At an energy of $10^{20}$~eV, the 90\% horizon scale is reduced to $H \sim 80$~\acrshort{Mpc}. 
As a consequence of the approach of the horizon, the number of sources potentially contributing will be limited to 
\begin{equation}
 N_{\text{s}} = n_{\text{s}}\times 4\pi \frac{H^{3}}{3} \simeq 21 \left(\frac{n_{\text{s}}}{10^{-5}\,\text{Mpc}^{-3}}\right)\times \left(\frac{H}{80\,\text{Mpc}}\right)^{3}
\end{equation}
on average, where $n_{\text{s}}$ is some effective source density.

The fraction of the flux actually contributed by each individual source at a given energy depends of course on its actual distance, intrinsic power, and precise attenuation due to intervening interactions of the accelerated cosmic rays. 
To study the combination of these effects, I did a series of Monte-Carlo simulations using a well-tested propagation code which was developed by D. Allard \cite{Allard+08,AllardPropagReview2011}. 
This Monte-Carlo code can be applied to models with various source spectra, compositions, powers, spatial distributions, and cosmological evolutions.

Various combinations are explored, under the requirement that the resulting propagated energy spectrum is compatible with current data. 
Among the possible models, four are used which include both conservative and extreme cases:
\begin{inparaenum}[i\upshape)]
 \item a model in which the UHECRs are composed only of protons, 
 \item a model in which UHECRs are composed solely of iron nuclei,
 \item a generic mixed composition model, and 
 \item the so-called low proton $E_{\max}$ model~\cite{Allard+07,Allard+08}, which accounts for a possible evolution towards a heavier composition above 10~EeV~\cite{AugerComposition2011}.
\end{inparaenum}

For each of these four models, a set of parameters that have been found to fit the data are shown in Table~\ref{tab:modelParameters}. 
The source spectrum is assumed to be a power law of index $x$. 
These sources are assumed to be limited in their ability to accelerate particles by the particle rigidity, and so the source spectra are modeled with an exponential cutoff above energy $E_{\max}$ for protons, and $Z\times E_{\max}$ for nuclei of charge $Z$. 
For the pure iron model, the maximum energy of the nuclei is therefore actually 26 times the quoted value of $E_{\max}$. 
For models with a mixed composition, $\beta$ is a heuristic parameter that implements a bias towards heavier nuclei~\cite{AllardPropagReview2011}. 
The relative abundance of nucleus $i$ in the source composition is given by $\alpha_{i} = \alpha_{\text{GCR},i}\times A_{i}^{\beta - 1}$, where $\alpha_{\text{GCR},i}$ is relative abundance of species $i$ in low-energy Galactic cosmic rays \cite{DuvTha96}. 
It is interesting to compare this composition biasing, which is an ad hoc relationship introduced in the literature in order to fit the observed spectra, 
with the results found in chapter \ref{CHAPTER:UHECR Source Maximum Energy and the UHECR Spectrum}, where an abundance enhancement $\alpha_{i,\text{eff}} = \alpha_{i,o} \times Z_{i}^{\beta-1}$ is found. 
In the latter case $\beta$ was the spectral index of the distribution of source maximum energies.   

\begin{table}[htdp]
\begin{center}
\begin{tabular}{|c|c|c|c|}
\hline
Model & $x$ & $\beta$ & $E_{\max}$ (EeV) \\
\hline
pure p & 2.5 & 0 & 110 \\
pure Fe & 2.3  & 0 & 13.85 \\
mixed & 2.3 & 2.3 & 316 \\
low-p $E_{\max}$ & 1.6 & 2.3 & 4 \\
\hline
\end{tabular}
\caption[Model Parameters]{\label{tab:modelParameters} Parameters chosen for the four models used in our Monte-Carlo simulations. $x$ is the power-law index of the source spectrum.  $\beta$ is a composition parameter (see text), and $E_{\max}$ is 
the maximum energy at the source for protons.}
\end{center}
\end{table}%

The quoted values correspond to the parameters that best fit the Auger data.
However, somewhat different values obtained by fitting both the HiRes and Telescope Array data were also used, and it was found that this had no significant impact on the results regarding source statistics.
The values obtained for $E_{\max}$ in each scenario are adjusted, together with the source spectral index and composition enhancement, 
to reproduce the observed cutoff in the UHECR spectrum without introducing unobserved features in the spectrum around the maximum proton energy. 
The choice of $10^{20.5}$~eV for the maximum proton energy in the mixed-composition model is arbitrary. Any value larger that this (i.e. well above the GZK cutoff) would produce essentially the same results.

In all cases, it is assumed that there is no evolution of the intrinsic source power and/or density as a function of redshift. This was done in order not to increase the number of free parameters, but 
to investigate the influence of such an evolution, a fifth model was added, corresponding to the mixed composition model with strong cosmological evolution. The cosmological evolution was included using an \gls{SFR} \cite{HopBea06} or an \gls{FR-II} \cite{Wall+05} evolution model, 
where the source power depends on redshift. This dependence of the source power on redshift requires a different source spectral index to fit the observed data of $x = 2.1$ and 1.8 for the SFR and FR-II models respectively.

\begin{figure*}[ht!]
\centering
\subfigure[Mixed Model]{\label{fig:HistoMixed}\includegraphics[width=0.480\textwidth]{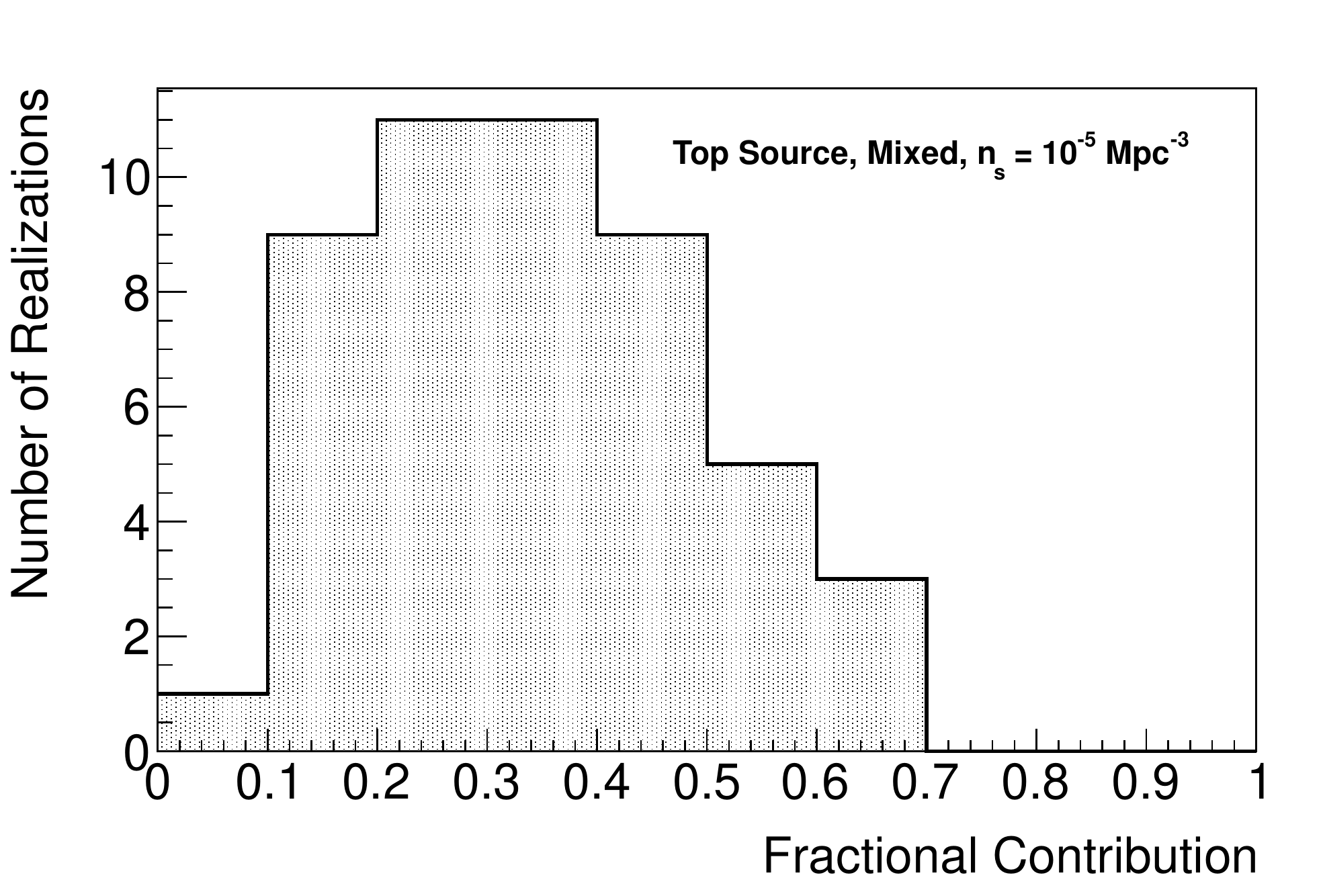}}
\subfigure[Low-$p~E_{\max}$ Model]{\label{fig:HistoLowPE}\includegraphics[width=0.480\textwidth]{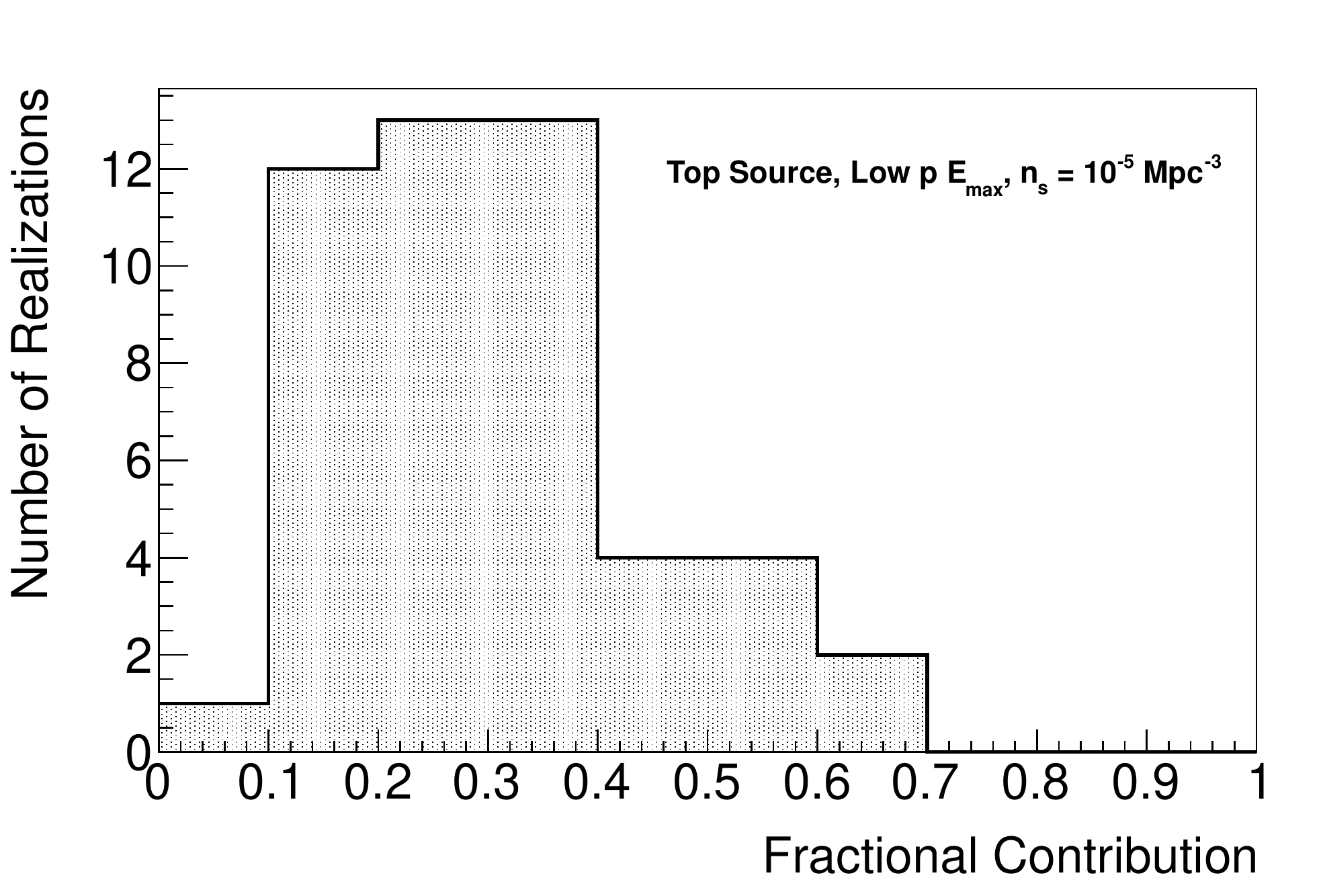}}\\
\subfigure[$p$-Only Model]{\label{fig:HistoP}\includegraphics[width=0.480\textwidth]{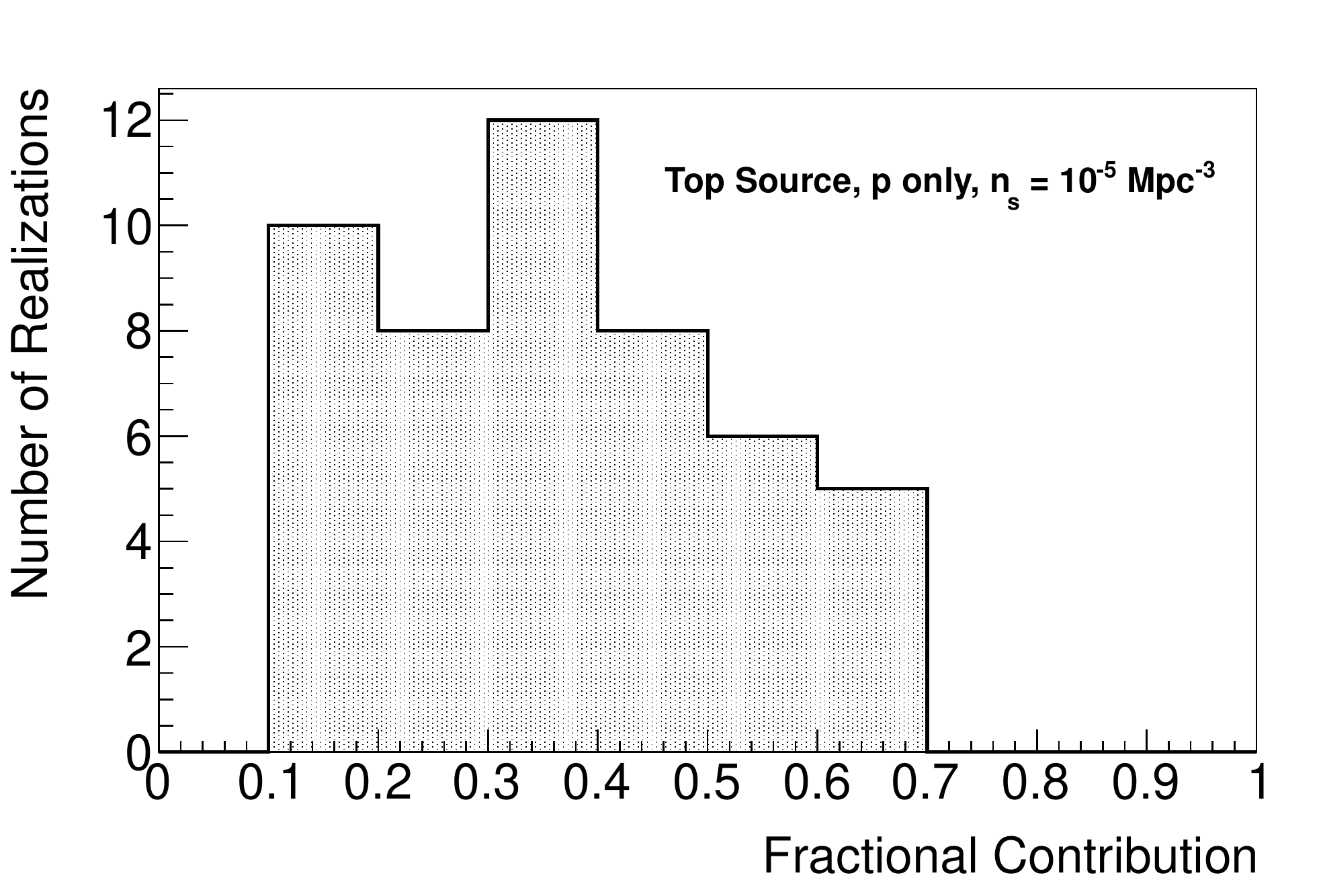}}
\subfigure[Fe-Only Model]{\label{fig:HistoFe}\includegraphics[width=0.480\textwidth]{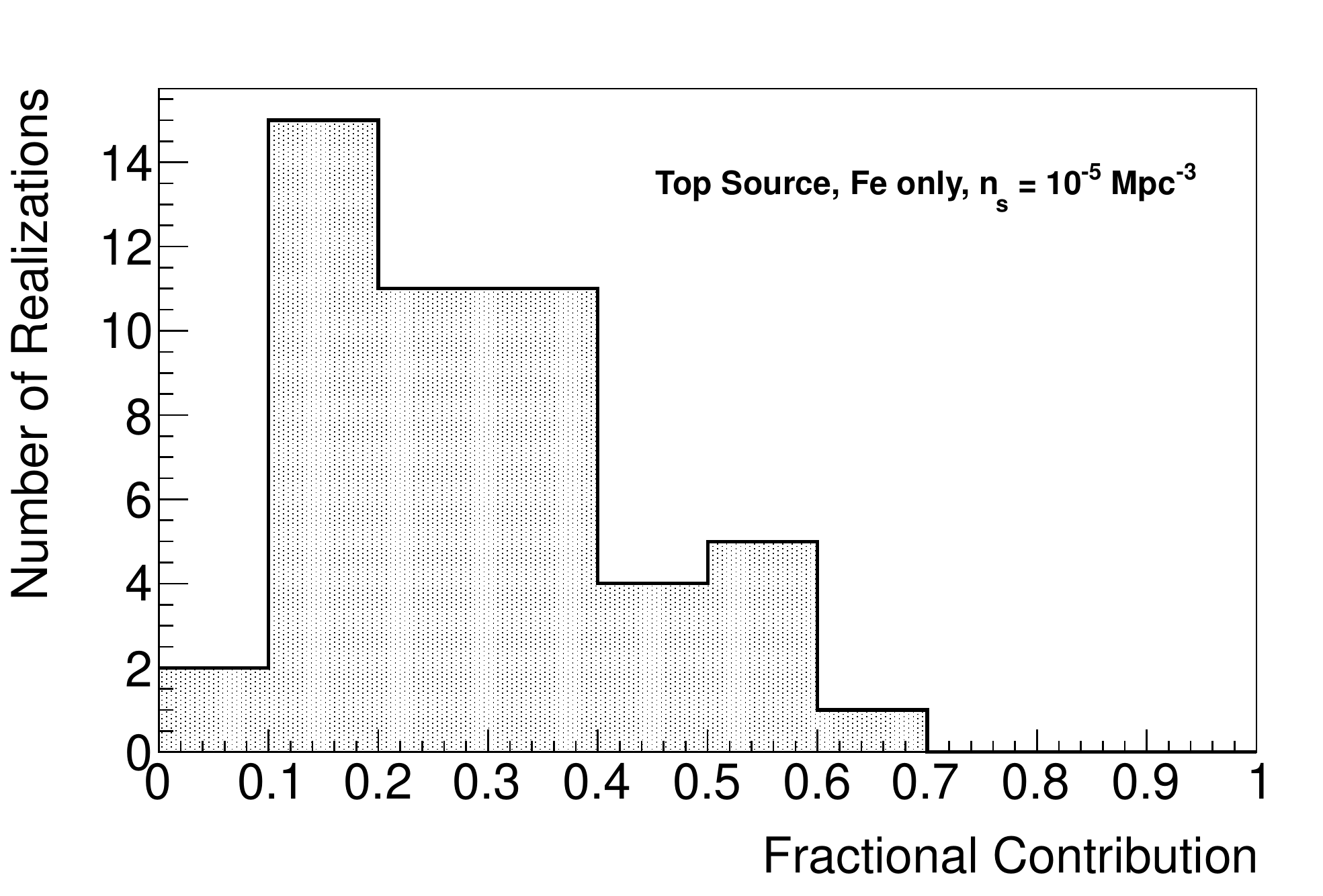}}
\caption[Histograms of Source Flux Fraction by Realization]{Histograms of the flux fraction contributed by the top (most contributing) source in each realization.
A histogram is shown for each of the four models (as per Table \ref{tab:modelParameters}) with a source density of  $n_{\text{s}}~=~10^{-5}\,\text{Mpc}^{-3}$ and $E_{\min}=100$~EeV.
}
\label{fig:S1Histo:4Models}
\end{figure*}

The contribution of each source to the total UHECR flux also depends on its luminosity. 
The default assumption made in the literature and here is that the sources are standard candles -- i.e., that each source has the same power.
The effect of a distribution of intrinsic source luminosities is also studied here by considering the same models, but assigning to each source a luminosity following a chosen probability distribution. 
A $\log_{10}$-normal distribution with a $\sigma$ of 1, a $\log_{10}$-normal distribution with a $\sigma$ of 2, and a power law distribution with an index of -2 are considered.


For each model, three different scenarios where built corresponding to three typical choices of source density: $n_{\text{s}} = 10^{-4}\,\text{Mpc}^{-3}$, $10^{-5}\,\text{Mpc}^{-3}$, and $10^{-6}\,\text{Mpc}^{-3}$. 
The lowest density corresponds to an average of $\sim 4$ sources within 100~Mpc, a situation in which only a few extreme sources are able to accelerate UHECRs up to and beyond 100~EeV.
The highest density is representative of the AGN density in the local universe, which is an upper limit for scenarios involving AGNs as UHECR sources. 

For each of these scenarios, a particular realization of the source configuration with the assumed density is constructed. 
This is done by drawing a sub-sample of the galaxies in the flux-limited 2MRS catalog~\cite{Huchra+12} up to a limiting radius of $R_{\text{limit}} = \left(180/2\pi\right) n_{\text{s}}^{-1/3} \leq 700~\text{Mpc}$, 
thereby mimicking the local source distribution. A continuous source distribution is assumed beyond $R_{\text{limit}}$. 
In the case of a model with a distribution of intrinsic source luminosities, 
each source is assigned a random luminosity according to the chosen distribution.

Hundreds of thousands of UHECRs are then propagated in the Monte Carlo simulation. The UHECRs are emitted by each of the sources in the realization according to their assumed spectrum and composition. 
The particles reaching Earth are collected into a data set which is sufficiently large to avoid Poissonian fluctuations. 
From this data set, the fraction of events which come from each source in the realization is determined by binning each event according to source and energy after propagation. 
This fraction depends on the energy of the UHECRs, since fewer and fewer sources contribute at higher and higher energy, and
the evolution of this fraction as a function of energy can be studied by considering only UHECR events above a given minimum energy threshold, $E_{\min}$.

As the actual universe is a definite, but a priori unknown realization of the underlying astrophysical scenario, the cosmic variance associated with various choices of source distances must be explored.
To this end, the above procedure was repeated for 50 different source configurations with the same density within the same astrophysical model, and the overall distribution of results was analyzed.

\section{Results}

A major goal of UHECR studies, if not the major goal, is to isolate sources in the sky and study their individual properties. 
Whether this can be achieved observationally depends i) on the apparent angular size of the sources, as seen from Earth after propagation through the intergalactic and interstellar magnetic fields, 
ii) on the number of sources visible in the sky and their respective weight, and iii) on the statistics collected by the detectors.

The goal of this study is to address the second point quantitatively, and to do that the fraction of UHECRs contributed by all the sources is
determined for each of the scenarios described in the previous section. The sources are then sorted by apparent luminosity.

\subsection{The Fractional Contribution of the Brightest Sources}

\fig\ref{fig:S1Histo:4Models} shows histograms of the contribution of the brightest source in the sky for four chosen astrophysical scenarios as a fraction of the total flux.
The relatively large spread corresponds to the mentioned cosmic variance, with the contribution of the brightest source in a given realization depending solely on the spatial configuration of the sources in 
that realization. 
It can be seen that, depending on the realization, the brightest source can contribute as much as 68\% of the total flux above $E_{\min} = 100$~EeV or as low as 10\%, with a standard deviation of $\sim 15$\%.

In order to compare different scenarios to one another, the value of the median of the distributions is used as the typical value to be expected for a given scenario. 
This means that the actual contribution of the source would be higher in half of the realizations and lower in the other half. 
A 68\% probability interval around the median is determined to quantify the spread in the distributions. 
This probability interval is calculated by counting the number of scenarios in each fractional contribution, to avoid making any assumption about
the limiting distribution (i.e.\ assuming that the distribution of the fraction contribution is Gaussian, etc.). 
The interval is calculated symmetrically, so that 32\% of the total number of realizations lay above and 32\% lay below, within the bin resolution.

\begin{figure*}[ht!]
\centering
\subfigure[Low-$p~E_{\max}$ Model]{\label{fig:3models3sources:lowp}\includegraphics[width=0.480\textwidth]{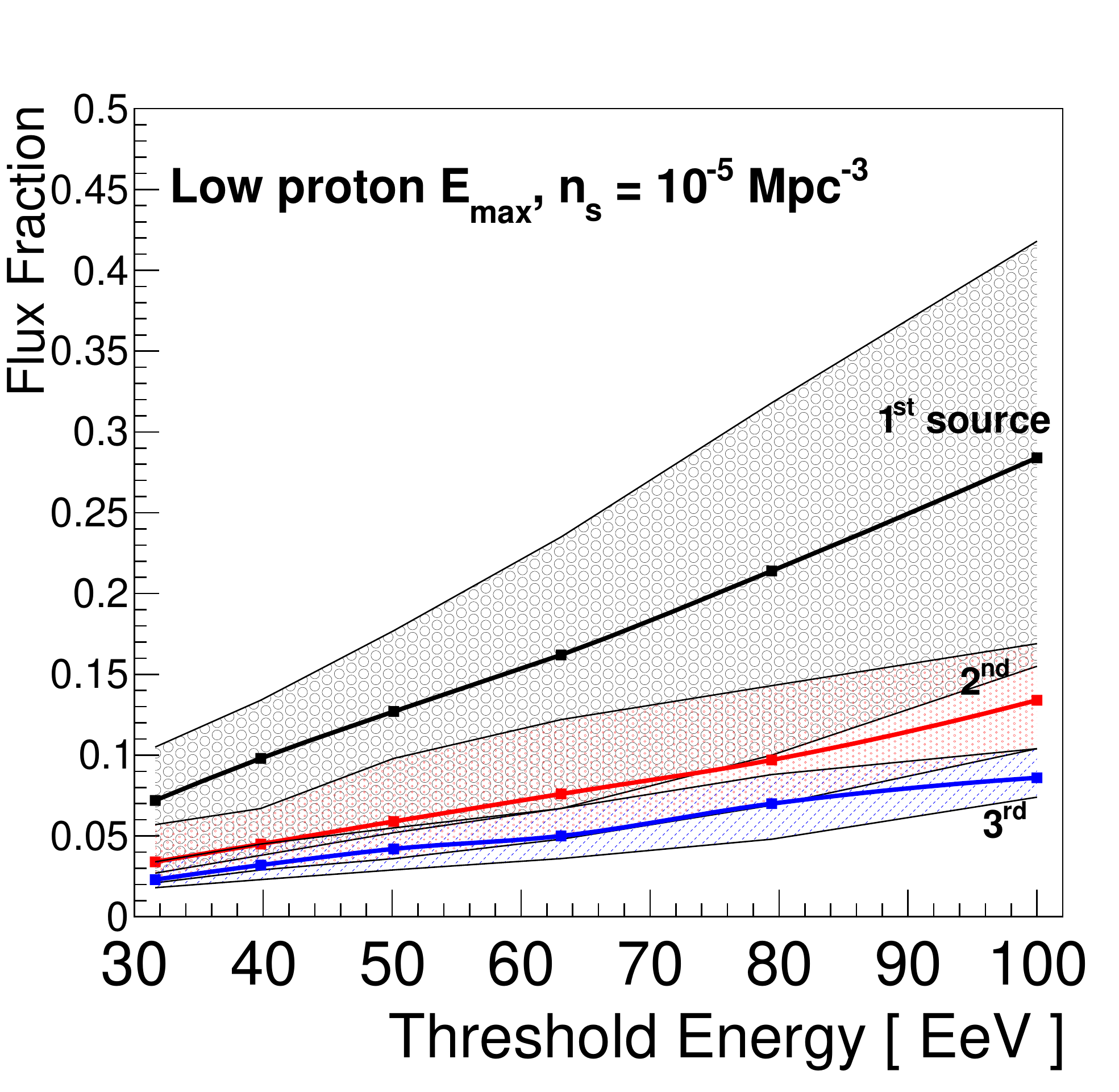}}
\subfigure[$p$-Only Model]{\label{fig:3models3sources:ponly}\includegraphics[width=0.480\textwidth]{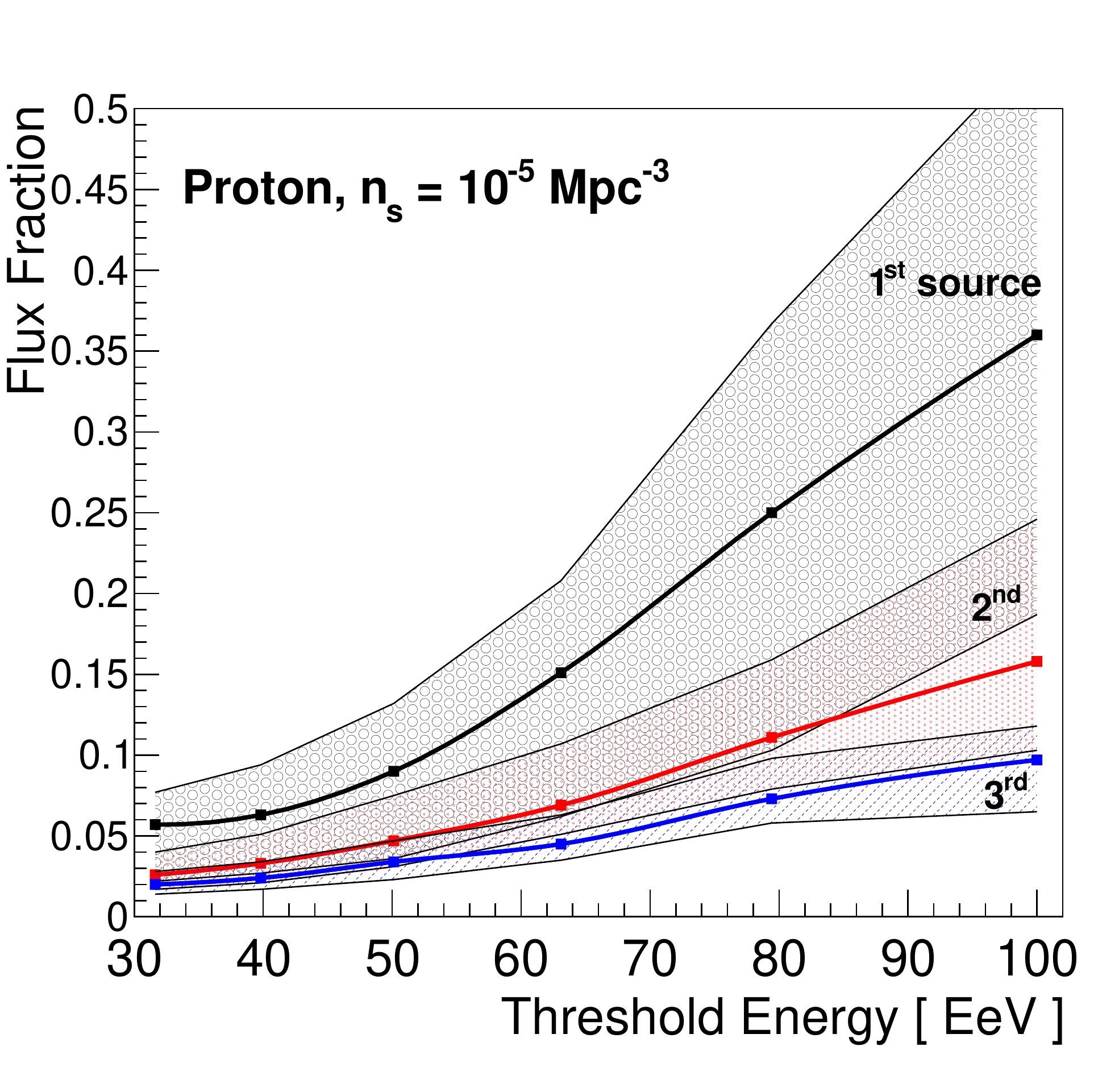}}
\subfigure[Fe-Only Model]{\label{fig:3models3sources:feonly}\includegraphics[width=0.490\textwidth]{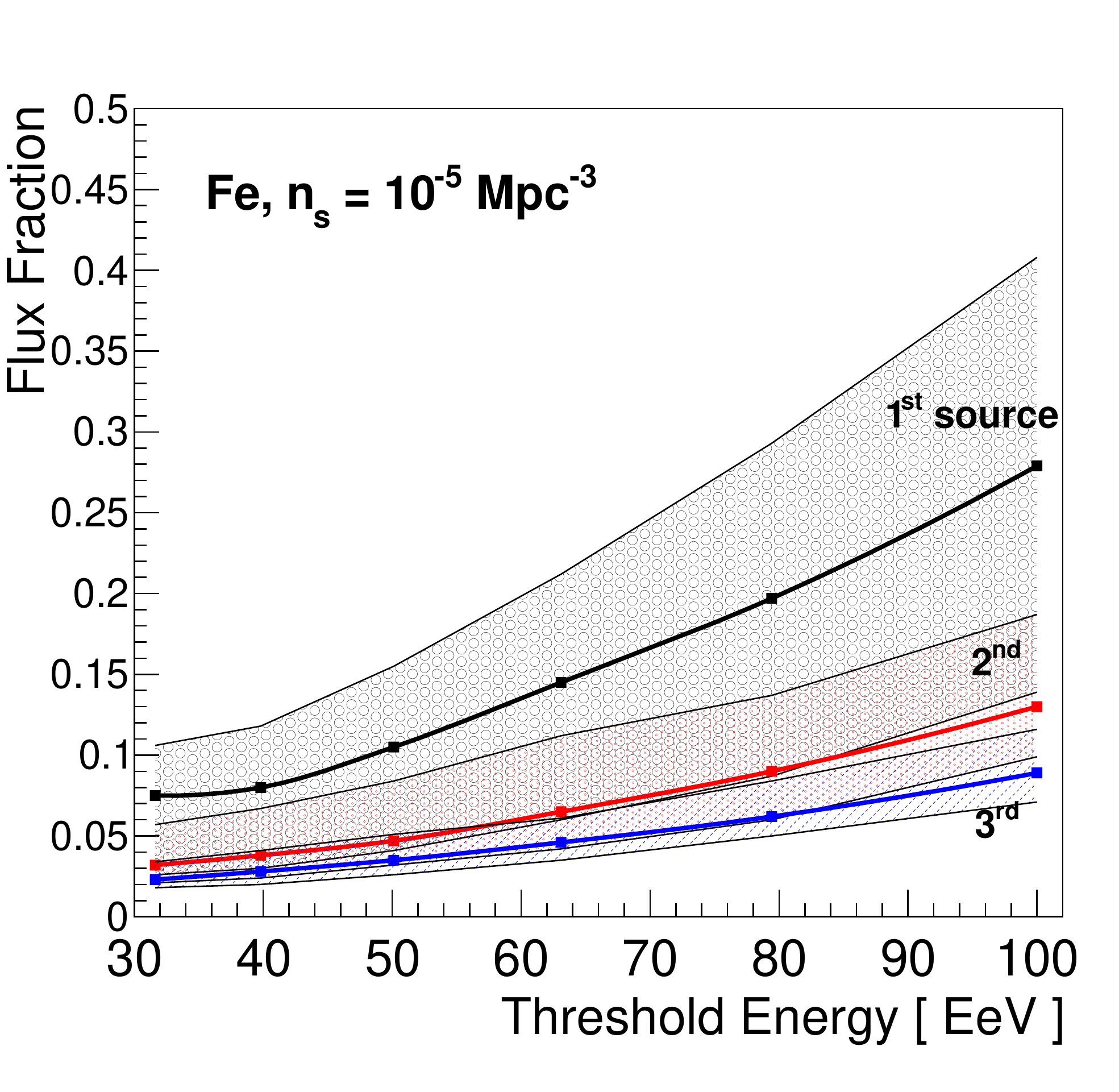}}\\
\caption[Median Flux for 3 Models]{Median flux as a percentage of the total for the three brightest sources in the sky, 
shown for the Low-$p~E_{\max}$, $p$-Only, and Fe-Only models (as given in Table \ref{tab:modelParameters}) 
using a source density of $n_{\text{s}}= 10^{-5}\,\text{Mpc}^{-3}$. Plots for these models with a source density of $10^{-4}$, and $10^{-6}\,\text{Mpc}^{-3}$
are shown in Figs.~\ref{fig:3models3sources:10minus4MPC} and \ref{fig:3models3sources:10minus6MPC} in the appendix.
}
\label{fig:3models3sources} 
\end{figure*}

In \fig\ref{fig:3models3sources}, the evolution of the fractional contribution of the three brightest sources as a function of threshold energy $E_{\min}$ is shown 
for the four models of Table~\ref{tab:modelParameters}, with a source density $n_{\text{s}} = 10^{-5}\,\text{Mpc}^{-3}$.
As expected, the contribution of the brightest sources increases with increasing energy, which is the direct consequence of the GZK effect. 

\begin{figure*}[ht!]
\centering
\subfigure[$n_{\text{s}} = 10^{-4}\,\text{Mpc}^{-3}$]{\label{fig:Mixedmodels3sources:4}\includegraphics[width=0.490\textwidth]{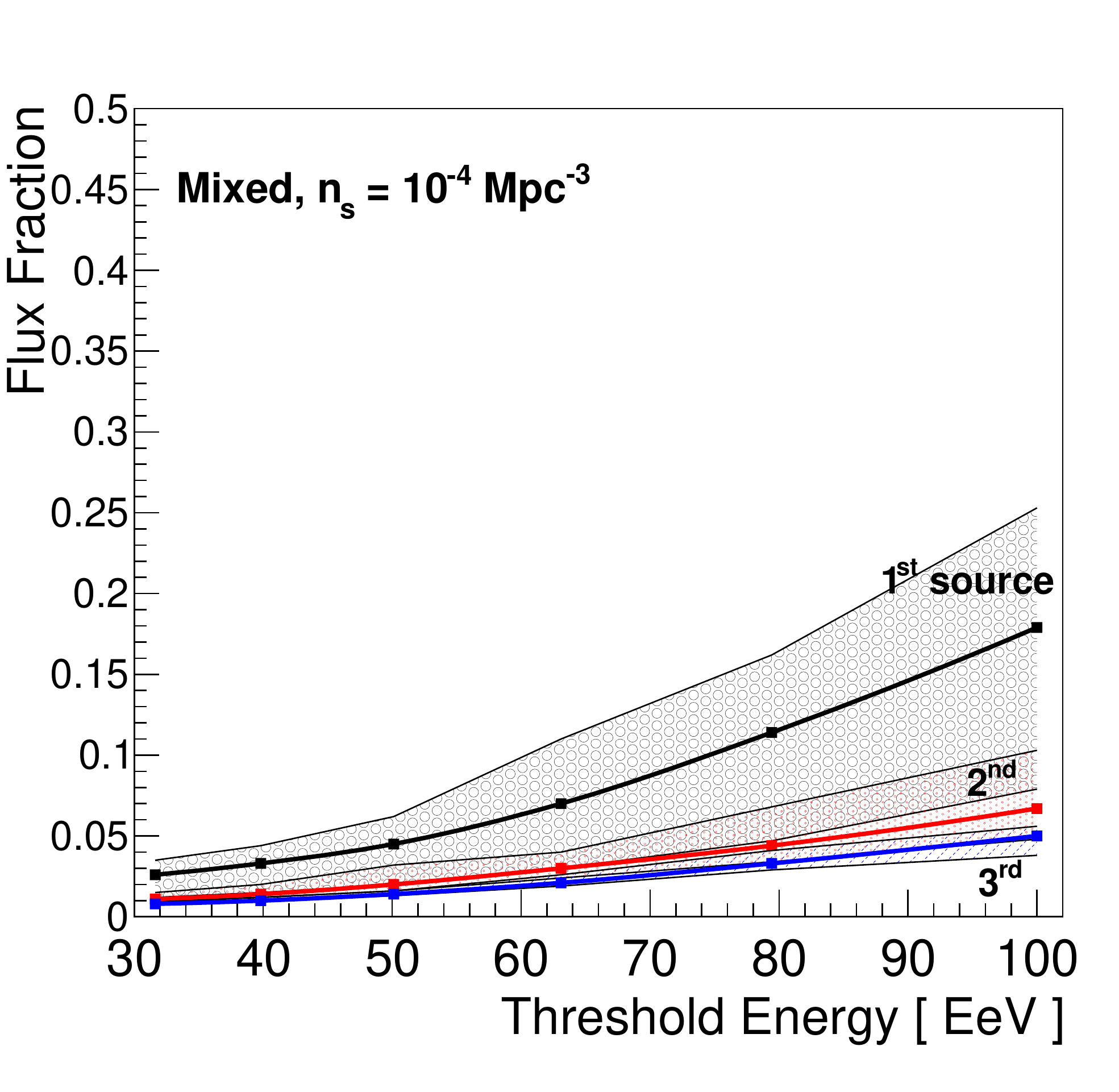}}
\subfigure[$n_{\text{s}}=10^{-5}\,\text{Mpc}^{-3}$]{\label{fig:Mixedmodels3sources:5}\includegraphics[width=0.490\textwidth]{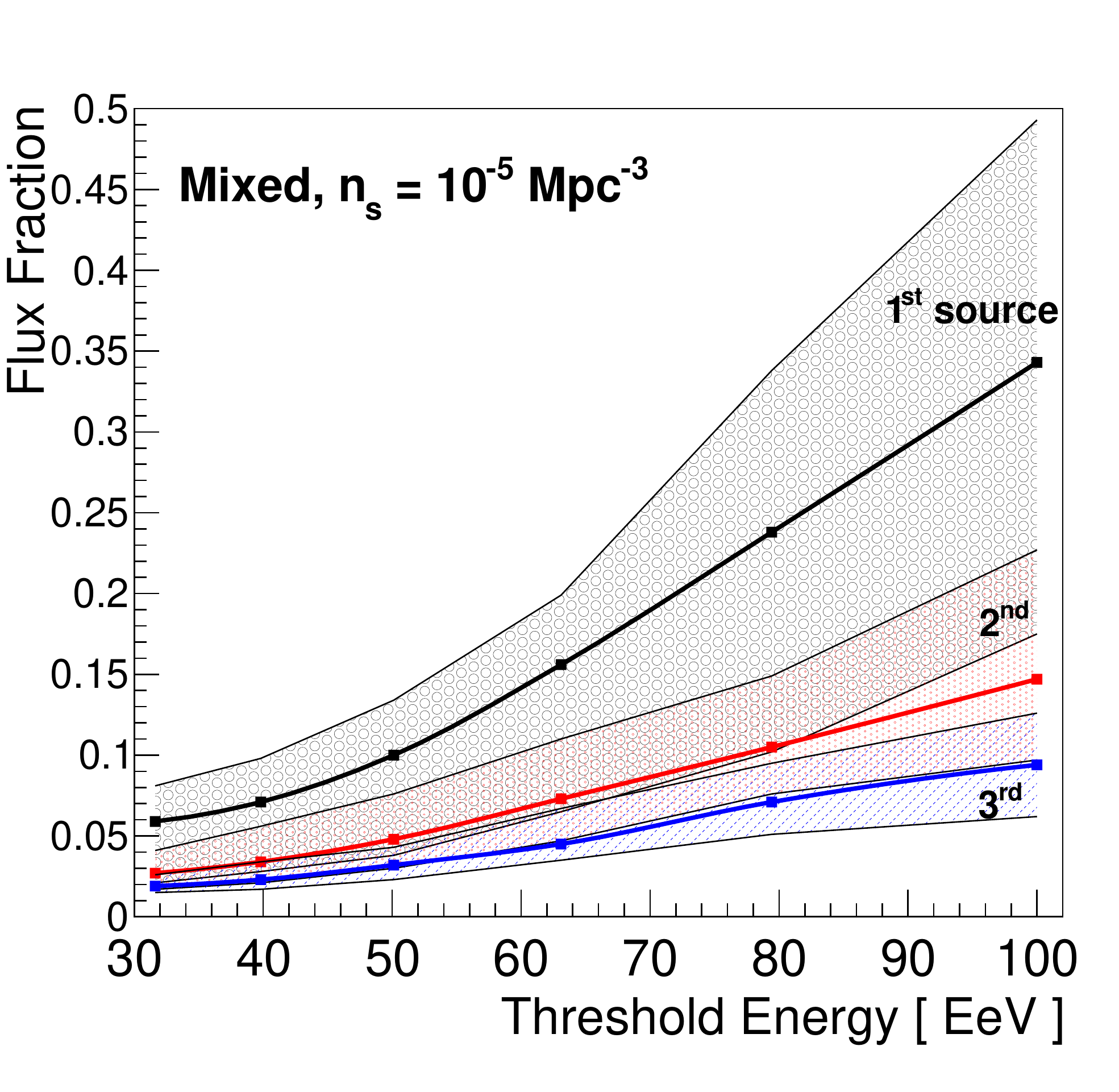}}
\subfigure[$n_{\text{s}}=10^{-6}\,\text{Mpc}^{-3}$]{\label{fig:Mixedmodels3sources:6}\includegraphics[width=0.490\textwidth]{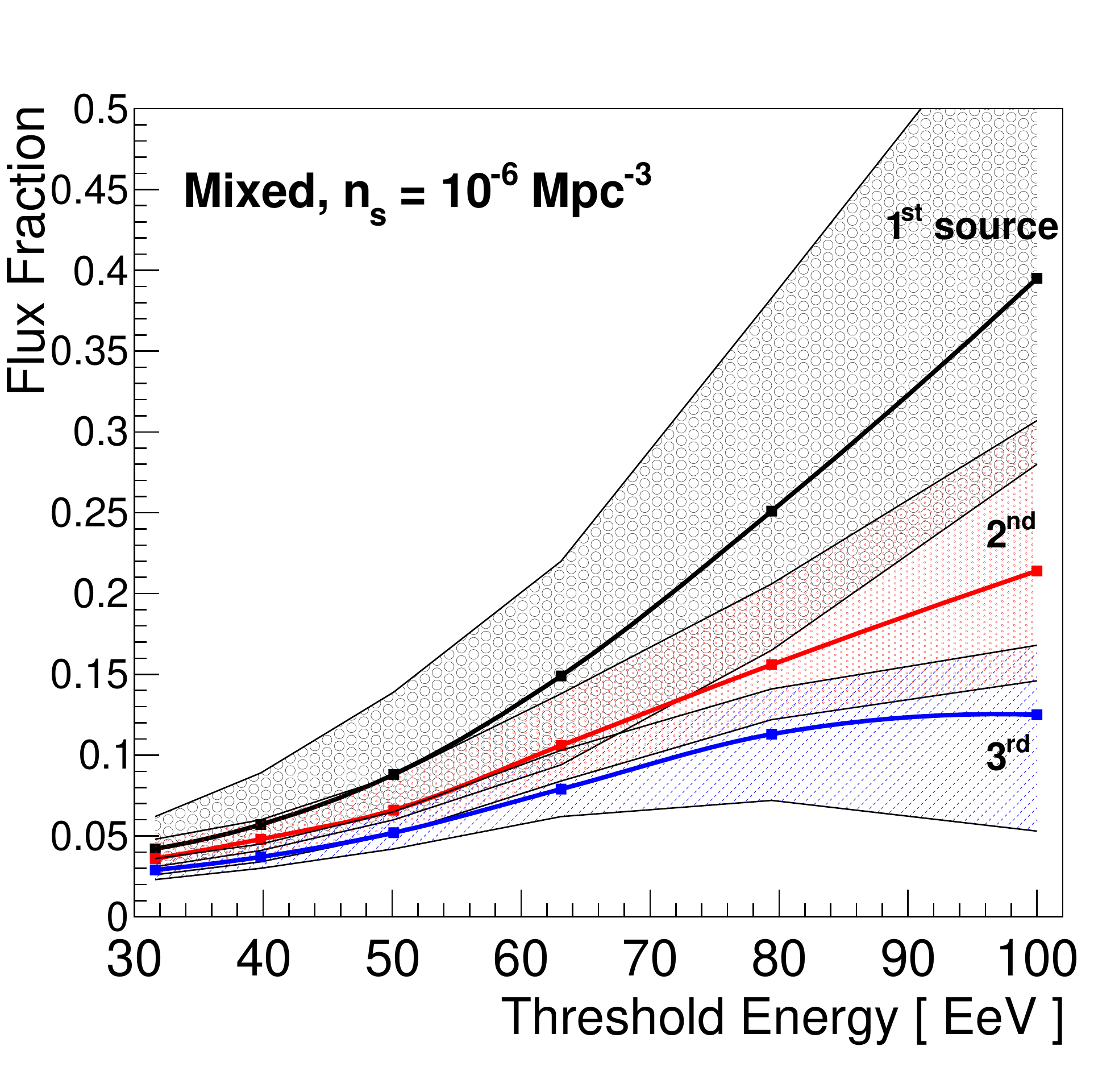}}\\
\caption[Median Flux for 3 Source Densities]{Median flux as a percentage of the total for the brightest 3 sources in the sky, shown for the mixed-composition model (parameters given on Table \ref{tab:modelParameters}) 
with source densities $n_{\text{s}}= 10^{-4}, 10^{-5}, \text{and}\,10^{-6}\,\text{Mpc}^{-3}$.}
\label{fig:Mixedmodels3sources} 
\end{figure*}

In \fig\ref{fig:Mixedmodels3sources}, the same results are shown for the mixed composition model with three different densities: $n_{\text{s}} = 10^{-6}$, $10^{-5}$, 
and $10^{-4}\,\text{Mpc}^{-3}$. As can be seen, the fractional contribution of the second brightest source is typically a factor of 2--2.5 lower than that of the brightest source, 
and the contribution of the third brightest source is another factor of two lower.
This hierarchy is clear when looking at the median of the distributions, 
but two or three sources may contribute roughly equally to the UHECR flux in individual realizations, as suggested by the 68\% probability region shown in the plots.

\begin{figure}[t]
\centering
\includegraphics[width=0.90\textwidth]{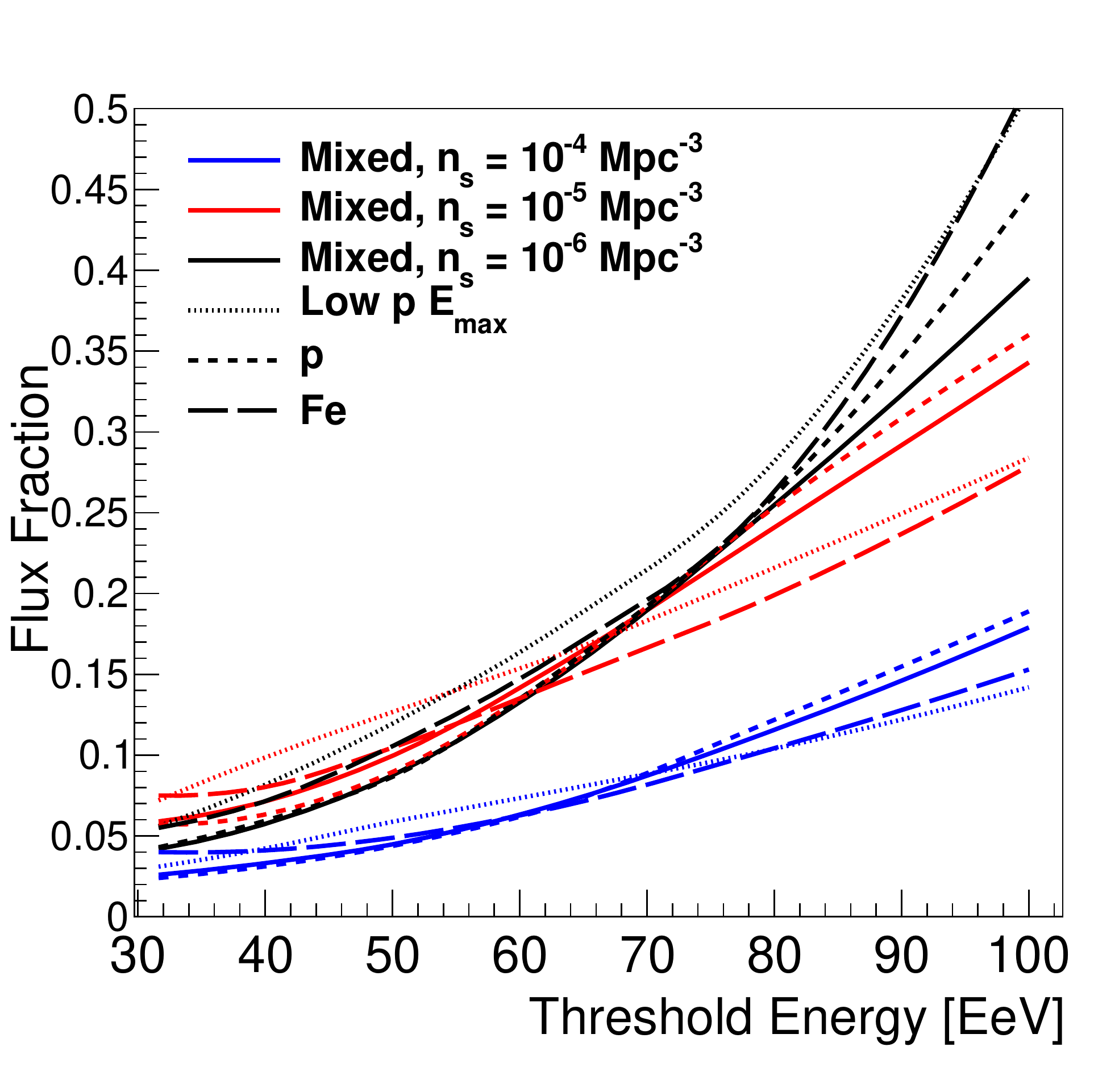}
\caption[Median over 4 Models and 3 Densities]{
 \label{fig:3Densities4Models} Median flux as a percentage of the total for the brightest source in the sky, 
shown for all four models of Table \ref{tab:modelParameters}) and for source densities of $n_{\text{s}} = 10^{-4}, 10^{-5}, \,\text{and}\,10^{-6}\,\text{Mpc}^{-3}$.}
\end{figure}

To compare the variation of the fractional contribution of the brightest source between models and source densities, the median for each scenario is plotted as a function of $E_{\min}$ in \fig\ref{fig:3Densities4Models}. 
At a given density, the difference between the contributions of the brightest source in the four models is relatively moderate, of the order of a 20\% relative variation at $E_{\min}=100$~EeV. 
However, a large difference can be seen for each model between the three source densities. 
At $E_{\min} = 100$~EeV, the typical fractional contribution is a factor of two larger at $n_{\text{s}} = 10^{-5}\,\text{Mpc}^{-3}$ than at $n_{\text{s}} = 10^{-4}\,\text{Mpc}^{-3}$, 
and another 50\% more at $n_{\text{s}} = 10^{-6}\,\text{Mpc}^{-3}$.

As would be expected, the domination of a few sources in the UHECR flux increases as the source density decreases, due primarily to the fewer number of sources overall. 
As can be seen, however, this effect is reduced at low energy. 
This is because the contribution of the most nearby sources, for which the actual density makes a difference (compared to a continuous distribution of sources), is reduced as the GZK horizon recedes. 
At the lowest energies considered here, $\sim 30$~EeV, the horizon scale is much larger than the distance between neighboring sources and larger than the radius $R_{\text{limit}}$, beyond which a continuous source distribution is assumed. 
As a consequence, the increase of the fractional contribution of the brightest source with energy is stronger for lower densities.

\subsection{The Influence of a Luminosity Distribution}

\begin{figure}[ht!]
\centering
\includegraphics[width=0.90\textwidth]{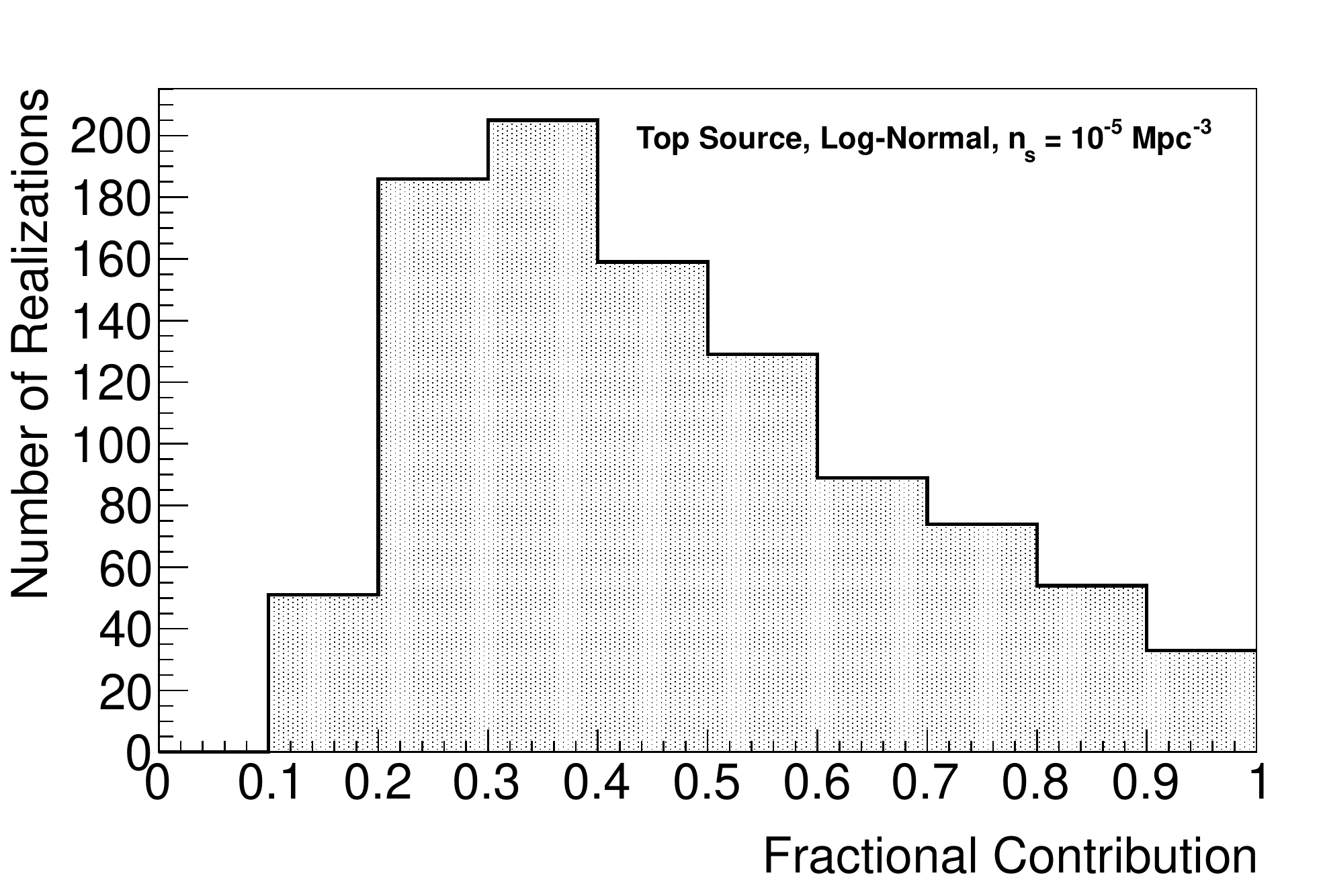}
\caption[Histogram of the Flux Assuming a Distribution of Source Luminosities]{Histogram of the flux contributed by the brightest source in each realization for a mixed-composition model (see Table \ref{tab:modelParameters}) with a source density of $n_{\text{s}}=10^{-5} \,\text{Mpc}^{-3}$ and $E_{\min}=100$~EeV. 
In this case, the individual sources have been given an intrinsic luminosity according to a log$_{10}$-normal distribution with $\sigma=1$.}
\label{fig:S1Histo:LogNLum}
\end{figure}

In \fig\ref{fig:S1Histo:LogNLum}, the same histogram as in \fig\ref{fig:HistoMixed} is shown, but with the source luminosity now distributed according to a log$_{10}$-normal distribution with a $\sigma$ of 1. 
The resulting distribution has a much larger cosmic variance, with the contributions of the brightest source being larger than 50\% in slightly more than one-third of the cases and reaching up to almost 100\% of the flux in a few rare instances.

\begin{figure}[ht!]
\centering
\includegraphics[width=0.90\textwidth]{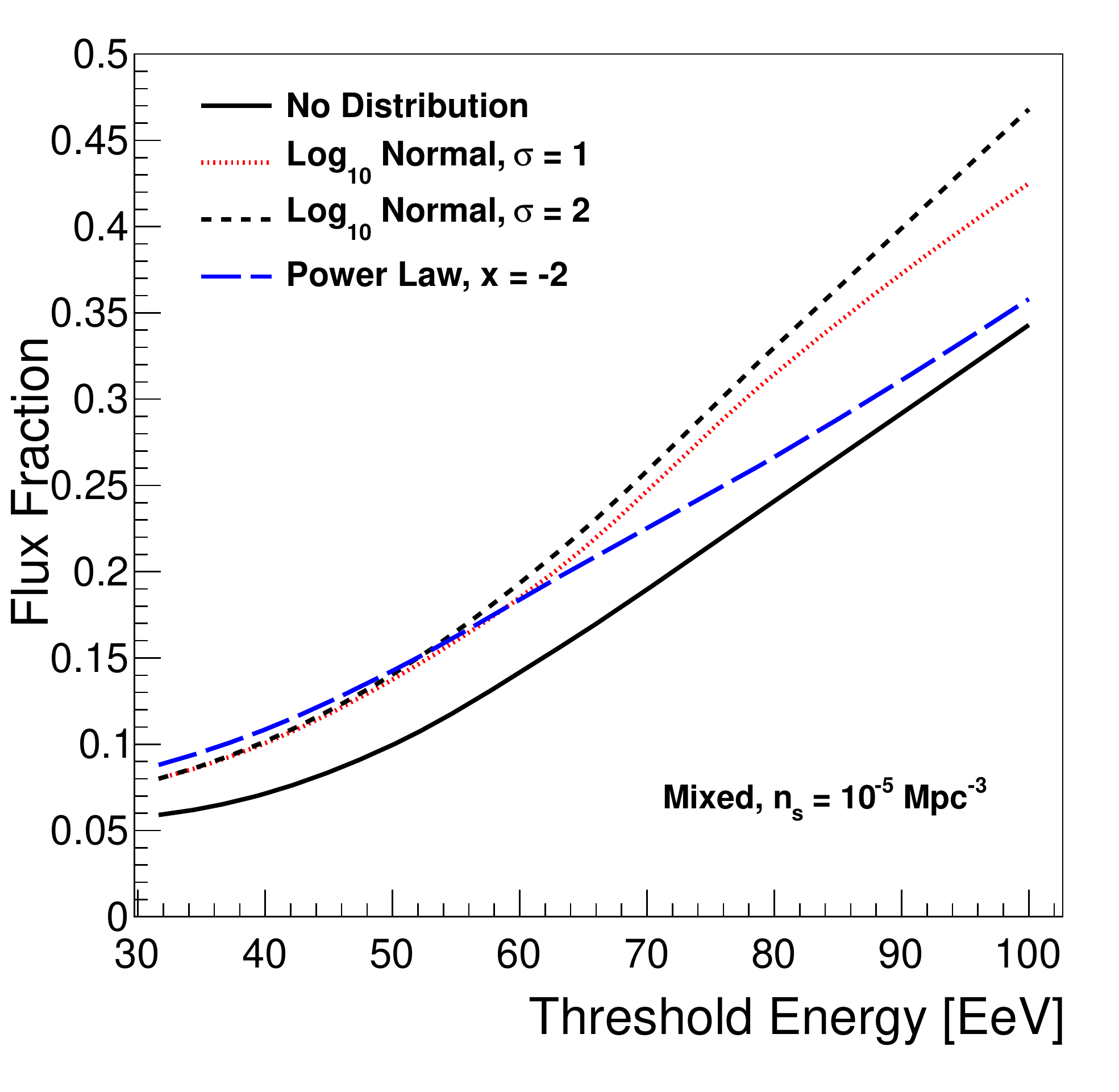}
\caption[Median with Source Luminosity Distribution]{Median flux fraction for the brightest source in the sky, shown for a mixed-composition model with a source density of $n_{\text{s}} = 10^{-5}\,\text{Mpc}^{-3}$. 
The median is shown for a uniform source luminosity and for source luminosities distributed according to either a log$_{10}$-normal distribution or a power law, as described in the text.}
\label{fig:MixedModelsLumDistributions}
\end{figure}

The influence of different luminosity distribution scenarios, as mentioned in section~\ref{sec:method}, is shown in \fig\ref{fig:MixedModelsLumDistributions}. 
In this figure the evolution with energy of the median fractional contribution of the brightest source for a mixed composition model with $n_{\text{s}} = 10^{-5}\,\text{Mpc}^{-3}$ is plotted. 
The contribution increases in every case, passing from 34\% at 100 EeV in the case of standard candles to up to 47\% for the log$_{10}$-normal, $\sigma = 1$, scenario. 
This is because upward fluctuations, where one of the most nearby sources happens to be brighter than average, extend the distribution towards higher fractional contributions. 
Downward fluctuations are more limited, however, and simply switch the ordering of the source brightnesses.

\subsection{The Influence of Source Evolution}

The previous plots show the results obtained assuming that there is no evolution of the intrinsic source power and/or density as a function of redshift, or in other words, time. 
Since the brightest sources are always nearby, within a few tens of Mpc (even at low energy, where the GZK horizon is further away), 
the intrinsic power of these sources should not be expected to be different in any significant way on the corresponding timescale, whatever the evolution scenario. 
However, in all the source evolution scenarios considered, the source power and/or density is higher at earlier times, so that more distant sources contribute more to the flux than in the case with no evolution. 
This enhances the cosmic ray flux at low energy with respect to that at high energy, which results in a change of the propagated spectrum obtained from a given source spectrum. 
Source evolution scenarios thus require harder source spectra in order to fit the observed UHECR data.

\begin{figure}[ht!]
\centering
\includegraphics[width=0.90\textwidth]{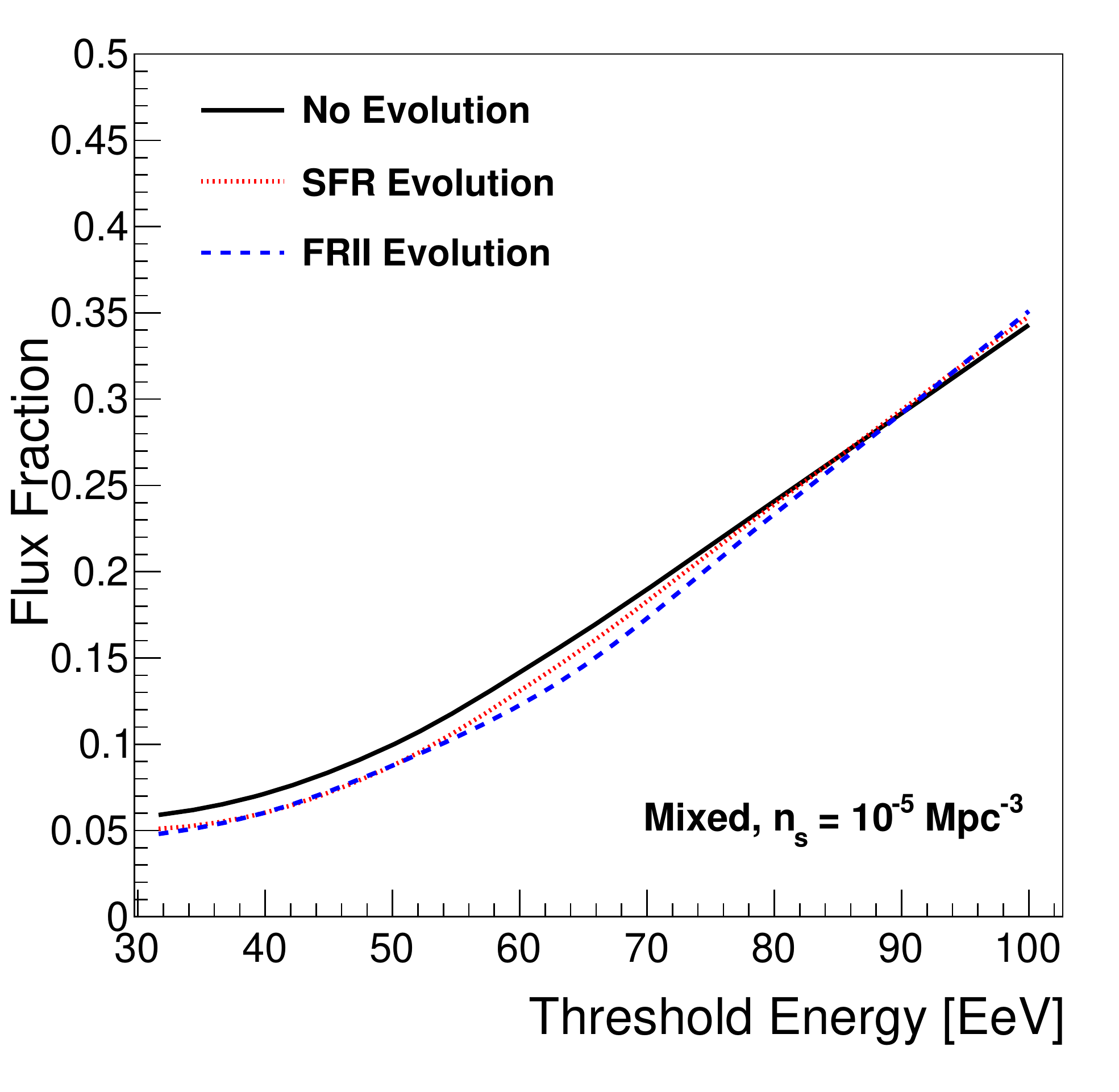}
\caption[Median Flux with Source Evolution]{Median flux fraction of the brightest source, shown for the mixed-composition model with a source density of $n_{\text{s}}= 10^{-5} \text{Mpc}^{-3}$, 
for three source evolution models: no evolution with time, SFR-like evolution, and FR-II-like evolution.}
\label{fig:MixedmodelSourceEvolution}
\end{figure}

Because this might in principle change the above results, the fractional contribution of the brightest sources was also computed assuming that the source power or number follows the evolution of the star formation rate \cite[SFR,][]{HopBea06} 
or that of the FR-II radio galaxies~\cite{Wall+05}. As can be seen in \fig\ref{fig:MixedmodelSourceEvolution}, the effect is negligible. 
This is because the relative contribution of the different sources is dictated by the GZK horizon structure, which is not modified by the source evolution. 
The results given in the last section, computed for a no-evolution scenario, are thus robust in this respect.

\subsection{Number of Sources Contributing to the UHECR Flux}

\begin{figure*}[ht!]
\centering
\subfigure[Mixed Model]{\label{fig:NumFifty:Mixed3den}\includegraphics[width=0.6\textwidth]{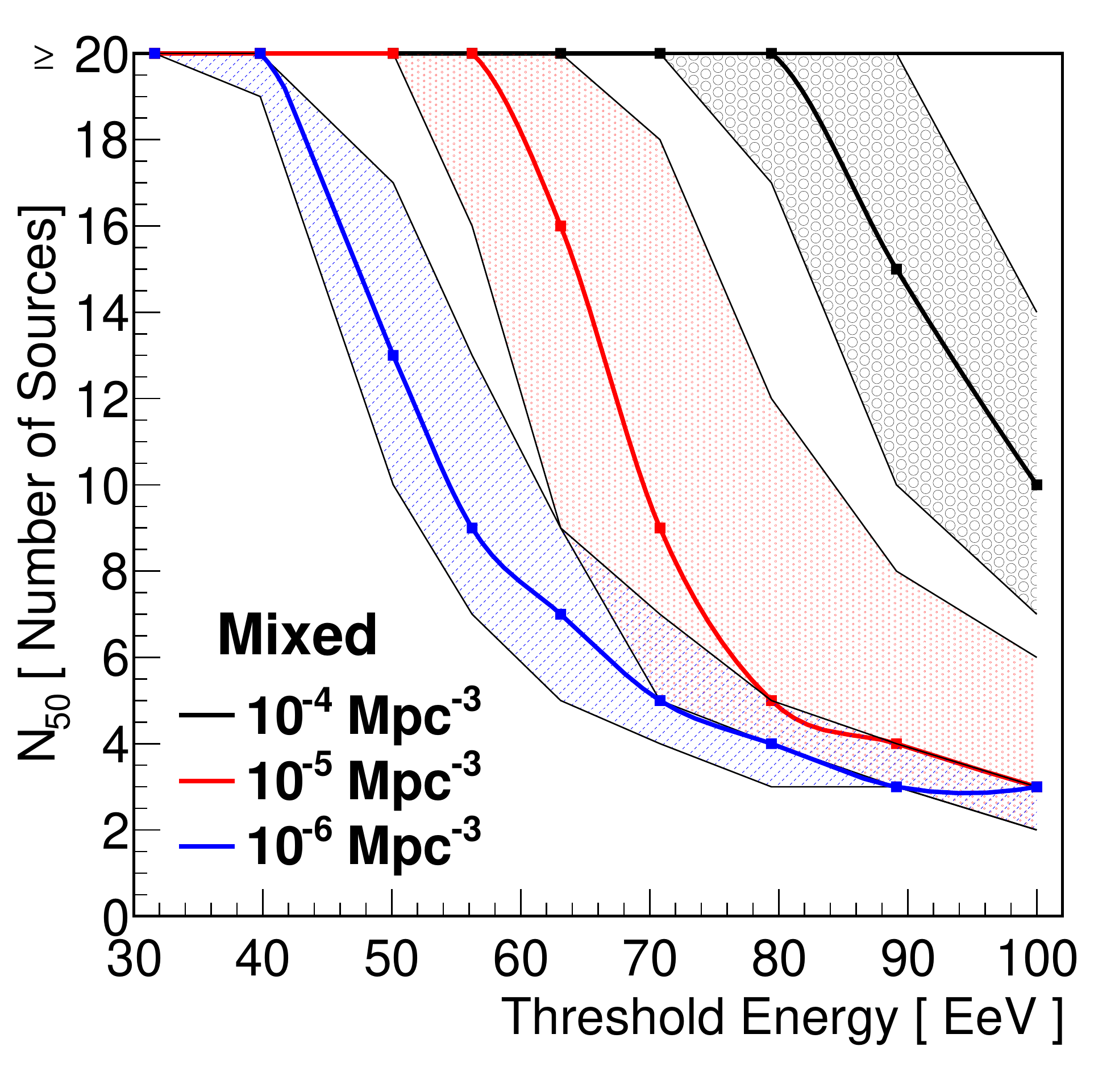}}\\
\subfigure[Various Models]{\label{fig:NumFifty:allModels}\includegraphics[width=0.6\textwidth]{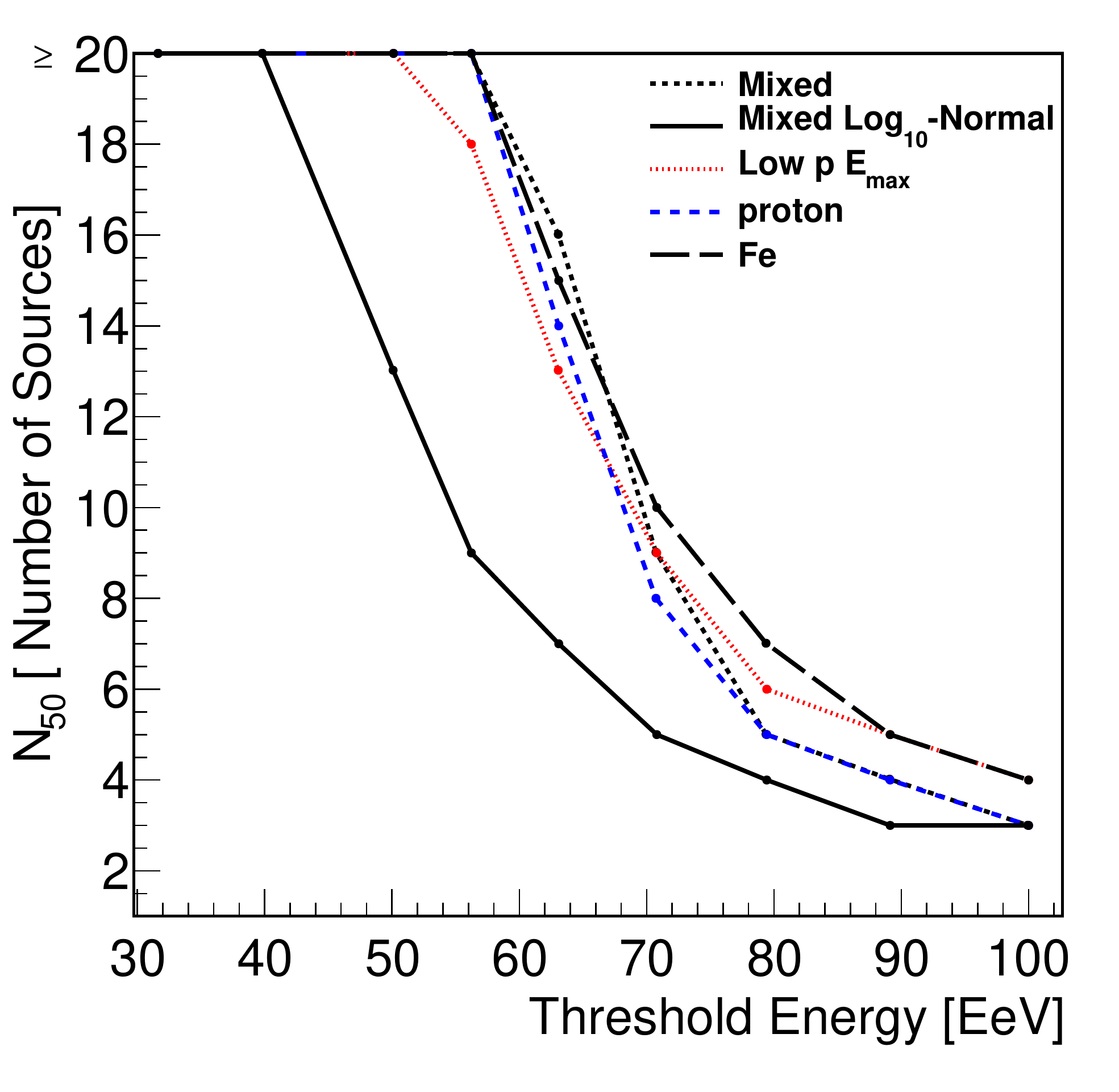}}
\caption[Number of UHECR Sources]{The number of sources, starting from the highest percentage contributor, which provide at least $50\%$ of the total flux, $N_{50\%}$, as a function $E_{\min}$, 
the minimum event energy. In figure \ref{fig:NumFifty:Mixed3den},  $N_{50\%}$ is shown for a mixed-composition model with source densities of $n_{\text{s}}=10^{-4}, 10^{-5}, \text{and}\,10^{-6}\,\text{Mpc}^{-3}$. 
Figure \ref{fig:NumFifty:allModels} shows $N_{50\%}$ for a source density of $n_{\text{s}}=10^{-5}\,\text{Mpc}^{-3}$ for each model, and, in addition, 
a mixed-composition model with source luminosities distributed according to a log$_{10}$-normal distribution with $\sigma=1$. 
The shaded contours show the region in which 68\% of all realizations lie, and is omitted in the second figure for clarity.}
\label{fig:NumFifty}
\end{figure*}

Figs.~\ref{fig:3models3sources} and~\ref{fig:Mixedmodels3sources}, show the contribution of the three brightest sources for various models. 
It is also interesting to determine how many sources are expected to make up more than, say, 50\% of the flux in each scenario. 
This number, denoted by $N_{50\%}$, is shown in \fig\ref{fig:NumFifty} for the mixed composition model and three values of the source density, 
as well as for each of the models of Table~\ref{tab:modelParameters} with a density $n_{\text{s}} = 10^{-5}\,\text{Mpc}^{-3}$.

The strongest influencing parameter is again the source density. For $n_{\text{s}} = 10^{-5}\,\text{Mpc}^{-3}$, $N_{50\%}$ is between three and four at $E = 100$~EeV. 
At 80~EeV, $N_{50\%}$ moderately increases to between four and seven. 
However, below 80~EeV, it rapidly increases as a result of the quickly receding horizon to reach more than 20 sources needed to make up more than 50\% of the flux. 
This is a direct demonstration of the GZK effect. The dramatic decrease in the overall UHECR spectrum above 60~EeV is indeed due to a dramatic reduction of the number of contributing sources in that energy range.

The source density influences $N_{50\%}$ as expected, in that 
fewer sources contribute a large fraction of the flux at lower densities. Only a handful of sources make up 50\% of the flux down to 60~EeV for $n_{\text{s}} = 10^{-6}\,\text{Mpc}^{-3}$, 
instead of 80~EeV for $n_{\text{s}} = 10^{-5}\,\text{Mpc}^{-3}$. For source densities as high as $n_{\text{s}} = 10^{-4}\,\text{Mpc}^{-3}$, $N_{50\%}$ is greater than ten even at 100~EeV.

Finally, for a given model, 
a distribution of source luminosities results in a lower value of $N_{50\%}$ compared to the same scenario assuming standard candle sources, 
as is shown in \fig\ref{fig:NumFifty:allModels}. 
This is in line with previous finding that a scenario with a distribution of luminosities 
is effectively similar to a standard-candle version of the same model with a lower source density, as expected.

\section{Discussion}

It has been shown here that above $E > 3\, 10^{19}~\text{eV}$ the number of sources which contribute to the UHECR flux can be expected to strongly decrease, down to only a few sources at the highest energies. 
This decrease is due to the energy-loss length of protons and heavy nuclei during propagation from their sources to the Earth, i.e.~the GZK effect. 
To quantify this effect, results for the fractional contribution of the brightest sources in the UHECR sky as a function of minimum threshold energy have been shown. 
Because the exact contribution is dependent on the spatial configuration of the closest sources, the median value over a set of realizations and for three choices of source density is studied. 
Several UHECR source scenarios have been considered with respect to composition and energy spectrum. The choice of source parameters was motivated by previous studies of UHECR propagation, so as to fit the data.

For a mixed-composition model with a source density of $n_{\text{s}} = 10^{-5}~\text{Mpc}^{-3}$, 
it was found that above $E = 100$~EeV the brightest UHECR source in the sky can be expected to contribute $34^{+15}_{-17}\%$ 
of the total flux, and the brightest three sources contribute more than 50\% of the total flux (actually 58\% in this case). 
In the previous number, the value of 34\% is the median of the values obtained for different realizations, and the range from 17\% to 49\% contains 68\% of all realizations. 
For lower source densities, the UHECR sky at the highest energies is dominated by even fewer sources. Scenarios with $n_{\text{s}} = 10^{-6}~\text{Mpc}^{-3}$ 
would typically result in three or four sources making up more than 50\% of the flux down to 80~EeV, and can leave only one or two sources contributing half of the total flux above 100~EeV.

The results presented in this chapter conservatively assume that the UHECR sources are standard candles. 
Whatever they may be, however, it is likely that they distribute over a range of intrinsic luminosities. 
This can only increase, on average, the weight of the most luminous sources in the overall UHECR sky and strengthen the effect investigated here. 
This effect was quantified for choice of several (hopefully representative) source luminosity distributions, 
finding that the contribution of the brightest source increases to $43^{+26}_{-16}\%$ for a log-normal distribution with $\sigma = 1$ and a source density $n_{\text{s}} = 10^{-5}~\text{Mpc}^{-3}$, 
at 100~EeV (the quoted range has the same interpretation as above). In such a scenario two sources largely dominate the observed flux at this energy.

Such low numbers for the count of sources contributing at the highest energies justify the hope that individual sources can be isolated in the sky by future UHECR detectors. 
At an energy where less than a handful of sources are responsible for the observed flux, with negligible ``background'' from other, 
much less intense sources, distinct regions of the sky may be populated essentially by UHECR events coming from a single source. 
This will be the case even with relatively large deflections due to intervening magnetic fields, if the deflections remain smaller than the angular distance between sources.
This angular distance will increase considerably on average when the source number is reduced to $\mathcal{O}$(1).

The results presented here illustrate the importance of concentrating on the highest energies, right \emph{inside} the GZK cutoff, 
in order to take full advantage of the GZK effect and the associated reduction of the number of visible sources. 
In particular, \fig\ref{fig:NumFifty} makes clear that a dramatic change in this number occurs between 50~EeV and 80~EeV. 
This suggests that a significant increase in the clustering signal can be expected if a new generation of detectors can be used to push the current statistics achieved at 50~or 60~EeV up to 80~or 100~EeV. 
Considering the very low number of contributing sources, one may loosely say that the sources ``isolate themselves'' at $10^{20}$~eV, as the GZK horizon removes sources that would otherwise overlap due to magnetic deflections.
This represents an observational challenge, since it implies increasing the statistics and detection power in a range of energy where the flux is severely reduced.

In this chapter the focus was on the source number and its evolution with energy. Given the low number of contributing sources suggested by this study, 
it is clear that any failure to detect a significant clustering signal at the highest energies will put strong constraints on the angular deflections of UHECRs and thus on the magnetic 
field and/or the UHECR composition. 
However, the present study clearly shows that the absence of a strong anisotropy signal and the inability of current detectors to isolate UHECR sources in the sky at 50 or 60~EeV does \emph{not} imply that this will not be possible at 100~EeV.

This can be seen as support for a new generation of detectors.
Increasing the acceptance of the detectors sufficiently to collect a statistically significant number of UHECR events deeply into the GZK cutoff is likely to require new observational set-ups.
This can be achieved by using space-based detectors, such as JEM-EUSO, or by exploring new detection techniques. This work has in fact been included as part of the JEM-EUSO science case.
 
Finally, since an order-of-magnitude gain in acceptance may be to the detriment of the precision of the measurements, 
it is important to investigate the effect of an imperfect energy resolution. With poor energy resolution, a cut in the UHECR energy, as assumed above, cannot be strictly applied. 
This is because some lower energy events will be (mis-)reconstructed at higher energy. These spillover events will contaminate the energy range where very few sources contribute to the overall 
flux with UHECRs from additional sources within the more distant GZK horizon which exists at lower energy. Since the UHECR spectrum decreases rapidly as energy increases, 
a small fraction of events reconstructed with an upward fluctuation of the estimated energy can represent a significant fraction of the events attributed to a higher energy bin.

\begin{figure}[t!]
\centering
\includegraphics[width=0.90\textwidth]{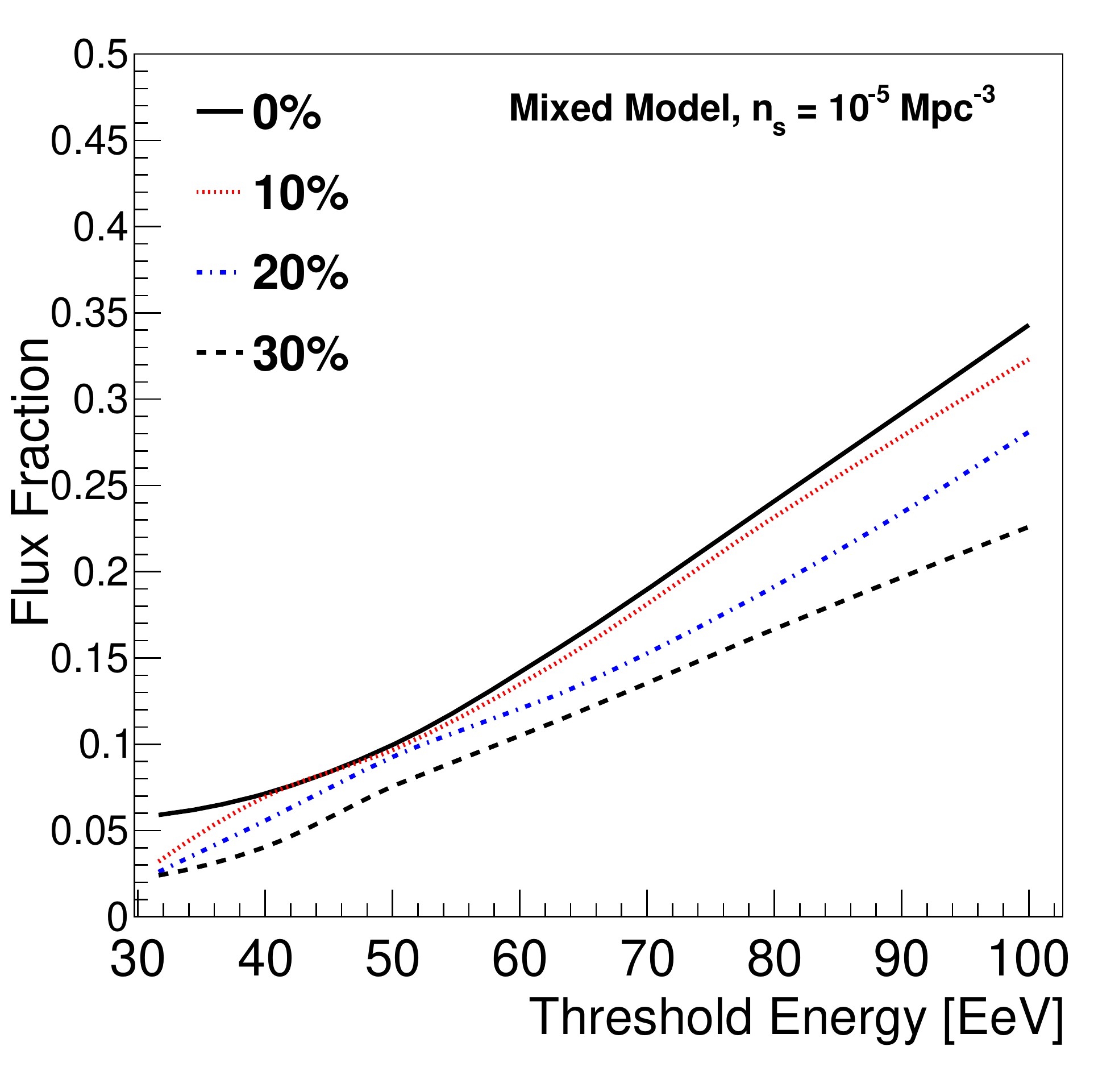}
\caption[Energy Resolution Effect on the Median Flux]{ \label{fig:MixedmodelEnergyRes} Median flux fraction of the brightest source in the sky as a function of minimum \emph{reconstructed} energy, 
shown for a mixed-composition model with a source density of $n_{\text{s}} = 10^{-5}\,\text{Mpc}^{-3}$. 
The top curve is the reference corresponding to a perfect energy reconstruction, and the other three curves correspond to detectors with an assumed Gaussian energy resolution of 10\%, 20\%, and 30\%.}
\end{figure}

To illustrate this effect, a Gaussian detector response was implemented when binning each UHECR event with respect to $E_{\text{min}}$. The analysis was then done using the effectively reconstructed energies, instead of the true UHECR energy. 
The results are shown in \fig\ref{fig:MixedmodelEnergyRes} for a detector with 10\%, 20\%, and 30\% energy resolution, overlaid with the results for a perfect detector. 
In the mixed-composition model with source density $n_{\text{s}} = 10^{-5}\,\text{Mpc}^{-3}$, this fraction goes from 24\% for perfect resolution at 80~EeV to respectively 23\%, 19\%, and 17\%, for a 10\%, 20\%, and 30\% energy resolution.
At 100~EeV, the reduction is from 34\% to 32\%, 28\%, and 22\%, respectively.

For current ground based observatories, this effect could play an important role in hiding the anisotropy which we would otherwise naturally expect in the UHE region.
The ability of a given detector to actually isolate the brightest sources in the sky will thus depend on its energy resolution. 
The energy dependence of the detector acceptance will also play a role. For instance, for detectors with a larger acceptance at higher energy, 
the above effect will be reduced to some extent by the fact that lower energy events have a lower probability of being detected at all. 
This should thus be modeled for each experiment, given their individual performances.

In sum, regardless of magnetic deflections and experimental limitations, the existence of the GZK effect implies that only a handful of sources will contribute to the UHECR flux above 80~EeV or so. 
The work presented here is not a ``discovery'' of this fact, but rather a detailed analysis and quantification of ``handfulness'' of the number of sources. 
The argument to be taken from these results is that, even though the anisotropy patterns observed by current experiments at 60~EeV are not as enlightening as had long been hoped, 
the quest to understand UHECRs through anisotropy studies should not stop now, and that expanding our observational capabilities at $10^{20}$~eV should give us key information. 
The isolation of the first UHECR source in the sky would be an important achievement, allowing us to constrain currently unknown astrophysical parameters, such as source density, 
maximum energy at the source, individual source power, and the UHECR energy budget, with corresponding constraints on the nature of the sources and acceleration mechanisms.

      \printbibliography[heading=subbibliography]
       \end{refsection}

   \begin{refsection}
  \chapter{Perspective and Conclusion}

In chapter~\ref{CHAPTER:UHECR Source Maximum Energy and the UHECR Spectrum} a generic class of models for UHECR phenomenology was studied, in which
the sources accelerate protons and nuclei with a power-law spectrum having the same index, but with
different values for the maximum proton energies. The maximum proton energy was assumed to be distributed according to a power-law. 
For energies above the maximum energy of some sources, but sufficiently lower than the maximum proton energy it was found that such models are equivalent to single-type source models, with a larger effective power law index and a heavier composition at the source.
The resulting enhancement of the abundance of nuclei was calculated, and typical values of a factor 2–10 for Fe nuclei were found. At the highest energies, the heavy nuclei enhancement ratios become larger, and
the granularity of the sources must also be taken into account. 

These results show that the effect of a distribution of maximum energies among sources must be taken into account in order to understand both the energy
spectrum and the composition of UHECRs, as measured on Earth. While including a distribution of maximum energies in UHECR models introduces new free parameters, 
these parameters can be related to the properties of UHECR source candidates. At this time such a study remains for future work.
As an example, if it is assumed that Active Galactic Nuclei (AGN) are the main source of UHECRs, the maximum energy distribution can be 
related to the AGN X-ray Luminosity Function (XLF). The XLF can then be inserted into a cosmic ray propagation simulation in order to study
both the resulting composition and the spectral index at the source needed to fit the observed spectrum. This could result in a deeper understanding of the UHECR spectrum and composition by relating
them to the properties of well-known astrophysical objects. 

Parallel to this, the observational search for the source of UHECRs is ongoing.
So far, a lack of a strong anisotropy or clear correlation with known astrophysical objects has raised questions about the ability to identify sources and study their properties through UHECRs.
This may be especially true in the case of a heavy composition at the highest energies, as suggested by composition results from the Pierre Auger Observatory, or strong magnetic deflections.
Even if the composition is heavy or magnetic deflections are high, however, UHECR sources could still be isolated in the sky if
\begin{inparaenum}[i\upshape)]
    \item the flux is dominated by a limited number of sources,
    \item the apparent angular size of the source objects in the sky is generally smaller than the separation between sources, and
    \item the statistics collected by the detector(s) are high enough.
     \end{inparaenum}
The first two items are properties of the physical phenomena, which are what they are. The last point, on the other hand, is dependent on experimental efforts, and represents a means by which we can learn something about the UHECR sky,
even if this information takes the form of being unable to distinguish the sources. 
From an experimental point of view, then, a large increase of exposure in next generation UHECR observatories is of central importance for many astrophysical questions related to UHECR.    

It is still legitimate to question, however, what we expect to be able to observe in this next generation of experiments. 
Due to the GZK effect, the horizon for a given source to contribute to the UHECR flux decreases with increasing energy, and
the GZK effect can thus become a useful phenomenon, which effectively reduces the number of UHECR sources in the sky with increasing energy.
This idea was quantified in chapter~\ref{CHAPTER:UHECR Source Statistics} by studying the contribution of individual sources to the UHECR flux, and it was found that the more than half of the UHECR flux can be attributed to only a few sources 
at energies above $\sim8~10^{19}~$eV. 

The next step is to study the apparent angular size of sources. This work is ongoing, and is being realized by extending the study presented in chapter~\ref{CHAPTER:UHECR Source Statistics} to include
the propagation of the UHECRs through Galactic and extra-Galactic magnetic fields and the drawing of realistic sky maps. 
UHECR source models, and the actual properties of current and future experiments, such as total exposure and angular resolution, can also be included in such a study in order to make realistic predictions about 
the observational power of UHECR Observatories within a given scenario. The preliminary results from this ongoing study have been presented at the 33rd International Cosmic Ray Conference \cite{ParizotBlaksleyICRC} and now form part of the JEM-EUSO science case.
Those results, and the work presented in this thesis, give strong reason to believe that the near future will continue to be an exciting period in the UHECR field
and that we will continue to understand more about the highest energy particles in the Universe.


     \printbibliography[heading=subbibliography]
       \end{refsection}

\part{Appendices}
\appendix
  \renewcommand\thesection{\Alph{section}}
  \begin{refsection}
\chapter{DAQ-User Guide}
 \label{Chapter:DAQuserguide}
 
Midas is a data-acquisition framework written in C++ which provides services and communication between sub-applications which control various aspects 
of the data acquisition \cite{MIDAS}. Midas provides the capability to collect data from multiple sources, linked to different computers, while analyzing and storing the collected data
online or offline on separate machines. The frontend codes, which control data readout, and the analysis codes, which sort incoming data, must be provided by the user. 

This short guide describes how to use the Data Acquisition System (DAQ), created for photomultiplier tube sorting and calibration at APC, and which is based on the Midas framework.
The hardware of this system was described in detail in chapter~\ref{CHAPTER:PMT Sorting}.
The focus in this chapter is on describing the routines written for the DAQ and how to use the system. A full, more general, guide for the Midas system can be found online \cite{MIDASwebsite}. 

The DAQ described here can, in principle, read out any combination of CAMAC and/or VME ADCs.
In this case, the frontend code runs on a MVME 30002 VME processor board, which is connected directly into the VME back-plate. CAMAC control functions are provided by the DCAMLIB library written by D. Kryn \cite{DCAMLIB}.
The processor board thus acts as an independent frontend computer which is connected to a backend (the working desktop computer) which runs the control and data analysis functions. 
Midas itself can run on Unix systems (Linux), Mac, and Windows, but this guide assumes that the system is running on Linux.

Section~\ref{sec:Introduction to MIDAS} presents an overview of the DAQ and the sub-processes on which it depends. Starting the DAQ system is also described at the same time.
A guide to using the system, including taking runs, and viewing data is given in section~\ref{sec:Using the DAQ}. 
After that, the frontend programs are presented in section~\ref{sec:The DAQ Frontends}, and the analysis routines are covered in section~\ref{sec:The DAQ Analyzer}.
A few useful commands for using the online database are briefly mentioned in section~\ref{sec:Working with the Online Database}.

\section{Introduction to the DAQ}
\label{sec:Introduction to MIDAS}


A Midas ``experiment'' is a collection of variables, acquisition processes, and analysis routines designed for a certain hardware configuration. These are written by the user, and 
handle the actual hardware access, read-out, and event analysis, while the process control and services are provided by the Midas framework. 

The available experiments and their respective directories are defined in the \emph{exptab} file. The exptab file is found by default in the [HOME]/online directory, where [HOME] refers to 
the users home or working directory. The [HOME] directory is defined by the individual Midas installation.
In the exptab file, each experiment is defined by an entry of the type:
\begin{verbatim}
 [Experiment Name]  [Path/to/Experiment/Directory] [UserName]
\end{verbatim}
and the contents of each experiment are defined by the online database files which are saved in the experiment directory. There is no need for the actual frontend or backend programs to 
live in the experiment directory. 

The actual directories and names of the C1205 data acquisition system (here called the DAQ) at APC will be used throughout this guide.
An existing experiment can easily be adapted by modifying and re-compiling the frontend and analysis codes. The DAQ experiment is defined as:
\begin{verbatim}
 C1205_Frontend  /home/blaksley/online/C1205_Frontend blaksley
\end{verbatim}
Any time a MIDAS process is started it will prompt the user for the experiment number which is assigned based on the order of experiments in the exptab file.
Currently this experiment is assigned number 10 as it is the tenth Midas experiment defined in the exptab file.
 
\subsection{The Basic Processes: Starting the DAQ}
\label{sec:Processes}
Starting the data acquisition is probably the most complicated part of its use. Once the system is running, however, it is very stable and can be left running indefinitely.
To make it easier to turn the system on, a start-up script can be written which will automatically start all
the processes which run on the backend computer. 

If the processes are started individually, then the experiment to which they should connect must be chosen from the experiment list or specified as the experiment name with the \emph{-e} option when running the command. 
Each process must be run in a independent terminal window.
The basic Midas processes are: 
\begin{itemize}
\item \textbf{backend}: These processes control the overall DAQ and provide other services. Each process is described below.
\begin{itemize}
\item \textbf{mserver}: This program provides the network services and connection between the frontend and the backend computer (if not the same). 
The network communications protocols used by Midas automatically account for the endianness of each system.
The mserver is run in a terminal with the command:
   \begin{verbatim}
    cd /home/blaksley/packages/midas/linux/bin/
    ./mserver 
   \end{verbatim}
\item \textbf{mhttpd}: This program provides access to the web control interface. The mhttpd program takes a port number with the \emph{-p} option, and is run in a terminal as:
   \begin{verbatim}
    cd /home/blaksley/packages/midas/linux/bin/     
    ./mhttpd -p 9081 -e C1205_Frontend
   \end{verbatim}
\item \textbf{Control Interface}: The web control interface gives full control over the data acquisition system, and can be accessed by opening any web browser on the computer
running mhttp and going to the address:
   \begin{verbatim}
   http://localhost:9081
   \end{verbatim}
The port number on the address (here 9081) and of the mhttpd process must match. See section \ref{sec:Using the DAQ} for how to use the basic functions of the control interface. 
The control can also be accessed remotely by pointing any web browser to the appropriate host name and port number.                                                                   
\item \textbf{mlogger}: This program provides storage of analyzed data. It is not needed if data storage is not required, and is run in a terminal with the command
   \begin{verbatim}
    cd /home/blaksley/packages/midas/linux/bin/     
    ./mlogger -e C1205_Frontend                                                                         
   \end{verbatim}
\item \textbf{Analyzer}: This program sorts, analyzes, and histograms the data sent from the frontend. It is written by the user according to their hardware and sorting requirements. 
 Without an analyzer, incoming data can only be stored by mlogger, not viewed online. The main analyzer in the DAQ is run in a terminal by the command:
   \begin{verbatim}
     cd /home/blaksley/online/C1205_frontend/c1205analyzer    
    ./analyzerb -e C1205_Frontend                                                                              
   \end{verbatim}
Any number of separate analysis routines can be connected to the same experiment at once. The actual DAQ analyzer will be discussed in more detail in section~\ref{sec:The DAQ Analyzer}.
\end{itemize}
 \item \textbf{The frontend(s)}: The main frontend program controls the CAMAC and VME modules and collects the resulting data. 
It is run on the MVME processor board, which is connected to the working computer through a standard Ethernet connection.
The remote connection to the MVME board is through \emph{ssh}. To connect type:
\begin{verbatim}
 ssh blaksley@apcdc4
\end{verbatim} 
once connected to the processor board, the frontend code can be run by navigating to the frontend directory.
Add \emph{-h}, followed by the host name of the backend, to the run command in order to direct the frontend to look for the experiment on the backend computer rather than the local machine.
The run command for the frontend is:
\begin{verbatim}
 cd /home/blaksley/online/C1205_Frontend
 ./C1205_Frontend -h apcdecpc3 -e C1205_Frontend
\end{verbatim} 
A screen as in \fig\ref{fig:DAQ_frontend} will be shown once the frontend is running. This screen shows the current status of the data sources controlled by the frontend program.
For the DAQ, there are two more frontend programs which are run on the backend computer. These programs read out the NIST photodiode and control the X-Y movement respectively.
The NIST photodiode read out is connected by serial cable (RS232) and the frontend is run as:
\begin{verbatim}
 cd /home/blaksley/online/NIST_slowcontrol
 ./scfe -e C1205_Frontend
\end{verbatim}
The X-Y movement is connected to the backend computer by USB and the frontend control is started as :
\begin{verbatim}
 cd /home/blaksley/online/XYmovement_Midas
 ./XYmovement_scfe -e C1205_Frontend
\end{verbatim} 

\begin{figure}
\begin{center}
 \includegraphics[width=0.95\textwidth]{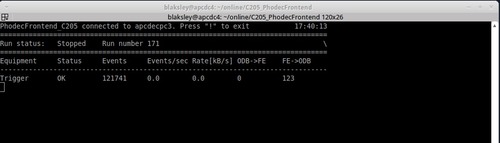}
\caption[DAQ Frontend Screen]{The frontend process screen. The process is run in a terminal through an \emph{ssh} session connecting to the frontend computer. 
The screen displays the actual status of the frontend, along with number of events collected in the current run and 
the event rate.}
\label{fig:DAQ_frontend}
\end{center}
\end{figure}

\item \textbf{Roody}: The Roody program interfaces the analyzer code with ROOT, allowing online viewing of histogramed data. Roody is run in a terminal with the command:
   \begin{verbatim}
     cd /home/blaksley/packages/fixed_roody/bin/     
    ./roody -Hlocalhost                                                                                 
   \end{verbatim}
The \emph{-Hlocalhost} option tells Roody to look on the local machine for analyzer output. Roody can also be connected to another computer if the analysis code is being run there, and Roody can also open
saved root files. Details on how to use Roody are given in section \ref{sec:UseDAQ}. The \emph{fixed\_roody} directory contains a version of Roody which has been modified to run 
on 64-bit systems.
\end{itemize}

Each of these commands have a help function listing their options which can be accessed with the terminal command:
\begin{verbatim}
 ./<processname> -help
\end{verbatim}

\section{Using the DAQ}
\label{sec:Using the DAQ}

\begin{figure}
\begin{center}
 \includegraphics[width=1.0\textwidth]{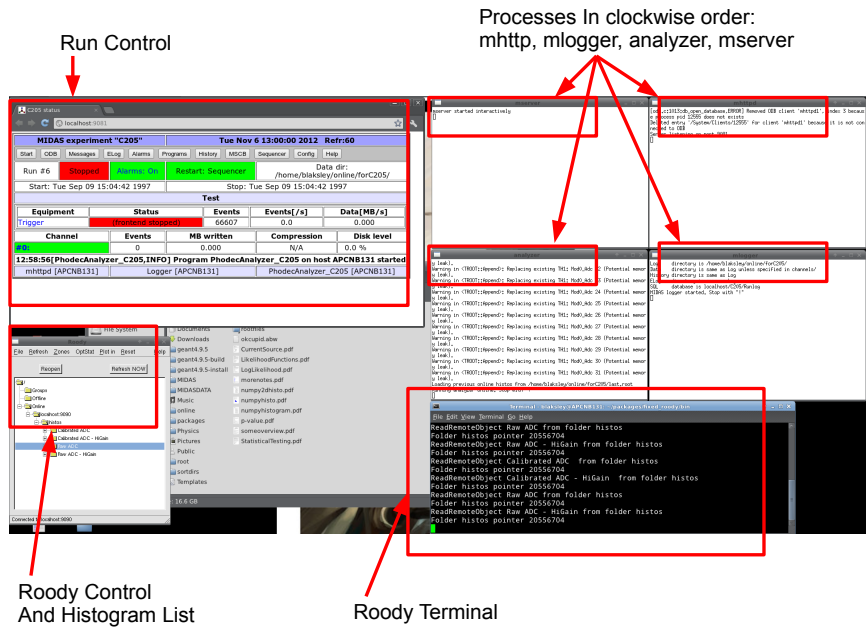}
\caption[DAQ Backend Processes]{An overview of the processes running on the backend computer.  The terminal containing the frontend display is not shown. The ``Run Control'' window is where options and controls for running the DAQ can be found. 
The Analyzer, mhttpd, mserver, and mlogger processes each run in their own terminals which will display messages relating to their current status. Roody is an interface between 
the Analyzer and ROOT, allowing online viewing of the histogramed data. 
Each process is described in section \ref{sec:Processes}.}
\label{fig:BackendOverview_labeled}
\end{center}
\end{figure}

\label{sec:UseDAQ}
Once the backend of the DAQ is running several windows will be open as shown in \fig\ref{fig:BackendOverview_labeled}.
The mhttpd, mlogger, mserver, and analyzer are all opened in independent terminals, and so
each can be shutdown and restarted independently. This is a key point of the DAQ, as if a process dies it will not freeze the system. All the other processes will continue, and the 
control interface will indicate that one of the processes has died. The only requirement to continue the acquisition correctly is to restart the process which was killed.
For example, the mlogger window can be closed if there is no need to save data to the hard-drive. The system will run as normal and if at any time data needs to be saved it is only necessary to restart the logger and 
connect it to the correct experiment.

\begin{figure}
\begin{center}
 \includegraphics[width=1.0\textwidth]{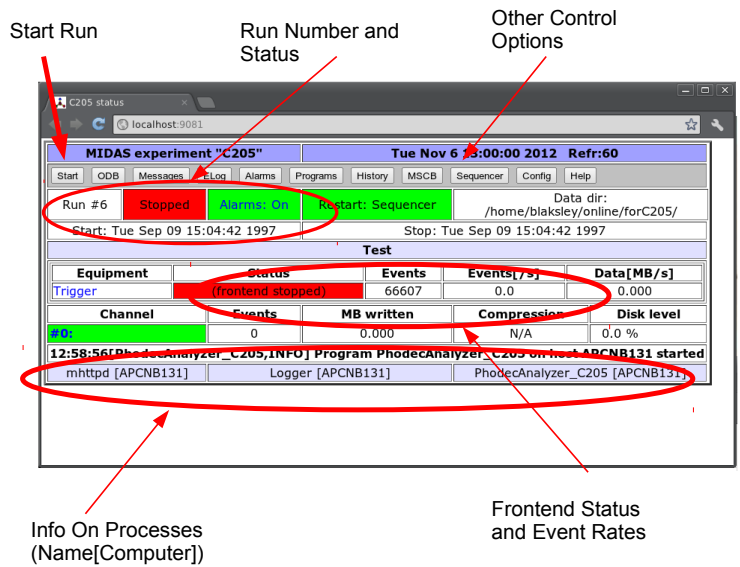}
\caption[DAQ Control Interface]{An overview of the control interface, opened here in the Chromium web browser. All of the immediate controls are located on the main screen. Various menus containing other options are along the top of the page. 
Details concerning their function can be found in sections~\ref{sec:The DAQ Analyzer} and \ref{sec:The DAQ Frontends}, or in the Midas documentation. The current run status is shown in the center of the page along with the directory to which data is being saved. The bottom of 
the page lists all Midas services and the computer on which they are running, in addition to any status log entries, alarms, or warning messages.}
\label{fig:Backend_Control_overviewLabeled}
\end{center}
\end{figure}

The control interface for data acquisition can be accessed through any web browser. The interface is shown in \fig\ref{fig:Backend_Control_overviewLabeled}.
Every soft option (not written into the frontend or analyzer code) is accessible from the menus along the top of the web control. Along the bottom of the interface, the name and host computer of every connected process is shown. 
This list will typically contain one mhttp, one mlogger, and one analyzer entry. 
If the mlogger or analyzer is not connected and running, then events from the frontend will not be sorted or stored, respectively. It is possible to run the DAQ with the logger and not the analyzer, and
in this case the raw events will be stored to the hard drive, but not sorted or histogramed.

The status of the frontend equipment is shown in the center of the control screen. If the frontend is running and connected properly, then the status entry will be green. ``Stopped'' or ``Running'' indicates whether the frontend is currently taking data or
not. In the top left is the run control and status; here the run number and current condition are displayed. 

\emph{It is important to note that the web control interface acts exactly like a web page. When  navigating the menus, the back button may be used at any time and all the current options will be saved. 
The interface refreshes automatically every few seconds.
If runs are quickly started and stopped, however, it is advised that the refresh button on the browser be used to ensure that the current status of the system is displayed.}  

\subsection{Starting and Stopping}

\begin{figure}
\begin{center}
 \includegraphics[width=1.0\textwidth]{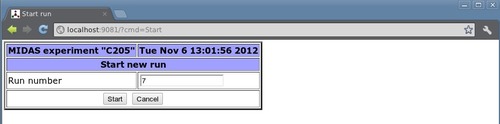}
\caption[The DAQ Start Screen]{The start run screen}
\label{fig:Backend_RunStart_small}
\end{center}
\end{figure}

\begin{figure}
\begin{center}
 \includegraphics[width=1.0\textwidth]{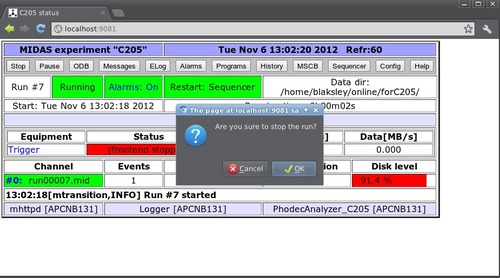}
\caption[DAQ Run State]{The control screen in the run state (in the background) and the stop run prompt.}
\label{fig:Backend_Stopping_small}
\end{center}
\end{figure}

The top-left most button labeled \emph{start} in \fig\ref{fig:Backend_Control_overviewLabeled} will start the run. 
On clicking this button a window will be shown as in \fig\ref{fig:Backend_RunStart_small}. If the run number entered already exists (as a stored data file in the current mlogger directory), then
the start of the run will abort and a notification will be given. The system will automatically increment the run number at each start. Once the system is running, the system status will be shown in the run status and frontend status fields. 
This can be seen in the background of \fig\ref{fig:Backend_Stopping_small}. Once the system is in the running state the \emph{start} button is replaced by a \emph{stop} and \emph{pause} button.
Both buttons will bring up a confirmation window as can be seen in \fig\ref{fig:Backend_Stopping_small}.

As previously mentioned, the control interface acts like a web page and is only refreshed every few seconds, or if a process sends certain types of updates. 
Therefore, if in doubt, the refresh button should be used to be sure that the displayed information is the most current.

\subsection{Viewing Spectra Online}
\label{sec:Roody}

\begin{figure}
\begin{center}
 \includegraphics[width=1.0\textwidth]{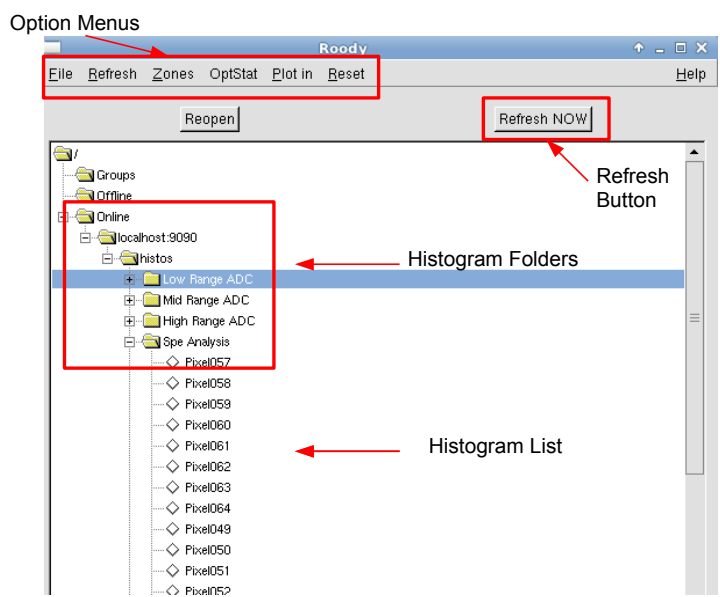}
\caption[Roody Interface]{The main Roody screen. Along the top are menus for plotting and refresh options. The program interface is point and click, and
 the available histograms are listed by folder. Any histogram in this list can be drawn according the current options by double clicking on the entry.}
\label{fig:Roody_1_small}
\end{center}
\end{figure}


Online spectra are displayed by Roody, which was introduced in section \ref{sec:Processes}. Roody is simply an interface between the analyzer code, which collects and sorts data from the frontend, and ROOT, which is used to 
display the histograms. Once Roody is connected to a given experiment, it will give access to all of the histograms defined in the analyzer. Because Roody is basically ROOT, Roody works exactly like ROOT. This means that every ROOT option, such as 
fitting of histograms and peak finding functions, is available. Roody is also the least stable part of Midas. It is also very easy to stop and restart, however, and if a problem occurs, Roody can be killed by clicking in the Roody terminal
and hitting \textbf{Ctrl+C}. Roody can then re-run it by hitting the \textbf{Up-Arrow} key and \textbf{Enter} in the terminal. The histograms of the current run will \emph{not} be lost as they are stored in shared memory until a new run is started. 

The main Roody window is shown in \fig\ref{fig:Roody_1_small}. Here a list of folders can be seen. The Offline folder contains any histograms which have been loaded from a file. 
This is a very useful aspect of Roody, in that it allows later viewing of histogramed data which has been saved to a ROOT file.
The Online folder contains histograms from the most recent run of every experiment
which is connected to Roody, and is subdivided by source computer (here localhost port 9090). 
If Roody is run as described in section \ref{sec:Processes}, then there will be only the online data source, which will be opened automatically. 

Within the data source folder are sub-folders corresponding to different sets of histograms, as defined in the analyzer program. For the C$\_$1205 DAQ there are 2 sub-folders. 
The first three folders, named ``Low Range ADC'', ``Mid Range ADC'', and ``High Range ADC'',  contain one histogram for each charge range for each QDC, labeled by module and channel number. 
Each spectrum is a  direct histogram of the converted QDC values.
These histograms are useful for debugging or in cases for which only a few QDC channels are being used.
The second folder, named ``Spe Analysis'' maps each QDC channel to a certain pixel according to a user defined map. The mapping is found in the analyzer parameters within the online database (ODB).
This was done to simplify working with the M64 PMT, so that each spectrum corresponds to an absolutely defined 
pixel within the PMT. The spectra in this folder will also display the results of single photoelectron spectrum analysis at the end of run, which will be discussed in section~\ref{sec:The DAQ Analyzer}.

\begin{figure}
\begin{center}
 \includegraphics[width=.8\textwidth]{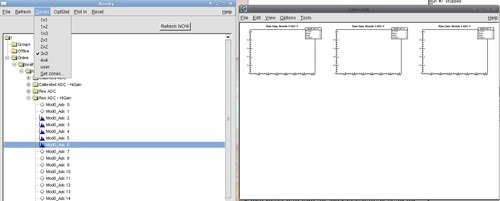}
\caption[Roody Histograms]{Roody, showing a drawn ROOT canvas and the zones menu. Here the histograms are being plotted in a single canvas which has been divided into 9 (3$\times$3) pads. }
\label{fig:Roody_Zones_small}
\end{center}
\end{figure}

\begin{figure}
\begin{center}
 \includegraphics[width=1.0\textwidth]{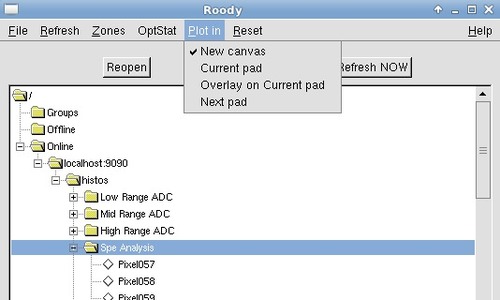}
\caption[Roody Plot Options]{The plot options menu. ``New Canvas" will draw every histogram in a new ROOT canvas. ``Current Pad" is the default and will replace the current canvas with the new histogram. 
The ``Overlay on Current Pad" option allows histograms to be drawn over top of each other on the same plot, and the ``Next Pad" option allows each selected histogram to be drawn in a sequence of pads (in a split canvas for example). }
\label{fig:Roody_Plot_In_small}
\end{center}
\end{figure}

Clicking on any histogram entry will draw it in the current Canvas. By default this will replace the currently displayed histogram, but the draw option can be changed through the menu, as shown in \fig\ref{fig:Roody_Plot_In_small}. 
If several histograms need to be viewed at once then each can be drawn in a new canvas by selecting the \emph{New Canvas} option. 
To put several histograms in the same canvas, as seen in \fig\ref{fig:Roody_Zones_small}, the number of rows and columns must be set in the ``Zones" menu and the \emph{Next Pad} option must be selected as the draw option.

The displayed histograms are updated every few seconds, according to selected refresh rate in the drop-down ``Refresh'' menu. 
The currently displayed histogram can be refreshed by clicking the \emph{Refresh Now} button. 
If the automatic refresh is off and you start a new run, the histograms will not be updated until the \emph{Refresh Now} button is hit. 
\emph{Therefore, always hit the Refresh button after opening a new histogram to sure that it is current.}   

A ROOT fit panel can be opened and the drawing options of the current histogram can be changed from the canvas menu. Histograms can also be saved to PDF, PS, etc. 
It is possible to zoom in on the current histogram by clicking and dragging across a range on the $X$ or $Y$ axis, or through the canvas menu, which can be opened by right-clicking on the canvas.
The integral, mean value, and standard deviation\footnote{Labeled as RMS in root, for historical reasons, but calculated as the standard deviation.}
displayed in the upper-right corner of the canvas are calculated for the displayed range.

\subsection{Viewing Saved Events}
The events collected by the DAQ are saved by the logger in Midas format (.mid) to the directory shown in the control interface. The data are saved unanalyzed, and event by event. 
This allows them to be later sorted by different analysis routines as long as the event structure is known.

If the DAQ analyzer code is run on an event file, it will create a *.root file of the sorted histograms, which can then be viewed in Roody, or in the ROOT TBrowser.
To run the analyzer in this way, the event file must first be copied to the directory where the analyzer code is located:
\begin{verbatim}
 cd /home/blaksley/online/C205_Frontend/c1205analyzer    
\end{verbatim}
and the analyzer must then be run in a terminal with the options:
\begin{verbatim}
./analyzer -i <filename1> <filename2> ... -o <filename>  
\end{verbatim}
The \emph{-i} option indicates the input file name(s)  which may contain a $\%05d$ to be replaced by the run number. Up to ten input files can be specified in one \emph{-i} statement.  

The \emph{-o} option gives the output file name. The extension may be *.mid (Midas binary), *.asc (ASCII), *.root (ROOT) or *.rz (HBOOK). If the name contains a $\%05d$, one output file is generated for each run. 
Use \emph{OFLN} as
an output file name to create a HBOOK shared memory instead of a file. More options for the analyzer codes can be seen by adding the \emph{-help} option to the run command.
Once the analyzer is finished, the output ROOT file can be opened by running Roody and selecting ``Open Data File" from the ``File" menu. 

\section{The DAQ Frontends}
\label{sec:The DAQ Frontends}
In the DAQ system there are three separate frontend routines. Two of these run directly on the backend computer, while the third runs on the MVME processor board located in the VME crate. 
The two frontend routines running on the backend computer control the NIST photodiode read out and the X-Y movement respectively. Here a brief overview will be given, but for a detailed understanding the 
user is invited to look at the codes themselves.
In Midas the data sources can be grouped into four categories:
\begin{inparaenum}[i\upshape)]
 \item periodic sources,
 \item polled sources,
 \item slow control sources, and
 \item sources which generate interrupts. 
\end{inparaenum}

Periodic sources are read out at a constant rate, which is defined as a parameter of the frontend code. The read-out rate can also be modified 
online through the online database entry of the respective frontend. The NIST photodiode falls into this category, being read out at a rate of 10 Hz. At each read-out cycle, the 
appropriate command is issued to the LaserStar through RS232 and the response is read by the frontend code. Because the read-out rate is low, the frontend itself decodes the data word from the LaserStar and saves it to disk.

Slow control sources are those which do not generate data \emph{per se}, but rather set some control value. An example of slow control hardware are high voltage power supplies or X-Y movements. The slow control frontend routine implements
a user defined command logic which issues predefined orders in set situations. In the case of the X-Y movement, for example, the slow control creates a \emph{demand} and a \emph{measured} variable for each degree of freedom.
If the user demanded value is modified, the program requests the current status of the hardware. The response of the hardware is stored in to the measured value. 
If the measured value differs from the demanded value, then the slow control routine attempts to set the 
new value in the hardware according to the defined control logic. The measured and demanded values are also compared at regular intervals, according to the settings defined by the slow control frontend. 
The advantage of the slow control system in Midas is that each demand and measured parameter is stored in the online database and so can be accessed by other 
parts of the system such as the run sequencer. 

ADC, QDC, and \acrshort{TDC} hardware is operated as a polled source or interrupt data source. Interrupt data sources are more difficult to handle, and, as none are used in the DAQ, they will not be discussed further.
Polled sources operate on some external trigger and their status is requested by the system at regular intervals.
In the DAQ, the QDCs are sent a CAMAC command which causes the module 
to return a ``yes'' if converted values are ready to be read. The frontend polls the QDCs continuously for a set time before breaking to allow housekeeping processes, after which the polling loop is 
re-entered. 

If the poll returns a ``yes'' then the routine enters the read cycle of the QDC bank. This reads the data words for each QDC channel sequentially using single 24-bit CAMAC transfers. 
At every cycle each QDC gives a header word, 16 data words, a overflow word, and an event separator word. Each of these is sent directly to the backend without being decoded.
The timing of the polling function is calibrated during the initialization of the QDC frontend. For efficiency measurements, the repetition rate of the trigger must be
low enough that the modules are never busy when a new trigger is issued. For the configuration of the DAQ this is a maximum read out rate of $\sim2~$kHz.

\section{The DAQ Analyzer}
\label{sec:The DAQ Analyzer}
The analyzer routine is responsible for histogramming and analyzing data received from the frontend. 
In the DAQ, the QDC frontend gives one event per charge integration gate, with each event composed of raw 24-bit data words directly from the QDC.
Each frontend connected to the DAQ system places events into \emph{banks}. 
One bank is defined in the QDC frontend for each QDC module, giving a total of four QDC banks. Each bank receives 
nineteen 24-bit words per event.

The analyzer is split into sub-routines called \emph{modules}. A bank list in the main analyzer routine defines
both the data sources and the output for which the analyzer looks. At the same time, each module defines which banks it requires and outputs. This allows a complex interconnection between
modules, with one module taking data from a frontend source, operating on it, and then placing the result into a new bank which can be used by another module.
In addition to multiple modules, several different main analyzer routines can be connected to the same experiment at once.    
In the DAQ, there are (when this was written) 6 modules within the core analyzer, and each of these will be discussed in turn.

Each module receives raw data words from the frontend source and then decodes these to extract channel, range, and count information. 
By processing the data offline (on the backend) the data collection is not slowed by the need to decode the raw data words.
In addition, the raw events, saved by the logger, are isolated from any decoding errors, and this also serves to make the frontend routines as general as possible.
The data word decoding is specific to the hardware and so the hardware manuals should be referenced for any questions concerning it. 

\subsection{Simple Charge Spectra Routines}
The first three modules in the analyzer handle the incoming QDC data in the simplest possible way. For each event
the data words in the four QDC banks are decoded to retrieve the channel, range, and count information. 
The C1205 QDC has three independent charge ranges, and each of the three analyzer modules handles one charge range.
These are respectively named ``Low Range ADC'', ``Mid Range ADC'', and ``High Range ADC''.

Within each module, if the range in the decoded event block corresponds to the module's range, then the event is histogramed in the spectrum corresponding to module given by the bank and the channel number given in each data word.
Each spectrum from these modules is therefore labeled according to the the QDC module and channel. The histograms are defined with bins centered on integer values, between the minimum and maximum 
QDC output codes. For the C1205 the output codes are 14 bits in two's complement notation.

\subsection{Single Photoelectron Spectra Analysis Routine}
The fourth routine, called ``Spe Analysis'', decodes incoming QDC data in the same way as the previous three modules.
Unlike the previous modules, however, this module is specialized for taking single photoelectron spectra, and so handles only event data from the low-charge range of the QDC.
Each spectrum in this module corresponds to a defined MAPMT pixel, and the mapping between QDC channel and pixel is defined as a look-up table in the module online database.

Each histogram is defined with bins centered on integer values. The \emph{Sumw2} ROOT function is called during the creation of the histograms to insure that bin errors are properly handled.
At the end of each run this module performs an analysis of each single photoelectron spectra to extract the gain and number of single photoelectron counts. The module also accesses the NIST photodiode power
in order to calculate the absolute efficiency. This analysis is performed in the following steps, looping over every defined pixel (spectra):
\begin{itemize}
 \item The end time of the run is recorded using the system clock. The stop time is compared to the run start time to determine the total time of the run. 
 This gives the total run time with a resolution on the order of a microsecond including both the network delay and the PC clock resolution.

 \item The average power measured by the NIST photodiode during the run is retrieved by searching the analyzer shared memory for the histogram of the photodiode readings created by the photodiode analysis module.
 The average power is determined by taking the mean from this histogram. 

 \item Each spectra is smoothed $n$ times using the 353QH algorithm \cite{Friedman:1974vj} (built into ROOT TH1 class). The number of smoothing iterations is defined by the user, with a value of $n=6$ used in the analysis presented in chapter~\ref{CHAPTER:EUSOBalloonMeasurements}.

 \item After smoothing, the histogram is passed to two routines which search for peaks and valleys within the spectrum. 
The valley search operates by looking for a bin containing a number of counts lower than or equal to both the preceding and following bin.
The peak search operates similarly, looking for a bin with a counts greater than or equal to both the preceding and following bin.  
This method naturally requires that the difference in contents between bins is statistically significant, and so relies on the 
spectrum smoothing to reduce statistical fluctuations.

The valley search begins from the first peak, which is found by looking just below the spectrum mean. For single photoelectron
spectra, the mean of the total spectra is just above the pedestal mean, as the pedestal dominates ($\sim99\%$ of total events) over the single photoelectron peak. The first valley found therefore corresponds to the 
valley between the single photoelectron peak and the pedestal.

\item Once the valley between the single photoelectron peak and the pedestal is found, the mean of the single photoelectron peak is determined by taking the mean of
the spectrum above the valley. The mean of the pedestal is found by taking the mean of the spectrum below the valley. The uncertainty on both mean values is calculated at the same time, and the position of 
both mean values are marked on the spectrum for later viewing.

\item The number of single photoelectron counts is determined by taking the integral of the spectrum above the valley. The uncertainty is also calculated and returned.

\item An extrapolation of the single photoelectron peak below the valley is estimated using a 3rd order polynomial. 
The polynomial is found by fitting the backside of the single photoelectron peak in the range from $\mu-(2/3)d$ to $\mu - (1/3)d$, where $\mu$ is the mean of the single photoelectron peak and $d$ is the distance between $\mu$ and the 
valley. The integral of the polynomial function between the valley and the mean of the pedestal is then used to estimate the number of single photoelectron counts below the valley. 

\item The number of single photoelectron counts and the means of the single photoelectron peak and pedestal are handed to functions which calculate the gain, the quantum efficiency, and the uncertainty on each.
\begin{itemize}
\item The gain, in QDC counts, is given by the difference of the two means.
In order to convert the gain in QDC counts into a numeric gain, the analysis routine looks up the QDC conversion constant corresponding to that QDC channel in a table held in the online database. The values
in this table are the result of the QDC characterization measurements discussed in chapter~\ref{CHAPTER:PMT Sorting}. The total uncertainty on the gain is also calculated, accounting for the uncertainty on the means and the error on the QDC conversion.
\item The efficiency is calculated according to \eq\ref{eq:EfficiencyCalculation}) using the power measured by the NIST photodiode during the run. 
Each parameter which appears in \eq\ref{eq:EfficiencyCalculation}) (the calibrated attenuation $\alpha$, the Planck constant, light wavelength, etc.) is stored in the online database of the analyzer module.
\end{itemize}
\item Every result is saved to user-readable files. One text file is created for each pixel (spectrum), along with a results summary file, a file containing statistics from the run, and a detailed log of the analysis.
These files are saved inside a folder named run$\_\%05d$, where $d$ is the run number. The folder is placed in the same location as the .mid file containing the logged events, in the directory defined in
the mlogger parameters.
\end{itemize}

This analysis routine was purposefully developed to avoid using a fit to determine the shape or location of the single photoelectron peak. 
The single photoelectron peak itself is correctly described by a Polya distribution, and the relative height of the pedestal and single photoelectron
peak is determined by the Poisson distribution. Due to this, analytic models of PMT
response have several free parameters which are model dependent and degenerate. The spectrum is also statistically dominated by the pedestal. This presents a difficult fitting problem for an automatic routine.
At the same time, a simple combined two-Gaussian fit of the pedestal and single photoelectron peak
does not correctly describe the spectrum, and often gives unphysical results. Either of these methods therefore requires a strong cross-check (that is looking at each spectrum) to ensure that the results are correct. 
Due to the number of spectra generated for even a single 64-pixel MAPMT, this approach was not preferred.

At the same time, searching for a valley was found to be extremely effective, giving highly reliable results when the single photoelectron spectra are clean.
The valley-searching method generally fails completely if the spectrum itself does not show a good valley or has some other artifact, which can be considered as an advantage, as unphysical results are obvious.
The decision was therefore made to use this technique, so that the analysis results can be trusted without the need to systematically check every single photoelectron spectra.


\subsection{Automatic Centering Routine}

\begin{figure}
\begin{center}
 \subfigure[AutoCentering Module]{ \label{fig:AutoCentering_parameters}\includegraphics[width=0.48\textwidth]{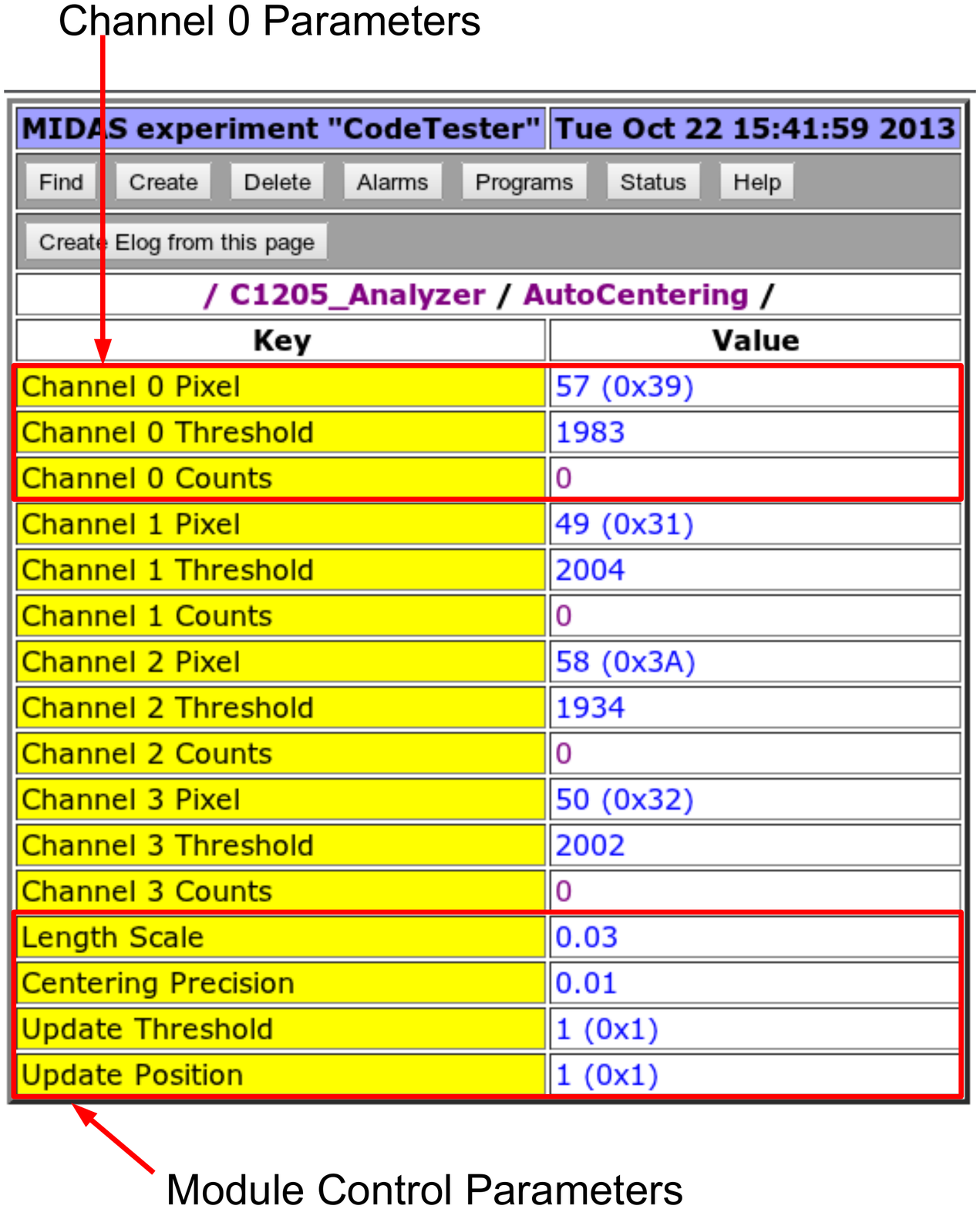}}
 \subfigure[Position Scanning Module]{\label{fig:PositionScanning_parameters}\includegraphics[width=0.48\textwidth]{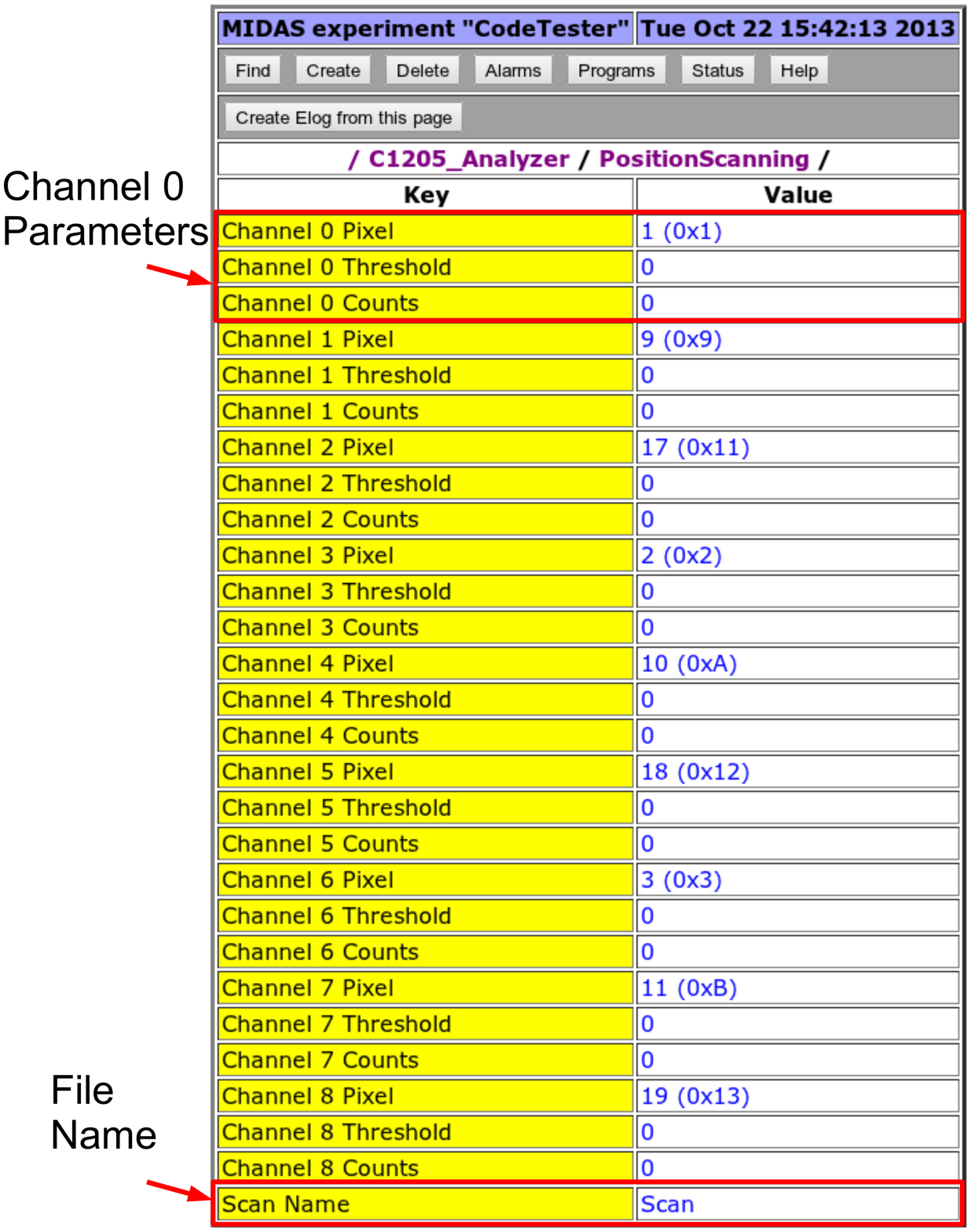}}
\caption[Analyzer Module Parameters]{ Screen captures of the online database parameters which control the ``AutoCentering'' and ``Position Scanning'' modules of the DAQ analyzer.
In each module a series of channels are requested by pixel number, as defined by the single photoelectron spectrum analysis pixel map. A threshold is defined for each channel, and 
the number of counts above this threshold is given at the end of each run. The other parameters of each module are described in the text.
\label{fig:module_parameters}}
\end{center}
\end{figure}

The ``AutoCentering'' routine interfaces the QDC with the slow control of the X-Y movement in order to find the cross between four pixels. 
This module is based on the single photoelectron analysis routine and takes number of parameters as shown in \fig\ref{fig:AutoCentering_parameters}. 
The module defines four \emph{centering channels}, each of which is tied to a pixel defined by the user. The pixel assignment can be changed between runs, but the analyzer must be restarted in order to correctly redefine the histograms.
The program finds the QDC channel corresponding to the requested pixel by looking in the pixel mapping used by the single photoelectron analysis routine. 
The four centering channel histograms are filled for each event as in the previous modules. 

At the end of a run, the number of counts in the each centering channel spectrum over the defined threshold is found. The threshold is in QDC counts and can be set by the user. If the \emph{Update Threshold} flag is set to 1,
then the module attempts to find the valley between the pedestal and the single photoelectron spectrum as done in the single photoelectron analysis module.

From the number of counts in each spectrum, a correction to the position of the light spot is calculated according to
\begin{equation}
 \Delta X =\frac{\left(N_{1} + N_{3}\right) - \left(N_{2} + N_{4}\right)}{\left(N_{1} + N_{2} + N_{3} + N_{4}\right)}
\end{equation}
and
\begin{equation}
 \Delta Y = \frac{\left(N_{1} + N_{2}\right) - \left(N_{3} + N_{4}\right)}{\left(N_{1} + N_{2} + N_{3} + N_{4}\right)}
\end{equation} 
The orientation of the four centering channels is defined as shown in \fig\ref{fig:CenteringDiagram} (from the top-left, starting at channel 0) with the positive $X$ direction to the right facing the photocathode, and the positive $Y$ direction 
towards the bottom.

If the \emph{Update Position} flag is set to 1, then the current position of the X-Y movement is incremented by an amount given by $ l \Delta X$ and $ l \Delta Y$, where $l$ is set by the \emph{Length Scale} parameter.
The AutoCentering module continues to update the position until both  $\Delta X$ and $\Delta Y$ are less than the \emph{Centering Precision} parameter.

This module is designed to be used with a run script which creates a loop of around 10-20 runs. 
If the length scale is set to the pixel size in the first few runs, then the initial position of the light spot need only be in one of the four centering channels. 
The length scale can then be reduced with each iteration, giving a convergence to within a few percent after $\sim10$ runs if the single photoelectron spectra are clean.

\subsection{Position Scanning Routine}


The ``Position Scanning'' module is an adaptation of the automatic centering module to the scanning of the photocathode. The parameters of the position scanning module are shown in \fig\ref{fig:PositionScanning_parameters}.
This module takes a number of user-defined pixels. As in the automatic centering routine, the corresponding QDC channel for the desired pixel is found by the module using the pixel map, and the 
QDC events are histogramed in each spectra accordingly.

At the end of each run, the number of counts in each \emph{scanning channel} over the threshold set for that channel is returned and amended to a text file.
The current position of the X-Y movement is also saved. The name of the text file is given by the \emph{Scan Name} parameter.
Coupled with an appropriate run script, this module allows collecting the number of single photoelectron counts in 
a number of pixels as a function of light spot position and was used for the pixel scanning results presented in chapter~\ref{CHAPTER:EUSOBalloonMeasurements}.

\section{Working with the Online Database}
\label{sec:Working with the Online Database}

Within Midas, parameters for each experiment are stored in a central structure called the Online DataBase (ODB). The ODB
contains run and analysis parameters, logging channel information, slow control values, status and performance data, and any other information which is defined by the user.
The parameters of the analysis routines presented in the last sections are held in the ODB. The ODB can be navigated and modified in the mhttp control web page, or it
can also be directly edited using the \emph{odbedit} routine. Odbedit can be run with the command:
\begin{verbatim}
  cd /home/blaksley/packages/midas/linux/bin/ 
  ./odbedit -e C1205_Frontend
\end{verbatim}
The odbedit program acts as a terminal-based browser. Information within the ODB is stored as series of \emph{keys} using a directory-like structure. Navigation through
the ODB directories uses the \emph{cd} command, and acts like typical Unix directory navigation.

Several useful odbedit commands include \emph{cleanup}, \emph{load}, and \emph{save}.
The \emph{cleanup} command deletes hanging clients, that is frontend or backend programs which are no longer present in the experiment, but which continue to have a ODB entry.
The \emph{save $<$filename$>$} command saves the ODB from the current directory to the file specified in \emph{$<$filename$>$}, given as an absolute path.
Similarly, the \emph{load $<$filename$>$} command loads parameters into the ODB from the specified file name.

These last two commands can be used to load different sets of analysis parameters, switching between different pixel maps, for example.
A set of four pixel maps, each corresponding to a different PMT within the EUSO-Balloon EC design can be loaded into the ODB using the command:
\begin{verbatim}
load "/home/blaksley/online/C1205_Frontend/ECPMT-[PMT]-PixelMapping.odb"
\end{verbatim}
where [PMT] is A, B, C, or D. This also loads the appropriate orientation of four center pixels into the AutoCentering analysis routine.
 
\section{Conclusion}
This appendix has presented a quick-start guide to the using the DAQ. 
Each sub-process of the data acquisition system was introduced and the most relevant details of the analysis routines were described.
These programs serve as a strong base on which further, more specialized, routines can be created. 
The power of the DAQ and the Midas framework on which it is based
is the complete flexibility it offers. It is highly recommended that the user consult the Midas 
documentation in order to understand the overall Midas system, and look at the code itself to understand how the DAQ frontend and analyzer function in detail.

 \printbibliography[heading=subbibliography]
  \end{refsection}
 \chapter{Figures and Tables}
   \label{appendix:chapter:Figures}

\section{Plots from the Calibration of the QDCs}
\label{app:sec:QDCmodule}
\begin{figure}[H]
  \centering
  \includegraphics[width=0.90\textwidth]{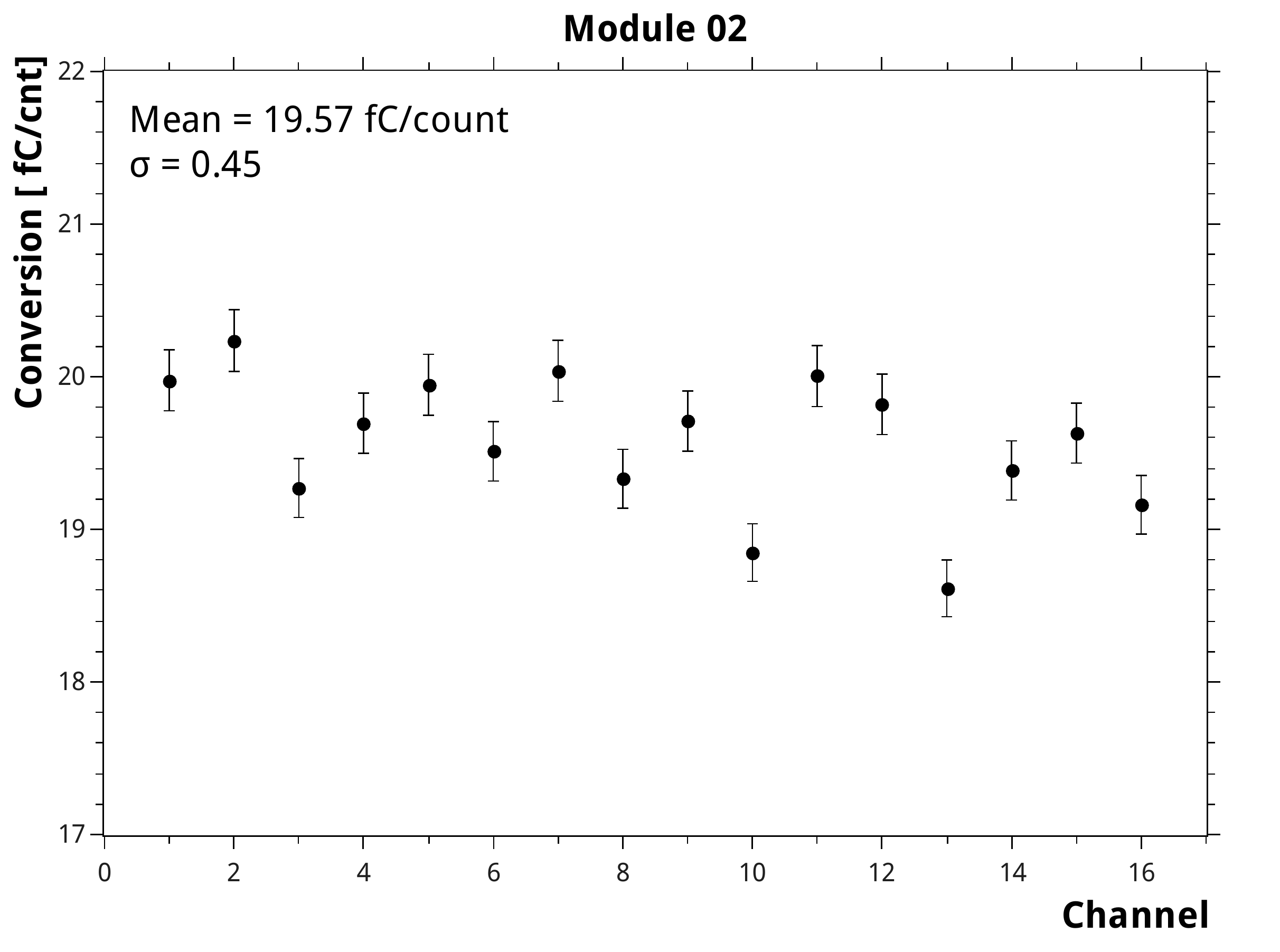}
  \caption[]{\label{fig:QDCMod2_overview}  A plot showing the results for the charge resolution for all 16 channels of QDC 2 (C1205$~\# 78$). Each QDC channel is measured independently, as described in the Section~\ref{sec:QDCCharacterization}.
The abscissa is the channel number within the module (numbered from 1 to 16), and the ordinates are the resolution in fC per count returned.}
 \end{figure}

\begin{figure}[H]
  \centering
  \includegraphics[width=0.90\textwidth]{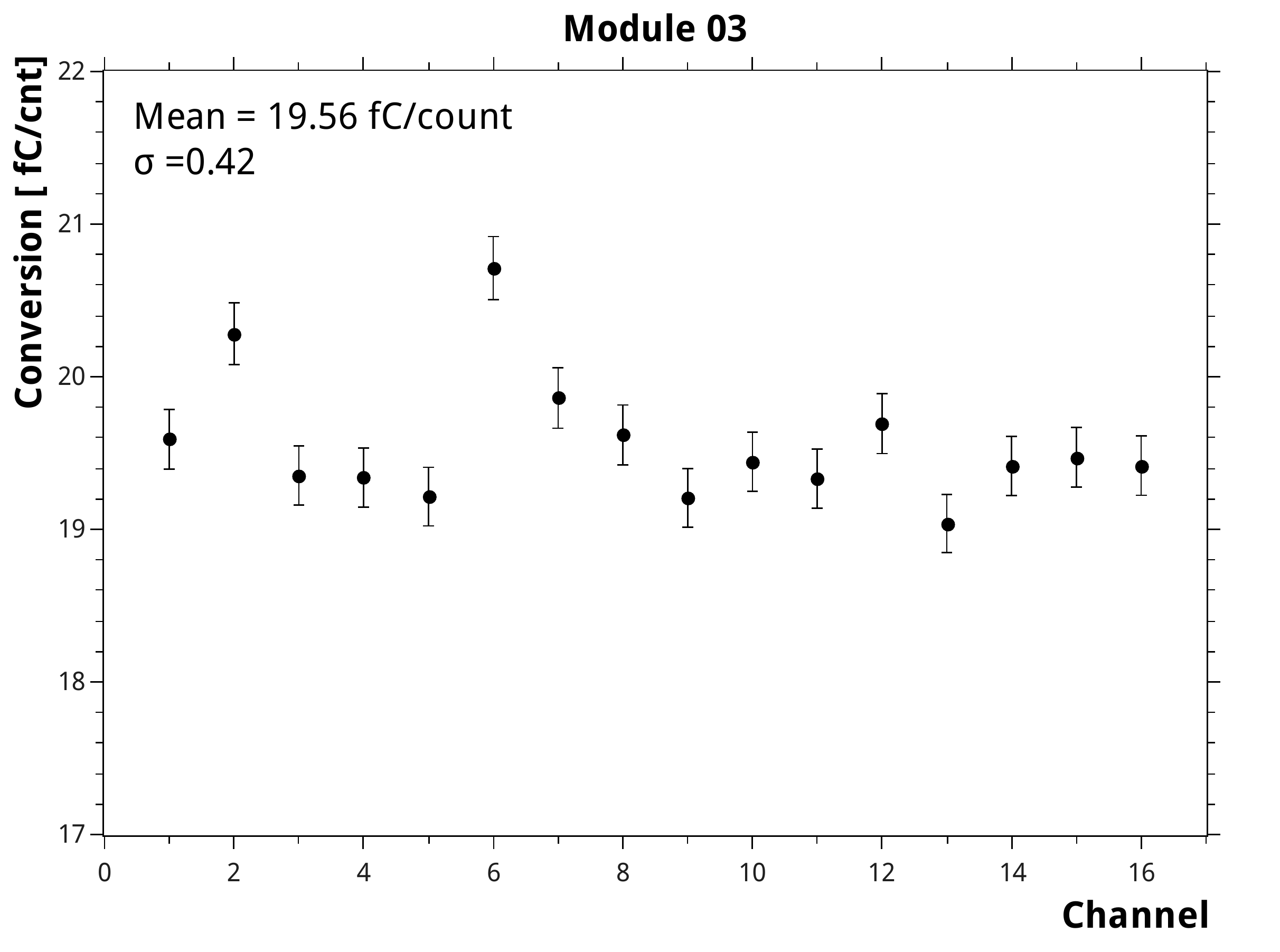}
  \caption[]{\label{fig:QDCMod3_overview}A plot showing the results for the charge resolution for all 16 channels of QDC 3 (C1205$~\# 95$). Each QDC channel is measured independently, as described in the Section~\ref{sec:QDCCharacterization}.
The abscissa is the channel number within the module (numbered from 1 to 16), and the ordinates are the resolution in fC per count returned.}
 \end{figure}

\begin{figure}[H]
  \centering
  \includegraphics[width=0.90\textwidth]{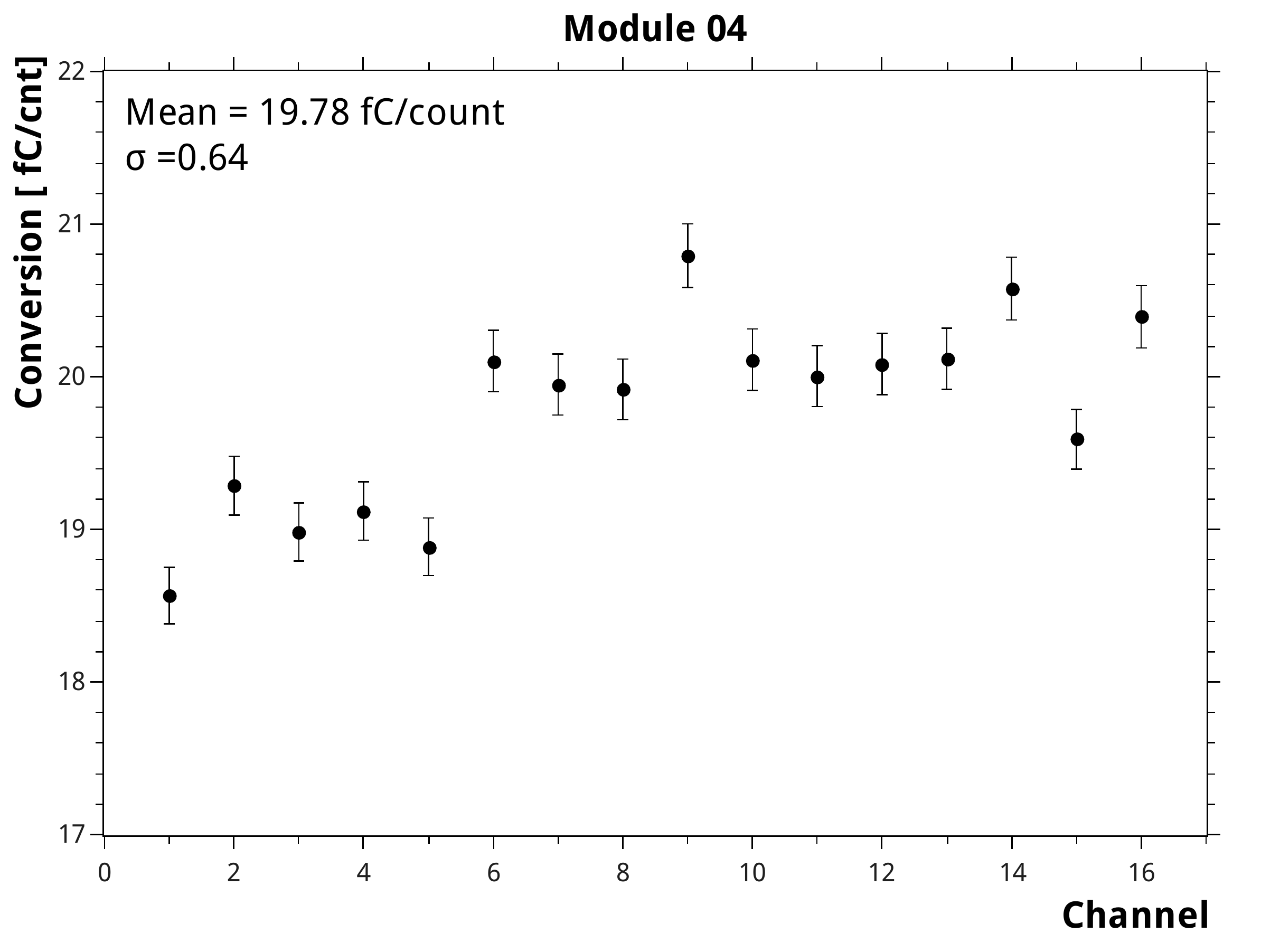}
  \caption[]{\label{fig:QDCMod4_overview} A plot showing the results for the charge resolution for all 16 channels of QDC 4 (C1205$~\# 87$). Each QDC channel is measured independently, as described in the Section~\ref{sec:QDCCharacterization}.
The abscissa is the channel number within the module (numbered from 1 to 16), and the ordinates are the resolution in fC per count returned.}
 \end{figure}

\section{Extra Plots for UHECR Source Statistics}

\begin{figure}[H]
\centering
\subfigure[Low-$p~E_{\max}$ Model]{\includegraphics[width=0.480\textwidth]{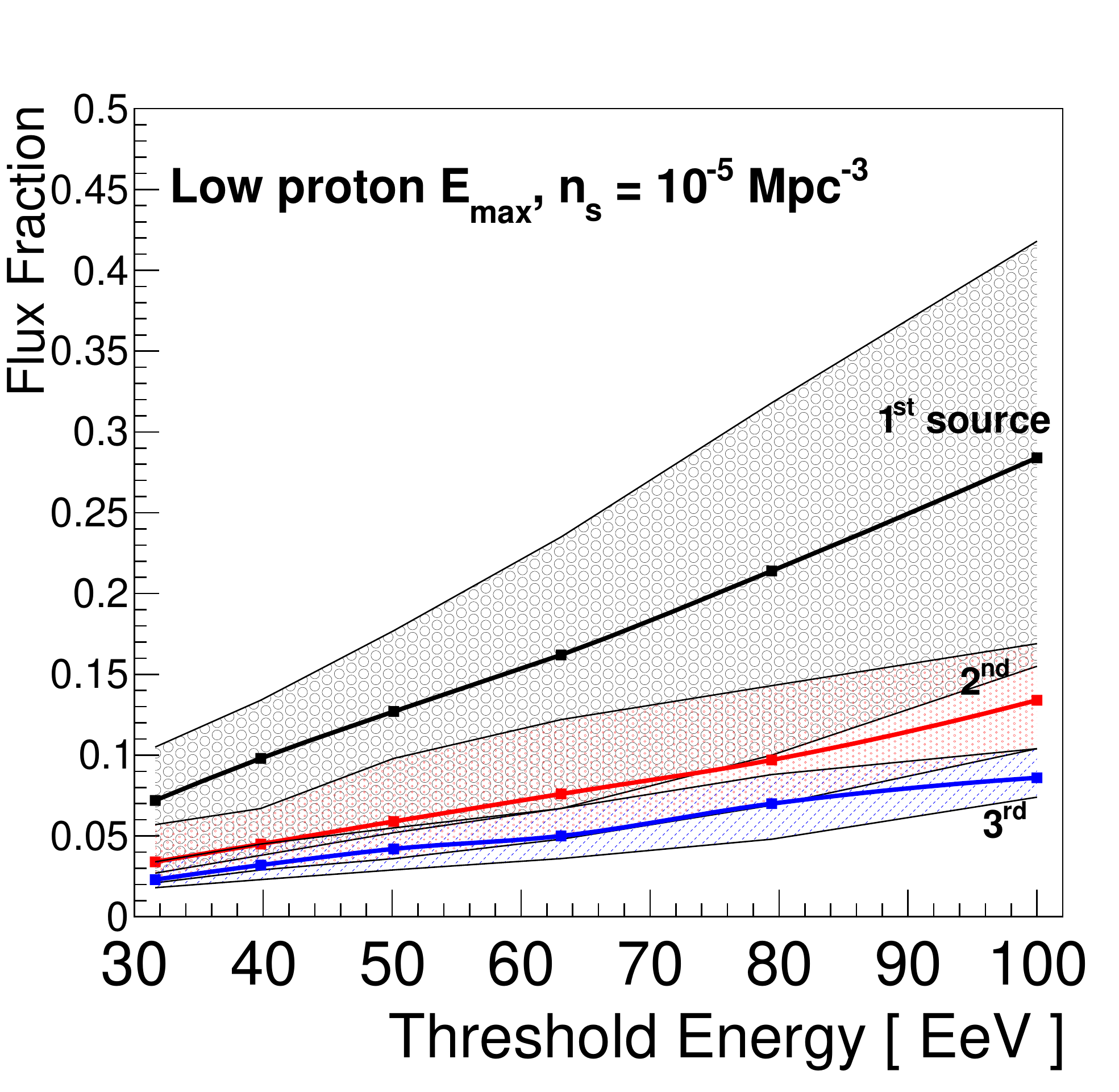}}
\subfigure[$p$-Only Model]{\includegraphics[width=0.480\textwidth]{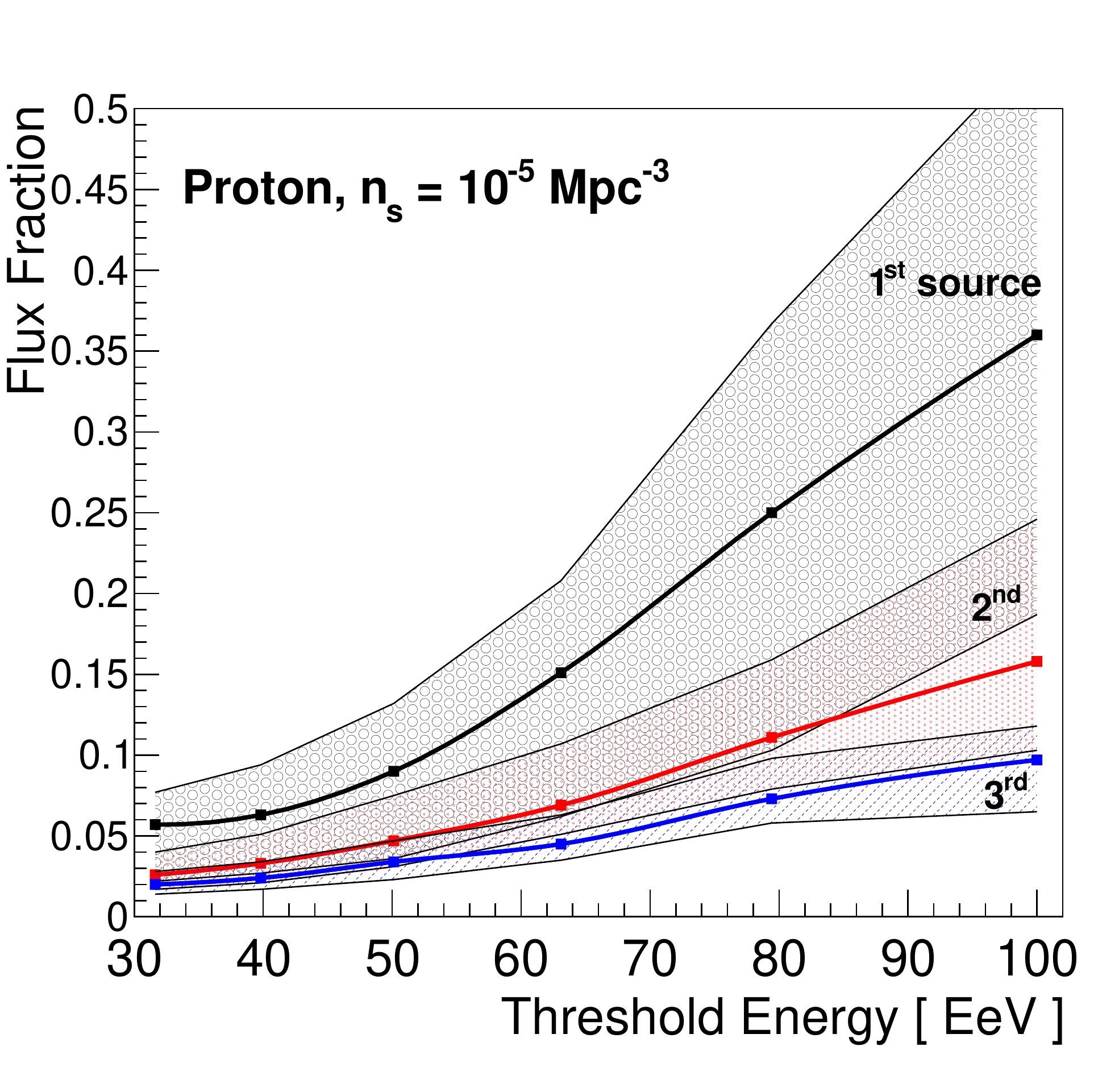}}
\subfigure[Fe-Only Model]{\includegraphics[width=0.490\textwidth]{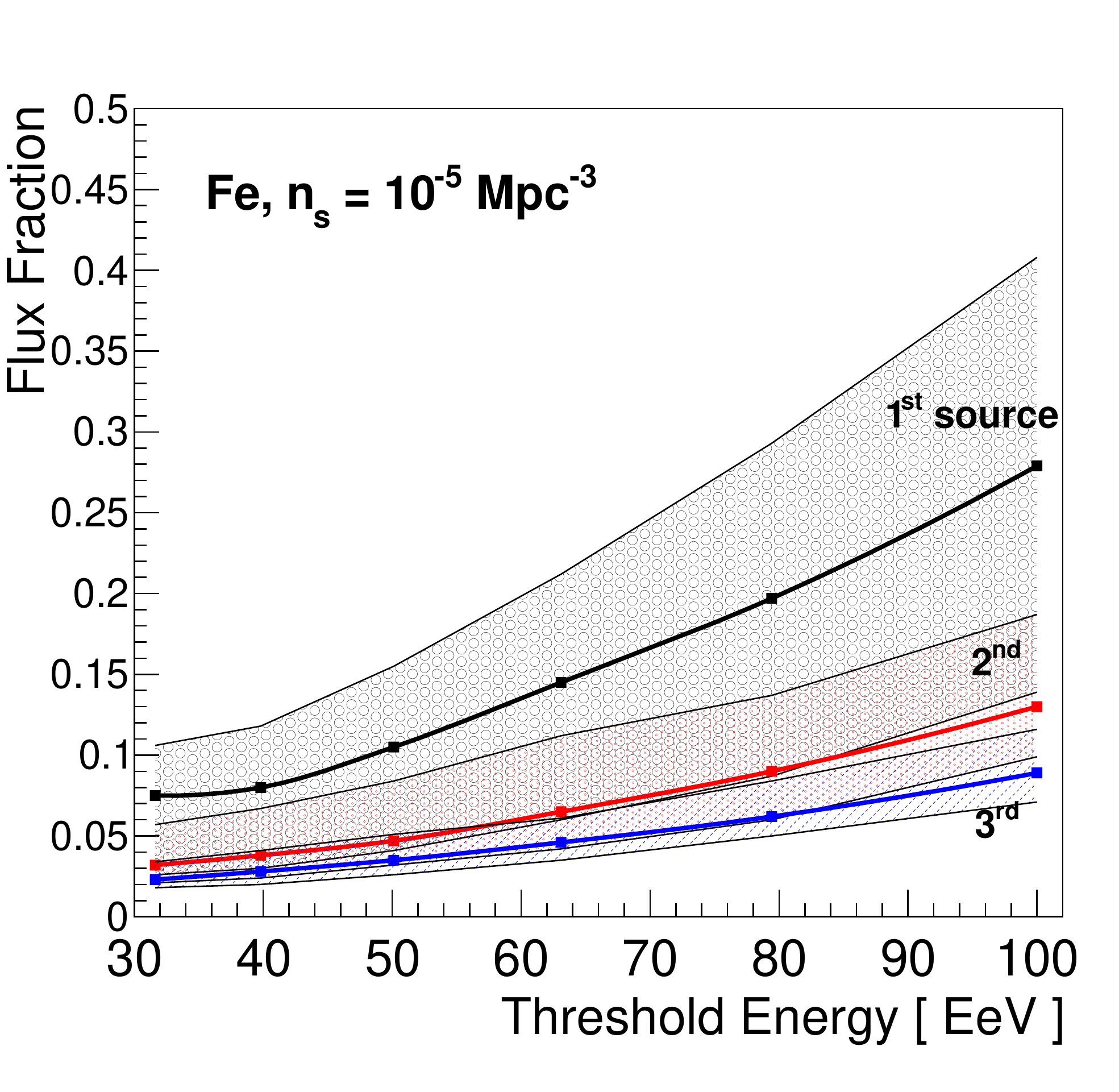}}\\
\caption[Median Flux for 3 Models]{Median flux as a percentage of the total for the three brightest sources in the sky, 
shown for the Low-$p~E_{\max}$, $p$-Only, and Fe-Only models (as given in Table \ref{tab:modelParameters}) 
using a source density of $n_{\text{s}}= 10^{-4}\,\text{Mpc}^{-3}$. 
}
\label{fig:3models3sources:10minus4MPC} 
\end{figure}

\begin{figure}[H]
\centering
\subfigure[Low-$p~E_{\max}$ Model]{\includegraphics[width=0.480\textwidth]{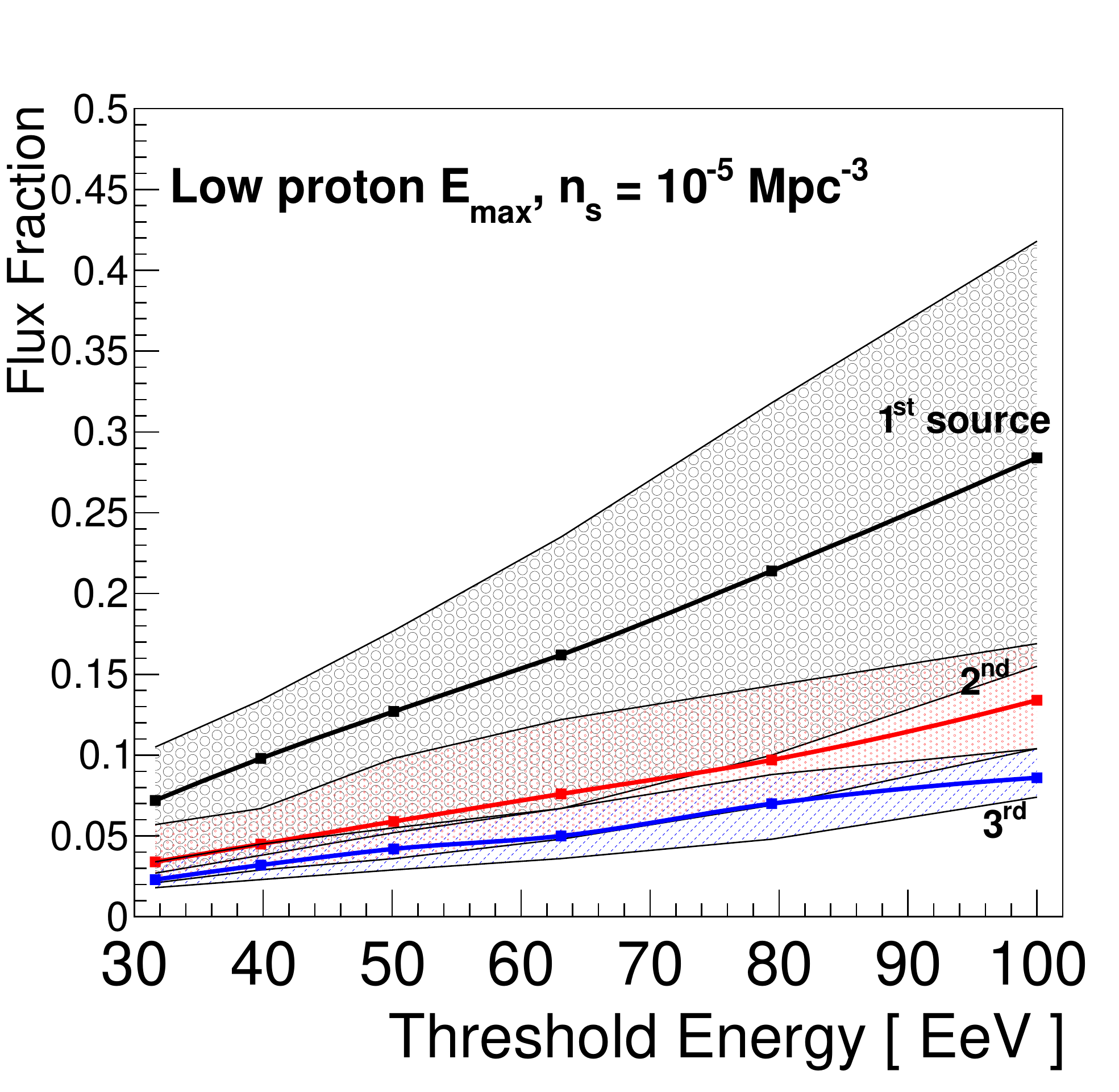}}
\subfigure[$p$-Only Model]{\includegraphics[width=0.480\textwidth]{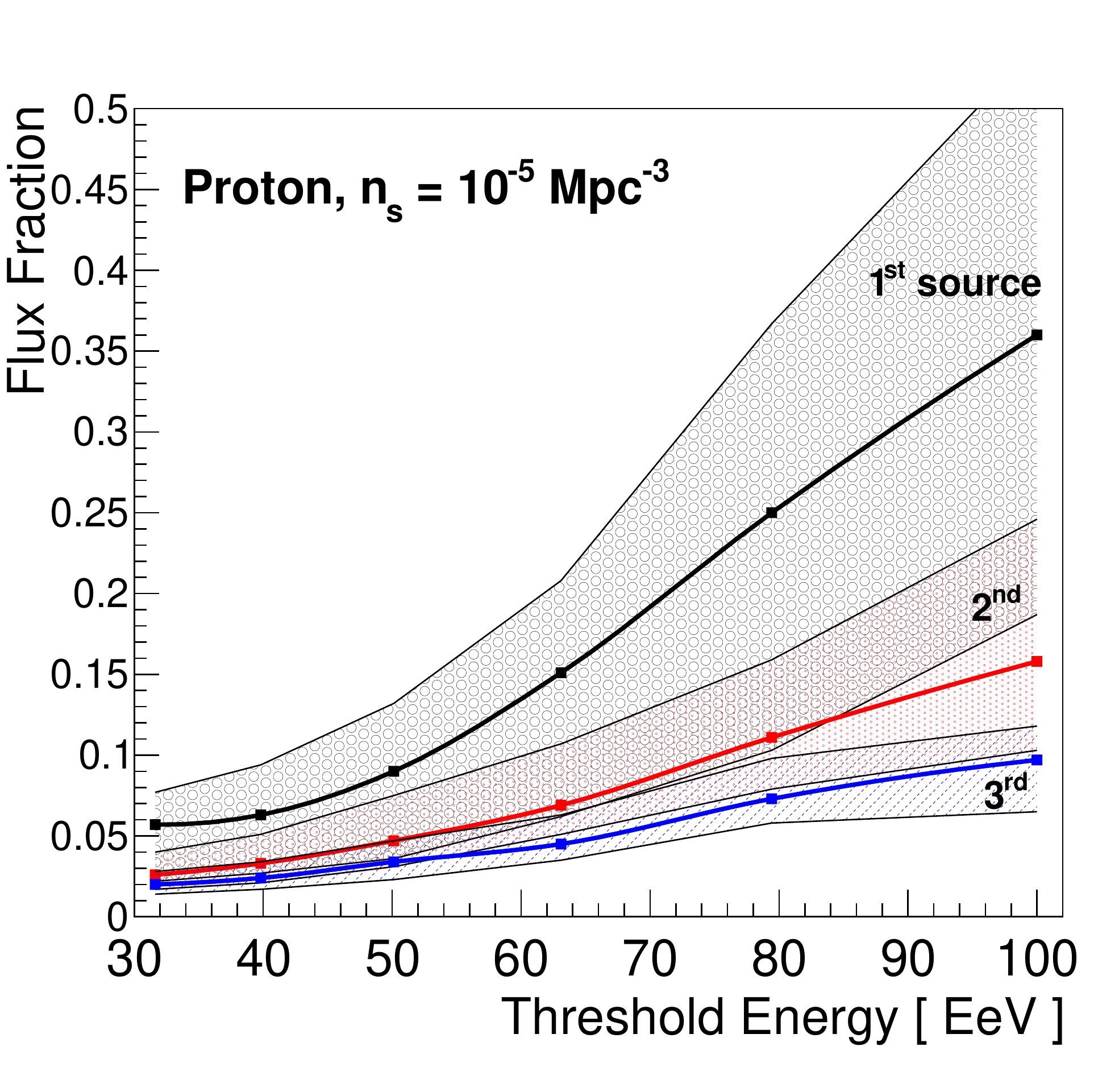}}
\subfigure[Fe-Only Model]{\includegraphics[width=0.490\textwidth]{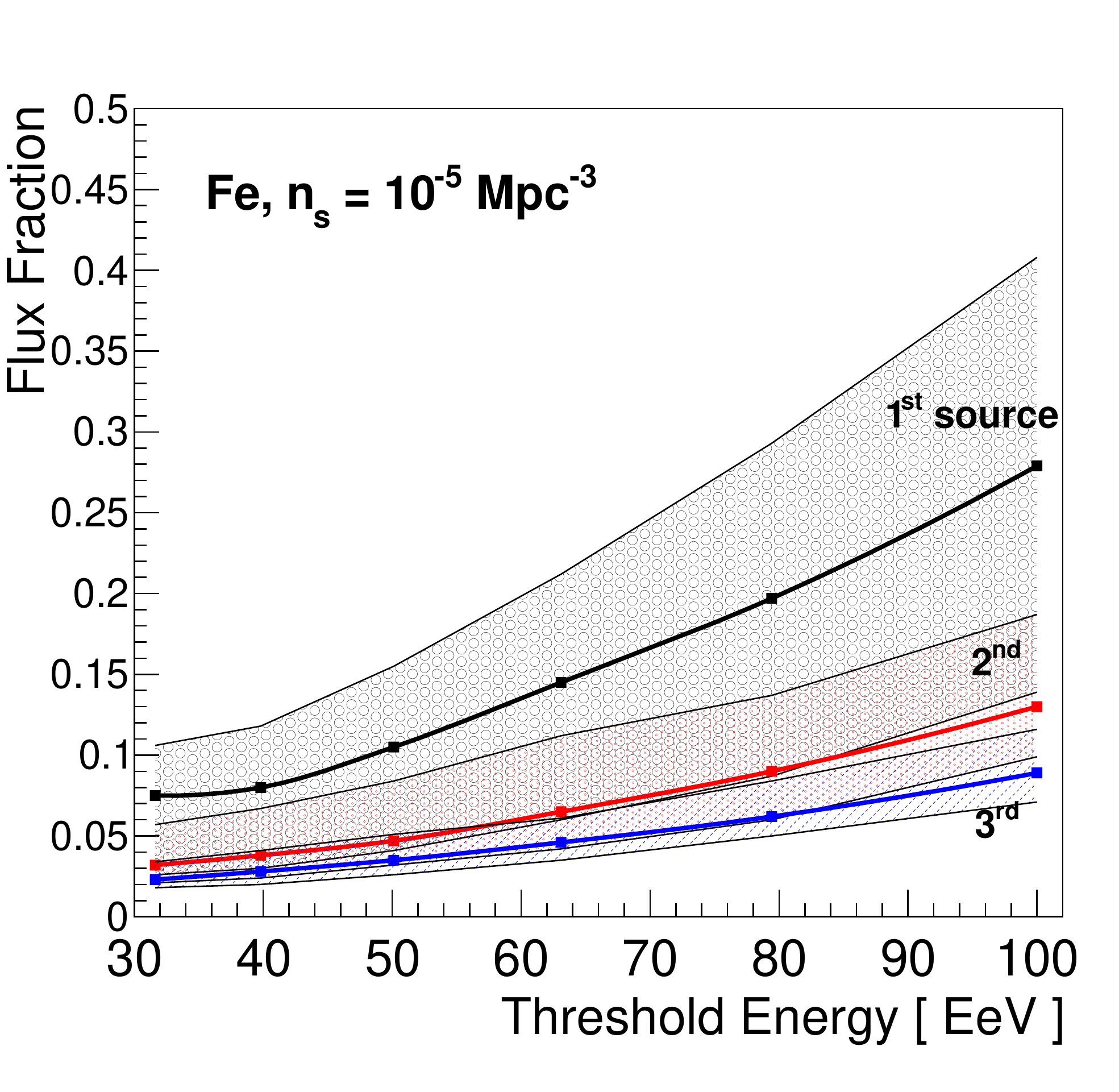}}\\
\caption[Median Flux for 3 Models]{Median flux as a percentage of the total for the three brightest sources in the sky, 
shown for the Low-$p~E_{\max}$, $p$-Only, and Fe-Only models (as given in Table \ref{tab:modelParameters}) 
using a source density of $n_{\text{s}}= 10^{-6}\,\text{Mpc}^{-3}$.
}
\label{fig:3models3sources:10minus6MPC} 
\end{figure}

\section{Efficiency Tables for EUSO-Balloon EC109}
 \label{app:sec:EfficiencyTables}

Tables of the Efficiency for each PMT of EC 109. No table is given for other ECs for brevity.

\begin{table}
\begin{center}
\begin{tabulary}{1.0\textwidth}{LCCC || LCCC}
\multicolumn{4}{l}{EC 109 PMT--A}\\
\toprule
Pixel & $\epsilon$ & $\delta \epsilon/\epsilon$ & P/V & Pixel & $\epsilon$ & $\delta \epsilon/\epsilon$ & P/V \\
\cmidrule(l){1-8}
01 & 0.252 & 0.028 & 1.7 & 33 & -- & -- &0.0 \\
02 & 0.266 & 0.030 & 2.3 & 34 & 0.229 & 0.025 & 2.7 \\
03 & 0.261 & 0.029 & 2.3 & 35 & 0.231 & 0.026 & 2.4 \\
04 & 0.241 & 0.026 & 2.3 & 36 & 0.227 & 0.026 & 2.6 \\
05 & 0.253 & 0.029 & 2.0 & 37 & 0.230 & 0.026 & 2.6 \\
06 & 0.261 & 0.027 & 1.5 & 38 & 0.240 & 0.026 & 2.6 \\
07 & 0.247 & 0.026 & 1.9 & 39 & 0.247 & 0.026 & 2.2 \\
08 & -- & -- &1.5 & 40 & 0.236 & 0.027 & 1.5 \\
09 & -- & -- &0.0 & 41 & 0.255 & 0.027 & 2.3 \\
10 & 0.222 & 0.025 & 2.8 & 42 & 0.234 & 0.026 & 2.2 \\
11 & 0.229 & 0.026 & 2.9 & 43 & 0.238 & 0.026 & 2.4 \\
12 & 0.228 & 0.025 & 3.0 & 44 & 0.242 & 0.029 & 2.4 \\
13 & 0.229 & 0.026 & 2.5 & 45 & -- & -- &1.3 \\
14 & 0.228 & 0.025 & 2.9 & 46 & 0.245 & 0.028 & 2.5 \\
15 & 0.231 & 0.026 & 2.7 & 47 & 0.242 & 0.028 & 2.1 \\
16 & 0.229 & 0.034 & 1.6 & 48 & 0.246 & 0.031 & 1.6 \\
17 & 0.252 & 0.025 & 2.6 & 49 & -- & -- &1.5 \\
18 & 0.228 & 0.025 & 2.7 & 50 & 0.239 & 0.028 & 2.1 \\
19 & 0.228 & 0.025 & 2.6 & 51 & 0.241 & 0.021 & 2.3 \\
20 & 0.238 & 0.031 & 1.6 & 52 & 0.245 & 0.025 & 2.5 \\
21 & 0.239 & 0.026 & 2.5 & 53 & 0.251 & 0.027 & 2.2 \\
22 & 0.229 & 0.025 & 2.6 & 54 & 0.254 & 0.025 & 2.6 \\
23 & 0.256 & 0.033 & 1.1 & 55 & 0.241 & 0.025 & 2.5 \\
24 & 0.240 & 0.031 & 1.6 & 56 & 0.241 & 0.029 & 1.6 \\
25 & 0.245 & 0.026 & 2.4 & 57 & -- & -- &0.5 \\
26 & 0.242 & 0.026 & 2.4 & 58 & -- & -- &1.0 \\
27 & 0.229 & 0.025 & 2.9 & 59 & 0.276 & 0.034 & 1.5 \\
28 & 0.228 & 0.027 & 2.1 & 60 & 0.266 & 0.031 & 1.8 \\
29 & 0.238 & 0.028 & 2.1 & 61 & 0.251 & 0.026 & 2.2 \\
30 & 0.244 & 0.025 & 2.4 & 62 & 0.257 & 0.028 & 1.9 \\
31 & 0.241 & 0.026 & 2.4 & 63 & 0.250 & 0.030 & 1.7 \\
32 & 0.245 & 0.031 & 1.6 & 64 & -- & -- &1.0 \\
\bottomrule
\end{tabulary}
\caption[EC109--PMTA Efficiency Results]{ \label{tab:EC109--PMTA EffResults}
The efficiency  $\epsilon$ measured for EC 109 PMT--A. Pixel 51 was used as the reference pixel. $P/V$ denotes the peak to valley ratio, which can be used as a figure of merit.
Blank entries are those pixels which did not pass quality cuts, as described in section \ref{sec:Efficiency and Gain Results}
}
\end{center}
\end{table}

\begin{table}
\begin{center}
\begin{tabulary}{1.0\textwidth}{LCCC || LCCC}
\multicolumn{4}{l}{EC 109 PMT--B}\\
\toprule
Pixel & $\epsilon$ & $\delta \epsilon/\epsilon$ & P/V & Pixel & $\epsilon$ & $\delta \epsilon/\epsilon$ & P/V \\
\cmidrule(l){1-8}
01 & 0.276 & 0.025 & 2.7 & 33 & 0.248 & 0.025 & 2.9 \\
02 & 0.272 & 0.025 & 3.6 & 34 & 0.249 & 0.029 & 1.6 \\
03 & 0.274 & 0.025 & 3.6 & 35 & 0.240 & 0.025 & 2.9 \\
04 & 0.286 & 0.026 & 2.5 & 36 & 0.229 & 0.025 & 2.8 \\
05 & 0.251 & 0.025 & 3.0 & 37 & 0.233 & 0.026 & 2.1 \\
06 & 0.280 & 0.026 & 2.5 & 38 & 0.238 & 0.025 & 2.8 \\
07 & 0.271 & 0.027 & 2.2 & 39 & 0.232 & 0.025 & 3.1 \\
08 & -- & -- &1.1 & 40 & 0.245 & 0.026 & 2.4 \\
09 & 0.267 & 0.025 & 3.3 & 41 & 0.264 & 0.025 & 2.7 \\
10 & 0.245 & 0.025 & 4.3 & 42 & 0.240 & 0.027 & 2.5 \\
11 & 0.246 & 0.025 & 4.2 & 43 & 0.230 & 0.025 & 3.3 \\
12 & 0.250 & 0.025 & 3.8 & 44 & 0.239 & 0.026 & 2.5 \\
13 & 0.248 & 0.025 & 3.4 & 45 & 0.224 & 0.026 & 2.4 \\
14 & 0.249 & 0.025 & 2.9 & 46 & 0.237 & 0.025 & 2.7 \\
15 & 0.254 & 0.028 & 2.7 & 47 & 0.241 & 0.026 & 2.8 \\
16 & 0.247 & 0.026 & 1.7 & 48 & 0.228 & 0.026 & 2.0 \\
17 & 0.272 & 0.025 & 3.1 & 49 & 0.268 & 0.025 & 3.1 \\
18 & 0.244 & 0.025 & 3.2 & 50 & 0.232 & 0.026 & 2.9 \\
19 & 0.247 & 0.025 & 3.4 & 51 & 0.237 & 0.029 & 2.2 \\
20 & 0.235 & 0.025 & 3.7 & 52 & 0.236 & 0.028 & 2.4 \\
21 & 0.239 & 0.025 & 3.4 & 53 & 0.227 & 0.026 & 2.5 \\
22 & 0.227 & 0.021 & 3.8 & 54 & 0.229 & 0.025 & 3.0 \\
23 & 0.241 & 0.025 & 2.6 & 55 & 0.245 & 0.025 & 2.7 \\
24 & -- & -- &1.1 & 56 & 0.244 & 0.028 & 2.0 \\
25 & 0.263 & 0.025 & 3.1 & 57 & 0.297 & 0.028 & 1.7 \\
26 & 0.232 & 0.025 & 3.4 & 58 & 0.273 & 0.029 & 2.1 \\
27 & 0.225 & 0.025 & 3.0 & 59 & 0.288 & 0.030 & 1.8 \\
28 & 0.247 & 0.032 & 1.8 & 60 & 0.274 & 0.028 & 2.1 \\
29 & 0.237 & 0.026 & 2.5 & 61 & 0.267 & 0.027 & 2.1 \\
30 & 0.238 & 0.026 & 3.2 & 62 & 0.290 & 0.030 & 1.9 \\
31 & 0.237 & 0.025 & 2.7 & 63 & 0.298 & 0.033 & 2.0 \\
32 & 0.241 & 0.025 & 2.5 & 64 & -- & -- &0.0 \\
\bottomrule
\end{tabulary}
\caption[EC 109--PMTB Efficiency Results]{ \label{tab:EC109--PMTB EffResults}
The efficiency  $\epsilon$ measured for EC 109 PMT--B. Pixel 22 was used as the reference pixel. $P/V$ denotes the peak to valley ratio, which can be used as a figure of merit.
Blank entries are those pixels which did not pass quality cuts, as described in section \ref{sec:Efficiency and Gain Results}
}
\end{center}
\end{table}

\begin{table}
\begin{center}
\begin{tabulary}{1.0\textwidth}{LCCC || LCCC}
\multicolumn{4}{l}{EC 109 PMT--C}\\
\toprule
Pixel & $\epsilon$ & $\delta \epsilon/\epsilon$ & P/V & Pixel & $\epsilon$ & $\delta \epsilon/\epsilon$ & P/V \\
\cmidrule(l){1-8}
01 & -- & -- &1.209 & 33 & 0.336 & 0.035 & 1.4 \\
02 & 0.385 & 0.034 & 1.7 & 34 & 0.293 & 0.027 & 2.3 \\
03 & 0.359 & 0.030 & 1.9 & 35 & 0.276 & 0.028 & 1.7 \\
04 & 0.332 & 0.026 & 2.0 & 36 & 0.283 & 0.025 & 2.4 \\
05 & 0.360 & 0.030 & 1.8 & 37 & 0.280 & 0.025 & 2.5 \\
06 & 0.349 & 0.027 & 1.6 & 38 & 0.288 & 0.025 & 2.4 \\
07 & 0.330 & 0.026 & 2.1 & 39 & 0.285 & 0.026 & 2.7 \\
08 & -- & -- &1.1 & 40 & 0.313 & 0.025 & 2.2 \\
09 & -- & -- &0.0 & 41 & 0.296 & 0.027 & 2.1 \\
10 & 0.293 & 0.025 & 3.1 & 42 & 0.275 & 0.026 & 1.8 \\
11 & 0.295 & 0.025 & 3.2 & 43 & 0.282 & 0.027 & 2.2 \\
12 & 0.285 & 0.025 & 2.8 & 44 & 0.278 & 0.025 & 2.1 \\
13 & 0.286 & 0.025 & 3.1 & 45 & 0.288 & 0.031 & 1.9 \\
14 & 0.296 & 0.025 & 2.9 & 46 & 0.281 & 0.025 & 2.6 \\
15 & 0.299 & 0.025 & 3.1 & 47 & 0.289 & 0.025 & 2.4 \\
16 & 0.334 & 0.028 & 2.4 & 48 & 0.313 & 0.025 & 2.4 \\
17 & 0.325 & 0.025 & 2.4 & 49 & -- & -- &1.0 \\
18 & 0.296 & 0.025 & 2.7 & 50 & 0.273 & 0.028 & 1.4 \\
19 & 0.291 & 0.025 & 2.7 & 51 & 0.297 & 0.029 & 1.6 \\
20 & 0.292 & 0.027 & 1.8 & 52 & 0.286 & 0.025 & 2.7 \\
21 & 0.290 & 0.025 & 2.7 & 53 & 0.273 & 0.028 & 2.1 \\
22 & 0.285 & 0.021 & 3.5 & 54 & 0.290 & 0.024 & 3.1 \\
23 & 0.280 & 0.027 & 1.8 & 55 & 0.291 & 0.025 & 2.8 \\
24 & 0.312 & 0.025 & 2.5 & 56 & 0.311 & 0.025 & 2.4 \\
25 & 0.303 & 0.025 & 2.5 & 57 & -- & -- &0.5 \\
26 & -- & -- &0.9 & 58 & -- & -- &0.6 \\
27 & 0.284 & 0.026 & 2.6 & 59 & -- & -- &1.2 \\
28 & 0.278 & 0.026 & 2.3 & 60 & 0.324 & 0.028 & 1.4 \\
29 & 0.277 & 0.026 & 2.3 & 61 & 0.327 & 0.025 & 2.5 \\
30 & 0.293 & 0.025 & 2.3 & 62 & 0.320 & 0.025 & 2.5 \\
31 & 0.287 & 0.025 & 2.5 & 63 & 0.314 & 0.026 & 2.4 \\
32 & 0.315 & 0.025 & 2.5 & 64 & 0.330 & 0.029 & 1.7 \\
\bottomrule
\end{tabulary}
\caption[EC 109--PMTC Efficiency Results]{ \label{tab:EC109--PMTC EffResults}
The efficiency  $\epsilon$ measured for EC 109 PMT--C. Pixel 22 was used as the reference pixel. $P/V$ denotes the peak to valley ratio, which can be used as a figure of merit.
Blank entries are those pixels which did not pass quality cuts, as described in section \ref{sec:Efficiency and Gain Results}
}
\end{center}
\end{table}

\begin{table}
\begin{center}
\begin{tabulary}{1.0\textwidth}{LCCC || LCCC}
\multicolumn{4}{l}{EC 109 PMT--D}\\
\toprule
Pixel & $\epsilon$ & $\delta \epsilon/\epsilon$ & P/V & Pixel & $\epsilon$ & $\delta \epsilon/\epsilon$ & P/V \\
\cmidrule(l){1-8}
01 & 0.300 & 0.028 & 1.9 & 33 & 0.299 & 0.024 & 2.5 \\
02 & 0.296 & 0.025 & 2.6 & 34 & 0.295 & 0.028 & 2.3 \\
03 & 0.295 & 0.025 & 2.7 & 35 & 0.269 & 0.025 & 3.3 \\
04 & 0.293 & 0.027 & 2.2 & 36 & 0.262 & 0.024 & 3.2 \\
05 & 0.287 & 0.025 & 2.6 & 37 & 0.276 & 0.027 & 2.2 \\
06 & 0.328 & 0.035 & 1.5 & 38 & 0.269 & 0.024 & 3.1 \\
07 & 0.314 & 0.031 & 1.8 & 39 & 0.270 & 0.024 & 3.1 \\
08 & -- & -- &0.8 & 40 & 0.293 & 0.025 & 2.5 \\
09 & 0.296 & 0.025 & 3.3 & 41 & 0.298 & 0.025 & 2.9 \\
10 & 0.281 & 0.024 & 3.7 & 42 & 0.263 & 0.025 & 3.2 \\
11 & 0.283 & 0.024 & 3.6 & 43 & 0.269 & 0.025 & 3.5 \\
12 & 0.278 & 0.024 & 3.6 & 44 & 0.274 & 0.025 & 2.8 \\
13 & 0.262 & 0.025 & 3.3 & 45 & 0.263 & 0.025 & 3.2 \\
14 & 0.257 & 0.024 & 2.9 & 46 & 0.261 & 0.024 & 3.5 \\
15 & 0.269 & 0.025 & 1.9 & 47 & 0.276 & 0.024 & 3.3 \\
16 & -- & -- &1.2 & 48 & 0.301 & 0.025 & 2.5 \\
17 & 0.296 & 0.024 & 3.1 & 49 & 0.316 & 0.028 & 2.2 \\
18 & 0.273 & 0.024 & 3.8 & 50 & 0.285 & 0.025 & 2.9 \\
19 & 0.272 & 0.024 & 4.2 & 51 & 0.277 & 0.021 & 2.8 \\
20 & 0.271 & 0.024 & 3.5 & 52 & 0.262 & 0.024 & 3.3 \\
21 & 0.266 & 0.024 & 3.5 & 53 & 0.272 & 0.025 & 3.1 \\
22 & 0.264 & 0.025 & 3.3 & 54 & 0.269 & 0.024 & 3.2 \\
23 & 0.277 & 0.025 & 3.0 & 55 & 0.286 & 0.024 & 3.2 \\
24 & -- & -- &1.3 & 56 & 0.292 & 0.025 & 2.3 \\
25 & 0.295 & 0.024 & 3.0 & 57 & -- & -- &0.8 \\
26 & 0.278 & 0.025 & 2.8 & 58 & -- & -- &1.0 \\
27 & 0.269 & 0.024 & 3.3 & 59 & 0.319 & 0.029 & 1.5 \\
28 & 0.280 & 0.028 & 2.4 & 60 & 0.315 & 0.027 & 1.8 \\
29 & 0.271 & 0.025 & 2.9 & 61 & 0.308 & 0.027 & 2.1 \\
30 & 0.266 & 0.025 & 2.7 & 62 & 0.296 & 0.025 & 1.9 \\
31 & 0.264 & 0.024 & 3.1 & 63 & 0.307 & 0.026 & 2.3 \\
32 & 0.314 & 0.026 & 2.5 & 64 & -- & -- &0.0 \\
\bottomrule
\end{tabulary}
\caption[EC109--PMTD Efficiency Results]{ \label{tab:EC109--PMTD EffResults}
The efficiency  $\epsilon$ measured for EC 109 PMT--D. Pixel 51 was used as the reference pixel. $P/V$ denotes the peak to valley ratio, which can be used as a figure of merit.
Blank entries are those pixels which did not pass quality cuts, as described in section \ref{sec:Efficiency and Gain Results}}
\end{center}
\end{table}


\section{Former EUSO-Balloon PDM Layout}
 \label{app:sec:EfficiencyTables}

\begin{figure}[H]
\centering
\subfigure[The PDM efficiency Map]{\includegraphics[width=0.8\textwidth]{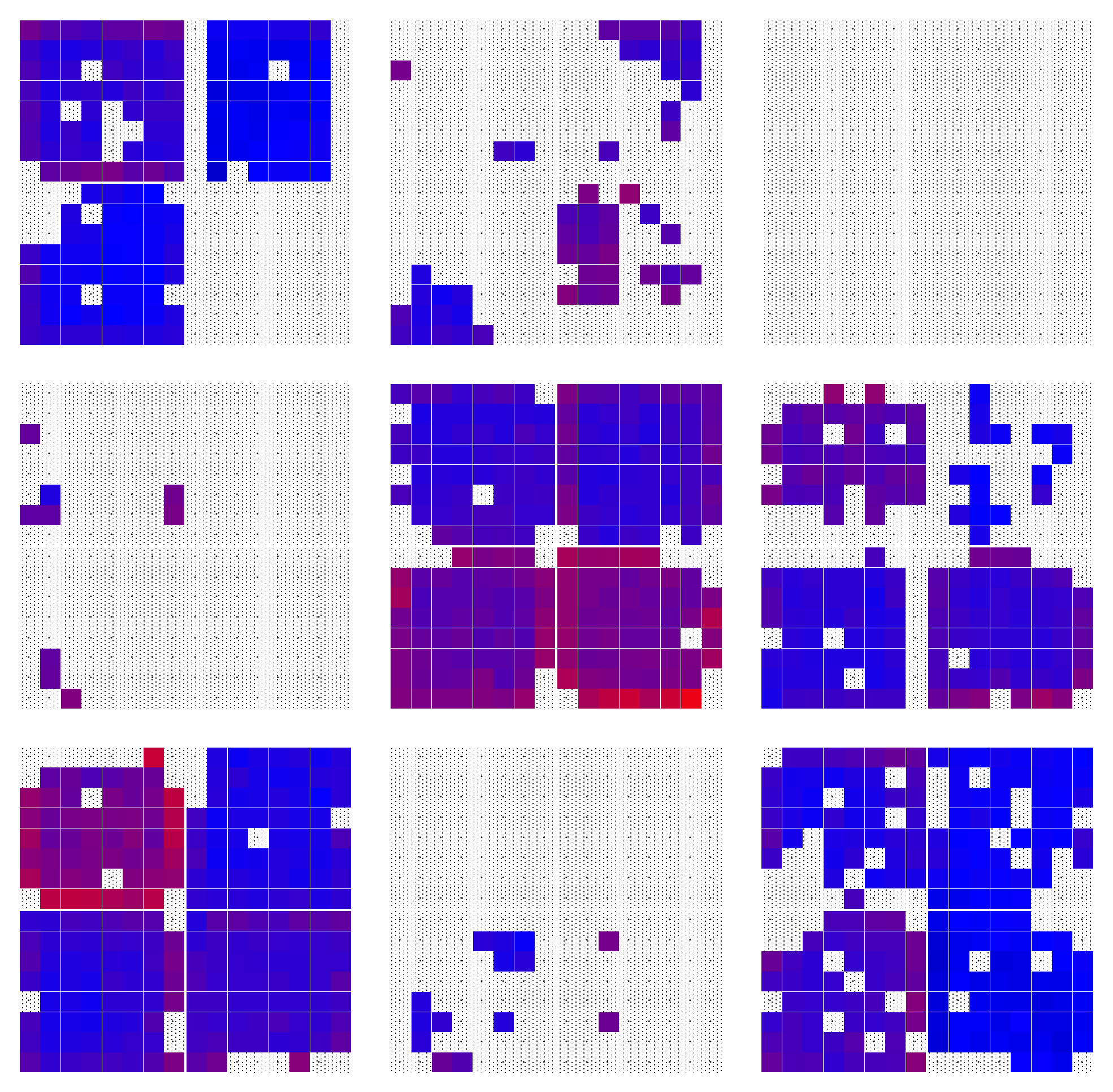}}
\subfigure[The PDM layout]{\label{fig:PDMlayout:old} \includegraphics[width=0.4\textwidth]{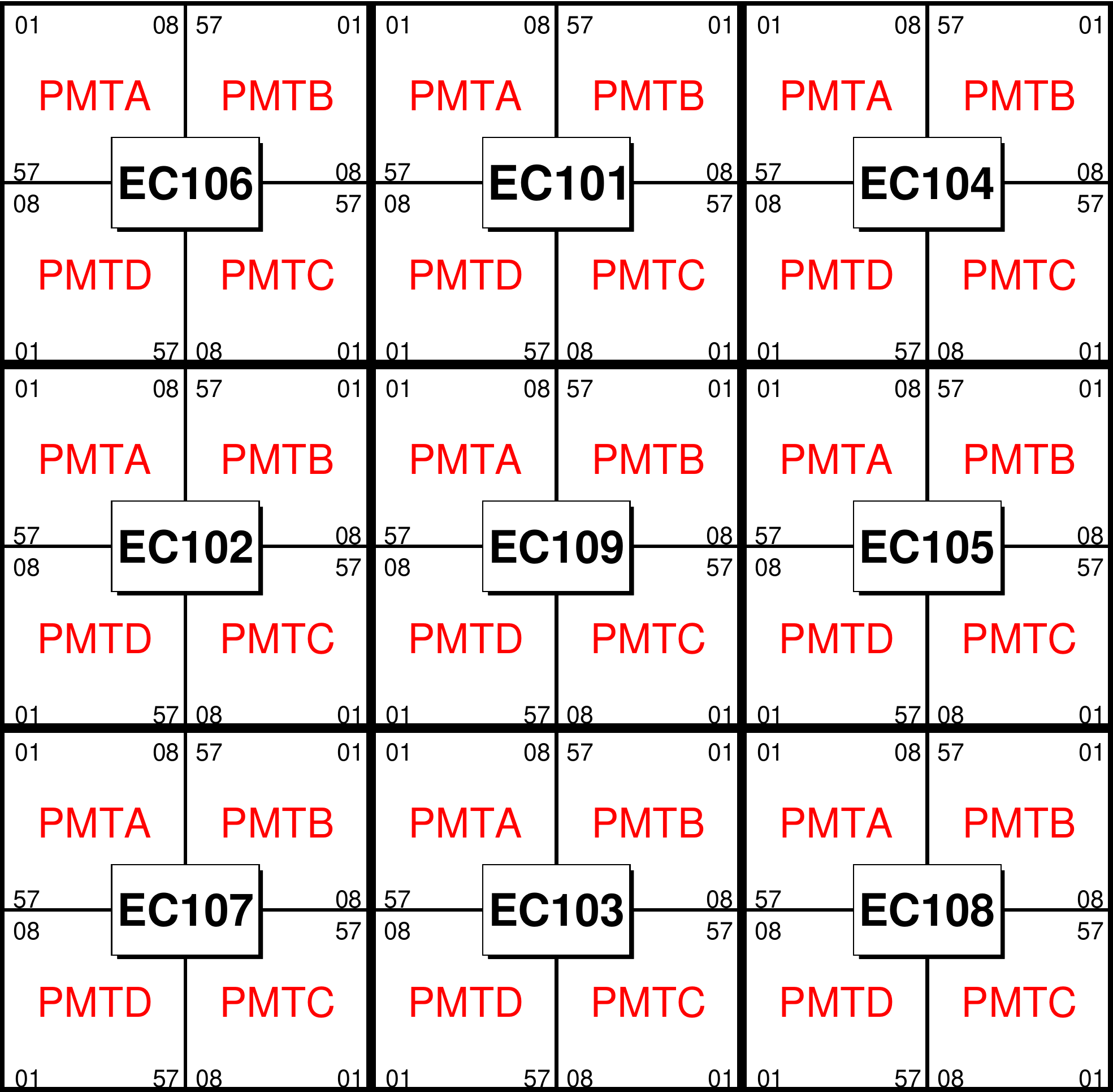}}\\\

\caption[EUSO-Balloon PDM Version 1]{
A map of the efficiency for the entire EUSO-Balloon PDM, using the original layout with the first nine ECs. 
The layout of the PDM can be seen in \fig\ref{fig:PDMlayout:old}. 
The color scale in this plot is the same as in \fig\ref{fig:EC109_EffPlot}.
Grayed-out squares indicate pixels for which no result is given.
EC 104 (the top-left corner) was not measured as it is still being assembled. 
No result is given for PMT-C of EC 106 (top-left), PMT-B and PMT-C of EC 102 (middle-left), and  PMT-A and PMT-B of EC 103 (bottom-middle) as these PMTs did not 
give any spectra which passed the quality cuts. It is emphasized again that this is only because the QDC is not as sensitive as the ASIC. 
}
\label{fig:3models3sources:10minus4MPC} 
\end{figure}
 \chapter{Photographs of Experimental Setups and Equipment}
    \label{appendix:chapter:Photographs}

\begin{figure}[p]
\centering
  \subfigure[]{\label{pic:BlackBox:1} \includegraphics[angle=0,width=1.0\textwidth]{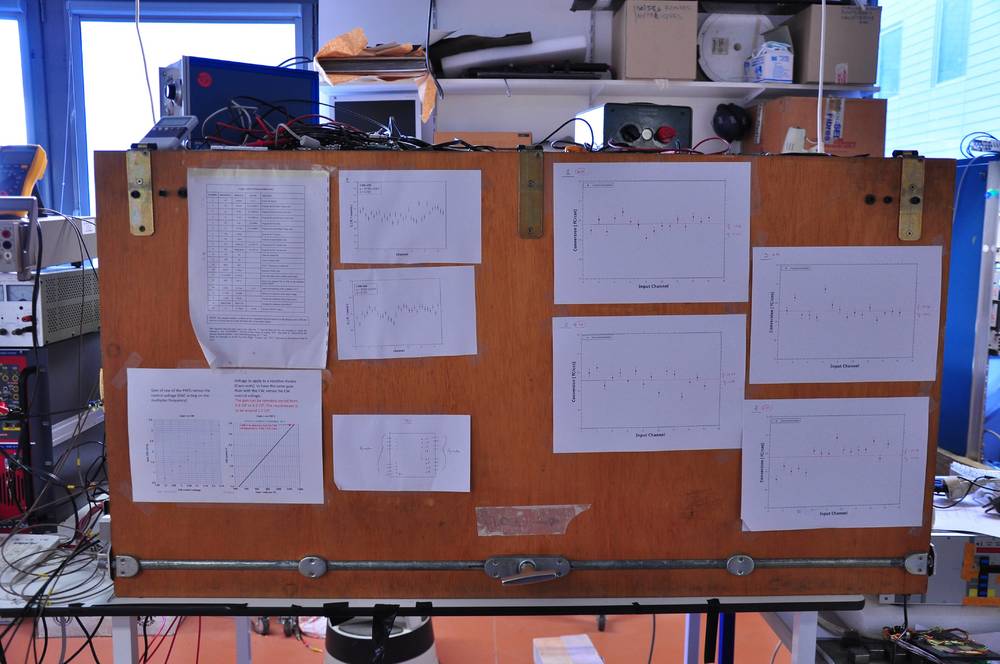}}\\
   \subfigure[]{\label{pic:BlackBox:2} \includegraphics[angle=0,width=1.0\textwidth]{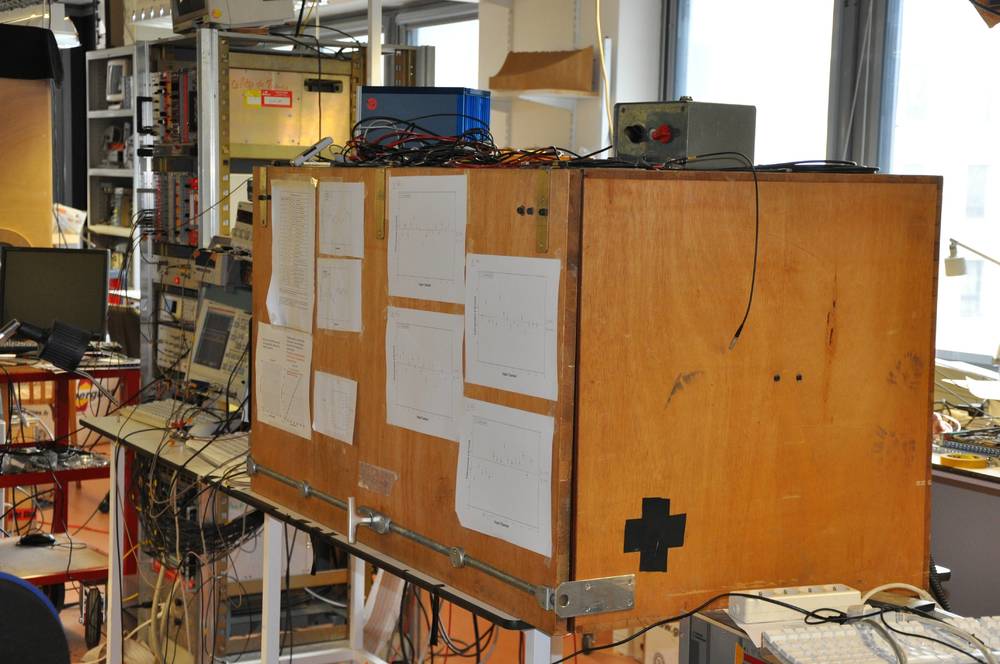}}
\caption[Photographs of the Black Box I]{ \label{pic:BlackBox:Outside} Several photographs of the outside of the black box. \fig\ref{pic:BlackBox:1} shows the front-side of the black box. The hinges and door handle of the black box can
be clearly seen. 
In \fig\ref{pic:BlackBox:2}, the black box is shown at an angle, and the tower housing the data acquisition hardware can be seen in the background. }
 \end{figure}

\begin{figure}[p]
\centering
 \includegraphics[angle=0,width=1.0\textwidth]{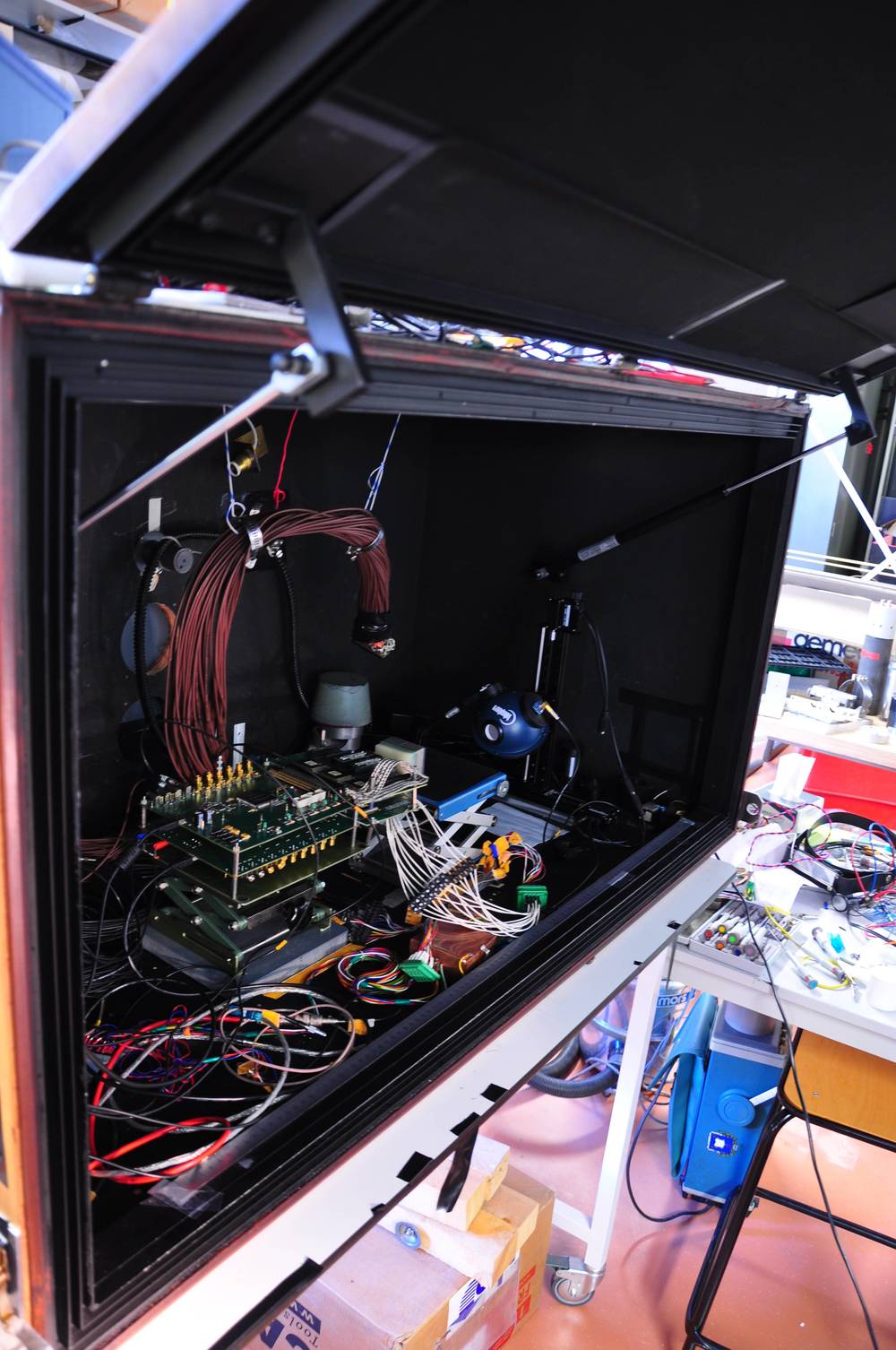} 
\caption[Photographs of the Black Box II]{ \label{pic:BlackBox:Open} A photograph looking into the black box. The baffles around the door ensure a light-tight seal, and the door pumps hold the door open.
The integrating sphere can be seen inside the box, along with a EUSO-Balloon EC, the ASIC board, and the LAL EC-ASIC test board. The numerous white wires leaving the EC go to the high voltage. The large hanging cable bundle
contains the 64 LEM00 connections to the QDC-based data acquisition system.}
 \end{figure}

\begin{figure}[p]
\centering
  \includegraphics[angle=0,width=1.0\textwidth]{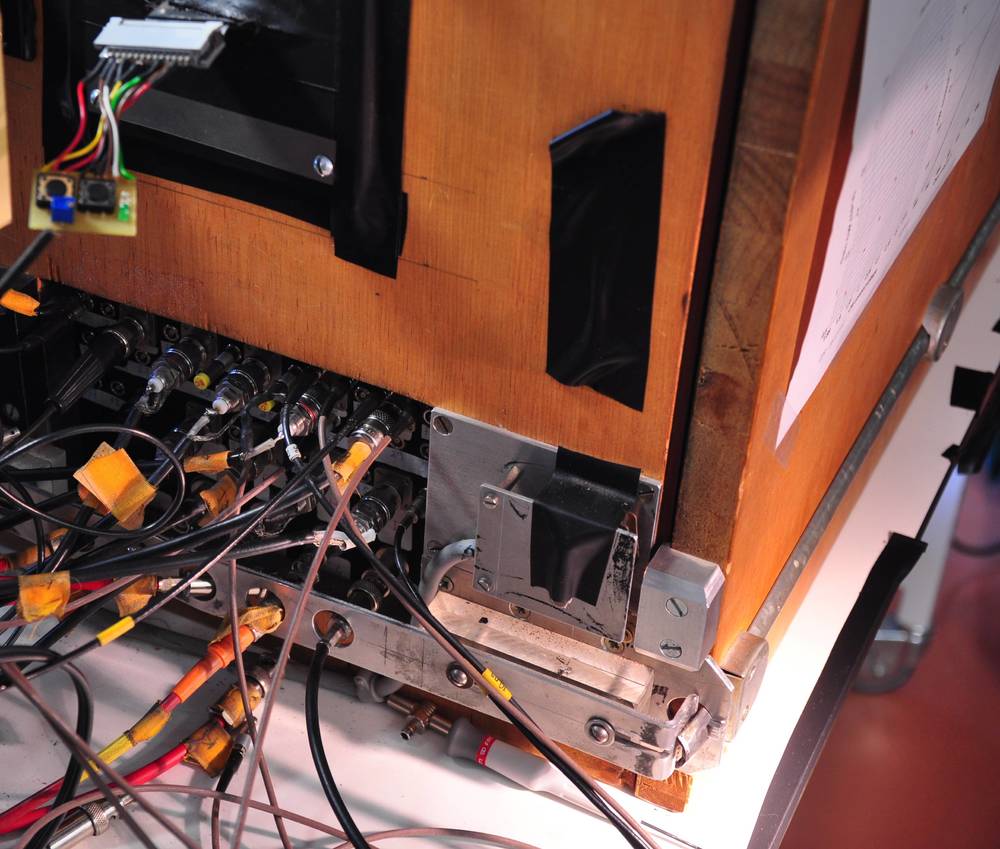}
\caption[Photographs of the Black Box III]{ \label{pic:BlackBox:LockandCables} A close-up photograph of the locking mechanism and cable feed-through connectors on the black box. The lock mechanism 
prevents the black box from being opened while the SHV cables are connected, ensuring that PMTs are unpowered anytime the box is opened to daylight. }
 \end{figure}

\begin{figure}[p]
\centering
\includegraphics[angle=0,width=1.0\textwidth]{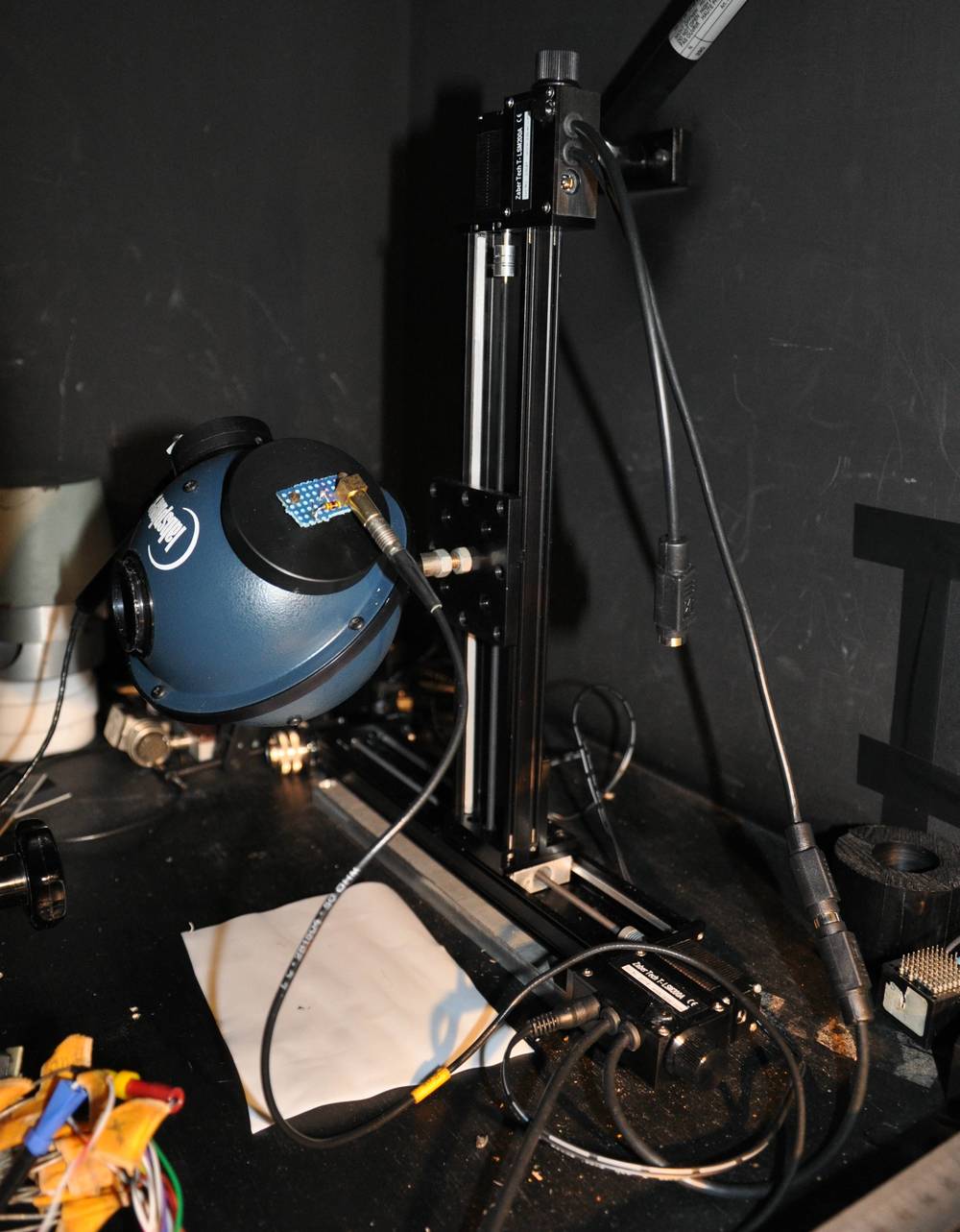}
\caption[A Photograph of the Integrating Sphere]{ \label{pic:IntegratingSphere:1}  A photograph of the integrating sphere, mounted on an X-Y stage. A single LED is attached to the integrating sphere port in the foreground. The port in the background 
holds the NIST photodiode, and the exit of the sphere is towards the left of the photograph.}
 \end{figure}

\begin{figure}[p]
\centering
  \subfigure[]{\label{fig:64Cables_outside} \includegraphics[angle=270,width=1.0\textwidth]{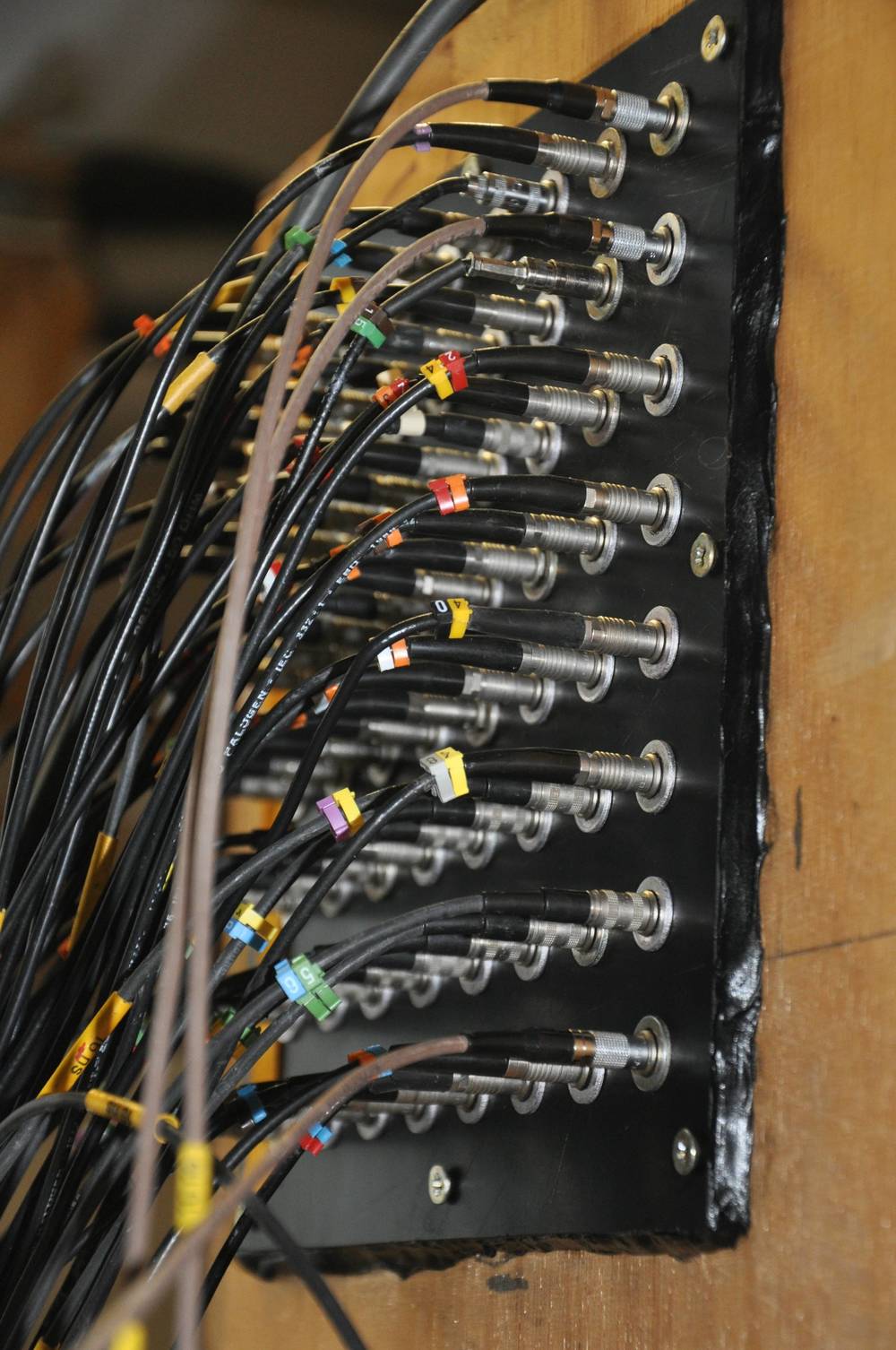}}
 \subfigure[]{\label{fig:64Cables_inside} \includegraphics[angle=90,width=1.0\textwidth]{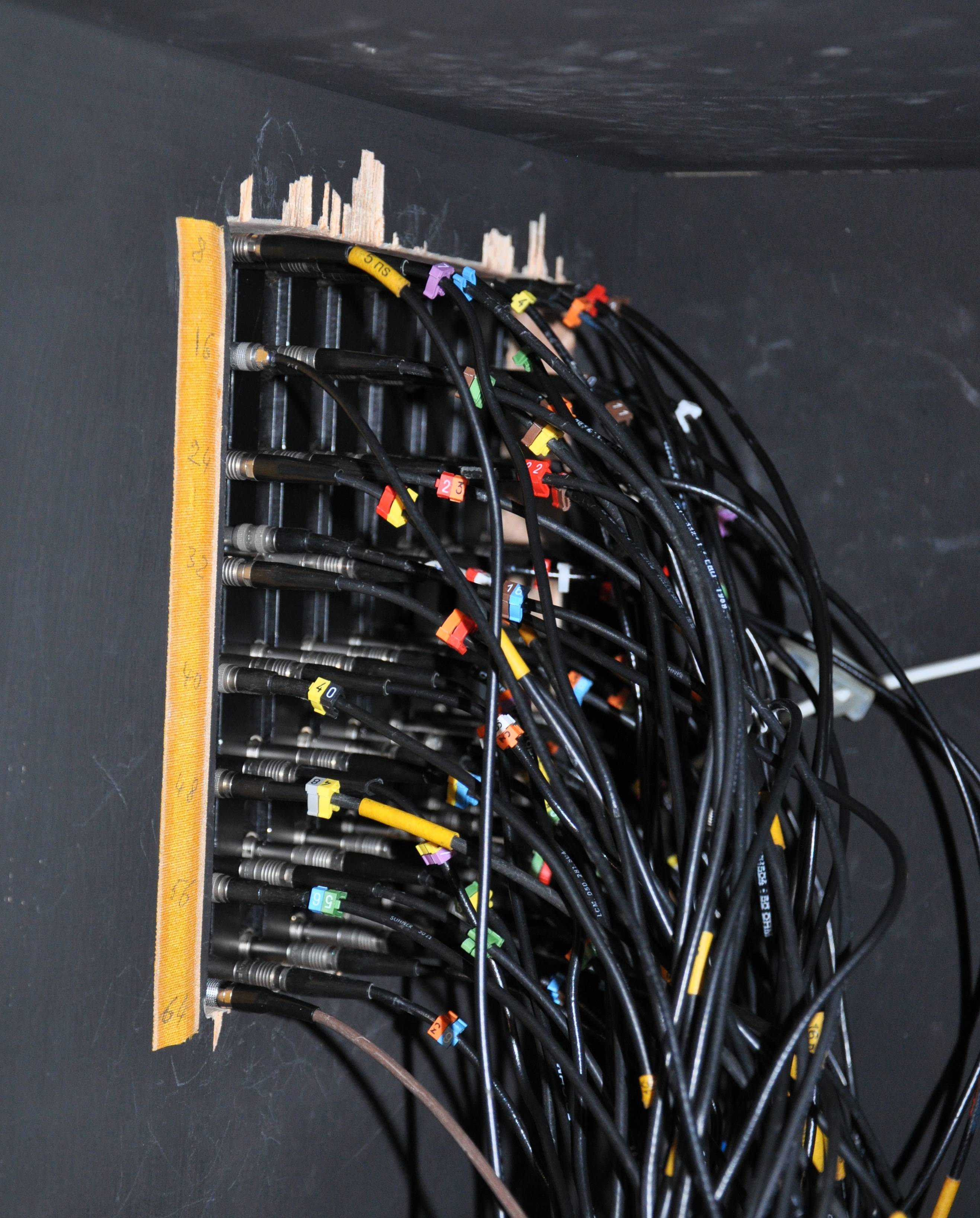}}
\caption[Photographs of the 64 LEM00 Cable Feed-Throughs]{ \label{pic:64cables} Photographs of the 64 LEM00 cable feed-throughs (rotated by $90^{\circ}$), one connection per pixel. These connectors are placed in a plastic plate, which is larger in size than the 
hole in the black box. The outside edge of the plate is also covered with putty to insure that no light leaks through. \fig\ref{fig:64Cables_inside} shows the plate from inside the black box, while \fig\ref{fig:64Cables_outside} shows the 
plate from the outside.}
 \end{figure}

\begin{figure}[p]
\centering
 \includegraphics[angle=90,width=1.0\textwidth]{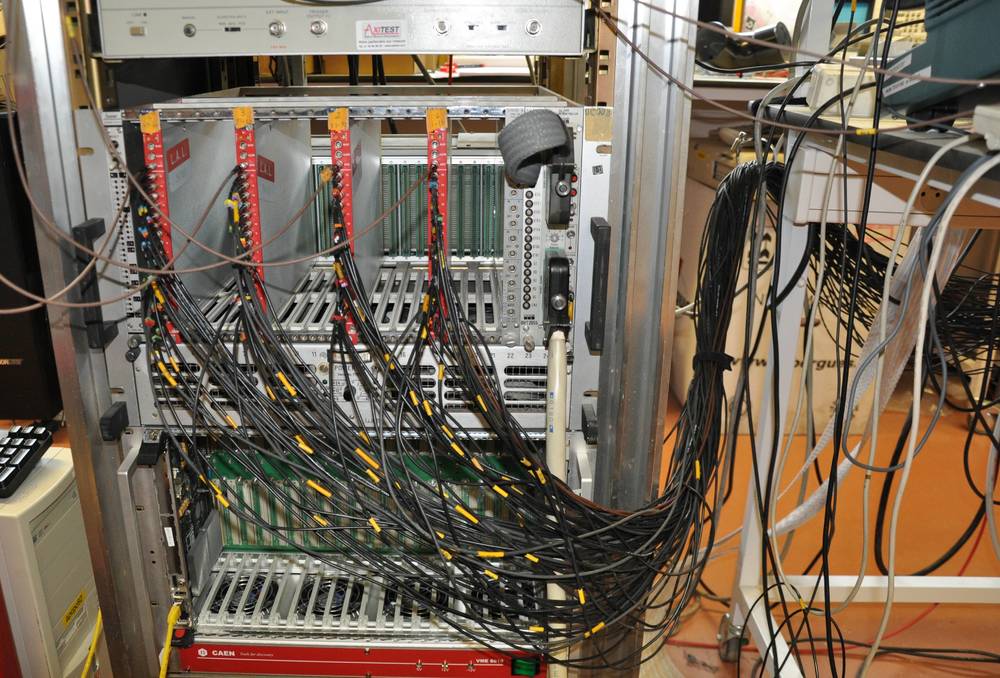}
\caption[Overview of the CAMAC Crate]{ \label{pic:CAMACCrate_Overview} A photograph of the CAMAC and VME crate (rotated by $90^{\circ}$). The 64 LEM00 cables from the black box can be seen coming from the right of the photograph, and go to the 64 QDC channels. 
The four QDC modules are housed in the CAMAC crate, which is placed directly above the VME crate. The CAMAC crate controller connects through the thick gray cable to the VME crate through a CBD 8210 VME to CAMAC board.
The VME, and by extension the CAMAC,
is controlled by a Motorola MVME processor board, which can be seen inside the VME crate on the left side. The MVME board is connected to the backend computer through the yellow Ethernet cable. }
 \end{figure}

\begin{figure}[p]
\centering
  \subfigure[]{\label{fig:QDCandControllerCloseUp} \includegraphics[angle=90,width=0.9\textwidth]{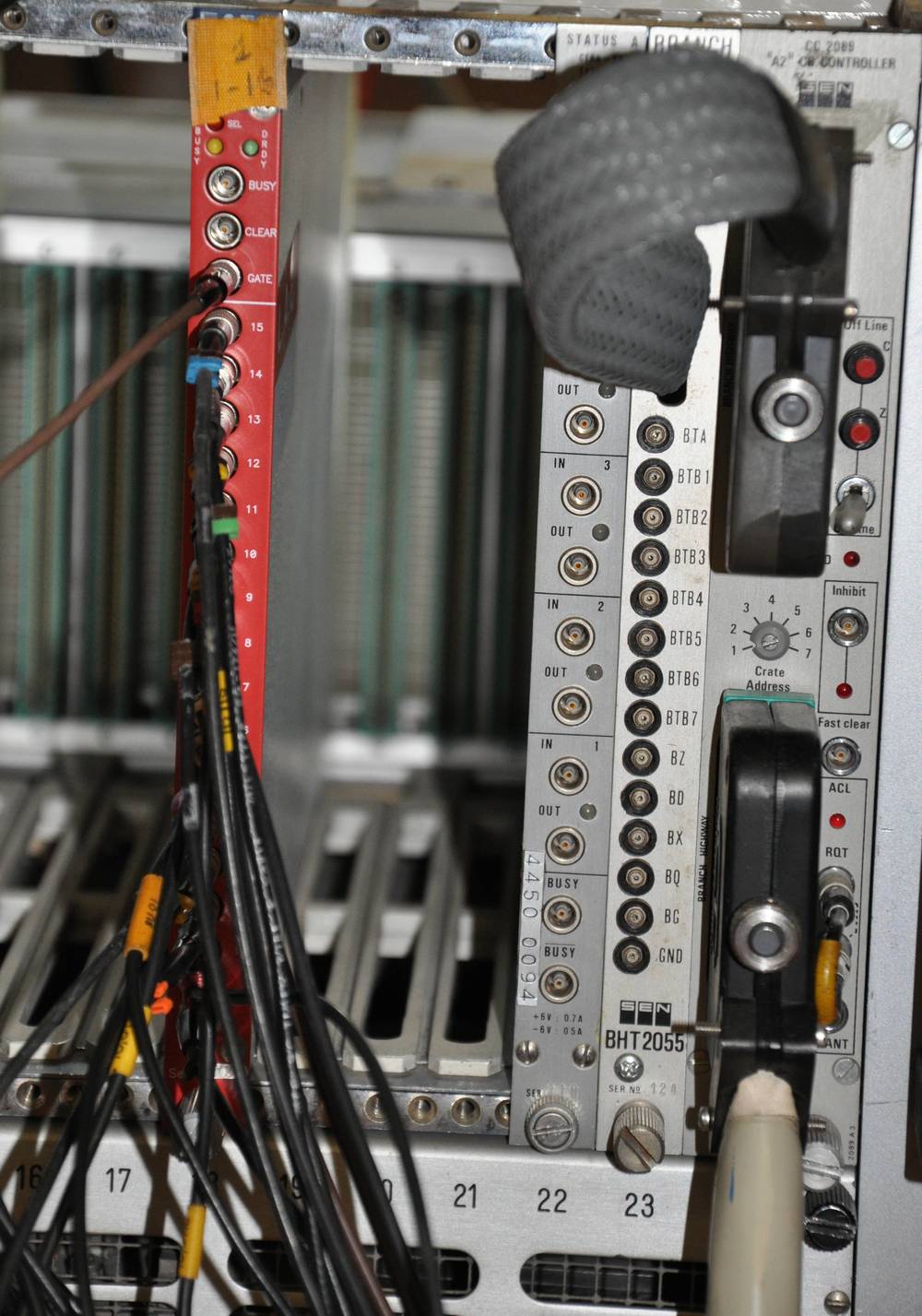}}
  \subfigure[]{\label{fig:NIMmodules_CloseUp} \includegraphics[angle=90,width=0.9\textwidth]{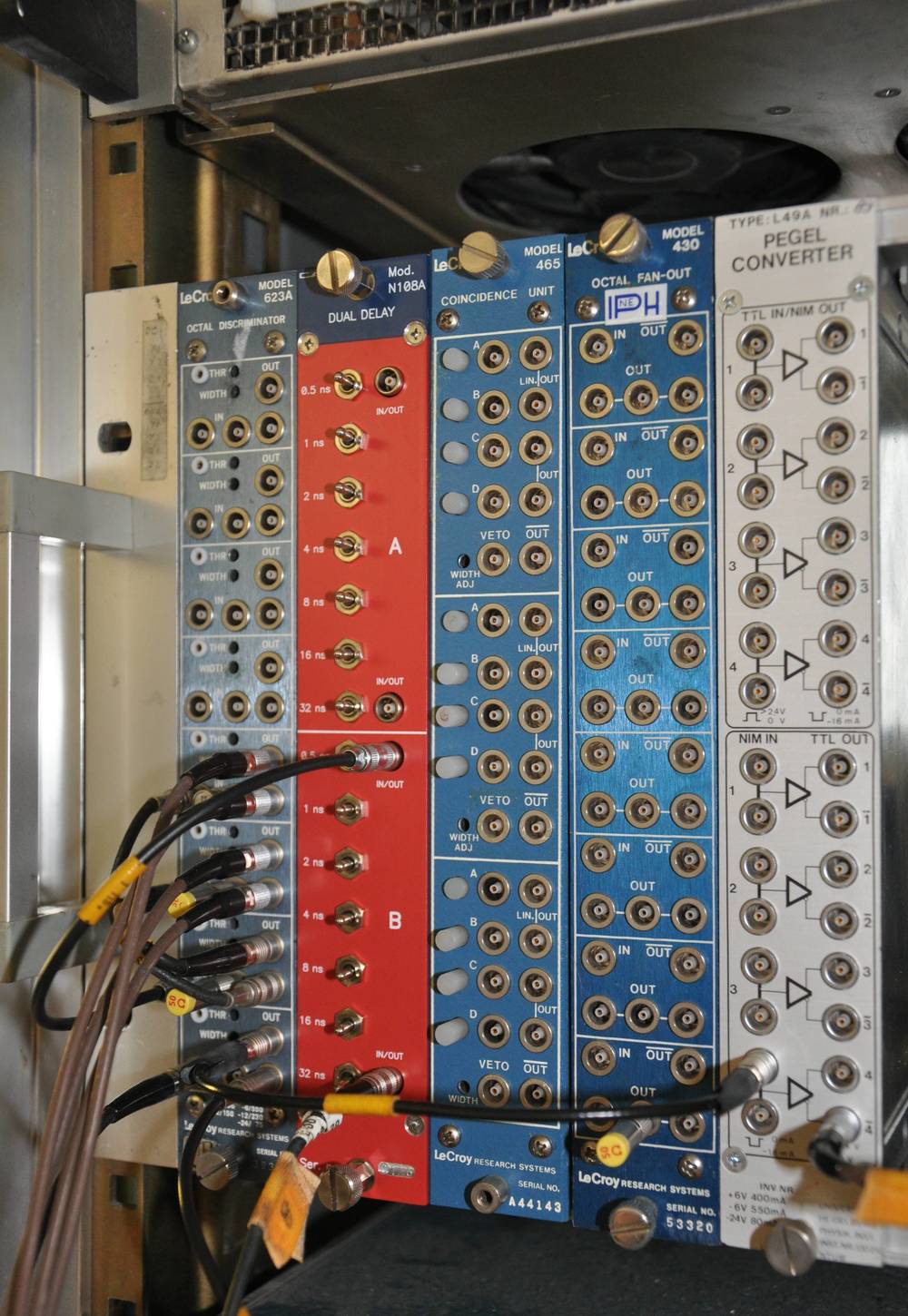}}
\caption[A Close-up of the QDCs]{ \label{pic:QDCandControllerCloseUp} Close-up photographs (rotated by $90^{\circ}$) of the CAMAC crate controller (\fig\ref{fig:QDCandControllerCloseUp}) and a section of the NIM crate (\fig\ref{fig:NIMmodules_CloseUp}).
In \fig\ref{fig:QDCandControllerCloseUp}, the branch terminator can be seen next to the CAMAC crate controller, in slot 23. The CERN status register, which was used for the QDC calibration, is in slot 22. On the left of the photograph one of the four 
C1205 QDC modules can be seen in slot 18. The first LEM00 cable is the gate input and the bottom 16 cables are the inputs from 16 MAPMT pixels. The busy and clear input of the QDC are not used in this setup.
The close-up of the NIM crate shows a discriminator unit on the far left, followed by a delay line (red). The input from the pulse generator is cascaded through the discriminator in order to create four gate pulses (one for each QDC) with the 
same delay relative to the LED signal. The output to the LED is sent through the PEGEL convertor on the far right of the photograph and then on to the LED driver. 
}
 \end{figure}

\begin{figure}[p]
\centering
  \subfigure[]{\label{fig:Balloon-CWHVPS} \includegraphics[angle=0,width=1.0\textwidth]{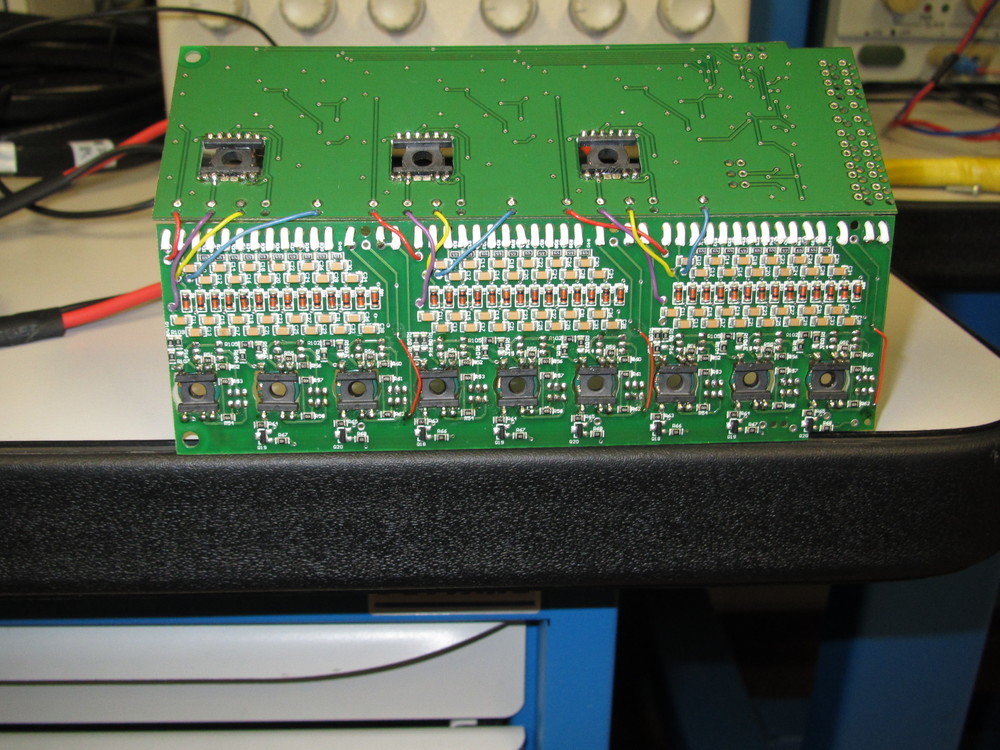}}\\
 \subfigure[]{\label{fig:Balloon-CWHVPS-switches} \includegraphics[angle=0,width=1.0\textwidth]{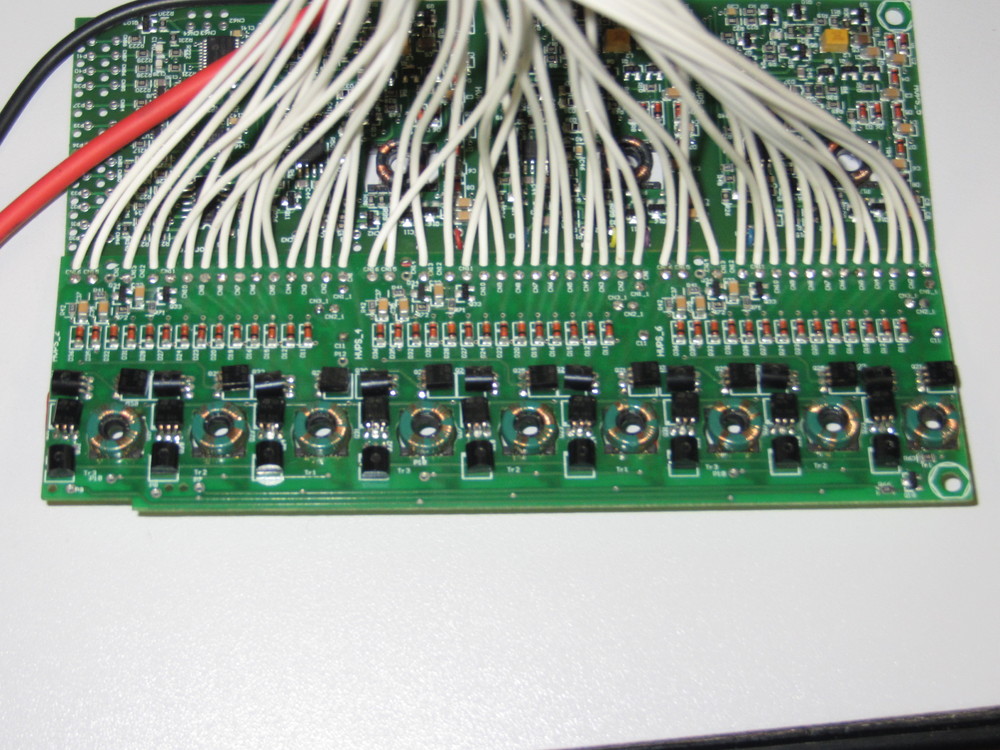}}
\caption[Photographs of EUSO-Balloon CW-HVPS]{ \label{pic:EUSOBalloonCWHVPS} Two photographs of one EUSO-Balloon flight-model Cockcroft-Walton high voltage power supply board. The high voltage circuits themselves
are shown in \fig\ref{fig:Balloon-CWHVPS}, and the ladder structure of the three Cockcroft-Walton circuits can be clearly seen. 
\fig\ref{fig:Balloon-CWHVPS-switches} shows the switches, three per high voltage circuit.
}
 \end{figure}


\begin{figure}[p]
\centering
 \includegraphics[angle=0,width=1.0\textwidth]{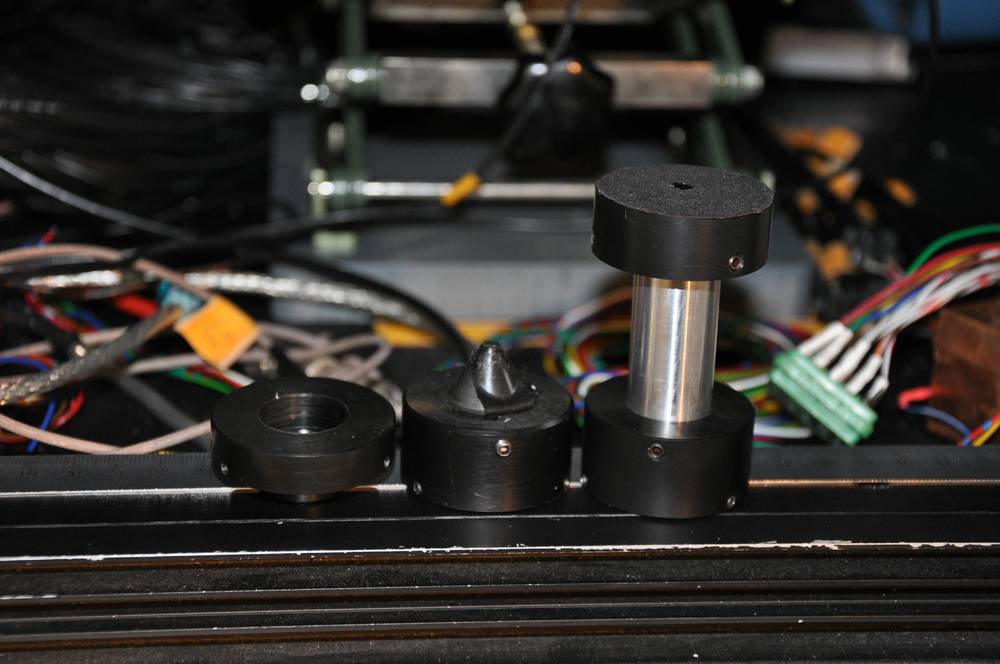}
\caption[Photograph of Collimators]{ \label{pic:ThreeCollimators} A photograph of the three collimators used through the work presented in this thesis. The first, on the left, is a diaphragm with an opening of 3 mm. The center collimator
was used in the calibration of the PMTs for the air fluorescence measurement presented in chapter~\ref{CHAPTER:AIRFLUOR}. This collimator has an exit pinhole of 0.3 mm, an entrance diameter of
1 mm, and a length of 23.5 mm.
The collimator on the right was used for the pixel by pixel measurements in chapter~\ref{CHAPTER:EUSOBalloonMeasurements}. This collimator has an exit pinhole with a diameter of 0.3 mm, an entrance 
diameter of 1 mm, and a length of 64 mm. }
 \end{figure}

\begin{figure}[p]
\centering
 \includegraphics[angle=90,width=1.0\textwidth]{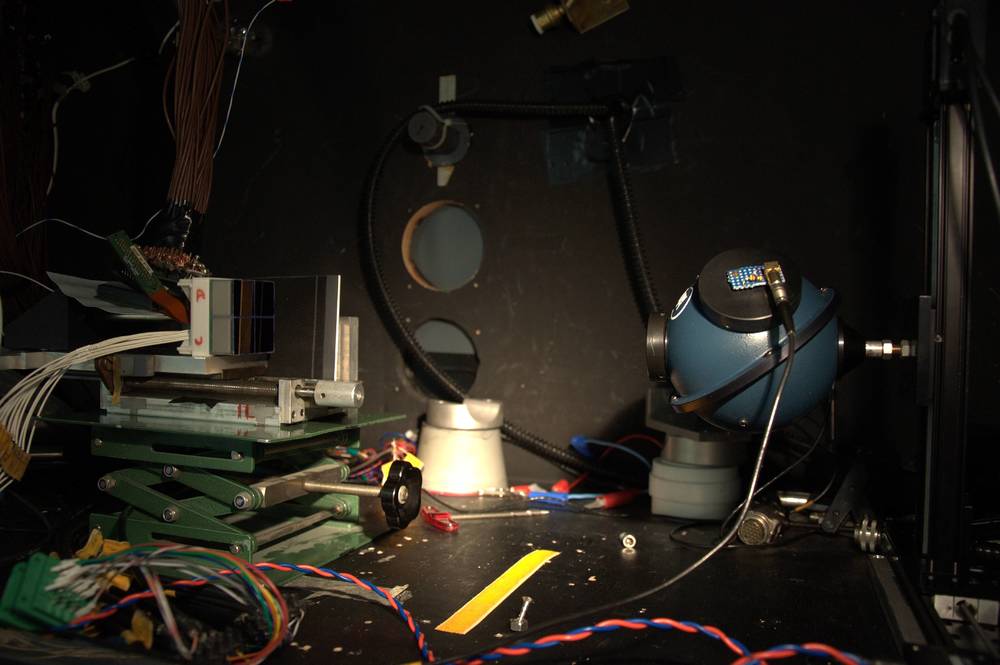}
\caption[Photograph of EC under Uniform Illumination I]{ \label{pic:ECUniformIllumination:Closeup} A photograph of one EUSO-Balloon EC in the black box. Here the EC is setup to be uniformly illuminated, and is about 40 cm from the 
exit of the integrating sphere. The 3 mm diaphragm, shown in \fig\ref{pic:ThreeCollimators}, is attached to the port of the sphere. The 14 white wires coming from the EC unit go to the HVPS, which is connected through the large green plug seen in the foreground.
The anode pin connector of one PMT is connected to the bundle of 64 LEM00 cables, which take the anode signals to the QDCs.}
 \end{figure}

\begin{figure}[p]
\centering
 \includegraphics[angle=90,width=1.0\textwidth]{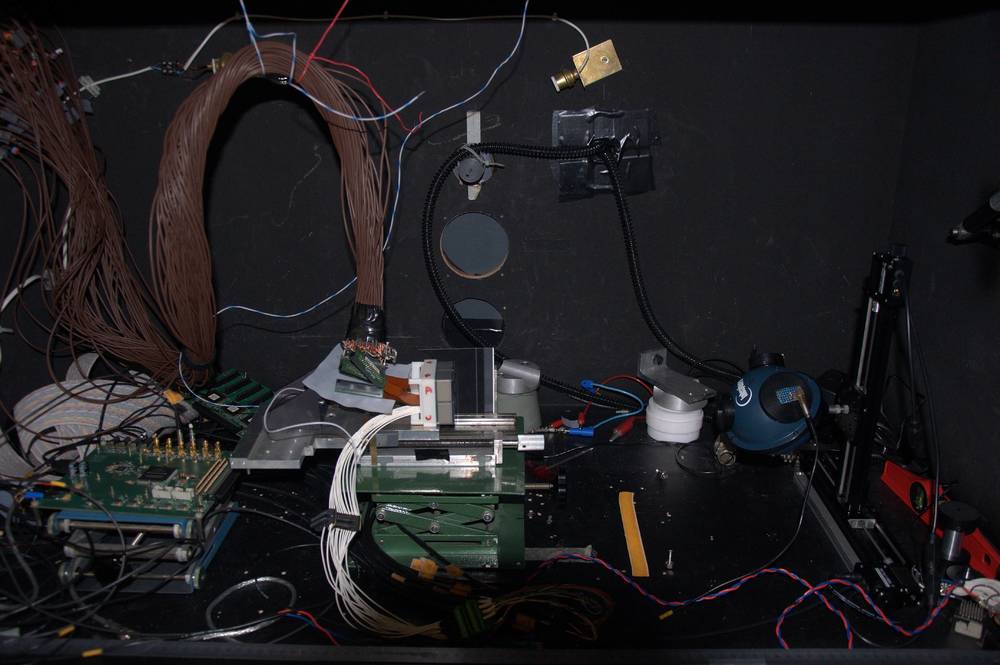}
\caption[Photograph of EC under Uniform Illumination II]{ \label{pic:ECUniformIllumination:Overview} Another photograph, again showing one EUSO-Balloon EC setup for uniform illumination. The ASIC test hardware can also be
seen in the foreground and background.}
 \end{figure}

\renewcommand*{\glspostdescription}{}
\glossarystyle{listgroup}
\printglossaries

\end{document}